\documentclass[longauth,bibyear]{aa} 

\usepackage{lscape}
\usepackage{float}
\usepackage{graphicx}
\usepackage{txfonts}
\begin{document} 

\title{An ALMA survey of submillimetre galaxies in the COSMOS field: Physical properties derived from energy balance spectral energy distribution modelling}

   \author{O.~Miettinen\inst{1}, I.~Delvecchio\inst{1}, V.~Smol\v{c}i\'{c}\inst{1}, M.~Aravena\inst{2}, D.~Brisbin\inst{2}, A.~Karim\inst{3}, B.~Magnelli\inst{3}, 
M.~Novak\inst{1}, E.~Schinnerer\inst{4}, M.~Albrecht\inst{3}, H.~Aussel\inst{5}, F.~Bertoldi\inst{3}, P.~L.~Capak\inst{6,7}, C.~M.~Casey\inst{8}, C.~C.~Hayward\inst{9,10}, 
O.~Ilbert\inst{11}, H.~T.~Intema\inst{12}, C.~Jiang\inst{13,2,14}, O.~Le F{\`e}vre\inst{11}, H.~J.~McCracken\inst{15}, A.~M.~Mu{\~n}oz Arancibia\inst{16}, F.~Navarrete\inst{3}, 
N.~D.~Padilla\inst{17,18}, D.~A.~Riechers\inst{19}, M.~Salvato\inst{20}, K.~S.~Scott\inst{21}, K.~Sheth\inst{22}, and L.~A.~M.~Tasca\inst{11}}

   \institute{Department of Physics, Faculty of Science, University of Zagreb, Bijeni\v{c}ka cesta 32, HR-10000 Zagreb, Croatia \\ \email{oskari@phy.hr} \and 
N\'ucleo de Astronom\'{\i}a, Facultad de Ingenier\'{\i}a, Universidad Diego Portales, Av. Ej\'ercito 441, Santiago, Chile \and 
Argelander-Institut f\"{u}r Astronomie, Universit\"{a}t Bonn, Auf dem H\"{u}gel 71, D-53121 Bonn, Germany \and 
Max-Planck-Institut f\"{u}r Astronomie, K\"{o}nigstuhl 17, 69117 Heidelberg, Germany \and 
AIM Unit\'e Mixte de Recherche CEA CNRS, Universit\'e Paris VII UMR n158, Paris, France \and 
Infrared Processing and Analysis Center, California Institute of Technology, MC 100-22, 770 South Wilson Ave., Pasadena, CA 91125, USA \and 
Spitzer Science Center, California Institute of Technology, Pasadena, CA 91125, USA \and 
Department of Astronomy, The University of Texas at Austin, 2515 Speedway Blvd Stop C1400, Austin, TX 78712, USA \and 
Center for Computational Astrophysics, Flatiron Institute, 162 Fifth Avenue, New York, NY 10010 \and 
Harvard-Smithsonian Center for Astrophysics, 60 Garden Street, Cambridge, MA 02138, USA \and
Aix Marseille Universit\'e, CNRS, LAM (Laboratoire d'Astrophysique de Marseille), UMR 7326, 13388, Marseille, France \and 
Leiden Observatory, Leiden University, PO Box 9513, 2300 RA Leiden, The Netherlands \and
CAS Key Laboratory for Research in Galaxies and Cosmology, Shanghai Astronomical Observatory, Nandan Road 80, Shanghai 200030, China \and 
Chinese Academy of Sciences South America Center for Astronomy, 7591245 Santiago, Chile \and 
Sorbonne Universit\'es, UPMC Universit\'e Paris 6 et CNRS, UMR 7095, Institut d'Astrophysique de Paris, 98 bis Boulevard Arago, 75014 Paris, France \and 
Instituto de F\'{\i}sica y Astronom\'{\i}a, Universidad de Valpara\'{\i}so, Av. Gran Breta\~{n}a 1111, Valpara\'{\i}so, Chile \and 
Instituto de Astrof\'{\i}sica, Pontificia Universidad Cat\'olica de Chile, Avda. Vicu\~{n}a Mackenna 4860, Santiago, Chile \and
Centro de Astro-Ingenier\'{\i}a, Pontificia Universidad Cat\'olica de Chile, Avda. Vicu\~{n}a Mackenna 4860, 782-0436 Macul, Santiago, Chile \and 
Astronomy Department, Cornell University, 220 Space Sciences Building, Ithaca, NY 14853, USA \and
Max-Planck-Institut f\"{u}r extraterrestrische Physik, Garching bei M\"{u}nchen, D-85741 Garching bei M\"{u}nchen, Germany \and 
North American ALMA Science Center, NRAO, 520 Edgemont Rd, Charlottesville, VA, USA 22903 \and 
NASA Headquarters, 300 E. St. SW, Washington, DC 20546, USA}

   \date{Received ; accepted}

\authorrunning{Miettinen et al.}
\titlerunning{Physical properties of ALMA SMGs in COSMOS}

\abstract {Submillimetre galaxies (SMGs) represent an important source population in the origin and cosmic evolution of the most massive galaxies. Hence, it is imperative to place firm constraints on the fundamental physical properties of large samples of SMGs.} {We determine the physical properties of a sample of SMGs in the COSMOS field that were pre-selected at the observed-frame wavelength of $\lambda_{\rm obs}=1.1$~mm, and followed up at $\lambda_{\rm obs}=1.3$~mm with the Atacama Large Millimetre/submillimetre Array (ALMA).}{We used the {\tt MAGPHYS} model package to fit the panchromatic (ultraviolet to radio) spectral energy distributions (SEDs) of 124 of the target SMGs, which lie at a median redshift of $z = 2.30$ (19.4\% are spectroscopically confirmed). The SED analysis was complemented by estimating the gas masses of the SMGs by using the $\lambda_{\rm obs}=1.3$~mm dust emission as a tracer of the molecular gas component.}{The sample median and 16th--84th percentile ranges of the stellar masses, obscured star formation rates, dust temperatures, and dust and gas masses were derived to be $\log(M_{\star}/{\rm M}_{\sun})=11.09^{+0.41}_{-0.53}$, ${\rm SFR}=402^{+661}_{-233}$~${\rm M}_{\sun}~{\rm yr}^{-1}$, $T_{\rm dust}=39.7^{+9.7}_{-7.4}$~K, $\log(M_{\rm dust}/{\rm M}_{\sun})=9.01^{+0.20}_{-0.31}$, and $\log(M_{\rm gas}/{\rm M}_{\sun})=11.34^{+0.20}_{-0.23}$, respectively. The $M_{\rm dust}/M_{\star}$ ratio was found to decrease as a function of redshift, while the $M_{\rm gas}/M_{\rm dust}$ ratio shows the opposite, positive correlation with redshift. The derived median gas-to-dust ratio of $120^{+73}_{-30}$ agrees well with the canonical expectation. The gas fraction ($M_{\rm gas}/(M_{\rm gas}+M_{\star})$) was found to range from 0.10 to 0.98 with a median of $0.62^{+0.27}_{-0.23}$. We found that $57.3\%$ of our SMGs populate the main sequence (MS) of star-forming galaxies, while $41.9\%$ of the sources lie above the MS by a factor of greater than three (one source lies below the MS). These super-MS objects, or starbursts, are preferentially found at $z \gtrsim 3$, which likely reflects the sensitivity limit of our source selection. We estimated that the median gas consumption timescale for our SMGs is $\sim 535$~Myr, and the super-MS sources appear to consume their gas reservoir faster than their MS counterparts. We found no obvious stellar mass--size correlations for our SMGs, where the sizes were measured in the observed-frame 3~GHz radio emission and rest-frame UV. However, the largest 3~GHz radio sizes are found among the MS sources. Those SMGs that appear irregular in the rest-frame UV are predominantly starbursts, while the MS SMGs are mostly disk-like.} {The physical parameter distributions of our SMGs and those of the equally bright, 870~$\mu$m selected SMGs in the ECDFS field (the so-called ALESS SMGs) are unlikely to be drawn from common parent distributions. This might reflect the difference in the pre-selection wavelength. Albeit being partly a selection bias, the abrupt jump in specific SFR and the offset from the MS of our SMGs at $z \gtrsim 3$ might also reflect a more efficient accretion from the cosmic gas streams, higher incidence of gas-rich major mergers, or higher star formation efficiency at $z \gtrsim 3$. We found a rather flat average trend between the SFR and dust mass, but a positive ${\rm SFR}-M_{\rm gas}$ correlation. However, to address the questions of which star formation law(s) our SMGs follow, and how they compare with the Kennicutt-Schmidt law, the dust-emitting sizes of our sources need to be measured. Nonetheless, the larger radio-emitting sizes of the MS SMGs compared to starbursts is a likely indication of their more widespread, less intense star formation activity. The irregular rest-frame UV morphologies of the starburst SMGs are likely to echo their merger nature. The current stellar mass content of the studied SMGs is very high, so they must quench to form the so-called red-and-dead massive ellipticals. Our results suggest that the transition from high-$z$ SMGs to local ellipticals via compact, quiescent galaxies (cQGs) at $z \sim 2$ might not be universal, and the latter population might also descend from the so-called blue nuggets. However, $z \gtrsim 4$ SMGs could be the progenitors of higher redshift, $z \gtrsim 3$ cQGs, while our results are also consistent with the possibility that ultra-massive early-type galaxies found at $1.2 \lesssim z \lesssim 2$ experienced an SMG phase at $z \leq 3$.}

\keywords{Galaxies: evolution -- Galaxies: formation -- Galaxies: starburst -- Galaxies: star formation -- Submillimetre: galaxies}

   \maketitle
%

\section{Introduction}

The second half of the 1990s witnessed a major progress in studies of massive galaxy formation and evolution when the first extragalactic 
submillimetre continuum surveys were performed, thus opening a new window into the extragalactic sky (\cite{smail1997}; \cite{hughes1998}; \cite{barger1998}; \cite{eales1999}). 
Most notably, these surveys led to the discovery of a new population of strongly star-forming dusty galaxies, now generally known as submillimetre galaxies or SMGs 
(see \cite{casey2014} for a review). 

Studies of SMGs over the past few tens of years have provided valuable insights into their properties. These include the redshift distribution (e.g. \cite{chapman2005}; \cite{aretxaga2007}; \cite{wardlow2011}; \cite{yun2012}; \cite{smolcic2012}; \cite{simpson2014}; \cite{zavala2014}; \cite{miettinen2015a}; \cite{chen2016a}; \cite{strandet2016}; \cite{simpson2017}; \cite{danielson2017}; \cite{brisbin2017}), spatial clustering and environment (e.g. \cite{ivison2000}; \cite{blain2004}; \cite{aravena2010}; \cite{hickox2012}; \cite{miller2015}; \cite{chen2016b}; \cite{wilkinson2017}; \cite{smolcic2017a}), merger incidence (e.g. \cite{conselice2003}), and circumgalactic medium (\cite{fu2016}). Regarding the intrinsic physical characteristics of SMGs, the properties studied so far include the sizes and morphologies (e.g. \cite{swinbank2010}; \cite{menendez2013}; \cite{aguirre2013}; \cite{targett2013}; \cite{chen2015}; \cite{simpson2015}; \cite{ikarashi2015}; \cite{miettinen2015b}; \cite{hodge2016}; \cite{miettinen2017b},c), panchromatic spectral energy distributions (SEDs; e.g. \cite{michalowski2010}; \cite{magnelli2012}; \cite{swinbank2014}; \cite{dacunha2015}; \cite{miettinen2017a}), stellar masses (e.g. \cite{dye2008}; \cite{hainline2011}; \cite{michalowski2012}; \cite{targett2013}), gas masses (e.g. \cite{greve2005}; \cite{tacconi2006}, 2008; \cite{engel2010}; \cite{ivison2011}; \cite{riechers2011}; \cite{bothwell2013}; \cite{huynh2017}), gas kinematics (e.g. \cite{alaghband2012}; \cite{hodge2012}; \cite{carilli2013}; \cite{olivares2016}), and active galactic nucleus (AGN) incidence (\cite{alexander2003}, 2005; \cite{laird2010}; \cite{johnson2013}; \cite{wang2013}). The role played by SMGs in a broader context of galaxy formation and evolution has also been investigated through models (e.g. \cite{baugh2005}; \cite{fontanot2007}; \cite{dave2010}; \cite{gonzalez2011}; \cite{hayward2013}) and observational approach (e.g. \cite{swinbank2006}; \cite{toft2014}; \cite{simpson2014}). 

Owing to observational and theoretical efforts, some of the main findings that have emerged are that SMGs are predominantly found in a $\lesssim3$~Gyr old universe (redshift $z\gtrsim2$), they are very massive in their stellar and molecular gas content, where the masses of both components can be of the order of hundred billion solar masses ($M_{\star}\simeq M_{\rm gas}\sim10^{11}$~M$_{\sun}$), and that their star formation rate (SFR) can reach astonishingly high values of a few or more solar masses per day, or $\gtrsim1\,000$~M$_{\sun}$~yr$^{-1}$ in more conventional units. The vigorous star formation activity of SMGs is traditionally viewed as a gas-rich, major merger-driven phase (e.g. \cite{tacconi2008}; \cite{engel2010}), but disk instabilities occurring in gas-rich, Toomre-instable disks, whose high gas fractions can be sustained by accretion from cosmic filaments (e.g. \cite{dekel2009a}; \cite{ceverino2010}), could also be behind the very high SFRs of SMGs; the relative importance of these processes remains to be quantified. Nevertheless, the high SFRs in conjunction with the compact sizes of SMGs (e.g. the dust-emitting region is typically a few kpc across in full width at half maximum (FWHM)), can cause even the system-wide, or galaxy-integrated SFR surface densities to reach the Eddington limit of $\Sigma_{\rm SFR}^{\rm Edd}\sim10^3$~M$_{\sun}$~yr$^{-1}$~kpc$^{-2}$ for a radiation pressure supported starburst disk (e.g. \cite{thompson2005}). 

In a bigger context of cosmological galaxy evolution, a compelling picture has emerged, which indicates that the SMGs that lie at high redshifts of $z\gtrsim3$ have the potential to quench their star formation and become the compact, quiescent galaxies (cQGs) seen at $z\sim2$, and which can further grow in size to evolve into the high stellar mass ($M_{\star}\geq10^{11}$~M$_{\sun}$) elliptical galaxies we see at the current epoch (e.g. \cite{lilly1999}; \cite{swinbank2006}; \cite{cimatti2008}; \cite{fu2013}; \cite{toft2014}; \cite{simpson2014}). Besides the intrinsic physical properties of SMGs being consistent with the aforementioned evolutionary connection, SMGs are also sometimes found to be physically associated with growing groups and clusters of galaxies, which further supports the idea that SMGs are protoellipticals, whose present-day matured versions are preferentially found near the centres of galaxy clusters. Hence, observational studies of SMGs are motivated by their potential to provide strong constraints on models of massive galaxy formation and evolution.

However, many previous SMG studies were based on fairly small samples, in which case final conclusions are inevitably 
limited. To improve our understanding of SMGs, and quantify their role in the overall massive galaxy evolution, their key physical properties (e.g. $M_{\star}$ and SFR) need to be determined for large, well-selected, and homogeneous source samples. A powerful technique to reach this goal is to construct and analyse the multiwavelength SEDs for a large sample of SMGs, which however is feasible only if sufficient amount of continuum imaging data over the whole electromagnetic spectrum, from the ultraviolet (UV) and optical to radio, are available. This is the core science theme of the present paper.

In this paper, we study a large sample of SMGs detected with the Atacama Large Millimetre/submillimetre Array (ALMA) in 
the Cosmic Evolution Survey (COSMOS; \cite{scoville2007})\footnote{{\tt http://cosmos.astro.caltech.edu}.} deep field. 
Thanks to the exceptionally rich multiwavelength coverage of COSMOS, we can examine the key physical properties of these SMGs 
by fitting their densely-sampled panchromatic SEDs. The layout of this paper is as follows. In Sect.~2, we describe our SMG sample, and the employed 
observational data sets. The SED analysis and its results, and the dust-based gas mass estimates are presented in Sect.~3. Section~4 is devoted to discussion, 
and in Sect.~5 we summarise the results and present our conclusions. A multiwavelength photometry table is provided in 
Appendix~A, the SED plots are shown in Appendix~B, and the derived physical parameters are tabulated in Appendix~C. 

The cosmology adopted in the present work corresponds to a spatially flat (the curvature parameter $k=0$) $\Lambda$CDM (Lambda cold dark matter) universe with the present-day dark energy density parameter $\Omega_{\Lambda}=0.70$, and total (dark plus luminous baryonic) matter density parameter $\Omega_{\rm m}=0.30$, both in units of the critical density. The Hubble constant is set at $H_0=70$~km~s$^{-1}$~Mpc$^{-1}$, which corresponds to a dimensionless Hubble parameter of $h=H_0/100=0.7$. A Chabrier (2003) Galactic-disk initial mass function (IMF) is assumed in the calculation of $M_{\star}$ and SFR. 

\section{Data}

\subsection{Source sample}

\subsubsection{Parent source sample: The ASTE/AzTEC 1.1~mm selected sources followed up with ALMA at 1.3~mm} 

Our target SMGs were originally identified by the $\lambda_{\rm obs}=1.1$~mm blank-field continuum survey over a continuous area of 0.72~deg$^2$ or 36.7\% of 
the full $1\fdg4\times1\fdg4$ COSMOS field (centred at $\alpha_{2000.0}\simeq 10$~hr and $\delta_{2000.0}\simeq 2\fdg2$) carried out with the 144~pixel AzTEC bolometer camera (\cite{wilson2008}) on the 10~m Atacama Submillimetre Telescope Experiment (ASTE; \cite{ezawa2004}) by Aretxaga et al. (2011). The angular resolution (FWHM) of the observed AzTEC 1.1~mm continuum map was $34\arcsec$, and had an average $1\sigma$ noise level of 1.26~mJy~beam$^{-1}$. The 129 brightest AzTEC sources (the detection signal-to-noise ratio cut at S/N$_{\rm 1.1\, mm}\geq4$; $S_{\rm 1.1\, mm}^{\rm AzTEC}\geq3.5$~mJy) were followed up with dedicated ALMA pointings at $\lambda_{\rm obs}=1.3$~mm and $\sim1\farcs6 \times 0\farcs9$ angular resolution with a $1\sigma$ root-mean-square (rms) noise of $\sim0.1$~mJy~beam$^{-1}$ by M.~Aravena et al. (in prep.) (Cycle~2 ALMA project 2013.1.00118.S; PI: M.~Aravena). Among these 129 target sources, 33 were resolved into two or three components, and in total we detected 152 ALMA sources at an S/N$_{\rm 1.3\, mm}\geq5$ ($S_{\rm 1.3\, mm}^{\rm ALMA}\gtrsim0.5$~mJy). This detection S/N$_{\rm 1.3\, mm}$ threshold yields a sample that is expected to be free of spurious sources, that is the sample reliability reaches a value of $\sim100\%$ (M.~Aravena et al., in prep.). 

The multicomponent sources are called AzTEC/C1a, C1b, etc., in order of decreasing 1.3~mm flux density. The corresponding full sample multiplicity fraction is $26\% \pm 4\%$ (at $\sim1\farcs6 \times 0\farcs9$ resolution and $S_{\rm 1.1\, mm}^{\rm AzTEC}\geq3.5$~mJy), where the quoted uncertainty represents the Poisson error on counting statistics. Our ALMA follow-up survey, which together with the source catalogue are described in detail by M.~Aravena et al. (in prep.), allowed us to accurately pin down the positions of the single-dish AzTEC sources. This way, we could reliably identify the multiwavelength counterparts of the target SMGs (\cite{brisbin2017}; \cite{miettinen2017b}), which is a crucial step towards the panchromatic SED analysis. As described in Brisbin et al. (2017), 98 ($64.5\% \pm 6.5\%$) of our SMGs were found to have a well-defined counterpart in the COSMOS2015 multiwavelength catalogue (\cite{laigle2016}). For 37 sources ($24.3\% \pm 4.2\%$), a deblending technique was required owing to confusion by a nearby source(s), and the observed-frame optical--mid-infrared (IR) photometry for these sources was manually extracted. 
Seventeen ALMA sources ($11.2\% \pm 2.7\%$) do not have a detected counterpart at optical to mid-IR wavebands, and hence no photometric redshift based on these short-wavelength data could be derived for these sources (for details, see \cite{brisbin2017}).

\subsubsection{Cleaning the submillimetre galaxy sample from active galactic nucleus contamination}

If a galaxy hosts an AGN, the nuclear emission can affect some of the galaxy properties derived through SED fitting. In particular, 
the AGN continuum emission can affect the derived stellar population parameters, and 
the AGN radiation field can heat the surrounding dusty torus, which can boost the observed mid-IR flux densities. 
If such an AGN component cannot be properly taken into account in the SED analysis to quantify its contribution to the SED, 
as is the case in the present work (see Sect.~3.1.1), AGN-host galaxies need to be discounted from an analysis of SED properties.

As described by Miettinen et al. (2017b), three of our target SMGs (AzTEC/C24b, 61, and 77a) were detected with the Very Long Baseline Array (VLBA) at a high, $16.2\times7.3$ square milliarcsecond resolution at $\nu_{\rm obs}=1.4$~GHz (N.~Herrera Ruiz et al., in prep.), which indicates the presence of a radio-emitting AGN or 
a very compact nuclear starburst (or both in symbiosis) in these SMGs. As also described in more detail by Miettinen et al. (2017b), the bright VLBA detection AzTEC/C61 ($S_{\rm 1.4\, GHz}^{\rm VLBA}=11.1$~mJy) was also detected in the X-rays (see \cite{civano2016} for the \textit{Chandra} COSMOS Legacy Survey), its observed-frame 3~GHz radio brightness temperature is $T_{\rm B}^{\rm 3\, GHz}>10^{4.03}$~K, and its radio spectrum between the observed-frame 1.4~GHz and 3~GHz is slightly inverted 
($S_{\nu}\propto \nu^{0.11}$). All these properties indicate that AzTEC/C61 harbours an AGN, and hence we removed it from the present sample. 
Although the other two VLBA detected SMGs, AzTEC/C24b ($S_{\rm 1.4\, GHz}^{\rm VLBA}=134.2$~$\mu$Jy; $T_{\rm B}^{\rm 3\, GHz}=75.2\pm10.4$~K) and AzTEC/C77a 
($S_{\rm 1.4\, GHz}^{\rm VLBA}=332.6.2$~$\mu$Jy; $T_{\rm B}^{\rm 3\, GHz}>243.9$~K), were not detected in the X-rays, we excluded them from 
the present analysis because the VLBA detection of these SMGs points towards the presence of buried AGN activity. At least in the case of AzTEC/C77a, 
this is indeed suggested by the sanity-checked SEDs presented in Appendix~A.

There are also seven other SMGs in our sample that were detected in the X-rays, but which were not detected with the VLBA. 
These are AzTEC/C11, 44b, 45, 56, 71b, 86, and 118 (for details, see \cite{miettinen2017b}, and references therein). 
All these seven X-ray detected SMGs were removed from our sample. 

Finally, we note that a non-detection of hard X-ray emission does not rule out the possibility of a heavily obscured AGN being present 
in some of our remaining SMGs, but its contribution to the observed SED is likely to be energetically unimportant. 
To quantify the potential AGN incidence among our SMGs, we checked how many of our X-ray non-detected SMGs 
satisfy the AGN selection criteria of Donley et al. (2012; their Eqs.~(1) and (2)), and are hence classified as AGN dominated 
in the near and mid-IR (see \cite{brisbin2017} for the relevant photometric data). We found that the percentage of such AGNs 
in our sample of X-ray non-detected SMGs is about 6.5\%. Hence, the vast majority of our target SMGs are purely star-forming galaxies (SFGs). 
  
\subsubsection{The final submillimetre galaxy sample}

In a nutshell, after considering the number of sources for which we possess optical to mid-IR photometry (and hence photometric redshifts; \cite{brisbin2017}), and excluding the SMGs that are potentially hosting an AGN that can affect the photometry, we end up with a final sample of 124 SMGs ($81.6\% \pm 7.3\%$ of the initial ALMA sample) that we analyse further in the present work. The deboosted ALMA 1.3~mm flux densities of these sources are in the range of $S_{\rm 1.3\, mm}=0.52-7.24$~mJy, which correspond to $S_{\rm 850\, \mu m}=2.3-32$~mJy for a dust emissivity index of $\beta=1.5$. Hence, all the target sources fulfil the definition of SMGs as galaxies having $S_{\rm 850\, \mu m}\geq1$~mJy (\cite{coppin2015}; \cite{simpson2017}; \cite{danielson2017}).

\subsection{Multiwavelength photometric data}

Because our SMGs lie within the COSMOS field, they benefit from an exceptionally rich arsenal 
of panchromatic observations that were collected across the electromagnetic spectrum, 
all the way from the X-ray regime to the radio bands. 

In the following subsections, we describe the COSMOS data sets employd here, and a selected compilation of mid-IR to radio flux densities of our SMGs 
is provided in Table~\ref{table:fluxes}. For multiwavelength image montages, we refer to Brisbin et al. (2017).

\subsubsection{From the observed-frame near-ultraviolet and optical to mid-infrared: Stellar and warm dust emissions}

In the present study, we employed the band-merged COSMOS2015 photometric catalogue, which contains extensive ground and 
space-based photometric data from the near-ultraviolet (UV) and optical to the mid-IR wavelength bands 
(\cite{laigle2016})\footnote{See {\tt http://cosmos.astro.caltech.edu/page/photom}.}. 

The deep $u^*$-band (central wavelength $\lambda_{\rm cen}=0.382$~$\mu$m) observations were 
carried out with the MegaCam imaging camera (\cite{boulade2003}) mounted on the Canada-France-Hawaii Telescope (CFHT). Most of the COSMOS2015 photometry data were obtained 
using the Subaru Prime Focus Camera (Suprime-Cam) mounted on the 8.2-metre Subaru telescope (\cite{miyazaki2002}; \cite{taniguchi2007}, 2015). From these data, we used those obtained with the six broad-band filters: $B$ ($\lambda_{\rm cen}=0.446$~$\mu$m), $g^+$ ($\lambda_{\rm cen}=0.478$~$\mu$m), $V$ ($\lambda_{\rm cen}=0.548$~$\mu$m), 
$r$ ($\lambda_{\rm cen}=0.629$~$\mu$m), $i^+$ ($\lambda_{\rm cen}=0.768$~$\mu$m), and $z^{++}$ ($\lambda_{\rm cen}=0.911$~$\mu$m). The intermediate and narrow-band Subaru data were not used here because their central wavelengths ($\lambda_{\rm cen}=0.426-0.824$~$\mu$m for the intermediate-band filters; $\lambda_{\rm cen}=0.712$~$\mu$m and 0.815~$\mu$m for the two narrow bands; see Table~1 in \cite{laigle2016}) are comparable to those of the broad-band filters, they pass only a small portion of the spectrum, and they can be sensitive to observed-frame optical spectral line features, which are not taken into account in the SED method used here. Moreover, we used data obtained with the Subaru/Hyper Suprime-Cam (HSC; \cite{miyazaki2012}) in its HSC-$Y$ band ($\lambda_{\rm cen}=0.979$~$\mu$m). 

Near-infrared imaging of the COSMOS field is being done by the UltraVISTA survey (\cite{mccracken2012}; \cite{ilbert2013}) in the $Y$ ($\lambda_{\rm cen}=1.021$~$\mu$m), $J$ ($\lambda_{\rm cen}=1.253$~$\mu$m), $H$ ($\lambda_{\rm cen}=1.645$~$\mu$m), and $K_{\rm s}$ ($\lambda_{\rm cen}=2.154$~$\mu$m) bands\footnote{The data products are produced by TERAPIX; see {\tt http://terapix.iap.fr}.}. The COSMOS2015 UltraVISTA data used in the present work correspond to the data release version 2 (DR2). The Wide-field InfraRed Camera (WIRCam; \cite{puget2004}) on the CFHT was also used for $H$ ($\lambda_{\rm cen}=1.631$~$\mu$m) and $K_{\rm s}$-band ($\lambda_{\rm cen}=2.159$~$\mu$m) imaging, and these data were used in case the source was not covered by UltraVISTA imaging. Longer wavelength near-IR and mid-IR observations were obtained with the \textit{Spitzer} Space Telescope's Infrared Array Camera 
(IRAC; 3.6--8.0 $\mu$m; \cite{fazio2004}), and still longer IR wavelengths were observed with 
the Multiband Imaging Photometer for \textit{Spitzer} (MIPS; 24--160~$\mu$m; \cite{rieke2004}). Most of these data were taken 
as part of the COSMOS \textit{Spitzer} survey (S-COSMOS; \cite{sanders2007}). However, the IRAC 3.6~$\mu$m and 4.5~$\mu$m data used here 
(and tabulated in the COSMOS2015 catalogue) were taken by the \textit{Spitzer} Large Area Survey with 
Hyper Suprime-Cam (SPLASH)\footnote{{\tt http://splash.caltech.edu}.} during the warm phase of the mission when cryogenic cooling was no longer available on-board \textit{Spitzer} (PI: P.~Capak; see \cite{steinhardt2014}). 
Of the \textit{Spitzer}/MIPS observations we used the 24~$\mu$m mid-IR data from the catalogue of 
Le Floc'h et al. (2009) that are also included in the COSMOS2015 fusion catalogue.

\subsubsection{Far-infrared to millimetre data: Colder dust emission}

In the present study, we used the far-IR (100~$\mu$m, 160~$\mu$m, and 250~$\mu$m) to submm (350~$\mu$m and 500~$\mu$m) \textit{Herschel} (\cite{pilbratt2010})\footnote{\textit{Herschel} is an ESA space observatory with science instruments provided by European-led Principal Investigator consortia and with important participation from NASA.} continuum observations, which were performed as part of the Photodetector Array Camera and Spectrometer (PACS) Evolutionary Probe (PEP; \cite{lutz2011}) and the \textit{Herschel} Multi-tiered Extragalactic Survey (HerMES\footnote{{\tt http://hermes.sussex.ac.uk}.}; \cite{oliver2012}) legacy programmes. Because the beam sizes (FWHM) of the \textit{Herschel} data are large, namely $6\farcs7$, $11\arcsec$, $18\arcsec$, $25\arcsec$, and $36\arcsec$ at 100, 160, 250, 350, and 500~$\mu$m, respectively, the \textit{Herschel} flux densities were extracted by using the ALMA 1.3~mm and \textit{Spitzer} 24~$\mu$m sources as positional priors as explained in Brisbin et al. (2017; cf.~\cite{magnelli2012}). For closely separated multicomponent ALMA sources ($<0.5\times {\rm FWHM}({Herschel})$), we first extracted a total, systemic \textit{Herschel} flux density, and then used the relative ALMA flux densities of the individual components to estimate their fractional contribution to the total \textit{Herschel} flux density.

From the ground-based single-dish telescope data, we used the deboosted ASTE/AzTEC 1.1~mm flux densities reported by Aretxaga et al. (2011; their Table~1), a study from which our parent SMG sample was drawn. In case the AzTEC source was resolved into two or three components in our ALMA imaging, we used the relative ALMA 1.3~mm flux densities of the components to estimate their contribution to the AzTEC 1.1~mm emission. For 13 sources studied here, we could also obtain the 450~$\mu$m and 850~$\mu$m photometry obtained through observations with the James Clerk Maxwell Telescope (JCMT)/Submillimetre Common User Bolometer Array 2 (SCUBA-2; \cite{holland2013}) by Casey et al. (2013). The beam FWHMs at these two wavelengths were $7\arcsec$ and $15\arcsec$. For 16 of our sources, we used the 870~$\mu$m data obtained by the Large APEX BOlometer CAmera (LABOCA; \cite{siringo2009}) survey of the inner 0.75~deg$^2$ of the COSMOS field (F.~Navarrete et al., in prep.; see also \cite{smolcic2012}). The effective angular resolution (FWHM) of these data was $27\farcs6$. Moreover, five of our analysed AzTEC sources (plus two candidate AGN-host SMGs) were detected at 1.2~mm with the Max-Planck Millimetre Bolometer Array 2 (MAMBO-2; \cite{kreysa1998}) at $11\arcsec$ resolution (\cite{bertoldi2007}). As in the case of the AzTEC 1.1~mm data, we used the relative ALMA flux densities of the multicomponent sources to estimate their fractional contribution to the SCUBA-2, LABOCA, and MAMBO-2 flux densities. 

Besides our ALMA 1.3~mm data, some of our SMGs benefit from additional interferometrically observed (sub-)mm flux densities. These include the 890~$\mu$m flux densities measured with the Submillimetre Array (SMA) by Younger et al. (2007, 2009) for seven of our ALMA sources. The SMGs AzTEC/C5, C17, and C42 were observed with the ALMA Band~7 during the second early science campaign to search for [\ion{C}{II}] or [\ion{N}{II}] line emission (Cycle~1 ALMA project 2012.1.00978.S; PI: A.~Karim). The Common Astronomy Software Applications ({\tt CASA}; \cite{mcmullin2007}) package\footnote{{\tt https://casa.nrao.edu}.} was used to construct the continuum images from the line-free channels at $\lambda_{\rm obs}=870$~$\mu$m, 857~$\mu$m, and 994~$\mu$m, respectively. To determine the flux densities, we used the National Radio Astronomy Observatory (NRAO) Astronomical Image Processing System ({\tt AIPS}) software package\footnote{\tt http://www.aips.nrao.edu.} to make two-dimensional elliptical Gaussian fits (the {\tt AIPS} task {\tt JMFIT}). For AzTEC/C5, C17, and C42, we derived $S_{\rm 870\, \mu m}=14.12\pm0.25$~mJy, $S_{\rm 857\, \mu m}=7.63\pm0.17$~mJy, and $S_{\rm 994\, \mu m}=6.52\pm0.28$~mJy (a combination of two Gaussians fitted to the two C42 components separated by only $0\farcs7$ (see \cite{miettinen2017b}), and valid for our aggregate SED). We also used the Plateau de Bure Interferometer (PdBI) 3~mm flux density for AzTEC/C5 from Smol\v{c}i\'{c} et al. (2011; $S_{\rm 3\, mm}=0.30\pm0.04$~mJy). For AzTEC/C17, we adopted the 3.6~mm PdBI and 7.1~mm Very Large Array (VLA) flux densities from Schinnerer et al. (2008); the former value is only of $\sim2\sigma$ significance ($S_{\rm 3.6\, mm}=0.2\pm0.09$~mJy), and the latter represents a $3\sigma$ upper limit ($S_{\rm 7.1\, mm}\leq0.15$~mJy). Finally,
for AzTEC/C6a and C6b, we used the ALMA Band~7 (870~$\mu$m) flux densities from Bussmann et al. (2015; their sources HCOSMOS02-Source0 and Source1), and the NOrthern Extended Millimetre Array (NOEMA) 1.8~mm flux densities from Wang et al. (2016; their sources 131077 and 130891). 

\subsubsection{Radio data: Non-thermal synchrotron and thermal free-free emissions}

To sample the radio regime of the source SEDs, we employed the 325~MHz and 610~MHz observations taken by the Giant Meterwave Radio Telescope (GMRT)-COSMOS survey (A.~Karim et al., in prep.). The synthesised beam sizes of the employed mosaics were $10\farcs76 \times 9\farcs49$ at 325~MHz, and $6\farcs5 \times 6\farcs5$ at 610~MHz, while the typical $1\sigma$ rms noises at these frequencies were 78~$\mu$Jy~beam$^{-1}$ and 50.6~$\mu$Jy~beam$^{-1}$, respectively.  
We used the {\tt BLOBCAT} source extraction software (\cite{hales2012}) to create a preliminary catalogue of GMRT sources, and which was complemented by closer eye inspection of the mosaics guided by the ALMA positions. Altogether, 
we assigned a 325~MHz counterpart for 66, and a 610~MHz counterpart for 62 out of our parent sample of 152 ALMA sources. A $3\sigma$ upper flux density limit was set for those sources that were not detected in the GMRT mosaics. 

We also used the VLA radio continuum imaging data at 1.4~GHz (\cite{schinnerer2007}, 2010), and at 3~GHz taken by the VLA-COSMOS 3~GHz Large Project (\cite{smolcic2017b}). As described by Miettinen et al. (2017b), 
$76\%\pm7\%$ of our parent sample of ALMA SMGs (115/152) have a 3~GHz counterpart, and the corresponding flux densities used in the present study can be found in their Table~C.1. 

\begin{figure}[!htb]
\centering
\resizebox{0.9\hsize}{!}{\includegraphics{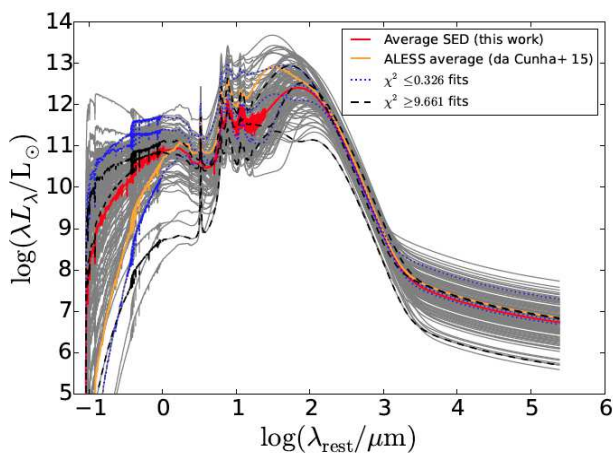}}
\caption{Best-fit panchromatic (UV–radio) rest-frame SEDs of all the 124 analysed SMGs (grey lines). The three best SED fits ($\chi^2 \leq 0.326$) are highlighted by the blue dotted lines, while the three worst SEDs ($\chi^2 \geq 9.661$) are shown by the black dashed lines. The red line shows the average SED 
of our SMGs, while the orange line shows the average {\tt MAGPHYS} SED of the 870~$\mu$m selected SMGs from da Cunha et al. (2015; cf.~their Fig.~11). See Sect.~4.8 for further discussion.}
\label{figure:averagesed}
\end{figure}

\section{Analysis and results}

\subsection{Spectral energy distributions from ultraviolet-optical to radio wavelengths}

\subsubsection{Method}

To derive the key physical properties of our SMGs, we constructed their UV-optical to radio SEDs using the data sets described in Sect.~2.2. 
The observational data were modelled using the Multiwavelength Analysis of Galaxy Physical Properties code {\tt MAGPHYS} 
(\cite{dacunha2008})\footnote{{\tt MAGPHYS} is publicly available, and can be retrieved at {\tt http://www.iap.fr/magphys/magphys/MAGPHYS.html}.}. 
Since the {\tt MAGPHYS} package has already been described in detail in several papers (e.g. \cite{dacunha2008}, 2010b; \cite{smith2012}; \cite{berta2013}; \cite{rowlands2014}; \cite{hayward2015}; \cite{smolcic2015}; \cite{dacunha2015}), we do not repeat it here, but just briefly mention that {\tt MAGPHYS} is built on a global energy balance between stellar and dust emissions: the UV-optical photons emitted by young stars are absorbed (and scattered) by dust grains in star-forming regions and more diffuse parts of the galactic interstellar medium (ISM), and the heated dust then reradiates this absorbed energy in the IR (e.g. \cite{devereux1990}). One potential caveat of the energy balance technique in the analysis of SMGs is that the visible (unobscured) stellar component can be spatially decoupled from the dust-emitting, obscured star-forming parts, in which case the UV-optical and far-IR to mm photometry might not be coupled in a way assumed in the SED modelling (\cite{simpson2017}; \cite{casey2017}; see also \cite{miettinen2017b} for SMG size comparisons).

In the present work, we used the new version of {\tt MAGPHYS}, which is optimised to fit the SEDs of $z > 1$ SFGs all the way from 
the UV to the radio regime. This SED fitting package is expected to be better suited to derive the physical properties of SMGs than the earlier 
versions of {\tt MAGPHYS} (see \cite{dacunha2015}; \cite{miettinen2017a}). The updated fitting tool has three key improvements over previous versions. First, it contains 
extended prior distributions of star formation history and dust optical thickness. Secondly, the absorption of UV photons by the intergalactic medium (IGM) is taken into account. 
Thirdly, the SED fit can be extended to the centimetre radio wavelengths. The latter is based on the assumption of a far-IR($42.5-122.5$~$\mu$m)-radio correlation with a $q_{\rm FIR}\propto \log(L_{\rm FIR}/L_{\rm 1.4\, GHz})$ parameter distribution centred at $q_{\rm FIR}=2.34$, which equals the mean value derived by Yun et al. (2001) for a sample of 1\,809 galaxies at $z\leq0.15$ detected with the \textit{Infrared Astronomical Satellite} (\cite{neugebauer1984}) at $S_{\rm 60 \, \mu m}\geq2$~Jy. The $q_{\rm FIR}$ parameter is assumed to have a scatter of $\sigma(q_{\rm FIR})=0.25$ to take possible variations into account. The thermal free-free emission in {\tt MAGPHYS} is fixed to have a spectral shape of $S_{\nu}^{\rm T}\propto \nu^{-0.1}$ (i.e. optically thin free-free emission). The non-thermal synchrotron emission is assumed to have a spectral shape of $S_{\nu}^{\rm NT}\propto \nu^{-0.8}$. The thermal radio emission is assumed to amount to 10\% ($f_{\rm T}=0.1$) of the total flux density at $\nu_{\rm rest}=1.4$~GHz. The possible contribution of an AGN to the radio emission is not taken into account. As demonstrated by, for example Miettinen et al. (2017a), some of the aforementioned assumptions might be invalid for individual SMGs.

The {\tt MAGPHYS} SED models used here assume that the interstellar dust is predominantly heated by the radiation produced by 
star formation activity, while the AGN contribution to the dust heating is not taken into account. 
However, as described in Sect.~2.1.2, we purified our sample from potential AGN hosts, and hence a lack of modelling 
the AGN emission is not expected to bias our results.  

Following da Cunha et al. (2015) and Miettinen et al. (2017a), the flux density upper limits in the SED fitting process 
were taken into account by setting the nominal value to zero, and using the upper limit value (here $3\sigma$) 
as the flux density error.

\begin{figure*}
\begin{center}
\includegraphics[width=0.2465\textwidth]{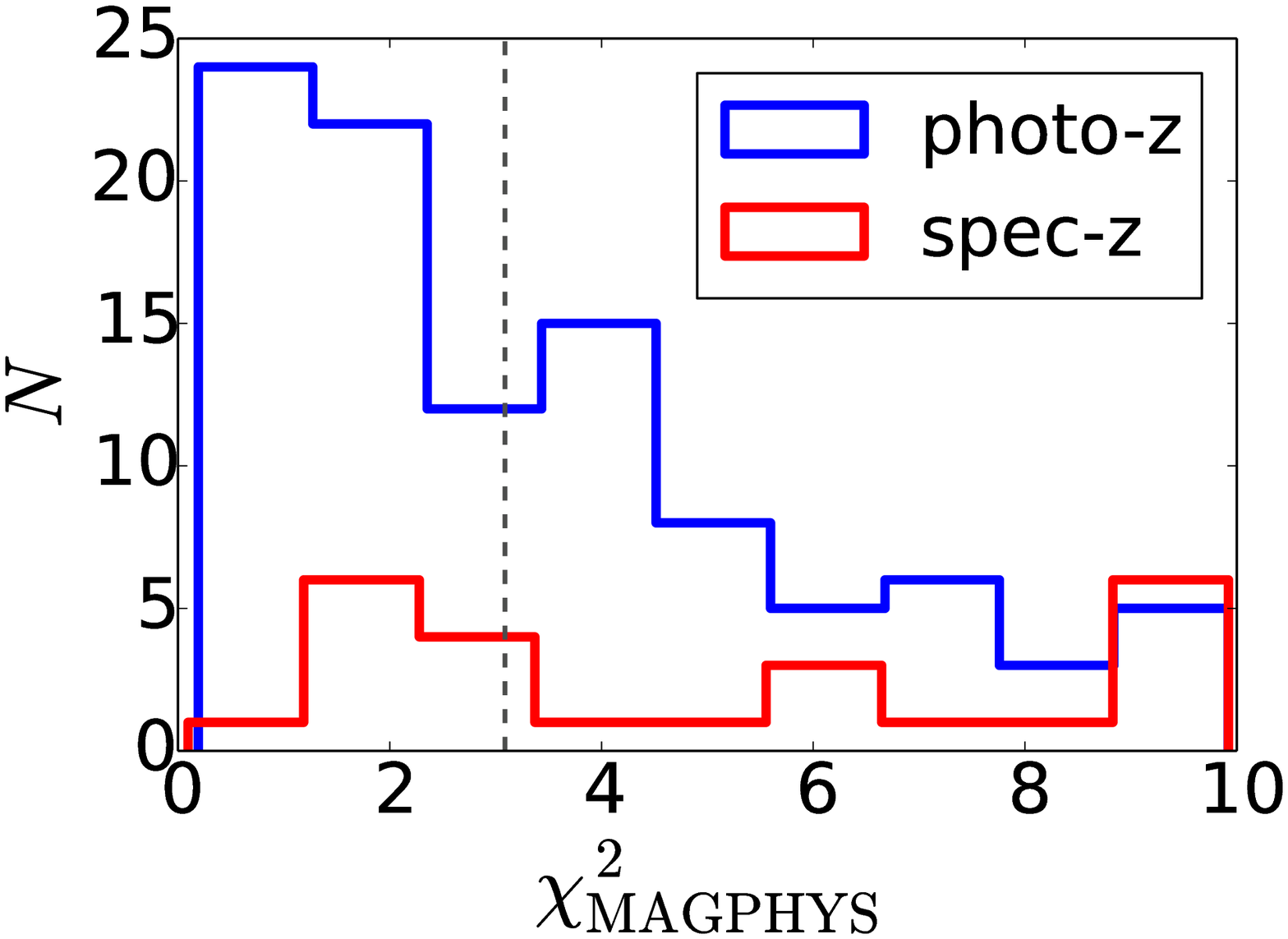}
\includegraphics[width=0.2465\textwidth]{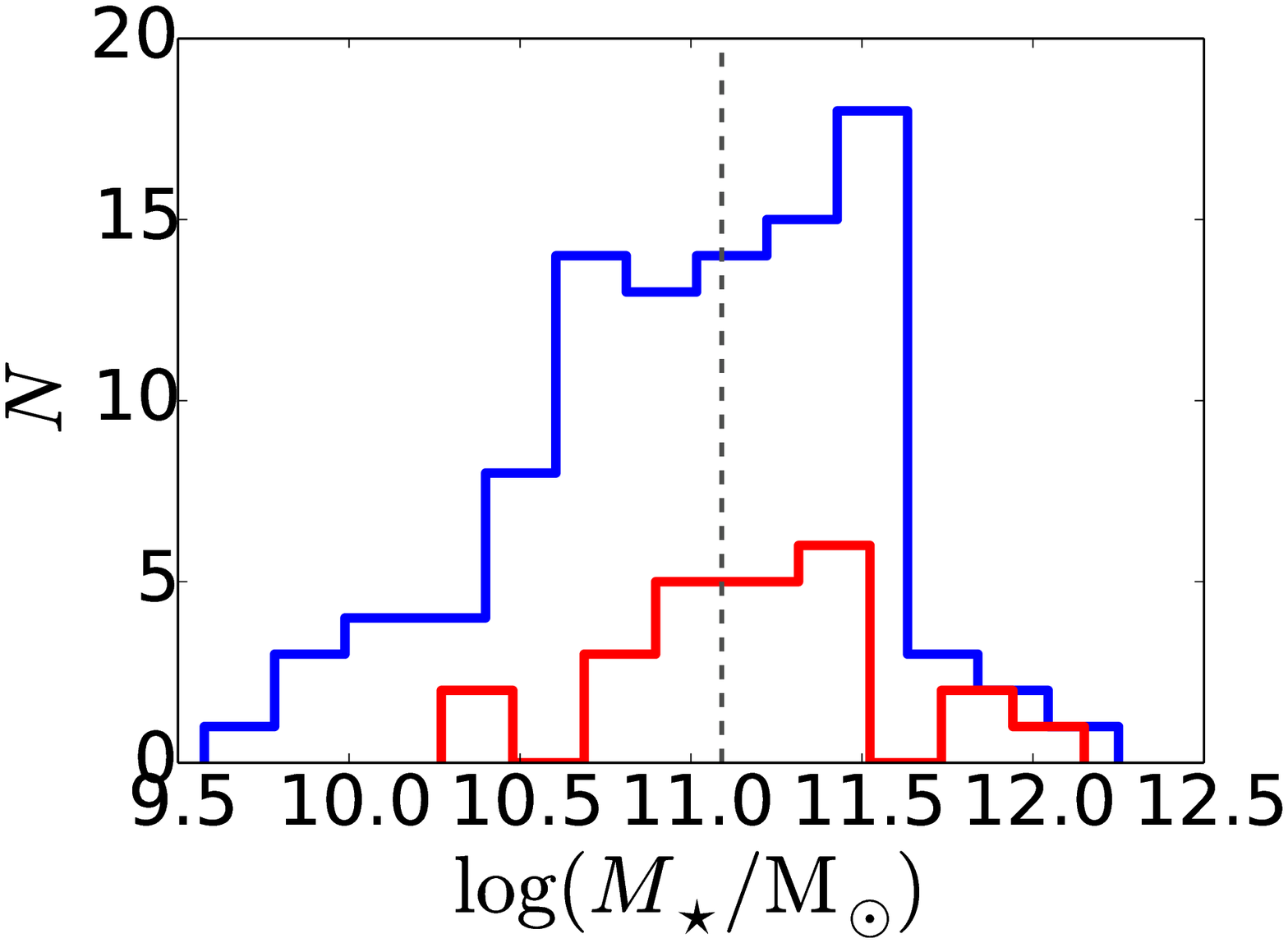}
\includegraphics[width=0.26\textwidth]{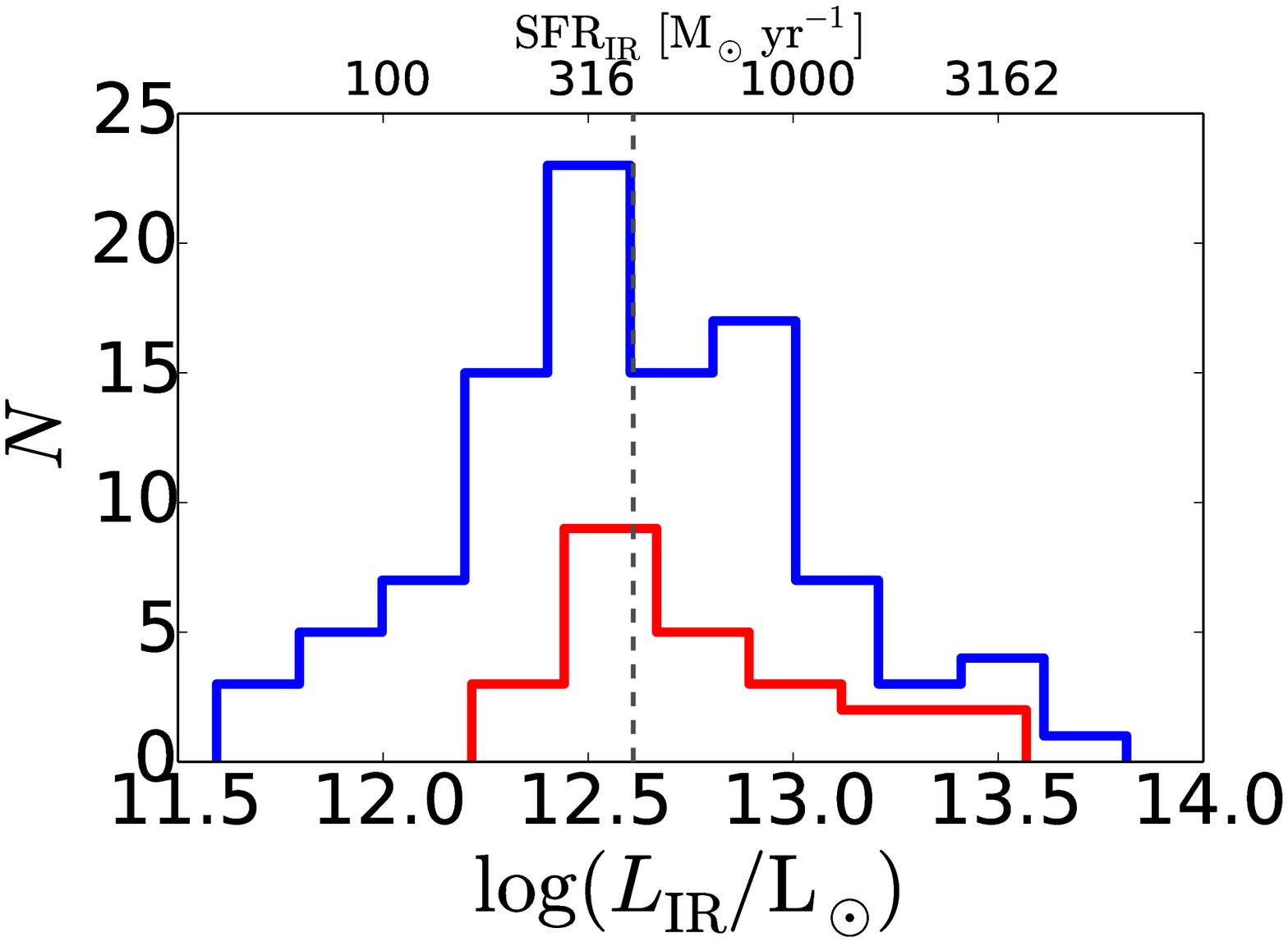}
\includegraphics[width=0.2465\textwidth]{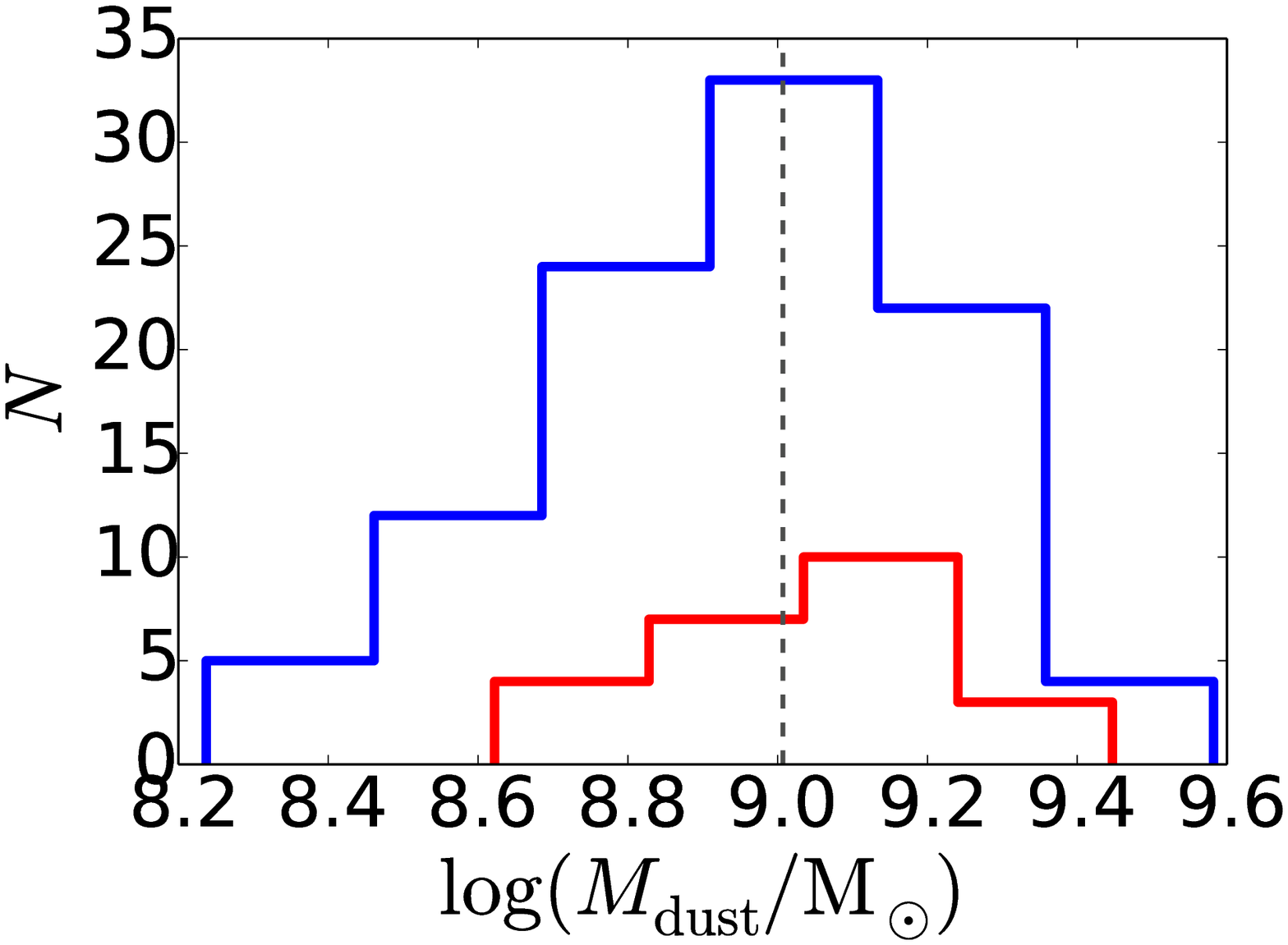}
\includegraphics[width=0.2465\textwidth]{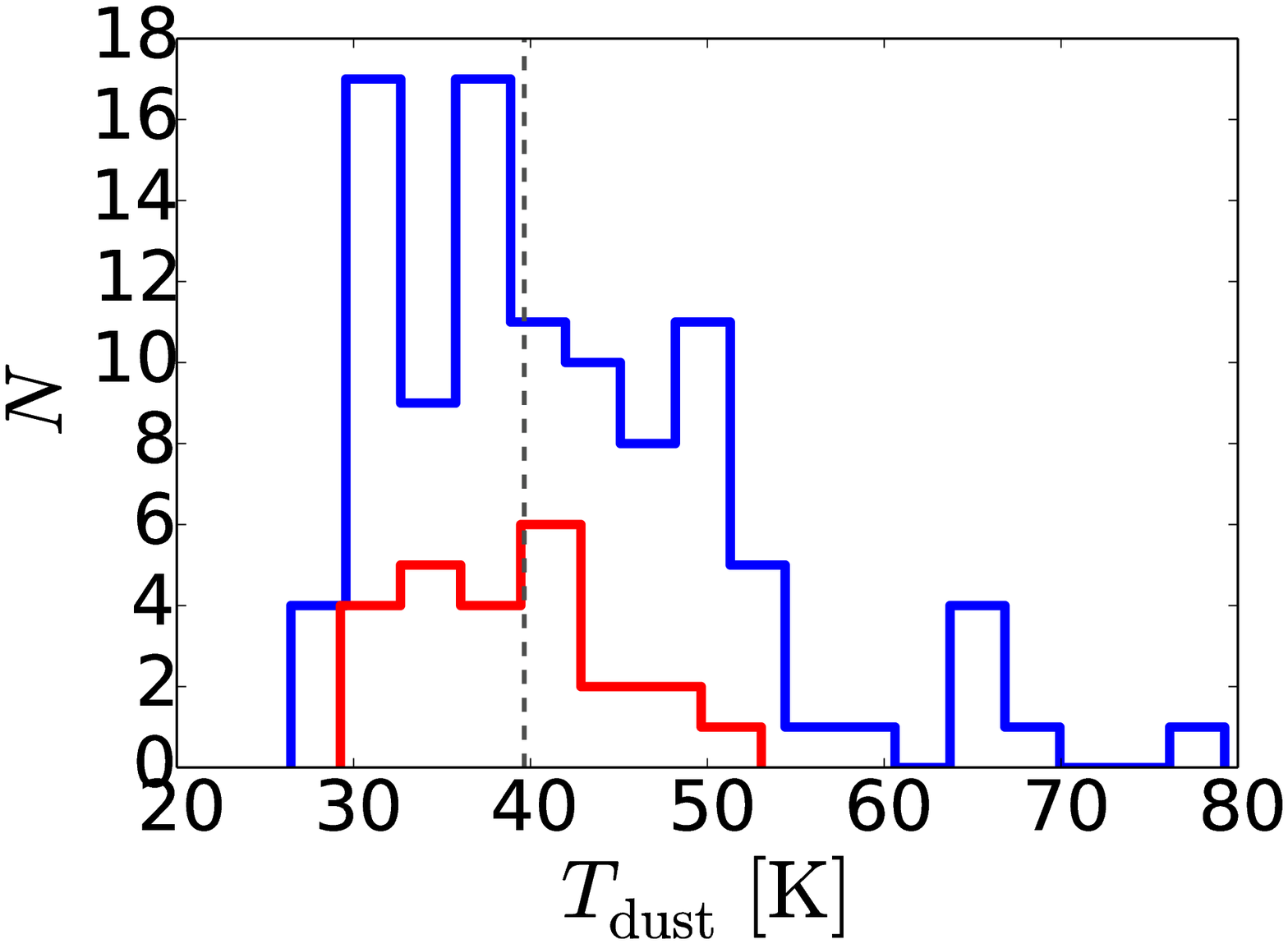}
\includegraphics[width=0.2465\textwidth]{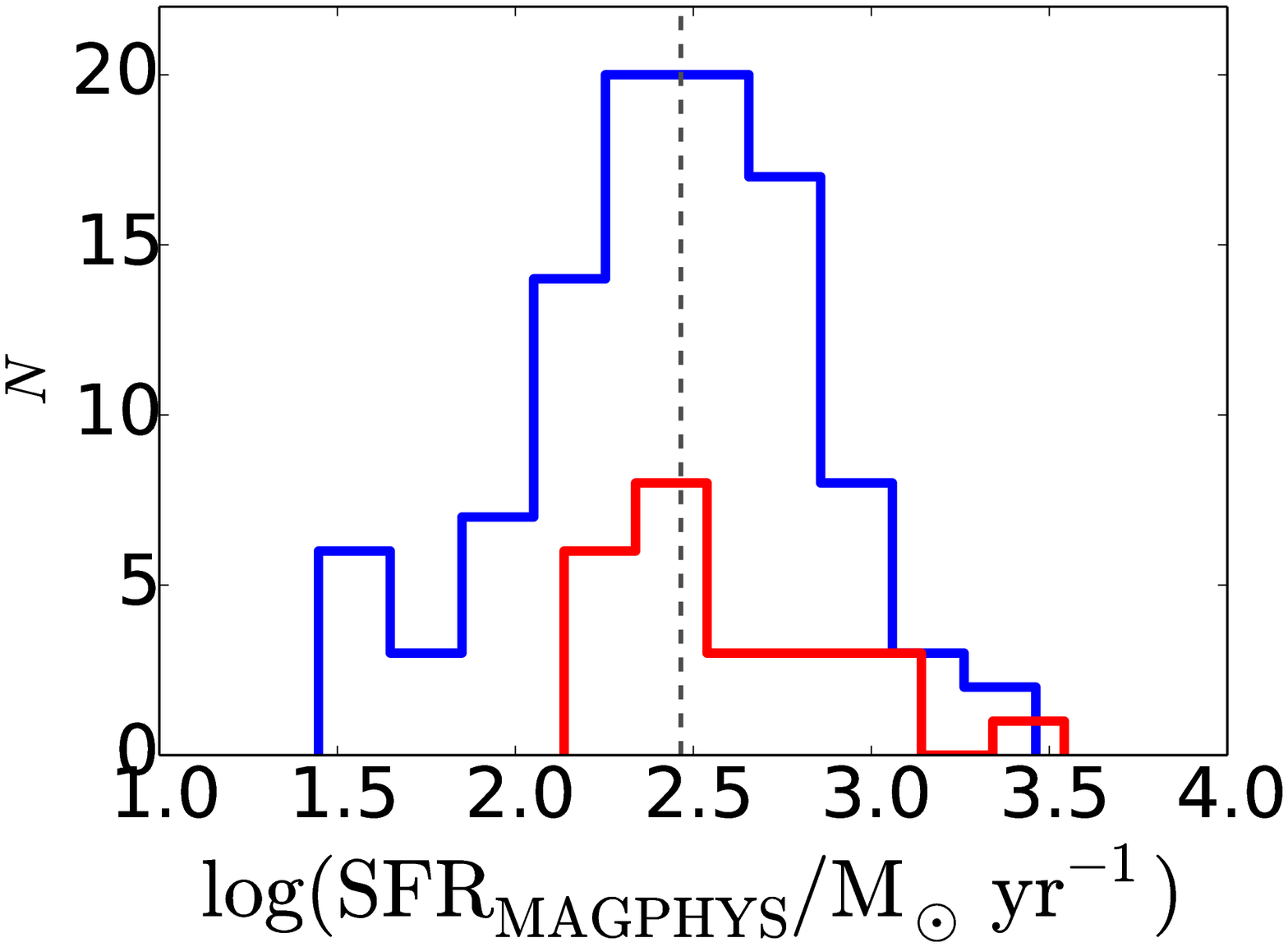}
\includegraphics[width=0.2465\textwidth]{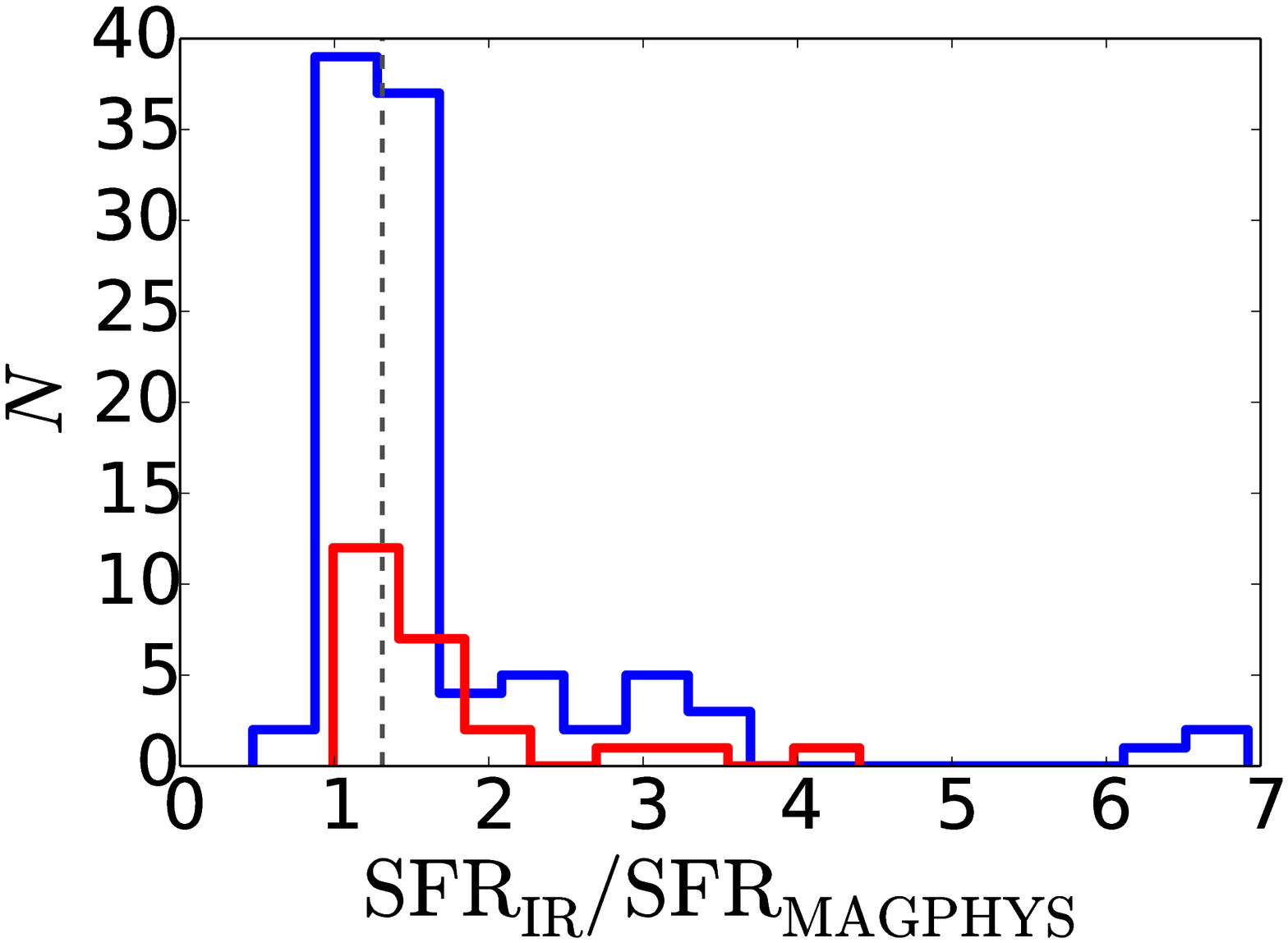}
\includegraphics[width=0.2465\textwidth]{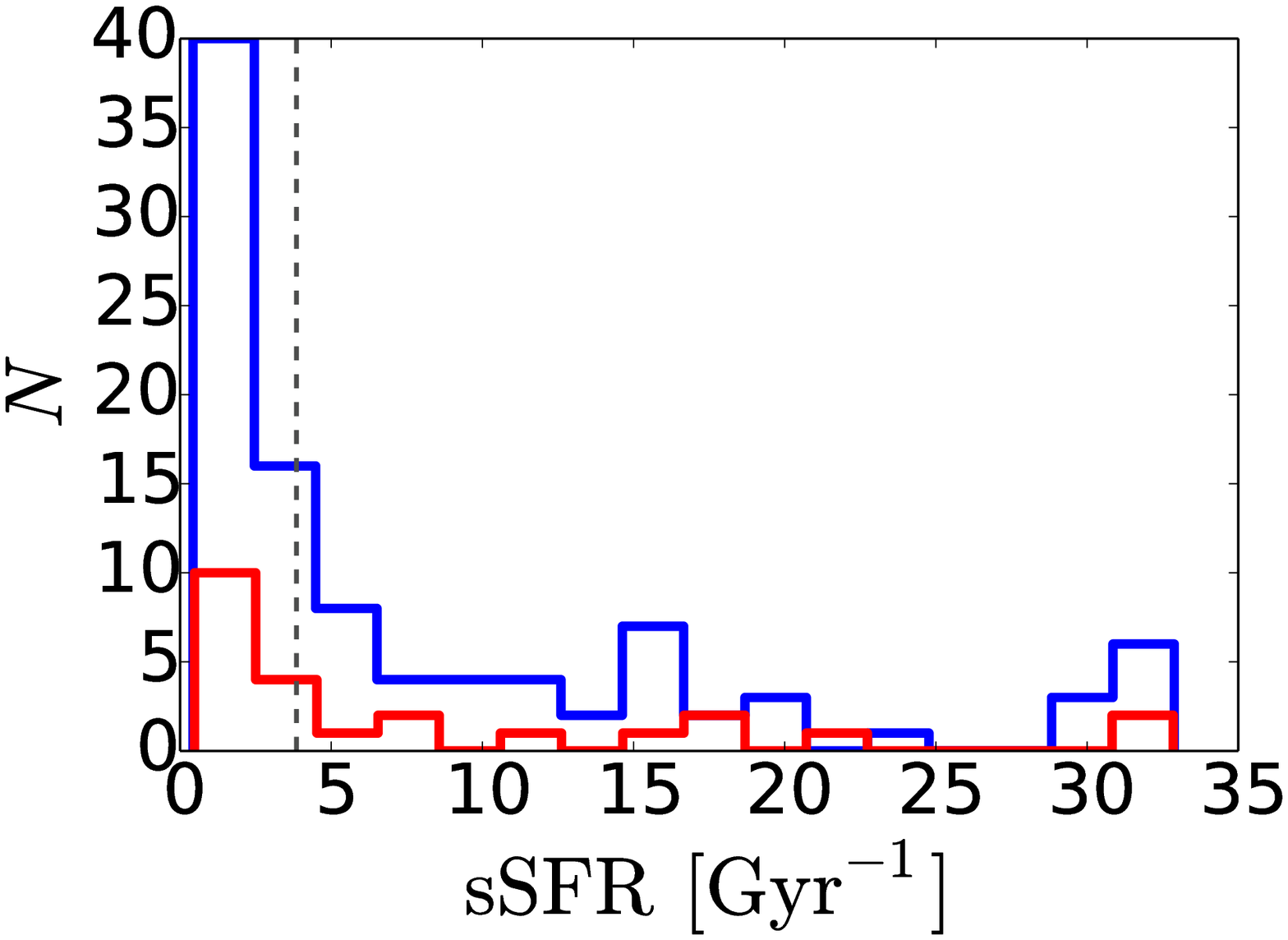}
\includegraphics[width=0.2465\textwidth]{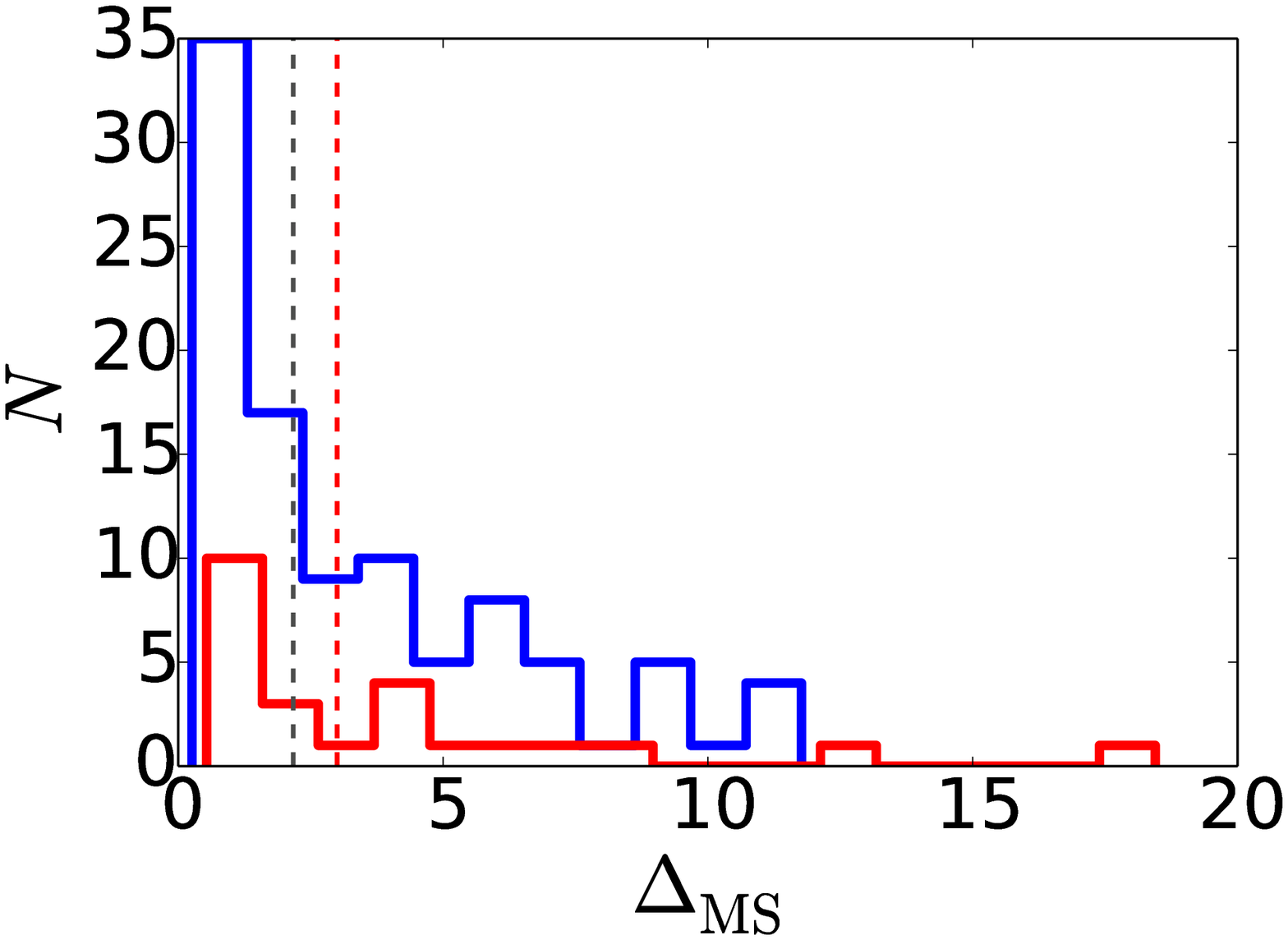}
\includegraphics[width=0.2465\textwidth]{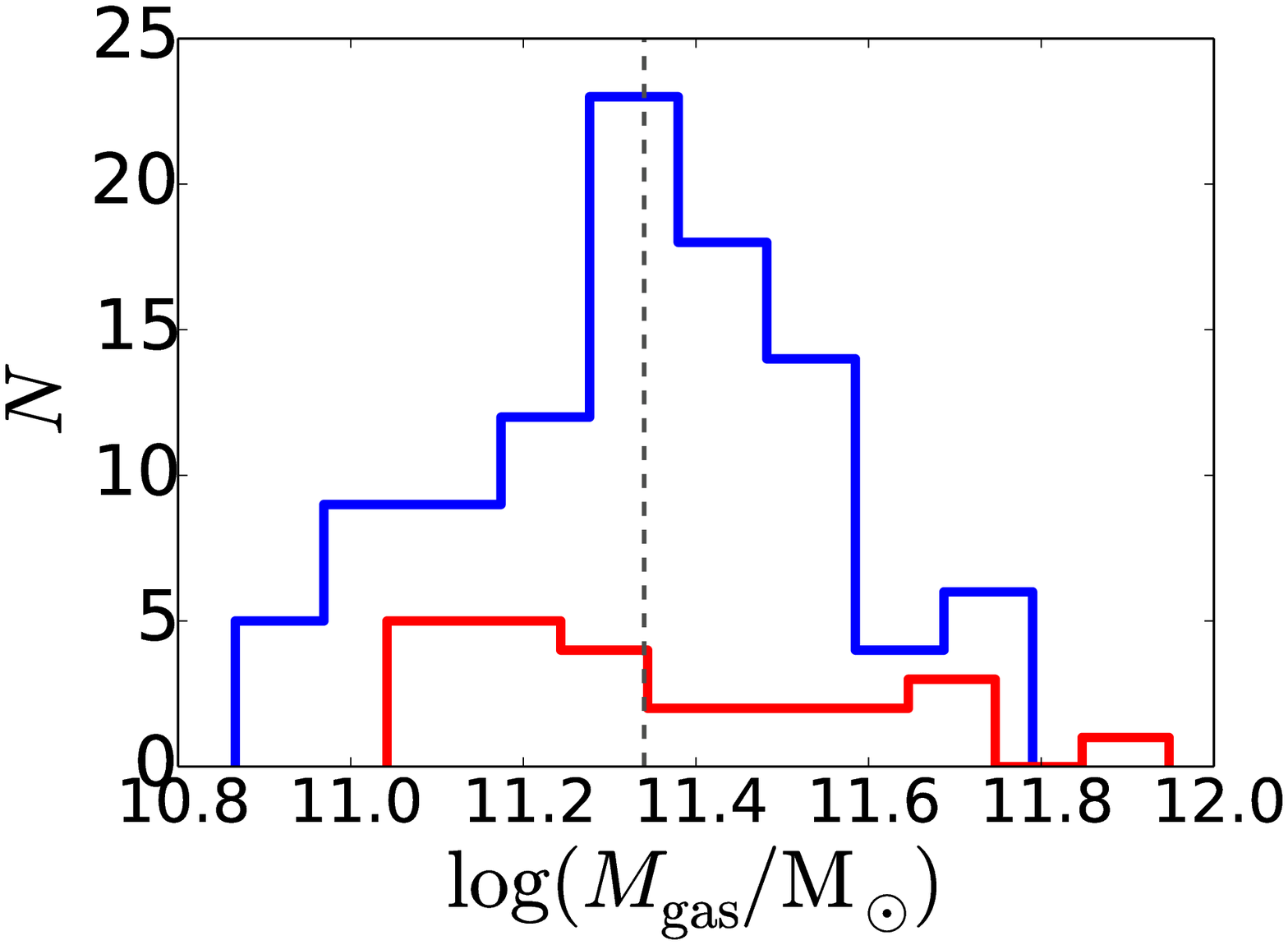}
\caption{Distributions of the {\tt MAGPHYS} SED results and other physical parameters, shown separately for the spectroscopically confirmed sources (red histogram) and 
the sources with photometric redshifts (blue histogram). The panels from top to bottom, left to right, show the $\chi^2$ (goodness of fit) values of the best-fit {\tt MAGPHYS} SEDs, stellar masses, total-IR luminosities (the upper $x$-axis shows the corresponding $L_{\rm IR}$-based SFRs), dust masses, dust temperatures, {\tt MAGPHYS}-derived SFRs ($\Delta t=100$~Myr), ratios between the $L_{\rm IR}$-based SFR and that directly output by {\tt MAGPHYS}, specific SFRs, starburstiness parameters (distance from the MS mid-line; see Sect.~4.1), and the ALMA 1.3~mm-based gas masses (Sect.~3.2). Base-10 logarithms are used for the masses, $L_{\rm IR}$, and {\tt MAGPHYS}-derived SFRs. The vertical dashed lines mark the full sample medians (see Table~\ref{table:sed2}), and the additional red vertical line in the starburstiness panel shows the upper boundary of the MS (i.e. $\Delta_{\rm MS}=3$, above which the source can be defined as a starburst galaxy).}
\label{figure:histograms}
\end{center}
\end{figure*}

\begin{table*}[!htb]
\caption{Statistics of the {\tt MAGPHYS} SED fitting results and the $M_{\rm gas}$ values.}
{\scriptsize
\begin{minipage}{2\columnwidth}
\centering
\renewcommand{\footnoterule}{}
\label{table:sed2}
\begin{tabular}{c c c c c c c c c c}
\hline\hline 
Parameter & $\chi^2$ & $\log(M_{\star}/{\rm M}_{\sun})$ & $\log(L_{\rm IR}/{\rm L}_{\sun})$ & SFR [${\rm M}_{\sun}~{\rm yr}^{-1}$] & sSFR [${\rm Gyr}^{-1}$] & $\Delta_{\rm MS}$ & $T_{\rm dust}$ [K] & $\log(M_{\rm dust}/{\rm M}_{\sun})$ & $\log(M_{\rm gas}/{\rm M}_{\sun})$ \\[1ex]
\hline
Min. & 0.095 & 9.58 & 11.59 & 39 & 0.4 & 0.3 & 26.5 & 8.24 & 10.87\\[1ex]
Max. & 9.922 & 12.25 & 13.81 & 6\,501 & 32.9 & 18.4 & 79.3 & 9.58 & 11.95\\[1ex]
Mean & $3.740\pm0.243$ & $11.04\pm0.04$ & $12.64\pm0.04$ & $722\pm84$ & $8.1\pm0.8$ & $3.5\pm0.3$ & $41.2\pm0.8$ & $8.96\pm0.02$ & $11.34\pm0.02$\\[1ex]
Median & $3.088^{+3.829}_{-1.984}$ & $11.09^{+0.41}_{-0.53}$ & $12.61^{+0.42}_{-0.38}$ & $402^{+661}_{-233}$\tablefootmark{a} & $3.9^{+12.7}_{-2.9}$ & $2.2^{+4.2}_{-1.3}$ & $39.7^{+9.7}_{-7.4}$ & $9.01^{+0.20}_{-0.31}$ & $11.34^{+0.20}_{-0.23}$\\ [1ex]
$\sigma_{\rm SD}$ & 2.696 & 0.50 & 0.42 & 938 & 9.4 & 3.3 & 9.3 & 0.27 & 0.21\\ [1ex]
\hline 
\end{tabular} 
\tablefoot{The columns are as follows: (1) statistical quantity; (2) $\chi^2$ (goodness of fit) of the best SED fit; (3) stellar mass; (4) IR luminosity calculated by integrating the SED over the rest-frame wavelength range of $\lambda_{\rm rest}=8-1\,000$~$\mu$m; (5) SFR calculated using the ${\rm SFR}(L_{\rm IR})$ relationship of Kennicutt (1998); (6) specific SFR ($={\rm SFR}/M_{\star}$); 
(7) ratio of SFR to that of a MS galaxy of the same redshift and stellar mass (i.e. offset from the MS); (8) luminosity-weighted dust temperature (see Eq.~(8) in \cite{dacunha2015}); (9) dust mass; (10) gas mass. Base-10 logarithms are used for the masses and $L_{\rm IR}$. The five rows list the minimum, maximum, mean, median, and standard deviation values. The reported uncertainty of the mean is the standard error of the mean ($\sigma_{\rm SD}/\sqrt{N}$, where $N$ is the sample size), while that of the median represents the 16th--84th percentile range (68\% confidence interval).\tablefoottext{a}{The median value of the {\tt MAGPHYS}-derived SFR is $291^{+377}_{-162}$~${\rm M}_{\sun}~{\rm yr}^{-1}$} (see Fig.~\ref{figure:histograms}).}
\end{minipage} }
\end{table*}

\subsubsection{Spectral energy distribution results and physical parameters}

The individual source SEDs are shown in Fig.~\ref{figure:seds}, 
while in Fig.~\ref{figure:averagesed} we plot the best-fit 
SEDs of all the 124 analysed SMGs along with the average SED. In the latter figure, the three best and worst SED fits are highlighted to illustrate the effect of the $\chi^2$ of the fit. For comparison, the average {\tt MAGPHYS} SED of the 870~$\mu$m selected SMGs from da Cunha et al. (2015) is also overplotted in Fig.~\ref{figure:averagesed} (see Sect.~4.8 for discussion). In Fig.~\ref{figure:seds2}, we show the SEDs of those SMGs that are potentially harbouring an AGN, but which are not analysed further here. All the individual SED parameters are listed in Table~\ref{table:sed}, while in Table~\ref{table:sed2} we tabulate the sample statistics, such as the mean and median values. The sample distributions are illustrated as histogram plots in Fig.~\ref{figure:histograms}, separately for the spectroscopically confirmed sources (24/124) and the sources whose redshfits were derived using photometric techniques (100/124).

As can be seen in Fig.~\ref{figure:seds}, in some cases the best-fit model is inconsistent with the upper flux density limits 
(e.g. the far-IR flux density upper limits for AzTEC/C42). At least some of the discrepancies could be the result of an incorrectly assigned 
upper flux density limit (we assumed $3\sigma$ upper limits). Also, the upper flux density limits in {\tt MAGPHYS} are not rigorously treated as such, 
that is the best-fit is not forced to lie below them (similarly, some of the uncensored data points can also lie below or above the best fit, 
such as in the case of AzTEC/C37). As a consistency check, we refit the SEDs of those sources that have some of the upper flux density limits below 
the best fit by ignoring these upper limits. For example, we removed the observed-frame 24~$\mu$m upper limit for AzTEC/C2a, C2b, C49, and C51b, 
the upper limits between the rest-frame wavelengths of $\lambda_{\rm rest}\sim1$~$\mu$m and $\sim10$~$\mu$m for AzTEC/C22b, all the upper limits except the shortest wavelength value for AzTEC/C66, 
and all the upper limits except the four shortest wavelength values for AzTEC/C87. The best-fit model SEDs (and hence the corresponding physical properties) were found to be practically identical 
to those where the upper limits were taken into account, which demonstrates that our best {\tt MAGPHYS} SED fits are heavily weighted by the large number of uncensored data points.

The $\chi^2$ values of our SED fits range from 0.095 to 9.922. The {\tt MAGPHYS} SED fits are often considered acceptable if 
the corresponding $\chi^2$ values satisfy a threshold probability of $P_{\rm thresh}<1\%$ for the observed data to be consistent with 
the model (\cite{smith2012}; \cite{poudel2016}; see also \cite{hayward2015}). Following the analysis by Smith et al. (2012; their Appendix~B) 
and Poudel et al. (2016), the typical number of photometric bands in our SEDs, $N_{\rm bands}=33$, suggests that the aforementioned threshold 
is reached above a $\chi^2$ value of $>14$. Because even the worst of our SED fits has a lower $\chi^2$ value, namely $\chi^2\simeq10$, 
we consider all of our SED fits acceptable.

As shown in Fig.~\ref{figure:histograms}, the spectroscopically confirmed sources exhibit a similar range of SED $\chi^2$ values as the sources whose redshifts are photometric. We note that the spec-$z$ values of the former group of sources are in very good agreement with their photo-$z$ solutions (\cite{brisbin2017}), the mean (median) ratio between the two being 1.04 (0.99). This gives us confidence that the redshifts, and hence SEDs of our sources with only photo-$z$ values available are generally accurate.

The physical parameters given in Tables~\ref{table:sed} and \ref{table:sed2} are $M_{\star}$, total-IR luminosity ($L_{\rm IR}$, defined as the integral under the best-fitting SED from the rest-frame $8-1\,000$~$\mu$m; \cite{sanders1996}), SFR, the specific SFR (defined by ${\rm sSFR}\equiv{\rm SFR}/M_{\star}$; \cite{guzman1997}), a ratio between the SFR and that of a main-sequence (MS) galaxy of the analogue $M_{\star}$ ($\Delta_{\rm MS}\equiv {\rm SFR}/{\rm SFR}_{\rm MS}$; see Sect.~4.1), an average, luminosity-weighted dust temperature ($T_{\rm dust}$; see Eq.~(8) in \cite{dacunha2015} for the definition), dust mass ($M_{\rm dust}$), and the gas mass ($M_{\rm gas}$; described in Sect.~3.2). We adopted the median of the likelihood distribution as an estimate of each {\tt MAGPHYS}-based parameter, and the quoted uncertainties in these parameters tabulated in Table~\ref{table:sed} represent the 68\% confidence interval or the 16th--84th percentile range of the corresponding likelihood distribution. The true uncertainty budgets, particularly owing to systematics from the assumptions used in the SED modelling (e.g. the IMF), are certainly higher than the quoted formal error bars. 

The stellar masses derived from {\tt MAGPHYS} refer to the current stellar mass content, rather than the total stellar mass ever formed, and hence the mass returned to the ISM via stellar losses is accounted for (\cite{dacunha2015}; footnote~19 therein). Regarding the reliability of the stellar masses derived from {\tt MAGPHYS}, we note that Micha{\l}owski et al. (2014) found that {\tt MAGPHYS} can recover the stellar masses of simulated SMGs to better than a factor of two with a very mild bias (systematic overestimate of $\sim0.1$~dex). The authors used the older version of {\tt MAGPHYS}, while the new version of the code we used is expected to be better suited for SMGs (Sect.~3.1.1), and hence can yield even more accurate stellar masses. For comparison, the average uncertainty of the derived stellar masses is $\pm0.106$ in log solar units, which is similar to the aforementioned overestimation factor.

\begin{figure}[!htb]
\centering
\resizebox{0.9\hsize}{!}{\includegraphics{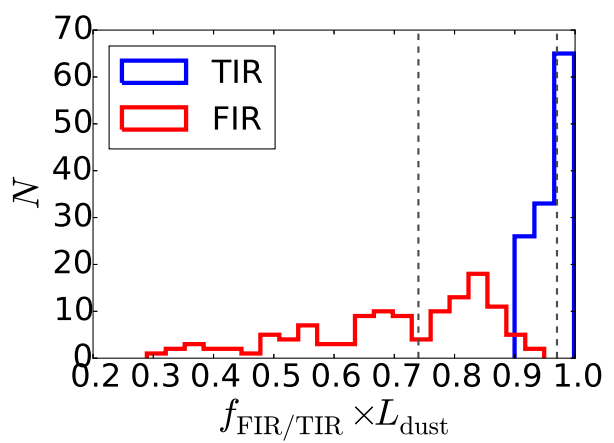}}
\caption{Distributions of the proportion of the {\tt MAGPHYS}-based dust luminosities ($L_{\rm dust}^{\rm 3-1\,000~\mu m}$) that are emitted in the far-IR ($42.5-122.5$~$\mu$m; $f_{\rm FIR}$; red histogram) and total-IR regimes ($8-1\,000$~$\mu$m; $f_{\rm TIR}$; blue histogram). The vertical dashed lines show the sample median values ($f_{\rm FIR}^{\rm median}=0.74$ and $f_{\rm TIR}^{\rm median}=0.97$).}
\label{figure:fractions}
\end{figure}

We also note that instead of $L_{\rm IR}$, {\tt MAGPHYS} gives the total dust IR luminosity ($L_{\rm dust}$ over $\lambda_{\rm rest}=3-1\,000$~$\mu$m) as an output parameter. In Fig.~\ref{figure:fractions}, we show the distributions of the proportions of $L_{\rm dust}$ that emerge in the far-IR ($42.5-122.5$~$\mu$m; e.g. \cite{helou1985}) and total-IR ($8-1\,000$~$\mu$m) ranges. As can be seen, the former quantity spans a wide range of values from $f_{\rm FIR}=0.29$ to 0.95 with a median of 0.74, while the $f_{\rm TIR}$ distribution is much narrower, $f_{\rm TIR}=0.9-1.0$, with a median of 0.97. Hence, the typical situation among our SMGs is that $L_{\rm IR}\simeq 1.3 \times L_{\rm FIR}$. 

Another issue regarding the derived IR luminosities is the possible AGN contribution to the dust heating. Although the potential AGN-hosts were removed from our final sample (Sect.~2.1.2), it is still possible that some of the remaining sources are subject to AGN heating (e.g. $\sim6.5\%$ of the sources satisfy the Donley et al. (2012) IR criteria for an AGN). Moreover, if the dust opacity of the AGN is very high, the mid-IR radiation could be reprocessed into far-IR continuum emission, and hence contribute to the observed far-IR luminosity. To quantitatively estimate the AGN contribution to $L_{\rm IR}$ of the analysed sources, we used the same method as we did in Delvecchio et al. (2017), that is the three-component SED fitting code {\tt SED3FIT}\footnote{The {\tt SED3FIT} code is publicly available at {\tt http://cosmos.astro.caltech.edu/page/other-tools}.} (\cite{berta2013}), which accounts for an additional AGN component (the other two components being the stellar and dust emission). The caveat is that {\tt SED3FIT} is based on the da Cunha et al. (2008) {\tt MAGPHYS} IR libraries, rather than the new high-$z$ {\tt MAGPHYS} model libraries we employed in the present analysis. Nevertheless, the mean (median) AGN contribution to $L_{\rm IR}$ was found to be only about 1.4\% (0.2\%), which strongly supports our earlier statement that our analysed SMGs tend to be either purely SFGs or SFGs with only a minor AGN heating contribution (Sects.~2.1.2 and 3.1.1).

\begin{figure}[!htb]
\centering
\resizebox{0.9\hsize}{!}{\includegraphics{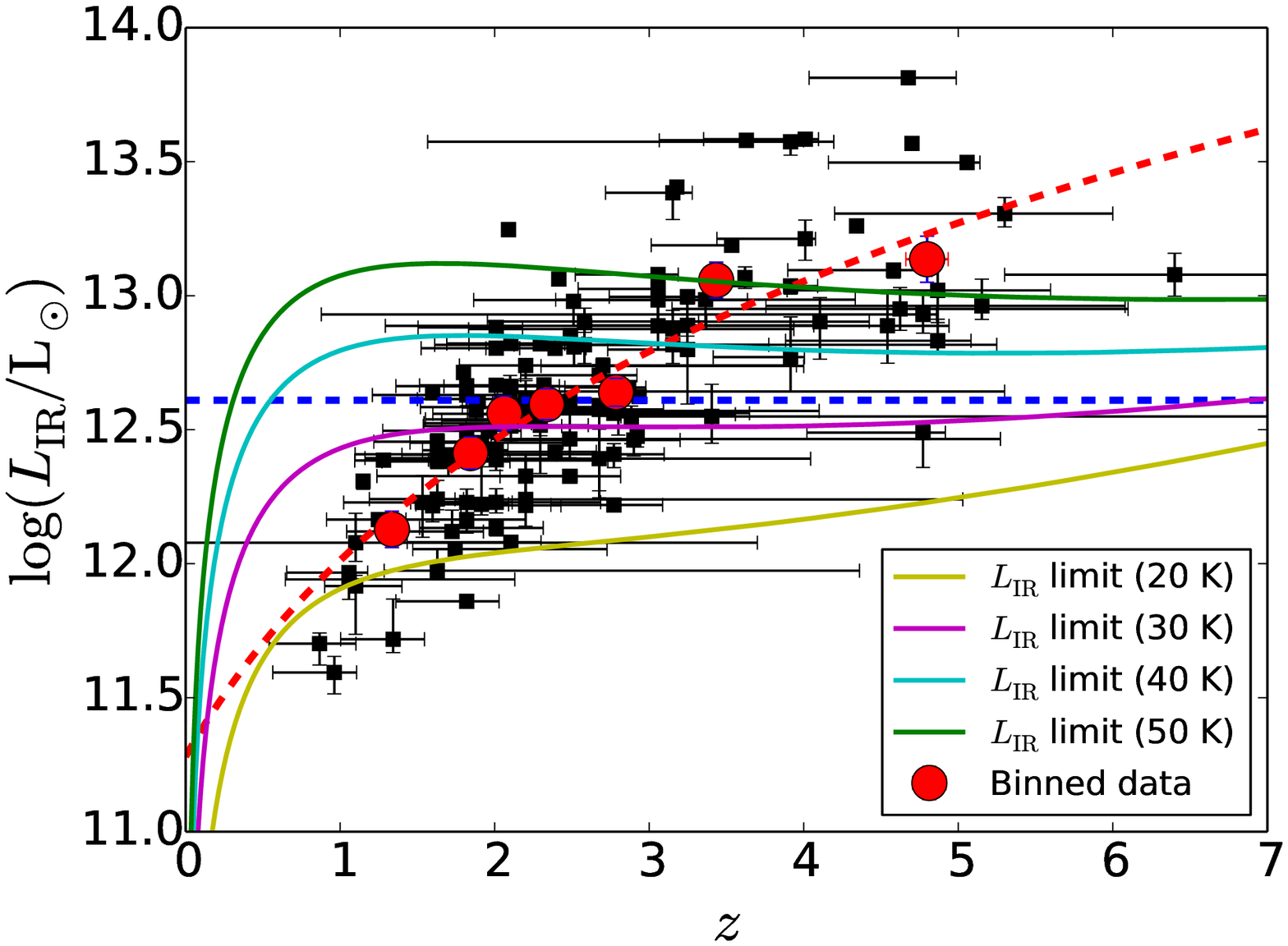}}
\resizebox{0.9\hsize}{!}{\includegraphics{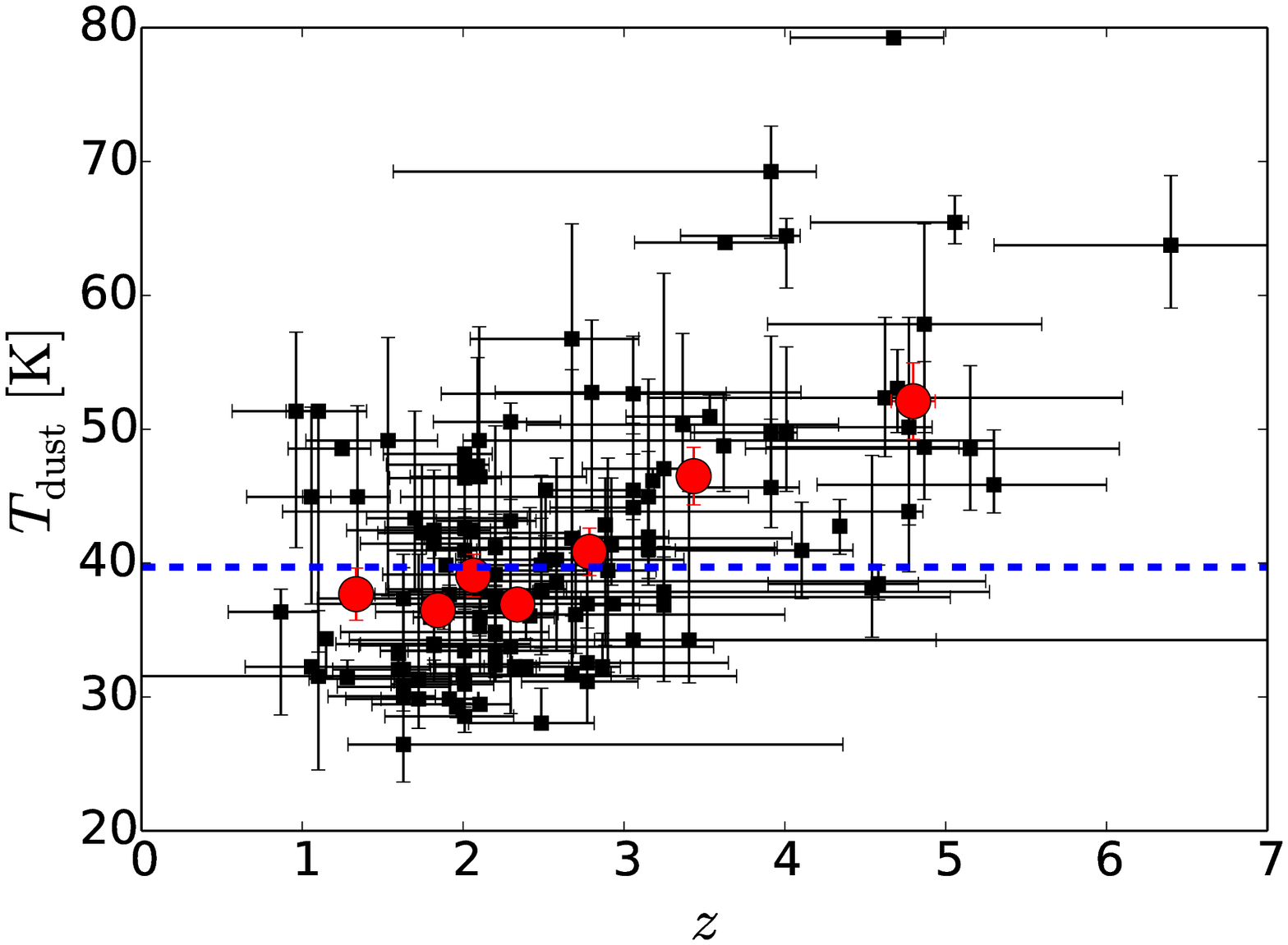}}
\caption{Infrared ($8-1\,000$~$\mu$m) luminosity (top) and $T_{\rm dust}$ (bottom) as a function of redshift. In both panels, the horizontal, blue dashed line shows the sample median ($\log(L_{\rm IR}/{\rm L}_{\sun})=12.61$ and $T_{\rm dust}=39.7$~K). The red filled circles represent the mean values of the binned data (each bin contains 18 SMGs, except the highest redshift bin, which contains 16 SMGs), with the error bars showing the standard errors of the mean values (mostly smaller than the size of the data points in the top panel). The red, dashed curve in the top panel indicates the best-fit function to the binned data, $\log(L_{\rm IR}/{\rm L}_{\sun})\propto~(1+z)^{0.09\pm0.01}$. The solid lines show the lower IR luminosity limits calculated for the $4\sigma$ flux density limit of 5~mJy at the initial AzTEC selection wavelength of 1.1~mm. These $L_{\rm IR}$ limits were computed assuming the local ULIRG template SEDs of Casey (2012), and are plotted at four different dust temperatures of 20~K (yellow), 30~K (magenta), 40~K (cyan), and 50~K (green).}
\label{figure:LIR}
\end{figure}

In the top panel in Fig.~\ref{figure:LIR}, we plot the values of $L_{\rm IR}$ as a function of redshift. The red, dashed curve overplotted in the figure represents the best-fit function to the binned, average data, and has a functional form of $\log(L_{\rm IR}/{\rm L}_{\sun})=(11.3 \pm 0.2)\times (1+z)^{0.09\pm0.01}$. This apparent behaviour of increasing $L_{\rm IR}$ with $z$ is a well-known selection effect. We note that beyond $z\sim3$, the average IR luminosities are much higher than the median luminosity, by factors of 2.8 and 3.4 for the two highest redshift bins ($\langle z \rangle=3.43$ and $\langle z \rangle=4.80$), respectively. The second highest redshift bin also lies above the best-fit curve by a factor of 1.37. For comparison, we also plot the IR luminosity detection limits, which correspond to the $4\sigma$ flux density limit of 5~mJy at the initial AzTEC selection wavelength of $\lambda_{\rm obs}=1.1$~mm (see Sect.~2.1.1; see also \cite{casey2014}; \cite{bethermin2015}). These $L_{\rm IR}$ limits, which are shown at four different representative dust temperatures of $T_{\rm dust}=20$, 30, 40, and 50~K, were computed assuming the local ultraluminous infrared galaxy (ULIRG; $L_{\rm IR}>10^{12}$~L$_{\sun}$) template SEDs of Casey (2012; see also \cite{caseyetal2012}). As illustrated in the top panel in Fig.~\ref{figure:LIR}, even our lowest redshift average data point lies above the 20~K IR luminosity limit, while the jump near $z\sim3$ can be understood as a selection bias if the sources at $z \gtrsim 3$ have higher dust temperatures ($\gtrsim40-50$~K; the cyan and green lines in the figure). In the bottom panel in Fig.~\ref{figure:LIR}, we plot the {\tt MAGPHYS}-inferred luminosity-weighted dust temperatures as a function of redshift. Although these $T_{\rm dust}$ values are not independent of the {\tt MAGPHYS}-derived dust luminosities (and hence $L_{\rm IR}$), they demonstrate the similar jump at $z\sim3$ as the IR luminosities in the top panel. More importantly, the two highest redshift bins have temperatures (46.5~K and 52.1~K) that are in good agreement with the 50~K IR luminosity limit plotted in the upper panel. This strongly supports the aforementioned remark that our $z\gtrsim3$ SMGs are likely biased towards warmer objects than at lower redshifts. In other words, our sample can lack sources, which are characterised by luminosity-weighted dust temperatures of $T_{\rm dust}\lesssim50$~K at $z\gtrsim3$.

Regarding the aforementioned selection effect related to the redshift evolution of $L_{\rm IR}$ and $T_{\rm dust}$, one might wonder whether it would influence the proportions of $L_{\rm dust}$ that emerge in the far-IR and total-IR ranges shown in Fig.~\ref{figure:fractions}. To explore this possibility, we divided our sample into two subsamples, one at $z\leq3$, and the other at $z>3$. For the former, lower-redshift sample the values of $f_{\rm FIR}$ were found to range from 0.33 to 0.95 with a mean (median) of 0.76 (0.79), while the values of $f_{\rm TIR}$ span a range of $0.91-1.0$ with both the mean and median being 0.97. For the $z>3$ subsample, the $f_{\rm FIR}$ values were derived to be $0.29-0.85$ with a mean (median) of 0.58 (0.59), while $f_{\rm TIR}$ was found to range from 0.90 to 0.99 with both the mean and median being 0.95. Hence, on average the $f_{\rm FIR}$ appears to be somewhat higher for the $z\leq3$ sources than for the $z>3$ sources, while the average $f_{\rm TIR}$ value is very similar for the two subsamples. We remind that the $f_{\rm TIR}$ values entered into our calculation of the total-IR luminosities, and are hence more relevant in our subsequent analysis.

The SFR reported in Table~\ref{table:sed} refers to a stellar mass range from $M_{\rm low}=0.1$~M$_{\sun}$ to $M_{\rm up}=100$~M$_{\sun}$, is averaged over the past $\Delta t=100$~Myr, and was calculated using the standard ${\rm SFR}(L_{\rm IR})$ relationship from Kennicutt (1998; here scaled to a Chabrier (2003) IMF)

\begin{equation} 
\label{eq:sfr}
{\rm SFR}=10^{-10}\times L_{\rm IR}[{\rm L}_{\sun}]\, {\rm M}_{\sun}~{\rm yr}^{-1}\,.
\end{equation}
This calibration relies on the starburst synthesis models of Leitherer and Heckman (1995), and it is based on the assumption of solar metallicity, and an optically thick ($\tau_{\rm dust}\gg1$) starburst region, in which case $L_{\rm IR}$ is a good proxy of the system's bolometric luminosity ($L_{\rm IR}\simeq L_{\rm bol}$), and hence a sound, calorimetric probe of the obscured, current stellar birth rate. A possible caveat is that the contribution to the dust heating by more evolved stellar populations (the cirrus component; e.g. \cite{helou1986}; \cite{lonsdale1987}; \cite{walterbros1996}) is not taken into account. If the cirrus ISM component heated by the more general galactic UV radiation field contributes to $L_{\rm IR}$, then the Kennicutt (1998) relationship overestimates the SFR. Another issue is the fact that some percentage of the UV photons can escape the starburst region without being absorbed, and hence are not reprocessed into IR photons (indeed, some of our SMGs are visible in the rest-frame UV images; \cite{miettinen2017b}). The {\tt MAGPHYS} code also gives the SFR as an output, and contrary to the aforementioned $L_{\rm IR}$ diagnostic, the model permits for the heating of the dust by older and longer-lasting stellar populations. 
We found that the ${\rm SFR}(L_{\rm IR})$ is somewhat higher on average than ${\rm SFR}_{\rm MAGPHYS}$: the ${\rm SFR}(L_{\rm IR})/{\rm SFR}_{\rm MAGPHYS}$ ratio was found to range from 0.47 to 6.92 with a median of $1.31^{+0.83}_{-0.17}$, where the $\pm$ errors represent the 16th--84th percentile range (see the corresponding panel in Fig.~\ref{figure:histograms}). If, instead of $\Delta t=100$~Myr, the aforementioned comparison is done by using the ${\rm SFR}_{\rm MAGPHYS}$ values averaged over the past $\Delta t=10$~Myr, the median ${\rm SFR}(L_{\rm IR})/{\rm SFR}_{\rm MAGPHYS}$ ratio is found to be $1.15^{+0.38}_{-0.27}$, which is consistent with the results obtained by da Cunha et al. (2015). Unless otherwise stated, in our subsequent analysis we use the SFR averaged over the past 100~Myr as calculated using Eq.~(\ref{eq:sfr}).

\subsection{Estimating the molecular gas mass} 

In the present work, we augment our SED-based physical parameter space by estimating the molecular gas masses of our target SMGs.
The mass of the molecular gas component of a galaxy is a key parameter to access many other important star formation-related 
parameters (e.g. gas fraction and gas consumption timescale), and hence to reach a better understanding of the galaxy's overall evolution. To our knowledge, only two of our target SMGs have a published CO tracer-based molecular gas mass estimate available (AzTEC/C5=AzTEC~1 (\cite{yun2015}) and AzTEC/C17=J1000+0234 (\cite{schinnerer2008})). For this reason, we estimate the gas masses of our SMGs using the long-wavelength ($\lambda_{\rm rest}\gtrsim250$~$\mu$m) dust continuum method of Scoville et al. (2016); see also Hildebrand (1983); Scoville (2013); Eales et al. (2012); Groves et al. (2015); Scoville et al. (2014, 2015); and Hughes et al. (2017). This method is based on the well-known fact that the Rayleigh-Jeans (R-J) tail of dust emission is generally optically thin ($\tau \ll 1$), and hence can be used as a direct probe of the total dust column density. However, the assumption of optically thin dust emission might not be valid for all SMGs, especially if the source is associated with a compact, nuclear starburst region, where the dust can be optically thick even at (sub-)mm wavelengths (cf.~Arp~220; e.g. \cite{scoville2017}, and references therein). In this case, the dust-based method would underestimate the true gas mass content. Nevertheless, for our ALMA 1.3~mm dust continuum measurements, we can write (cf.~Eq.~(16) in \cite{scoville2016})

\begin{equation}
\label{eq:ism}
\begin{aligned}
M_{\rm gas} &= 1.78\times10^{10}(1+z)^{-4.8}\frac{S_{\nu_{\rm obs}}}{{\rm mJy}} \left(\frac{\nu_{\rm obs}}{350\,{\rm GHz}}\right)^{-3.8} \\ 
& \qquad \times \frac{\Gamma_{\rm RJ}^{z=0}}{\Gamma_{\rm RJ}}\left(\frac{d_{\rm L}}{{\rm Gpc}}\right)^2\, {\rm M}_{\sun}\\
        & =  6.08 \times10^{10}(1+z)^{-4.8}\Gamma_{\rm RJ}^{-1}\frac{S_{\rm 1.3\, mm}}{{\rm mJy}}\left(\frac{d_{\rm L}}{{\rm Gpc}}\right)^2\, {\rm M}_{\sun}\,, 
\end{aligned}
\end{equation}
where the mass-weighted $T_{\rm dust}$ is assumed to have a constant value of 25~K (i.e. the dust mass is dominated by the cold component), $d_{\rm L}$ is the luminosity distance, and the function $\Gamma_{\rm RJ}$ is defined by

\begin{equation}
\label{eq:gamma}
\Gamma_{\rm RJ}(z,\,\nu_{\rm obs},\,T_{\rm dust})=\frac{h\nu_{\rm obs}(1+z)}{k_{\rm B}T_{\rm dust}}\frac{1}{e^{h\nu_{\rm obs}(1+z)/k_{\rm B}T_{\rm dust}}-1}\,,
\end{equation}
where $h$ is the Planck constant, and $k_{\rm B}$ the Boltzmann constant. The purpose of $\Gamma_{\rm RJ}$ is to correct for the deviation from the $S_{\nu}\propto \nu^2$ form of the R-J tail. The value of the unitless correction factor $\Gamma_{\rm RJ}^{z=0}$ is 0.7 at the reference frequency of 350~GHz. The gas masses derived using Eq.~(\ref{eq:ism}) are only weakly dependent on $T_{\rm dust}$ because the method is based on the R-J regime of the dust SED. For example, the value of $\Gamma_{\rm RJ}$ calculated using Eq.~(\ref{eq:gamma}) at the median redshift of the analysed SMGs ($z=2.30$) is 0.35, 0.44, 0.51, 0.57, and 0.61 at $T_{\rm dust}=20$, 25, 30, 35, and 40~K, respectively (cf.~Fig.~11 in \cite{scoville2016}).

In Eq.~(\ref{eq:ism}), the value of the dust emissivity index is fixed at $\beta=1.8$, which corresponds to a Galactic mean value of $\beta$ derived through observations with the \textit{Planck} satellite (\cite{planck2011}). For comparison, in the high-$z$ model libraries of {\tt MAGPHYS} we employed, the value of $\beta$ is fixed at 1.5 for the warm dust component (30--80~K), while that for the colder (20--40~K) dust is $\beta=2$. Equation~(\ref{eq:ism}) also assumes that the rest-frame 850~$\mu$m specific luminosity-to-gas mass ratio is $\alpha_{\rm 850\, \mu m}=L_{\rm 850\, \mu m}/M_{\rm gas}=(6.7 \pm 1.7) \times 10^{19}$~erg~s$^{-1}$~Hz$^{-1}$~${\rm M}_{\sun}^{-1}$ (see Eq.~(13) in \cite{scoville2016}). On the other hand, the 850~$\mu$m normalised dust opacity per unit total gas mass underlying Eq.~(\ref{eq:ism}) is $\kappa_{\rm 850\, \mu m}=\kappa_{\rm dust,\,850\, \mu m}\delta_{\rm dgr}=4.10\times10^{-3}$~cm$^2$~g$^{-1}$, where $\delta_{\rm dgr}$ is the dust-to-gas mass ratio. A canonical value of $\delta_{\rm dgr}=0.01$ for solar metallicity would imply $\kappa_{\rm dust,\,850\, \mu m}=0.41$~cm$^2$~g$^{-1}$, which is a factor of 1.878 lower than that assumed in our {\tt MAGPHYS} analysis ($\kappa_{\rm dust,\,850\, \mu m}=0.77$~cm$^2$~g$^{-1}$). It is also worth mentioning that the empirical calibration of Eq.~(\ref{eq:ism}) is partly based on a sample of 30 SMGs that lie at redshifts $z=1.44-2.96$ and have CO$(1-0)$ measurements available (\cite{scoville2016}). 

\begin{figure}[!htb]
\centering
\resizebox{0.9\hsize}{!}{\includegraphics{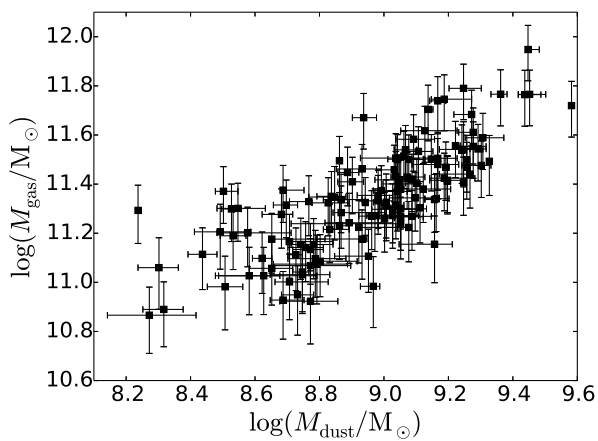}}
\caption{Molecular gas mass calculated using Eq.~(\ref{eq:ism}) plotted against the dust mass derived from {\tt MAGPHYS}.}
\label{figure:gasvsdust}
\end{figure}

Finally, we note that Eq.~(\ref{eq:ism}) was calibrated by Scoville et al. (2016) to yield the molecular gas mass rather than a total (atomic+molecular) gas mass of $M_{\rm gas}=M_{\rm HI}+M_{\rm H_2}$ as done in the Scoville et al. (2014) calibration. Indeed, for high-$z$ SMGs it is reasonable to assume that the gaseous ISM is mostly molecular, and hence $M_{\rm gas}\simeq M_{\rm H_2}$. The derived $M_{\rm gas}$ values are listed in Col.~(11) in Table~\ref{table:sed}, where the quoted formal uncertainties were propagated from the ALMA 1.3~mm flux density and $\alpha_{\rm 850\, \mu m}$ calibration constant uncertainties. Naturally, the true uncertainties in the derived gas masses are larger owing to the uncertain dust properties and calibration assumptions (see below). The values of $M_{\rm gas}$ are plotted as a function of dust mass in Fig.~\ref{figure:gasvsdust} to illustrate the clear positive correlation between the two quantities, which is built into Eq.~(\ref{eq:ism}).

Now, we turn our attention to the question how the values of $M_{\rm gas}$ compare with the CO-based gas masses available for AzTEC/C5 and C17. For the former SMG, Yun et al. (2015) derived a value of $M_{\rm gas}^{\rm CO}=(1.4\pm0.2)\times10^{11}$~M$_{\sun}$, while for the latter one Schinnerer et al. (2008) obtained a value of $M_{\rm gas}^{\rm CO}=2.6\times10^{10}$~M$_{\sun}$. Both of these values are based on CO$(J=4-3)$ measurements and the assumption that the CO-to-H$_2$ conversion factor is $\alpha_{\rm CO}=M_{\rm gas}/L_{\rm CO}'=0.8$~M$_{\sun}$~(K\,km~s$^{-1}$\,pc$^2$)$^{-1}$, where $L_{\rm CO}'$ is the CO$(1-0)$ line luminosity. Schinnerer et al. (2008) assumed that the gas is thermalised with $L_{\rm CO(1-0)}'=L_{\rm CO(4-3)}'$, while Yun et al. (2015) assumed that $L_{\rm CO(1-0)}'=2.174\times L_{\rm CO(4-3)}'$, which is based on the average SMG values compiled by Carilli \& Walter (2013). The values of $M_{\rm gas}$ we derived for AzTEC/C5 and C17, $M_{\rm gas}=(5.5\pm 1.4)\times10^{11}$~M$_{\sun}$ and $M_{\rm gas}=(3.8 \pm 1.0)\times10^{11}$~M$_{\sun}$, respectively, are $3.9\pm1.1$ and $14.6\pm 3.8$ times higher than the CO-based values. However, for a fair comparison, the aforementioned CO-based $M_{\rm gas}$ values should be scaled up by a factor of 8.125 to be consistent with a Galactic conversion factor of $\alpha_{\rm CO}=6.5$~M$_{\sun}$~(K\,km~s$^{-1}$\,pc$^2$)$^{-1}$ assumed by Scoville et al. (2016); the latter value includes a factor of 1.36 to take the contribution of helium (9\% by number) into account. In this case, we obtain the ratios $M_{\rm gas}/M_{\rm gas}^{\rm CO}=0.5\pm0.1$ and $M_{\rm gas}/M_{\rm gas}^{\rm CO}=1.8\pm0.5$ for AzTEC/C5 and C17, respectively. Hence, the two methods provide fairly similar (within a factor of two) gas masses. Besides some of the uncertain assumptions in the R-J dust continuum method (e.g. a uniform dust temperature of 25~K), the molecular gas masses derived from a single mid-$J$ transition ($J=4-3$ in our case) can suffer from significant uncertainties owing to the rotational level excitation effects. This highlights the need for larger CO$(1-0)$ surveys of SMGs (e.g.~\cite{huynh2017}). As tabulated in Col.~(10) in Table~\ref{table:sed2}, the $M_{\rm gas}$ values we estimated span from $7.4\times10^{10}$~M$_{\sun}$ to $8.9\times10^{11}$~M$_{\sun}$ with both the mean and median being $2.2\times10^{11}$~M$_{\sun}$. In an absolute sense, the major uncertainty factor in these dust-inferred gas mass estimates is the assumption of a high, Galactic $\alpha_{\rm CO}$ factor for all the sources, regardless of the redshift, metallicity, or the mode or level of star formation. We stress that if the true value of $\alpha_{\rm CO}$ is close to $0.8$~M$_{\sun}$~(K\,km~s$^{-1}$\,pc$^2$)$^{-1}$, as commonly assumed for ULIRGs and SMGs (e.g. \cite{downes1998}), then the dust-based gas masses can be overestimated by a factor of $\gtrsim8$.

\begin{figure}[!htb]
\centering
\resizebox{0.99\hsize}{!}{\includegraphics{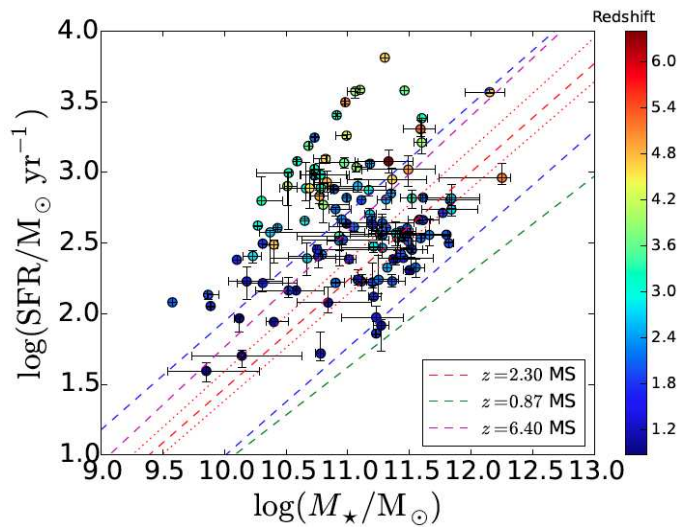}}
\caption{Main sequence diagram for our ALMA SMGs (a log-log plot of the SFR versus stellar mass). The data points are colour-coded with redshift as shown in the colourbar on the right. The red dashed line shows the mid-line position of the star-forming MS at the median redshift of the analysed SMGs ($z=2.30$) as given by Speagle et al. (2014), with the lower and upper blue dashed lines indicating a factor of three offset below and above the MS. The red, dotted lines indicate the MS mid-line boundaries of the interquartile redshift range of ${\rm IQR}(z)=3.15-1.91=1.24$. For comparison, the green and magenta dashed lines show the MS mid-lines at the lowest and highest redshifts of the analysed sources ($z=0.87$ and $z=6.40$), respectively.}
\label{figure:ms}
\end{figure}

\begin{figure}[!htb]
\centering
\resizebox{0.9\hsize}{!}{\includegraphics{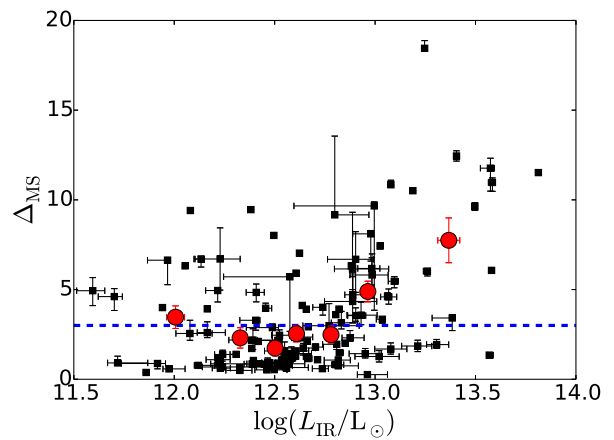}}
\caption{Starburstiness or the distance from the MS (parameterised as $\Delta_{\rm MS}={\rm SFR}/{\rm SFR}_{\rm MS}$) as a function 
of $\log(L_{\rm IR})$. The red filled circles show the binned averages as in Fig.~\ref{figure:LIR}. A MS border of $\Delta_{\rm MS}=3$ is 
indicated by a horizontal, blue dashed line.}
\label{figure:sblir}
\end{figure}

\section{Discussion}

\subsection{Star formation mode in the COSMOS ASTE/AzTEC submillimetre galaxies}

\subsubsection{The stellar mass -- star formation rate diagram: Comparison with the galaxy main sequence}

In Fig.~\ref{figure:ms}, we plot the $L_{\rm IR}$-based SFR values of our SMGs as a function of 
their stellar mass. The well-known, tight, and almost linear correlation between the SFR and $M_{\star}$ is the so-called MS of star-forming galaxies (e.g. \cite{brinchmann2004}; \cite{noeske2007}; \cite{elbaz2007}; \cite{daddi2007}; \cite{karim2011}; \cite{whitaker2012}; \cite{speagle2014}; \cite{salmon2015}), 
and it provides valuable insight into how galaxies convert their gaseous ISM into stars. 
To illustrate how our SMGs compare with the galaxy MS, in Fig.~\ref{figure:ms} we overlay 
the best fit functional form from Speagle et al. (2014), which is based on a compilation of 25 observational studies out to $z\sim6$ with different pre-selections (UV, optical, far-IR; see their Table~3), and is given by 

\begin{equation}
\label{eqn:ms}
\begin{aligned}
\log({\rm SFR}/{\rm M}_{\sun}~{\rm yr}^{-1}) &=(0.84-0.026\times \tau_{\rm univ})\log(M_{\star}/{\rm M}_{\sun})\\ 
& \qquad -(6.51-0.11\times \tau_{\rm univ})\, ,
\end{aligned}
\end{equation}
where $\tau_{\rm univ}$ is the age of the universe in Gyr. Hence, the MS evolves with redshift by shifting to higher SFRs at earlier cosmic times. We note that Speagle et al. (2014) adopted a Kroupa (2001) IMF in their calibration, which is very similar to a Chabrier (2003) IMF (see Sect.~3.1.1 in \cite{speagle2014} for discussion). 

In Fig.~\ref{figure:ms}, we plot Eq.~(\ref{eqn:ms}), that is the MS locus, at the median redshift of our analysed SMGs, $z=2.30$, and also at the lowest and highest redshifts of the analysed sources, namely at $z=0.87$ and $z=6.40$ (see \cite{brisbin2017} and \cite{miettinen2017b} for more details on the redshifts of our sources). To illustrate the midspread of our data, in Fig.~\ref{figure:ms} we plot the MS loci at the 75th and 25th percentiles of the redshifts ($z=3.15$ and $z=1.91$). Besides the mid-line locus, we also plot the factor of three (0.477~dex) lines below and above the MS at the sample median redshift of $z=2.30$. This illustrates the thickness (or scatter) of the MS, and includes both the intrinsic scatter and measurement uncertainties (see e.g. \cite{magdis2012}; \cite{dessauges2015}; \cite{mitra2017}, and references therein). The intrinsic MS scatter could reflect the physical evolution of galaxies undergoing the phases of gas compaction, depletion, replenishment, and quenching (\cite{tacchella2016}). Also, if the SFR-$M_{\star}$ scatter is (partly) a result of bursty star formation phases (\cite{sparre2017}), it could be linked to the fluctuations of the baryon accretion rate in the dark matter host halo (\cite{mitra2017}).

To quantify the offset from the MS mid-line, we calculated the ratio of the derived SFR to that expected for a MS galaxy of the same redshift and $M_{\star}$, that is $\Delta_{\rm MS}\equiv{\rm SFR}/{\rm SFR}_{\rm MS}$. The values of this ratio range from $0.3$ to $18.4$ with a median of $2.2^{+4.2}_{-1.3}$ (see column~(7) in Table~\ref{table:sed2}, and column~(8) in Table~\ref{table:sed}). Fifty-two out of the 124 SMGs ($41.9\%$ with a Poisson counting error of $5.8\%$) analysed here lie above the $\Delta_{\rm MS}=3$ border, the most significant outlier being AzTEC/C113. In keeping with the studies by da Cunha et al. (2015) and Miettinen et al. (2017a), we define these sources as starbursts, but we note that different definitions exist in the literature. For example, Elbaz et al. (2011) defined a galaxy to be a starburst if $\Delta_{\rm MS}={\rm sSFR}/{\rm sSFR}_{\rm MS} \geq 2$. Bauermeister et al. (2013) adopted a larger offset of $\Delta_{\rm MS}>4$ for their starburst galaxies, while Cowley et al. (2017a) set the MS-starburst border at $\Delta_{\rm MS}=10$ in their galaxy formation model.   

Seventy-one of our SMGs ($57.3\%\pm6.8\%$) have $1/3 < \Delta_{\rm MS} \leq3$, and hence lie within the MS. One of the sources, AzTEC/C107 at $z_{\rm phot}=5.15$, appears to lie just below the lower boundary of the MS, but the shape of the MS is uncertain at this high redshift, although we note that Tasca et al. (2015) found that the $\log({\rm SFR})-\log(M_{\star})$ relationship remains roughly linear up to $z=5$ (while the normalisation increases with redshift). In addition, the stellar mass of AzTEC/C107 is poorly constrained because it is based only on upper flux density limits at the rest-frame UV; if the true stellar mass of AzTEC/C107 is lower than the estimated value, it would move onto the MS. 

The aforementioned results are consistent with previous studies where some 
of the SMGs are found to be located on or close to the MS 
(at the high-$M_{\star}$ end), while a fair percentage of SMGs, especially the most luminous objects, are found to lie above the MS 
(e.g. \cite{magnelli2012}; \cite{michalowski2012}; \cite{roseboom2013}; \cite{dacunha2015}; \cite{koprowski2016}; \cite{miettinen2017a}). 
Our large, flux-limited sample strongly supports the view that SMG populations exhibit two different types of star formation modes, 
namely a steady conversion of gas into stars as in normal (non-starburst) star-forming disk galaxies, and star formation in a more violent, 
bursty event, which is likely triggered by major mergers. This type of bi-modality is also seen in hydrodynamic simulations 
(e.g. \cite{hayward2011}, 2012, and references therein).

To illustrate how the source IR brightness influences its position in the $M_{\star}$ -- SFR plane, we plot $\Delta_{\rm MS}$ against $\log(L_{\rm IR})$ in Fig.~\ref{figure:sblir}. The binned average data demonstrate how the fainter SMGs ($L_{\rm IR}\lesssim6\times10^{12}$~L$_{\sun}$ on average) are preferentially found within the MS, while the brightest SMGs ($L_{\rm IR}\gtrsim9\times10^{12}$~L$_{\sun}$ on average) typically lie above the MS, and hence are starbursts. This is consistent with the model of Hopkins et al. (2010), which predicts that at $z \sim 2-3$, merger-driven starbursts dominate sources with $L_{\rm IR} \gtrsim 6 \times 10^{12}$~L$_{\sun}$, while at lower IR luminosities, normal star-forming disk galaxies dominate.

\begin{figure*}
\begin{center}
\includegraphics[width=0.90\textwidth]{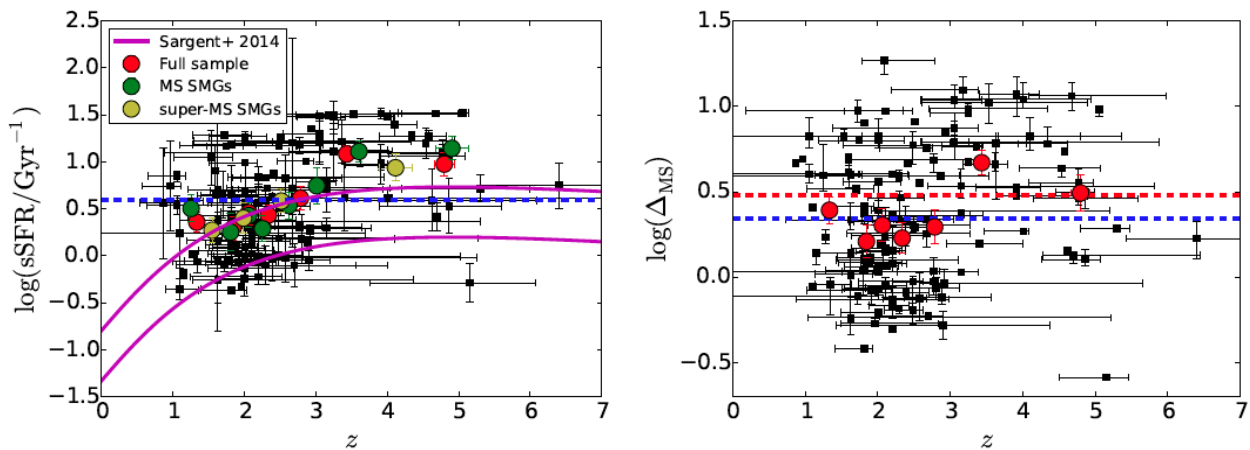}
\caption{Specific SFR (sSFR; left panel) and starburstiness (right panel) as a function of redshift. In both panels, the red filled circles represent the mean values of the binned data for the full sample as in Fig.~\ref{figure:LIR}. In the left panel, the green and yellow filled circles represent the mean values of the binned data for the MS SMGs and starburst SMGs, respectively. For the MS sample, each bin contains nine sources, except the highest redshift bin, which contains eight SMGs. For the starburst sample, each bin contains 13 sources. The blue horizontal dashed lines mark the sample median values of ${\rm sSFR}=3.9$~Gyr$^{-1}$ and $\Delta_{\rm MS}=2.2$. The magenta curves in the left panel represent the ${\rm sSFR}(M_{\star},\,z)$ relationship of Sargent et al. (2014; Eq.~(A1) therein), which is here plotted for our minimum and maximum stellar mass values (upper and lower curves, respectively). In the right panel, the red dashed line shows the upper boundary of the MS at $\Delta_{\rm MS}=3$.}
\label{figure:sSFR}
\end{center}
\end{figure*}

\subsubsection{Redshift evolution of the specific star formation rate and starburstiness}

In the left panel in Fig.~\ref{figure:sSFR}, we plot the sSFR, which reflects 
the strength of the current star formation activity with 
respect to the underlying galaxy stellar mass, as a function of redshift. 
As illustrated by the binned version of the data, 
the sSFR appears to increase as a function of redshift from $z\sim1$ to $z\sim3$ with a jump at $z\sim3$ (similar to that found by Miettinen et al. (2017a) for their sample of AzTEC SMGs in COSMOS), and then showing a plateau at $z\gtrsim3$. The jump at $z\sim3$ is likely caused by the selection bias discussed in Sect.~3.1.2 and illustrated in Fig.~\ref{figure:LIR}. 

In Fig.~\ref{figure:sSFR}, we also plot the stellar mass and redshift-dependent ${\rm sSFR}(M_{\star},\,z)$ relationship of Sargent et al. (2014; their Eq.~(A1)), which the authors derived using observational results from earlier studies out to $z<7$ (the majority of the data probed MS galaxies out to $z\sim3$). Our binned average data points at $z <3$, including the MS and starburst SMGs, are mostly consistent with the Sargent et al. (2014) relationship plotted for the lowest stellar mass in our sample (9.58 in log-10 solar units). At $z>3$, all binned averages lie above the Sargent et al. (2014) relationship (see above), but they exhibit a similar flattening towards higher redshifts as expected from the overplotted relationship. 

An increasing sSFR towards higher redshifts has been seen in several previous studies as well, which include large samples of SFGs selected from multiple fields over wide redshift ranges up to $z\sim6$ (e.g. \cite{feulner2005}; \cite{karim2011}; \cite{weinmann2011}; \cite{reddy2012}; \cite{tasca2015}; \cite{schreiber2015}; \cite{faisst2016}; \cite{koprowski2016}). Specifically, it has been found that the sSFR exhibits a very steep rise from $z=0$ to $z\sim2$, while it starts to plateau, or saturate at $z\gtrsim2$ (see also \cite{madau2014} for a review). 

A positive ${\rm sSFR}(z)$ evolution is likely to reflect the higher molecular gas masses and densities at earlier cosmic times (e.g. \cite{dutton2010}; \cite{saintonge2016}; \cite{tacconi2017}, and references therein). The physics behind this is likely governed by the specific cosmological accretion rate of baryons onto dark matter haloes, which is a steep function of redshift, namely $\dot{M}_{\rm halo} /M_{\rm halo} \propto (1+z)^{2.25}$ (\cite{neistein2008}; \cite{dekel2009a},b; \cite{bouche2010}). In the large, cosmological, smoothed particle hydrodynamics simulations by van de Voort et al. (2011), the cold-mode accretion rate density, which was defined to have a gas temperature of $T\lesssim 3.2\times 10^5$~K, peaks at $z \approx 3$, and then declines rapidly at lower redshifts. Assuming that sSFR tracks cold-mode accretion, the van de Voort et al. (2011) simulations agree with our finding of a jump in sSFR near $z\sim3$ (albeit being also a selection effect in the present study), and suggests only a short timescale (${\rm sSFR}^{-1}$) for the ISM to convert into stars. More specifically, the evolution of sSFR might be regulated by the angular momentum of the accreted gas: if the latter is higher at lower redshifts ($z\lesssim2$), the accreted gas tends to settle in the outer parts of the galactic disk, which results in a lower gas surface density of accreted gas, and hence lower sSFR (\cite{lehnert2015}). 

In the right panel in Fig.~\ref{figure:sSFR}, we plot the starburstiness as a function of redshift. The observed average behaviour is similar to that seen in the left panel, where the sSFR represents the normalisation of the MS at a given stellar mass, while the $\Delta_{\rm MS}$ in the right panel is the offset from the MS mid-line. On average, our SMGs at $z > 3$ have $\Delta_{\rm MS}>3$, while at $z <3$ the SMGs are typically consistent with the MS. As discussed above, the abrupt jump at $z\sim3$ is likely to reflect the sensitivity limits of our dust continuum data. Nevertheless, it is also possible that a boosted mode of star formation operates at $z \gtrsim 3$, and is possibly driven by mergers (\cite{khochfar2011}). A viable physical reason behind this trend is that when the fractional abundance of cold gas in galaxies decreases (Sect.~4.3), the occurrence of gas-rich major mergers that can trigger starbursts should also decrease.

\subsubsection{Do the different star formation modes exhibit different gas depletion times ?}

The gas depletion timescale, or the so-called Roberts time (\cite{kennicutt1994}), is given by 

\begin{equation}
\label{eqn:taudep}
\tau_{\rm dep}=\frac{M_{\rm gas}}{{\rm SFR}}\,.
\end{equation}
The Roberts time for gas depletion is built on the assumptions of 
a constant SFR and the simple closed-box model, in which no gas is either 
lost owing to galactic outflows or accreted by the galaxy. 

Using Eq.~(\ref{eqn:taudep}), we derived a wide range of gas depletion times for our full sample of SMGs, namely $\tau_{\rm dep}\sim30-5\,566$~Myr, with a median of $\sim535$~Myr. However, if part of the gaseous ISM is ejected by outflows with a rate comparable to the SFR, $\tau_{\rm dep}$ could be a factor of two shorter (e.g. \cite{tadaki2017}).

\begin{figure}[!htb]
\centering
\resizebox{0.9\hsize}{!}{\includegraphics{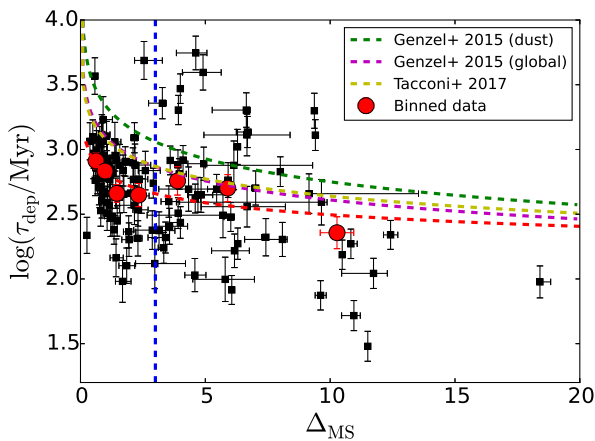}}
\caption{Gas depletion timescale as a function of $\Delta_{\rm MS}={\rm SFR}/{\rm SFR}_{\rm MS}$. The red filled circles represent the mean values of the binned data as in Fig.~\ref{figure:LIR}. A MS border of $\Delta_{\rm MS}=3$ is indicated by a vertical, blue dashed line. The red, dashed curve indicates the best-fit function to the binned data, $\log(\tau_{\rm dep}/{\rm Myr})=(2.82\pm0.04)\times \Delta_{\rm MS}^{-(0.053\pm0.016)}$. The green and magenta dashed curves show the relationships derived by Genzel et al. (2015) for the dust-based and CO plus dust-based (global) $\tau_{\rm dep}$ values, respectively. The yellow dashed line shows the best-fit relation from Tacconi et al. (2017), which almost overlaps with the global Genzel et al. (2015) relation.}
\label{figure:lifetime}
\end{figure}

To see how our data compare with the expectation that fainter, MS SMGs have a longer lifetime than their brighter, starburst counterparts (e.g. \cite{chen2016b}, and references therein), we plot the gas depletion timescale against $\Delta_{\rm MS}$ in Fig.~\ref{figure:lifetime}. The red, dashed curve overplotted in this figure represents the best-fit function to the binned, average data, and is given by 

\begin{equation}
\label{eqn:taudepeq}
\log(\tau_{\rm dep}/{\rm Myr})=(2.82\pm0.04)\times \Delta_{\rm MS}^{-(0.053\pm0.016)}\,.
\end{equation}
The Spearman rank-order correlation coefficient between the nominal binned data point values is $\rho=-0.71$, which is indicative of a fairly strong negative correlation. Although the slope in Eq.~(\ref{eqn:taudepeq}) is shallow and only of $3.3\sigma$ significance, for the MS sources ($\Delta_{\rm MS}<3$) we see a mean trend of decreasing $\tau_{\rm dep}$ when the distance from the MS mid-line increases. Above the MS ($\Delta_{\rm MS}>3$), our data points show more scatter, which is manifested by the nominal values of the last three binned averages, which do not lie on the red curve described by Eq.~(\ref{eqn:taudepeq}).

For comparison, in Fig.~\ref{figure:lifetime} we also overplot the relationships from Genzel et al. (2015), which they derived for a large sample of $\sim500$ SFGs at $z\sim0-3$ by using the \textit{Herschel} dust-based molecular gas depletion timescales and combined CO and dust data sets to calculate $\tau_{\rm dep}$ (see their Table~3 for the functional forms). However, a direct comparison with our result is complicated by the different methods of analysis applied by Genzel et al. (2015). For example, they derived the dust masses using the Draine \& Li (2007) dust models, and the dust masses were converted to gas masses using a metallicity-dependent dust-to-gas ratio. On the other hand, their CO-inferred gas masses were derived by scaling a Galactic $\alpha_{\rm CO}$ conversion factor of 4.36~M$_{\sun}$~(K\,km~s$^{-1}$\,pc$^2$)$^{-1}$ with a metallicity-dependent factor. Depending on the source, the SFR in Genzel et al. (2015) was calculated by using the Kennicutt (1998) IR indicator (similar to us), the rest-frame UV plus IR luminosities, or the UV-optical SED fits. Indeed, both the Genzel et al. (2015) relationships shown in Fig.~\ref{figure:lifetime} have a higher normalisation compared to our average fit, and they also have steeper slopes than derived here. Besides the different methods used in the analysis, these differences are likely caused by the fact that Genzel et al. (2015) focused on near-MS galaxies and their sample was much larger than ours (57\% of our SMGs lie within the MS). We also plot the best-fit relation derived by Tacconi et al. (2017), which was updated from the work by Genzel et al. (2015) by adding new CO data (making the CO-detected SFG sample size to be 650) and new dust observations. The best-fit Tacconi et al. (2017; see their Table~3) relationship plotted in Fig.~\ref{figure:lifetime} refers to the same Speagle et al. (2014) MS prescription as adopted here, and it basically overlaps with the global relation from Genzel et al. (2015).

For the MS SMGs, the estimated depletion times range from $\tau_{\rm dep}\sim96$~Myr to $\sim4.9$~Gyr, with a median of 644~Myr. For the super-MS SMGs, $\tau_{\rm dep}$ is found to lie in the range of about 30~Myr--5.6~Gyr, with a median of 407~Myr. We note that the three longest depletion times, $\sim3.9$, $\sim4.9$, and $\sim5.6$~Gyr, are found for SMGs at $\Delta_{\rm MS}=2.5$, 4.6, and 4.9, respectively. The estimated gas masses of these sources, $\sim(1.5-5.8)\times10^{11}$~M$_{\sun}$, are 0.7 to 2.7 times the sample median gas mass, and hence their very long depletion timescales are the result of their relatively low 
estimated SFRs, $\sim39-120$~M$_{\sun}$~yr$^{-1}$. However, the exceptionally long depletion time values suggest that either the molecular gas masses are overestimated for these sources, or the SFRs are underestimated, or both. On the other hand, the strongest starbursts with $\Delta_{\rm MS}>10$ systematically exhibit very short depletion times of only 30--220~Myr, in agreement with the results of B\'ethermin et al. (2015) for their similarly strong starbursts in COSMOS (see their Fig.~10). Hence, our results are broadly consistent with the expected picture of MS galaxies forming stars longer than super-MS galaxies. However, this picture is an oversimplification because we are not considering the gaseous inflows. The cold-mode accretion discussed in Sect.~4.1.2 can act as a source of continuous supply of fresh gas, and enable high SFRs ($>100$~M$_{\sun}$~yr$^{-1}$) for much longer periods of time.

\subsection{Relative mass contents of gas, dust, and stars, and their redshift evolution}

In this section, we compare the mass contents of the ISM (gas and dust) and stellar components. We stress that the gas masses, which were 
estimated from the dust continuum emission, can suffer from large uncertainties, particularly owing to the critical assumption of a uniform, 
Galactic CO-to-H$_2$ conversion factor (factor of $\gtrsim8$ uncertainty in $M_{\rm gas}$ owing to the uncertainty of $\alpha_{\rm CO}$ alone). 
On the other hand, the derived stellar masses have a systematic uncertainty factor of $\sim2$ owing to the uncertain (and possibly varying) 
stellar IMF (we assumed a Chabrier (2003) IMF). Moreover, regarding the redshift evolution we explore below, 
it should be kept in mind that our data appear to be subject to the selection effect illustrated in Fig.~\ref{figure:LIR}, that is the $z\gtrsim3$ 
sources tend to be warmer than the lower-redshift sources, which can also bias the corresponding ISM mass estimates.

\begin{figure*}
\begin{center}
\includegraphics[width=0.45\textwidth]{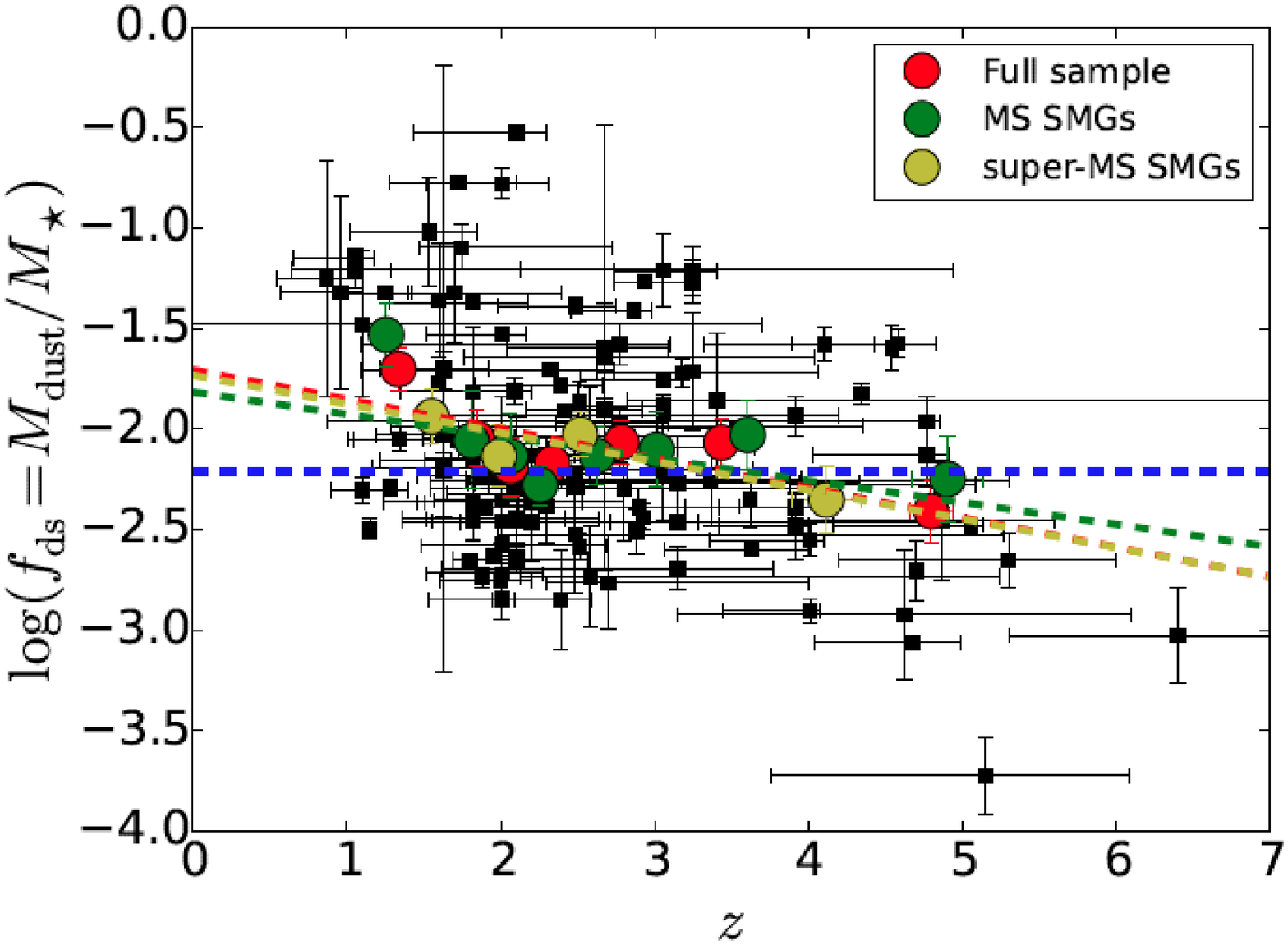}
\includegraphics[width=0.45\textwidth]{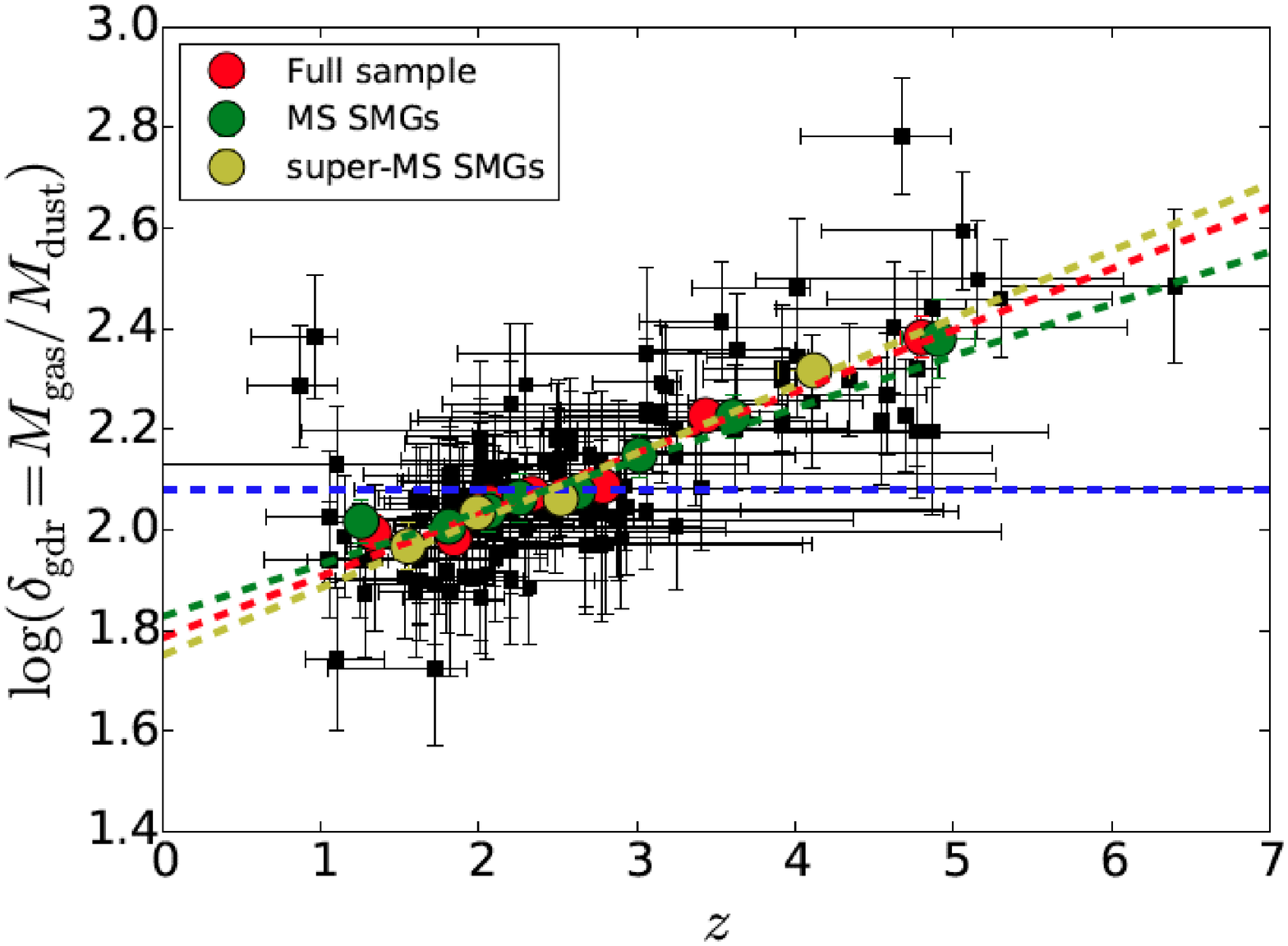}
\caption{Dust-to-stellar mass ratio (left) and gas-to-dust mass ratio (right) as a function of redshift. The red, green, and yellow filled circles show 
the binned averages as in the left panel in Fig.~\ref{figure:sSFR}. In the left panel, the blue horizontal dashed line marks 
the sample median $M_{\rm dust}/M_{\star}$ ratio of 0.006, while that in the right panel shows the median $M_{\rm gas}/M_{\rm dust}$ 
ratio of 120. The other dashed lines show the linear fits through the binned data points (see text for details).}
\label{figure:dusttostellar}
\end{center}
\end{figure*}

In Fig.~\ref{figure:dusttostellar}, we show the dust-to-stellar and gas-to-dust mass ratios as a function of redshift (left and right panel, respectively). The $f_{\rm ds}\equiv M_{\rm dust}/M_{\star}$ ratio, that is the specific dust mass, is found to span a wide range from $1.9\times10^{-4}$ to 0.30, with a mean (median) of 0.019 (0.006). The binned data suggest that, on average, $f_{\rm ds}$ decreases towards earlier epochs by factors of about 5.1, 5.2, and 2.6 for the full sample, MS SMGs, and super-MS SMGs over the redshift range studied here, respectively. For example, considering the full sample, the value of  $\langle f_{\rm ds}\rangle$ drops from 0.02 at $\langle z \rangle =1.34$ to $\langle f_{\rm ds}\rangle =0.004$ at $\langle z\rangle =4.80$.

The linear least squares fits through the binned data points overplotted in the left panel in Fig.~\ref{figure:dusttostellar} yielded $\log(f_{\rm ds})=-(0.15 \pm 0.06)\times z - (1.70\pm0.17)$, $\log(f_{\rm ds})=-(0.11 \pm 0.09)\times z - (1.82\pm0.25)$, and $\log(f_{\rm ds})=-(0.14 \pm 0.06)\times z - (1.73\pm0.14)$ for the full sample, MS SMGs, and super-MS SMGs, respectively. The corresponding Pearson correlation coefficients are $r=-0.76$, $r=-0.54$, and $r=-0.90$. 
Because the aforementioned slopes deviate from a flat trend by only $1.2\sigma-2.5\sigma$, the redshift evolutions cannot deemed to be statistically significant. By analysing the redshift dependencies of $M_{\rm dust}$ and $M_{\star}$ separately, we found that the trend shown in the left panel in Fig.~\ref{figure:dusttostellar} is mostly driven by a decreasing dust mass towards higher redshifts ($\log(M_{\rm dust}/{\rm M}_{\sun})\propto -(0.08\pm0.03)\times z$ with $r=-0.92$ for the full sample), while the average stellar mass is fairly constant as a function of redshift ($\log(M_{\star}/{\rm M}_{\sun})\propto (0.04\pm0.06)\times z$ with $r=0.43$ for the full sample). The observed increase of $M_{\rm dust}$ towards lower redshifts could be an indication of an elevated metal production, and hence more efficient dust formation, or changing relative proportions between heavy element enrichment, grain growth, and the dust destruction efficiency. As discussed by B\'ethermin et al. (2015, and references therein), the gas-phase metallicity, $Z_{\rm gas}$, decreases towards higher redshifts, and hence $M_{\rm dust}\propto Z_{\rm gas}\times M_{\rm gas}$ can also be expected to decrease as a function of redshift as we found. 

One potential caveat to our analysis of the redshift evolution of $M_{\rm dust}$ is the assumption of a fixed, redshift-independent dust opacity in the calculation of $M_{\rm dust}$. However, the dust properties, such as opacity, are likely to change as a function of metallicity (e.g. \cite{remy2014}; \cite{bate2014}). Following the above discussion, a linear dependence of the dust opacity on metallicity, $\kappa_{\rm dust}\propto Z_{\rm gas}$, would imply a lower opacity, and hence higher $M_{\rm dust}\propto \kappa_{\rm dust}^{-1}$ at higher redshifts as derived from optically thin dust emission (the other dust parameters being unchanged). On top of this effect, an increasing dust temperature towards higher redshifts would act to make the dust masses lower at higher redshifts. These complicating factors should be borne in mind when interpreting the aforementioned behaviour of $M_{\rm dust}(z)$.

Regarding the behaviour of the MS SMGs' $f_{\rm ds}$ shown in the left panel in Fig.~\ref{figure:dusttostellar}, one physical interpretation is that the dust-to-stellar mass ratio has not evolved much from the earliest SMGs to their main cosmic epoch at $z\sim2$ (a factor of $1.55$ increase from $\langle z\rangle =4.90$ to $\langle z\rangle =1.81$), but lower redshift ($z < 2$) SMGs start to have elevated dust-to-stellar mass ratios.  

B{\'e}thermin et al. (2015) found that their strong COSMOS starbursts ($\Delta_{\rm MS}>10$) exhibit higher values of $f_{\rm ds}$ than their MS sample (typically by a factor of five), and that the former population exhibits a postive evolution of $f_{\rm ds}$ with redshift, which is opposite to our super-MS SMGs' behaviour. On the other hand, the MS galaxies of B{\'e}thermin et al. (2015) showed an increase in $f_{\rm ds}$ up to $z\sim1$, and flattening towards higher redshifts. Similarly, Calura et al. (2017) found that $f_{\rm ds}$ of SFGs increases from $z\sim0$ to $z\sim1$, followed by a roughly flat $f_{\rm ds}(z)$ out to $z\sim2.5$. The flattening of $f_{\rm ds}(z)$ at $z>1$ found by B{\'e}thermin et al. (2015) and Calura et al. (2017) is broadly consistent with our result shown in Fig.~\ref{figure:dusttostellar}. We note that the stellar masses in B{\'e}thermin et al. (2015) were derived using the {\tt Le PHARE} SED code (\cite{arnouts1999}; \cite{ilbert2006}), while their dust masses were derived from the Draine \& Li (2007) dust model SEDs. The stellar masses in the analysis by Calura et al. (2017) were derived using the older version of {\tt MAGPHYS} (\cite{dacunha2008}), while their dust masses were calculated through fitting the source SEDs with a modified blackbody (MBB) function.

The derived gas-to-dust ratios range from $\delta_{\rm gdr}=53$ to 606, with a mean (median) value of 141 (120). To calculate the $\delta_{\rm gdr}$ ratio, we took into account the different values of the dust opacities used to derive our {\tt MAGPHYS}-based dust masses compared to the calculation of the gas masses (Sect.~3.2). As can be seen in the right panel in Fig.~\ref{figure:dusttostellar}, $\delta_{\rm gdr}$ appears to increase towards higher redshifts, and as illustrated by the red filled circles in the plot, the gas-to-dust ratio rises above the full sample median at $z\sim3$. A linear least squares fit through the binned data gives $\log(\delta_{\rm gdr})=(0.12\pm0.01)\times z + (1.79\pm0.04)$ with a Pearson $r$ of 0.98 for the full sample. For the MS and super-MS SMG populations we derived $\log(\delta_{\rm gdr})=(0.10\pm0.01)\times z + (1.83\pm0.03)$ with $r=0.98$ and $\log(\delta_{\rm gdr})=(0.13\pm0.02)\times z + (1.75\pm0.05)$ with $r=0.99$, respectively. Again, to unravel the correlation, we individually checked the behaviours of the gas and dust masses as a function of redshift, and we found that the $\delta_{\rm gdr}(z)$ trend is mostly driven by the aforementioned decreasing dust mass towards earlier times, while $M_{\rm gas}$ shows a jump at $z\sim3$, which explains the aforementioned rise in $\delta_{\rm gdr}$ at $z\sim3$. Because the gas masses were calculated from the ALMA 1.3~mm dust continuum flux densities, the jump at $z\sim3$ is likely to reflect the sensitivity limit discussed in Sect.~3.1.2 (Fig.~\ref{figure:LIR}). As mentioned above, the dust and gas masses depend on each other via the gas-phase metallicity, which drops towards higher redshifts, and consequently the $\delta_{\rm gdr}\propto Z_{\rm gas}^{-1}$ increases with redshift (\cite{bethermin2015}, and references therein). 

\begin{figure}[!h]
\centering
\resizebox{0.9\hsize}{!}{\includegraphics{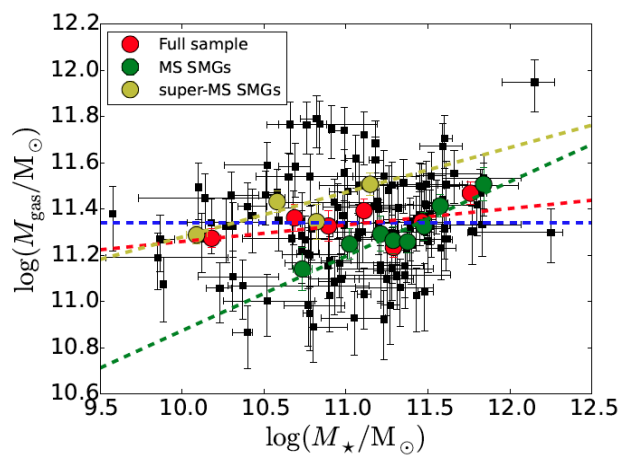}}
\caption{Gas mass plotted against the stellar mass. The red, green, and yellow filled circles show the binned mean values  as in the left panel in Fig.~\ref{figure:sSFR}. The horizontal dashed line represents the sample median gas mass of $M_{\rm gas}=2.2\times10^{11}$~M$_{\sun}$. The other dashed lines show the linear fits through the binned data points (see text for details).}
\label{figure:ismvsstar}
\end{figure}

Finally, in Fig.~\ref{figure:ismvsstar}, we show a plot of $M_{\rm gas}$ versus $M_{\star}$. The binned, full sample averages exhibit a hint of a mild, positive correlation, which could be a manifestation of an ongoing accretion of cold gas from the IGM onto our massive SMGs (e.g. \cite{saintonge2016}). However, a linear least squares fit through the mean values yields $\log(M_{\rm gas}/{\rm M}_{\sun})\propto (0.07\pm0.05)\times \log(M_{\star}/{\rm M}_{\sun}) +(10.55\pm0.59)$ ($r=0.51$), which indicates only a very weak, statistically insignificant ($1.4\sigma$) positive correlation. At least partly, this approximate average constancy of the gas mass as a function of $M_{\star}$ can be the result of our flux-limited sample selection, together with the fact that the gas masses were estimated from the observed-frame 1.3~mm flux density, which cause the derived $M_{\rm gas}$ values to span one decade (1.08~dex) from $10^{10.87}$ to $10^{11.95}$~M$_{\sun}$. 

On the other hand, when the sample is split into subsamples of MS and super-MS SMGs, the correlations are found to be stronger. For the former population we derived $\log(M_{\rm gas}/{\rm M}_{\sun})\propto (0.32\pm0.06)\times \log(M_{\star}/{\rm M}_{\sun}) +(7.65\pm0.72)$ ($r=0.94$), and for the latter we obtained $\log(M_{\rm gas}/{\rm M}_{\sun})\propto (0.19\pm0.06)\times \log(M_{\star}/{\rm M}_{\sun}) +(9.34\pm0.66)$ ($r=0.81$). Indeed, because the stellar mass is positively correlated with SFR (Sect.~4.1.1), and the SFR is higher at higher gas masses (Sect.~4.4), a positive correlation between $M_{\rm gas}$ and $M_{\star}$ is to be expected (see also e.g. \cite{sargent2014}; \cite{schinnerer2016}). 

The aforementioned $M_{\rm gas}$--$M_{\star}$ correlation is the strongest for our MS SMGs, which represent the majority ($\sim57\%$) of our source sample. This result is consistent with the view that for MS SMGs high SFRs can be sustained over long timescales owing to cold gas accretion from the filamentary streams of the cosmic web (e.g. \cite{keres2005}; \cite{dekel2009a}; \cite{brooks2009}; \cite{dave2010}). If SMGs reside predominantly in dark matter haloes of mass $M_{\rm halo}\sim10^{11.5}h^{-1}-10^{12}h^{-1}$~M$_{\sun}$ (e.g. \cite{cowley2017b}), accretion of cold, unshocked gas can indeed be expected (e.g. \cite{birnboim2003}; \cite{dekel2006}). However, the clustering measurements of SMGs by Chen et al. (2016b) suggest that SMGs tend to live in haloes more massive than a critical mass scale of $M_{\rm halo,\, crit}\sim10^{12}$~M$_{\sun}$ above which virial shock-heating of the inflowing gas emerges. On the other hand, the finding that the median $M_{\rm gas}/M_{\star}$ ratio for our full sample, MS SMGs, and starburst SMGs is about 1.6, 0.8, and 5.7, respectively, that is of order unity or higher, suggests that the gas could indeed flow cold towards the galactic disk, and hence act as a very efficient channel of providing gas to SMGs (e.g. \cite{khochfar2011}).

\subsection{Gas fraction and its redshift evolution}

Another useful parameter we can derive from the gas and stellar masses is the gas fraction 

\begin{equation}
f_{\rm gas}=\frac{M_{\rm gas}}{M_{\rm gas}+M_{\star}}=\frac{1}{1+({\rm sSFR}\times \tau_{\rm dep})^{-1}}\,,
\label{eq:gasfraction}
\end{equation}
where the gas depletion timescale is given by Eq.~(\ref{eqn:taudep}). According to Eq.~(\ref{eq:gasfraction}), $M_{\rm gas}\geq M_{\star}$ when $f_{\rm gas}\geq0.5$. For our full sample of SMGs, we derived the gas fractions in the range of $f_{\rm gas}=0.10-0.98$ with both the mean and median being $\langle f_{\rm gas}\rangle =0.62$. Hence, on average the gas mass estimated from the observed-frame 1.3~mm dust continuum emission (Sect.~3.2) exceeds the stellar mass, but we recall that some of our gas masses can be overestimated, which would explain some of the extreme gas fractions of near unity. If we split the sample into MS and super-MS objects, the values of $f_{\rm gas}$ are found to lie in the range of $0.24-0.89$ (mean 0.49, median 0.46) and $0.46-0.98$ (mean 0.81, median 0.85) for the two populations, respectively. Hence, the starburst SMGs have on average a factor of 1.65 times higher gas fraction than the MS SMGs.

\begin{figure}[!h]
\centering
\resizebox{0.9\hsize}{!}{\includegraphics{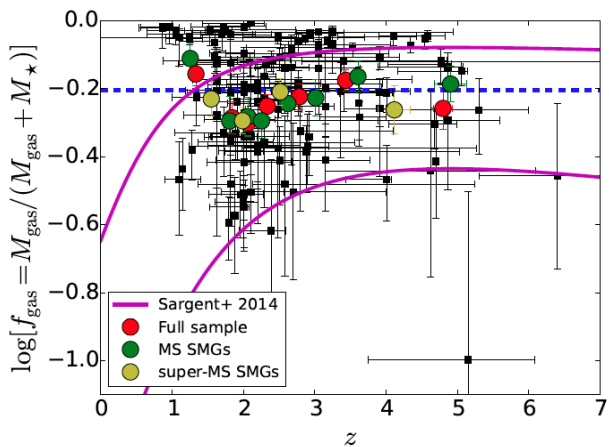}}
\caption{Gas fraction as a function of redshift. The red, green, and yellow filled circles show the binned averages as in the left panel in Fig.~\ref{figure:sSFR}. The blue horizontal dashed line marks the full sample median gas fraction of $f_{\rm gas}=0.62$. The magenta curves represent the $f_{\rm gas}(z)$ relationship from Sargent et al. (2014), which is here plotted for our minimum and maximum stellar mass values (see text for details).}
\label{figure:gasfraction}
\end{figure}

In Fig.~\ref{figure:gasfraction}, we plot the gas fraction against redshift. For comparison, we also plot the redshift evolution of $f_{\rm gas}$ for normal MS galaxies predicted by Eq.~(26) of Sargent et al. (2014), which is based on the positive correlation of SFR with both the stellar and molecular (H$_2$) gas masses. In Fig.~\ref{figure:gasfraction}, we show the $f_{\rm gas}$ curves for our minimum and maximum stellar mass values (the upper and lower curves, respectively). 
 
As can be seen in Fig.~\ref{figure:gasfraction}, most of our average data points tend to be bracketed by the maximum and minimum stellar mass $f_{\rm gas}(z)$ curves. The two exceptions are the lowest redshift bin of the full and MS samples ($\langle z \rangle=1.34$, $\langle f_{\rm gas} \rangle=0.70$ and $\langle z \rangle=1.26$, $\langle f_{\rm gas} \rangle=0.77$), both of which lie above our minimum stellar mass curve. The reason for these low-redshift outliers is likely to be the overestimated gas mass, and hence overestimated gas fraction. Apart from the lowest redshift bin, the average data points for the full and MS sample show an increase of $f_{\rm gas}(z)$ out to $z\sim3.5$, and flattening or even a mild decrease beyond that. The starburst SMGs exhibit a comparable average trend. Hence, our results are consistent with previous studies, which indicated that $f_{\rm gas}(z)$ plateaus at $z\gtrsim2-3$ (e.g. \cite{saintonge2013}; \cite{bethermin2015}; \cite{schinnerer2016}). We note that B\'ethermin et al. (2015), who estimated the gas masses from the dust masses by using a metallicity-dependent gas-to-dust ratio, found the flattening of $f_{\rm gas}(z)$ at $z>2$ for their MS galaxies only when they assumed a universal (i.e. no redshift evolution) fundamental metallicity relation (FMR) of Mannucci et al. (2010) to connect the gas-phase metallicity to $M_{\star}$ and SFR. However, similarly to our results, B\'ethermin et al. (2015) found that their strong starbursts have higher gas fractions, but follow the same increasing redshift evolution up to $z\sim2.4$ as their MS objects.

That we found such a wide range of gas fractions for our SMGs could, at least partly, reflect the different merger and star formation histories of the individual sources (e.g. \cite{geach2011}). For example, during a starburst event, the molecular gas reservoirs can be largely consumed in the star formation process, in which case $f_{\rm gas}$ decreases. The associated feedback, such as the galactic wind fuelled by supernovae and stellar winds, can also lead to a decrease in $f_{\rm gas}$ if a significant amount of cold gas is being ejected from the galaxy. On the other hand, $f_{\rm gas}$ can increase (again) if the galaxy accretes gas from the circumgalactic medium or the IGM (e.g. \cite{genzel2015}; \cite{tacchella2016}). 

\subsection{Exploring the scaling laws of star formation for the ALMA 1.3~mm detected submillimetre galaxies}

\begin{figure*}
\begin{center}
\includegraphics[width=0.45\textwidth]{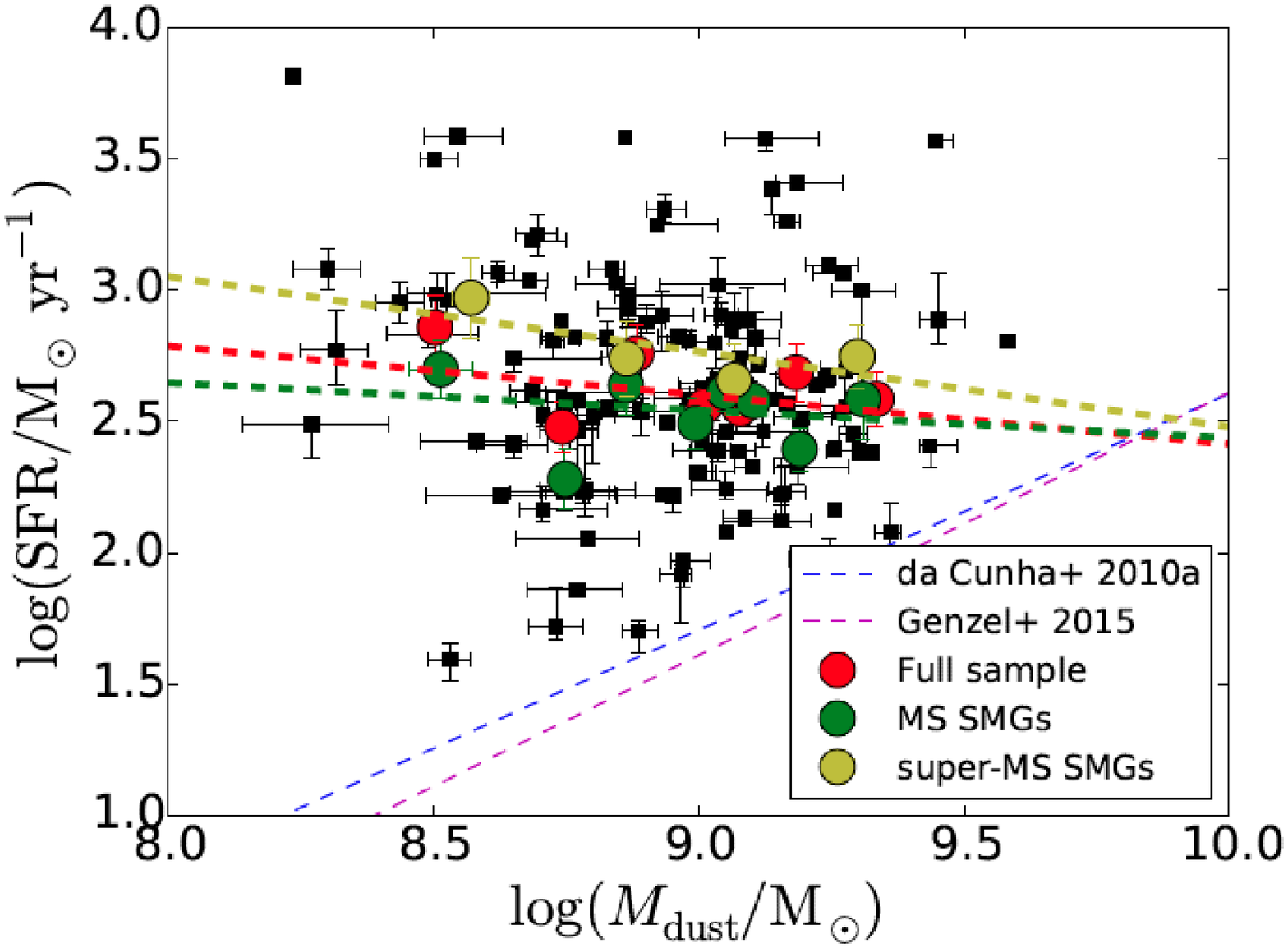}
\includegraphics[width=0.45\textwidth]{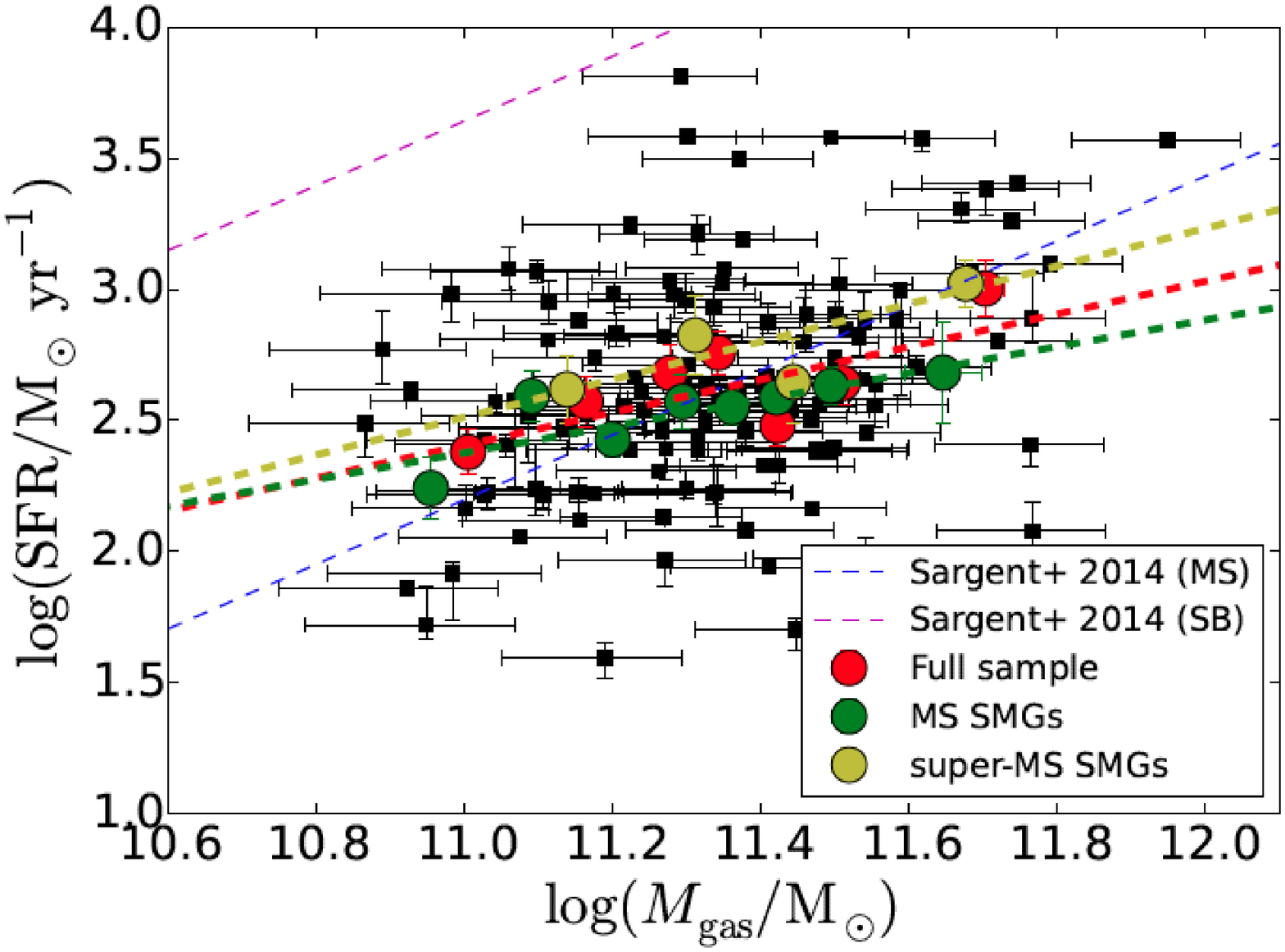}
\caption{Star formation rate as a function of dust mass (left panel) and gas mass (right panel; the so-called integrated K-S diagram). In both panels, the red, green, and yellow filled circles represent the mean values of the binned data as in the left panel in Fig.~\ref{figure:sSFR}. The blue dashed line in the left panel shows the da Cunha et al. (2010a) relationship, that is ${\rm SFR}[{\rm M}_{\sun}\,{\rm yr}^{-1}]=3.95\times10^{-7}\times (M_{\rm dust}/{\rm M}_{\sun})^{0.9}$. The magenta dashed line in the left panel shows the $M_{\rm dust}-T_{\rm dust}-{\rm SFR}$ scaling relation from Genzel et al. (2015; Eq.~(9) therein), which is here plotted at $T_{\rm dust}=25$~K. In the right panel, the blue and magenta dashed lines show the relationships calibrated for MS and starburst galaxies by Sargent et al. (2014; their Eq.~(4)). }
\label{figure:sk}
\end{center}
\end{figure*}

To explore how the dust and gas mass contents of our SMGs are related with their SFR, in Fig.~\ref{figure:sk} 
we show log-log plots of SFR versus $M_{\rm dust}$ (left panel) and $M_{\rm gas}$ (right panel). 

As can be seen in the left panel, the SFR appears to be fairly constant on average over the $M_{\rm dust}$ range explored here. This is true for the full sample, and separately for the MS SMGs and starburst SMGs. To quantify this, we fit the binned data using linear least squares fits. For the full sample, we derived the functional form $\log({\rm SFR}/{\rm M}_{\sun}\,{\rm yr}^{-1})\propto -(0.19\pm0.21)\times \log(M_{\rm dust}/{\rm M}_{\sun})$, while for the MS (starburst) SMGs the slope was derived to be $-0.11\pm0.23$ ($-0.29\pm0.20$).

It is interesting to remark that the aforementioned results are very different from the tight, positive ${\rm SFR}-M_{\rm dust}$ relationship derived by da Cunha et al. (2010a) for low-redshift Sloan Digital Sky Survey (SDSS; \cite{york2000}) galaxies over four orders of magnitude in both quantities (and where the analysis was based on {\tt MAGPHYS} SEDs, and the ${\rm SFR}$ was averaged over 100~Myr as in the present study). This relationship, $M_{\rm dust}[{\rm M}_{\sun}]=1.28\times10^7\times({\rm SFR}/{\rm M}_{\sun}\,{\rm yr}^{-1})^{1.11}$, is shown in its inverted form by the dashed blue line in the left panel in Fig.~\ref{figure:sk}. This positive trend is also physically understandable, if an increasing SFR, and hence an increasing supernova rate, leads to a more efficient enrichment of the ISM with metals and dust. However, if the dust production is dominated by the grain growth in the ISM (e.g. \cite{michalowski2016}), one might not necessarily expect a correlation between the dust mass content and SFR.

That we do not see a positive ${\rm SFR}-M_{\rm dust}$ correlation among our SMGs is likely a source selection effect (but see below). Because our initial source selection was based on the observed-frame 1.1~mm detections, and subsequent detections with ALMA at $\lambda_{\rm obs}=1.3$~mm, our sample is comprised of highly star-forming, dusty galaxies. Indeed, looking at Fig.~5 in da Cunha et al. (2010a), most of their target SDSS galaxies have SFRs of $\lesssim30$~M$_{\sun}$~yr$^{-1}$, with a relatively few sources extending to ${\rm SFR}\lesssim100$~M$_{\sun}$~yr$^{-1}$. Conversely, and as illustrated in the left panel in Fig.~\ref{figure:sk}, our SMG sample is probing the higher SFR regime of the ${\rm SFR}-M_{\rm dust}$ plane, which is unpopulated by the da Cunha et al. (2010a) sample. Another issue that might be causing the different behaviour of our SMGs in the ${\rm SFR}-M_{\rm dust}$ diagram is their potentially very different star formation histories compared to low-redshift galaxies. Nonetheless, our result is consistent with that of Hjorth et al. (2014), who found that the SMGs from Swinbank et al. (2014) tend to lie above the inverted da Cunha et al. (2010a) relationship (we note that Hjorth et al. (2014) plot the rest-frame 870~$\mu$m luminosity-based $M_{\rm dust}$ on the $y$-axis, and total-IR-based SFR on the $x$-axis in their Fig.~1). Hjorth et al. (2014) interpreted this finding under the scenario of the ${\rm SFR}-M_{\rm dust}$ relationship being a manifestation of an evolutionary sequence, where the dust is produced during the initial, rapid starburst phase, and owing to the highest attainable dust mass, a population of starbursting galaxies with lower dust masses than predicted by the da Cunha et al. (2010a) relationship for a given SFR is to be expected. Calura et al. (2017) also found that SMGs tend to lie above the inverted da Cunha et al. (2010a) relationship, and exhibit a flatter ${\rm SFR}-M_{\rm dust}$ relationship than seen locally (the authors employed the SMG data from Santini et al. (2010), where the SFRs were calculated from $L_{\rm IR}$, and dust masses were re-computed using a MBB SED fitting technique). Finally, as illustrated in Fig.~\ref{figure:sk}, the da Cunha et al. (2010a) relationship is similar to the $M_{\rm dust}-T_{\rm dust}-{\rm SFR}$ scaling relation from Genzel et al. (2015; their Eq.~(9)), which we show at $T_{\rm dust}=25$~K, that is at the dust temperature assumed in the gas mass calculation in Sect.~3.2. The idea behind the plotted Genzel et al. (2015) scaling relation is that the dust acts as a calorimeter by absorbing all the stellar UV photons, and re-radiates the absorbed energy in the optically thin IR regime at an average dust temperature (the emissivity index was assumed to be $\beta=1.5$). 
Hence, the da Cunha et al. (2010a) relationship can also be understood in terms of a simple scaling arising from dust-obscured star formation.

In the right panel in Fig.~\ref{figure:sk}, we plot the SFR against the gas mass. This type of a plot can be interpreted as the integrated Kennicutt-Schmidt (K-S) diagram (rather than the K-S diagram constructed from the SFR and gas mass surface densities; cf.~Fig.~6 in \cite{kennicutt1998}). The binned average data show a positive correlation between these two quantities. For the full sample, a linear least squares fit to the binned averages yielded $\log({\rm SFR}/{\rm M}_{\sun}\,{\rm yr}^{-1})=(0.63\pm0.28)\times \log(M_{\rm gas}/{\rm M}_{\sun})-(4.48\pm3.19)$, with $r=0.78$. For the MS SMGs, the slope and the intercept were derived to be $0.51\pm0.16$ and $-3.20\pm1.78$ ($r=0.81$), and for the starbursts they were found to be $0.72\pm0.26$ and $-5.42\pm2.97$ ($r=0.78$). The derived sub-linear slopes are $2.3\sigma-3.2\sigma$ above a zero slope, and hence the positive correlations are not statistically very significant. We note that because our gas masses were derived from the 1.3~mm flux densities (Eq.~(\ref{eq:ism})), and the SFRs were calculated from the total-IR luminosity, it is not surprising to see a positive correlation between the SFR and $M_{\rm gas}$, although such correlation is expected on the basis of the K-S -type scaling law. Indeed, while the average SFR for the full sample increases by a factor of 4.2 from its minimum to maximum value, the corresponding average $M_{\rm gas}$ increases by a comparable factor of $\sim5$. The shallow correlations we found can be partly understood to arise from the assumption of a uniform $T_{\rm dust}$ of 25~K in the calculation of $M_{\rm gas}$, while the sources with higher SFR are presumably associated with warmer dust. We also note that while the SFR appears to be roughly constant as a function of dust mass on average (Fig.~\ref{figure:sk}, left panel), the gas-to-dust ratio exhibits an increasing evolution as a function of redshift (Fig.~\ref{figure:dusttostellar}, right panel). These two behaviours also imply a rising trend of SFR as a function of gas mass. For comparison with our SMG data, we also plot the ${\rm SFR}-M_{\rm gas}$ relationships calibrated for MS and starburst galaxies by Sargent et al. (2014; their Eq.~(4)). Their MS relationship overlaps with our average data points, albeit with a steeper, somewhat super-linear slope of $1.235\pm0.046$. B\'ethermin et al. (2015) also found that their MS galaxies follow the Sargent et al. (2014) relationship, and this applied out to $z=4$ under the assumption of a broken FMR; for a universal FMR, their MS galaxies were shifted upwards from the Sargent et al. (2014) relationship. The strong starburst relationship of Sargent et al. (2014) has the same slope as their MS function, but it is offset towards higher SFRs (higher star formation efficiency (SFE)) with respect to our starburst SMG sequence (e.g. by a factor of 1.46 (0.16~dex) at $\log(M_{\rm gas}/{\rm M}_{\sun})=11$). Conversely, B\'ethermin et al. (2015) found that their strong starbursts lie within the $1\sigma$ confidence region of the Sargent et al. (2014) starbursts relationship, although the data were found to systematically lie below the mid-line of the relation. That we do not see two clearly separated scaling laws between our MS and starburst objects, but rather fairly similar ones for the two populations, is likely to reflect the critical assumption of a uniform, Galactic $\alpha_{\rm CO}$ conversion factor used in the calculation of our dust-based gas masses (see below).

In principle, the $M_{\rm dust}-M_{\rm gas}-{\rm SFR}$ comparison (more precisely the slopes) can provide insight into the actual underlying K-S-type star formation law, that is $\Sigma_{\rm SFR}\propto \Sigma_{\rm gas}^N$ (\cite{schmidt1959}; \cite{kennicutt1998}; see \cite{santini2014}; \cite{hjorth2014}; \cite{bethermin2015}). Hjorth et al. (2014) showed that under the assumption of a simple, closed-box chemical evolution model (dust produced by supernovae, while the contributions from asymptotic giant branch stars and dust grain growth in the ISM were not taken into account), the power-law slope of the $M_{\rm dust}-{\rm SFR}$ relationship is inversely proportional to the exponent of a global, integrated K-S star formation law of ${\rm SFR}\propto M_{\rm gas}^k$ (see also \cite{gall2011}). As discussed by the authors, if the global K-S index is $k=1.5$, one would expect a relationship of $M_{\rm dust}\propto {\rm SFR}^{2/3}$, while the da Cunha et al. (2010a) relation suggests a value of $k=0.9$ (Fig.~\ref{figure:sk}, left panel). As illustrated in the right panel in Fig.~\ref{figure:sk}, the integrated K-S law for MS galaxies from Sargent et al. (2014), which has a central role in the description of the gas component of SFGs in their 2-Star Formation Mode framework, has a slope ($1.235\pm0.046$) fairly similar to the aforementioned value of $k=1.5$, while our SMGs show a different behaviour, that is the nominal, sub-linear ${\rm SFR}-M_{\rm gas}$ slope ranges from 0.51 for 
the MS objects to 0.72 for starbursts, which could mean that they do not follow a traditional, K-S-type law. Of course, 
both the dust and gas-emitting size scales of our sources should be measured to examine how the SFR and mass surface densities are 
related to each other in the functional form of $\Sigma_{\rm SFR}\propto \Sigma_{\rm gas}^N$. For example, Bouch{\'e} et al. (2007) found that bright SMGs ($S_{\rm 850\, \mu m}\geq5$~mJy) lie at the high-$\Sigma_{\rm gas}$ end of a universal K-S relationship ($N\simeq 1.7$) out to $z=2.5$. There is, however, some observational evidence for distinct, bimodal star formation laws between normal disk galaxies and starburst SMGs with an elevated SFE (e.g. \cite{daddi2010}; \cite{genzel2010}), but the discrepancy might be (partly) caused by the different $\alpha_{\rm CO}$ conversion factors and CO level excitation assumptions adopted for the two populations (\cite{ivison2011}; \cite{narayanan2012}; \cite{liu2015}; \cite{scoville2016}). We note that the ${\rm SFR}-M_{\rm gas}$ relationship derived by Daddi et al. (2010) for normal disk galaxies (their Eq.~(1)) is fairly similar to the corresponding relationship from Sargent et al. (2014), with the slope of the former function being only a factor of 1.06 steeper. Also, the starburst sequence derived by Daddi et al. (2010) for their combined ULIRG and SMG data set is close to that of Sargent et al. (2014), with the same slope as the normal galaxy function but shifted upwards by 1.1~dex. The $\Sigma_{\rm SFR}\propto \Sigma_{\rm gas}^N$ relationship of a large subsample ($\sim40$) of the present SMGs will be presented by Miettinen et al. (2017c), who used high-resolution ($0\farcs2$) ALMA Cycle~4 observations (PI: O.~Miettinen) of $\lambda_{\rm obs}=870$~$\mu$m emission to calculate the source sizes, and hence the SFR and gas mass surface densities.

\begin{figure*}
\begin{center}
\includegraphics[width=0.45\textwidth]{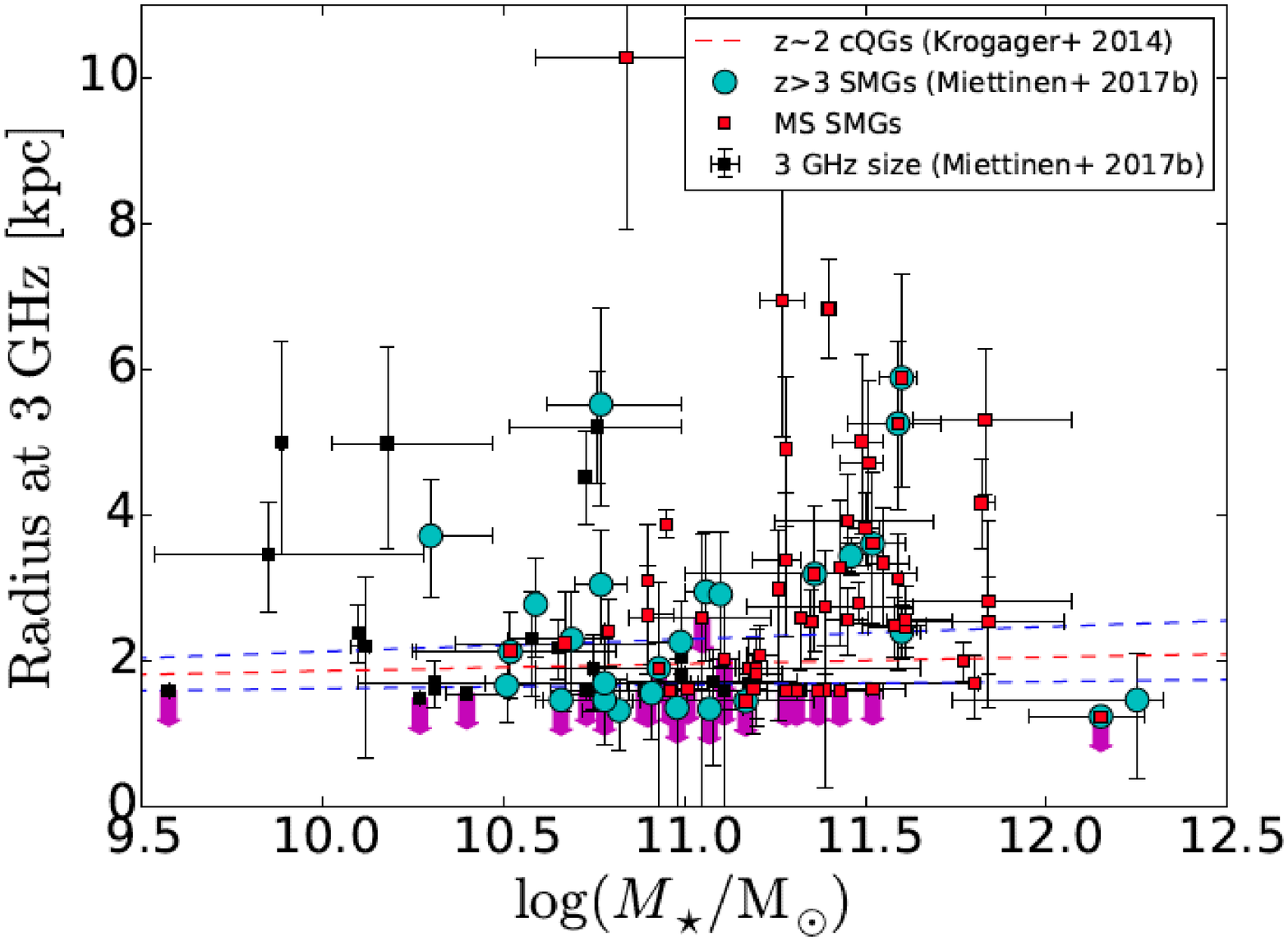}
\includegraphics[width=0.45\textwidth]{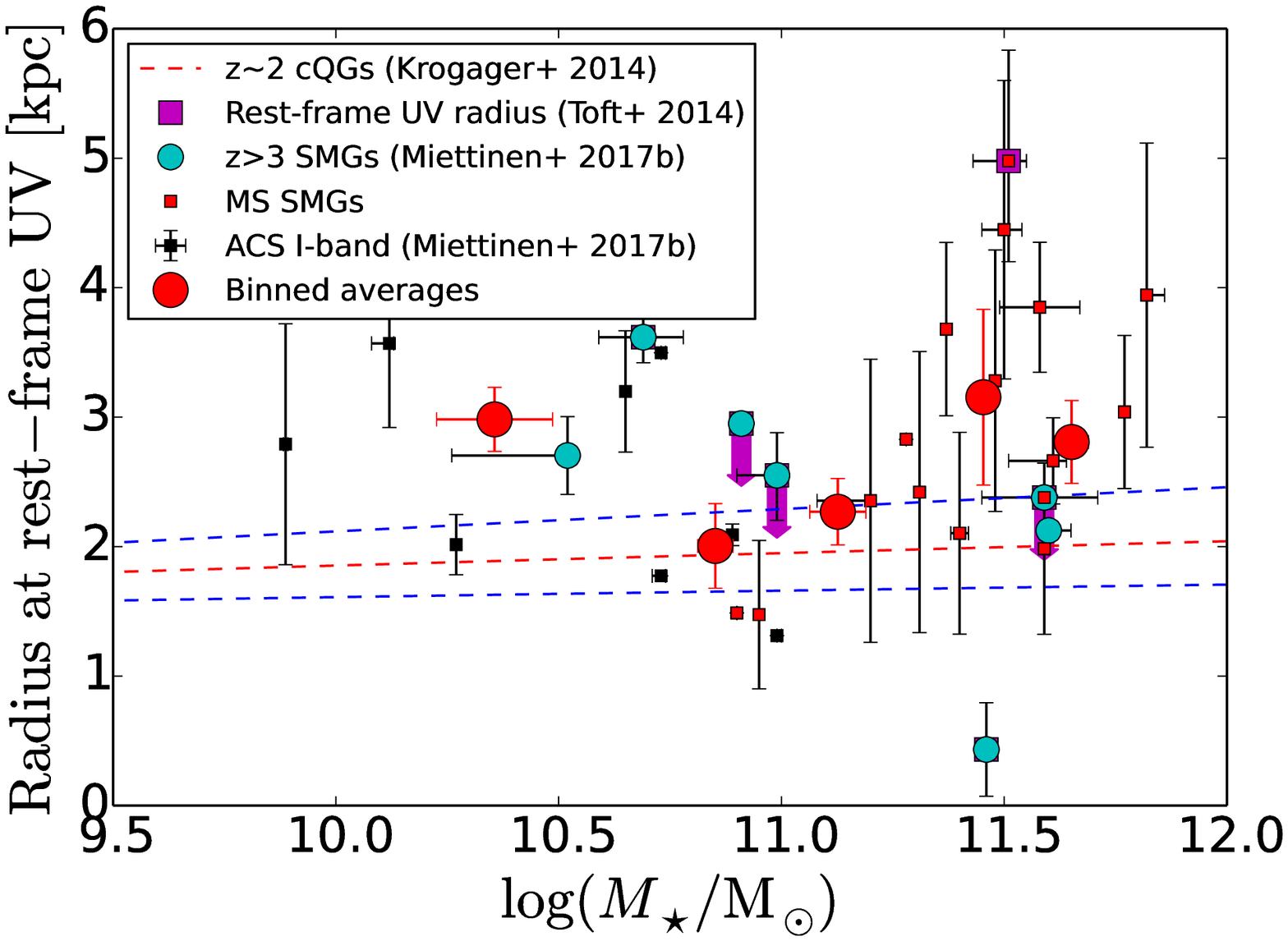}
\caption{Radii of our SMGs plotted against their stellar masses. In the left panel, we plot the observed-frame 3~GHz radio-emitting sizes (defined as half the deconvolved major axis FWHM) from Miettinen et al. (2017b). In the right panel, we plot the radii at the rest-frame UV, which are based on observations with the \textit{Hubble}/ACS in the $I$-band (see \cite{miettinen2017b}, and references therein). These data are supplemented by the rest-frame UV sizes measured by Toft et al. (2014; scaled to the present redshifts and cosmology). In both panels, the MS SMGs are highlighted by small red squares, while 
those SMGs that lie at $z>3$ are highlighted by filled cyan circles. The upper size limits are indicated by downwards-pointing magenta arrows. In the right panel, the large red circles represent the survival analysis-based mean values of the binned data (each bin is equally populated by six SMGs, except the highest stellar mass bin, which contains seven SMGs), with the error bars showing the standard errors of the mean values. The three dashed lines in both panels show the mass-size relationship of $z\sim2$ cQGs from Krogager et al. (2014), where the blue dashed lines represent the dispersion in the parameters (see text for details). We note that the plotting ranges are different in the left and right panels.}
\label{figure:size}
\end{center}
\end{figure*}

\subsection{Stellar mass -- size relationship}

The stellar masses derived in the present work can be compared with the observed-frame 3~GHz radio-emitting sizes derived for our SMGs by Miettinen et al. (2017b) to see if the two parameters exhibit any correlation. This comparison could be done for 98 SMGs for which both the {\tt MAGPHYS} SED fit and radio size (or its upper limit) could be derived. The stellar mass-radio size diagram is plotted in the left panel in Fig.~\ref{figure:size}. 

For comparison, in the right panel in Fig.~\ref{figure:size} we show a separate plot of the rest-frame UV sizes as a function of stellar mass. For 25 of our analysed SMGs, the UV size measurements are based on observations with the \textit{Hubble}/Advanced Camera for Surveys (ACS) in the $I$-band (F814W), and they are described in more detail in Miettinen et al. (2017b, and the references therein). These data are supplemented by the observed-frame near-IR (rest-frame UV) size measurements for six of our sources (AzTEC/C2a, C4, C5, C10b, C17, and C42) from Toft et al. (2014). The sizes from Toft et al. (2014) were scaled to our adopted cosmology, and revised redshifts were used whenever needed (see also \cite{miettinen2017b}). In both panels in Fig.~\ref{figure:size}, those SMGs that lie at $z>3$ are highlighted by filled, cyan circle symbols. Among the SMGs from Toft et al. (2014), only AzTEC/C10b lies at $z\lesssim 3$ (its redshift is estimated to be $z=2.90^{+0.30}_{-0.90}$; see \cite{brisbin2017} for details). 

The main purpose of the stellar mass-size comparison here is to see how our SMGs compare with the stellar mass-size relationship of cQGs at $z\sim2$, 
because the latter galaxy population is suggested to descent from high-redshift ($z>3$) SMGs (\cite{toft2014}). The three dashed lines shown in 
Fig.~\ref{figure:size} show the stellar mass-size (half-light radius) relationship and its dispersion for $z\sim2$ cQGs derived by Krogager et al. (2014). 
This is given by 

\begin{equation}
\label{eqn:krogager}
r_{\rm e}=\gamma \left(\frac{M_{\star}}{10^{11}~{\rm M}_{\sun}}\right)^{\beta}\,,
\end{equation}
where $\log(\gamma/{\rm kpc})=0.29\pm0.07$ and $\beta=0.53^{+0.29}_{-0.21}$ for their sample of cQGs in COSMOS, which is partly (41.2\%) composed of 
spectroscopically confirmed sources. We note that Krogager et al. (2014) used the Fitting and Assessment of Synthetic Templates ({\tt FAST}) code (\cite{kriek2009}) to derive their Chabrier (2003) IMF-based stellar masses.

The data points in the stellar mass -- radio size plane show a fairly large scatter, and no obvious correlation is visible between the two quantities. This is a possible manifestation of the finding that the radio emission from SMGs can partly originate in processes not linked to stellar evolution (high-mass star formation and supernova activity). For instance, this would be the case if two interacting galaxies have formed a radio-emitting bridge in between them (a so-called Taffy system; see \cite{miettinen2015b}, 2017b, and references therein). We also note that on average, the physical 3~GHz radio size was found to exhibit no evolution as a function of redshift (\cite{miettinen2017b}; Fig.~11 therein). However, a bimodality of stellar masses between MS and super-MS SMGs is seen in both panels in Fig.~\ref{figure:size}, so that the former population is preferentially found at $M_{\star}\gtrsim10^{11}$~M$_{\sun}$, while the latter type of SMGs tend to populate the $M_{\star}\lesssim10^{11}$~M$_{\sun}$ part of the diagram.

Twenty-one, or about $21\%\pm5\%$ of our sample of 98 SMGs have nominal masses and radio sizes that place them within the dispersion of the Krogager et al. (2014) relationship. Among these sources, only six lie at $z>3$, that is at cosmic epoch where the progenitors of the $z\sim2$ cQGs are suggested to be found (\cite{toft2014}). Owing to the large error bars of the radio sizes, the aforementioned overlap with the Krogager et al. (2014) relationship could be either stronger (up to 13 sources at $z>3$) or weaker. Another complicating factor is that the size parameter in the Krogager et al. (2014) relationship refers to a circularised half-light radius ($r_{\rm e}$) derived through S\'ersic profile fits, while in Fig.~\ref{figure:size} we have parameterised the radio radius as half the major axis FWHM. However, the simple conversion of ${\rm FWHM}=2 \times r_{\rm e}$ is strictly valid only for a circular Gaussian profile with a S\'ersic index of $n = 0.5$, while the mean (median) S\'ersic index derived by Krogager et al. (2014) is $\langle n \rangle = 3.50$ (3.08), which suggests an average (median) relationship of ${\rm FWHM}= 7.2\times10^{-4}\times r_{\rm e}$ (${\rm FWHM}= 2.8\times10^{-3}\times r_{\rm e}$; e.g. \cite{voigt2010}, their Eq.~(18)).

In the right panel in Fig.~\ref{figure:size}, which shows the rest-frame UV sizes as a function of stellar mass, we also plot the binned averages of the data. The latter values were calculated by using a Kaplan-Meier (K-M) survival analysis to take the upper size limits (left-censored data) into account. As can be seen in Fig.~\ref{figure:size}, the largest rest-frame UV sizes are found among the SMGs with the highest stellar masses, but the binned averages of the data do not reveal any obvious positive correlation. Only if the lowest stellar mass bin ($\log(M_{\star}/{\rm M}_{\sun})=10.36\pm0.13$) is not considered, we do see a hint of a rising trend in size as a function of mass, and where the two lowest mass bins are perfectly consistent with the Krogager et al. (2014) relationship. 

It is worth mentioning that in the present work, we revised some of the redshifts and stellar masses of the SMGs studied by Toft et al. (2014; see our Appendix~D). However, the redshift changes are minor, our values being $0.91-1.08$ (median 0.94) times the Toft et al. (2014) values, while our stellar masses are $0.26-3.63$ (median 1.64) times the values from Toft et al. (2014). The three unresolved sources (out of the six in total) from Toft et al. (2014) are formally consistent with the plotted mass-size relationship of the cQGs at $z\sim2$, while two sources clearly lie above it, and the very compact source AzTEC/C42 with $r_{\rm e}^{\rm UV}=0.43\pm0.36$~kpc lies below it. If we consider the $z>3$ SMGs plotted in the right panel in Fig.~\ref{figure:size}, four out of seven are formally consistent with the Krogager et al. (2014) relationship. Hence, our results are only broadly consistent with the conclusion drawn by Toft et al. (2014), namely that the $M_{\star}-r_{\rm e}^{\rm UV}$ distribution of $3 < z < 6$ SMGs is comparable to that of $z\sim2$ cQGs, which would support the $z>3$ SMG population to be composed of potential precursors of the latter population. We return to this topic in Sect.~4.9. One potentially important caveat in the above analysis is that the rest-frame UV size measurements of dust-obscured objects like SMGs can be subject to large uncertainties (see \cite{miettinen2017b}, and references therein). This is especially true if the source exhibits clumpy morphology in the rest-frame UV, which itself can be the manifestation of spatially differential dust obscuration. However, our main conclusions regarding 
the right panel of Fig.~\ref{figure:size} are based on the binned averages of the data points, which can partly compensate for the outliers caused 
by the uncertain rest-frame UV sizes.

\begin{figure*}
\begin{center}
\includegraphics[width=0.45\textwidth]{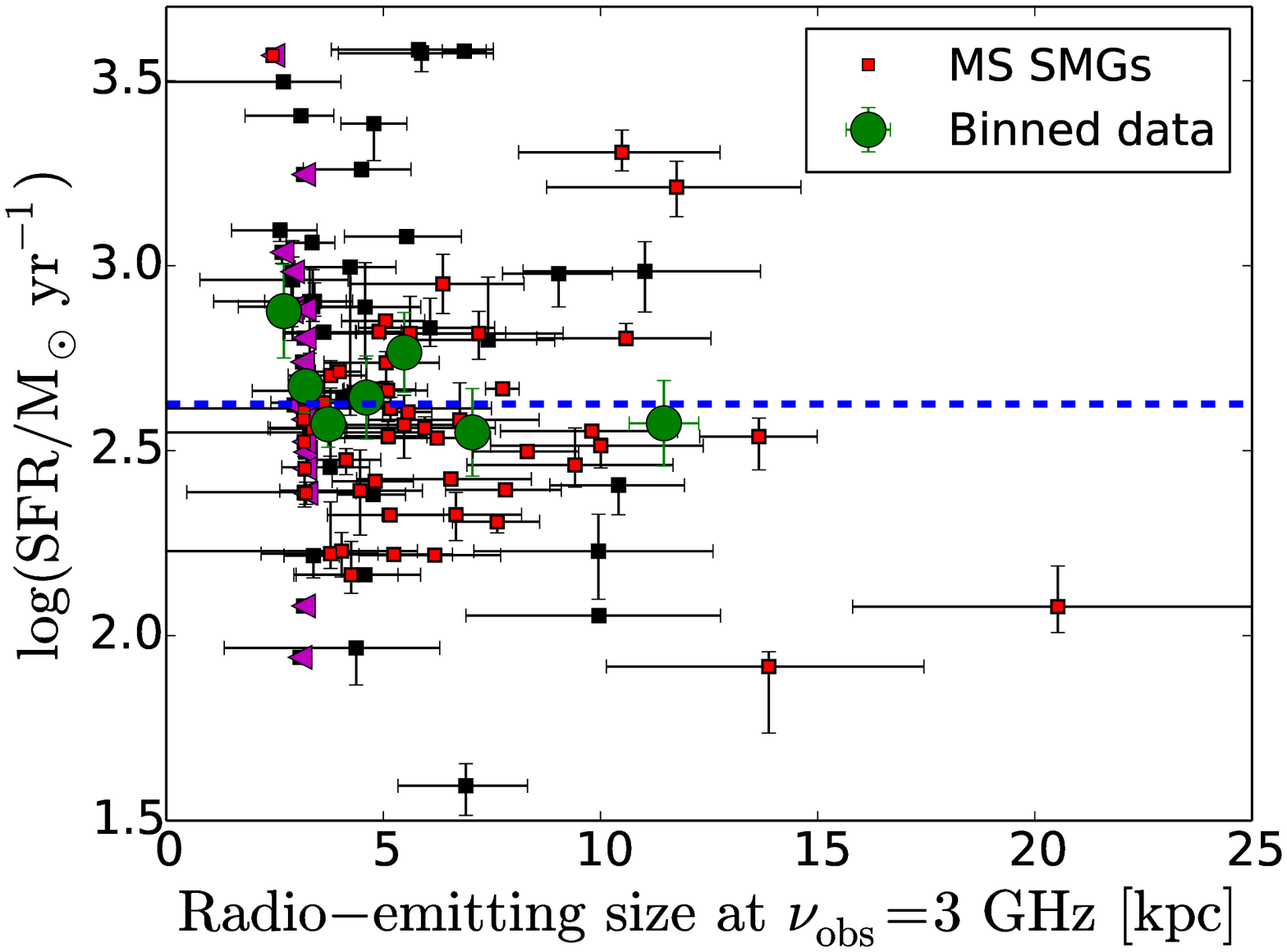}
\includegraphics[width=0.45\textwidth]{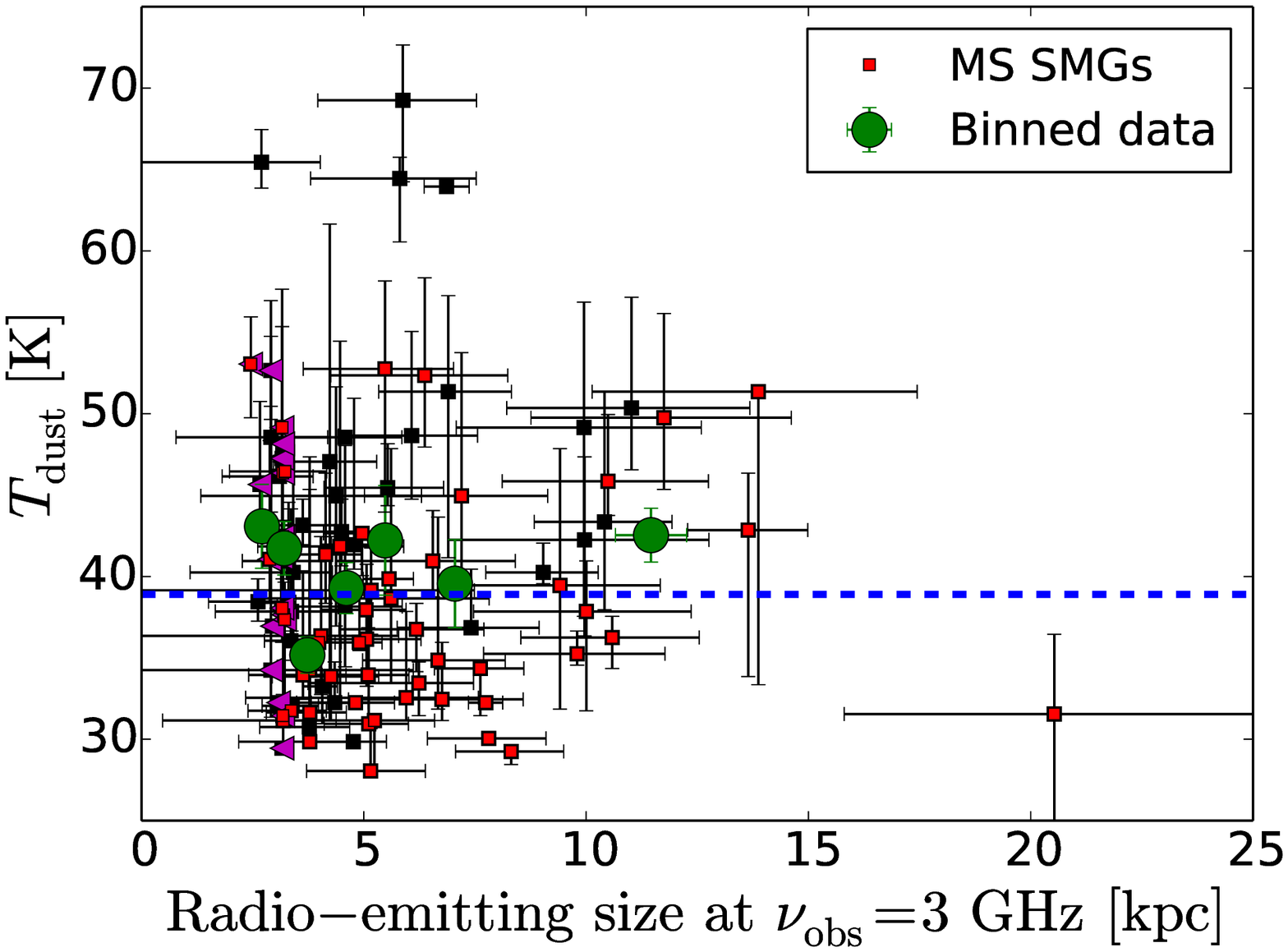}
\caption{Star formation rate (left) and $T_{\rm dust}$ (right) plotted as a function of the size of the radio-emitting region (FWHM of the deconvolved major axis in kpc) at $\nu_{\rm obs}=3$~GHz derived by Miettinen et al. (2017b). The MS SMGs are highlighted by red markers. The left-pointing magenta triangles indicate the upper size limits, while the green filled circles represent the survival-analysis based mean values of the binned data (each bin contains 14 SMGs), with the error bars showing the standard errors of the mean values. In both panel, the horizontal blue dashed line marks the median value of the $y$-axis parameter for the 98 analysed SMGs (${\rm SFR}=423$~${\rm M}_{\sun}~{\rm yr}^{-1}$ and $T_{\rm dust}=38.9$~K).}
\label{figure:corr}
\end{center}
\end{figure*}

\subsection{Comparison of the star formation rates, starburstiness levels, and dust temperatures with the radio-emitting sizes}

We now turn our attention to possible dependencies of the SFR, $\Delta_{\rm MS}$, and $T_{\rm dust}$ on the radio-emitting sizes of our SMGs derived by Miettinen et al. (2017b). As mentioned in Sect.~4.5, we can make these comparisons for a large subsample of 98 SMGs with both the SED and radio size available.

In Fig.~\ref{figure:corr}, we plot the SFR and $T_{\rm dust}$ as a function of the spatial extent (deconvolved FWHM of the major axis) of the radio-emitting region at the observed-frame frequency of 3~GHz. The binned average values plotted in this figure were derived using a K-M survival analysis of the left-censored data. No trend is discernible neither in SFR nor $T_{\rm dust}$ as a function of the radio size. However, the smallest radio-size bin ($2.7\pm0.1$~kpc) exhibits the highest SFR ($754^{+255}_{-190}$~M$_{\sun}$~yr$^{-1}$), and the highest $T_{\rm dust}$ value ($43.1\pm2.6$~K). In fact, the binned SFR and luminosity-weighted $T_{\rm dust}$ averages in both panels show the same behaviour. This is not surprising because the SFR was derived from the IR luminosity, which is strongly dependent on $T_{\rm dust}$ (for an optically thick source of a given size, $L_{\rm IR}\propto T_{\rm dust}^4$, while in the situation of optically thin dust emission, $L_{\rm IR}\propto M_{\rm dust}\times T_{\rm dust}^{4+\beta}$).

It has been suggested that local and low-redshift ULIRGs have warmer dust temperatures owing to their more compact sizes than their higher redshift, spatially more extended versions (e.g. \cite{elbaz2011}; \cite{rujopakarn2013}). Hence, one might expect that 
the dust temperature is higher if the star formation occurs in a spatially more compact region of the galaxy, 
but such a trend is not clearly visible in Fig.~\ref{figure:corr}. This might be an indication that our SMGs do not exhibit a geometry where a central starburst region is surrounded by a spatially more extended dust zone, but rather a geometric configuration where the heating sources (young stars) and dust are well mixed with a similar scaling, in which case $T_{\rm dust}$ is insensitive to the size (\cite{misselt2001}; \cite{safarzadeh2016}; see also \cite{miettinen2017b}). Ideally, the dust temperature and SFR would be compared with the size of the dust-emitting region, rather than with the radio-emitting size. The reason for this is that the observed spatial scale of radio emission from SMGs might not (always) be a good proxy of the zone of active high-mass star formation. For example, the radio-emitting region might appear puffed in merger-driven SMGs (see \cite{miettinen2017b}, and references therein). Hence, it seems possible that the absence of clear trends in Fig.~\ref{figure:corr} is caused by the SMG sizes being measured in the radio regime. Our ALMA Cycle~4 observations mentioned in Sect.~4.4 can be used to address these issues in our future work, that is we can investigate the dependencies of $T_{\rm dust}$ and $L_{\rm IR}$ (SFR) on the more appropriate dust-emitting sizes for a large subsample of the present SMGs.  

\begin{figure}[!h]
\centering
\resizebox{0.999\hsize}{!}{\includegraphics{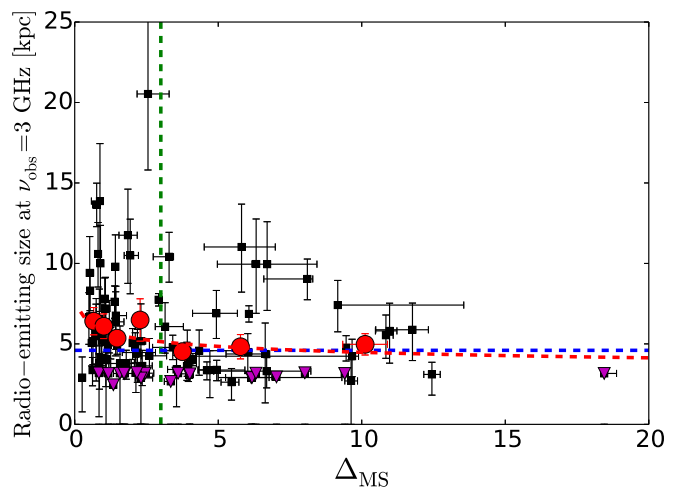}}
\caption{Radio-emitting size at $\nu_{\rm obs}=3$~GHz (deconvolved FWHM of the majoar axis in kpc) from Miettinen et al. (2017b) as a function of the deviation from the MS. The down-pointing magenta triangles indicate the upper size limits, while the red filled circles represent the binned averages as in Fig.~\ref{figure:corr}. The red, dashed curve represents a least squares fit to the mean data points (see text for details). The horizontal blue, dashed line marks the survival analysis-based median radio size of 4.6~kpc for the plotted sample. The vertical green, dashed line shows the upper boundary limit of the MS, that is $\Delta_{\rm MS}=3$.}
\label{figure:burstsize}
\end{figure}

As a final point, it is relevant to ask if the radio-emitting size exhibits any depedence on the distance from the MS. 
For this purpose, in Fig.~\ref{figure:burstsize} we plot the 3~GHz sizes as a function of the starburstiness parameter. 
The largest radio sizes are seen among the MS SMGs, and a least squares fit to the mean data points yields 

\begin{equation}
{\rm FWHM}_{\rm 3\, GHz}^{\rm major}[{\rm kpc}]=(5.83\pm0.29)\times \Delta_{\rm MS}^{-(0.115\pm0.043)}\,.
\end{equation}
Hence, there is a hint (at $2.7\sigma$ significance level) that the radio-emitting sizes of MS SMGs are typically larger than their starburst counterparts. The nominal values of the binned averages have a Spearman correlation coefficient of $\rho=-0.64$, which indicates the presence of a moderate negative trend. A similar result was obtained by Miettinen et al. (2017a) for their much smaller (by a factor of seven), partly overlapping sample of SMGs in COSMOS (see Fig.~10 therein). 

If the radio-emitting size really gets smaller the further the SMG lies above the MS, this might 
bring into question the scenario discussed by Miettinen et al. (2017b) where the large radio size is the result of galaxy interaction, because one would expect those systems to be associated with triggered starburst activity, and hence to have elevated $\Delta_{\rm MS}$ values. On the other hand, the trend shown in Fig.~\ref{figure:burstsize} could conform to the hydrodynamic simulations by Hayward et al. (2012), which suggest that the merger-driven starburst gets stronger as the (nuclear) separation between the interacting galaxies decreases. In this case, the radio-emitting region of the system should also decrease in size as the interacting galaxies approach the final coalescence. Figure~\ref{figure:burstsize} also provides a hint of an alternative view, namely that, on average, the MS SMGs possess a more extended radio-emitting disk, which could be an indication of their more widespread star formation compared to super-MS objects.

\subsection{Comparison of the physical properties between different galaxy morphologies}

In Miettinen et al. (2017b), we could assign a morphological classification for 30 of our 152 ALMA SMGs. 
The classifications were taken from the COSMOS morphology catalogues, 
and they are based on the $I$-band imaging with the \textit{Hubble}/ACS, 
which in our cases is probing the rest-frame UV (see \cite{miettinen2017b}, and references therein for details). 
In the present work, we could obtain a morphological class for 26 sources for which a {\tt MAGPHYS} SED could be 
derived, out of which 13 (50\%) are classified as disks, and the second half as irregulars. We stress that similar to the rest-frame 
UV sizes, the galaxy morphologies in the rest-frame UV can also be uncertain owing to the potential spatially varying dust obscuration. In particular, the differential dust obscuration could give the impression that the galaxy has an irregular morphology, although the real, underlying geometric confguration would be that of a smooth disk (cf.~Fig.~12 in \cite{miettinen2017b}). For comparison, Miettinen et al. (2017b) found that the 3~GHz radio sizes are similar for the different, aforementioned morphological classes (disks and irregulars).

\begin{figure*}
\begin{center}
\includegraphics[width=\textwidth]{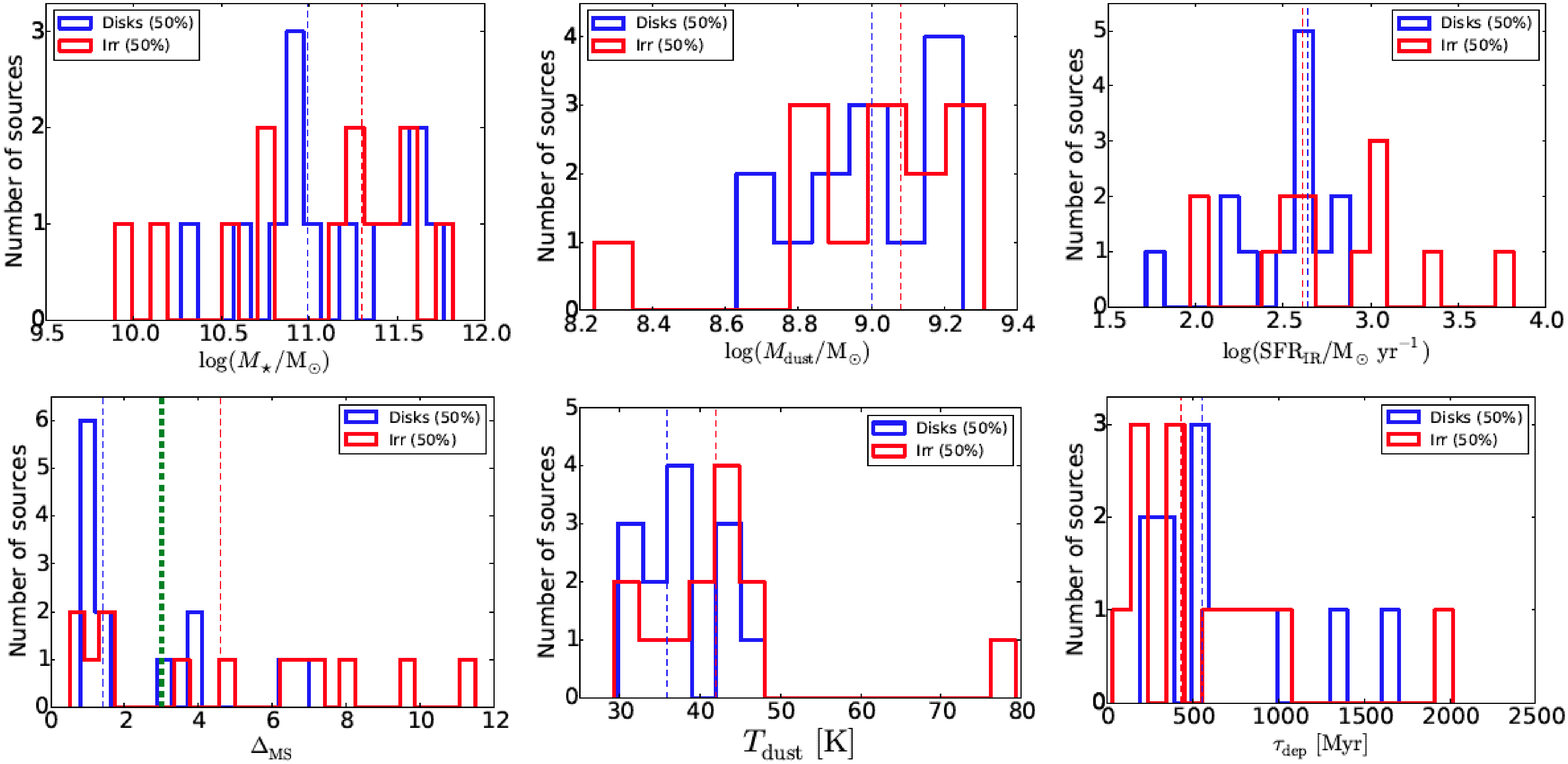}
\caption{Panels from left to right, top to bottom, show the distributions of $M_{\star}$, $M_{\rm dust}$, SFR, $\Delta_{\rm MS}$, $T_{\rm dust}$, and $\tau_{\rm dep}$ for our SMGs that are classified as disks and irregular galaxies (see text and \cite{miettinen2017b} for details). The blue and red vertical dashed lines show the sample medians (disks: $M_{\star}=9.8\times10^{10}$~M$_{\sun}$, $M_{\rm dust}=1.0\times10^9$~M$_{\sun}$, ${\rm SFR}=434$~M$_{\sun}$~yr$^{-1}$, $\Delta_{\rm MS}=1.4$, $T_{\rm dust}=36$~K, and $\tau_{\rm dep}=549$~Myr; irregulars: $M_{\star}=2.0\times10^{11}$~M$_{\sun}$, $M_{\rm dust}=1.2\times10^9$~M$_{\sun}$, ${\rm SFR}=404$~M$_{\sun}$~yr$^{-1}$, $\Delta_{\rm MS}=4.6$, $T_{\rm dust}=42$~K, and $\tau_{\rm dep}=429$~Myr). In the bottom left panel, the vertical green dashed line marks the upper boundary of the adopted MS definition, that is $\Delta_{\rm MS}=3$.}
\label{figure:morph}
\end{center}
\end{figure*}

In Fig.\ref{figure:morph}, we plot the distributions of $M_{\star}$, $M_{\rm dust}$, SFR, $\Delta_{\rm MS}$, $T_{\rm dust}$, and $\tau_{\rm dep}$ separately for disk-like SMGs and irregular systems. Overall, the stellar and dust mass distributions between the two morphological classes are comparable to each other, although we note that the lowest values of $M_{\star}$ and $M_{\rm dust}$ are found for irregular sources, and the stellar masses of disks show a peak near $10^{11}$~M$_{\sun}$, which causes the median $M_{\star}$ of the disks to be a factor of two lower than that of irregulars. The disks and irregulars show comparable median SFRs (434~M$_{\sun}$~yr$^{-1}$, and 404~M$_{\sun}$~yr$^{-1}$, respectively), but the latter sources have a tail towards higher SFRs. The disks appear to be predominantly found within the MS with a median $\Delta_{\rm MS}$ of 1.4, while irregulars are typically starbursts with a median $\Delta_{\rm MS}$ of 4.6. Irregulars are on average also warmer than disk-type SMGs (the median $T_{\rm dust}$ values are 42~K and 36~K, respectively), which is consistent with the Genzel et al. (2015) result that the dust temperature increases with distance from the MS. The overall distributions of the gas depletion times also appear to be fairly similar between disks and irregulars, with the disks having only a factor of 1.3 longer median $\tau_{\rm dep}$ than irregulars. 

To conclude, although the rest-frame UV morphologies of SMGs should be taken with reservation, it seems possible that the SMGs classified as irregulars 
are undergoing galaxy mergers, where the dynamical process(es) is triggering the starburst episode. Disks, on the other hand, are more likely to exhibit 
a more steady star formation, which makes the sources to appear as MS galaxies.    

\subsection{Comparison with the physical properties of the 870~$\mu$m selected sources in the Extended \textit{Chandra} Deep Field South (ECDFS)}

Following Miettinen et al. (2017a), we take our main comparison sample of SMGs to be the so-called ALESS SMGs studied by da Cunha et al. (2015). The ALESS SMGs were uncovered in the LABOCA 870~$\mu$m survey of the Extended \textit{Chandra} Deep Field South (ECDFS) or LESS survey by Wei{\ss} et al. (2009), and later followed up with $1\farcs6 \times 1\farcs2$ resolution Cycle 0 ALMA observations at 870~$\mu$m (\cite{hodge2013}; \cite{karim2013}). The median rms noise of the ALESS 870~$\mu$m data is $1\sigma=0.4$~mJy~beam$^{-1}$, which, assuming that $\beta=1.5$, corresponds to the same 1.3~mm sensitivity of $\sim0.1$~mJy~beam$^{-1}$ as our ALMA data have.

Besides the similar depths of our ALMA 1.3~mm survey and the ALESS survey, the three core arguments why we compare our results with those from da Cunha et al. (2015) are the following. First, the ECDFS is one of the best-studied extragalactic fields. Secondly, similar to da Cunha et al. (2015), we also used the new, high-$z$ version of {\tt MAGPHYS} to derive the SMG physical properties, which allows for a direct comparison between the results. Thirdly, the SMG sample of da Cunha et al. (2015) is relatively large; they analysed the 99 most reliable ($>3.5\sigma$ detection within the ALMA primary beam FWHM) SMGs detected in the ALESS 870~$\mu$m survey (the so-called MAIN sample; \cite{hodge2013}). 

As can be seen in Fig.~\ref{figure:averagesed}, the average {\tt MAGPHYS} SED of the ALESS SMGs is practically identical to ours at $\lambda_{\rm rest}>100$~$\mu$m. However, the largest discrepancies are found at wavelengths of $\lambda_{\rm rest}<1$~$\mu$m, and $\lambda_{\rm rest}\sim10-100$~$\mu$m. At least partly, this is likely to reflect the different sensitivities of the multiwavelength imaging surveys of COSMOS and ECDFS.

To make a fair comparison with the ALESS sample from da Cunha et al. (2015), we limit their sample to those SMGs that have ALMA 870~$\mu$m flux densities corresponding to our ALMA 1.3~mm flux density range in the analysed sample, that is $0.52~{\rm mJy}\leq S_{\rm 1.3\, mm} \leq7.24$~mJy. Under the assumption that $\beta=1.5$, this flux density range corresponds to $2.1~{\rm mJy}\leq S_{\rm 870\, \mu m} \leq29.5$~mJy. The number of MAIN ALESS SMGs that fall in this flux density range is 77 (\cite{hodge2013}), and da Cunha et al. (2015) performed SED fitting for all of them. By comparing the spectroscopic redshifts derived for these SMGs by Danielson et al. (2017) with the photometric redshifts from da Cunha et al. (2015), we found that 11 sources have a significant difference between their $z_{\rm spec}$ and $z_{\rm phot}^{\rm MAGPHYS}$ values ($\vert \Delta z/(1+z_{\rm spec})\vert =0.26-0.89$). However, only six out of these 11 $z_{\rm spec}$ values can be considered reliable (quality flag $Q=1$ or $Q=2$ in \cite{danielson2017}), while the remaining five $z_{\rm spec}$ values are only tentative because they are based on only one or two faint spectral features ($Q=3$; \cite{danielson2017}). Hence, we exluded the aforementioned six sources with poor $z_{\rm phot}^{\rm MAGPHYS}$ values, which makes the size of our ALESS comparison sample to be 71. The photometric redshifts of these SMGs, as derived by da Cunha et al. (2015), lie in the range of $z_{\rm phot}^{\rm MAGPHYS}=1.33-5.82$ with a median and 16th--84th percentile range of $z_{\rm phot}^{\rm MAGPHYS}=2.97^{+1.22}_{-0.99}$. This median redshift is a factor of 1.29 higher than that of our analysed SMGs ($z=2.30^{+1.42}_{-0.58}$). Seven sources in our ALESS comparison sample are AGN hosts (ALESS~11.1, 17.1, 57.1, 66.1, 70.1, 73.1, and 84.1), but the AGN emission is not believed to significantly affect the derived SED properties (\cite{dacunha2015}, and references therein). Moreover, one of the sources, ALESS~5.1, is potentially weakly lensed (\cite{hodge2016}).

For the aforementioned flux-limited ALESS sample, the median values and the 16th--84th percentile ranges of the physical parameters are 
$\log(M_{\star}/{\rm M}_{\sun})=10.92^{+0.46}_{-0.53}$, $\log(L_{\rm dust}/{\rm L}_{\sun})=12.66^{+0.26}_{-0.37}$, 
${\rm SFR}=407^{+417}_{-255}$~M$_{\sun}$~yr$^{-1}$, ${\rm sSFR}=5.8^{+9.0}_{-4.7}$~Gyr$^{-1}$, $T_{\rm dust}=43.0^{+5.0}_{-6.8}$~K, and $\log(M_{\rm dust}/{\rm M}_{\sun})=8.87^{+0.25}_{-0.29}$. These median values are 0.9--2.1 times those derived by da Cunha et al. (2015) for their full sample of 99 ALESS SMGs (see their Table~1), where the largest discrepancy is found for the {\tt MAGPHYS}-based sSFR. That the medians are mostly similar is not surprising because our comparison subsample is composed of 71 ALESS SMGs, which make 72\% of the full sample analysed by da Cunha et al. (2015). We note that da Cunha et al. (2015) defined their SFR to be averaged over the last 10~Myr, while the corresponding timescale in the present study is $\Delta t =100$~Myr. Hence, for a proper comparison with their results, we will here refer to our {\tt MAGPHYS}-based SFR($\Delta t=10$~Myr) values as well.

In what follows, we compare the physical properties of our SMGs with those of the aforementioned flux-limited ALESS sample. The ratios between our median $M_{\star}$, $L_{\rm dust}$, SFR, $T_{\rm dust}$, and $M_{\rm dust}$ values and those of the ALESS SMGs are provided in Table~\ref{table:comparison}. As can be seen in Table~\ref{table:comparison}, the median properties are in the same ballpark, the ratios (AzTEC/ALESS) between the different parameters ranging from 0.8 to 1.5. We applied the Welch's $t$-test (\cite{welch1947}) to the comparison of the samples with unequal population variance. The null hypothesis was that the mean values of the two independent samples are identical, under the assumption of a normally distributed parent population. The $t$-test statistics ($t$) and $p$-values for the comparisons of $M_{\star}$, $L_{\rm dust}$, ${\rm SFR}_{\rm MAGPHYS}(10\,{\rm Myr})$, $T_{\rm dust}$, and $M_{\rm dust}$ are given in Table~\ref{table:comparison}. In general, the highest $p$-values are found for those parameters whose medians are also similar to each other.

To quantify the sample comparison further, we also performed a two-sided Kolmogorov–Smirnov (K-S) test between the aforementioned physical parameters of our ALMA sources and the flux-limited ALESS sample. The null hypothesis was that these two samples are drawn from a common underlying parent distribution. The K-S test statistics ($D_{\rm KS}$) and $p$-values for the comparisons of the $M_{\star}$, $L_{\rm dust}$, ${\rm SFR}_{\rm MAGPHYS}(10\,{\rm Myr})$, $T_{\rm dust}$, and $M_{\rm dust}$ values are also given in Table~\ref{table:comparison}. For the dust luminosity values, besides the similar mean and median values of the samples, we also found a comparatively high K-S probability of $p=0.48$. This is expected, because our ALESS comparison sample was constructed from sources, which are equally bright to our ALMA sources. However, although the sample medians of other parameters are fairly similar to each other, the K-S test results suggest that the parameters are unlikely to be drawn from the same parent distribution  (the $p$-values range from $p=0.003$ to $p=0.22$). A caveat is that our ALESS comparison sample is smaller than our sample by a factor of 1.75, and hence the K-S tests presented here can be subject to small number statistics. This is indeed suggested by the shapes of the parameter distributions, which are fairly similar between our ALMA sources and the ALESS sources, except that we have more sources in certain parameter intervals, for example in the stellar mass range of $\sim10.6-11.6$ in log-10 solar units, at dust luminosities of $\gtrsim10^{13}$~L$_{\sun}$, in the dust temperature range of $\sim30-40$~K, and in the dust mass range of $\sim9-9.3$ in log-10 solar units. Nevertheless, at least part of the differences found here could be caused by the different selection wavelength ($\lambda_{\rm obs}^{\rm AzTEC}=1.1$~mm versus $\lambda_{\rm obs}^{\rm LESS}=870$~$\mu$m), which is also probing the different rest-frame wavelengths at the median source redshifts. Another factor, as mentioned above, might be the different depths of the optical--IR observations available in COSMOS and the ECDFS, which might be the reason behind the different stellar masses based on these short-wavelength data (see also \cite{miettinen2017a}). 

da Cunha et al. (2015) found that, at $z \simeq 2$, half of the ALESS SMGs (49\%) lie above the galaxy MS (i.e. $\Delta_{\rm MS}>3$), while the other half of the sources (51\%) are consistent with being at the high-$M_{\star}$ end of the MS, where their MS prescription was also adopted from Speagle et al. (2014). For the flux-density limited ALESS SMG sample analysed here (at a median redshift of $z_{\rm phot}=2.97$), these percentages are different: only $32.4\%\pm6.8\%$ of the sources are found to have $\Delta_{\rm MS}>3$, while $64.8\%\pm9.6\%$ lie within a factor of three of the MS, and two sources ($2.8\%\pm2.0\%$) lie below the MS. 

The percentages we found for our SMGs are more consonant with the full ALESS sample, that is $41.9\%\pm5.8\%$ are above the MS, and $57.3\%\pm6.8\%$ are consistent with the MS. If we base our analysis on the 10~Myr-averaged SFRs output by {\tt MAGPHYS} as da Cunha et al. (2015) did, we find that $39\% \pm 6\%$ of our SMGs lie at $\Delta_{\rm MS}>3$, $58\% \pm 7\%$ lie within the MS, and $3\% \pm 2\%$ of the sources fall below the MS ($\Delta_{\rm MS}<1/3$). These percentages are consistent with those derived from the Kennicutt (1998) $L_{\rm IR}-{\rm SFR}$ calibration.

\begin{table}[H]
\renewcommand{\footnoterule}{}
\caption{Comparison of the physical properties between our ALMA detected AzTEC SMGs and the equally bright ALESS SMGs.}
{\normalsize
\begin{minipage}{1\columnwidth}
\centering
\label{table:comparison}
\begin{tabular}{c c}
\hline\hline 
Parameter\tablefootmark{a} & Value\\
\hline 
$M_{\star}^{\rm AzTEC}/M_{\star}^{\rm ALESS}$ & $1.5^{+11.4}_{-1.3}$ \\ [1ex]
$L_{\rm dust}^{\rm AzTEC}/L_{\rm dust}^{\rm ALESS}$ & $0.9^{+4.9}_{-0.7}$ \\ [1ex]
${\rm SFR}^{\rm AzTEC}/{\rm SFR}^{\rm ALESS}$ & $0.8^{+8.9}_{-0.6}$\tablefootmark{b} \\ [1ex]
$T_{\rm dust}^{\rm AzTEC}/T_{\rm dust}^{\rm ALESS}$ & $0.9^{+0.4}_{-0.3}$ \\ [1ex]
$M_{\rm dust}^{\rm AzTEC}/M_{\rm dust}^{\rm ALESS}$ & $1.4^{+2.9}_{-1.0}$ \\[1ex]
\hline 
\multicolumn{2}{c}{$t$-test results\tablefootmark{c}}\\ [1ex]
\hline
$M_{\star}$ & $t=2.12$, $p=0.04$ \\ [1ex]
$L_{\rm dust}$ & $t=1.00$, $p=0.32$ \\ [1ex]
${\rm SFR}$ & $t=0.53$\tablefootmark{b}, $p=0.59$\tablefootmark{b} \\ [1ex]
$T_{\rm dust}$ & $t=-1.15$, $p=0.25$ \\ [1ex]
$M_{\rm dust}$ & $t=2.74$, $p=0.007$ \\ [1ex]
\hline
\multicolumn{2}{c}{K-S test results\tablefootmark{d}}\\ [1ex]
\hline
$M_{\star}$ & $D_{\rm KS}=0.15$, $p=0.22$ \\ [1ex]
$L_{\rm dust}$ & $D_{\rm KS}=0.12$, $p=0.48$ \\ [1ex]
${\rm SFR}$ & $D_{\rm KS}=0.17$\tablefootmark{b}, $p=0.13$\tablefootmark{b} \\ [1ex]
$T_{\rm dust}$ & $D_{\rm KS}=0.22$, $p=0.02$ \\ [1ex]
$M_{\rm dust}$ & $D_{\rm KS}=0.26$, $p=0.003$ \\ [1ex]
\hline
\end{tabular} 
\tablefoot{\tablefoottext{a}{The ratio between the sample medians, where the error bars were calculated from the corresponding 16th and 84th percentiles.}\tablefoottext{b}{The comparison was done between the {\tt MAGPHYS} output values averaged over 10~Myr.}\tablefoottext{c}{Results from a two-sided Welch's $t$-test between the two sets of physical properties. The $t$-statistic tests whether the sample mean values are different, and the corresponding $p$-value gives the probability for the the null hypothesis that the samples have identical mean values.}\tablefoottext{d}{Results from a two-sided K-S test between the two sets of physical properties. The maximum distance between the two cumulative distribution functions is given by the K-S test statistic $D_{\rm KS}$, while the corresponding $p$-value describes the probability that the two data sets are drawn from the same underlying parent distribution. }   }
\end{minipage} }
\end{table}

\subsection{The studied submillimetre galaxies in a wider context of massive galaxy evolution}

A multitude of physical properties we derived for our large sample of SMGs allow us to discuss these sources in a broader 
context of the evolution of massive galaxies. One important finding in this respect is that less than half ($42\%\pm6\%$) 
of our SMGs appear to lie above the MS (by a factor of greater than three), and can be considered starbursts. The remainder of the sample ($57\%\pm7\%$) 
appears to be composed of highly star-forming MS galaxies. Considering our sample and the $L_{\rm IR}$ selection effect it is subject to (Fig.~\ref{figure:LIR}), 
the dividing IR luminosity limit between MS SMGs and starburst SMGs is $L_{\rm IR}\sim9\times10^{12}$~L$_{\sun}$ (Fig.~\ref{figure:sblir}). 
The IR-based SFRs of the super-MS, starburst SMGs are $\sim39-6\,500~{\rm M}_{\sun}~{\rm yr}^{-1}$ with a median of $767~{\rm M}_{\sun}~{\rm yr}^{-1}$, 
while those for the MS SMGs were derived to be $\sim50-3\,700~{\rm M}_{\sun}~{\rm yr}^{-1}$ with a median of $346~{\rm M}_{\sun}~{\rm yr}^{-1}$. Clearly, the star formation activity is very intense in the MS SMGs as well. Figure~\ref{figure:sSFR} illustrates how our SMGs exhibit an increasing MS slope, or sSFR towards higher redshifts (left panel), and stronger level of starburstiness at earlier cosmic times (right panel). In particular, we found that, on average, our $z>3$ SMGs are found above the MS, while at $z<3$ the SMGs typically populate the MS. Although the abrupt jump in the starburstiness at $z\sim3$ shown in the right panel in Fig.~\ref{figure:sSFR} can be understood as a selection bias resulting from the sensitivity limit of our dust continuum data, it is also possible that it is partly reflecting the higher abundance of molecular gas reservoir available for star formation in higher redshift SMGs (see below), and the higher efficiency at which dense gas is converted into stars.

The aforementioned very high SFRs could be triggered by gas-rich galaxy mergers (e.g. \cite{noguchi1986}; \cite{mihos1996}), or by gravitational instabilities in very gas-rich galaxies for which the gas supply is maintained by continuous, cold gas accretion (\cite{narayanan2009}; \cite{dekel2009a}; \cite{engel2010}; \cite{wiklind2014}; \cite{narayanan2015}). As shown in Fig.~\ref{figure:lifetime}, the MS SMGs appear to exhaust their molecular gas reservoir through star formation slower on average than the starburst objects. In relation to this, we found that the MS SMGs exhibit a stronger positive correlation between their gas and stellar mass contents than the starburst SMGs (Fig.~\ref{figure:ismvsstar}), which is in line with the view that the MS objects can maintain their high observed SFRs over longer timescales as a result of being supplied with gas via cold accretion from the cosmic filaments. In Fig.~\ref{figure:gasfraction}, we can recognise an increasing trend in the gas fraction as a function of redshift, with a hint of flattening at $z\gtrsim3$, although the overall behaviour is raveled by the very high gas fractions of the lowest redshift bins, which might be a signature of the overestimated gas masses.

Related to the potentially different star formation properties of MS and starburst SMGs, 
it is still unsettled whether normal star-forming disks and starbursts follow different 
star formation scaling laws. This question is most often addressed by constructing the K-S -type diagram. 
In the integrated K-S diagram shown in Fig.~\ref{figure:sk} (right panel), we see the expected positive correlation 
between the SFR and $M_{\rm gas}$, although the MS and starburst relations are comparable to each other, 
both in their slopes and normalisations. Here, it is important to emphasise that 
the Scoville et al. (2016) gas mass estimator we used is calibrated for a single, Galactic CO-to-H$_2$ conversion factor, 
while the bi-modal star formation laws presented by some authors (e.g. \cite{daddi2010}) might, at least partly, 
be due to different assumptions about the $\alpha_{\rm CO}$ conversion factor for different types of galaxies. 
Hence, more information about the $\alpha_{\rm CO}$ values of our target SMGs is needed to 
study whether starbursts are more efficient star formers (in terms of the SFR/$M_{\rm gas}$ ratio) than the MS SMGs.

In Sect.~4.7, we demonstrated how those SMGs that are classified as irregulars on the basis of their rest-frame UV emission, and hence are potential mergers, tend to be starbursts. On an individual basis, there is persuasive evidence that one of our very high-redshift SMGs, AzTEC/C17 at $z_{\rm spec}=4.542$, is involved in an ongoing major merger, namely the broad CO$(4-3)$ line emission ($\sim10^3$~km~s$^{-1}$ to zero intensity), and its highly disturbed morphology in the rest-frame UV (\cite{schinnerer2008}). Another example is AzTEC/C4 that was resolved into two components at $0\farcs05$ resolution with ALMA at $\lambda_{\rm obs}=860$~$\mu$m by Iono et al. (2016). This indicates a mid-stage merger with a projected separation of $\sim1.5$~kpc between the interacting pair. At still higher resolutions of $0\farcs017 \times 0\farcs014$ and $0\farcs026 \times 0\farcs018$, the authors found that AzTEC/C2a and C5 both exhibit a double nucleus structure with a separation of $\sim200$~pc and $\sim150$~pc between the nuclei, respectively. This suggests that the latter two SMGs are observed near the final stage of merger. As discussed in the companion paper by Miettinen et al. (2017b), AzTEC/C22 and C42 are also found to be ongoing merger systems, where the dust-emitting components are separated by 13.8~kpc in the former one (the C22a and C22b components), and by 5.3~kpc in the latter SMG. 

On the other hand, in Sect.~4.6 we found evidence that the MS SMGs exhibit, on average, larger radio-emitting disks than their super-MS counterparts. This is a potential footprint of more widespread star-forming disks among the MS objects, where the star formation is governed by gravitational disk instabilities rather than violent merger events. Consistent with this, in Sect.~4.7 we found that disk-like morphologies in the rest-frame UV are most common among the MS SMGs. It is worth mentioning, however, that disky structures can re-form rapidly after a major merger event, or perhaps even survive the event to some degree (e.g. \cite{hayward2011}; \cite{hopkins2013}, and references therein). 

In what follows, we attempt to place our SMGs at two different redshift bins, at $z>3$ and $z\leq3$, in a context of massive galaxy 
evolution, the main emphasis being in the stellar mass assembly. The comparison samples of different galaxy populations discussed in the following two subsections are summarised in Table~\ref{table:pop}.

\subsubsection{Submillimetre galaxies within the first $\sim2$~Gyr of the universe ($z>3$)}

Thirty-six out of the 124 SMGs analysed here ($29\%$) lie at $z>3$, that is at the cosmic epoch where the progenitors of cQGs seen at $z\sim2$ are believed to exist (\cite{toft2014}). However, only five of our $z>3$ SMGs are spectroscopically 
confirmed. The cosmic time interval between $z>3$ and $z\sim2$, $\Delta t \gtrsim1.1$~Gyr, is much longer than 
the gas depletion timescale in our $z>3$ SMGs, which range from about 30~Myr to 760~Myr with a median of 218~Myr (see Sect.~4.1.3).

Following Ivison et al. (2016; their Sect.~4.3), and assuming that the duration of the starburst phase is 100~Myr, we roughly estimate that the comoving number density of our $z>3$ SMGs is $n_{\rm C}>1.8\times10^{-5}$~Mpc$^{-3}$, which is scaled by the ASTE/AzTEC survey area of 0.72~deg$^2$ from which our sources were initially selected. A lower limit to $n_{\rm C}$ is the result of neglecting the sample completeness corrections, which will be quantitatively studied elsewhere (M.~Aravena et al., in prep.). However, our estimated number density of $z>3$ SMGs exceeds that from Toft et al. (2014) by at least a factor of seven, but clearly these estimates are hampered by the uncertainties in the photometric redshifts of the sources (\cite{brisbin2017}). On the other hand, our comoving number density estimate is in better agreement with that of $z\sim2$, $M_{\star}>10^{11}$~M$_{\sun}$ cQGs derived by Toft et al. (2014), that is $(6.0\pm2.1)\times10^{-5}$~Mpc$^{-3}$.  

Toft et al. (2014) concluded that the similarity between the stellar mass-size relationships of $z\sim2$ cQGs and $z>3$ SMGs supports an evolutionary link between the former high stellar-density systems and the latter high-redshift, highly SFGs. As mentioned in Sect.~4.5, within the measurement uncertainties, there are six to thirteen $z>3$ SMGs in our sample whose stellar masses and radio sizes place them within the scatter of the stellar mass-size relationship of $z\sim2$ cQGs (\cite{krogager2014}). These sources represent about $22\%-48\%$ of all the $z>3$ SMGs plotted in the stellar mass-radio size plane in Fig.~\ref{figure:size}. The scatter seen in the mass-radio size distribution in 
Fig.~\ref{figure:size} is likely related to the finding that the spatial extent of radio emission is generally not a good proxy of that of the stellar distribution in SMGs (see \cite{miettinen2017b}), and hence no clear trend is visible between the size of the non-thermal synchrotron-emitting region and $M_{\star}$. Following Toft et al. (2014), we also constructed a rest-frame UV-emitting size distribution as a function of stellar mass for a subsample of our SMGs. For this purpose, we also re-analysed six of the Toft et al. (2014) SMGs that are common to ours. As shown in the right panel in Fig.~\ref{figure:size}, four out of the seven $z>3$ SMGs ($57\%$) are formally consistent with the $M_{\star}-R$ distribution of $z\sim2$ cQGs. In the case of rest-frame UV emission, however, the SMG size measurements can be hampered by strong dust obscuration, which can be spatially variable and lead to a clumpy source morphology. We conclude that the present analysis of the mass-size plane, and what it tells us about the role of $z>3$ SMGs in the formation of $z\sim2$ cQGs remains unsettled, and careful, dedicated stellar distribution measurement surveys of SMGs are required to better understand their link to the compact systems at $z\sim2$.

\begin{figure*}
\begin{center}
\includegraphics[width=0.45\textwidth]{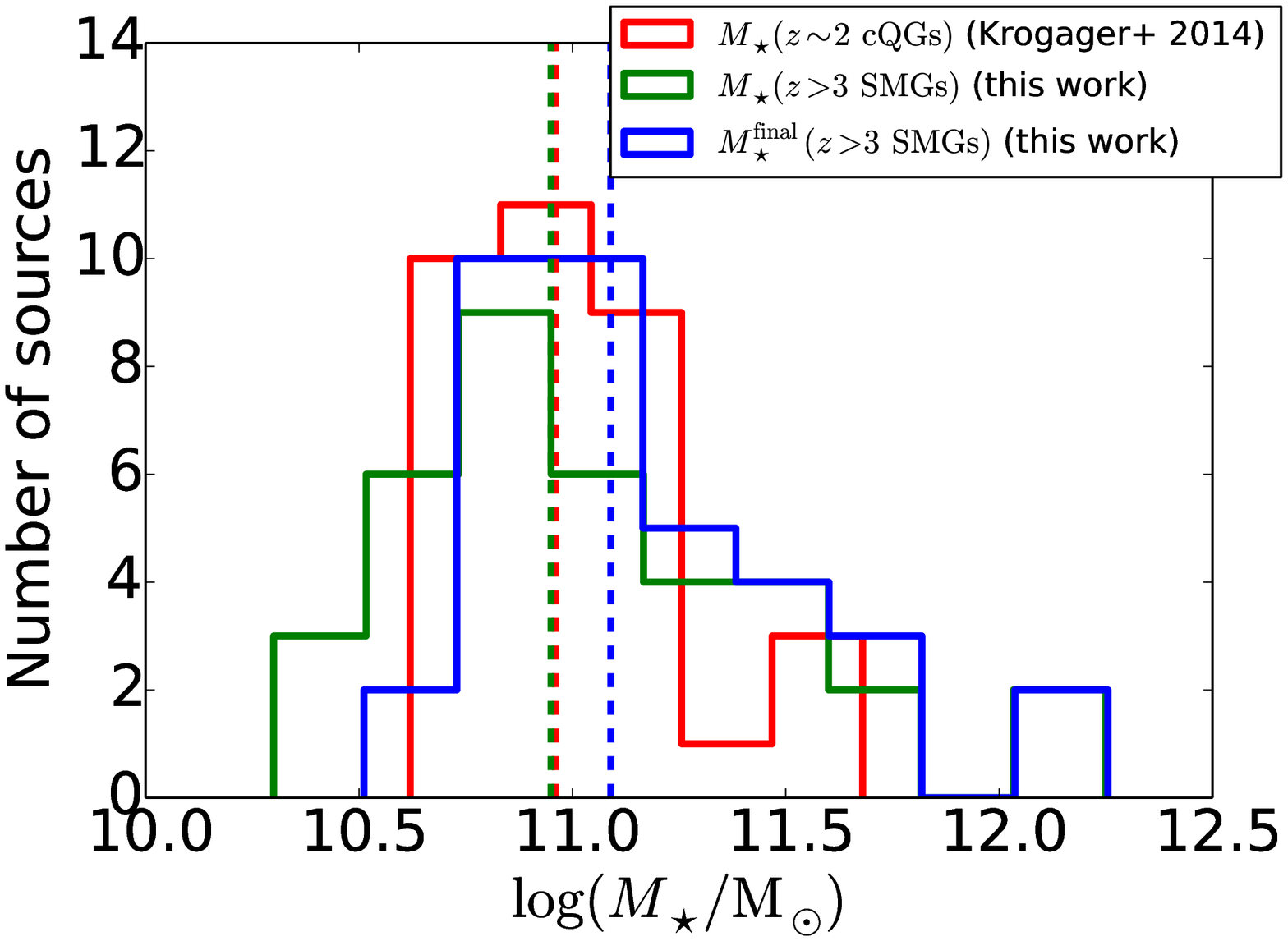}
\includegraphics[width=0.45\textwidth]{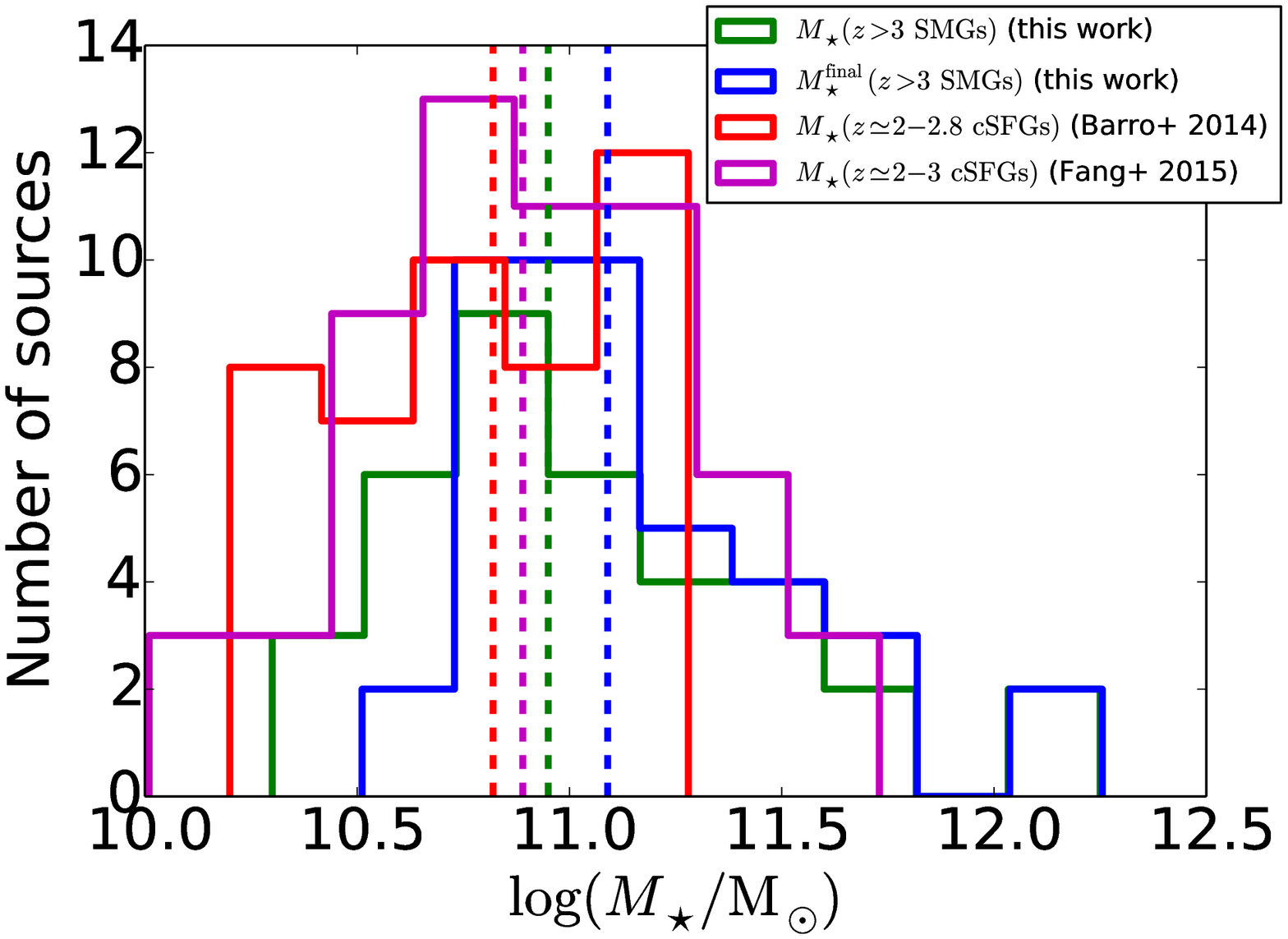}
\caption{\textbf{Left:} Stellar mass distribution of our $z>3$ SMGs (green histogram), and $z\sim2$ cQGs (\cite{krogager2014}; red histogram). The blue histogram shows the distribution of the estimated final stellar masses of our $z>3$ SMGs by assuming that $10\%$ of their putative gas mass content ($M_{\rm gas}$ derived in Sect.~3.2) is converted into stars by the end of the rapid star formation event. The vertical dashed lines mark the median mass values. \textbf{Right:} The green and blue histograms are as in the left panel, while the red and magenta histograms show the stellar mass distributions of cSFGs from Barro et al. (2014) and Fang et al. (2015), respectively.}
\label{figure:masshist}
\end{center}
\end{figure*}

As another quantitative test of the SMG--cQG evolutionary link, we follow the approach of Toft et al. (2014) and Miettinen et al. (2017a), and compare the current stellar mass distributions of our $z>3$ SMGs and $z\sim2$ cQGs, where the latter values are taken from Krogager et al. (2014; see our Table~\ref{table:pop}). The distributions are shown in Fig.~\ref{figure:masshist}. We also plot the distribution of the estimated final stellar masses of our $z>3$ SMGs, which were calculated by using the $M_{\rm gas}$ values derived in Sect.~3.2, and assuming that $10\%$ of the cold molecular gas reservoir is converted into stars ($M_{\star}^{\rm final}=M_{\star}+0.1\times M_{\rm gas}$) by the end of the rapid star formation episode, that is before quenching. This SFE is based on the hydrodynamic simulations by Hayward et al. (2011) as described by Toft et al. (2014), but observations of SMGs suggest that the SFE could be higher by a factor of a few or more (\cite{tacconi2006}; \cite{ma2016}). The final stellar mass distribution, with a median of $\log(M_{\star}/{\rm M}_{\sun})=11.09$, is shown by the blue histogram in Fig.~\ref{figure:masshist}. 

As illustrated in Fig.~\ref{figure:masshist}, the stellar masses of our $z>3$ SMGs are already high, with a median of 
$\log(M_{\star}/{\rm M}_{\sun})=10.95$, which is very similar to that of the $z\sim2$ cQG comparison sample. 
The aforementioned estimates of the final stellar masses suggest a median stellar mass growth by 38\% for 
our $z>3$ SMGs, which leads to a factor of 1.35 higher final median stellar mass than the median $M_{\star}$ of cQGs 
at $z\sim2$. This comparison does not take into account the stellar mass losses, such as the 
fact that the high-mass stars in $z>3$ SMGs die during the aforementioned time interval from $z>3$ to $z\sim2$, that is $\Delta t \gtrsim1.1$~Gyr. However, it is also not taken into account that the stellar mass returned to the ISM can become part of the gas supply from which new stars can form (e.g. \cite{leitner2011}).

A two-sided K-S test between our $M_{\star}^{\rm final}$ distribution and the Krogager et al. (2014) distribution of $M_{\star}$ values for $z\sim2$ cQGs yielded a K-S test statistic of $D_{\rm KS}=0.22$ and a $p$-value of 0.33. This suggests that our final stellar mass distribution of the $z>3$ SMGs and the stellar masses of $z\sim2$ cQGs might not be drawn from a common parent distribution, which is the result of the high-$M_{\star}^{\rm final}$ tail of our SMGs. 

Similarly to the mass-size relationship, the mass distribution analysis presented here is only roughly consistent 
with the proposed evolutionary connection between the $z>3$ SMG and $z\sim2$ cQG populations. However, we note that the source samples analysed here are fairly small (36 SMGs and 34 cQGs), and hence not necessarily well suited for a K-S test. Another obvious caveat here is related to the uncertainty in the gas mass estimates that we mentioned in Sect.~3.2. Nevertheless, if the gas mass estimates are close to the real values, a SFE higher than 0.1 would make the mass distribution discrepancy even larger. 

Above, we addressed a possible evolutionary link between $z>3$ SMGs and $z\sim2$ cQGs, but cQGs have also been found at higher redshifts, from $z\lesssim2.7$ (e.g. \cite{daddi2005}; \cite{kriek2006}; \cite{vandokkum2008}) out to $z\sim4$ (\cite{straatman2014}, 2015); see Table~\ref{table:pop}. For example, to reach the median stellar mass of the $z\sim2.3$ cQGs studied by van Dokkum et al. (2008), our $z>3$ SMGs should grow in their median $M_{\star}$ by a factor of 1.8, while our estimated median final stellar mass falls short by a factor of 1.32. In principle, this could be an indication of SFE being higher than $10\%$. Moreover, the very compact rest-frame far-IR dust-emitting sizes of SMGs, only $\sim1.4-3.1$~kpc in FWHM (\cite{ikarashi2015}; \cite{hodge2016}; \cite{simpson2017}; \cite{miettinen2017c}; see also \cite{miettinen2017b}), suggest that they could evolve into $z\sim2.3$ cQGs, which in the van Dokkum et al. (2008) sample are typically only a few kpc across. Similarly, Ikarashi et al. (2017) suggested that compact SMGs at $z \gtrsim 4$ have the potential to turn into cQGs seen at high redshifts of $z \gtrsim 3$.

Pushing even further back in time, we consider the Straatman et al. (2015) sample of $z\sim3.6$ cQGs, out of which 12.5\% are spectroscopically confirmed (see also \cite{straatman2014}). Owing to the median stellar age of 790~Myr for these high-$z$ cQGs, Straatman et al. (2014) suggested that they likely started to form their stars before $z=5$, and that this might have occurred in a dust-obscured SMG phase. Of the SMGs we have analysed in the present paper, 17 lie at $z>4$, and four at $z>5$. The median stellar masses of these two subpopulations are $\log(M_{\star}/{\rm M}_{\sun})=11.10$ and $\log(M_{\star}/{\rm M}_{\sun})=11.46$. Although some of the very high stellar masses we derived for our highest redshift SMGs can be overestimated owing to the non-detection at relevant optical to near-IR wavebands, it seems possible that they are already too massive (by a factor of $\sim3.8$ for the $z>5$ SMGs) to be good candidates for the progenitors of the typical cQGs identified by Straatman et al. (2014, 2015). 

It is important to note that besides $z>3$ SMGs, also other types of SFGs have been suggested to be the possible
progenitors of $z\sim2$ cQGs. These include the compact SFGs (cSFGs) at $z\gtrsim2$ to $z\sim4$ (e.g. \cite{barro2013}; \cite{patel2013}; \cite{stefanon2013}; \cite{barro2014}; \cite{fang2015}; \cite{vandokkum2015}; \cite{spilker2016}), and the passive, $VJL$ selected galaxies at $z=2.5-4$, which tend to be disk-dominated (\cite{fan2013}). Toft et al. (2014) speculated that the Barro et al. (2013) cSFGs at $z\lesssim3$ might represent a transition phase between some of the $z\gtrsim 3$ SMGs and $z\sim2$ cQGs. 

Fang et al. (2015) concluded that most of the progenitors of the cSFGs underwent gas-rich (wet) galaxy interactions in the past, and hence the question arises whether SMGs at $z>3$ might be a manifestation of such evolutionary stage. To address this question, in the right panel in Fig.~\ref{figure:masshist}, we plot the stellar mass distributions for our $z>3$ SMGs and their estimated final stellar masses as above, along with the cSFG mass distributions from Barro et al. (2014) and Fang et al. (2015). The median 
stellar mass of our $z>3$ SMGs is already 15\% to 35\% higher than that of the $z\sim2-3$ cSFGs before even considering the stellar mass growth of the SMGs, and hence it seems unlikely that these cSFGs would universally evolve from the $z>3$ SMGs. A two-sided K-S test between our estimated final mass distribution of the $z>3$ SMGs and the comparison cSFG mass distributions yielded very low probabilities of $p=4.4\times10^{-4}$ (\cite{barro2014}) and $p=0.016$ (\cite{fang2015}), which also suggest that the two are not linked. 

On the other hand, Barro et al. (2014) concluded that their $z\simeq2-2.8$ cSFGs are the natural progenitors of $z\sim2$ cQGs, and if we compare their median stellar mass with that from Krogager et al. (2014), we would expect a $\sim36\%$ median increase in stellar mass ($16\%$ for the Fang et al. (2015) cSFGs), and a rapid quenching of active star formation after that. If the latter turns out to be true, then it might be at odds with the suggested pathway from the $z>3$ SMGs to the cQG population at $z\sim2$.   

To close, the exact origin(s) of cQGs at $z\sim2$ is still unclear, and also other possible progenitor galaxy populations (other than SMGs) have been proposed in the literature (e.g. \cite{wellons2015}; \cite{belli2017}). Nonetheless, whether the $z>3$ SMGs are the progenitors of $z\sim2$ cQGs or not, the latter types of galaxies can later ($z<2$) grow in size via repeated gas-poor (dry), minor mergers, and undergo an inside-out metamorphosis to become the (central parts of the) giant ellipticals seen in the local universe (\cite{naab2006}; \cite{trujillo2007}; \cite{bournaud2007}; \cite{khochfar2009}; \cite{bezanson2009}; \cite{naab2009}; \cite{hopkins2009}; \cite{vandokkum2010}; \cite{oser2012}; \cite{vandesande2013}; \cite{toft2014}). 

\subsubsection{Submillimetre galaxies in the universe older than $\sim2$~Gyr ($z\leq3$)}

Following Miettinen et al. (2017a), we also discuss the possible role played by the $z\leq3$ SMGs in massive galaxy evolution. Of the analysed SMGs, the majority (88/124 or $71\%$) lie at $z\leq3$ (the lowest redshift source is AzTEC/C49 at $z_{\rm phot}=0.87^{+0.23}_{-0.33}$). 
As in Sect.~4.9.1, we estimate that the lower limit to the comoving number density of these sources is $n_{\rm C}>2.1\times10^{-4}$~Mpc$^{-3}$. 
Because massive cQGs have also been found at $z<2$, such as most of the $BzK$ selected sources from Daddi et al. (2005), and the spectroscopically confirmed sample of $0.9<z_{\rm spec}<1.6$ sources studied by Belli et al. (2014), one could think of a scenario where these galaxies are the evolutionary outcome of the SMG phase at $z\leq3$. 

\begin{figure}[!h]
\centering
\resizebox{0.999\hsize}{!}{\includegraphics{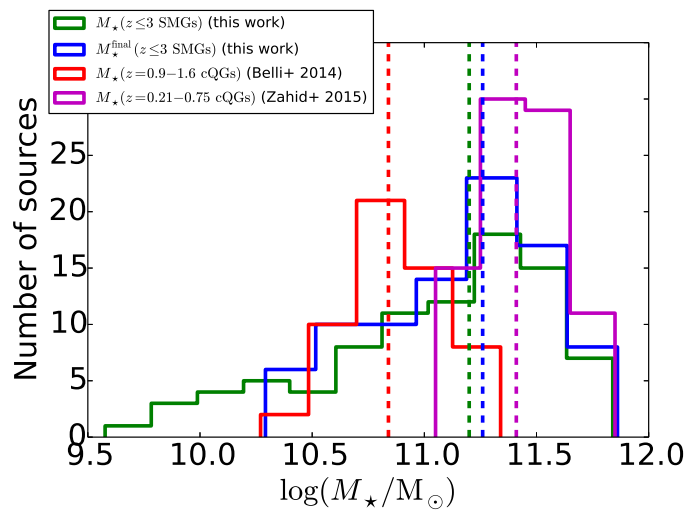}}
\caption{Stellar mass distribution of our $z\leq 3$ SMGs (green histogram), and their estimated final stellar mass distribution (blue histogram). Also shown are the stellar mass distributions of the cQGs from Belli et al. (2014; red) and Zahid et al. (2015; magenta). The vertical dashed lines mark the median mass values.}
\label{figure:masshist2}
\end{figure}

In Fig.~\ref{figure:masshist2}, we plot the stellar mass distribution of the Belli et al. (2014) sources, and the stellar masses and estimated final masses of our $z\leq3$ SMGs. As illustrated in the figure, the median stellar mass of our $z\leq3$ SMGs is $\log(M_{\star}/{\rm M}_{\sun})=11.20$. If we limit our sample to $1.6 < z \leq3$ SMGs (75 sources) to consider the potential precursors of the Belli et al. (2014) sources, the median stellar mass remains similar, $\log(M_{\star}/{\rm M}_{\sun})=11.21$. So, without even considering the further stellar mass growth of these sources, their median stellar mass is already a factor of 2.3 (0.37~dex) higher than that of the Belli et al. (2014) sample. Hence, it seems unlikely that the $z\leq3$ SMGs could be the progenitors of $z<2$ cQGs (see also \cite{miettinen2017a}). Instead, the cQGs found at intermediate redshifts might have evolved from massive ($\log(M_{\star}/{\rm M}_{\sun})>10$), cSFGs (\cite{barro2013}, 2014; \cite{belli2015}).

In Fig.~\ref{figure:masshist2}, we also plot the stellar mass distribution of those sources in COSMOS from Zahid et al. (2015) that fulfil the Barro et al. (2013) criterion for massive cQGs, that is $\log(M_{\star}/r_{\rm e}^{1.5})\geq 10.3$~M$_{\sun}$~kpc$^{-1.5}$. This sample is composed of 85 spectroscopically confirmed sources at a median redshift of $z_{\rm spec}=0.438$. The estimated final stellar mass of our $z\leq3$ SMGs falls short by a factor of 1.41 of the median stellar mass of these cQGs, and a significant size growth by a factor of $\sim2.5-4$ (assuming a circular Gaussian profile, and that the central, dust-obscured starburst region of an SMG quenches into a cQG) would be required to reach their sizes. Zahid et al. (2015) suggested that their intermediate-redshift cQGs represent the high-$M_{\star}$ tail of the normal QG population, and these sources might descend from compact, post-starburst (PSB) E+A galaxies (\cite{zahid2016}). Owing to the wide range of properties and environments of E+As (e.g. \cite{bekki2001}; \cite{tran2004}; \cite{pracy2009}), it is unclear what percentage (if any) of them could have evolved from SMGs.  

Nevertheless, the traditional SMGs that are predominantly found at $z<3$ have been shown to be promising candidates for the ancestors of the present-day passive ellipticals (e.g. \cite{smail2002}, 2004; \cite{swinbank2006}; \cite{simpson2014}). For example, Simpson et al. (2014) concluded that this is the case for their sample of 77 ALESS SMGs at a median redshift of $z=2.3$. On the other hand, the median stellar mass of the $z\leq3$ ALESS SMGs with reliable redshifts from da Cunha et al. (2015; 49 sources) is $\log(M_{\star}/{\rm M}_{\sun})= 10.61$, which is a factor of 3.9 (0.59~dex) lower than that of our $z\leq3$ SMGs. Also, a K-S test between the two mass distributions yields the values $D_{\rm KS}=0.36$ and $p=2.7\times10^{-4}$, which suggest that they are not draw from a common parent distribution. If we limit the da Cunha et al. (2015) sample to the sources that were selected from equally bright detections as ours, and which lie at $z\leq3$, we end up with 36 sources whose median stellar mass is $\log(M_{\star}/{\rm M}_{\sun})=10.86$, which is still a factor of 2.2 lower than ours. Hence, it remains unclear whether our 1.1~mm selected, 1.3~mm detected SMGs studied here could follow the same evolutionary path(s) to today's ellipticals as has been suggested for the 870~$\mu$m selected ALESS sources. 

Regarding the mass-size relationship analysed in Sect.~4.5, among our $z\leq3$ SMGs, we found 14 out of 71 sources (20\%) whose nominal stellar masses and radio sizes place them within the $z\sim2$ cQGs' mass-size relationship from Krogager et al. (2014), but within the uncertainties, this number (percentage) could be up to 34 sources (48\%). The percentages are similar (five to 11 sources or 21\%-46\%) in the stellar mass-UV size plane shown in the right panel in Fig.~\ref{figure:size}. This suggests that up to half of our $z\leq3$ SMG population might quench rapidly enough to transform into cQGs at $z\sim2$.

Wild et al. (2016) carried out a number density and stellar mass function study of $0.5<z<2$ PSB galaxies, and speculated that on the basis of these characteristics they might be the descendants of SMGs. Specifically, the characteristic stellar masses were found to be $\log(M_{\star}^{\rm Chabrier}/{\rm M}_{\sun})=10.15\pm0.20$, $10.53\pm0.07$, and $10.39\pm0.06$ at $0.5<z<1$, $1<z<1.5$, and $1.5<z<2$, with the corresponding number densities of about 1.6, 7.1, and 6.2 in units of $10^{-5}$~Mpc$^{-3}$ (see their Table~2). If we split our sample into redshift bins of $1<z \leq 3$, $1.5<z \leq 3$, and $2< z \leq 3$, we end up with 86, 78, and 53 sources, respectively. The median stellar masses and estimated comoving number densities of these sources are 11.20, 11.21, and 11.25 in log-10 solar units, and $>1.9$, $>1.2$, and $>0.7$ in units of $10^{-4}$~Mpc$^{-3}$. Our SMGs in these three redshift intervals have already median stellar masses higher by factors of 4.8--11.2 than the characteristic PSB masses from Wild et al. (2016), and the SMGs' number densities exceed those of the PSBs. Hence, our results clearly pose a challenge for the evolutionary link between SMGs and $z<2$ PSBs.  

Owing to the very high stellar masses of our $z\leq3$ SMGs, they might be more valid candidates for the progenitors of the ultra-massive, early-type galaxies (ETGs) seen at $z<2$. To address this possibility, we selected a spectroscopically confirmed ($z_{\rm spec}=1.242-1.910$) sample of 12 ETGs from Gargiulo et al. (2016; Tables~4 and 5 therein). The estimated median final stellar mass of our $1.91 < z\leq3$ SMGs is in fairly good agreement (higher by a factor of 1.32) with the median $M_{\star}$ of the dense ETGs from Gargiulo et al. (2016). If the dense, $z\sim1.4$ ETGs subsequently puff up in size via non-dissipative dry mergers towards the $z=0$ universe, they could well represent an evolutionary link between our $z\leq3$ SMGs and local massive ellipticals.

Lastly, an important aspect of the link between SMGs and the present-day giant ellipticals is to study the environment where the SMGs tend to sit, because today's massive, red-and-dead ellipticals are found to reside in rich galaxy clusters (e.g. \cite{dressler1980}). Smol{\v c}i{\'c} et al. (2017a) found evidence that AzTEC/C2a, C4, C5, C6a, C10b, C17, C18, and C42 are associated with galaxy overdensities; see also the earlier study of AzTEC/C6a or Cosbo-3 by Aravena et al. (2010). Wang et al. (2016) found that AzTEC/C6a and C6b belong to a $z=2.506$ protocluster, which is composed of 17 spectroscopically confirmed members. On the other hand, because SMGs at $1<z<2$ are not found to be clustered as strongly as their higher redshift counterparts, they are also not as promising candidates for being the progenitors of today's massive ellipticals found in galaxy clusters (\cite{wilkinson2017}; see also \cite{wild2016}). In a potential relation to this, Smail et al. (2014) found that the most active, submm-bright members of the $z = 1.62$ cluster Cl0218.3-0510 reside more in the outskirts of the system, rather than in the densest zone, which is populated by red-and-dead galaxies. The authors suggested that the bright dust emitters might evolve into some of the faint and less massive ellipticals seen at $z\sim0$. These issues will be addressed in a dedicated, forthcoming paper about the environments and potential galaxy overdensities associated with our SMGs.

\begin{table*}[!htb]
\caption{Summary of different galaxy populations discussed in Sect.~4.9.}
{\scriptsize
\begin{minipage}{2\columnwidth}
\centering
\renewcommand{\footnoterule}{}
\label{table:pop}
\begin{tabular}{c c c c c c c}
\hline\hline 
Population & $N_{\rm sample}$\tablefootmark{a} & $z$ & $M_{\star}$\tablefootmark{b} & $M_{\star}^{\rm final}$\tablefootmark{b,c} & $r_{\rm e}$\tablefootmark{d} [kpc] & Reference \\[1ex]
\hline
$z\leq3$ SMGs & 88 & $0.86-2.93$ (2.03) & $9.58-11.84$ (11.20) & $10.29-11.86$ (11.26) & \ldots & This work \\[1ex]
$z\leq3$ SMGs & 52 & $1.33-2.97$ (2.10) & $8.38-11.89$ (10.58) & \ldots & \ldots & \cite{dacunha2015} \\[1ex]
$z>3$ SMGs & 36 & $3.06-6.40$ (3.91) & $10.30-12.25$ (10.95) & $10.51-12.25$ (11.09) & \ldots & This work \\[1ex]
$z>3$ SMGs & 47 & $3.03-6.12$ (3.67) & $10.21-11.87$ (10.92) & \ldots & \ldots & \cite{dacunha2015} \\[1ex]
cSFGs & 59 & $2.03-2.98$ (2.35) & $10.01-11.73$ (10.89) & \ldots & $0.04-7.04$ (1.34) & \cite{fang2015} \\[1ex]
cSFGs & 45 & $1.97-2.79$ (2.43) & $10.20-11.28$ (10.82) & \ldots & $0.29-2.97$ (1.0) & \cite{barro2014} \\[1ex]
cQGs & 85 & $0.213-0.746$ (0.438) & $11.05-11.85$ (11.41) & \ldots & $1.18-9.04$ (3.99) & \cite{zahid2015} \\[1ex]
cQGs & 56 & $0.901-1.598$ (1.242) & $10.27-11.34$ (10.84) & \ldots & $0.73-7.18$ (1.96) & \cite{belli2014} \\[1ex]
cQGs & 7 & $1.39-2.47$ (1.76) & $10.44-11.60$ ($10.74-11.0$)\tablefootmark{e} & \ldots & $0.6-5.6$ (0.8)\tablefootmark{e} & \cite{daddi2005} \\[1ex]
cQGs & 34 & $1.84-2.28$ (2.08) & $10.62-11.68$ (10.96) & \ldots & $0.46-10.0$ (2.67) & \cite{krogager2014} \\[1ex]
cQGs & 9 & $2.02-2.56$ (2.34) & $10.75-11.47$ (11.21) & \ldots & $0.47-2.38$ (0.92) & \cite{vandokkum2008} \\[1ex]
cQGs & 16 & $3.46-4.05$ (3.61) & $10.60-11.25$ (10.88) & \ldots & $0.27-3.22$ (0.58) & \cite{straatman2015} \\[1ex]
ETGs & 12 & $1.242-1.910$ (1.396) & $11.03-11.59$ (11.19) & \ldots & $1.5-4.34$ (2.5) & \cite{gargiulo2016} \\[1ex]
\hline 
\end{tabular} 
\tablefoot{For each parameter, we give the range of values, and the median value is given in parenthesis.\tablefoottext{a}{Number of sources in the sample.}\tablefoottext{b}{The stellar masses are normalised to a Chabrier (2003) IMF, and given in units of $\log(M_{\star}/{\rm M}_{\sun})$.}\tablefoottext{c}{The estimated final stellar masses of our SMGs were calculated by assuming a SFE of $10\%$ (see text for details).}\tablefoottext{d}{Circularised effective radius.}\tablefoottext{e}{The stellar mass ranges were calculated from the range of values reported in Table~3 of Daddi et al. (2005), while the effective radii were read from their Fig.~12.}     }
\end{minipage} }
\end{table*}

\section{Summary and conclusions}

We studied the physical properties of a large, flux-limited sample of 124 SMGs in the COSMOS field. The target SMGs were originally pre-selected in a 1.1~mm dust continuum survey carried out with the ASTE/AzTEC bolometer, and then followed up with our ALMA 1.3~mm continuum imaging. Our main results are summarised as follows:

\begin{enumerate}
\item We used the new version of {\tt MAGPHYS} of da Cunha et al. (2008, 2015) to interpret the observed panchromatic SEDs of our SMGs, and complemented the analysis by estimating the gas masses of the sources from the observed-frame 1.3~mm dust continuum emission. For example, the median values and 16th--84th percentiles for the stellar mass, SFR, dust mass, and gas mass were derived to be $\log(M_{\star}/{\rm M}_{\sun})= 11.09^{+0.41}_{-0.53}$, ${\rm SFR}= 402^{+661}_{-233}$~${\rm M}_{\sun}~{\rm yr}^{-1}$, $\log(M_{\rm dust}/{\rm M}_{\sun})=9.01^{+0.20}_{-0.31}$, and $\log(M_{\rm gas}/{\rm M}_{\sun})=11.34^{+0.20}_{-0.23}$. 
\item We found that the dust-to-stellar mass ratio decreases as a function of redshift, while the gas-to-dust mass ratio exhibits a positive correlation with redshift. The median of the gas-to-dust ratio, $120^{+73}_{-30}$, is in good agreement with a canonical value of 100.
\item The dense gas fraction was found to span a huge range of values from 0.10 to 0.98 with a median of $0.62^{+0.27}_{-0.23}$. Hence, the gas mass estimated from dust emission is typically higher than the stellar mass content, but some of the dust-inferred gas masses can be overestimated owing to the uniform, Galactic CO-to-H$_2$ conversion factor assumed in 
the calculation. The redshift evolution of the gas fraction is broadly consistent with previous studies, that is it rises up to redshifts of $z\sim2-3$, and then shows a plateau or gentle decline.
\item Comparison with the galaxy MS calibration of Speagle et al. (2014) showed that $57.3\% \pm 6.8\%$ of our SMGs are consistent with being MS objects, while $41.9\% \pm 5.8\%$ of the sources can be classified as starbursts. We defined the starbursts as super-MS objects lying above the mid-line of the MS by more than a factor of three.
\item Starbursts are preferentially found at $z\gtrsim3$, beyond which the sSFR of our SMGs exhibits an abrupt jump. Although this is likely to reflect the sensitivity limit of our source selection, the higher sSFR values can be caused by a stronger cosmological gas accretion, higher major-merger rate, or higher efficiency in converting dense gas into stars at $z\gtrsim3$.
\item The gas depletion time was also found to exhibit a wide range of values from $\sim30$~Myr to $\sim5.6$~Gyr with a median of $\sim535$~Myr. We found evidence that the super-MS SMGs exhaust their gas reservoir faster than their MS counterparts (median depletion times are 407~Myr and 644~Myr, respectively).
\item Unlike low-redshift SFGs, our SMGs do not show a trend of increasing SFR as a function of dust mass. Instead, the SFR is fairly constant on average, regardless of whether we consider the full sample, MS SMGs, or the super-MS SMGs. This is likely to reflect our selection of highly star-forming dusty galaxies, where the low-$z$ trend is not visible. In contrast, the SFR appears to increase as a function of the estimated gas masses, that is we see a positive correlation 
in the integrated K-S plane. Supplementing the gas mass and SFR comparisons with the information about the rest-frame far-IR-emitting sizes of our SMGs will allow us to investigate whether the MS and starburst SMGs obey different star formation sequences, and how those are related to the K-S law (\cite{miettinen2017c}). 
\item No obvious trends were found between the SMG sizes measured in the radio and rest-frame UV and the stellar mass. At least partly the lack of correlation can be attributed to the finding that the radio emission from SMGs is not probing the spatial extent of the already formed stellar distribution, and that rest-frame UV radiation can be strongly affected by differential dust obscuration.
\item We found that the MS SMGs exhibit larger observed-frame 3~GHz radio sizes than their starburst counterparts. This might be a manifestation of the formation of stars being more widespread, and hence less intense in MS objects compared to compact starbursts. 
\item The SMGs that are classified as irregulars in the rest-frame UV appear to be predominantly starbursts, while those classified as disks are mostly found within the MS. We interpret this as an evidence of irregulars being merger systems where high-mass stars are being formed at a rapid rate.  
\item Owing to the high stellar and gas masses of the studied SMGs, it is clear that they must evolve over cosmic times into massive galaxies with no ongoing star formation activity. However, SMGs form a diverse group of galaxies, and it is not trivial to link them to other galaxy populations at lower redshifts. Our results are only in broad agreement with the suggested scenario where $z>3$ SMGs quench into very compact, spheroid-like galaxies found at $z \sim 2$, and hence SMGs at $z>3$ are unlikely to evolve into a single class of galaxies by $z \sim 2$. This conclusion is based 
on the mass-size relationship analysis, which however is hampered by the uncertainties in the sizes of SMGs as seen at different wavelengths. More accurate gas masses are also needed to better understand the expected final stellar masses of our $z>3$ SMGs, and how those compare with the masses of $z \sim 2$ cQGs. Another obvious caveat here is that only five of our $z>3$ SMGs are spectroscopically confirmed, and hence a dedicated spectroscopic follow-up is critical to obtain the accurate redshifts, and hence the physical properties and number densities of these sources. On the other hand, it seems possible that the $z \sim 2$ cQGs evolved from compact SFGs, or blue nuggets, rather than from more extreme SMGs. Our results do not support the view that these three types of galaxies would form an evolutionary sequence. Instead, our results agree best with the scenario where the highest redshift ($z \gtrsim 4$) SMGs quench their spatially compact star formation, and evolve into quiescent systems found at high redshifts of $z \gtrsim 3$. Regarding our $z \leq 3$ SMGs, we found some supporting evidence on the basis of their estimated stellar mass growth that they might form the ultra-massive, dense ellipticals seen at $z < 2$. 

The present study underscores the importance of large samples when trying to piece the SMG properties together and paint a coherent picture of massive galaxy evolution. Our forthcoming studies of the rest-frame far-IR-emitting sizes and galaxy environments of the present target SMGs will help to improve our understanding of their role in a wider context of the origin of massive galaxies. 
\end{enumerate}

\begin{acknowledgements}

We would like to thank the anonymous referee for providing comments and suggestions. 
This research was funded by the European Union's Seventh Framework programme 
under grant agreement 337595 (ERC Starting Grant, 'CoSMass'). M.~A. acknowledges 
partial support from FONDE-CYT through grant 1140099. A.~K. acknowledges support by the Collaborative Research Council 956,
sub-project A1, funded by the Deutsche Forschungsgemeinschaft (DFG). E.~S. acknowledges funding from the European Research Council 
(ERC) under the European Union's Horizon 2020 research and innovation programme (grant agreement No. 694343). 
D.~R. acknowledges support from the National Science Foundation under grant number AST-1614213 to Cornell University. 

This work was performed in part at the Aspen Center for 
Physics, which is supported by National Science Foundation grant PHY-1066293. This work was partially supported by 
a grant from the Simons Foundation. The Flatiron Institute is supported by the Simons Foundation. 
This paper makes use of the following ALMA data: ADS/JAO.ALMA\#2012.1.00978.S and 
ADS/JAO.ALMA\#2013.1.00118.S. ALMA is a partnership of ESO (representing its member states), NSF (USA) and NINS (Japan), 
together with NRC (Canada), NSC and ASIAA (Taiwan), and KASI (Republic of Korea), in cooperation with the Republic of Chile. 
The Joint ALMA Observatory is operated by ESO, AUI/NRAO and NAOJ. This paper is also partly based
on data products from observations made with ESO Telescopes at the La Silla Paranal Observatory under 
ESO programme ID 179.A-2005 and on data products produced by TERAPIX and the Cambridge Astronomy Survey Unit 
on behalf of the UltraVISTA consortium. This research has made use of NASA's Astrophysics Data System, 
and the NASA/IPAC Infrared Science Archive, which is operated by the JPL, California Institute of Technology, 
under contract with the NASA. This research has made use of the NASA/IPAC Extragalactic Database (NED) which is 
operated by the Jet Propulsion Laboratory, California Institute of Technology, under contract with the National Aeronautics 
and Space Administration. This research made use of Astropy, a community-developed core Python package for Astronomy 
(\cite{astropy2013}). We gratefully acknowledge the contributions of the entire COSMOS collaboration consisting of more 
than 100 scientists. More information on the COSMOS survey is available at {\tt http://cosmos.astro.caltech.edu}.

PACS has been developed by a consortium of institutes led by MPE (Germany) and including UVIE (Austria); 
KU Leuven, CSL, IMEC (Belgium); CEA, LAM (France); MPIA (Germany); INAF-IFSI/OAA/OAP/OAT, LENS, SISSA (Italy); 
IAC (Spain). This development has been supported by the funding agencies BMVIT (Austria), ESA-PRODEX (Belgium), 
CEA/CNES (France), DLR (Germany), ASI/INAF (Italy), and CICYT/MCYT (Spain). 

SPIRE has been developed by a consortium of institutes led by Cardiff University (UK) and including Univ. 
Lethbridge (Canada); NAOC (China); CEA, LAM (France); IFSI, Univ. Padua (Italy); IAC (Spain); 
Stockholm Observatory (Sweden); Imperial College London, RAL, UCL-MSSL, UKATC, Univ. Sussex (UK); 
and Caltech, JPL, NHSC, Univ. Colorado (USA). This development has been supported by national funding agencies: 
CSA (Canada); NAOC (China); CEA, CNES, CNRS (France); ASI (Italy); MCINN (Spain); SNSB (Sweden); STFC, UKSA (UK); and NASA (USA).

\end{acknowledgements}

\appendix

\section{SED plots}

The SEDs of the target SMGs are shown in Fig.~\ref{figure:seds}, while those of the potential AGN-hosts are 
shown in Fig.~\ref{figure:seds2}.

\begin{figure*}
\begin{center}
\includegraphics[width=0.2465\textwidth]{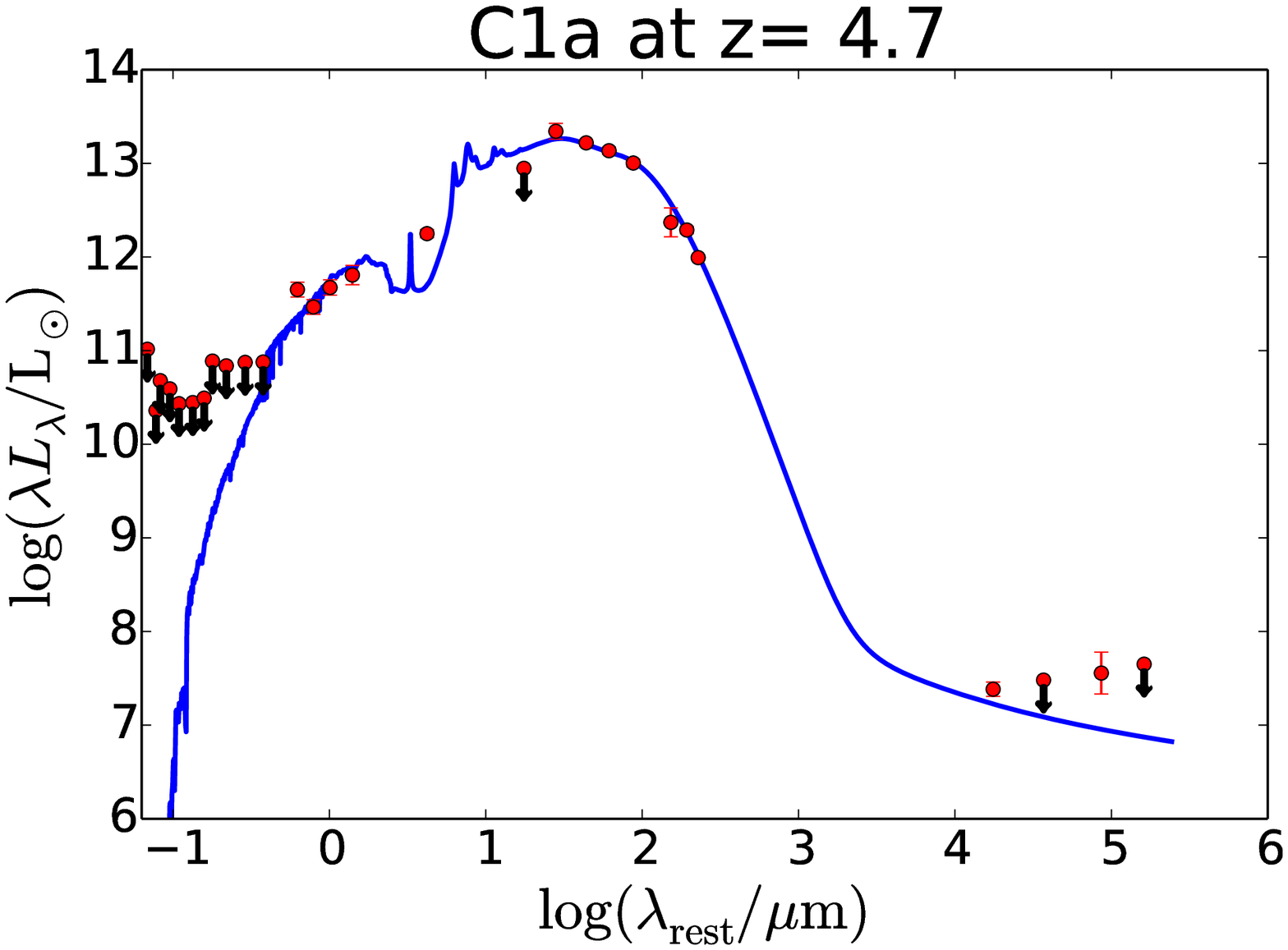}
\includegraphics[width=0.2465\textwidth]{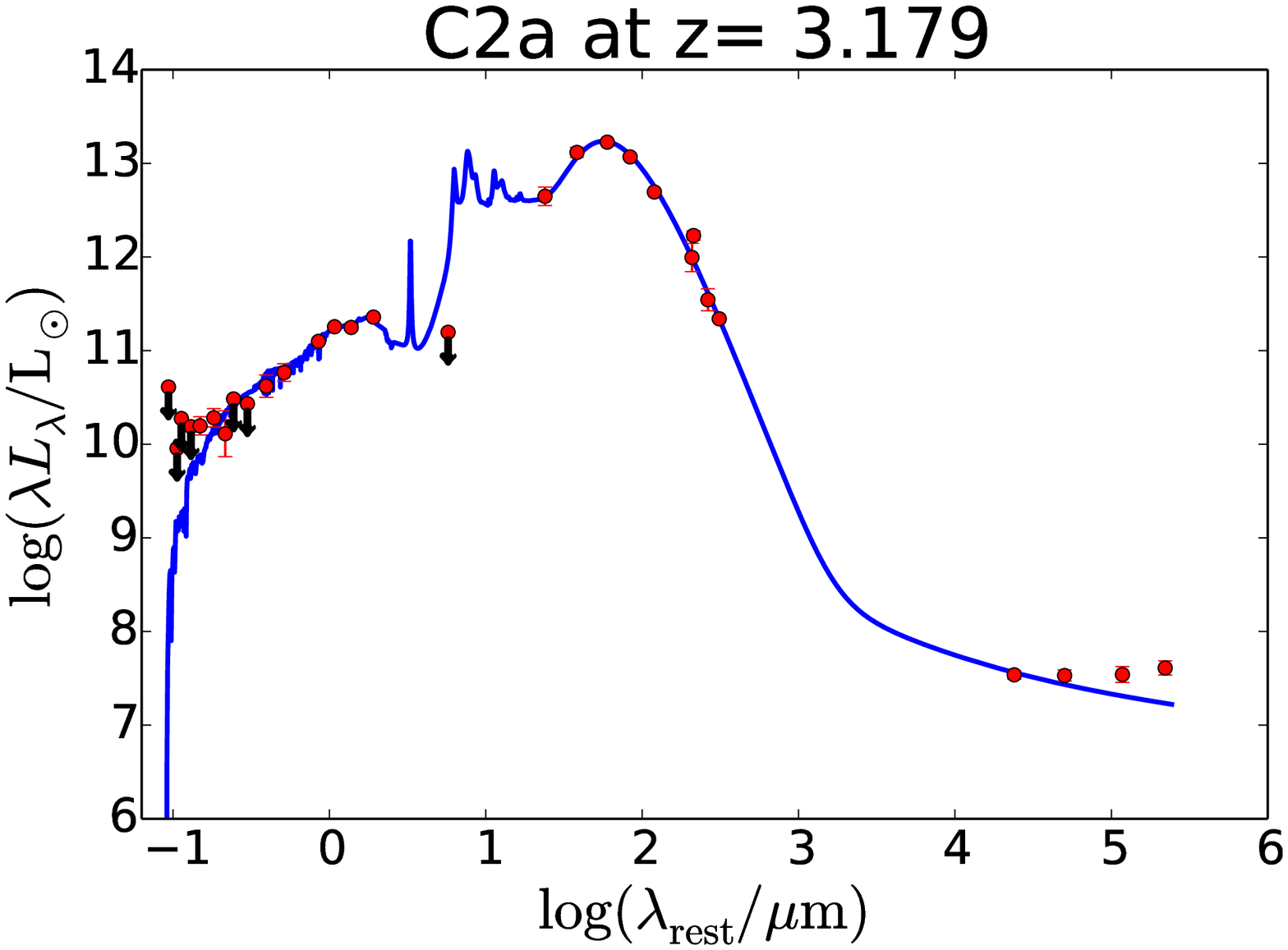}
\includegraphics[width=0.2465\textwidth]{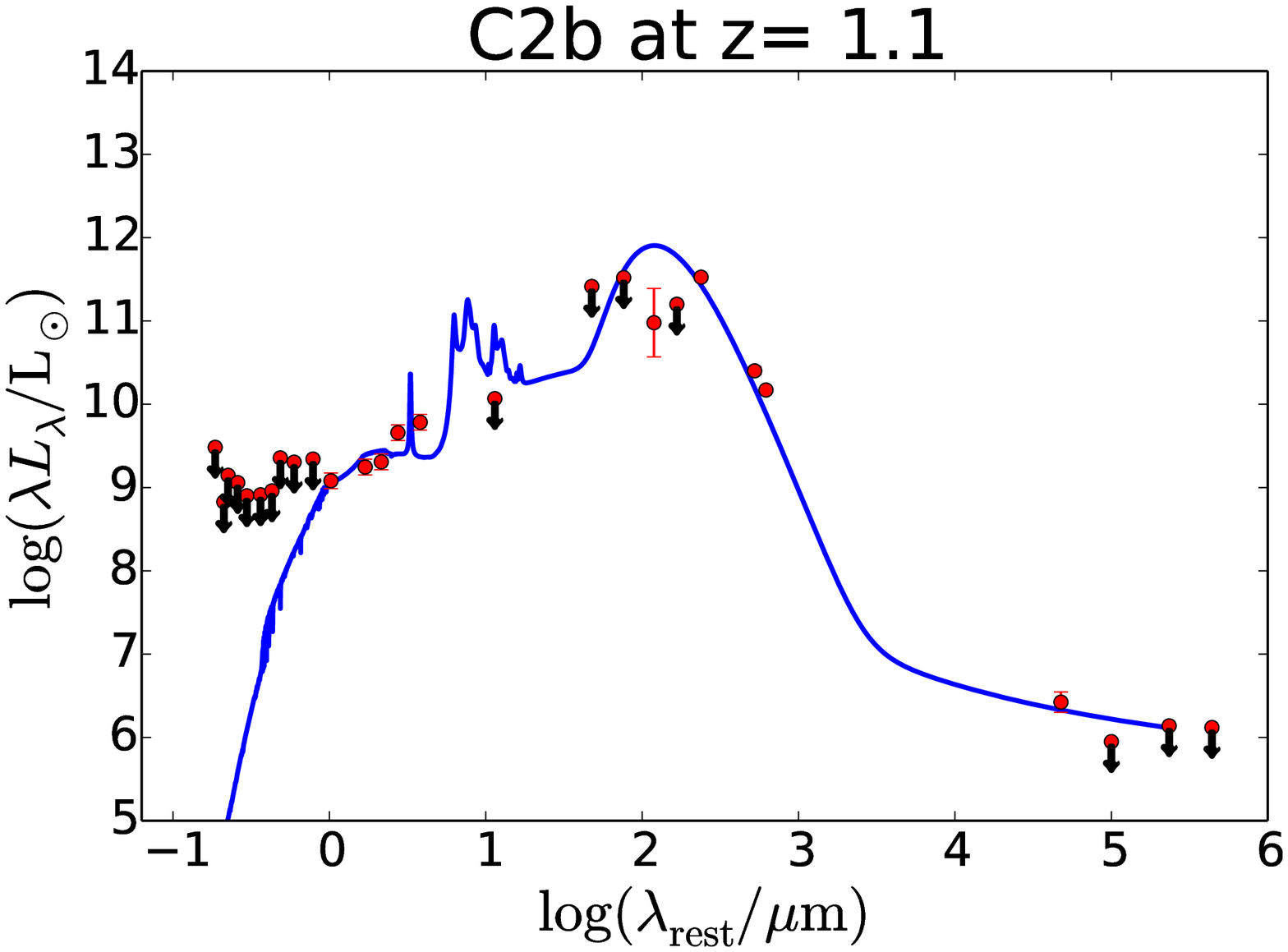}
\includegraphics[width=0.2465\textwidth]{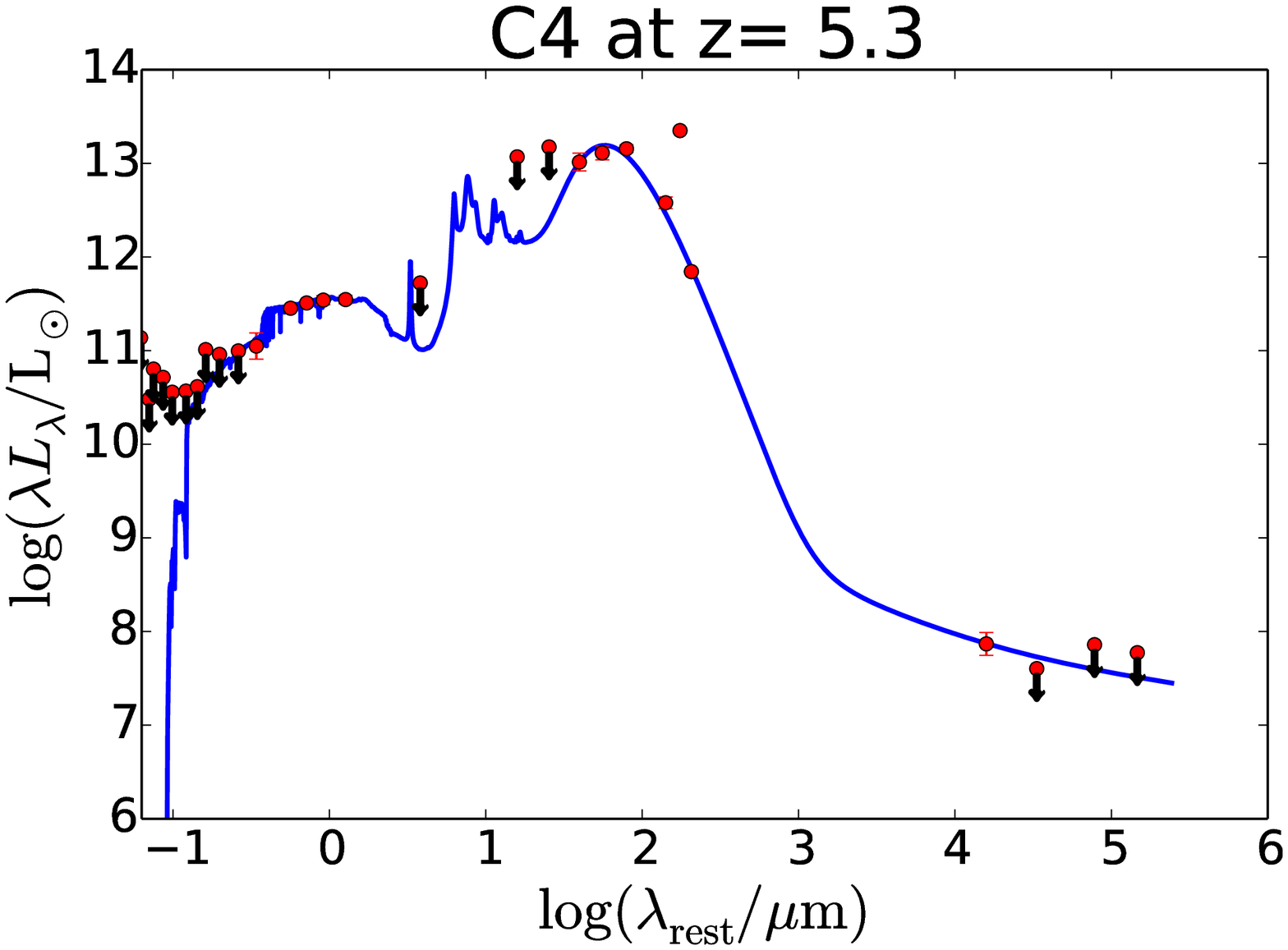}
\includegraphics[width=0.2465\textwidth]{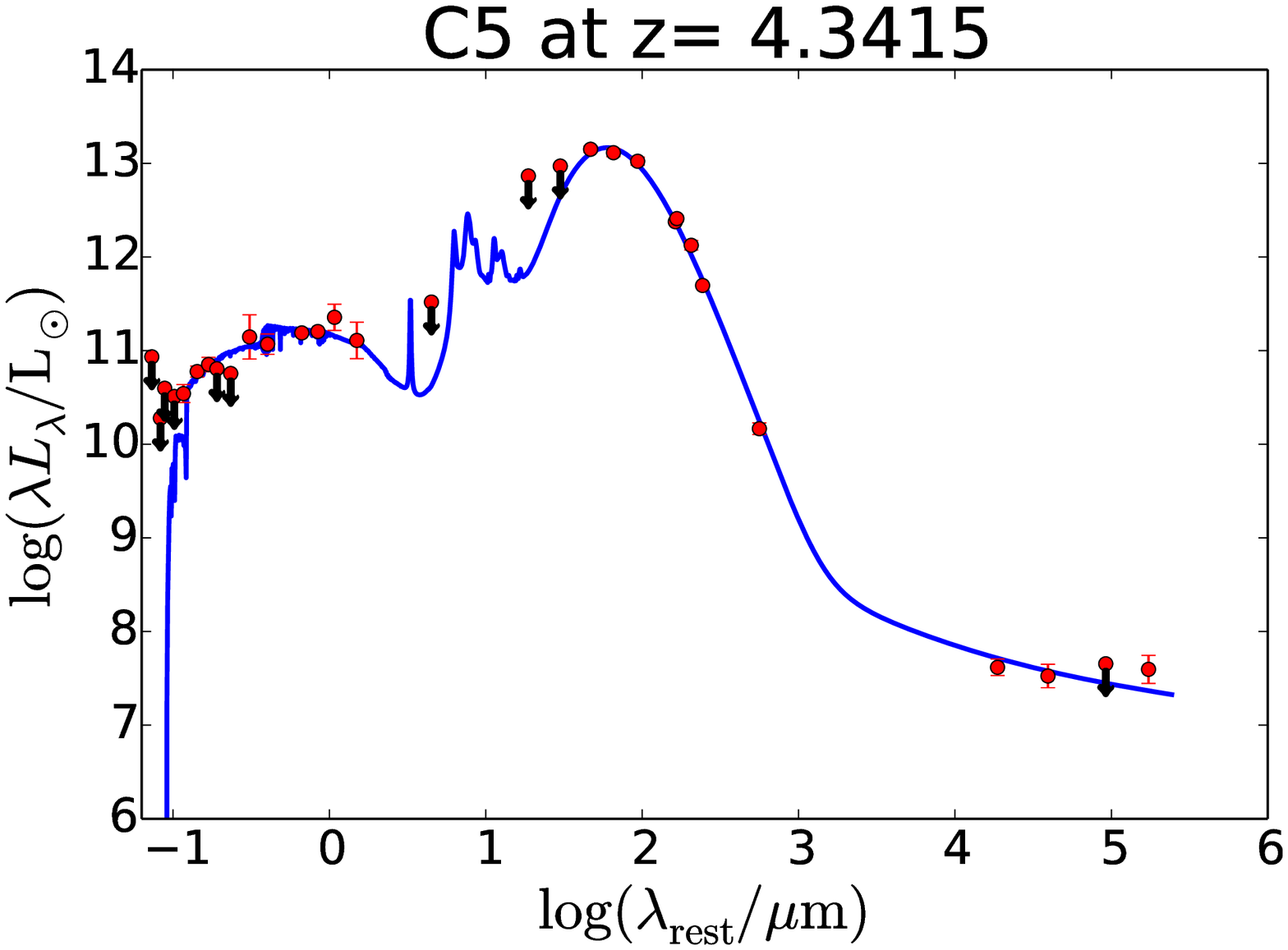}
\includegraphics[width=0.2465\textwidth]{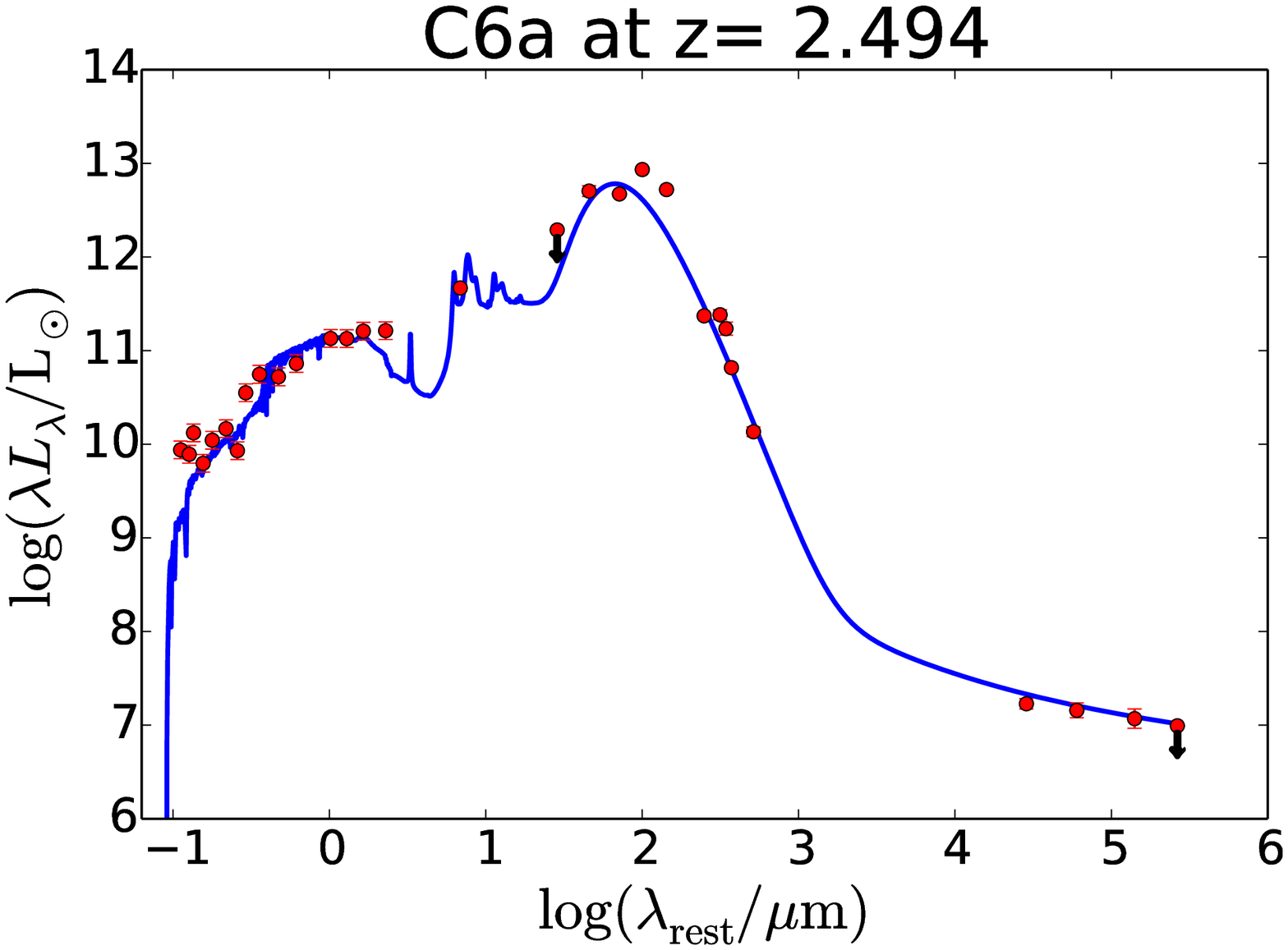}
\includegraphics[width=0.2465\textwidth]{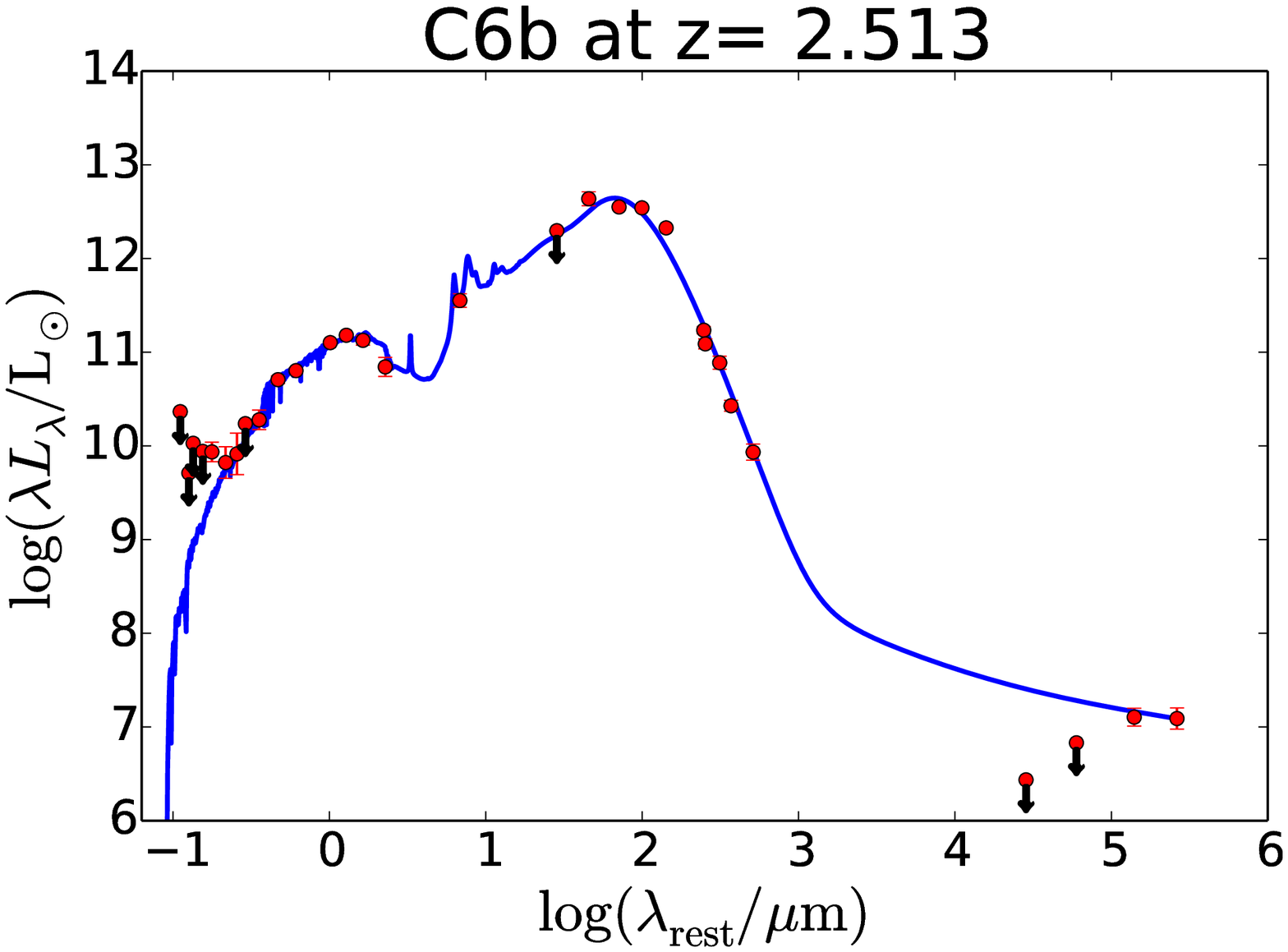}
\includegraphics[width=0.2465\textwidth]{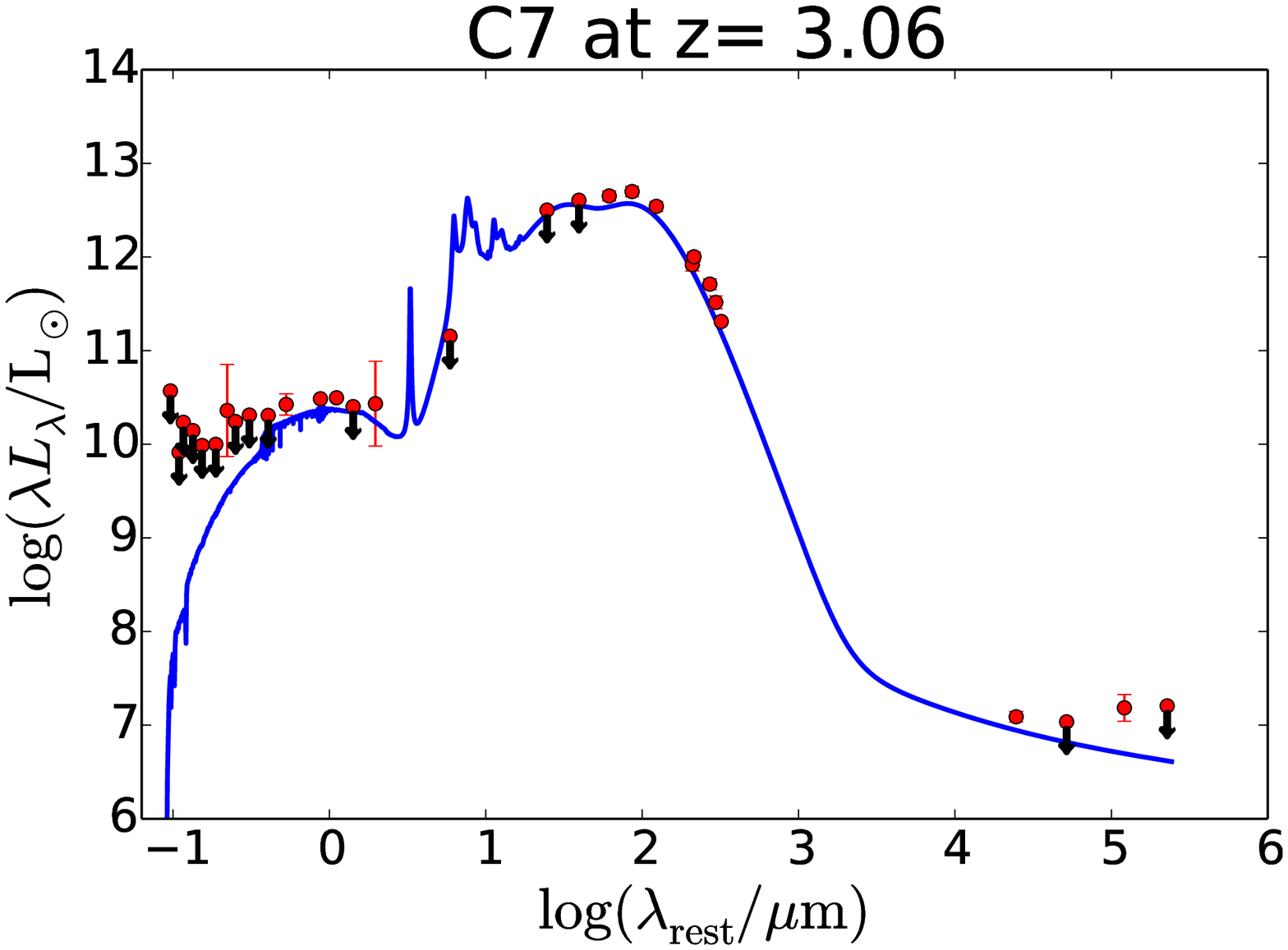}
\includegraphics[width=0.2465\textwidth]{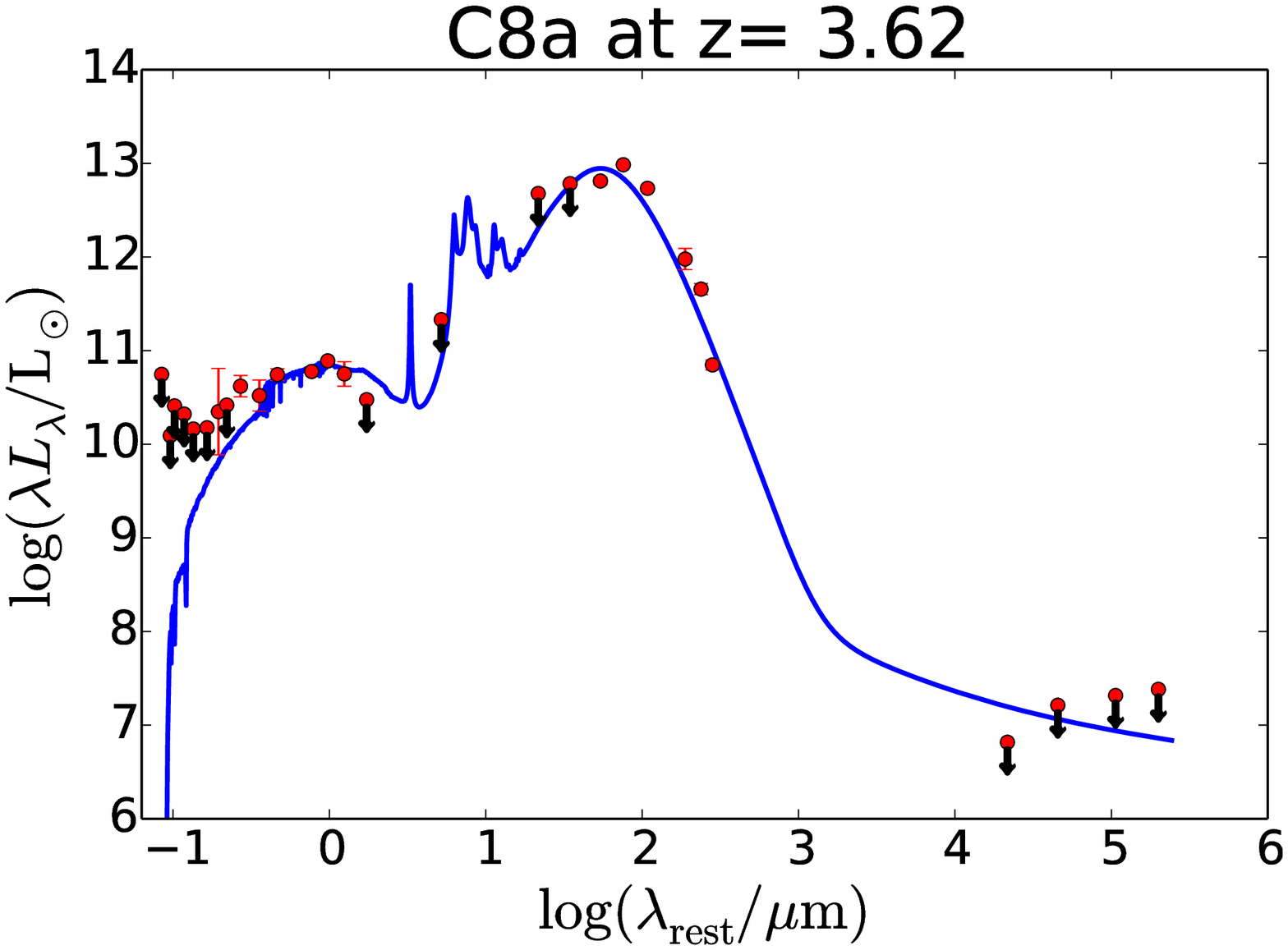}
\includegraphics[width=0.2465\textwidth]{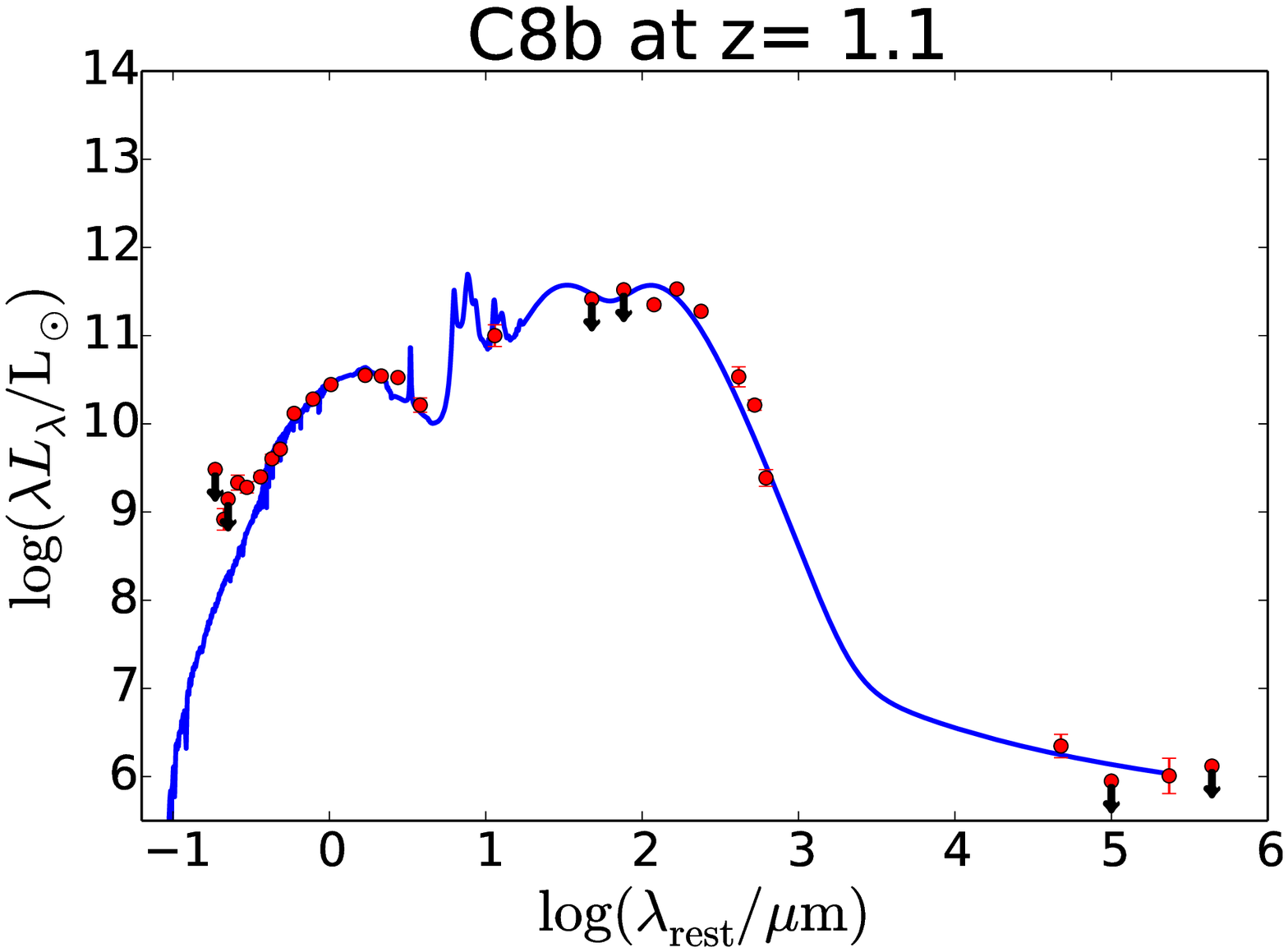}
\includegraphics[width=0.2465\textwidth]{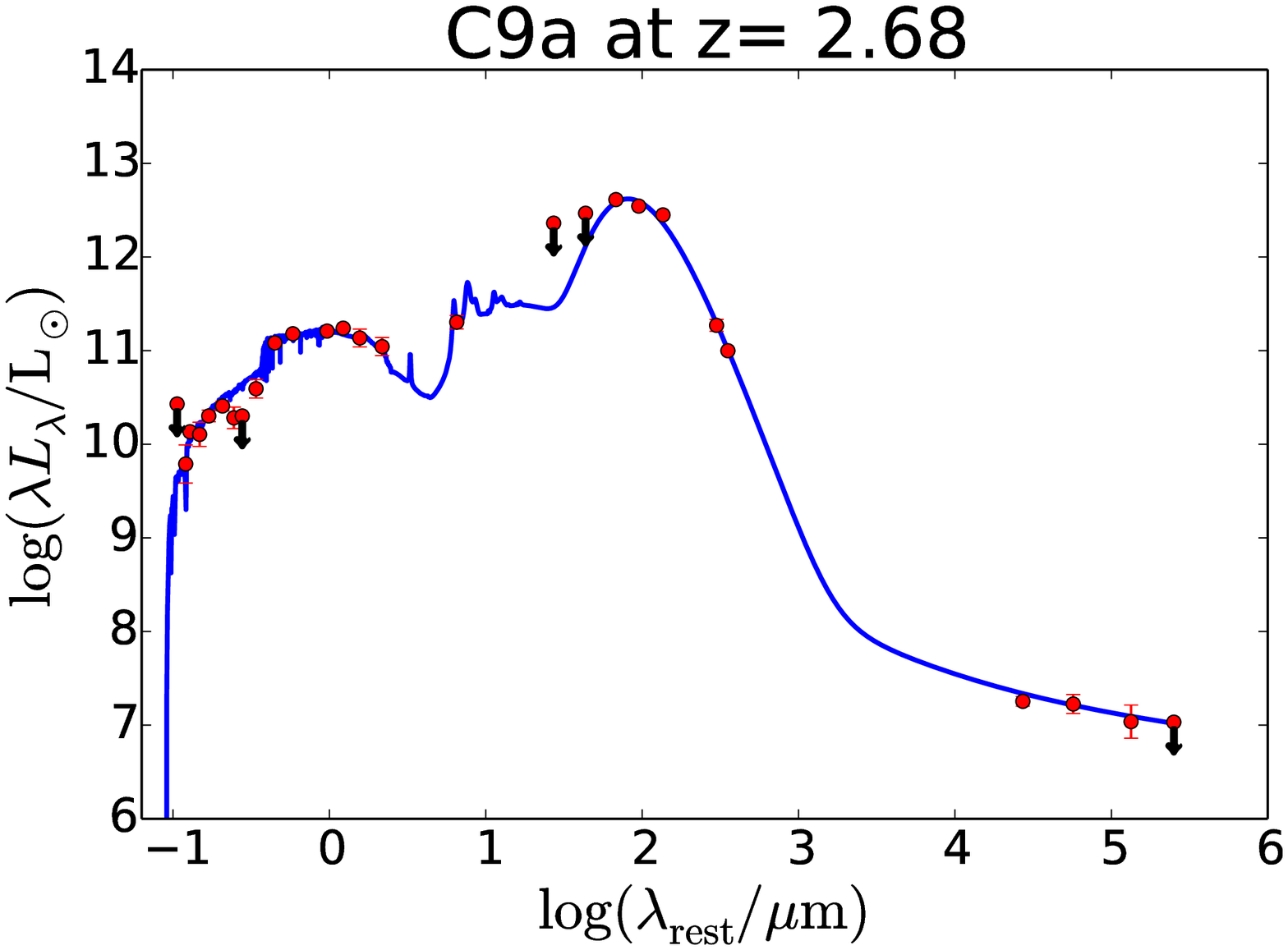}
\includegraphics[width=0.2465\textwidth]{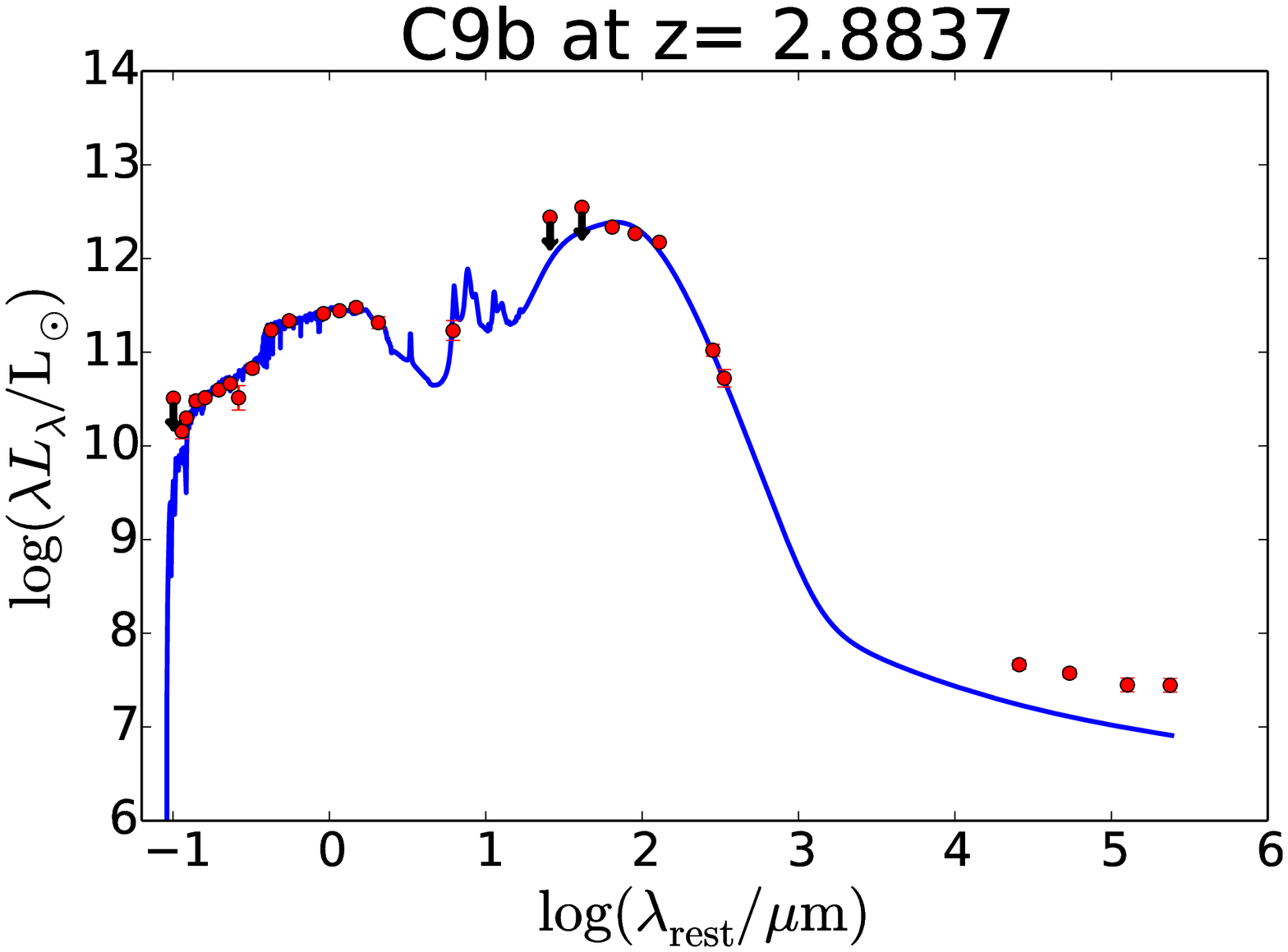}
\includegraphics[width=0.2465\textwidth]{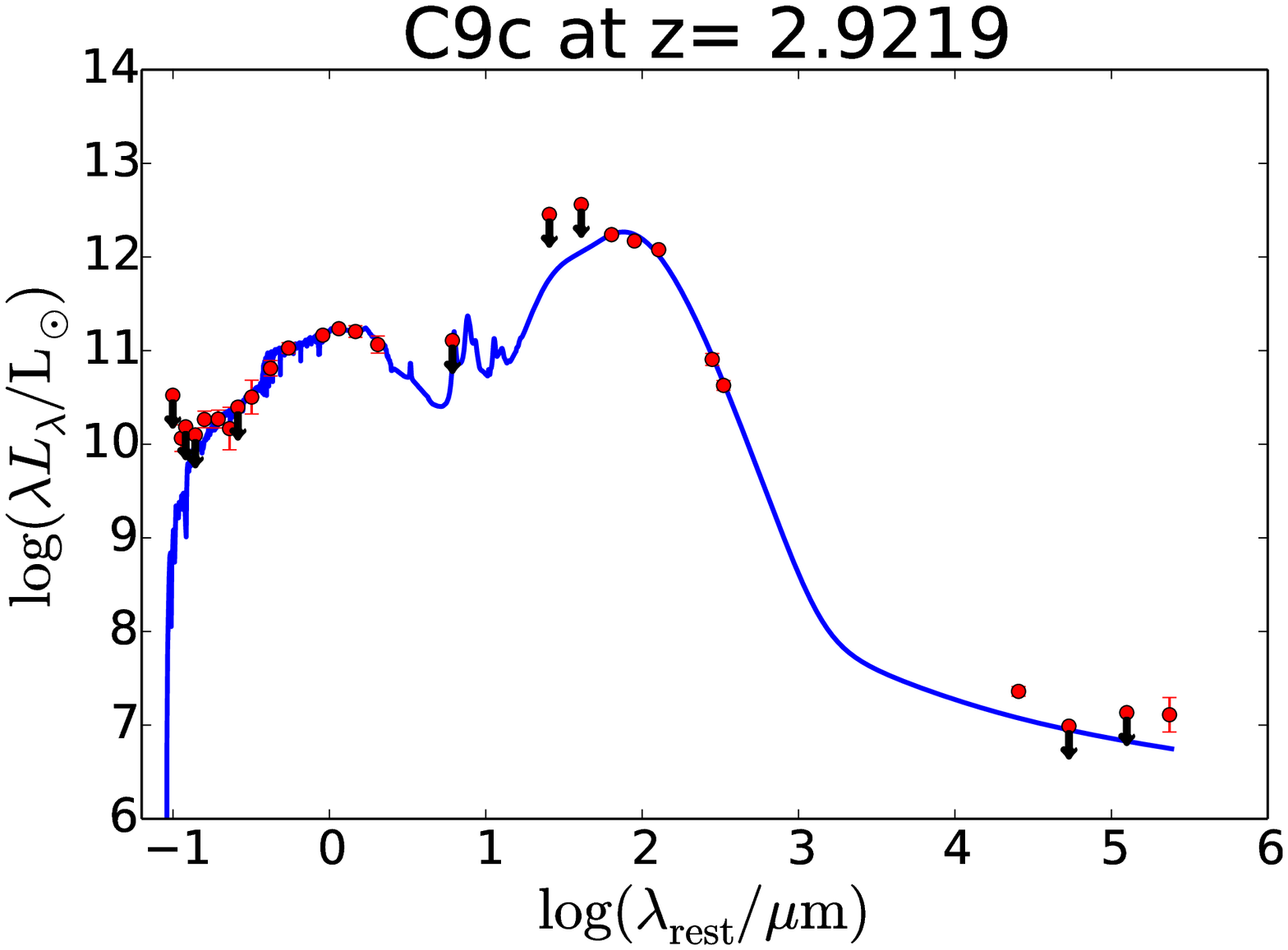}
\includegraphics[width=0.2465\textwidth]{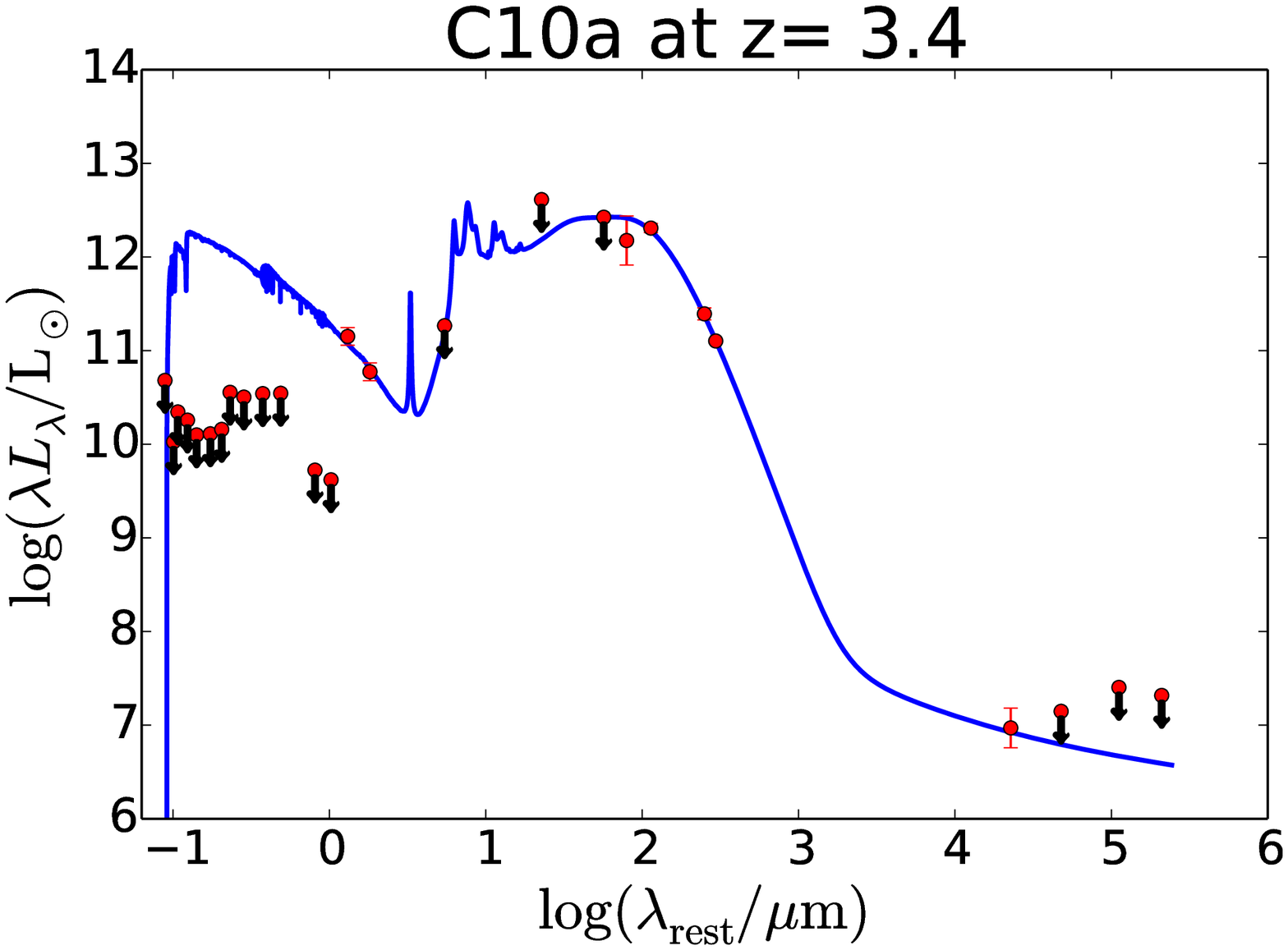}
\includegraphics[width=0.2465\textwidth]{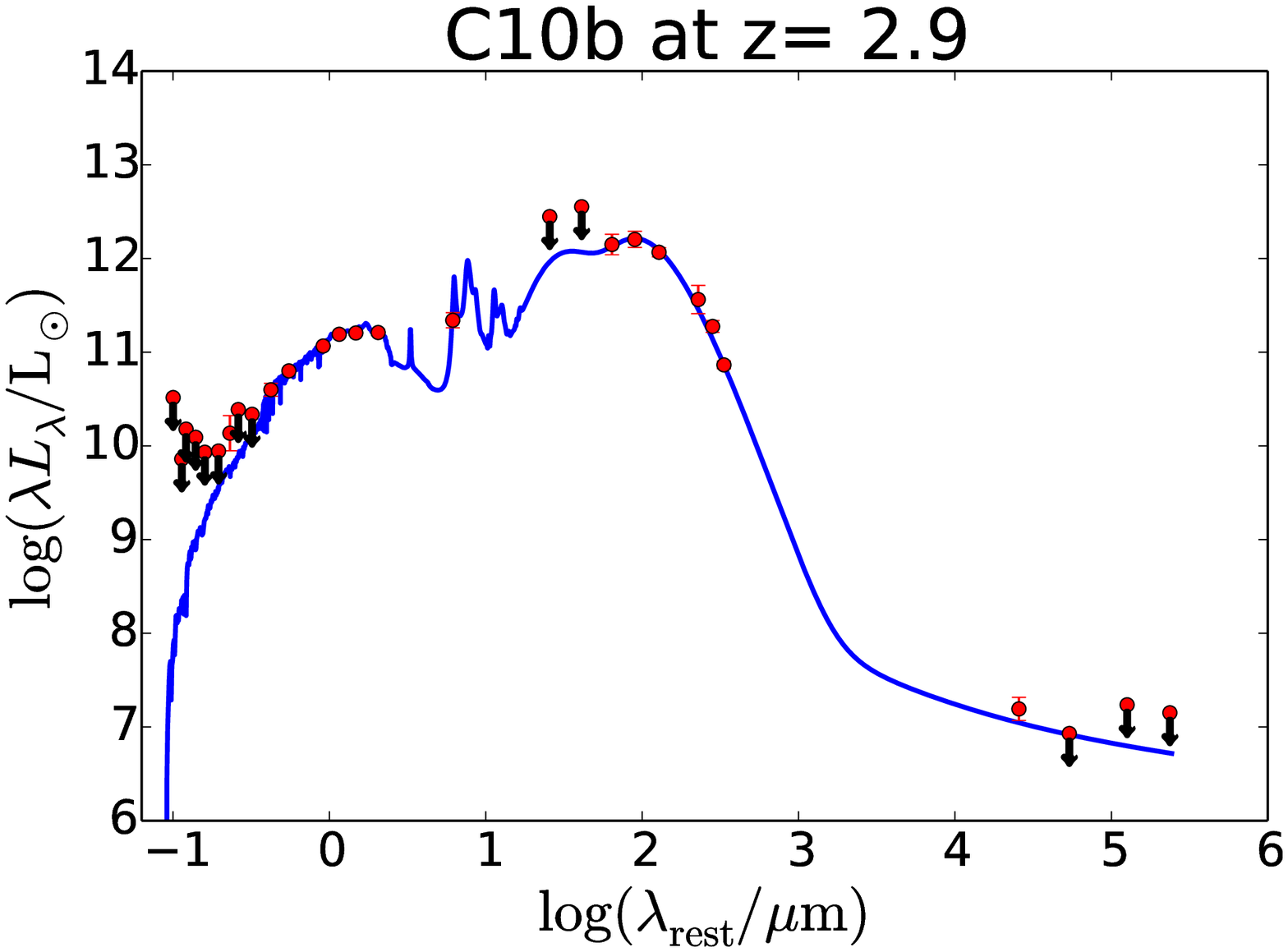}
\includegraphics[width=0.2465\textwidth]{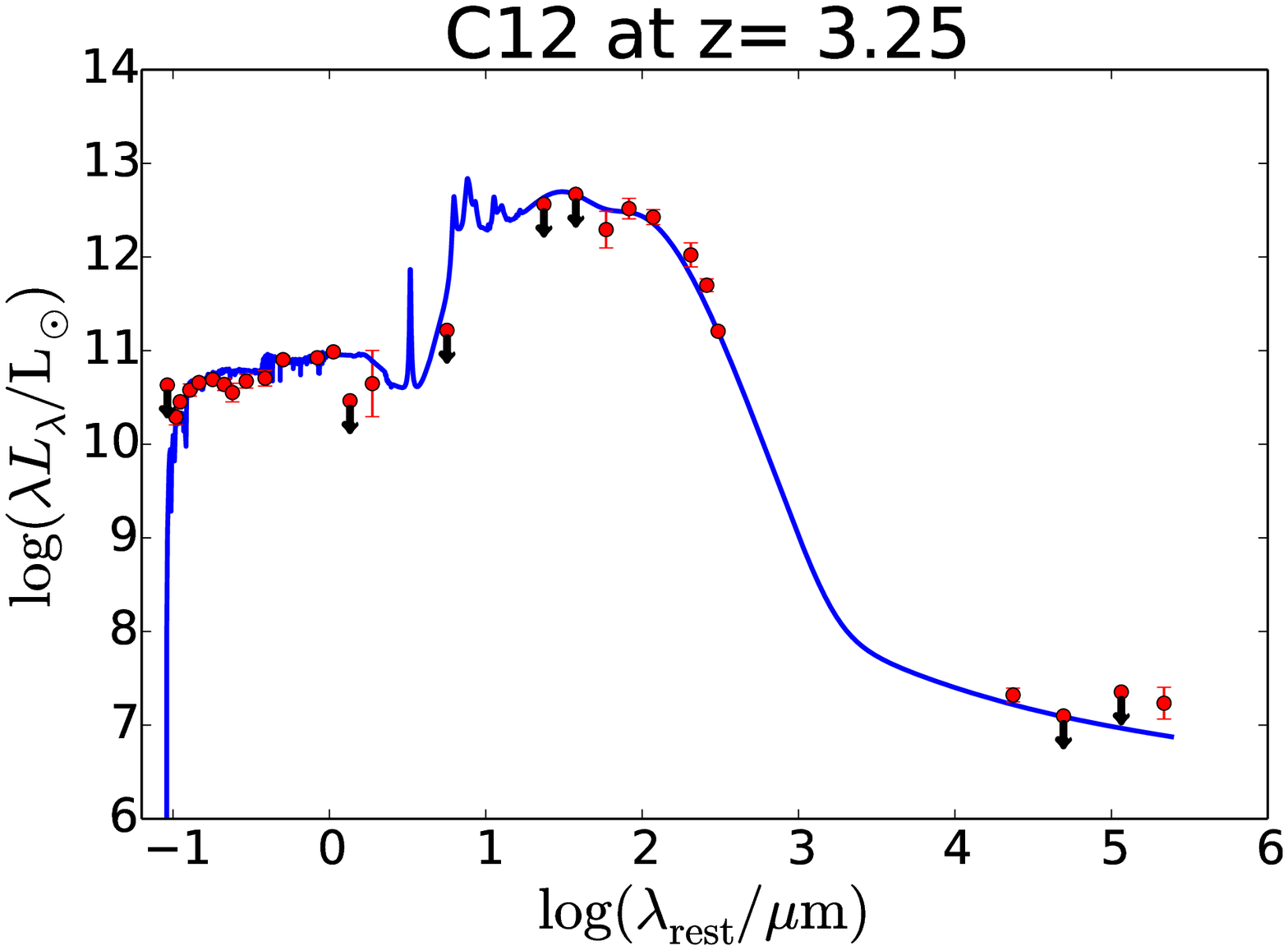}
\includegraphics[width=0.2465\textwidth]{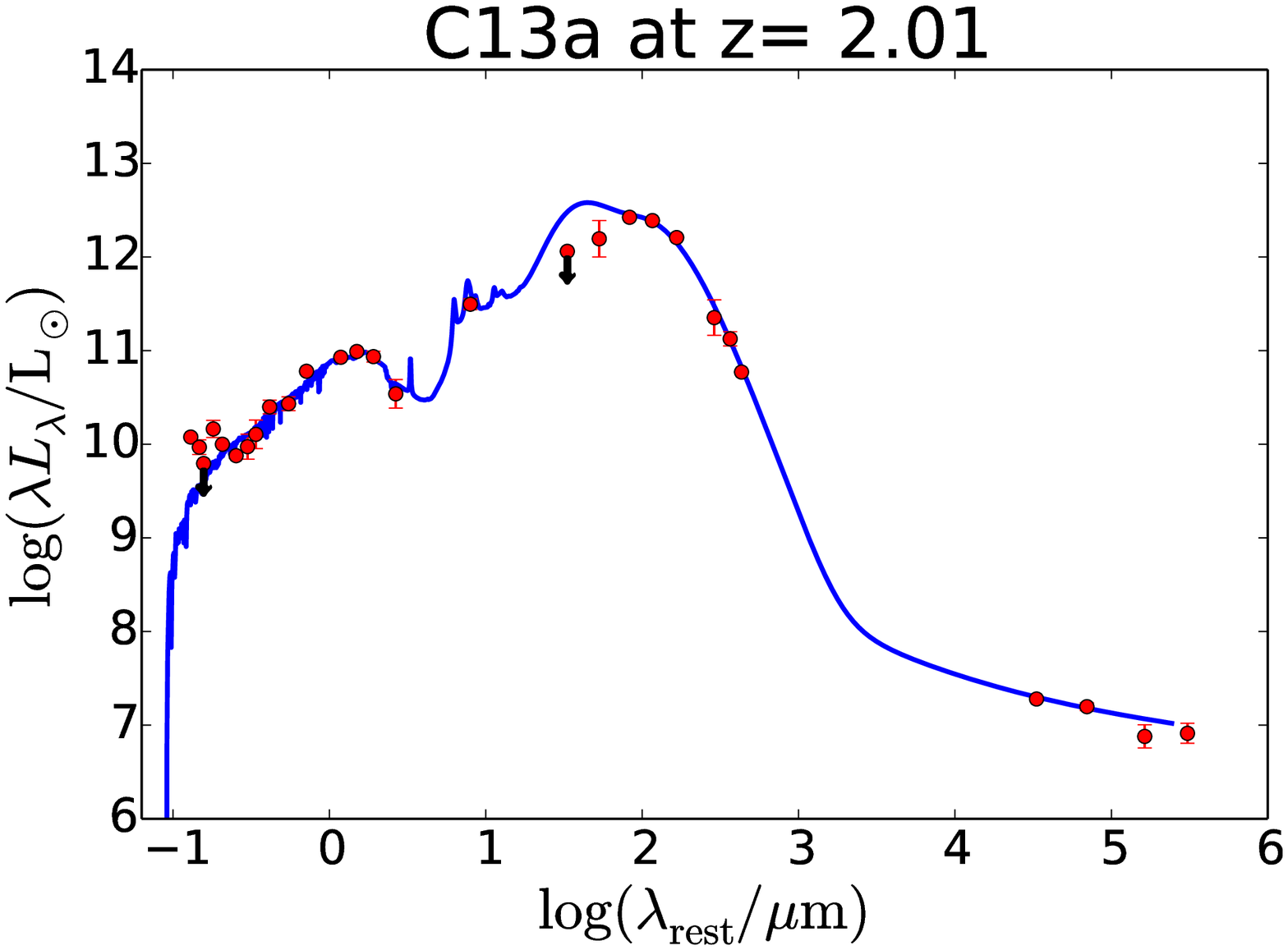}
\includegraphics[width=0.2465\textwidth]{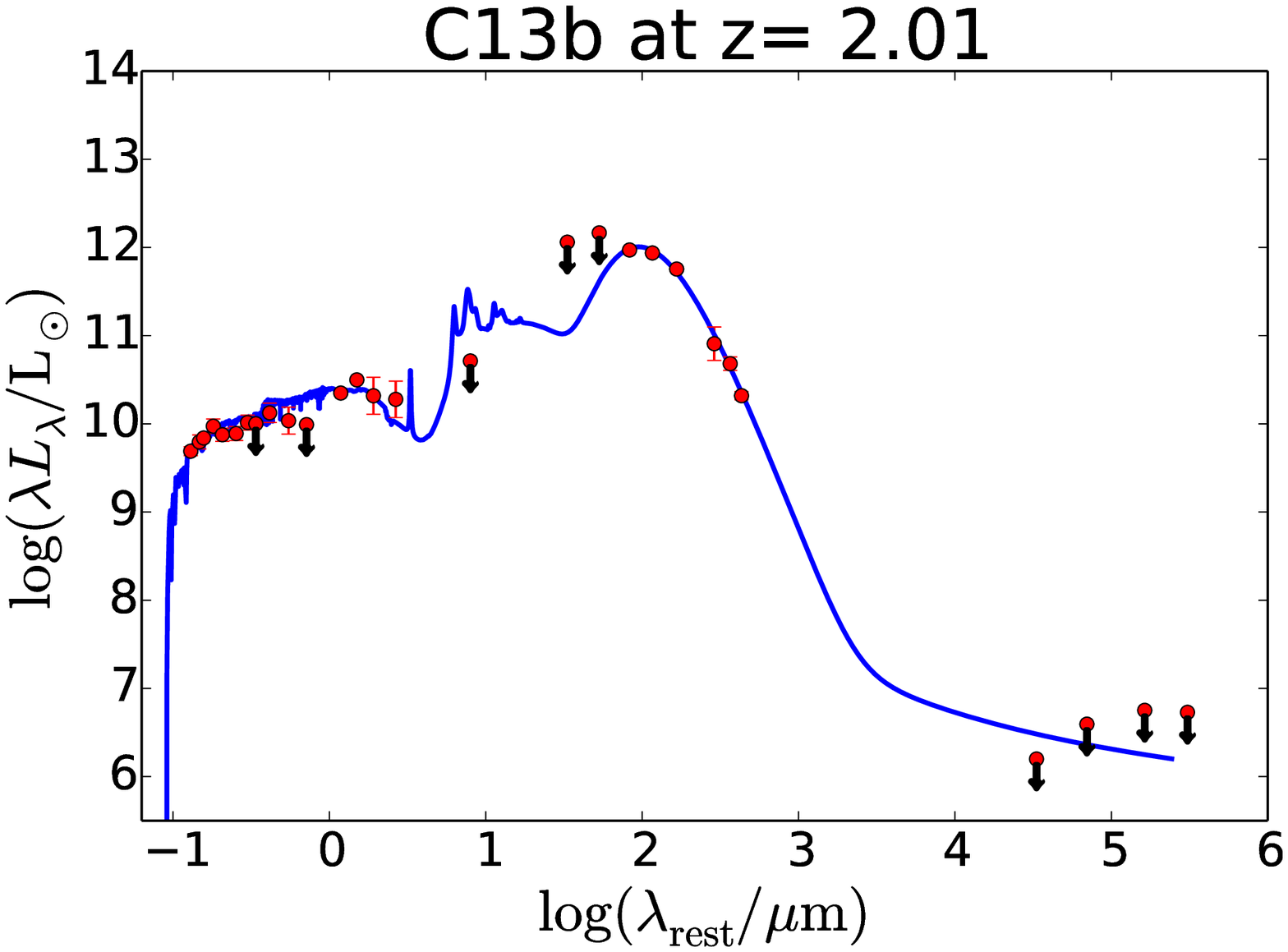}
\includegraphics[width=0.2465\textwidth]{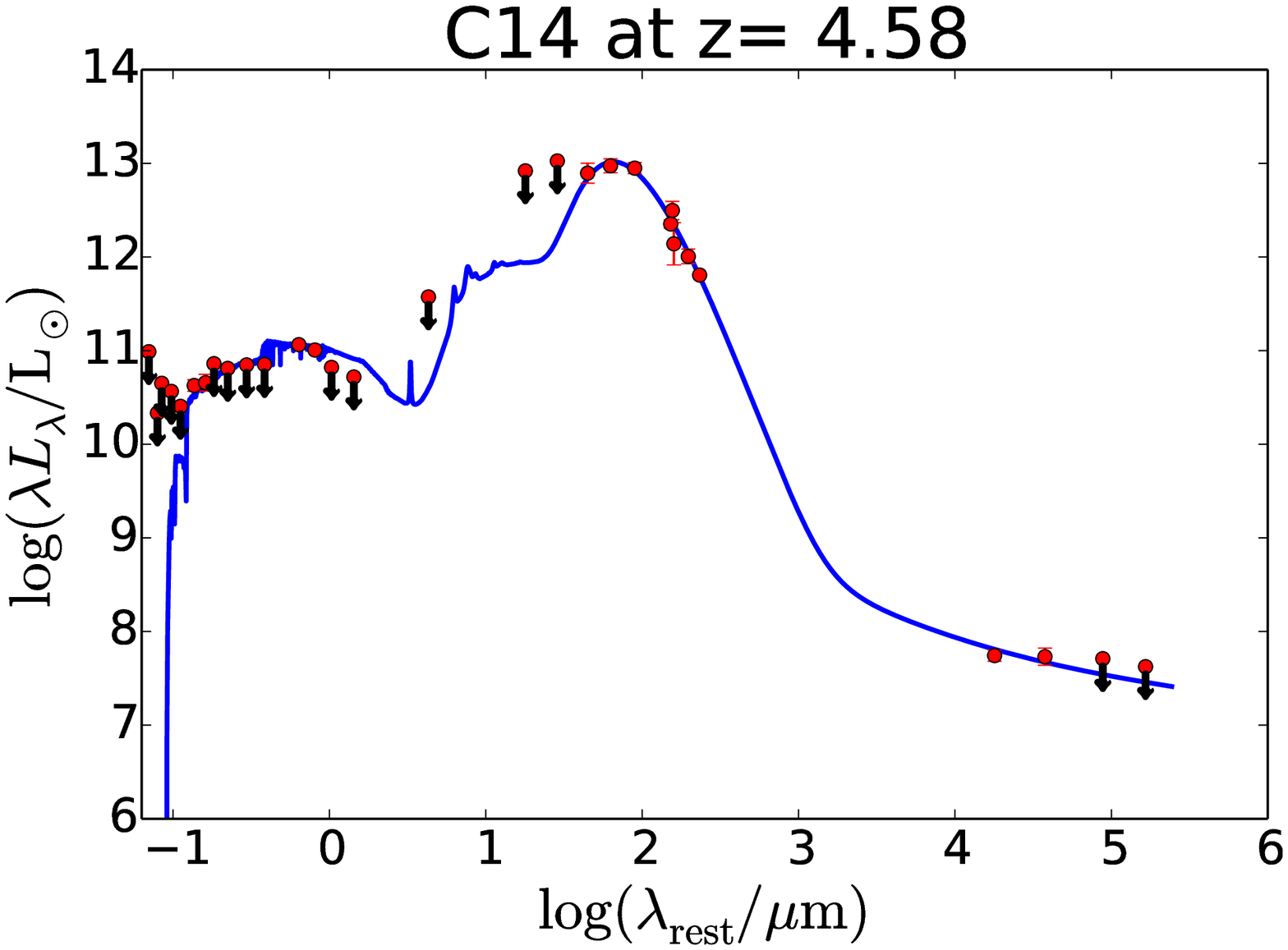}
\includegraphics[width=0.2465\textwidth]{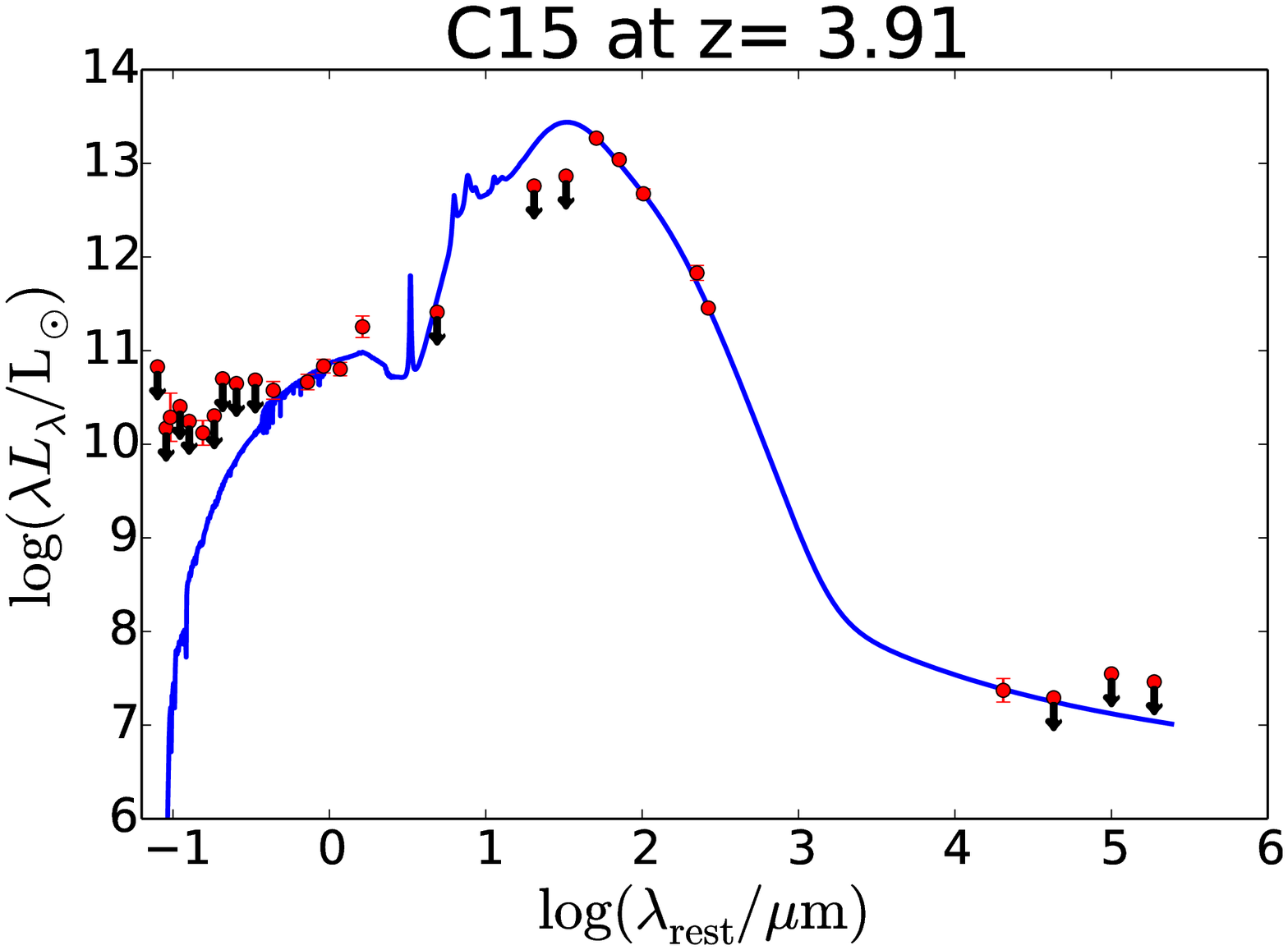}
\includegraphics[width=0.2465\textwidth]{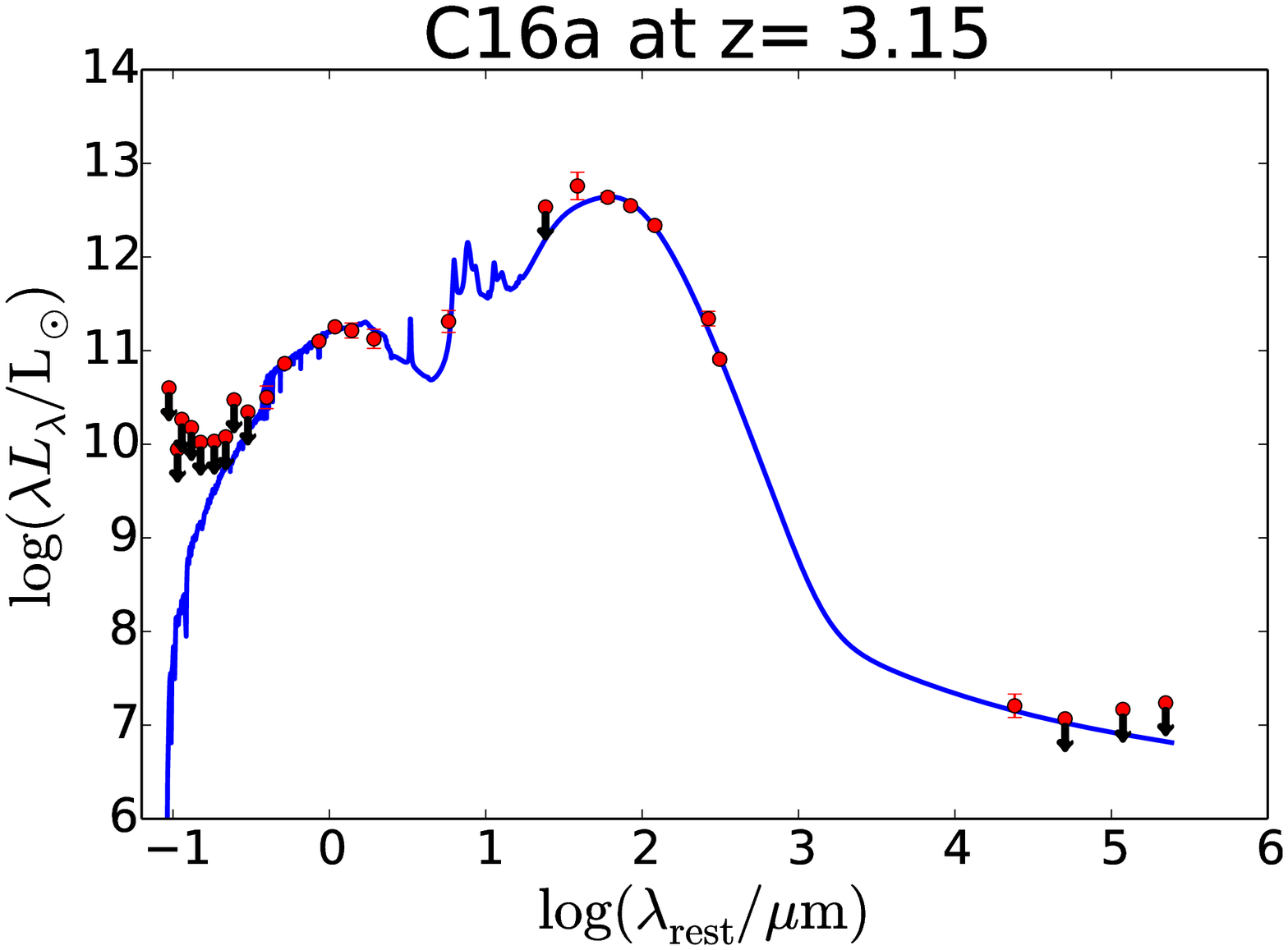}
\includegraphics[width=0.2465\textwidth]{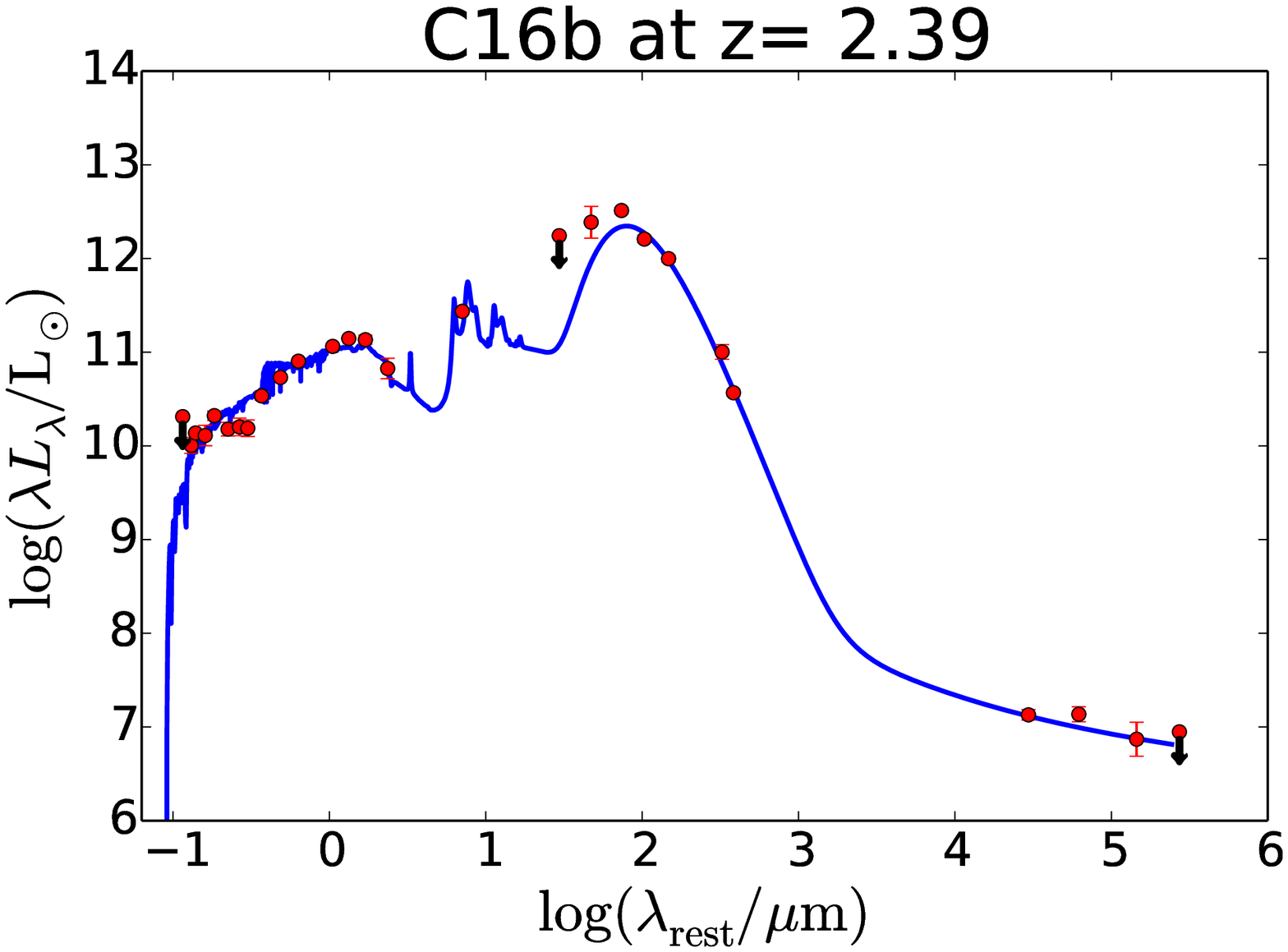}
\includegraphics[width=0.2465\textwidth]{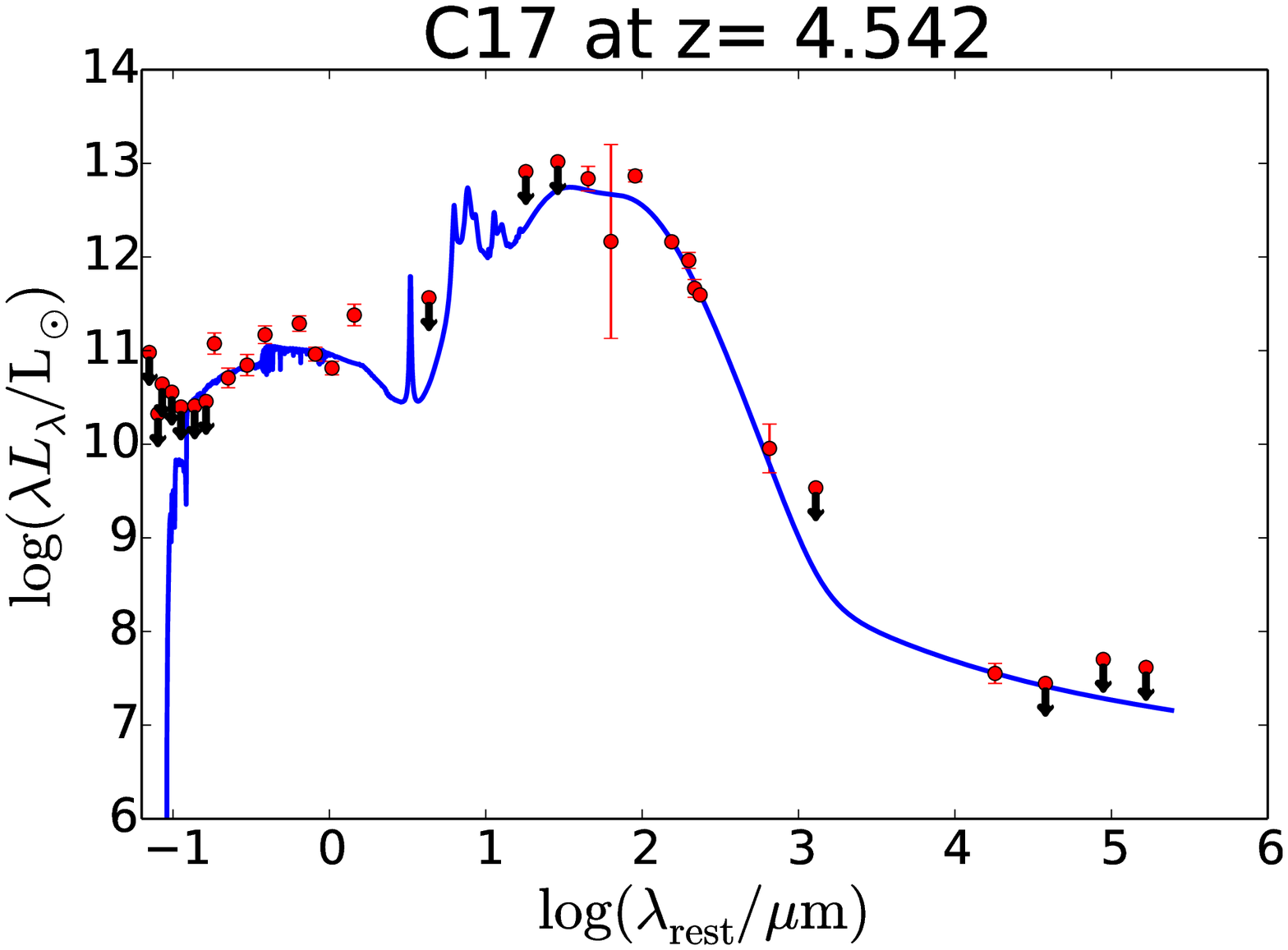}
\includegraphics[width=0.2465\textwidth]{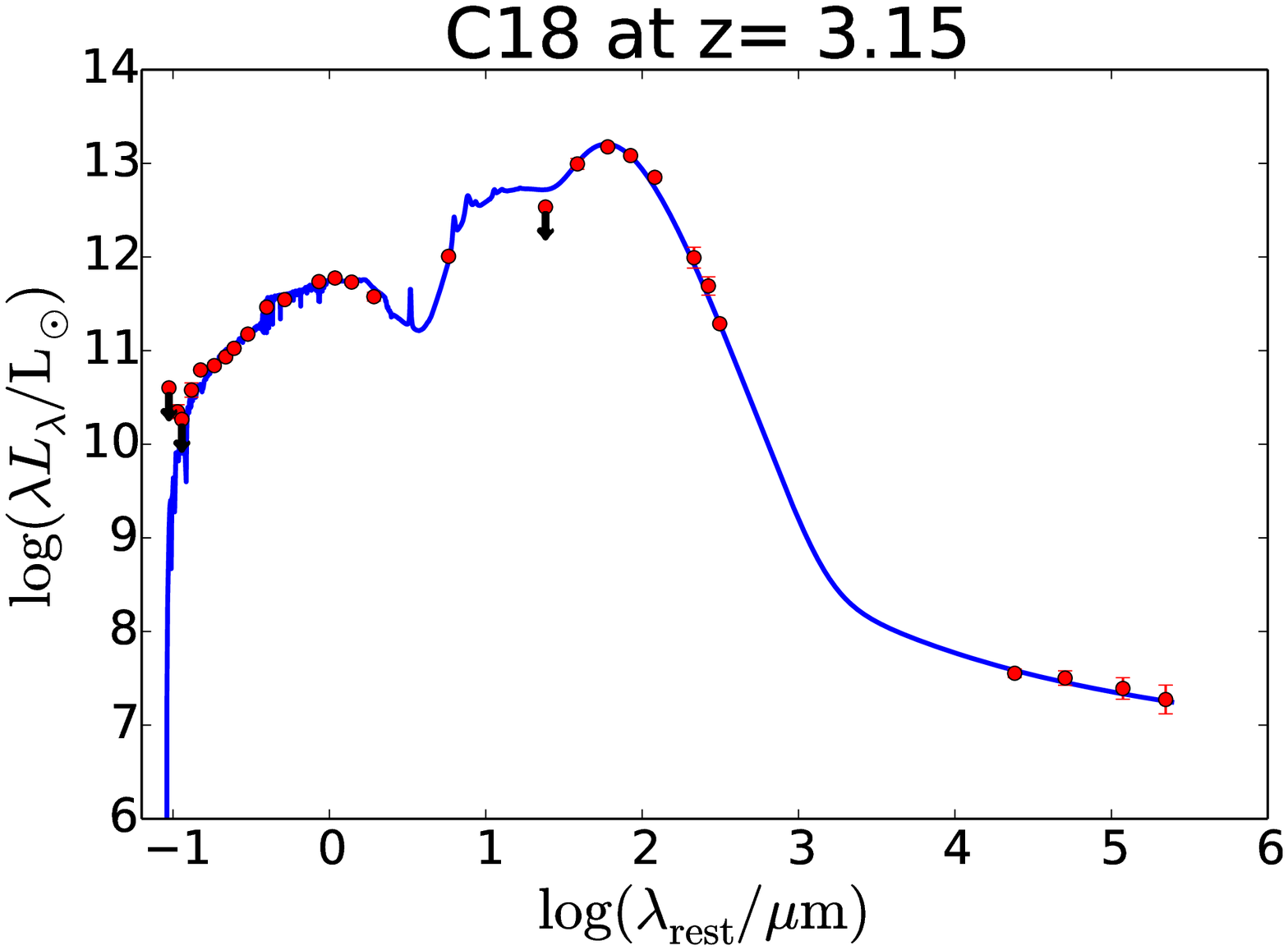}
\caption{Best-fit panchromatic (UV--radio) rest-frame SEDs of our SMGs. The source ID and redshift are shown on top of each panel. The red points with vertical error bars represent the observed photometric data, and those with downwards pointing arrows mark the $3\sigma$ upper flux density limits (taken into account in the fits). The blue line is the best-fit {\tt MAGPHYS} model SED from the high-$z$ library (\cite{dacunha2015}). In some cases the best-fit model is inconsistent with the upper flux density limits, which can be the result of invalid model assumptions (e.g. those in the radio regime; Sect.~3.1.1), or incorrectly assigned upper flux density limits (i.e. different from the $3\sigma$ limits; Sect.~3.1.2). }
\label{figure:seds}
\end{center}
\end{figure*}

\addtocounter{figure}{-1}
\begin{figure*}
\begin{center}
\includegraphics[width=0.2465\textwidth]{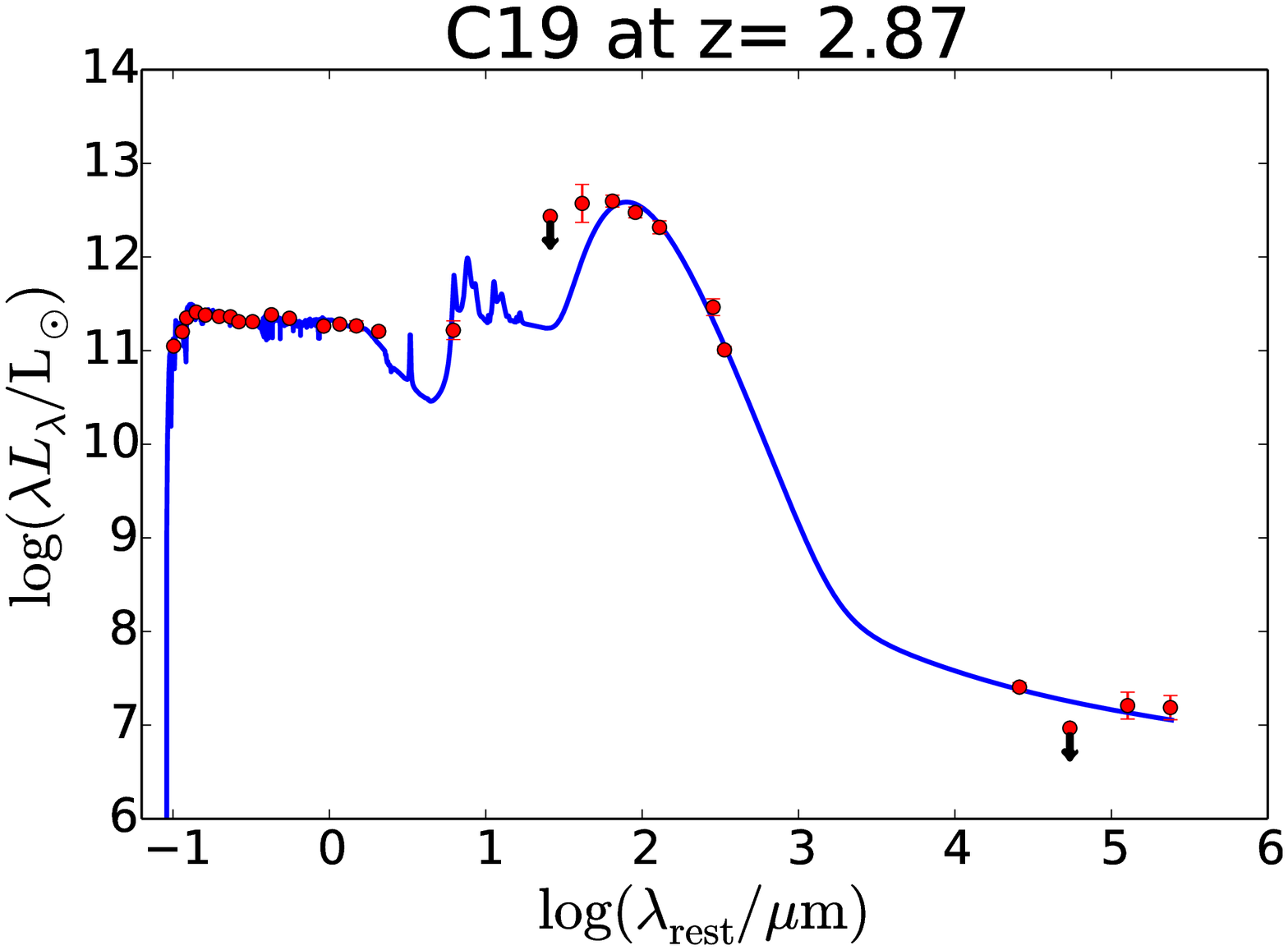}
\includegraphics[width=0.2465\textwidth]{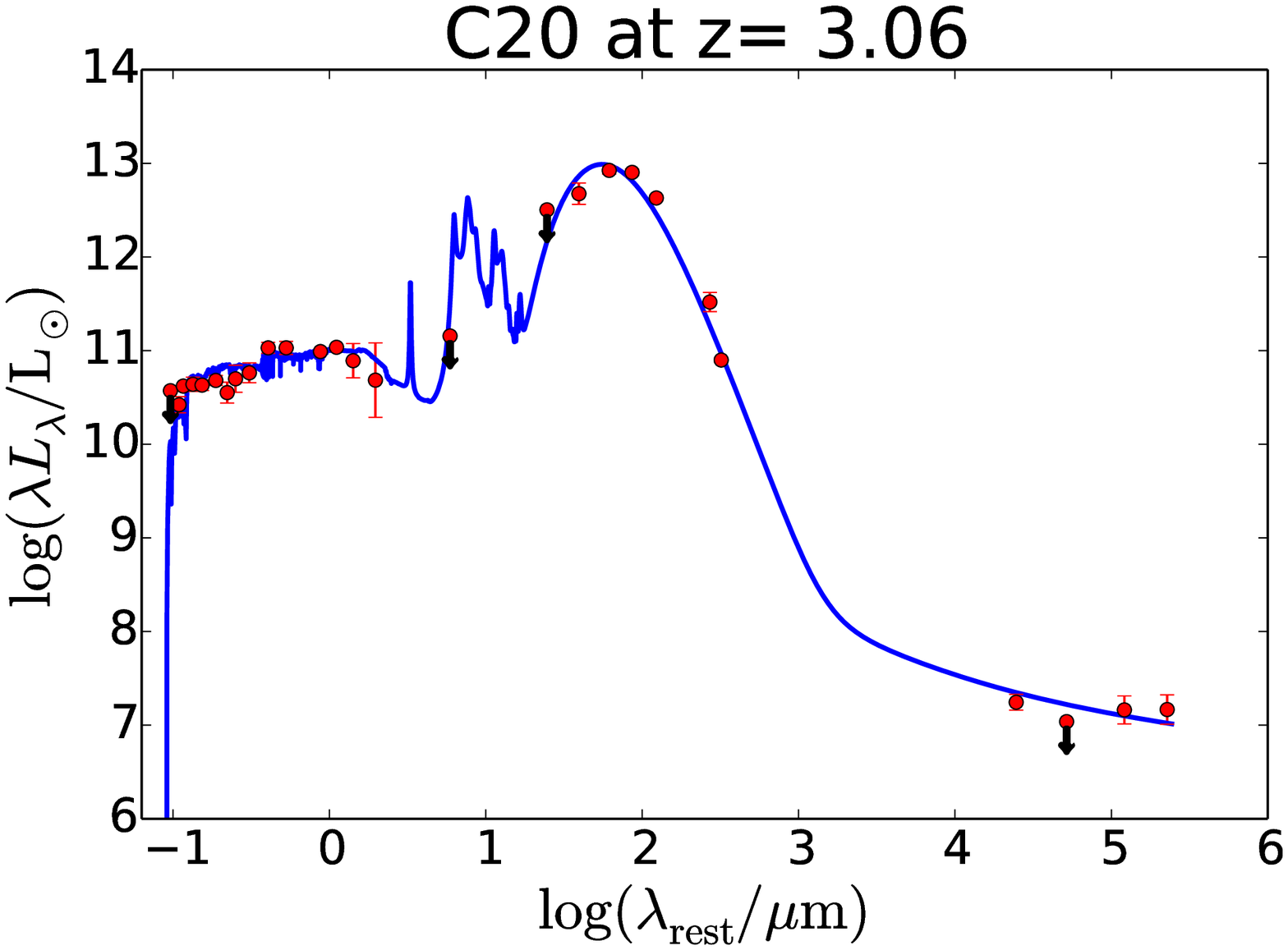}
\includegraphics[width=0.2465\textwidth]{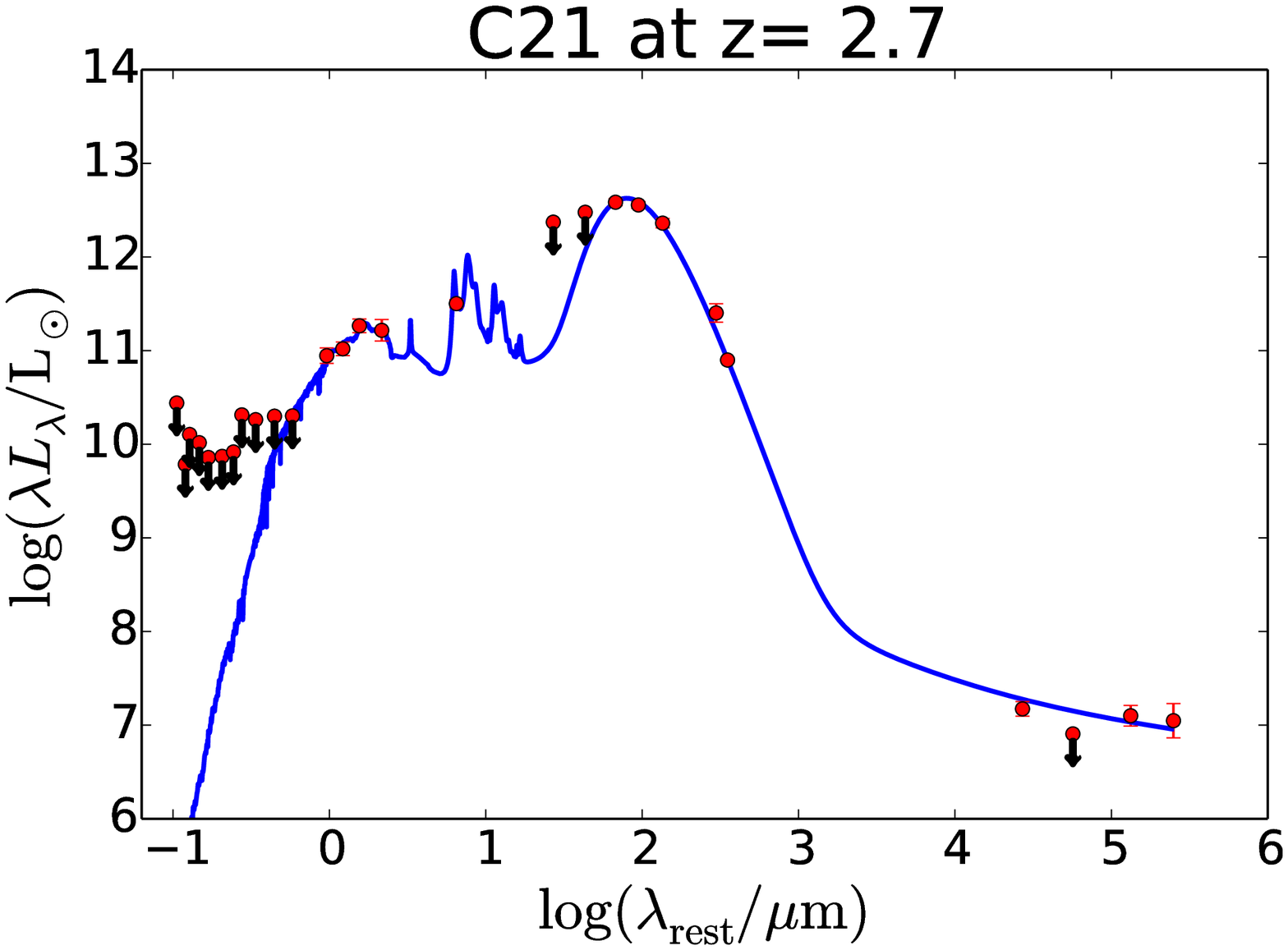}
\includegraphics[width=0.2465\textwidth]{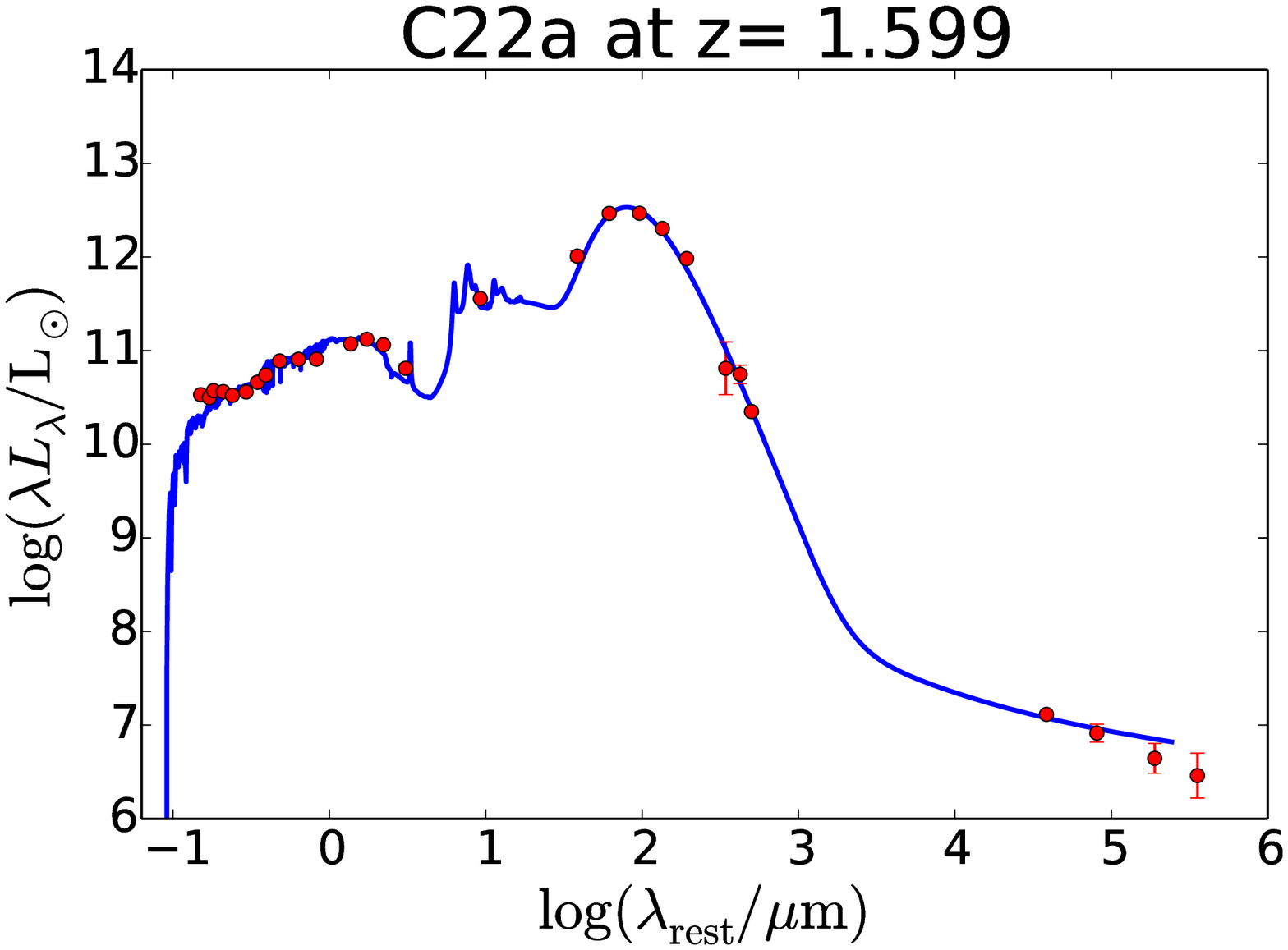}
\includegraphics[width=0.2465\textwidth]{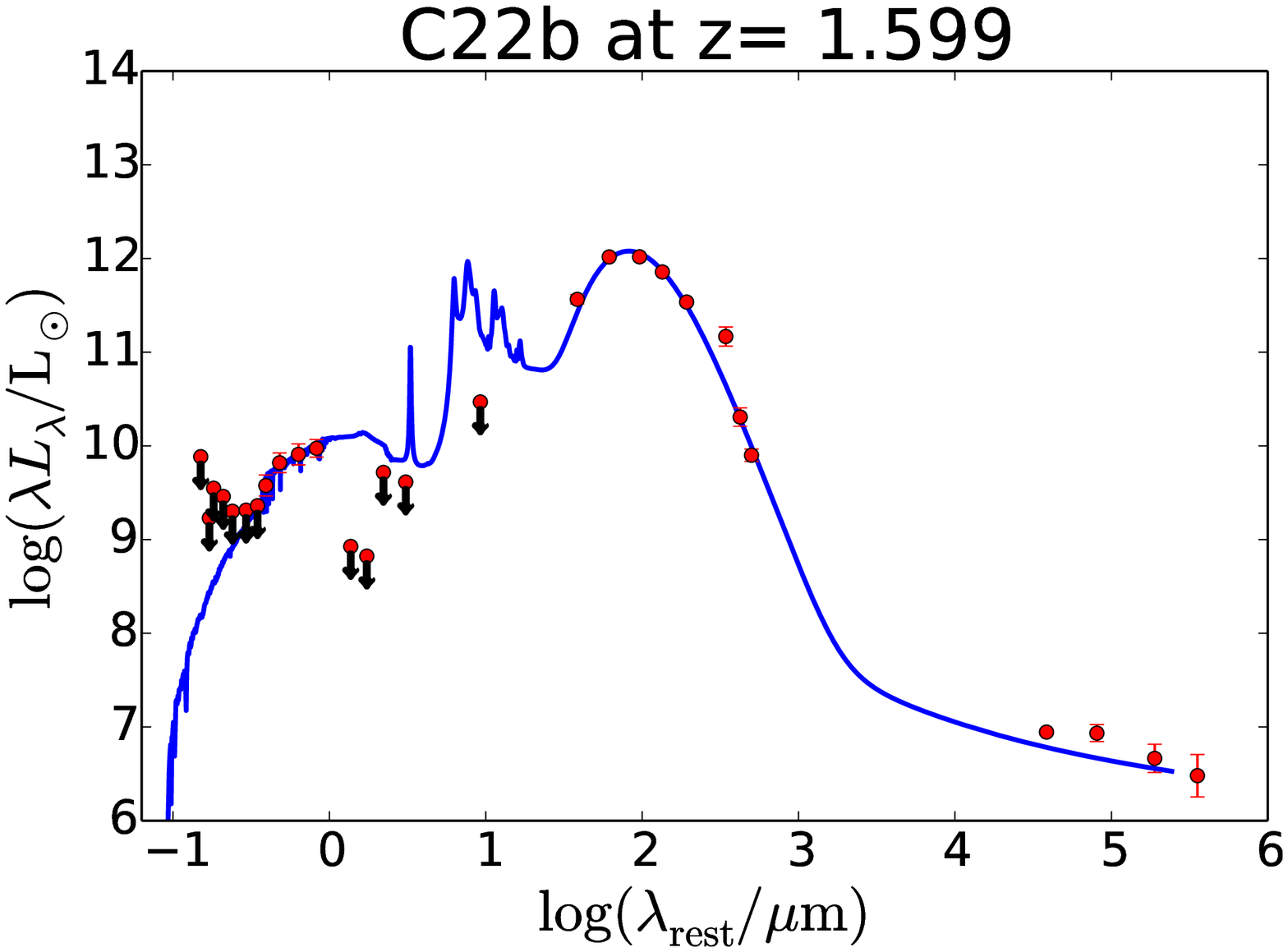}
\includegraphics[width=0.2465\textwidth]{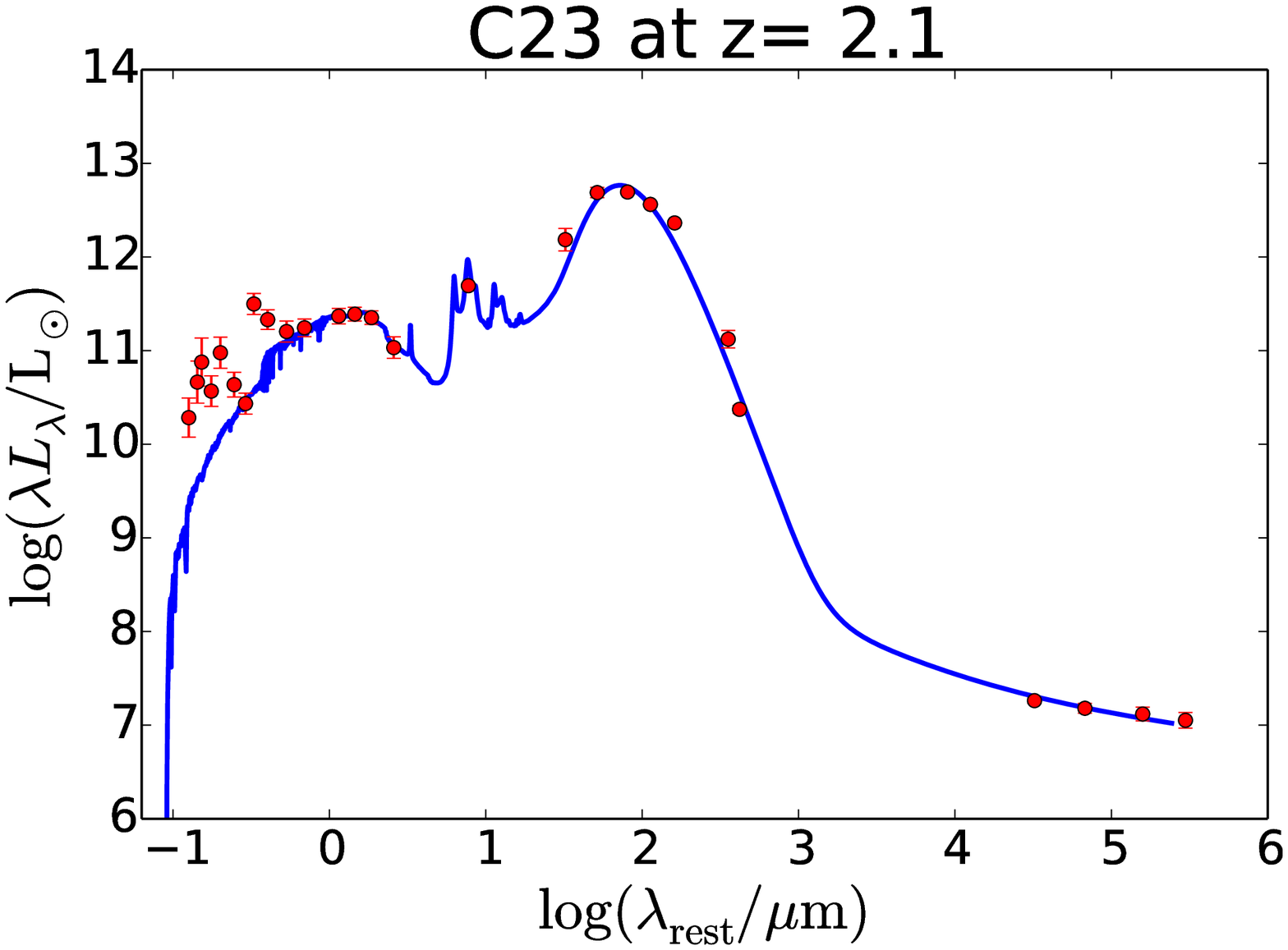}
\includegraphics[width=0.2465\textwidth]{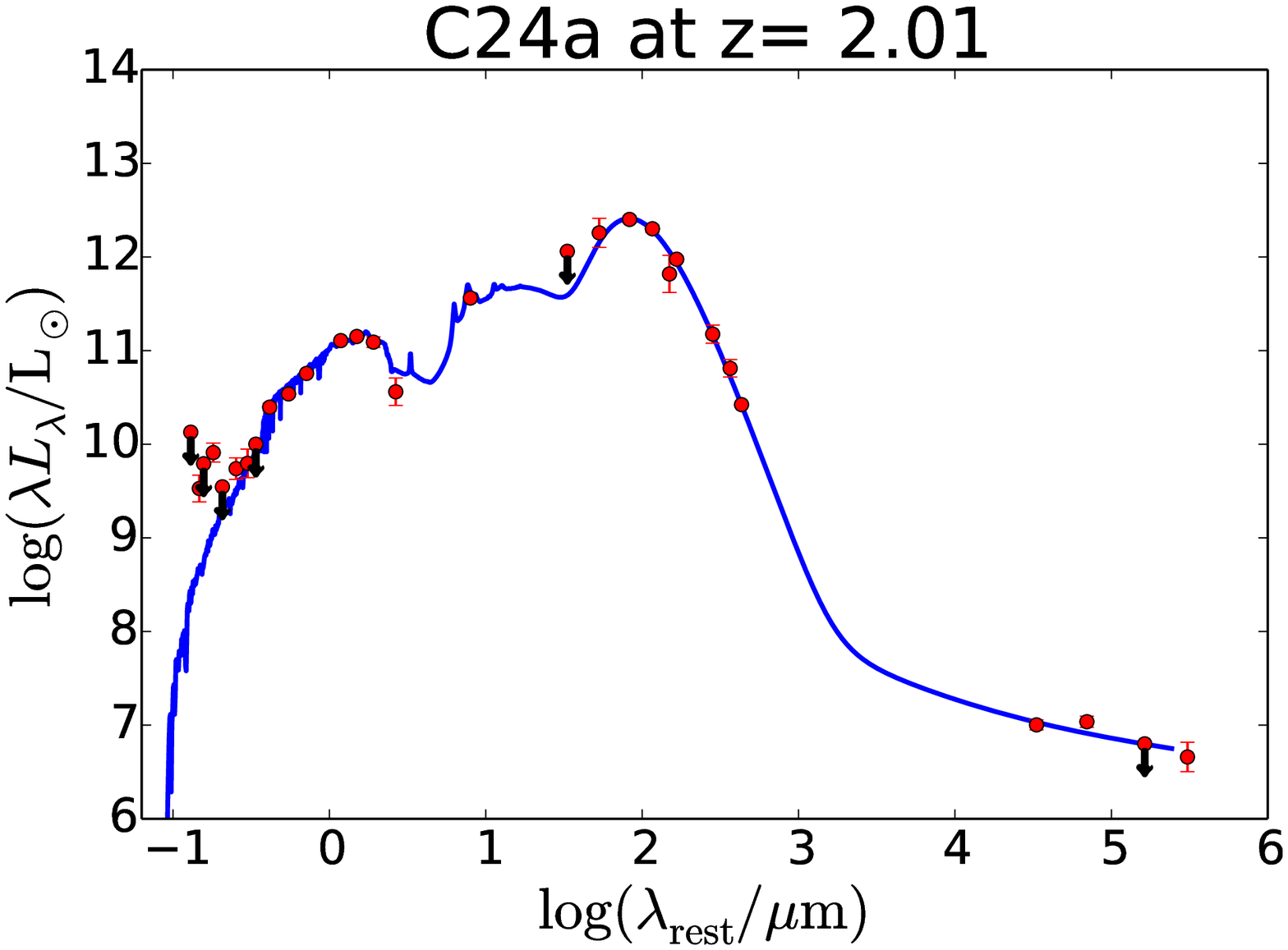}
\includegraphics[width=0.2465\textwidth]{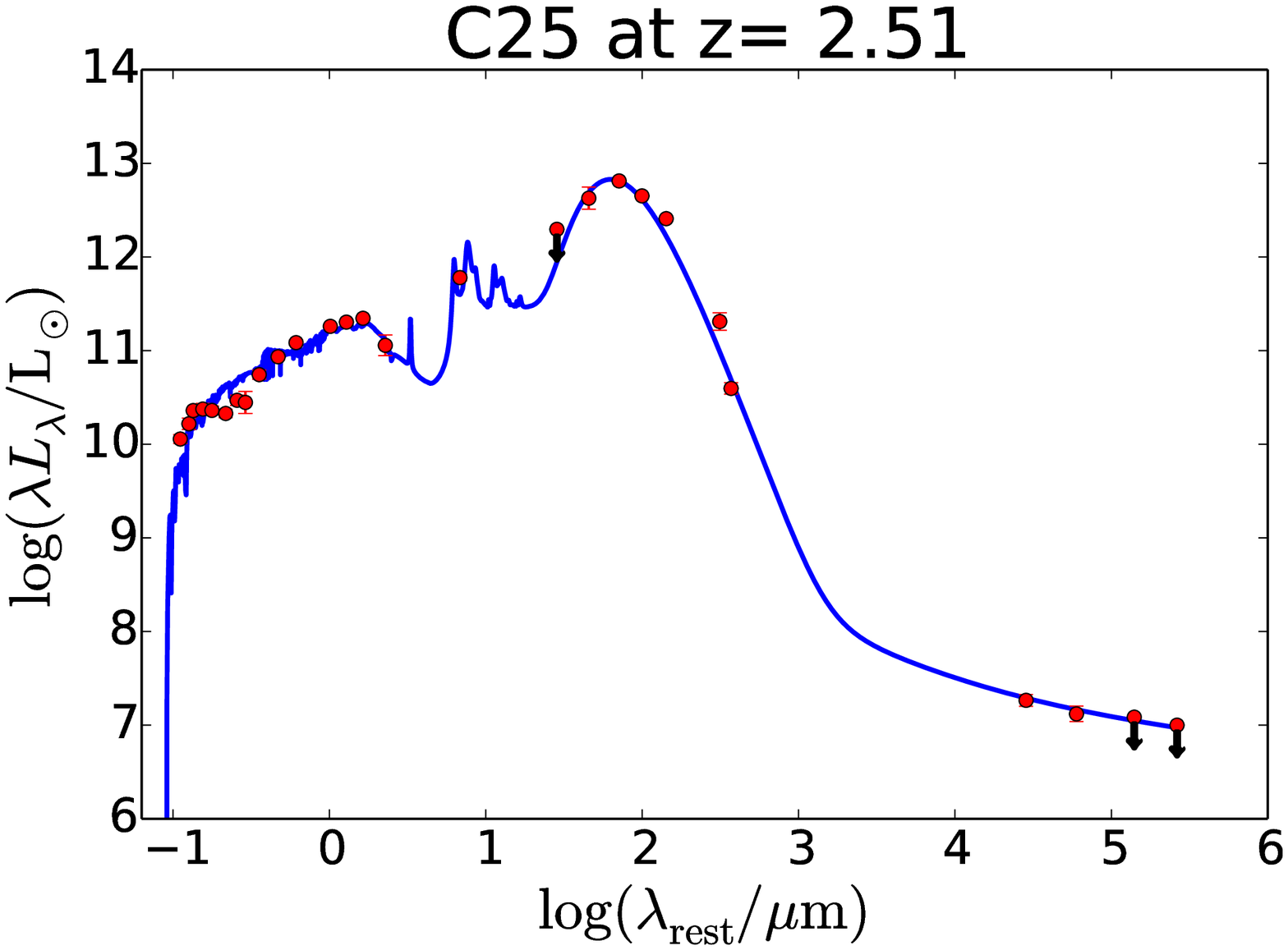}
\includegraphics[width=0.2465\textwidth]{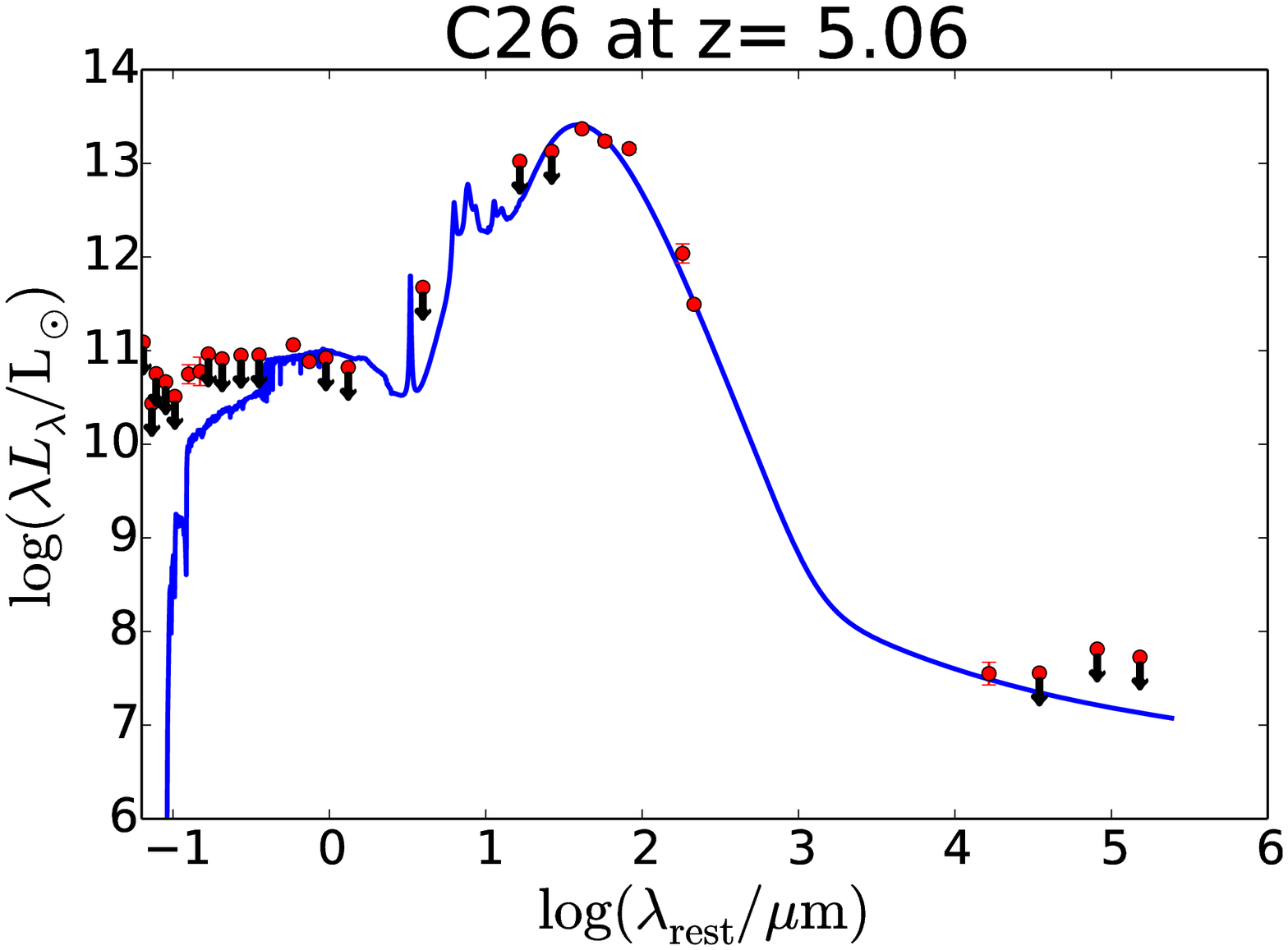}
\includegraphics[width=0.2465\textwidth]{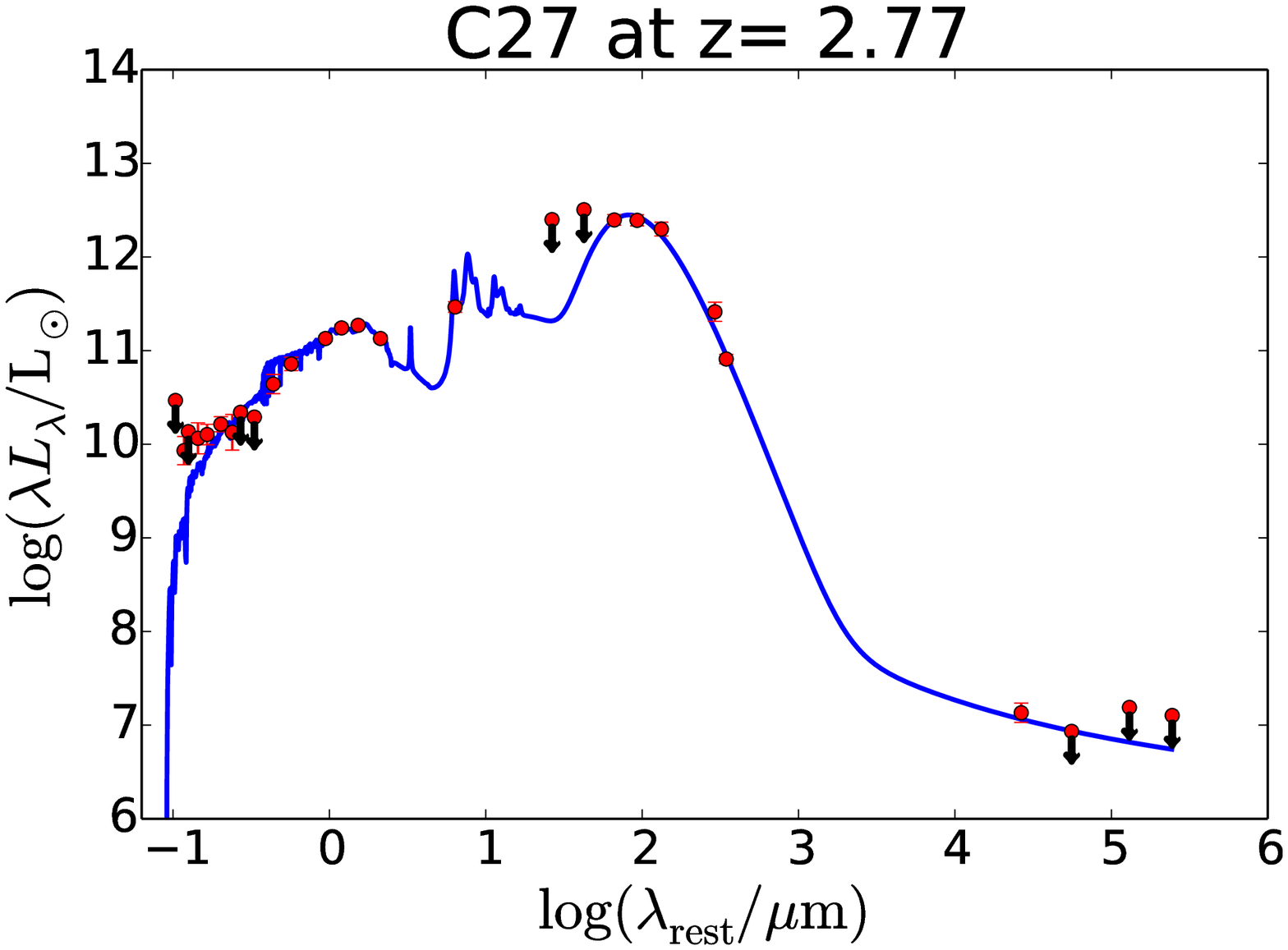}
\includegraphics[width=0.2465\textwidth]{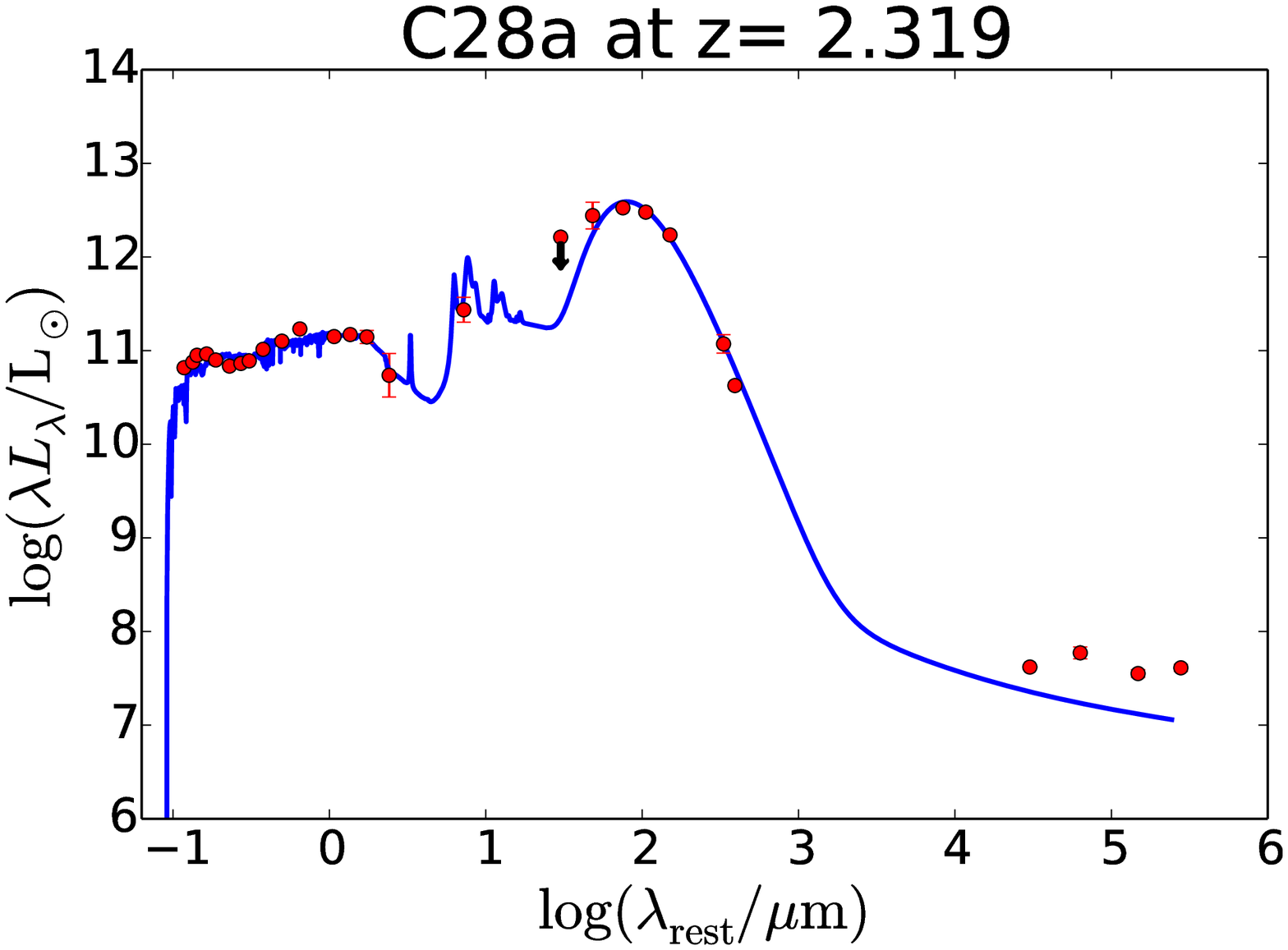}
\includegraphics[width=0.2465\textwidth]{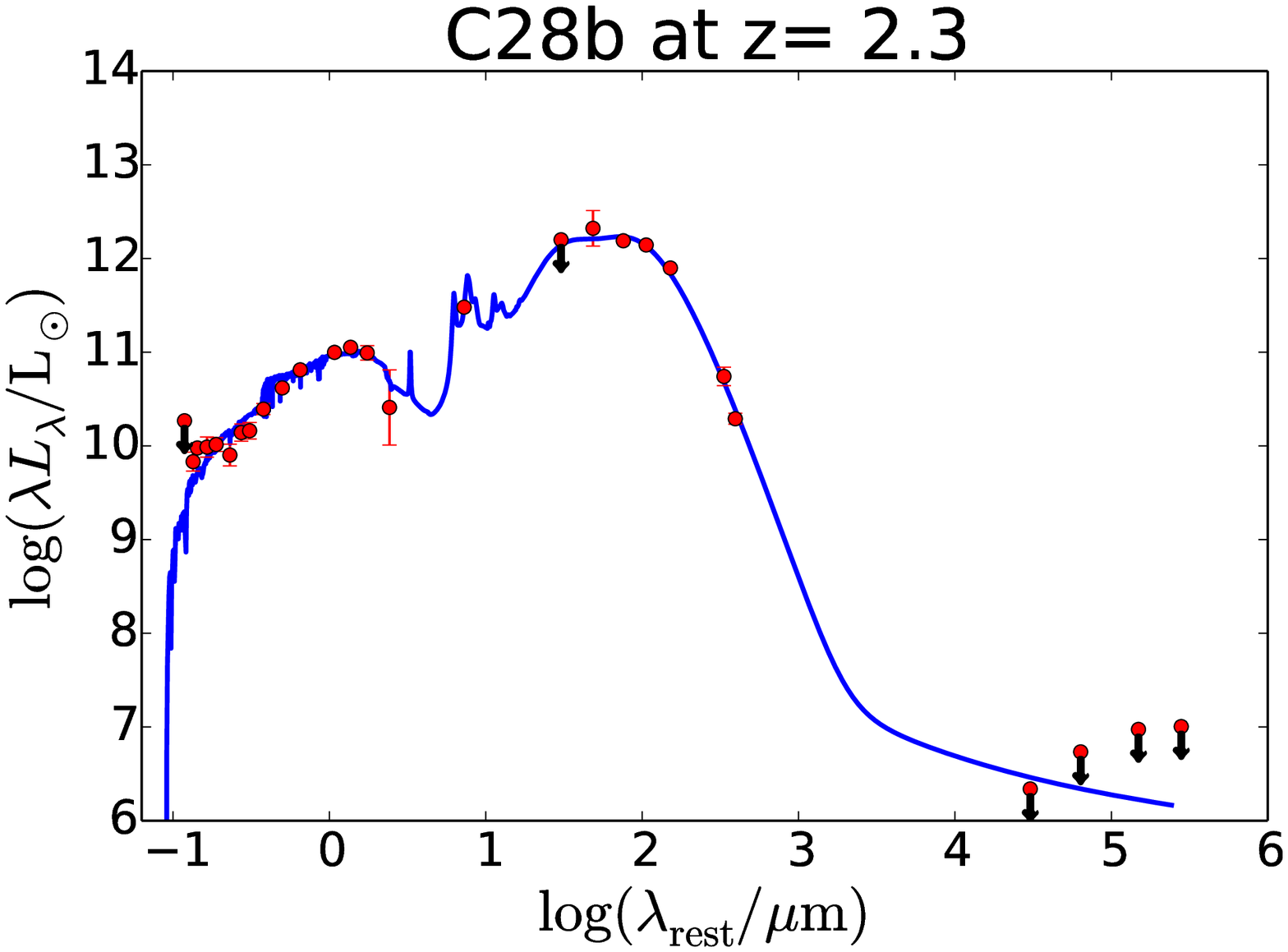}
\includegraphics[width=0.2465\textwidth]{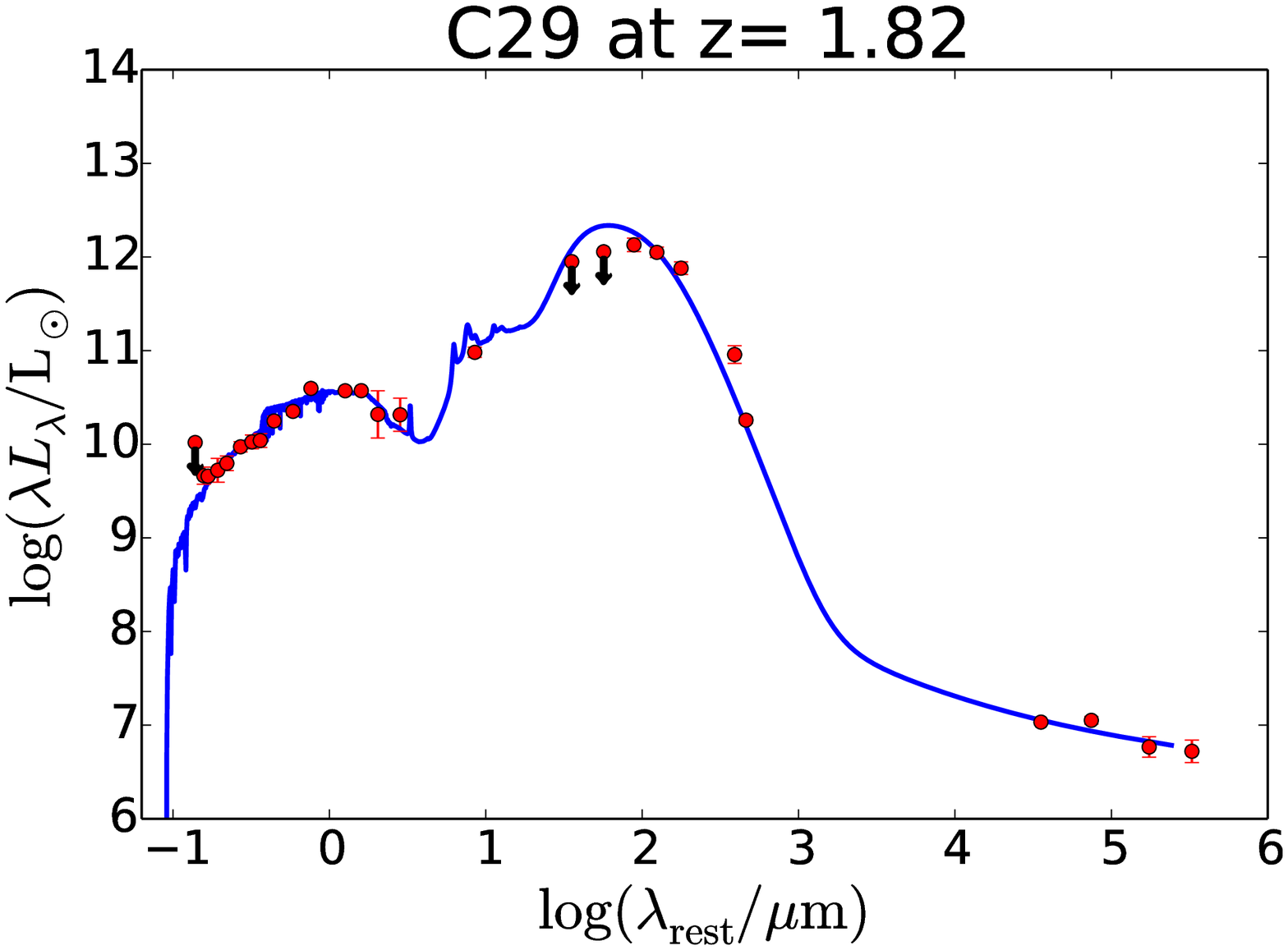}
\includegraphics[width=0.2465\textwidth]{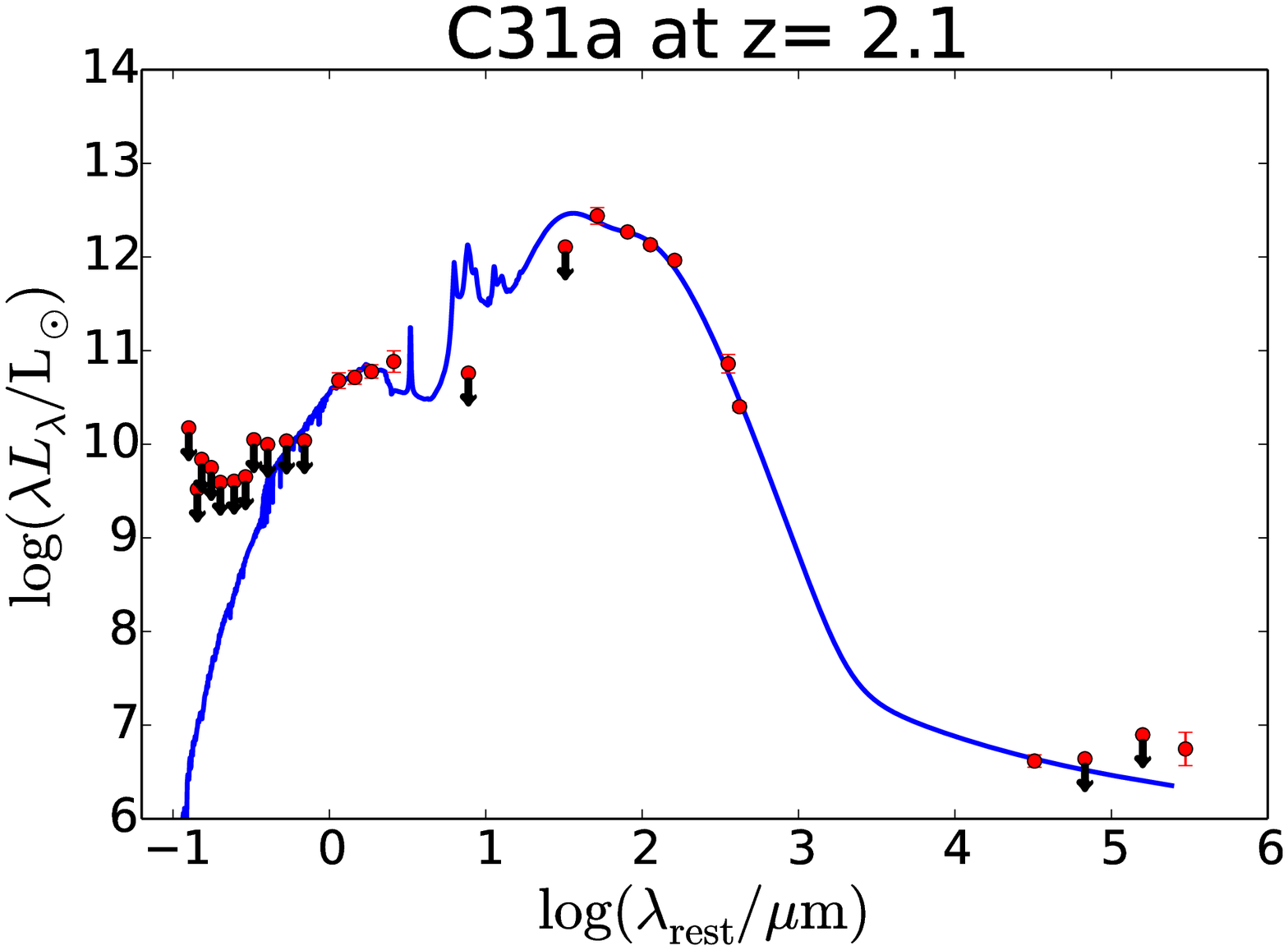}
\includegraphics[width=0.2465\textwidth]{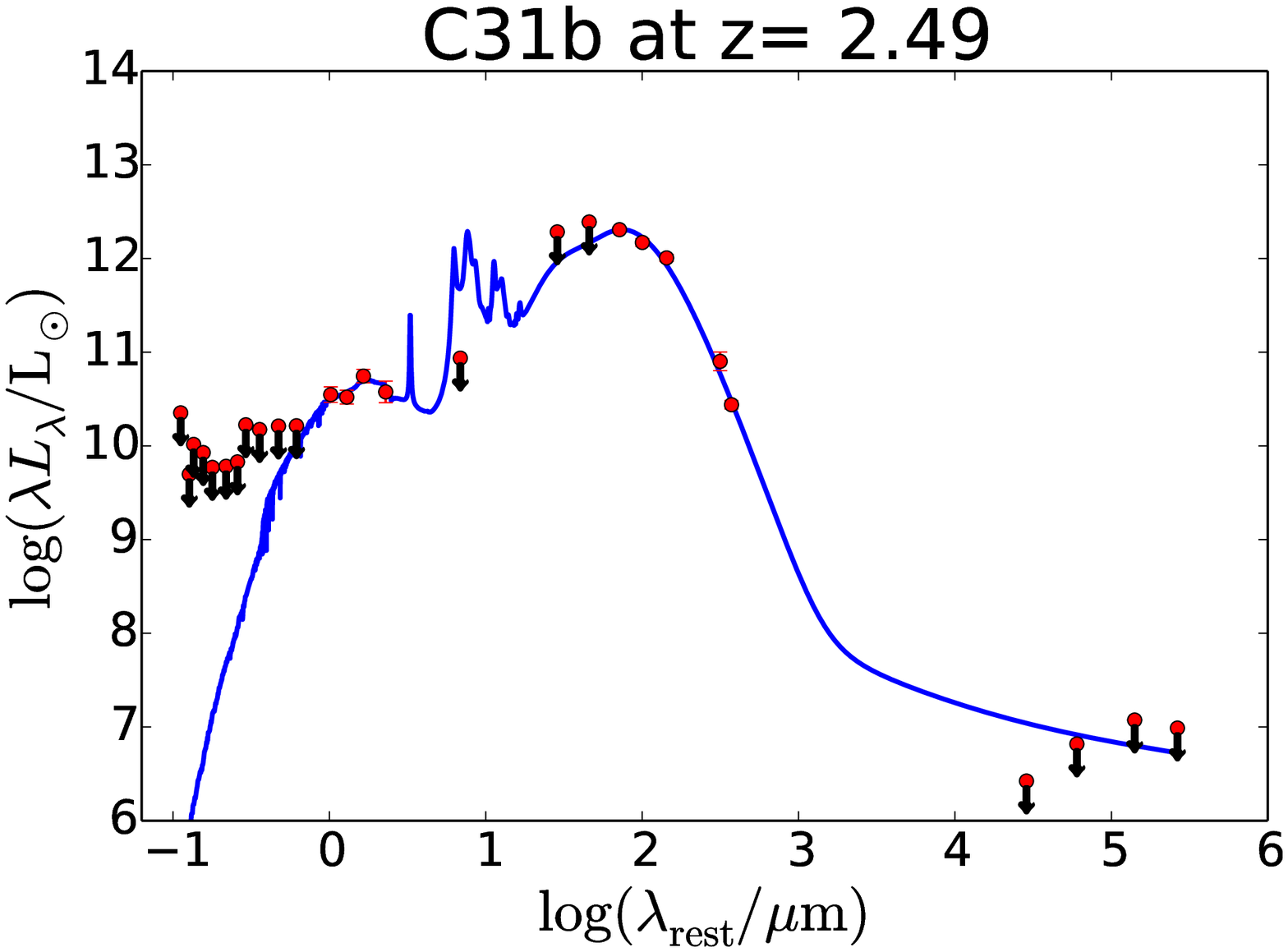}
\includegraphics[width=0.2465\textwidth]{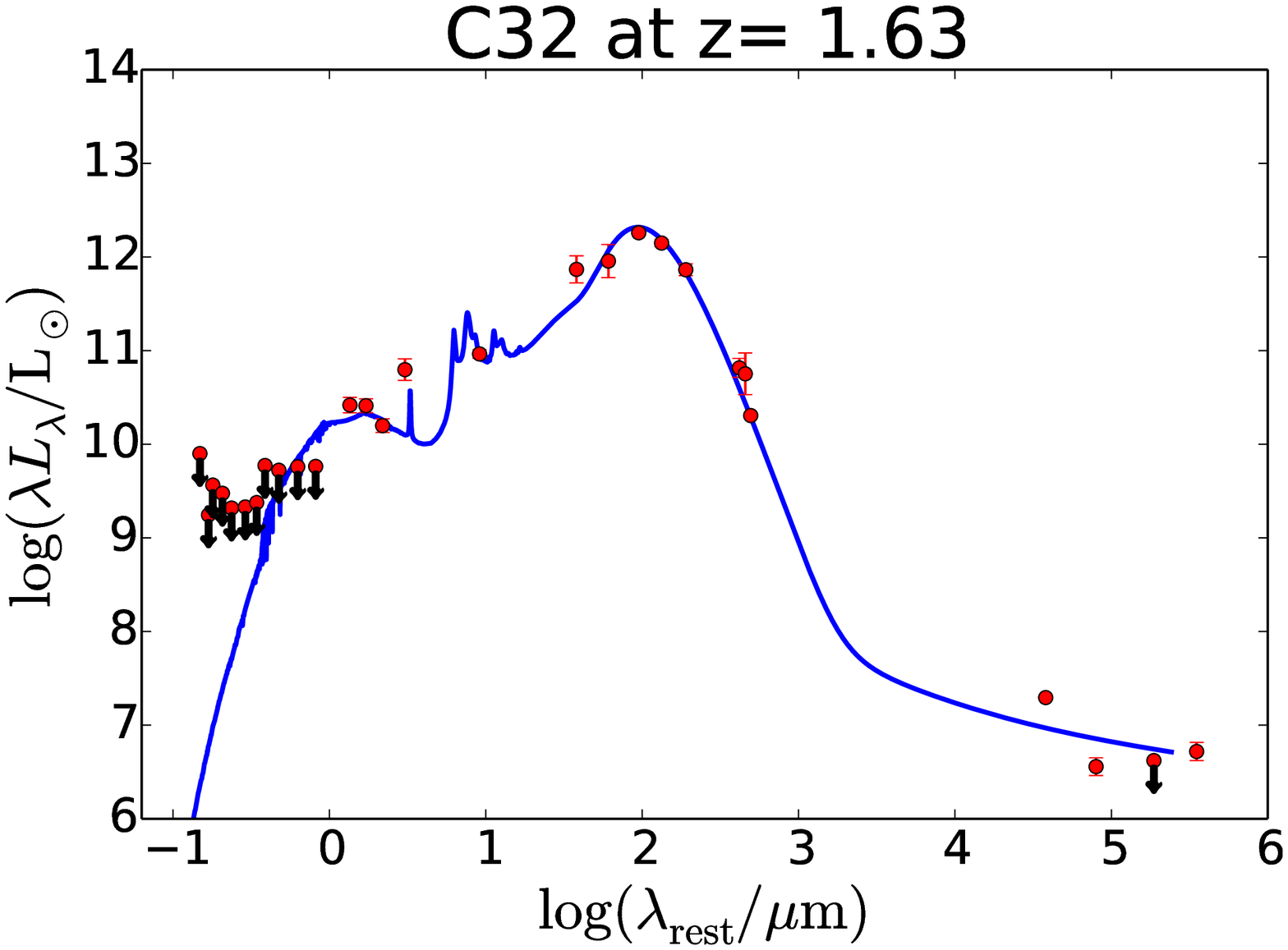}
\includegraphics[width=0.2465\textwidth]{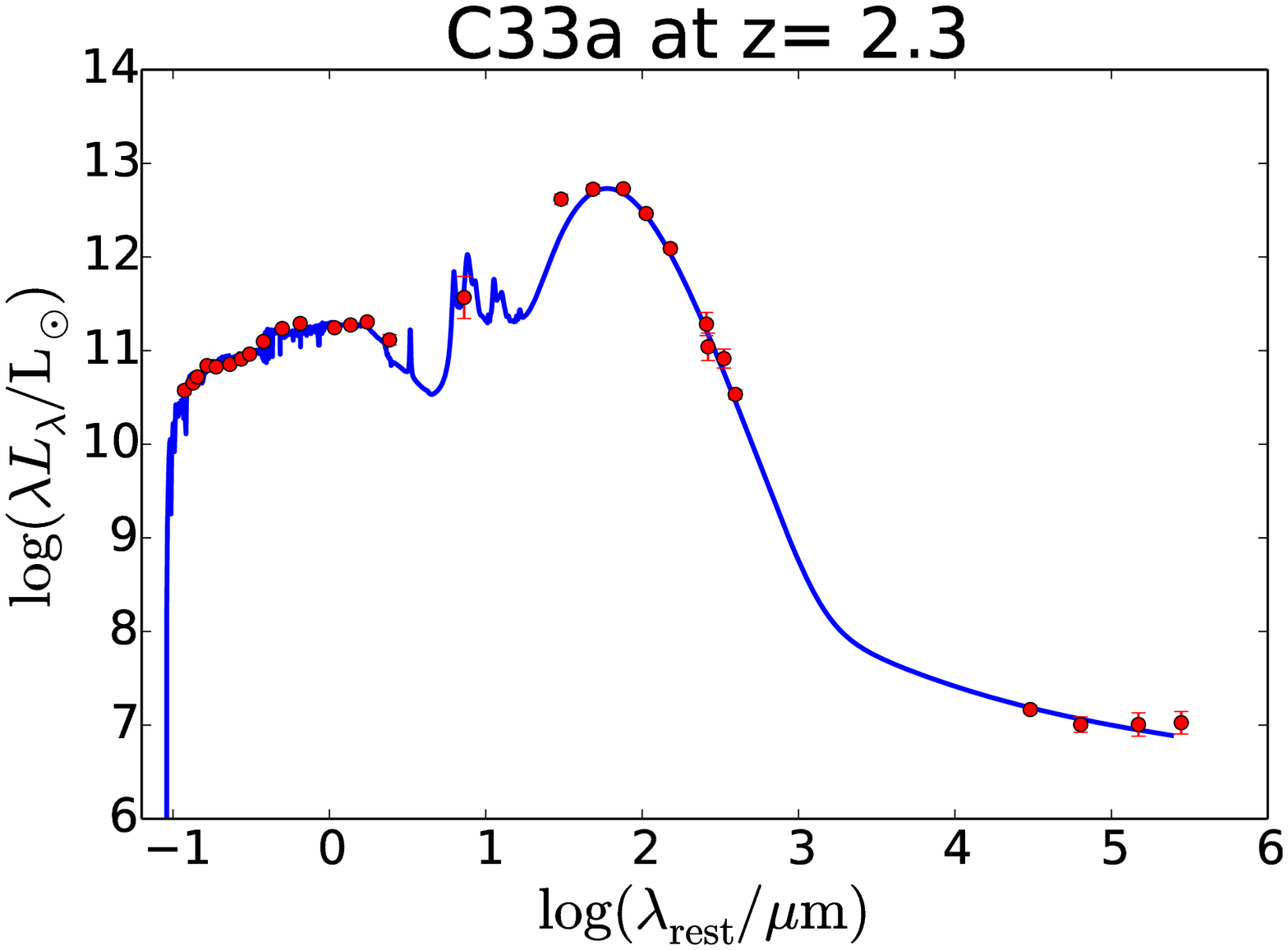}
\includegraphics[width=0.2465\textwidth]{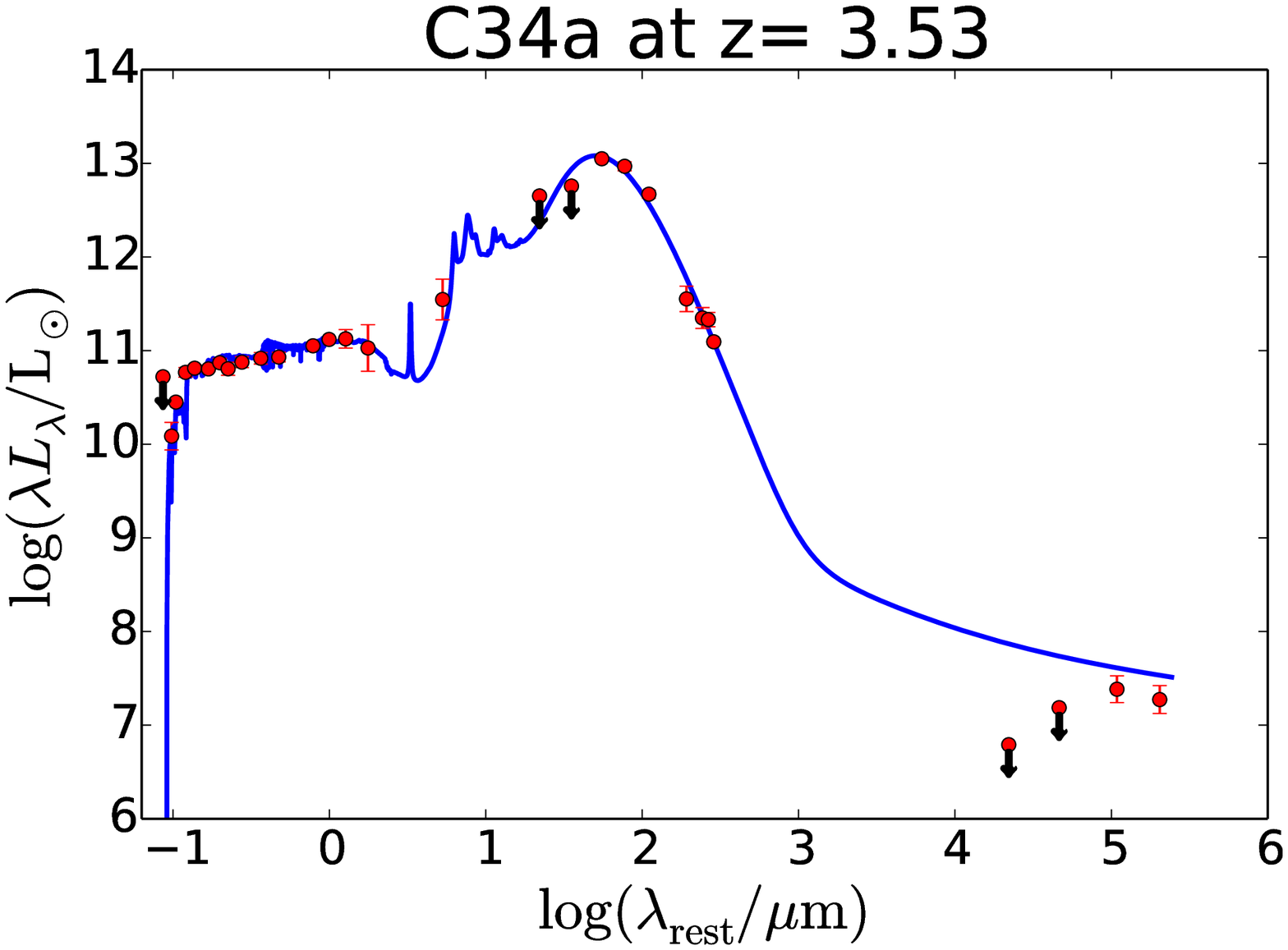}
\includegraphics[width=0.2465\textwidth]{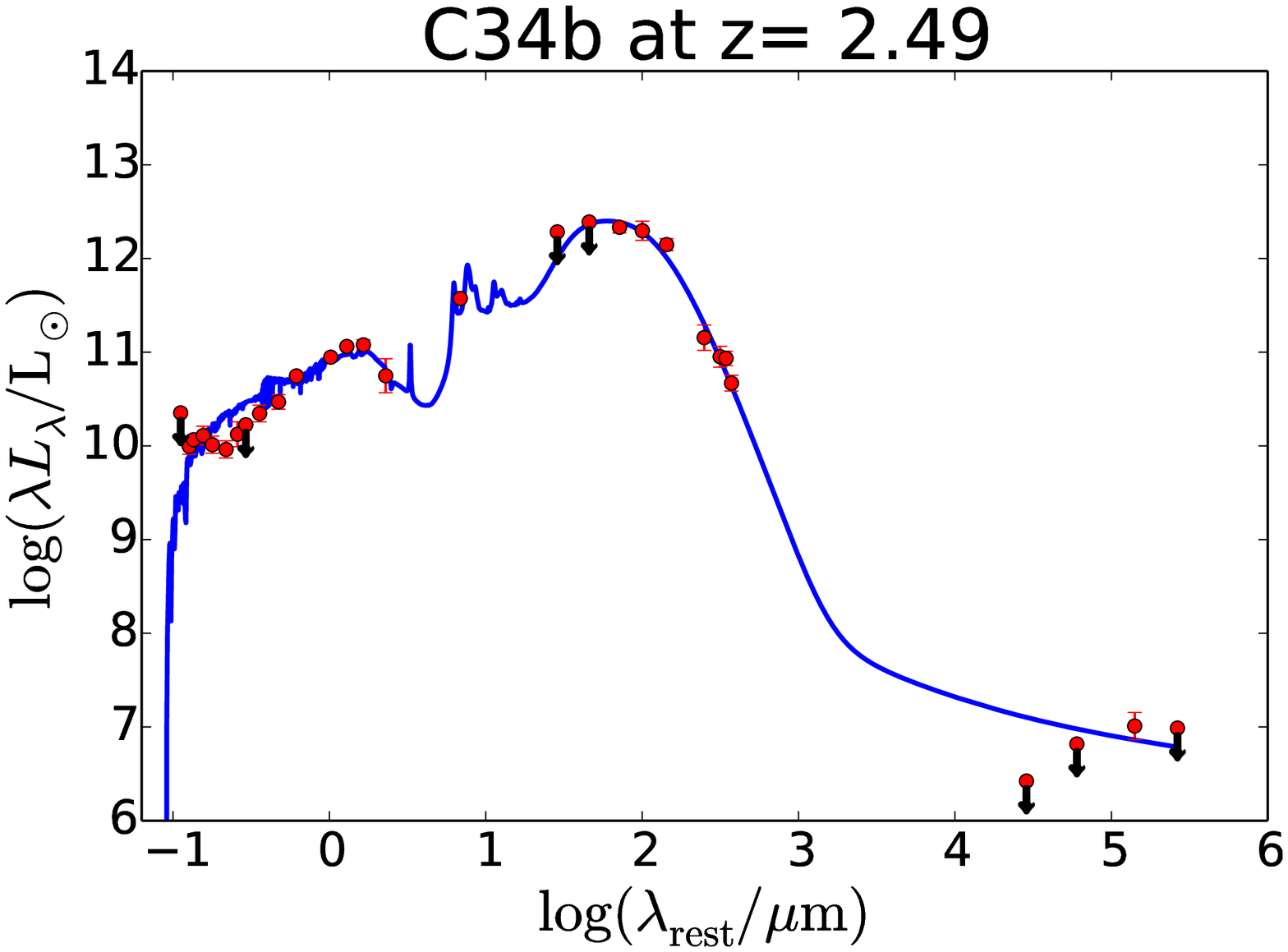}
\includegraphics[width=0.2465\textwidth]{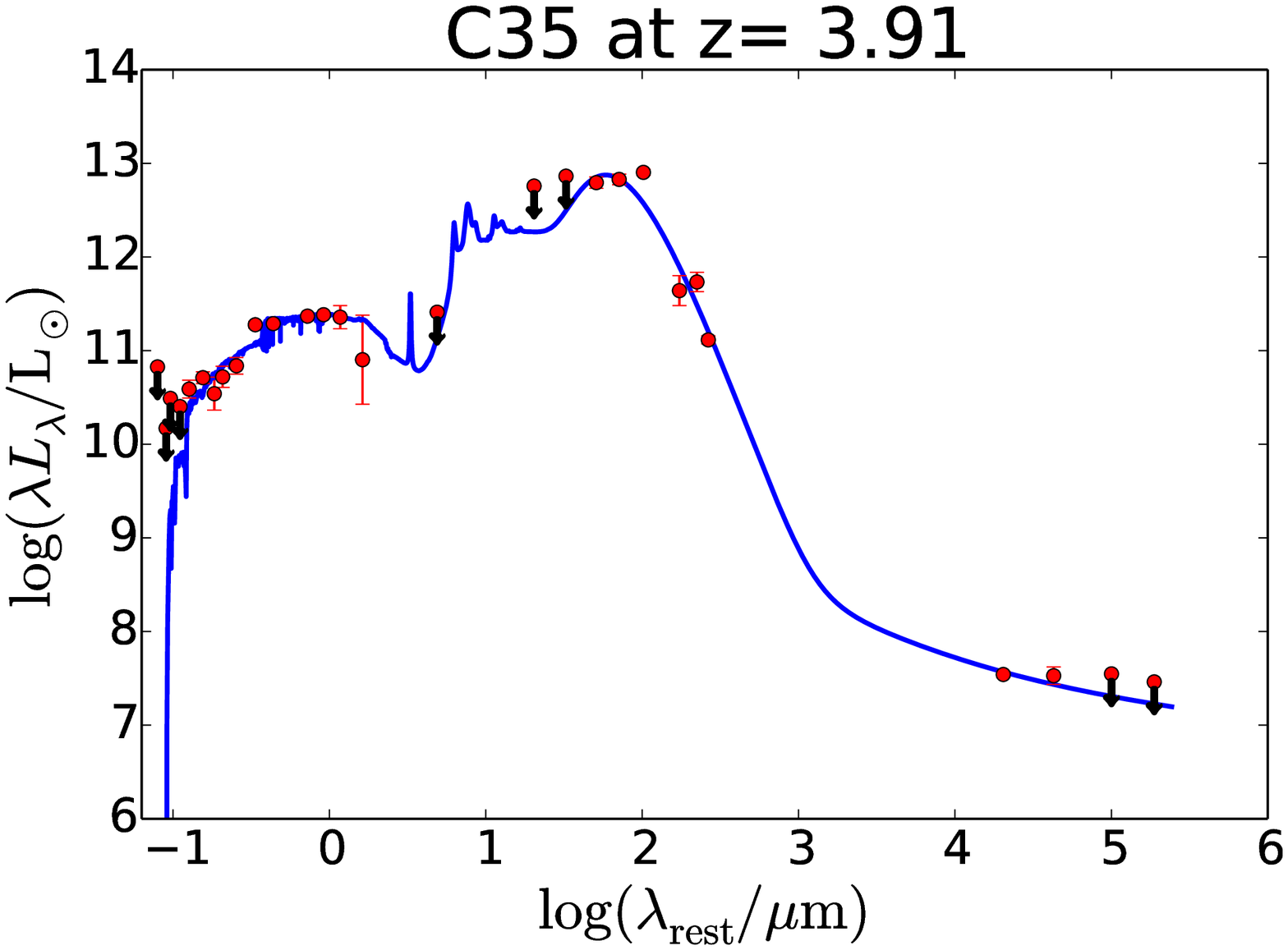}
\includegraphics[width=0.2465\textwidth]{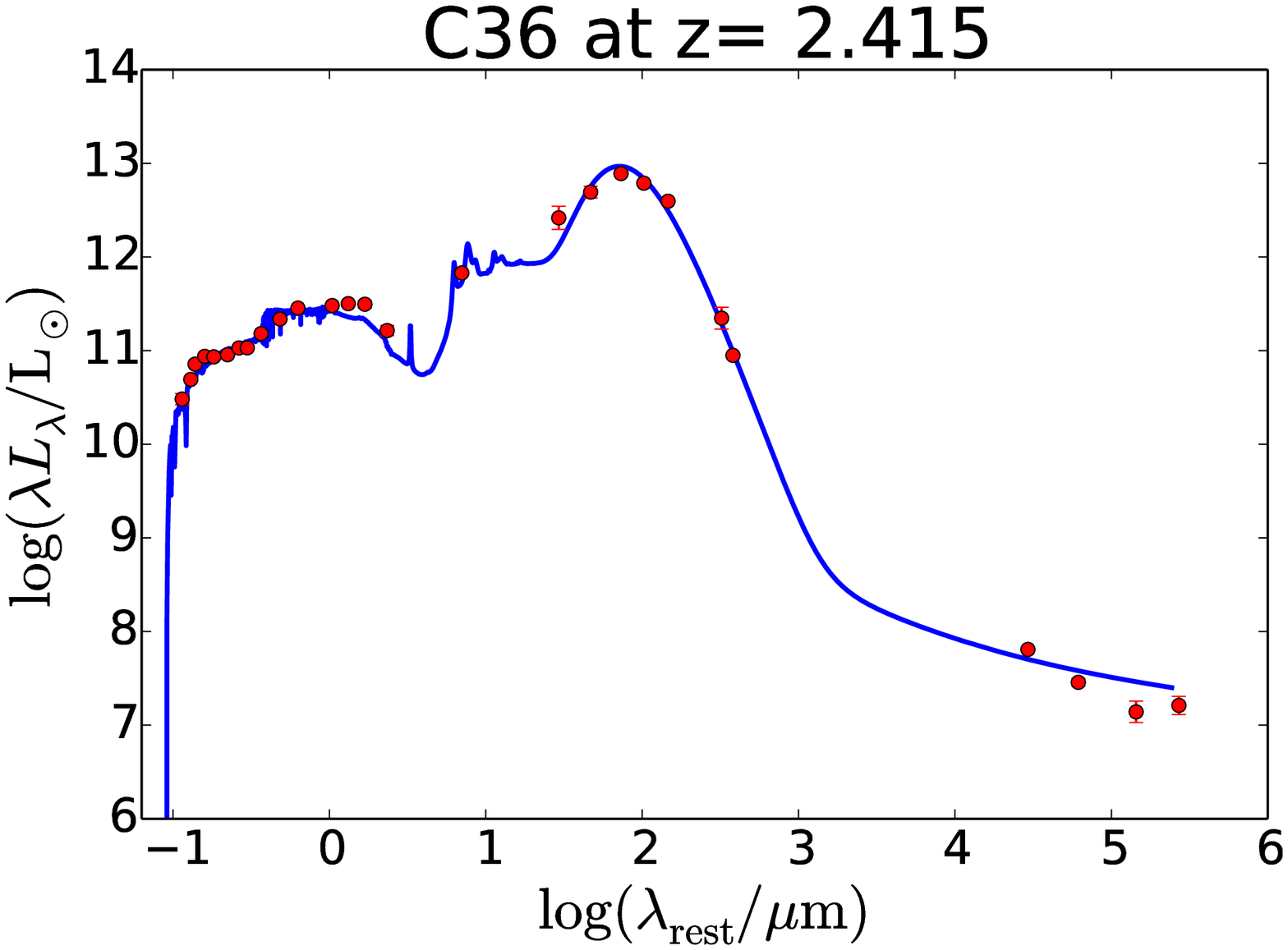}
\includegraphics[width=0.2465\textwidth]{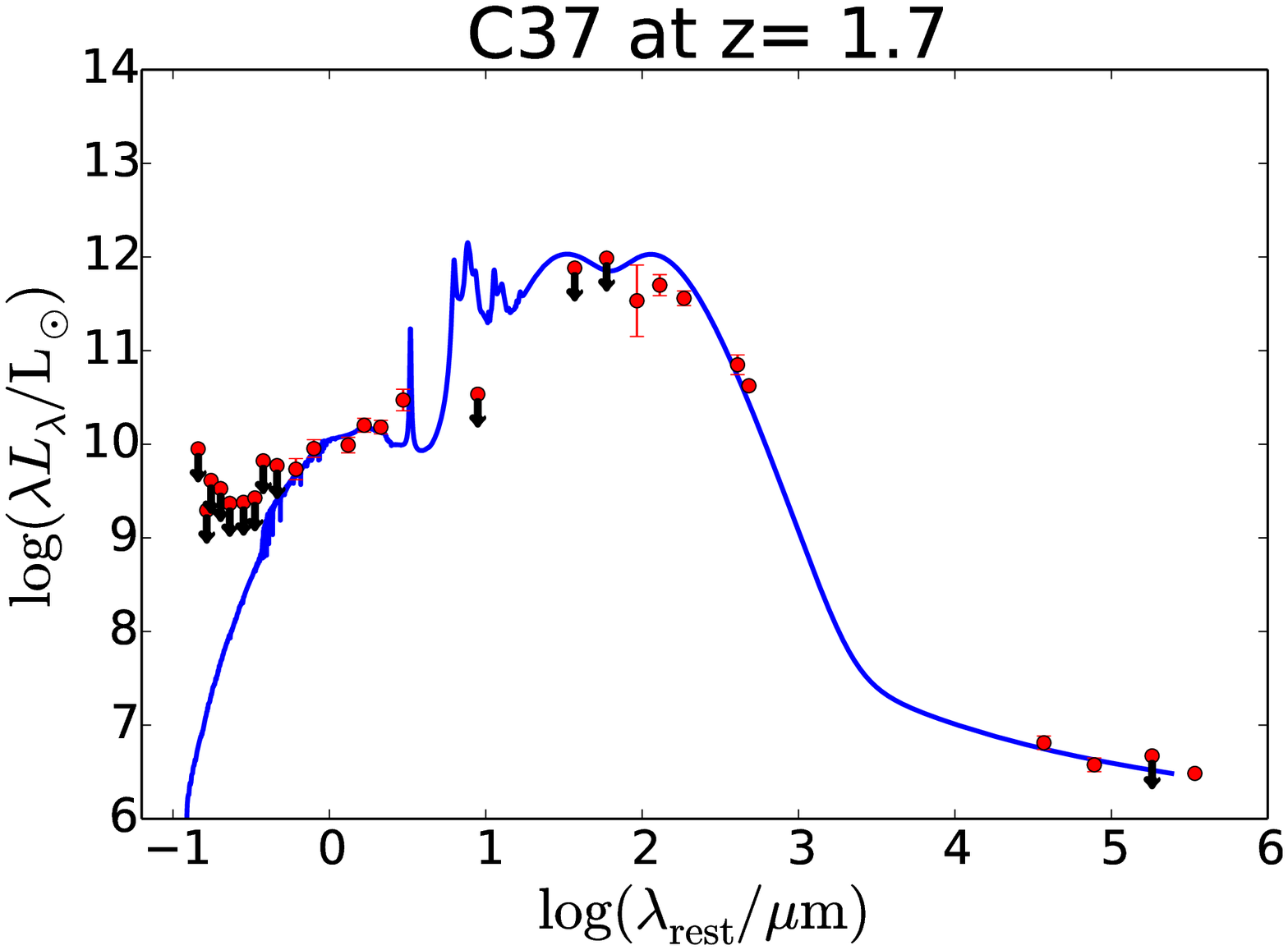}
\includegraphics[width=0.2465\textwidth]{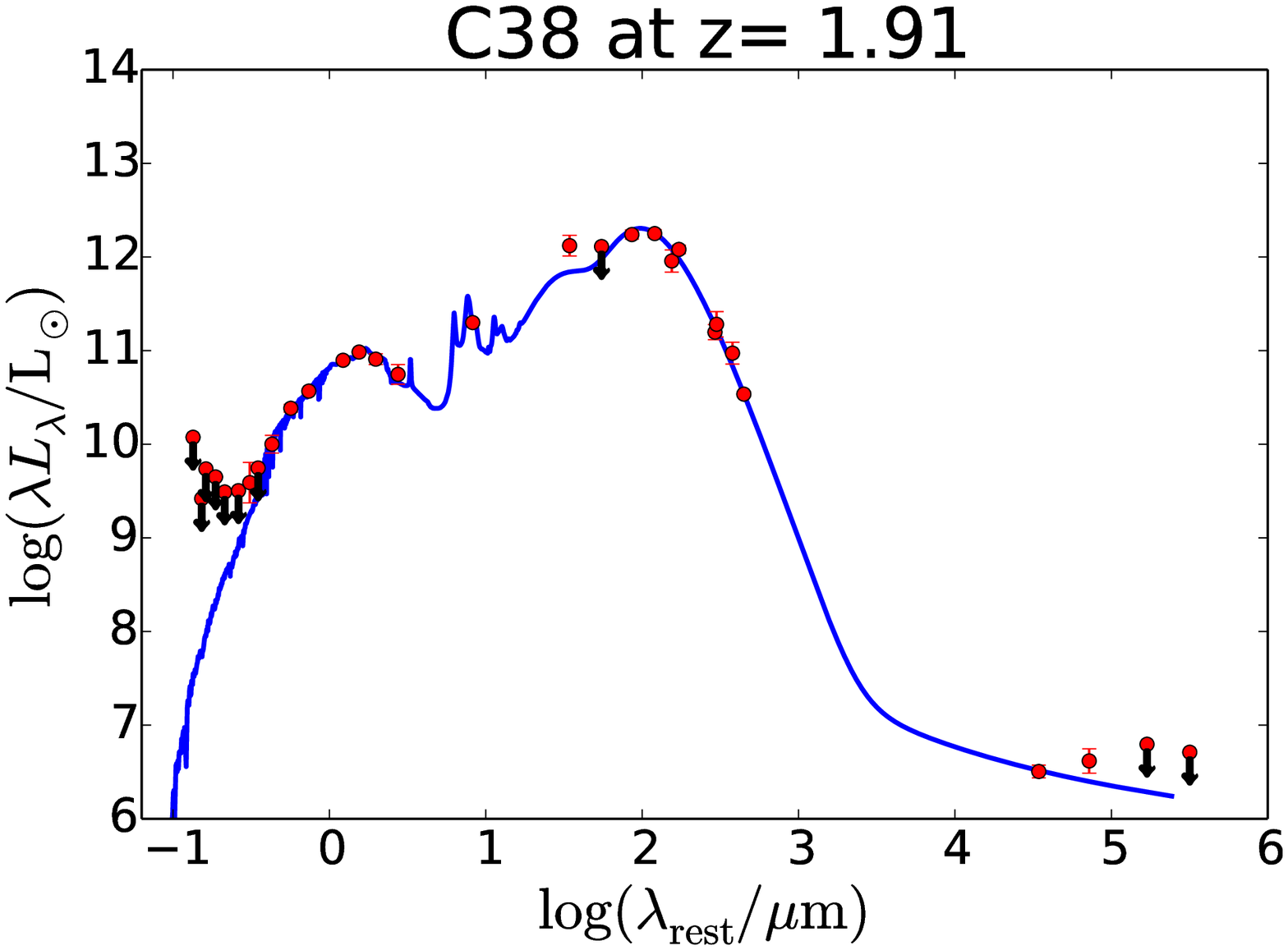}
\includegraphics[width=0.2465\textwidth]{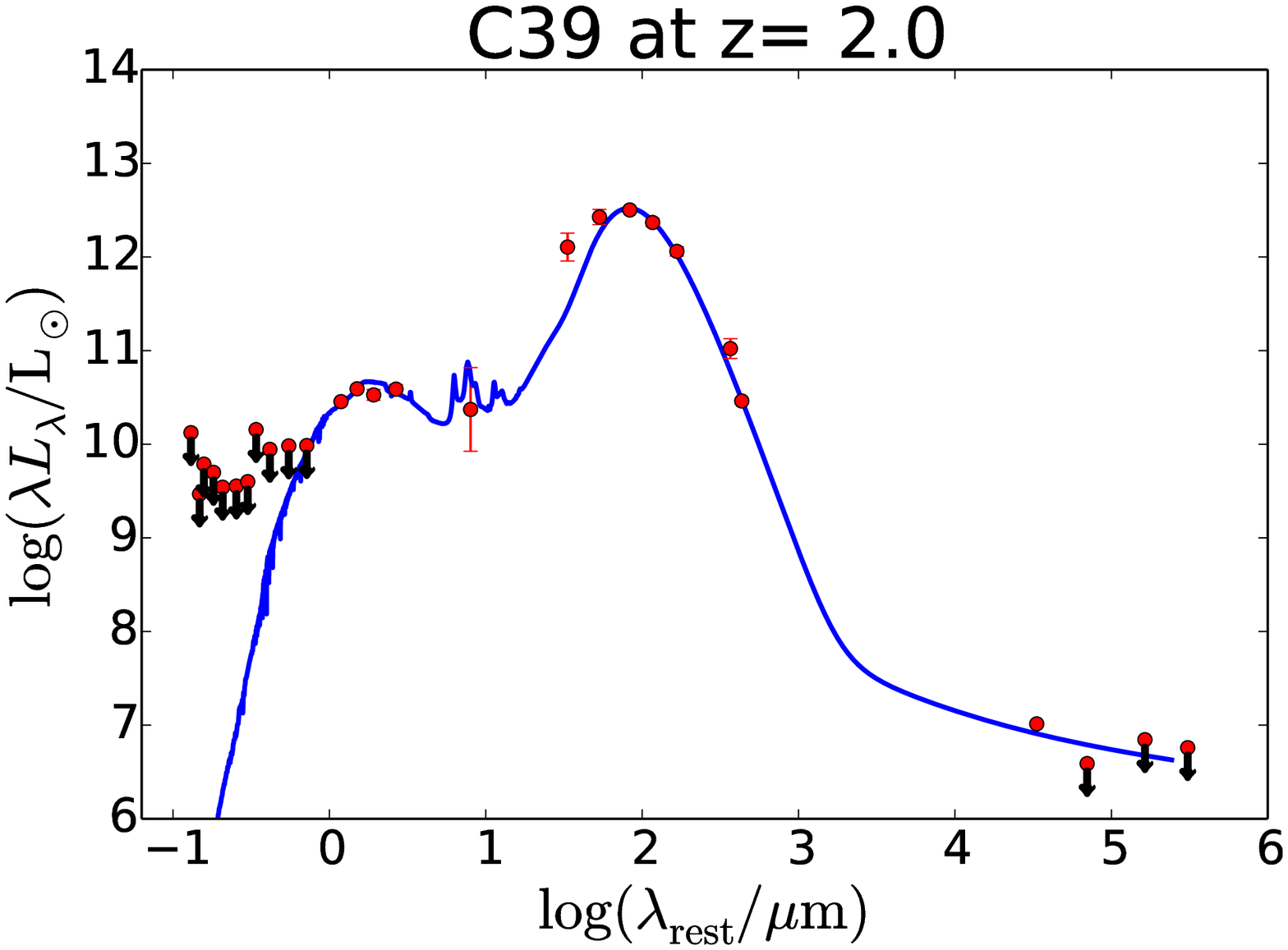}
\includegraphics[width=0.2465\textwidth]{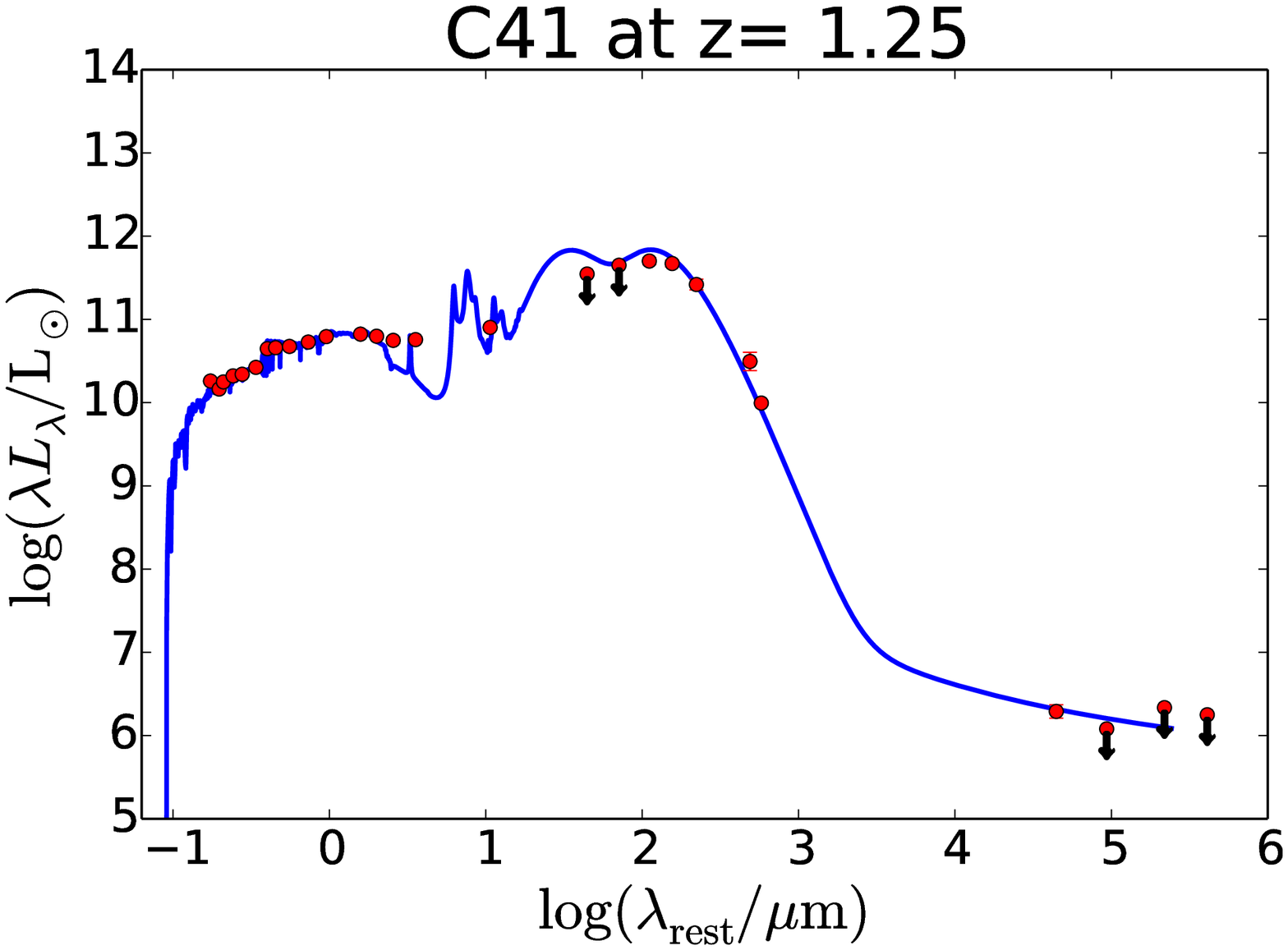}
\includegraphics[width=0.2465\textwidth]{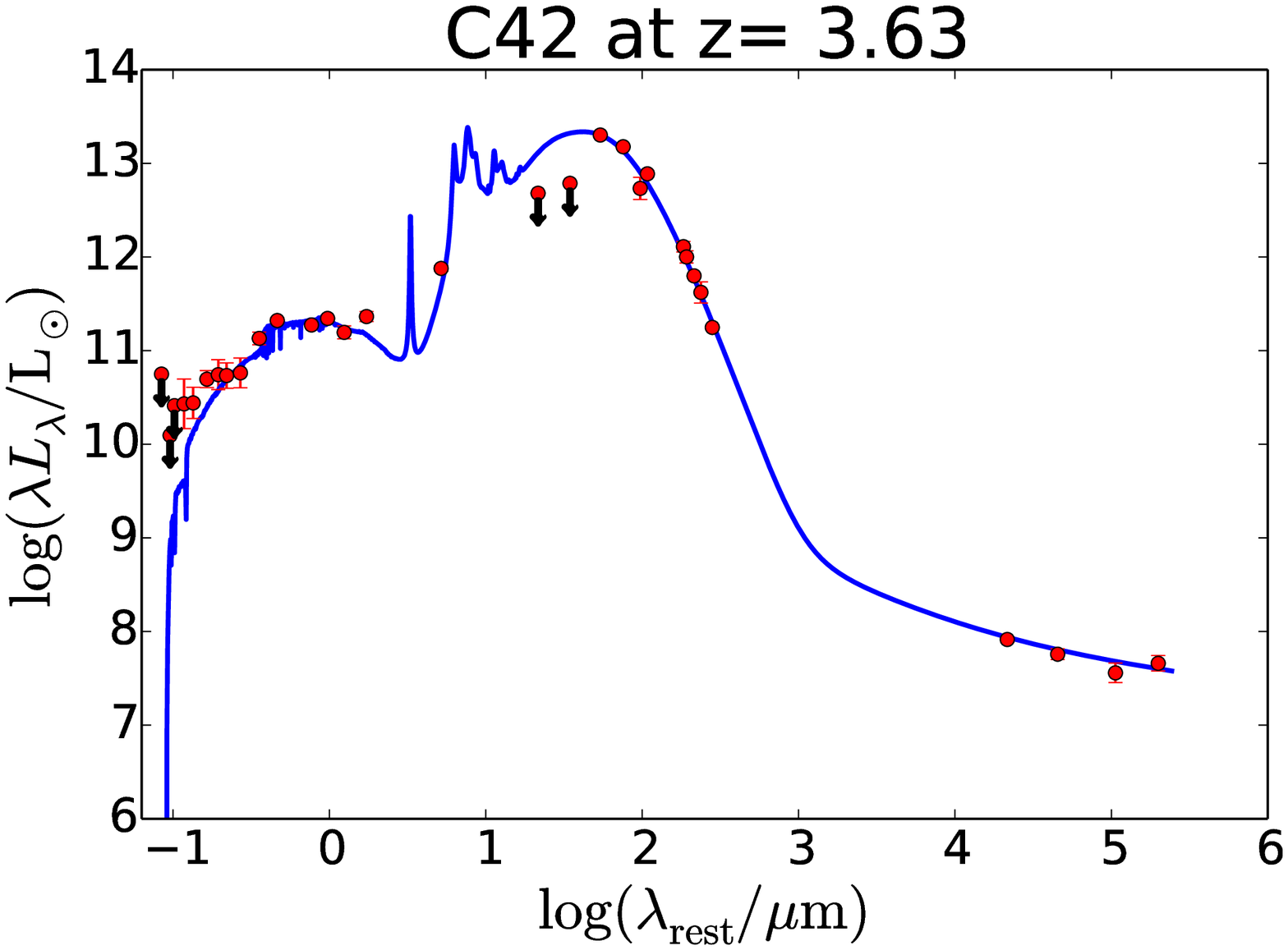}
\includegraphics[width=0.2465\textwidth]{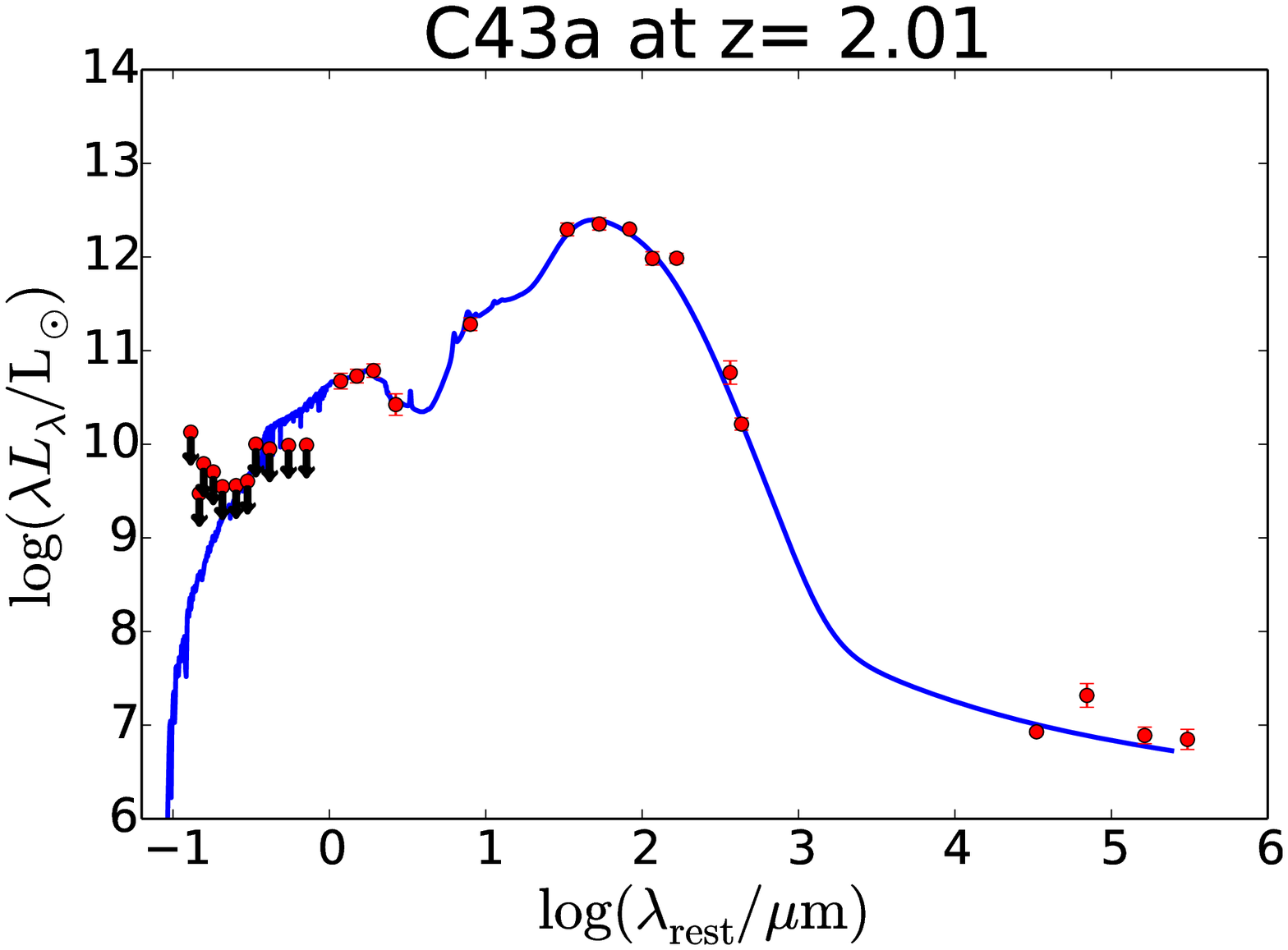} 
\includegraphics[width=0.2465\textwidth]{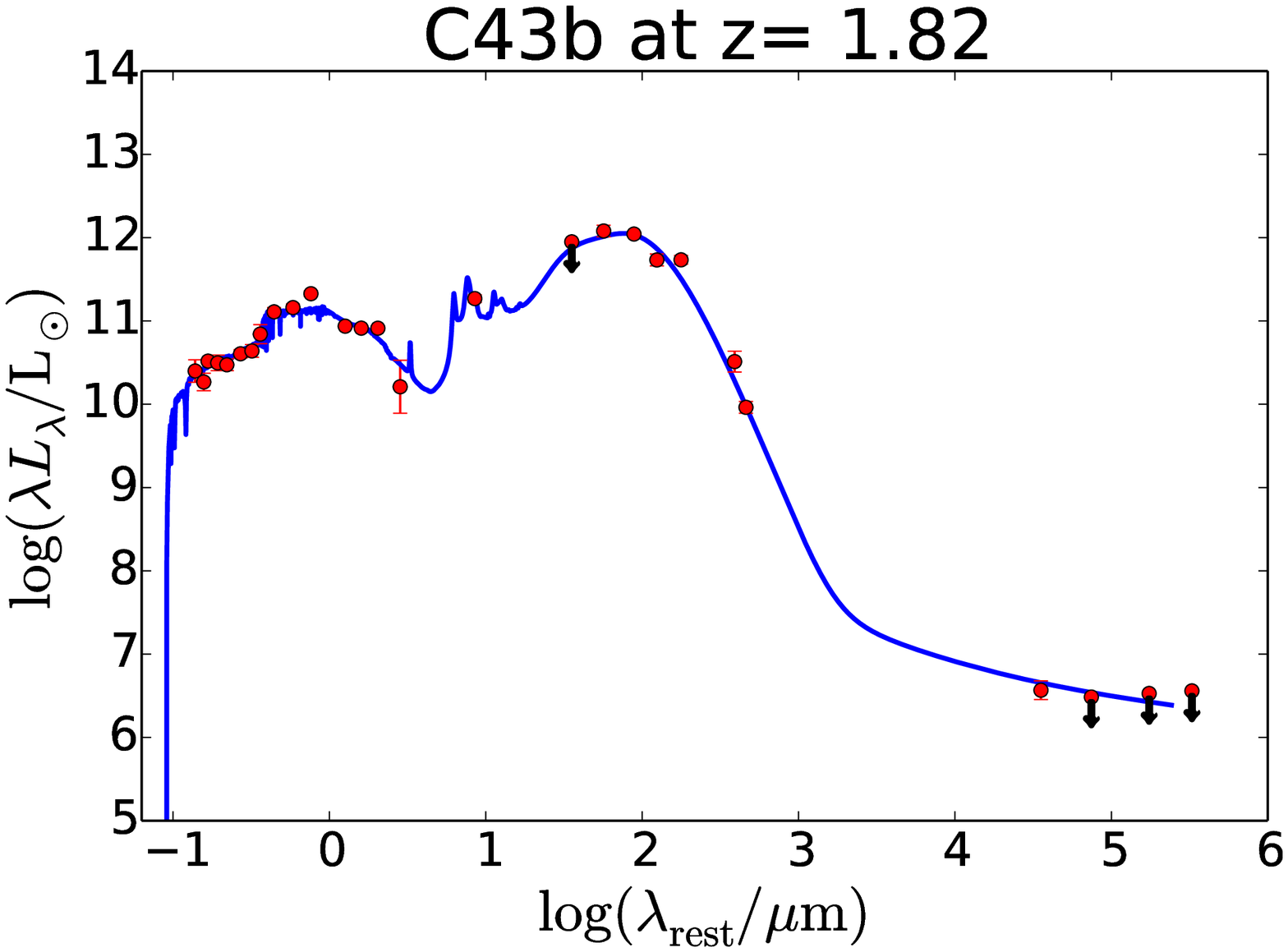}
\caption{continued.}
\label{figure:seds}
\end{center}
\end{figure*}

\addtocounter{figure}{-1}
\begin{figure*}
\begin{center}
\includegraphics[width=0.2465\textwidth]{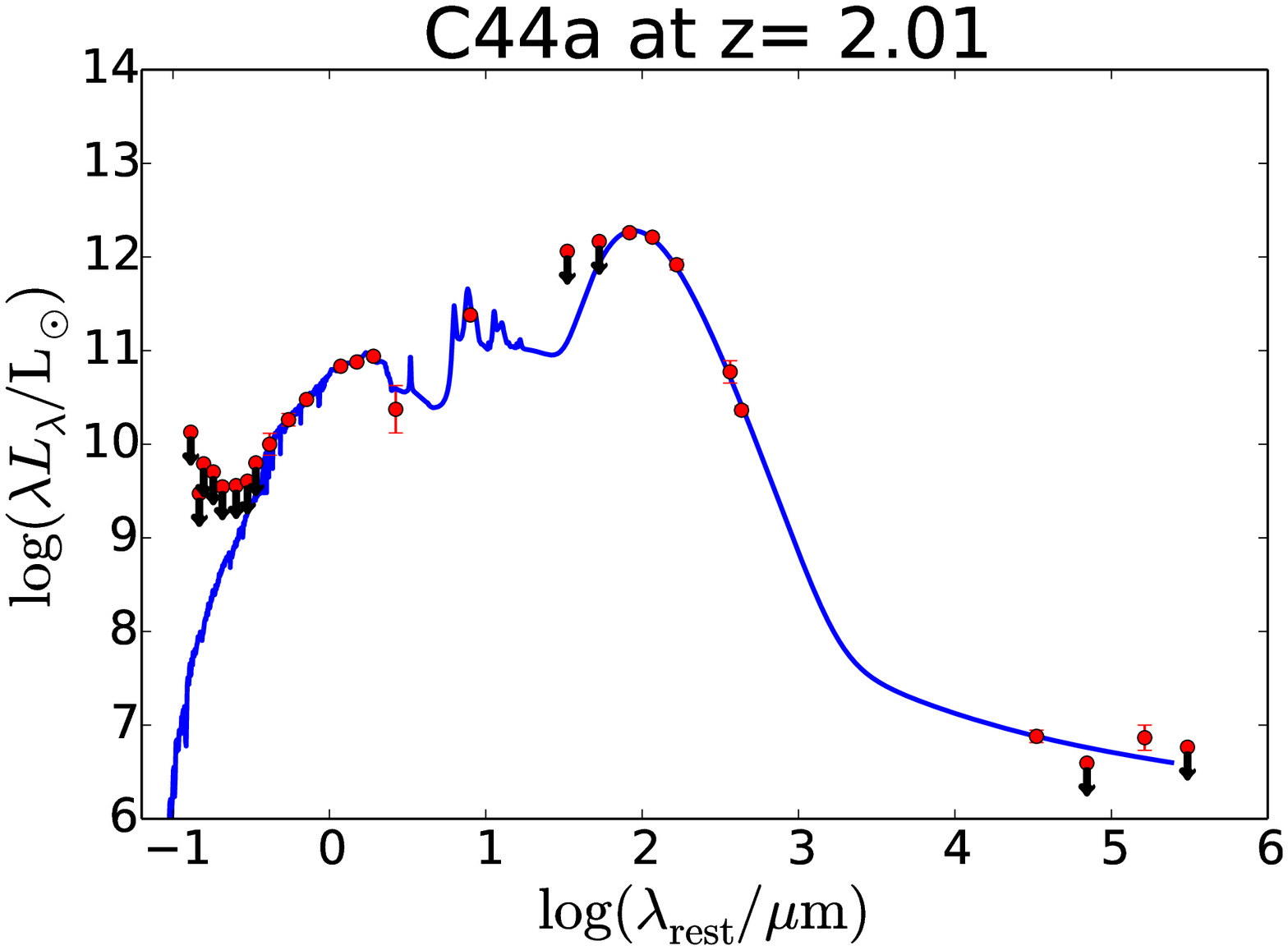}
\includegraphics[width=0.2465\textwidth]{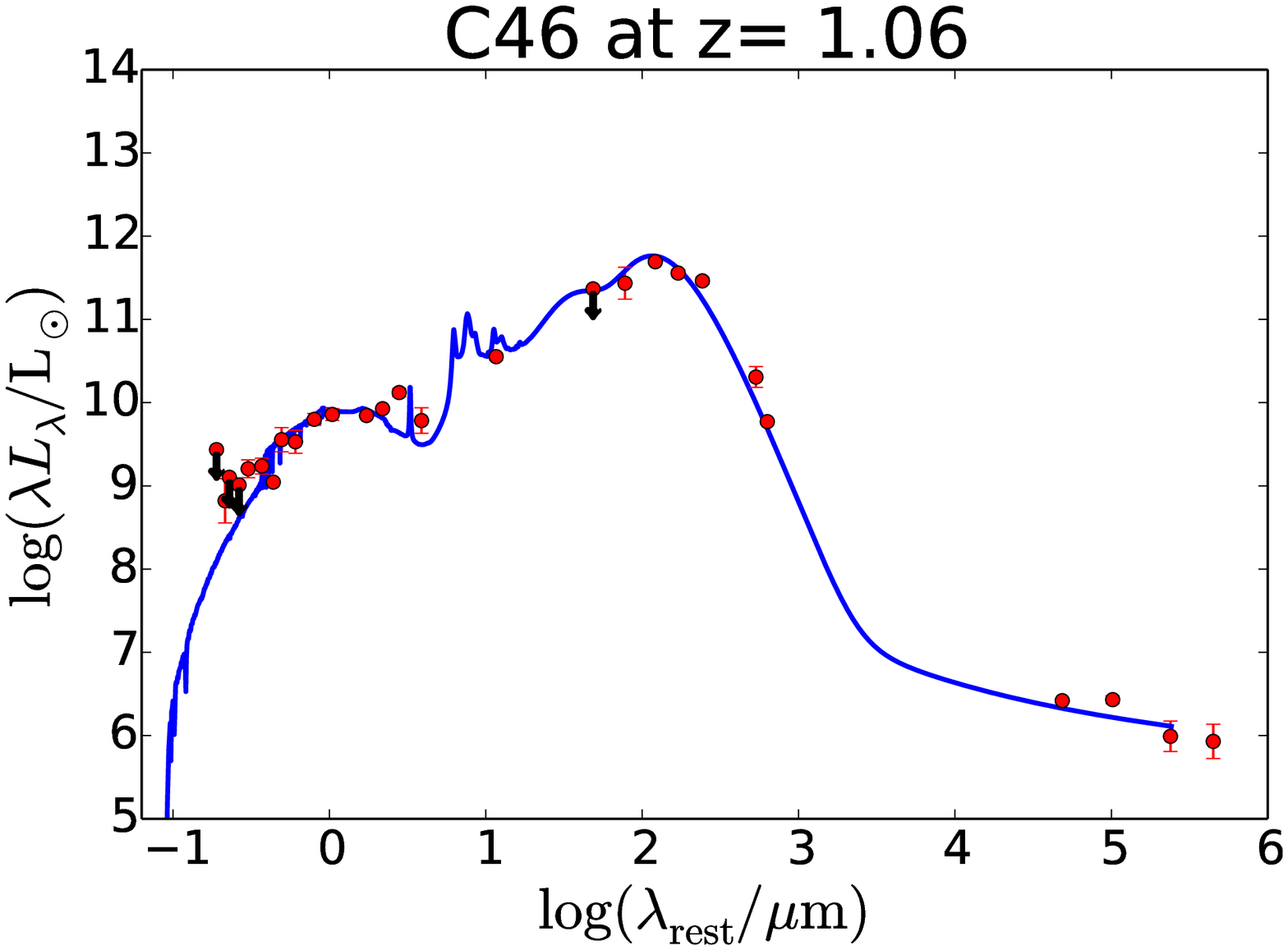}
\includegraphics[width=0.2465\textwidth]{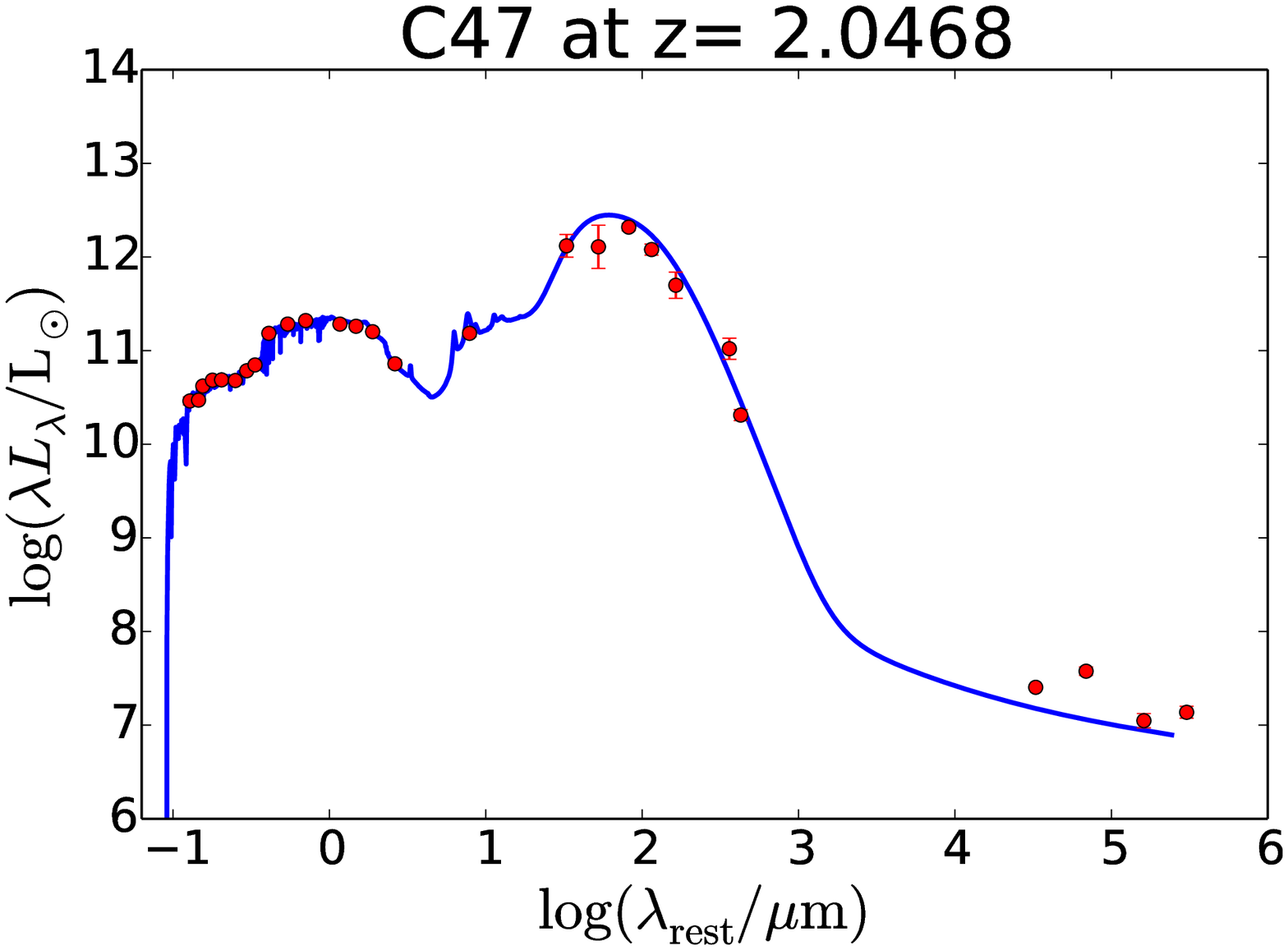}
\includegraphics[width=0.2465\textwidth]{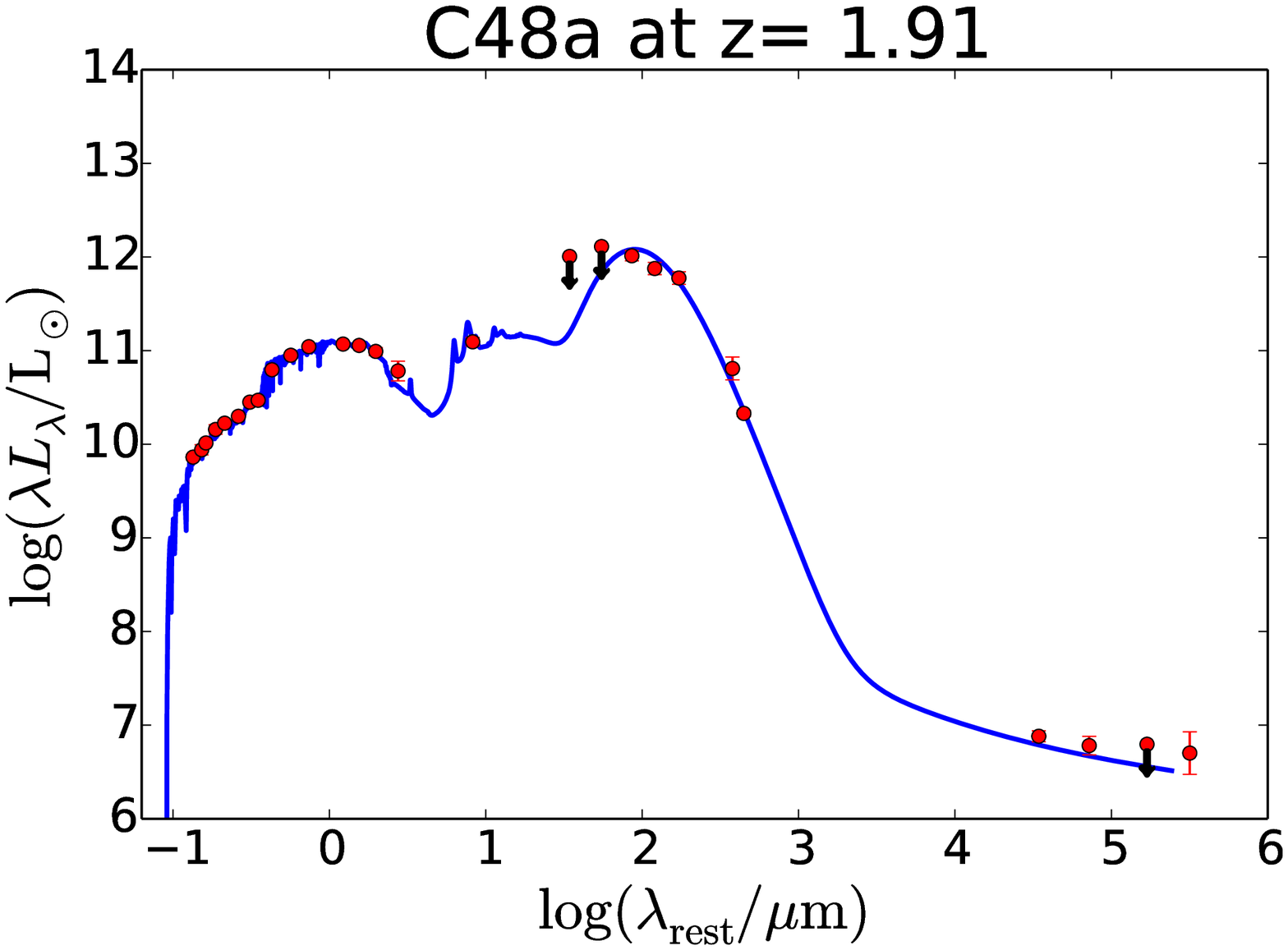}
\includegraphics[width=0.2465\textwidth]{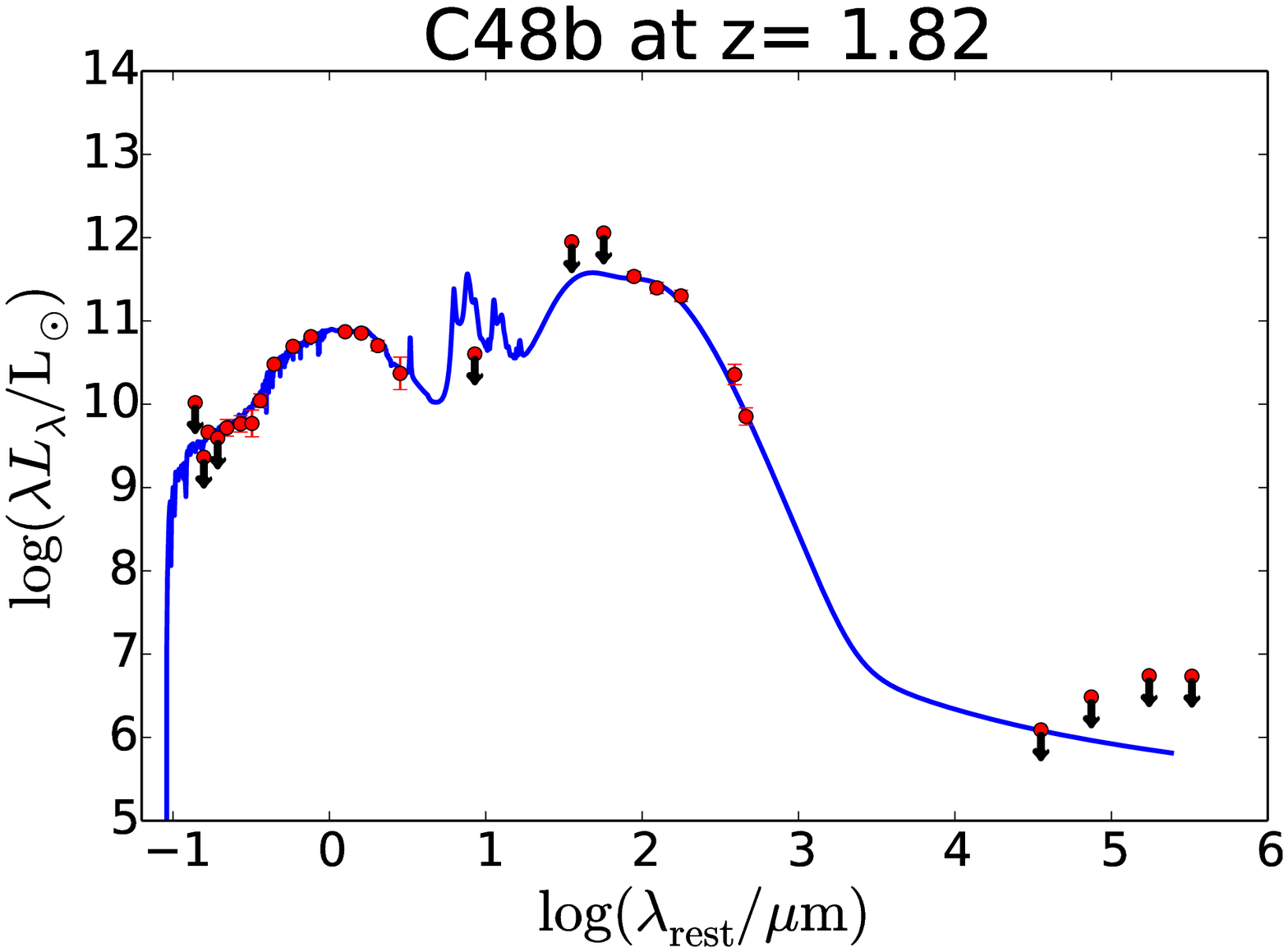}
\includegraphics[width=0.2465\textwidth]{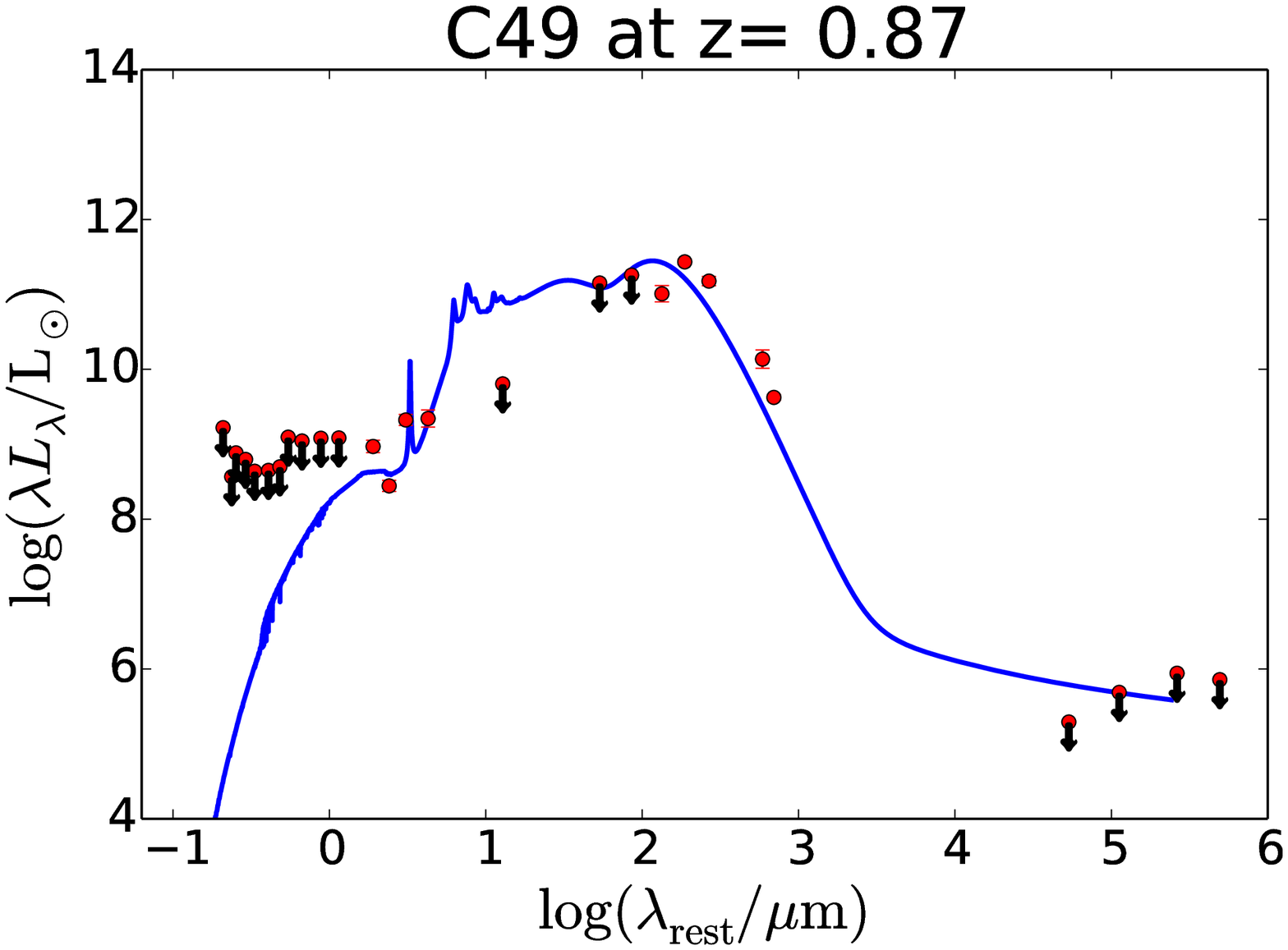}
\includegraphics[width=0.2465\textwidth]{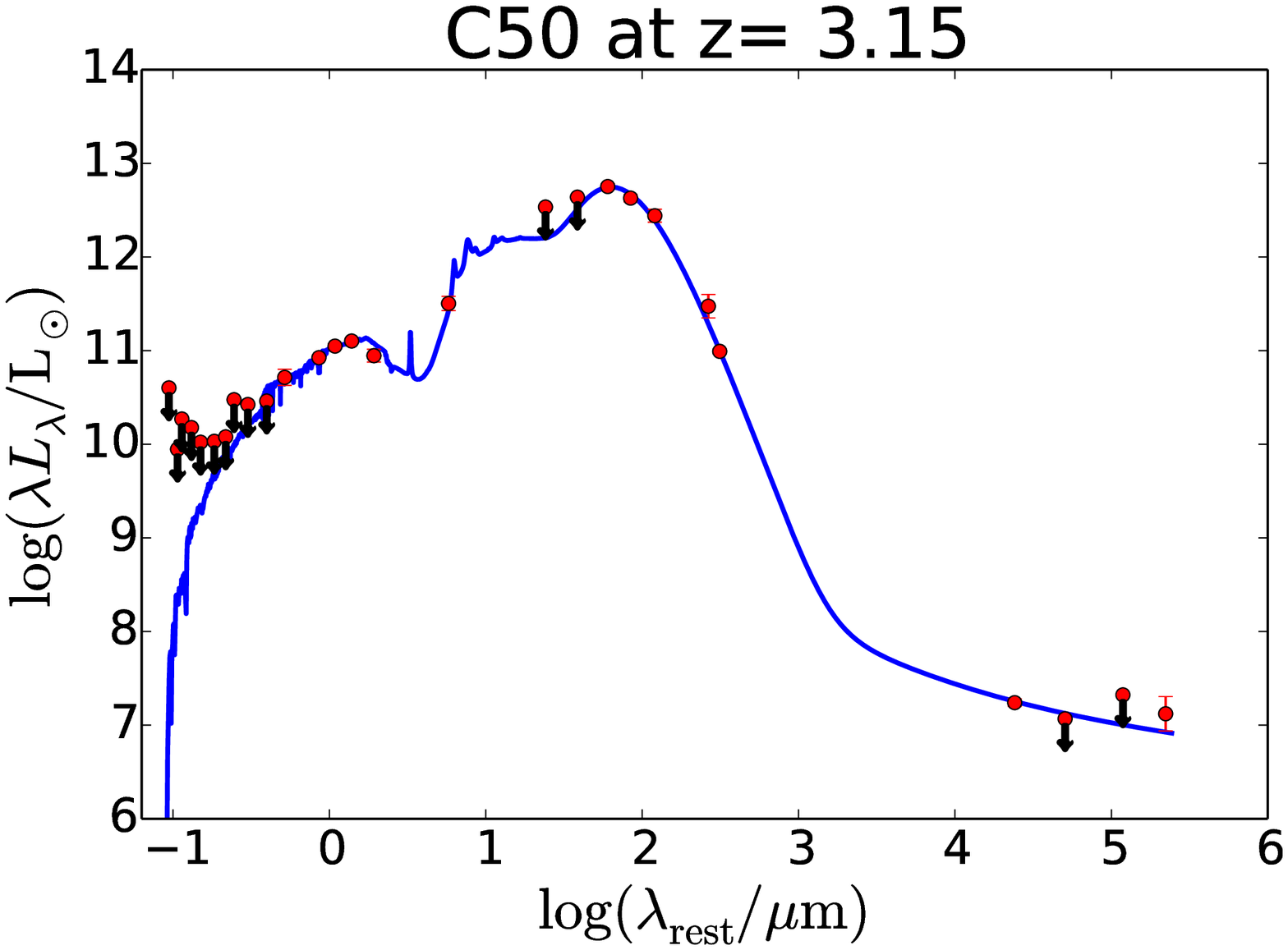}
\includegraphics[width=0.2465\textwidth]{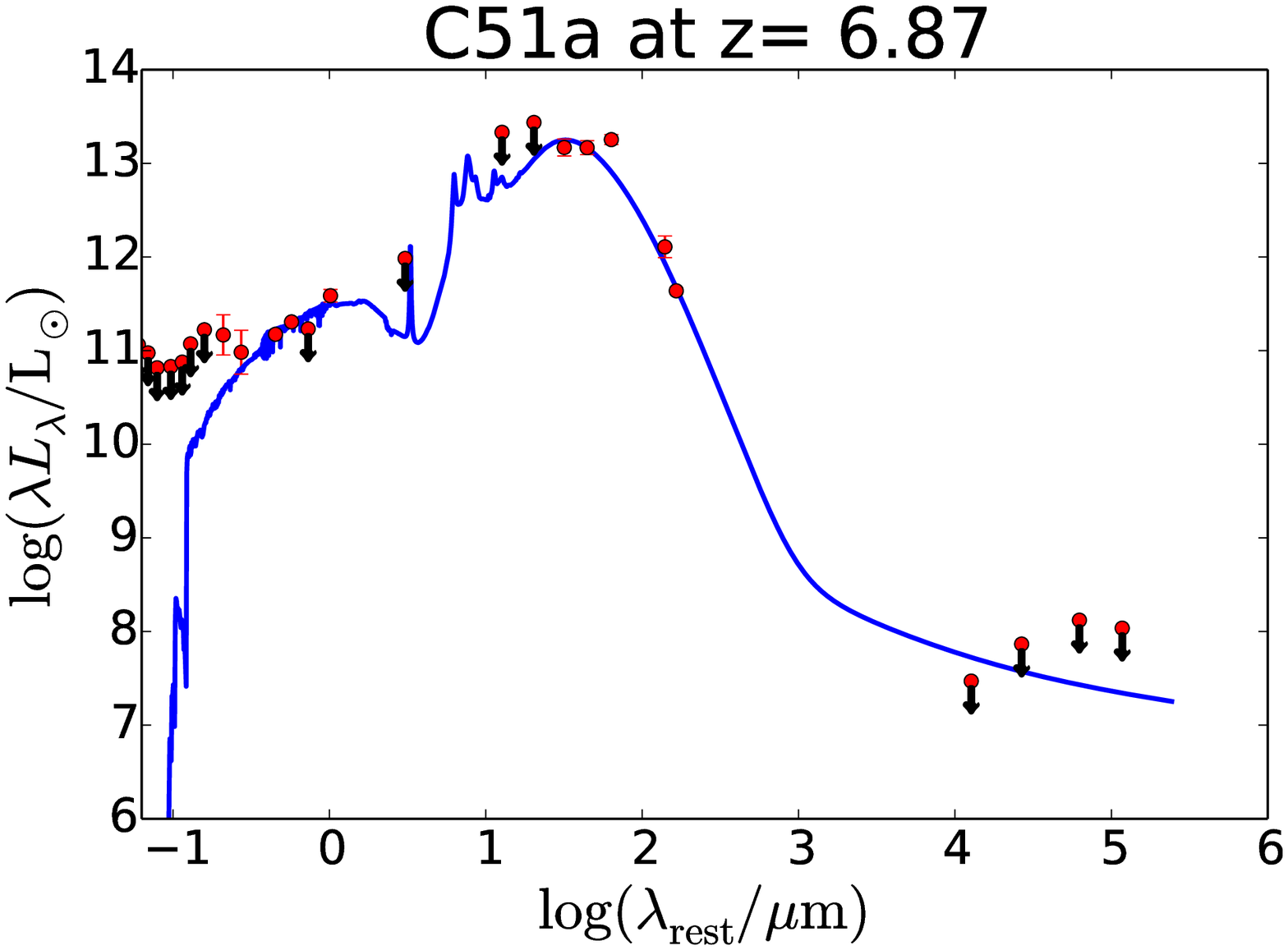}
\includegraphics[width=0.2465\textwidth]{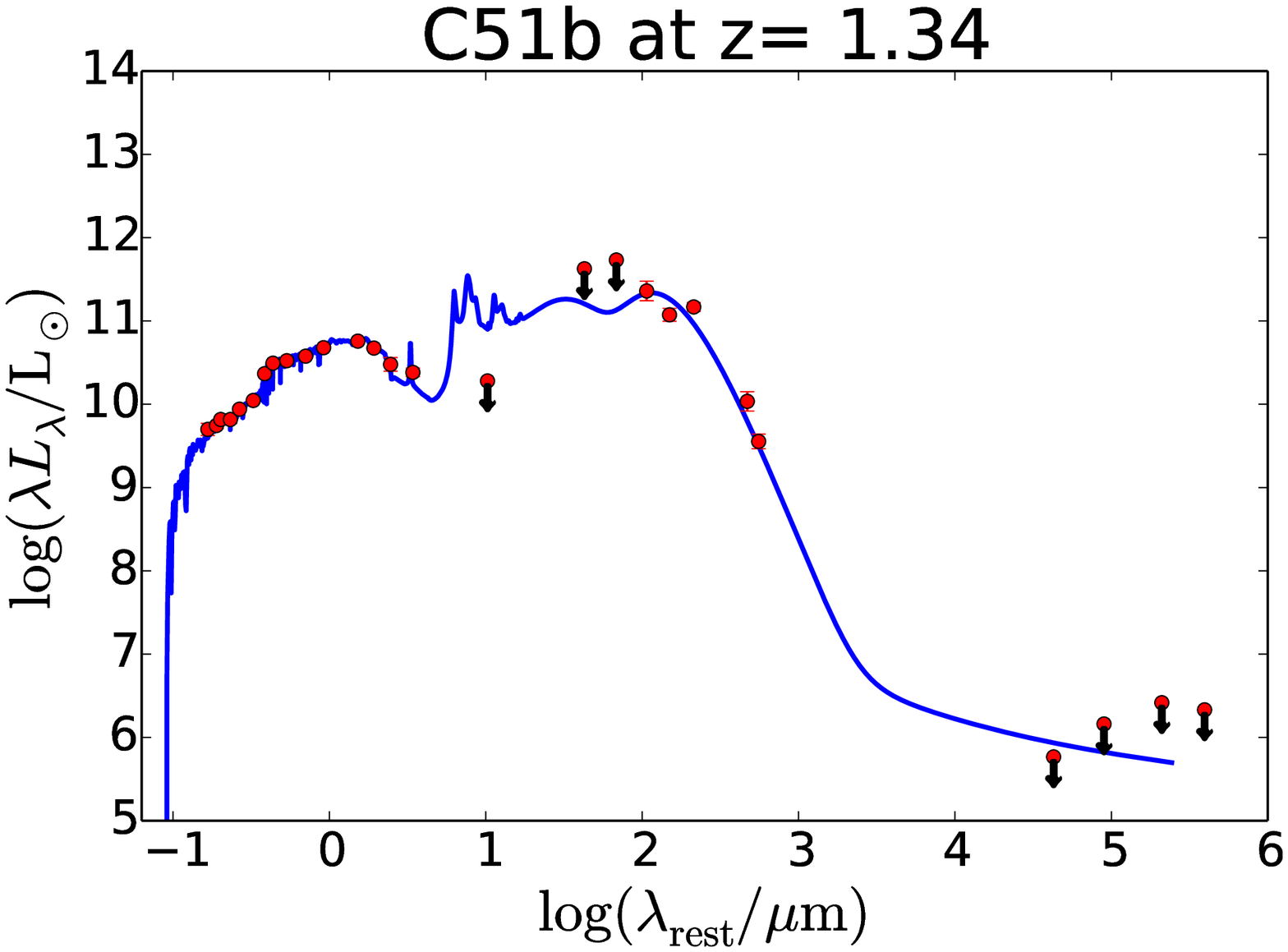}
\includegraphics[width=0.2465\textwidth]{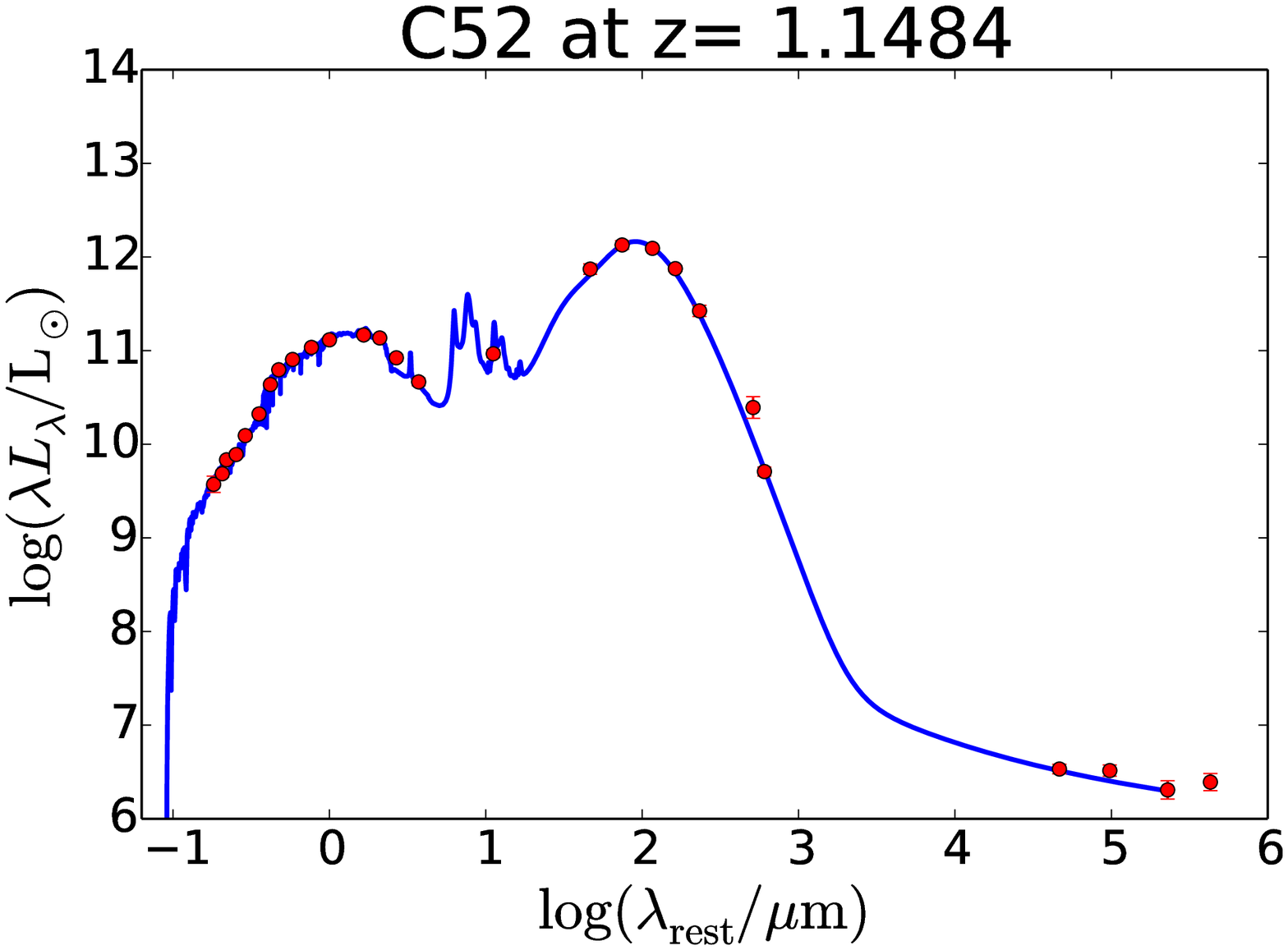}
\includegraphics[width=0.2465\textwidth]{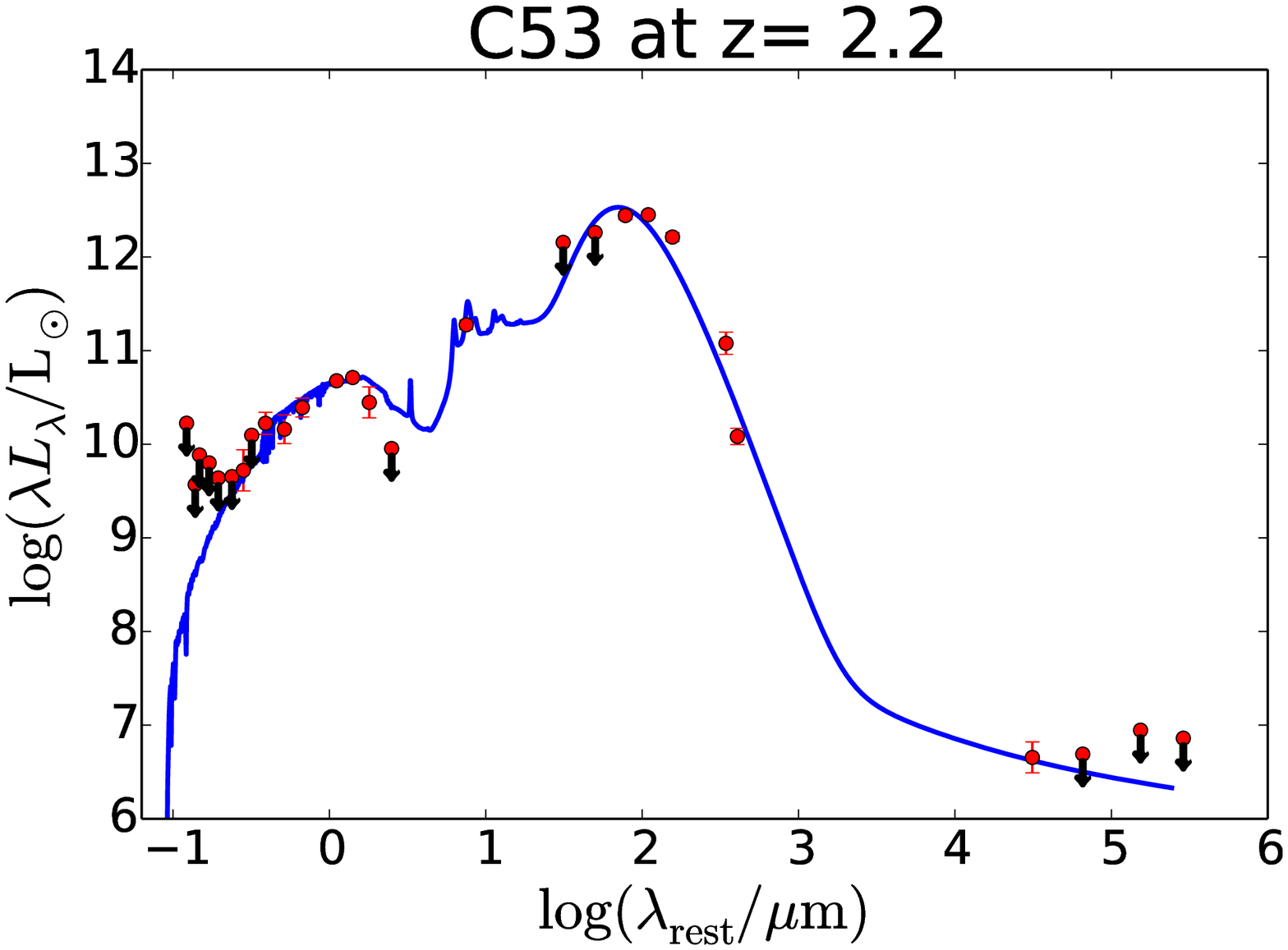}
\includegraphics[width=0.2465\textwidth]{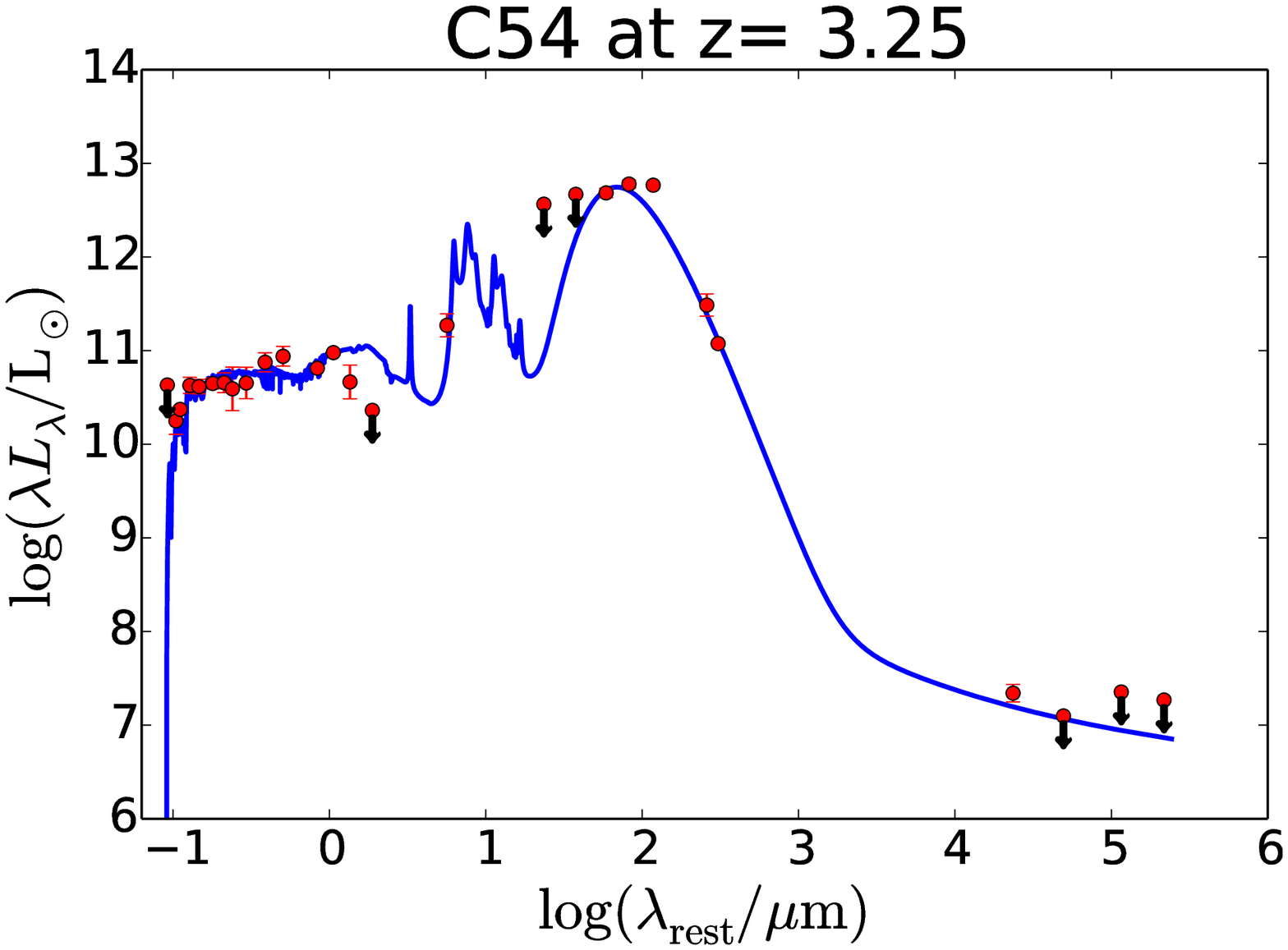}
\includegraphics[width=0.2465\textwidth]{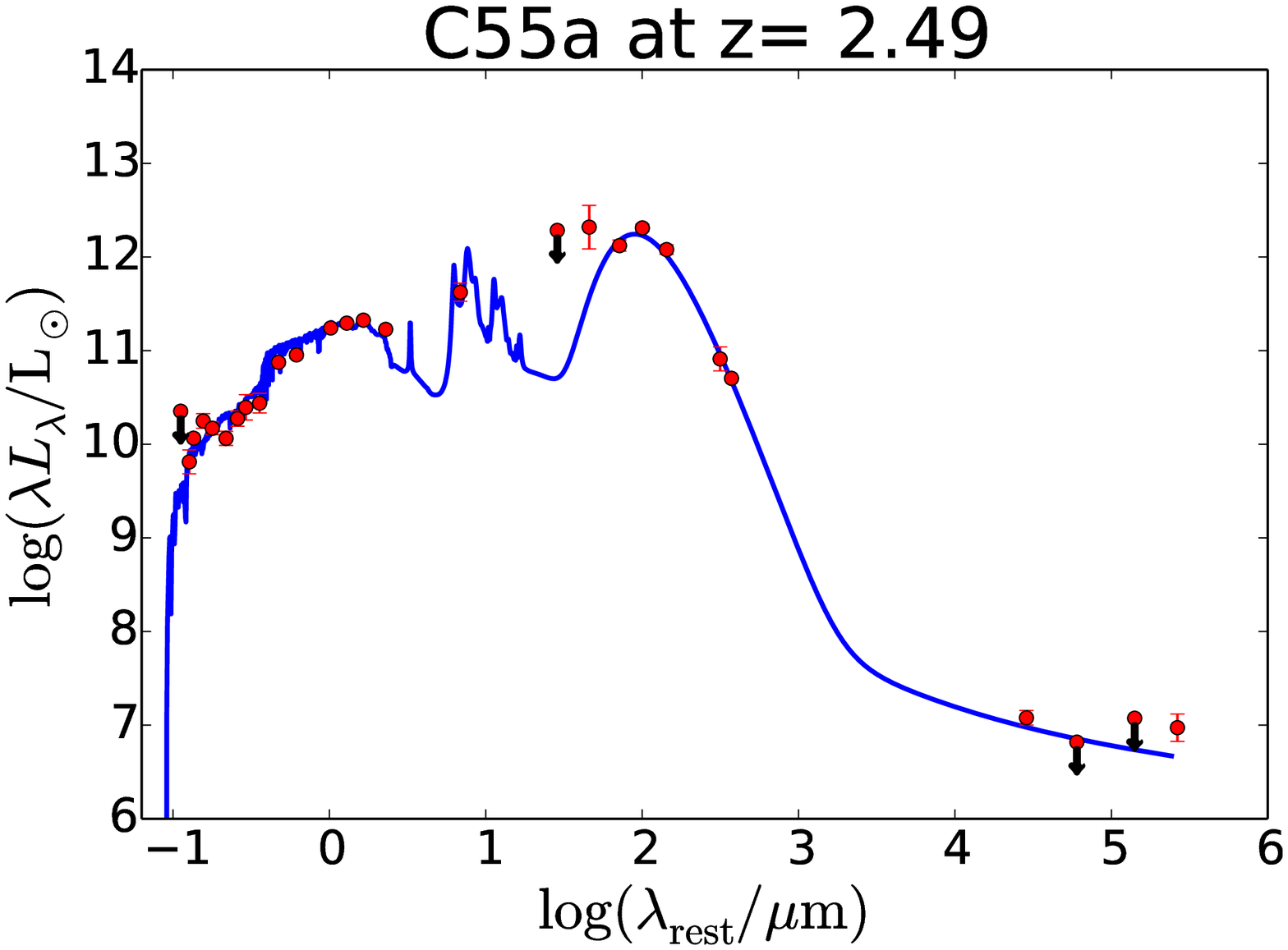}
\includegraphics[width=0.2465\textwidth]{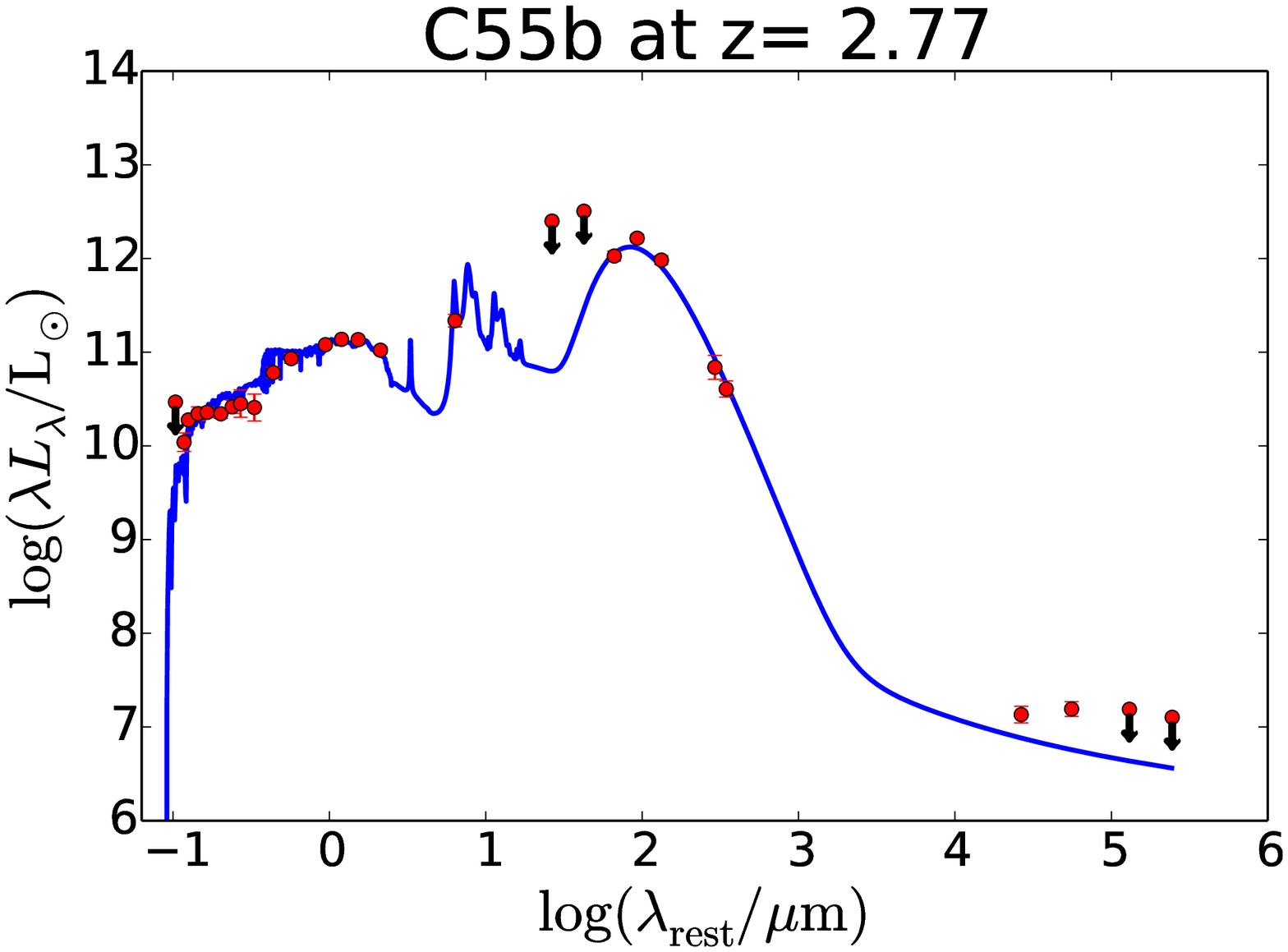}
\includegraphics[width=0.2465\textwidth]{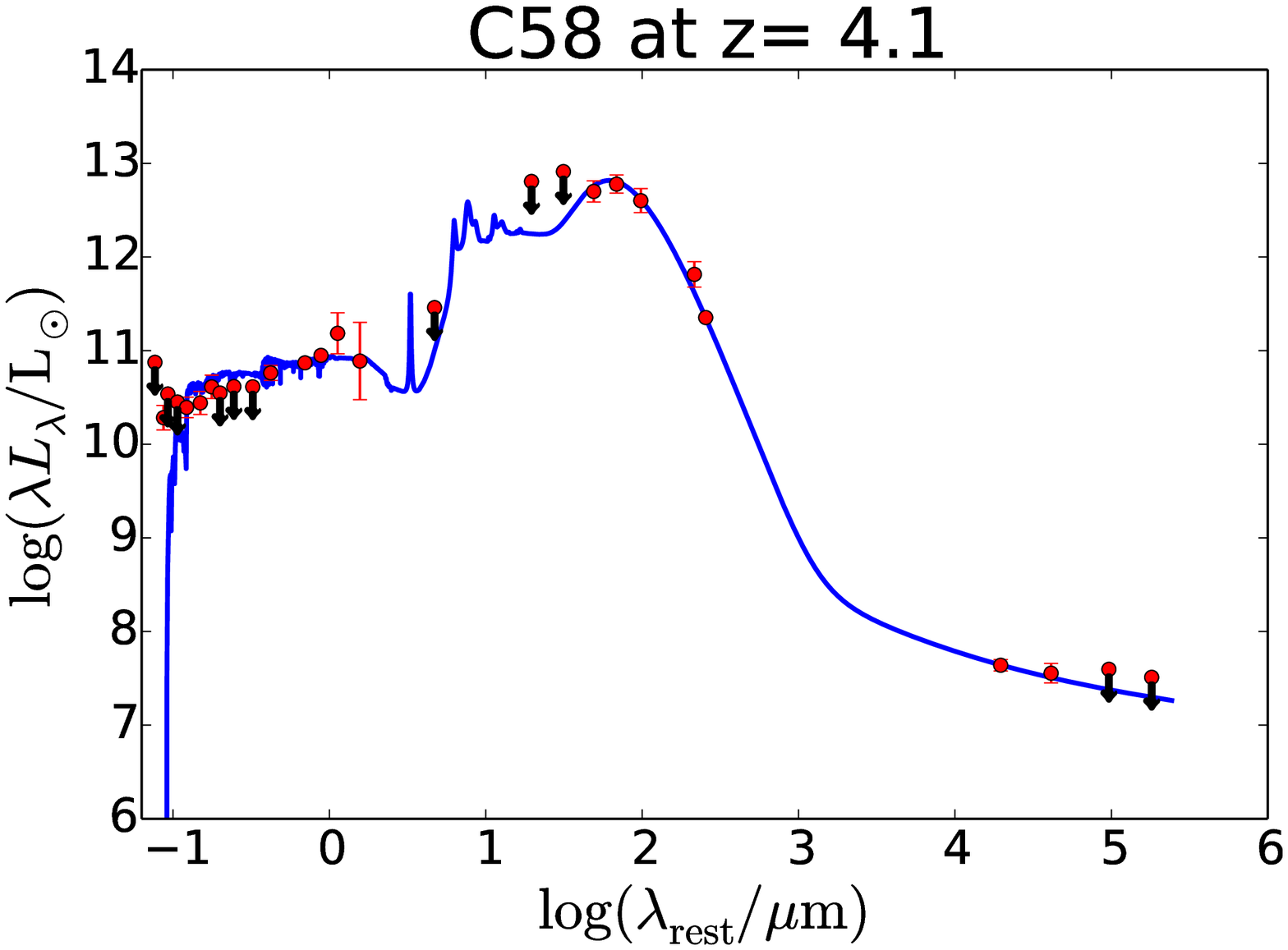}
\includegraphics[width=0.2465\textwidth]{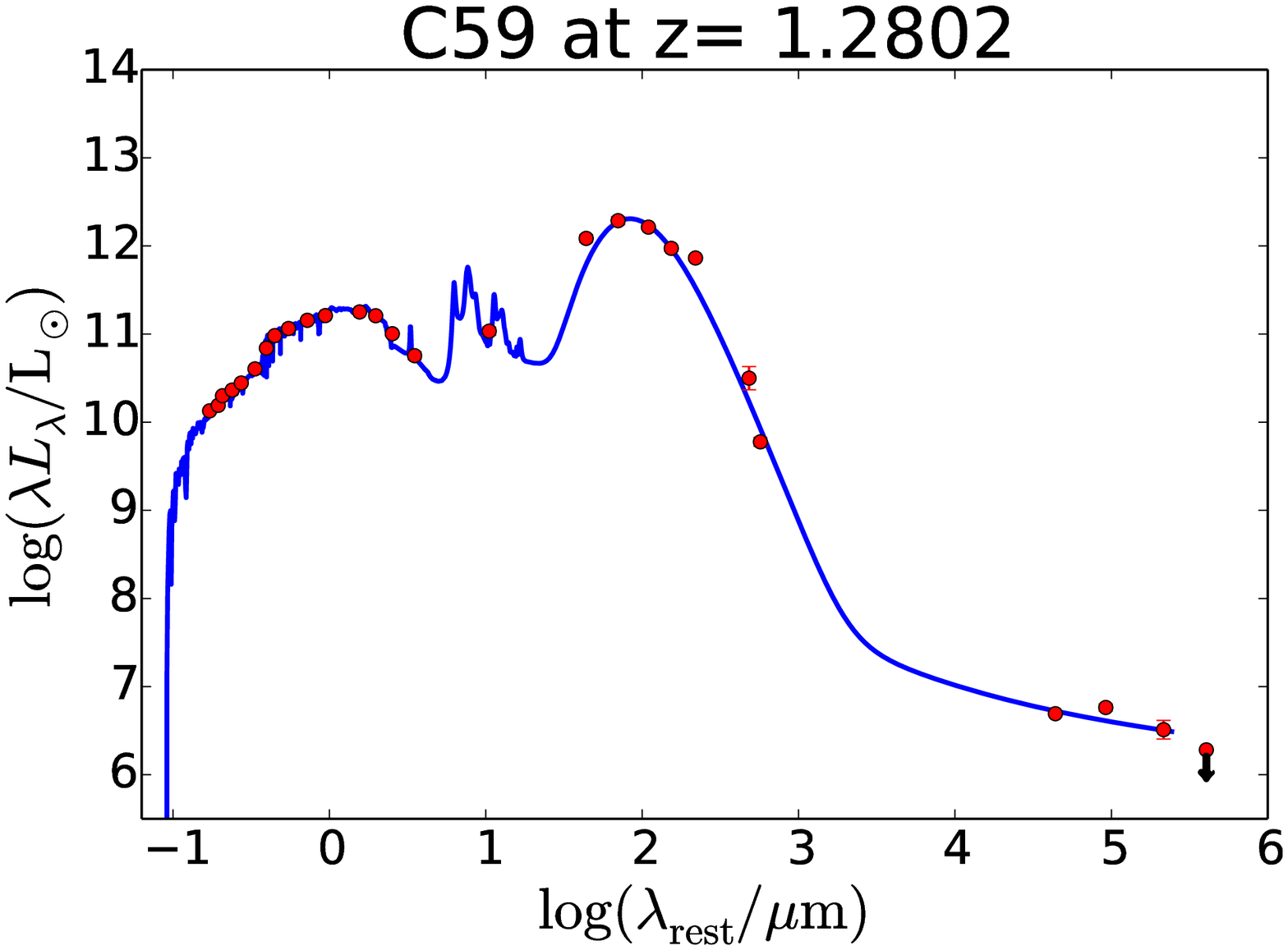}
\includegraphics[width=0.2465\textwidth]{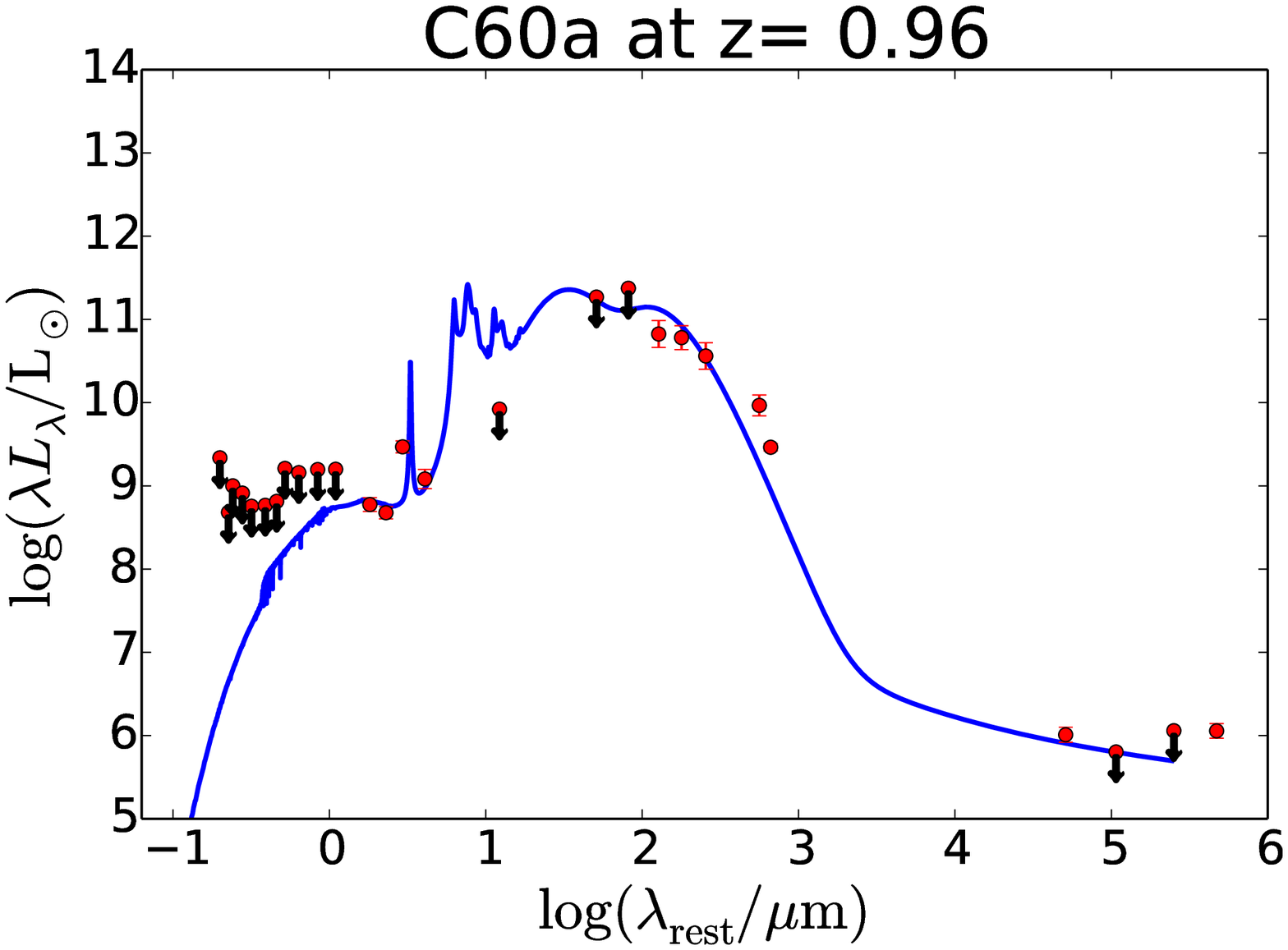}
\includegraphics[width=0.2465\textwidth]{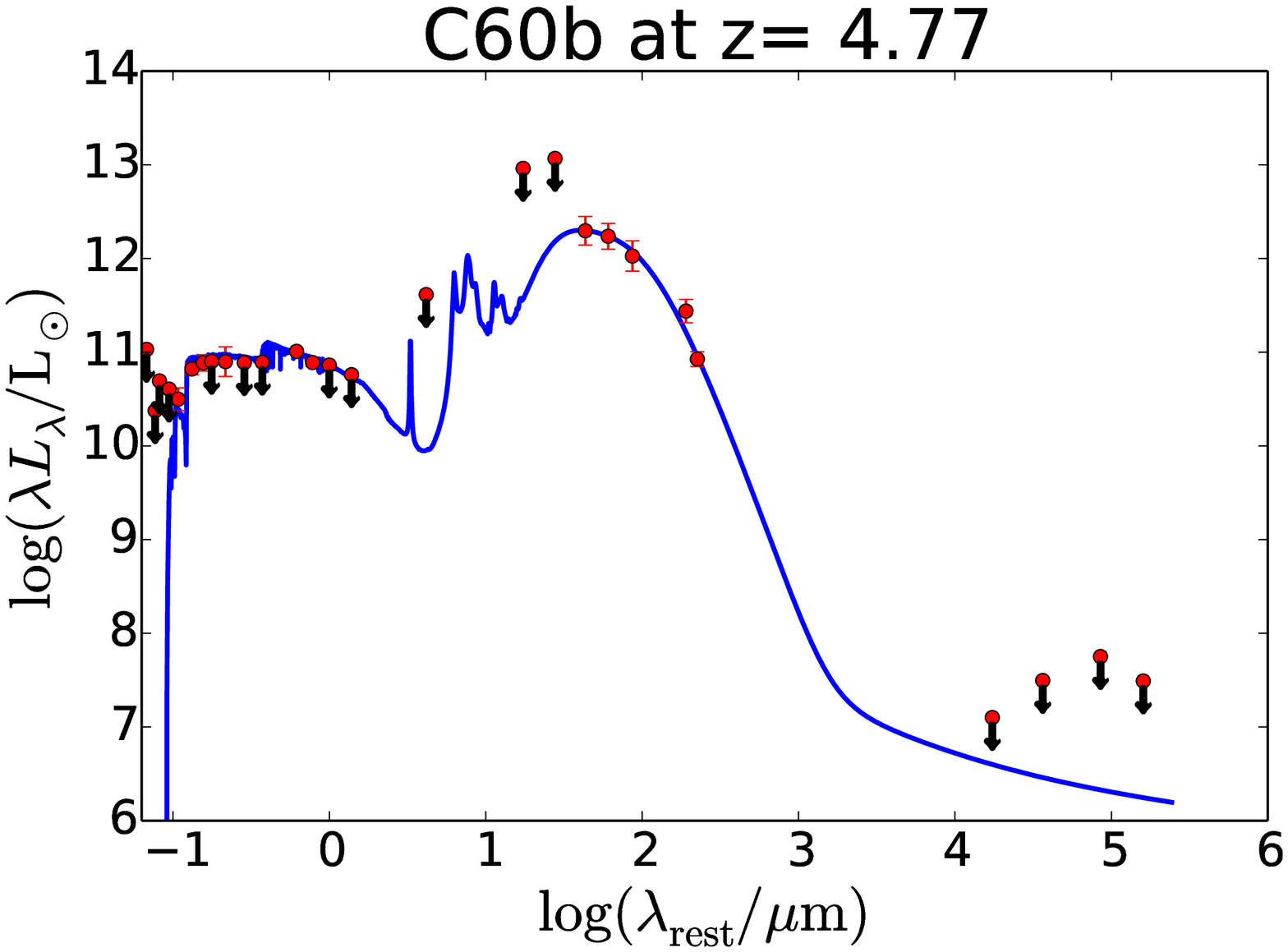}
\includegraphics[width=0.2465\textwidth]{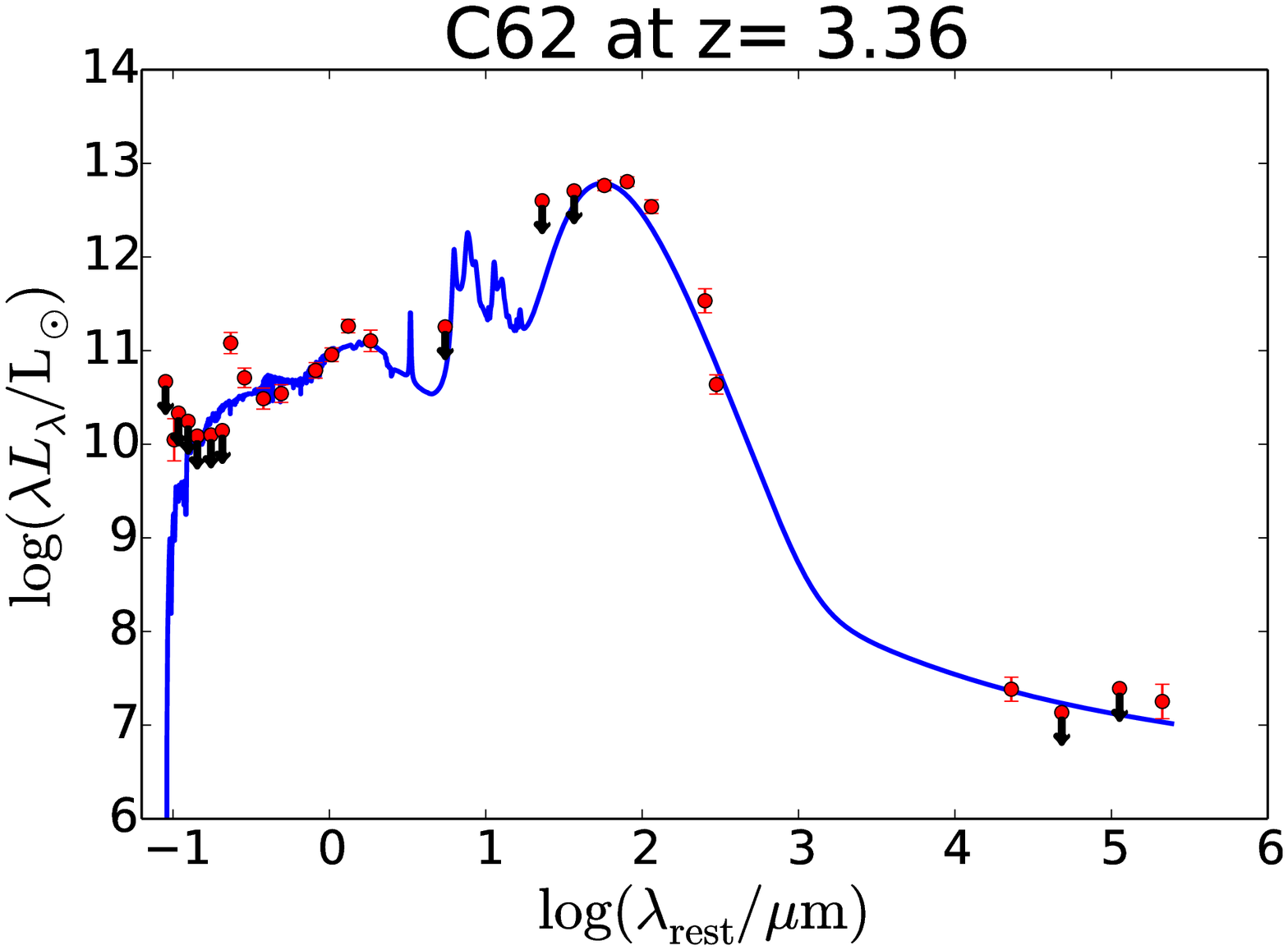}
\includegraphics[width=0.2465\textwidth]{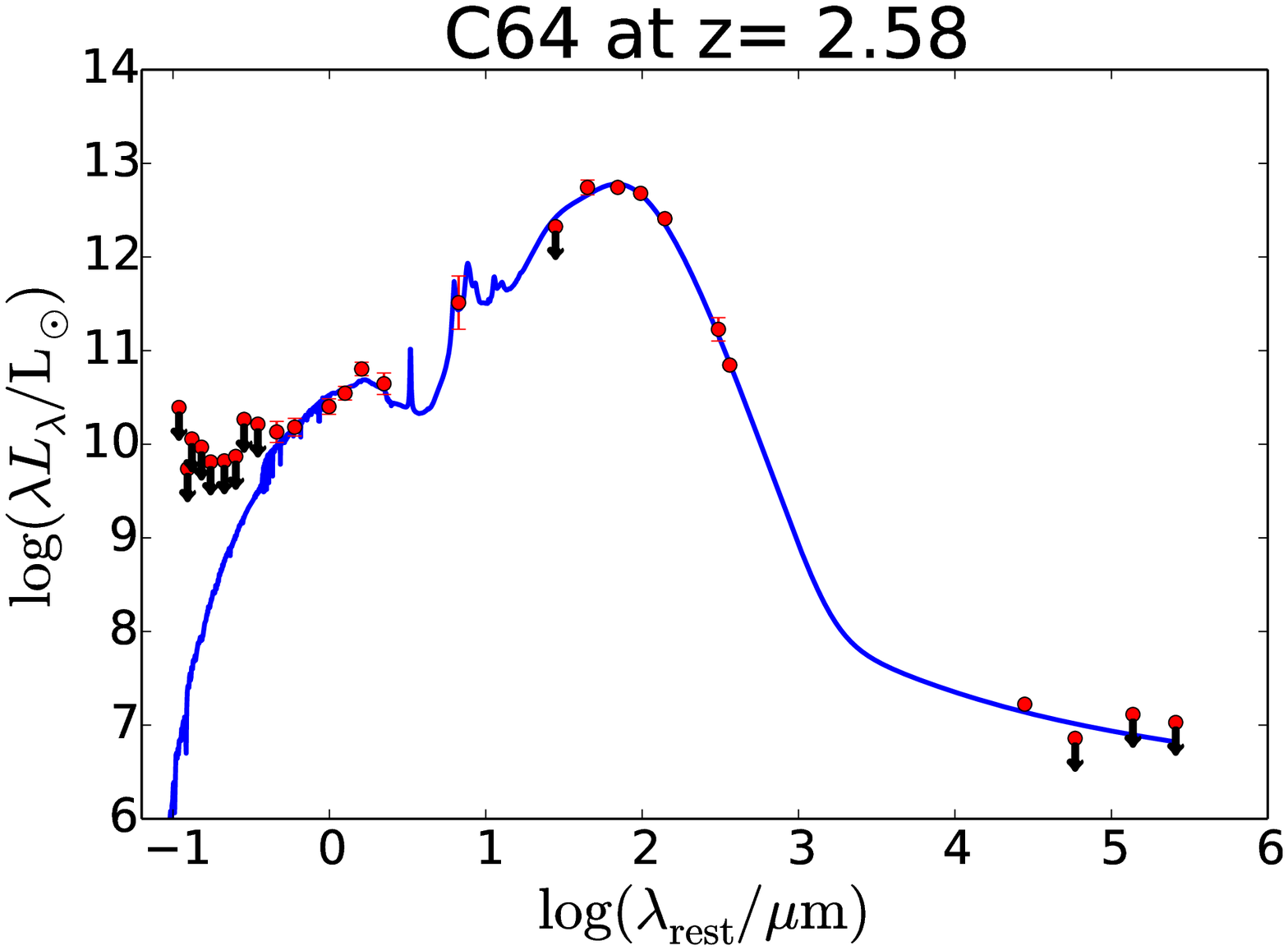}
\includegraphics[width=0.2465\textwidth]{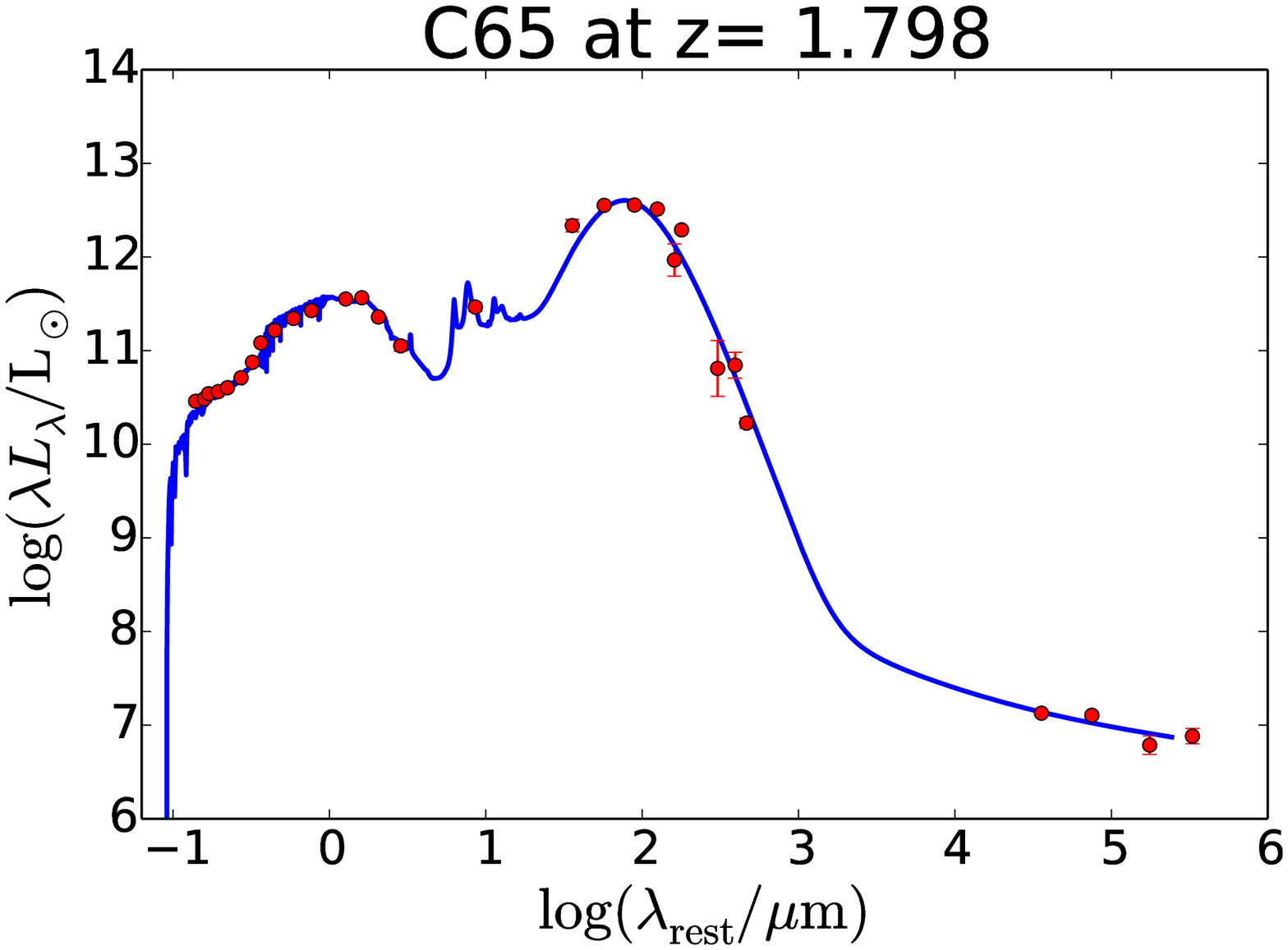}
\includegraphics[width=0.2465\textwidth]{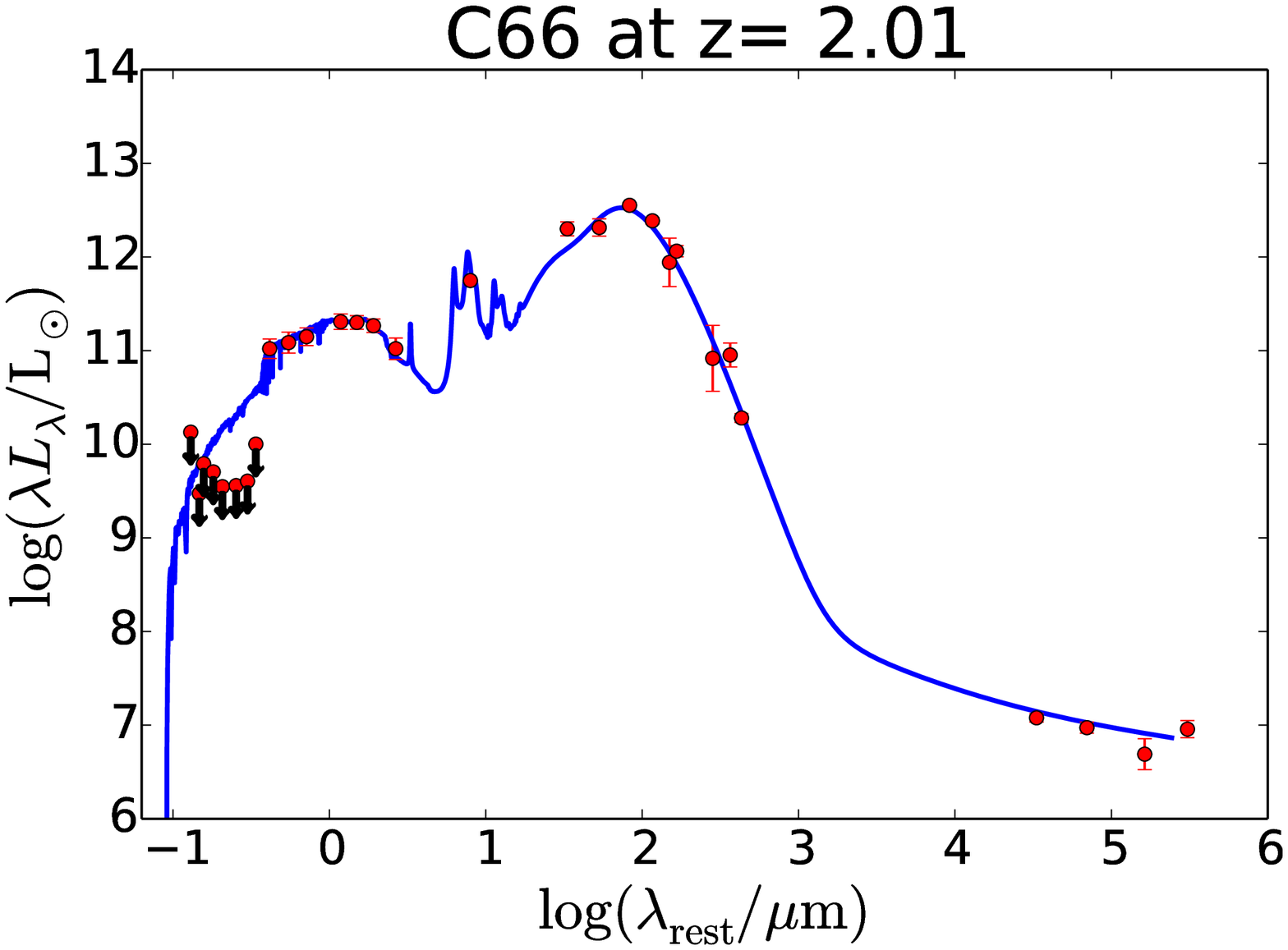}
\includegraphics[width=0.2465\textwidth]{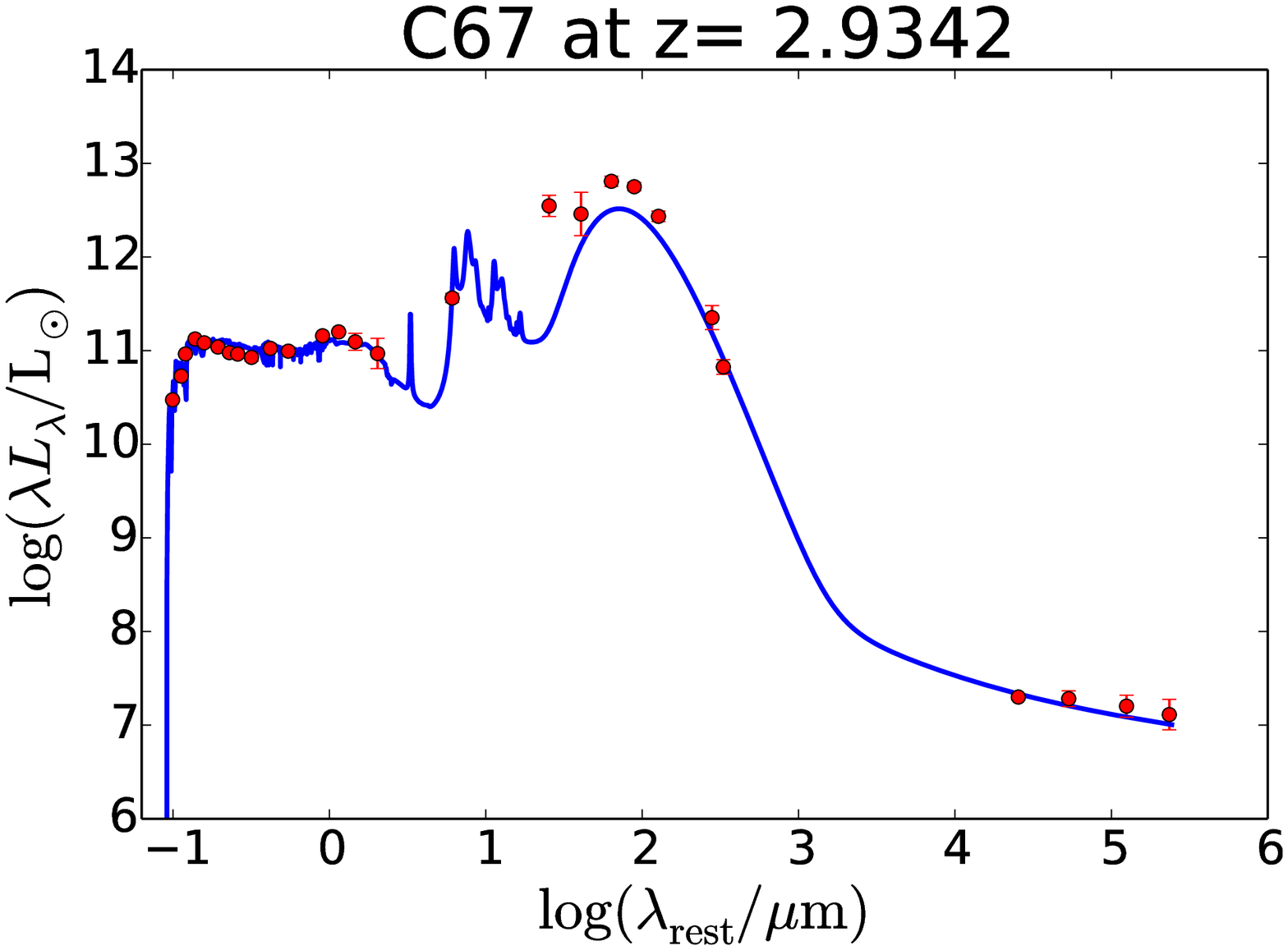}
\includegraphics[width=0.2465\textwidth]{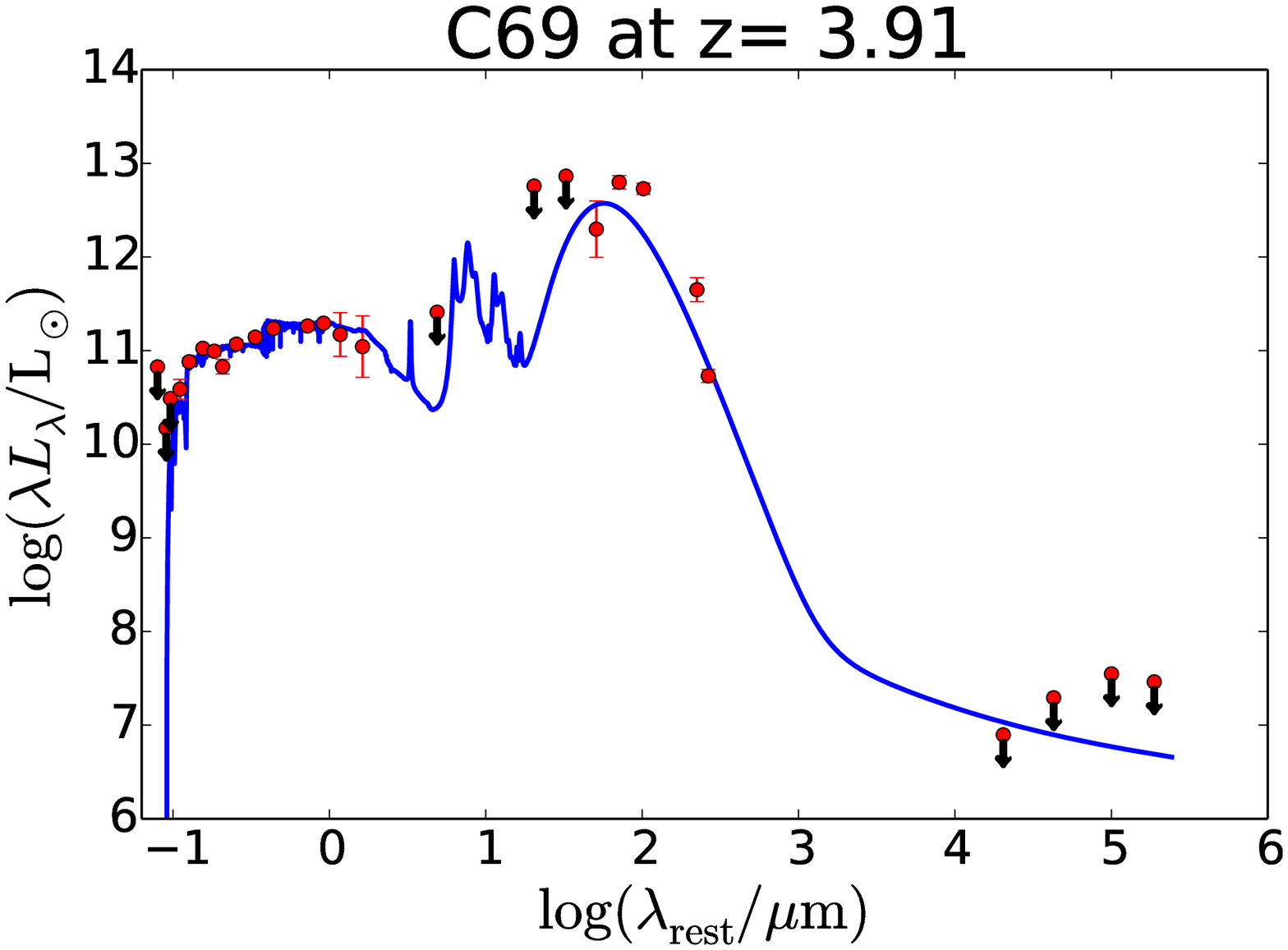}
\includegraphics[width=0.2465\textwidth]{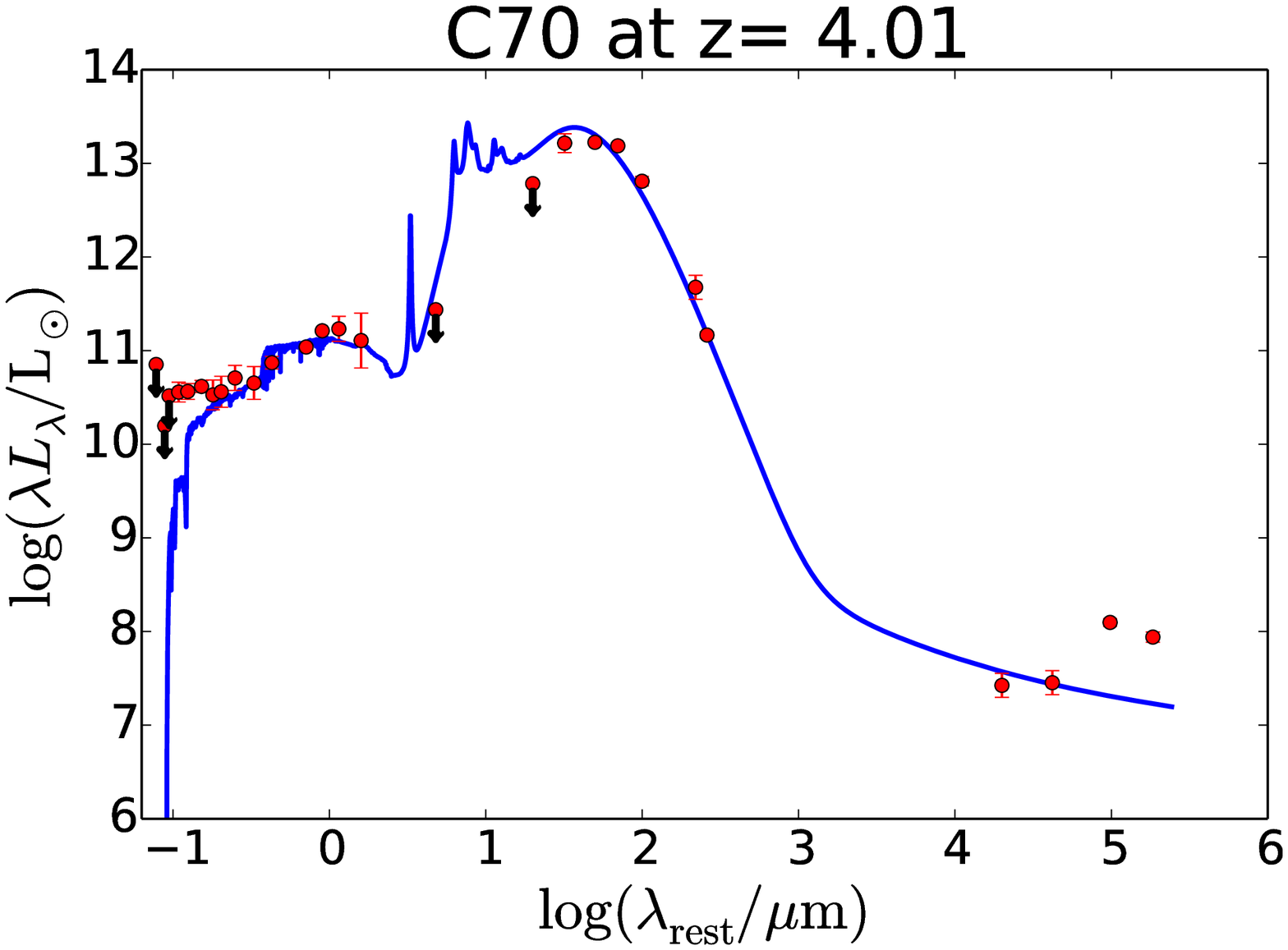}
\includegraphics[width=0.2465\textwidth]{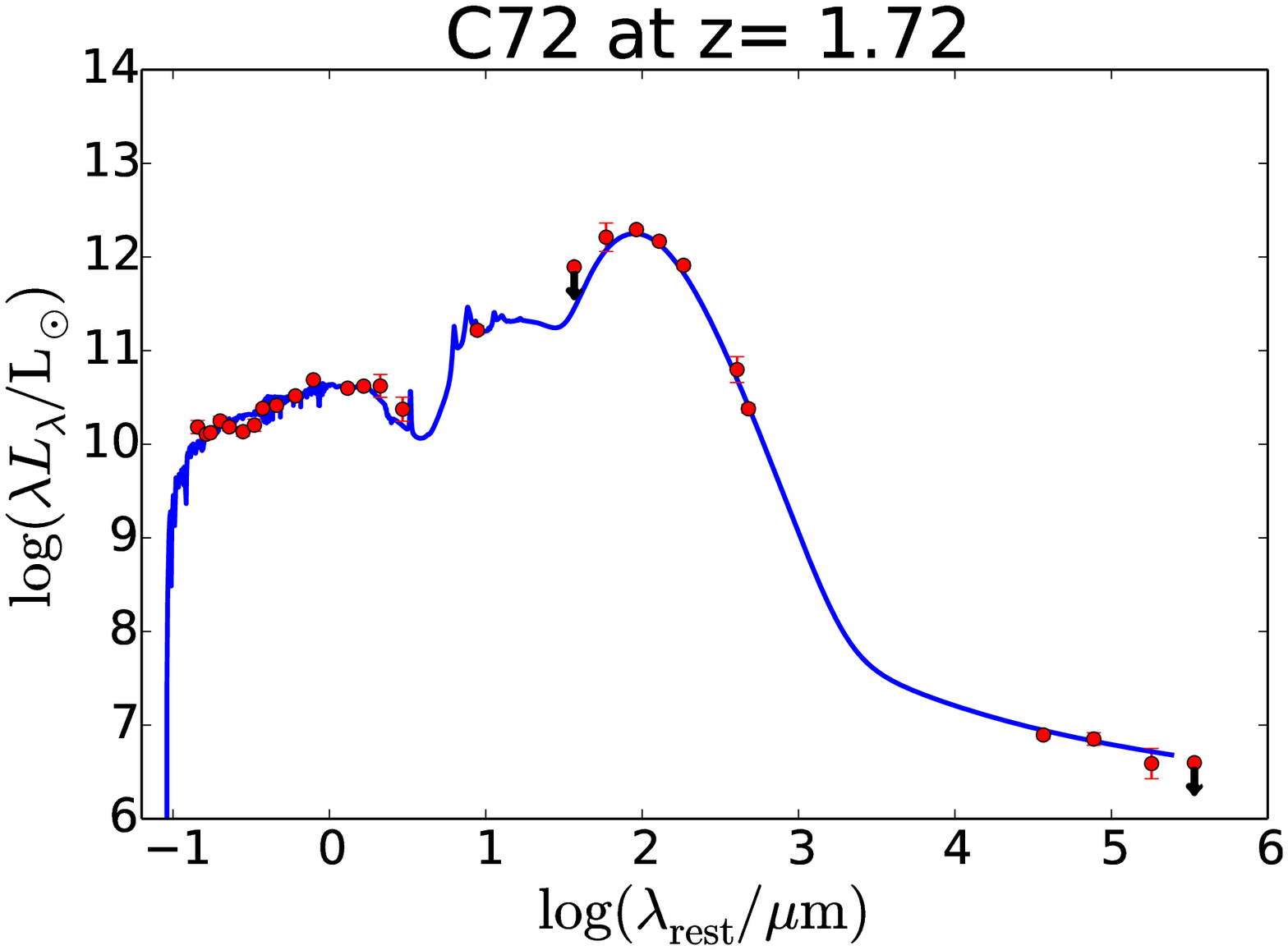}
\includegraphics[width=0.2465\textwidth]{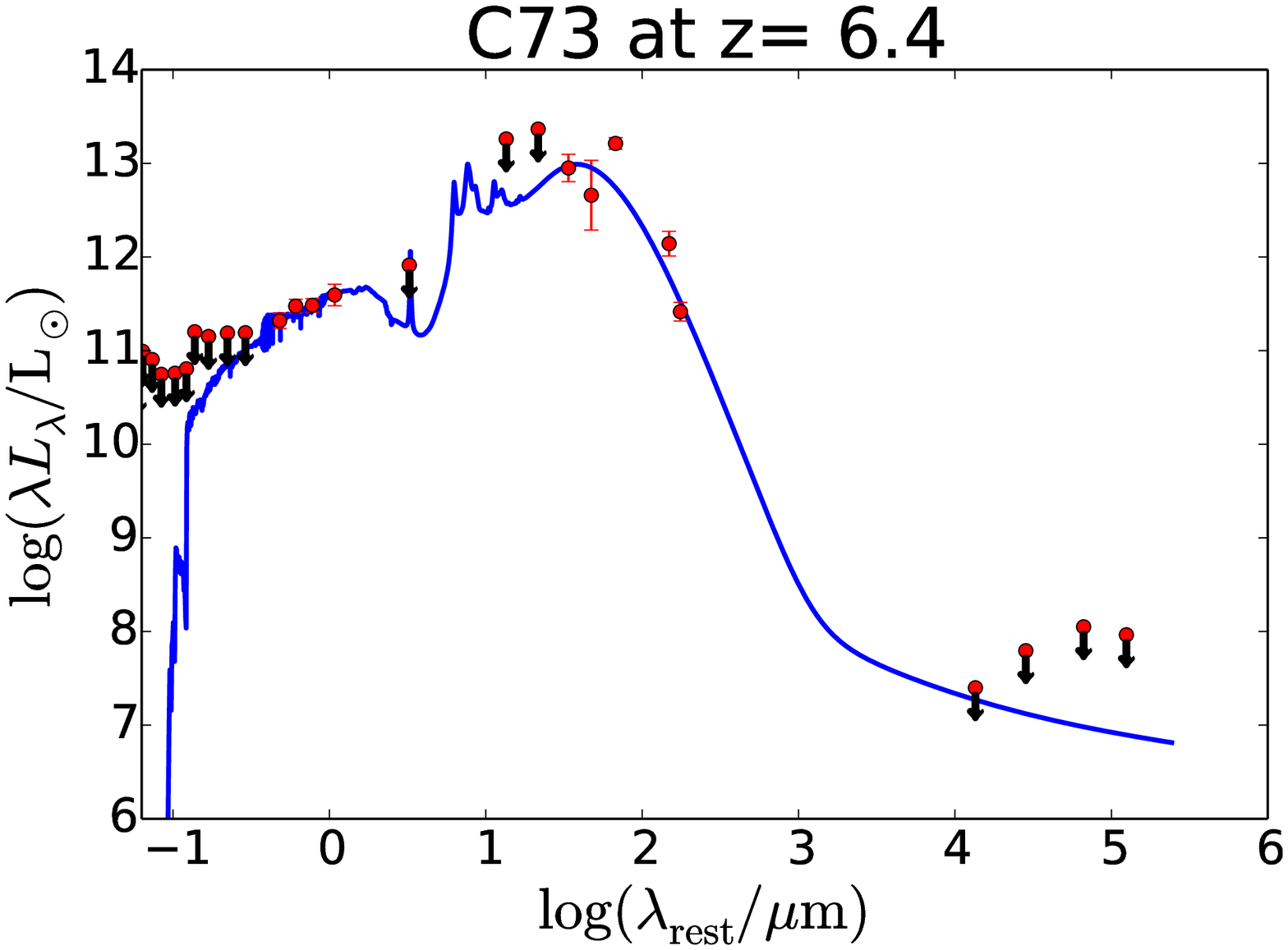} 
\includegraphics[width=0.2465\textwidth]{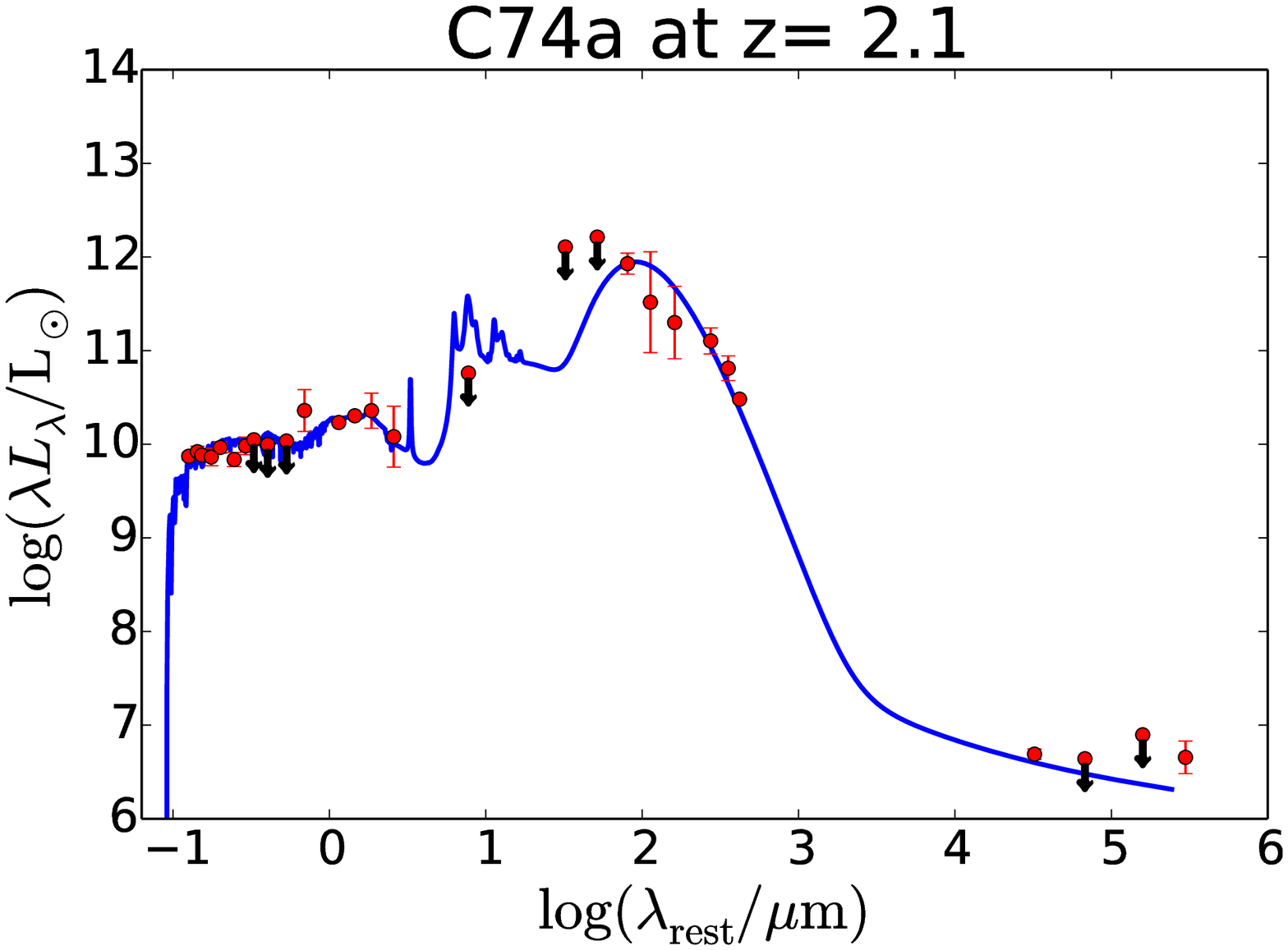}
\caption{continued.}
\label{figure:seds}
\end{center}
\end{figure*}

\addtocounter{figure}{-1}
\begin{figure*}
\begin{center}
\includegraphics[width=0.2465\textwidth]{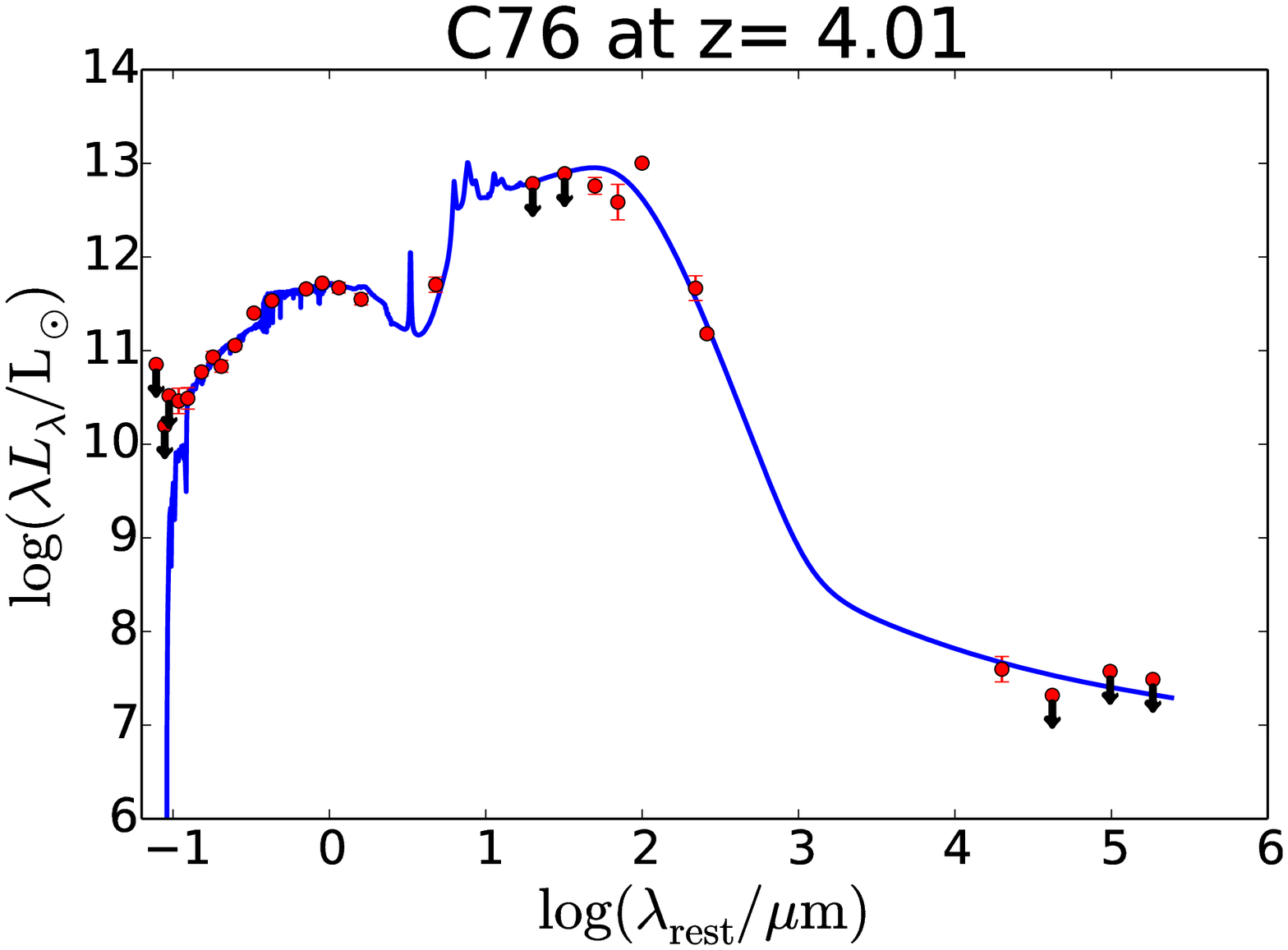}
\includegraphics[width=0.2465\textwidth]{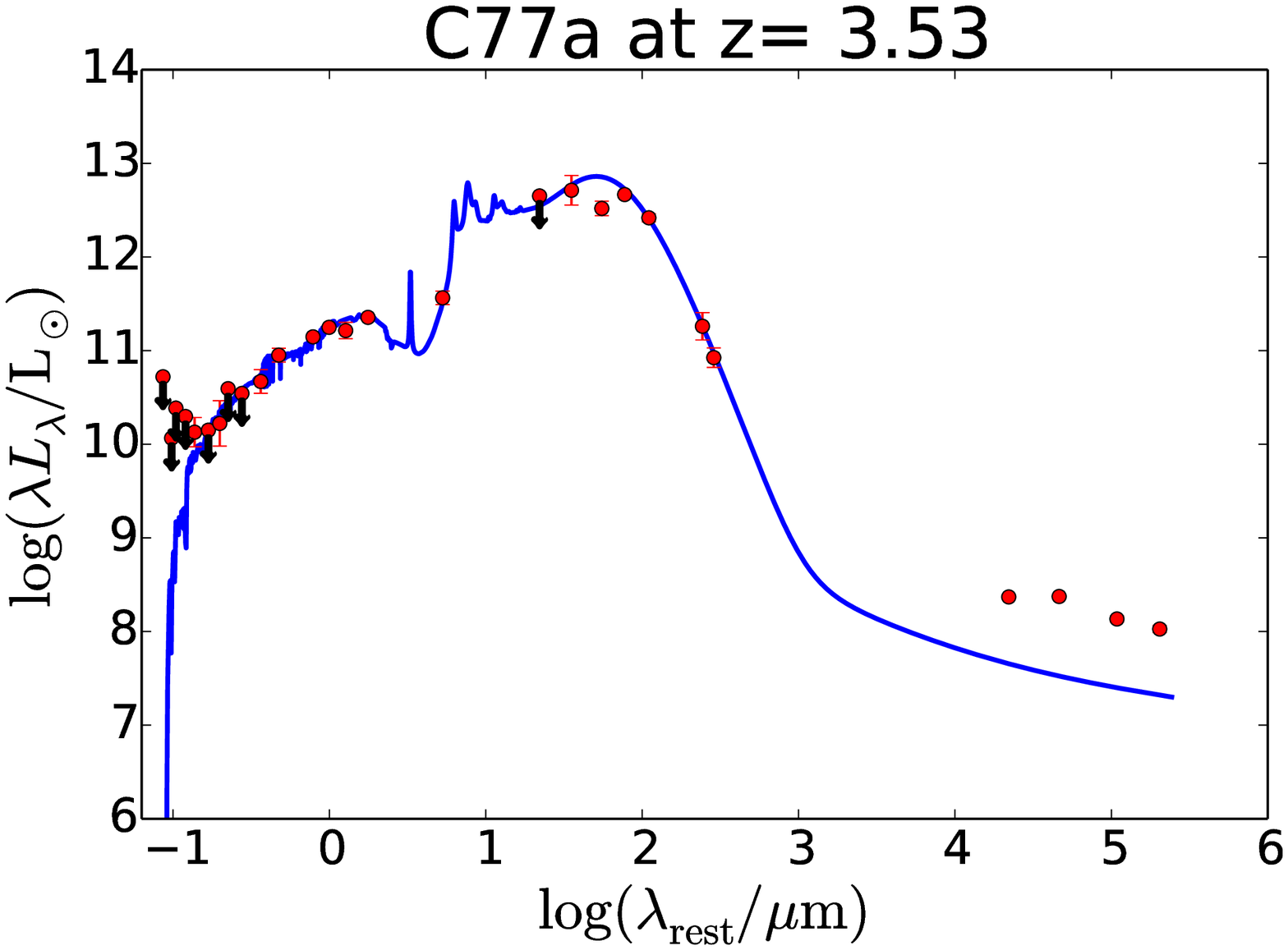}
\includegraphics[width=0.2465\textwidth]{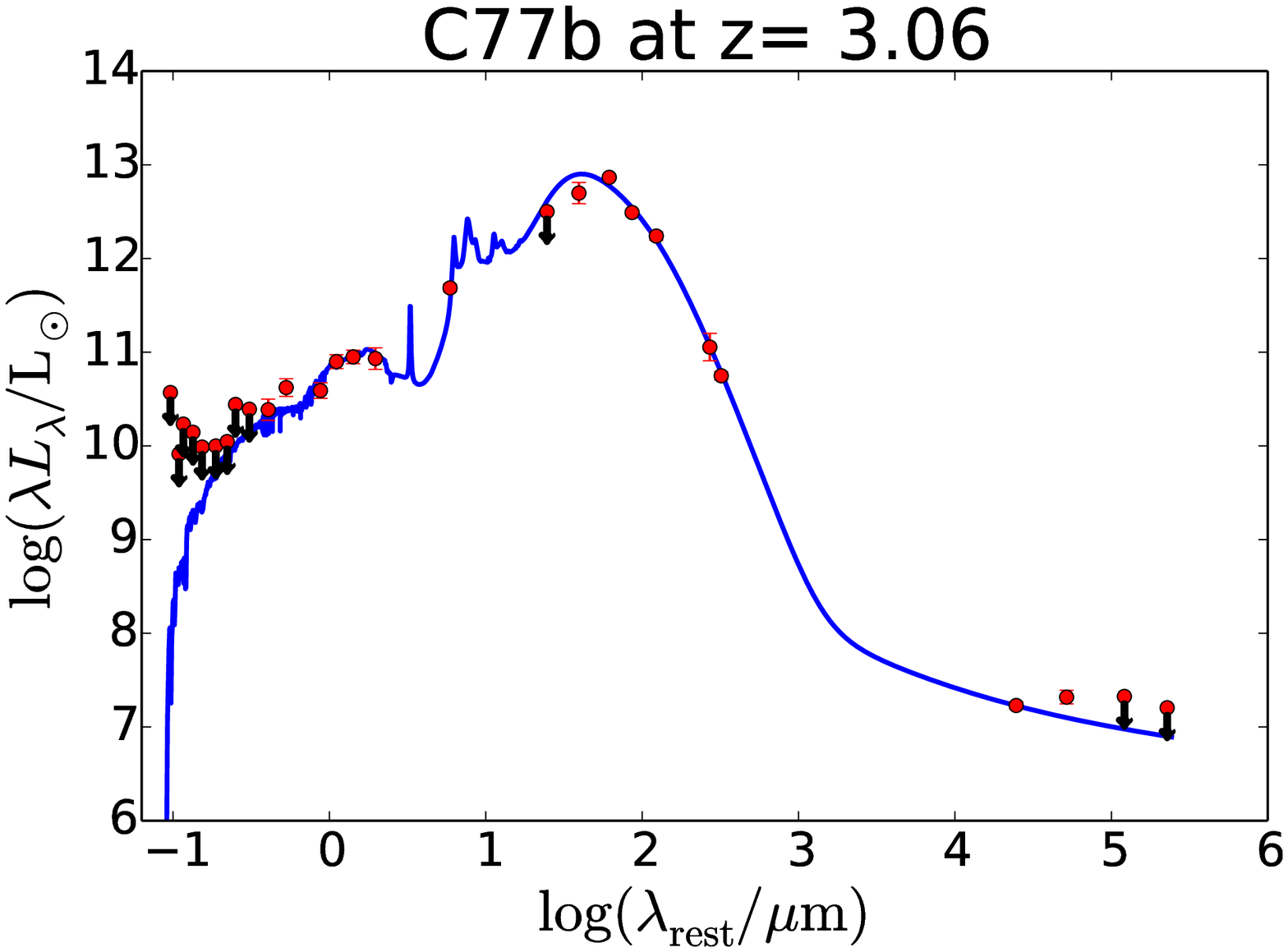}
\includegraphics[width=0.2465\textwidth]{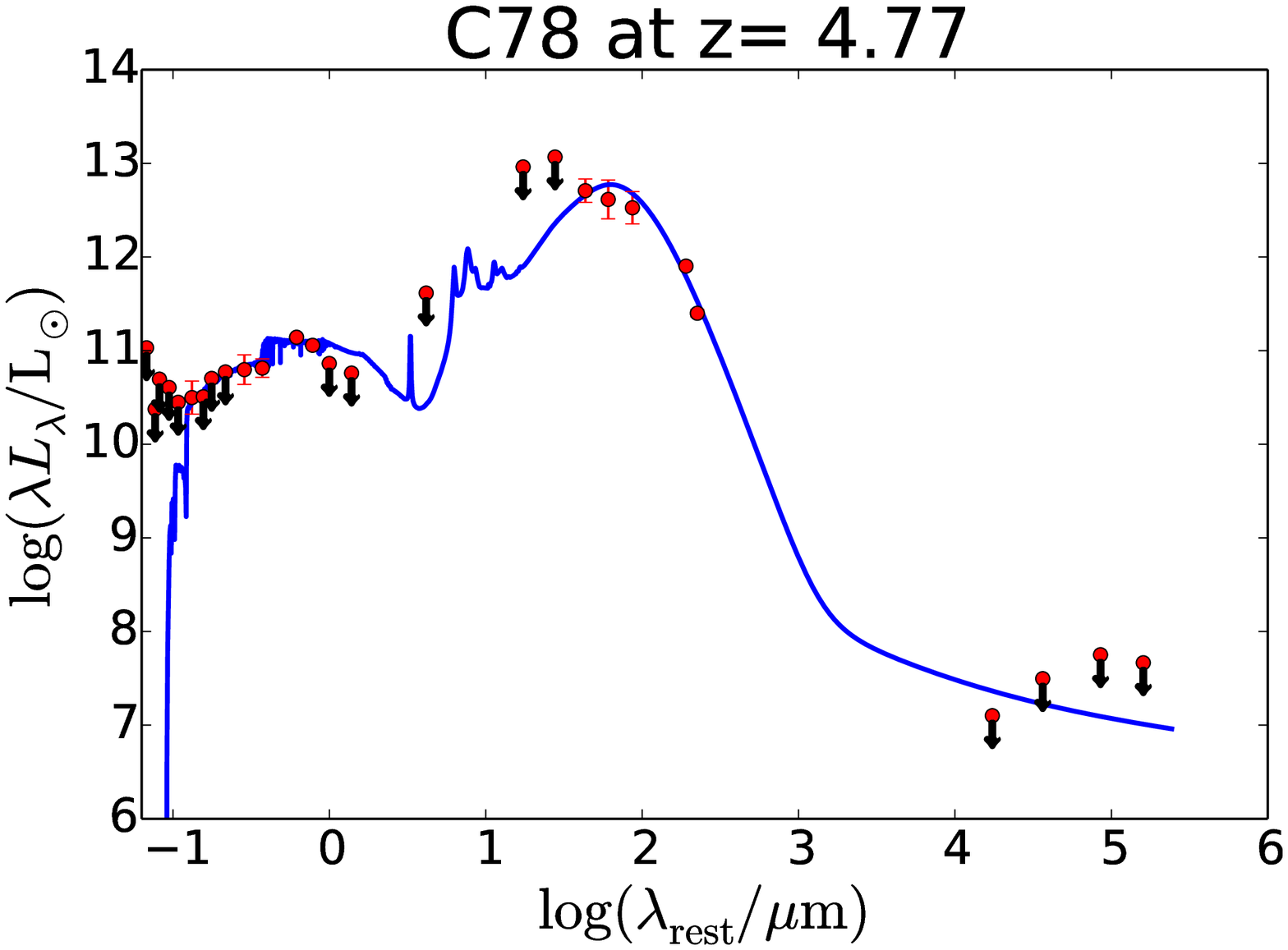}
\includegraphics[width=0.2465\textwidth]{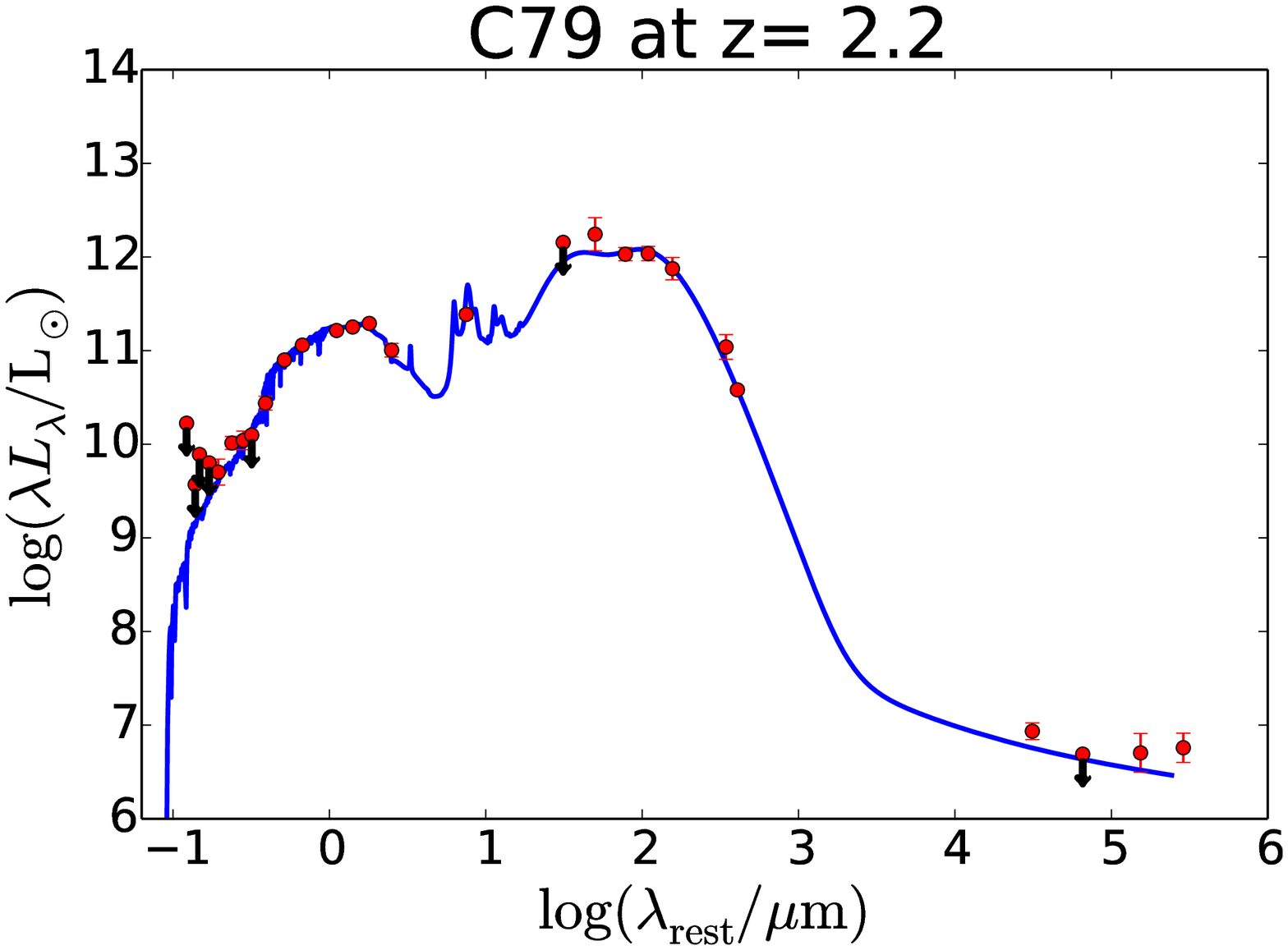}
\includegraphics[width=0.2465\textwidth]{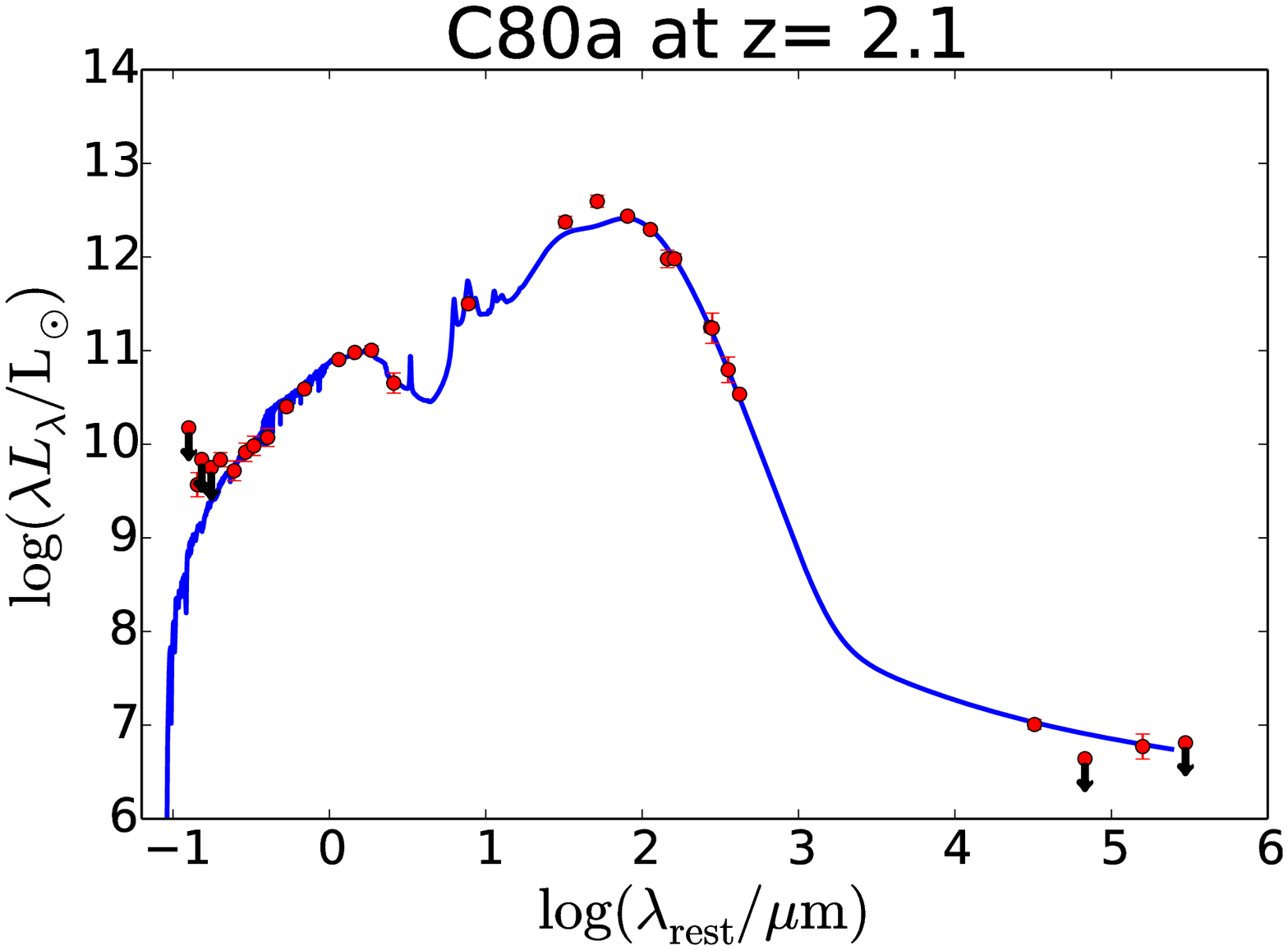}
\includegraphics[width=0.2465\textwidth]{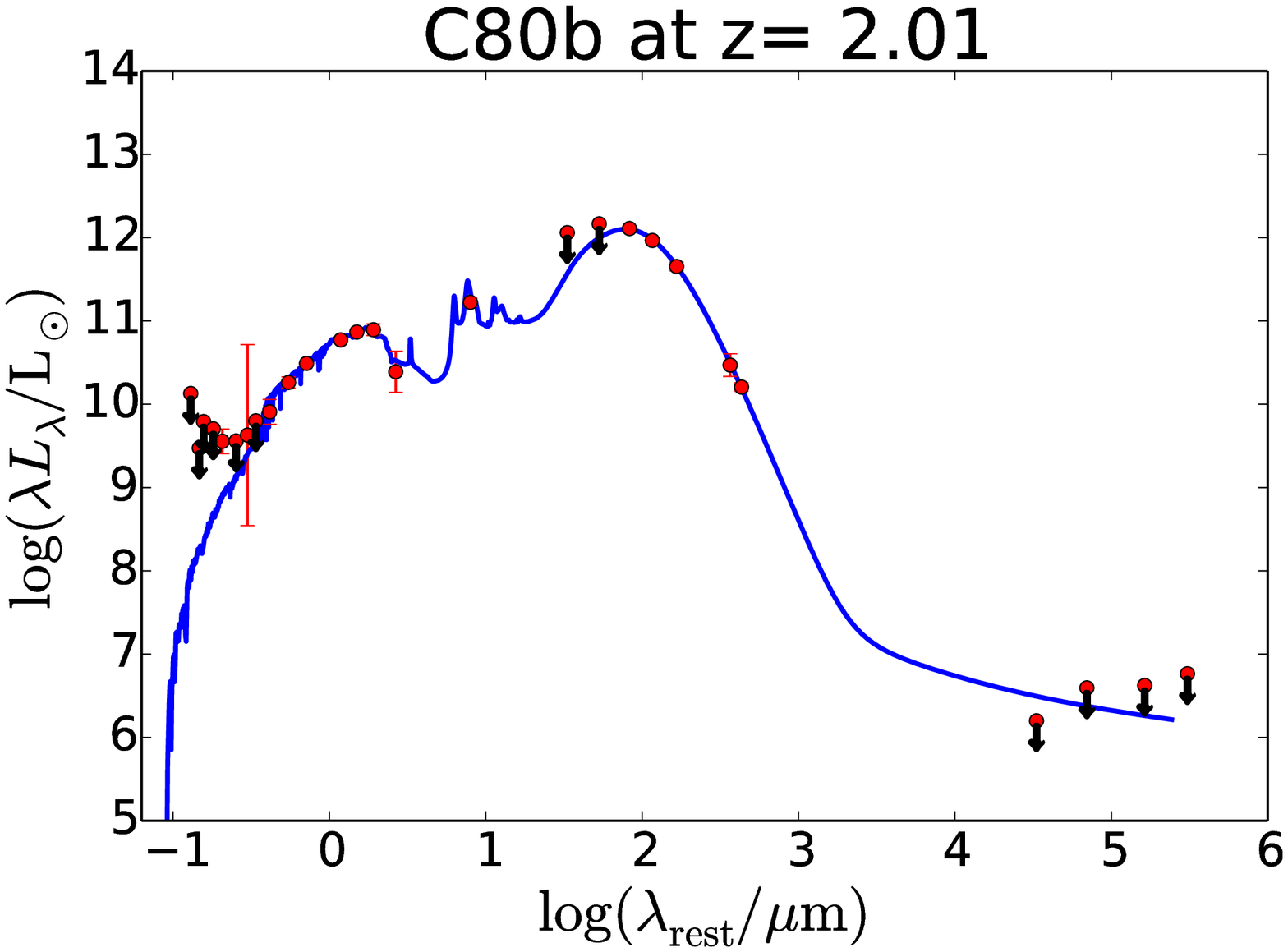}
\includegraphics[width=0.2465\textwidth]{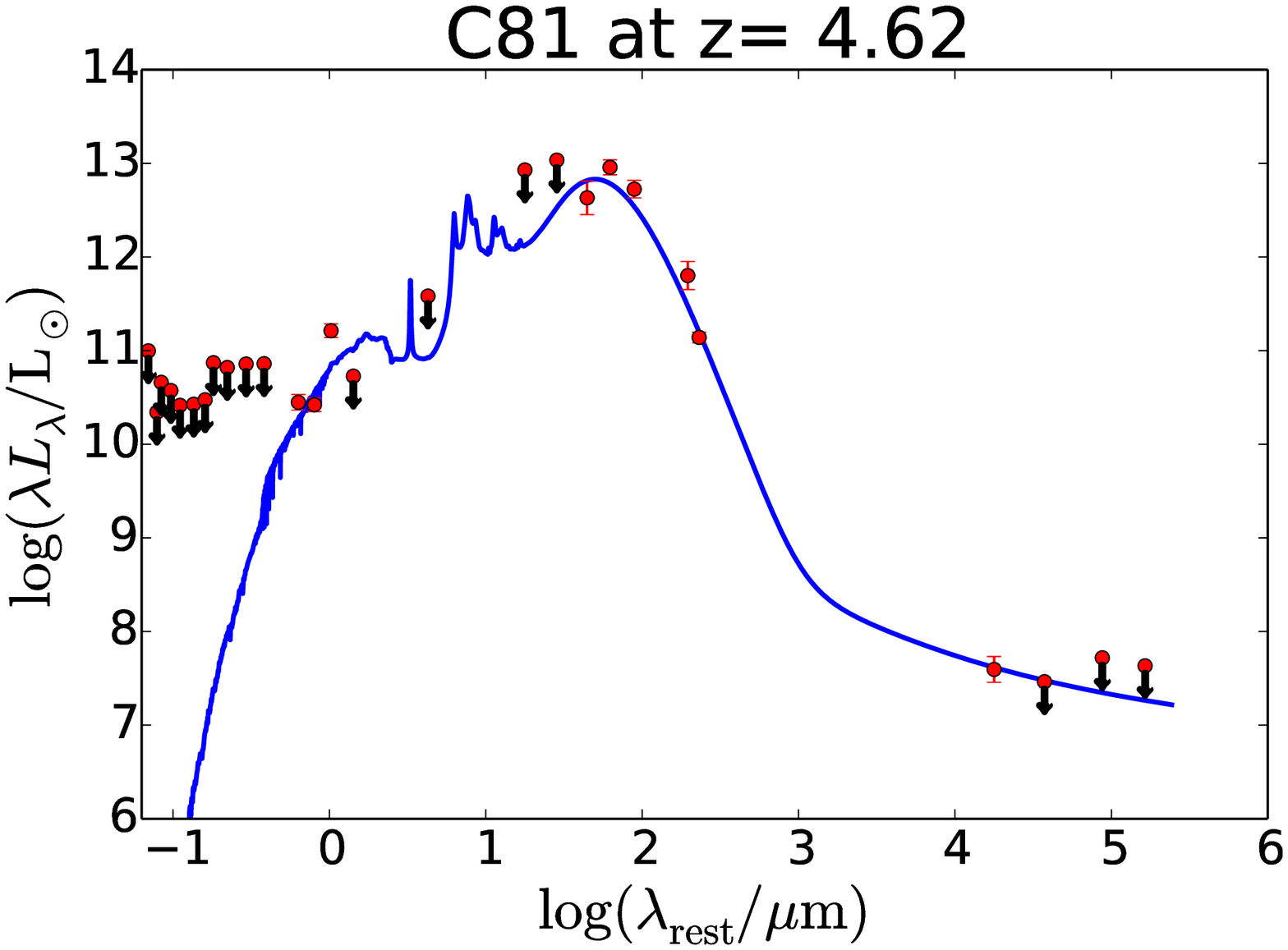}
\includegraphics[width=0.2465\textwidth]{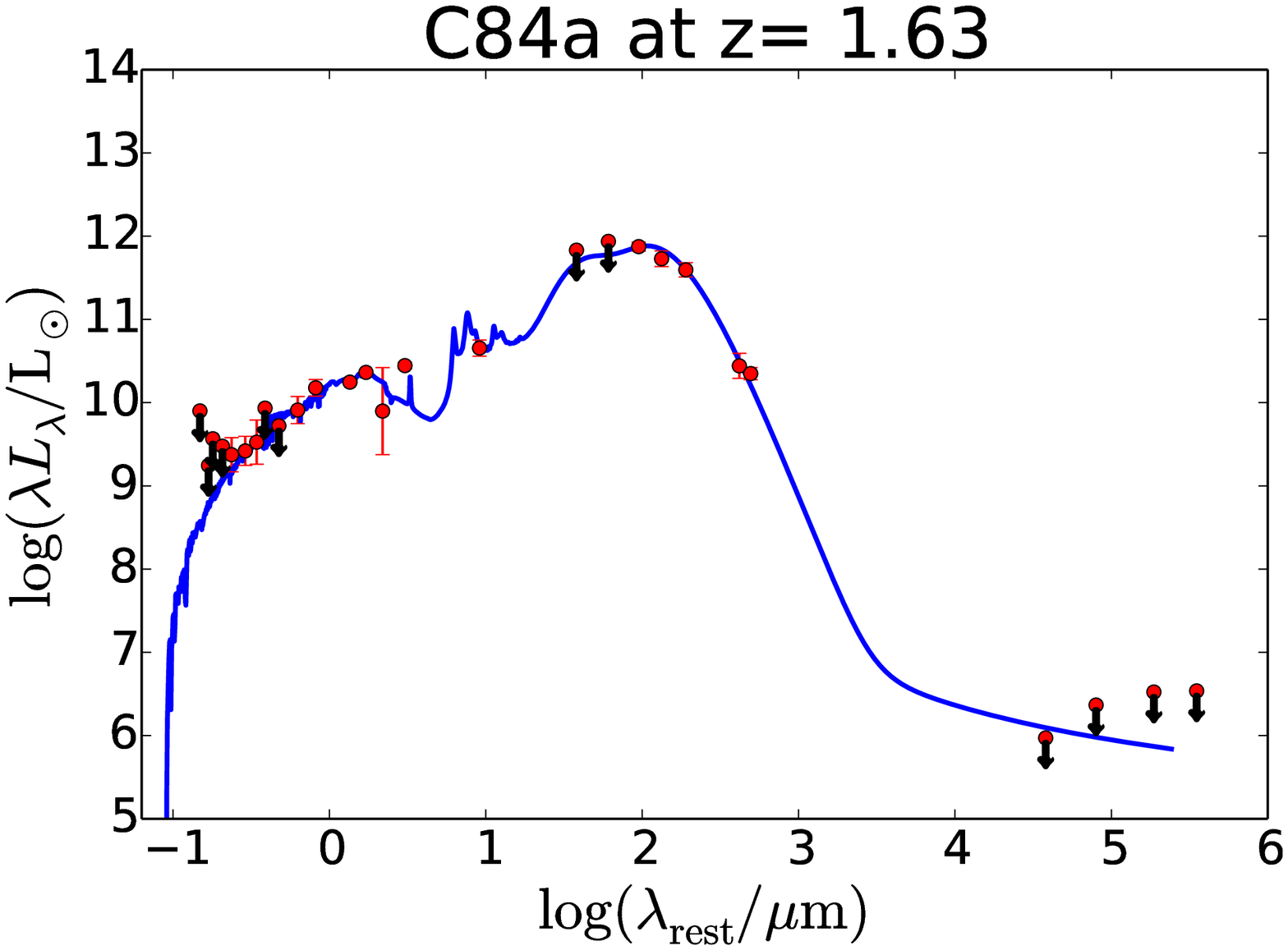}
\includegraphics[width=0.2465\textwidth]{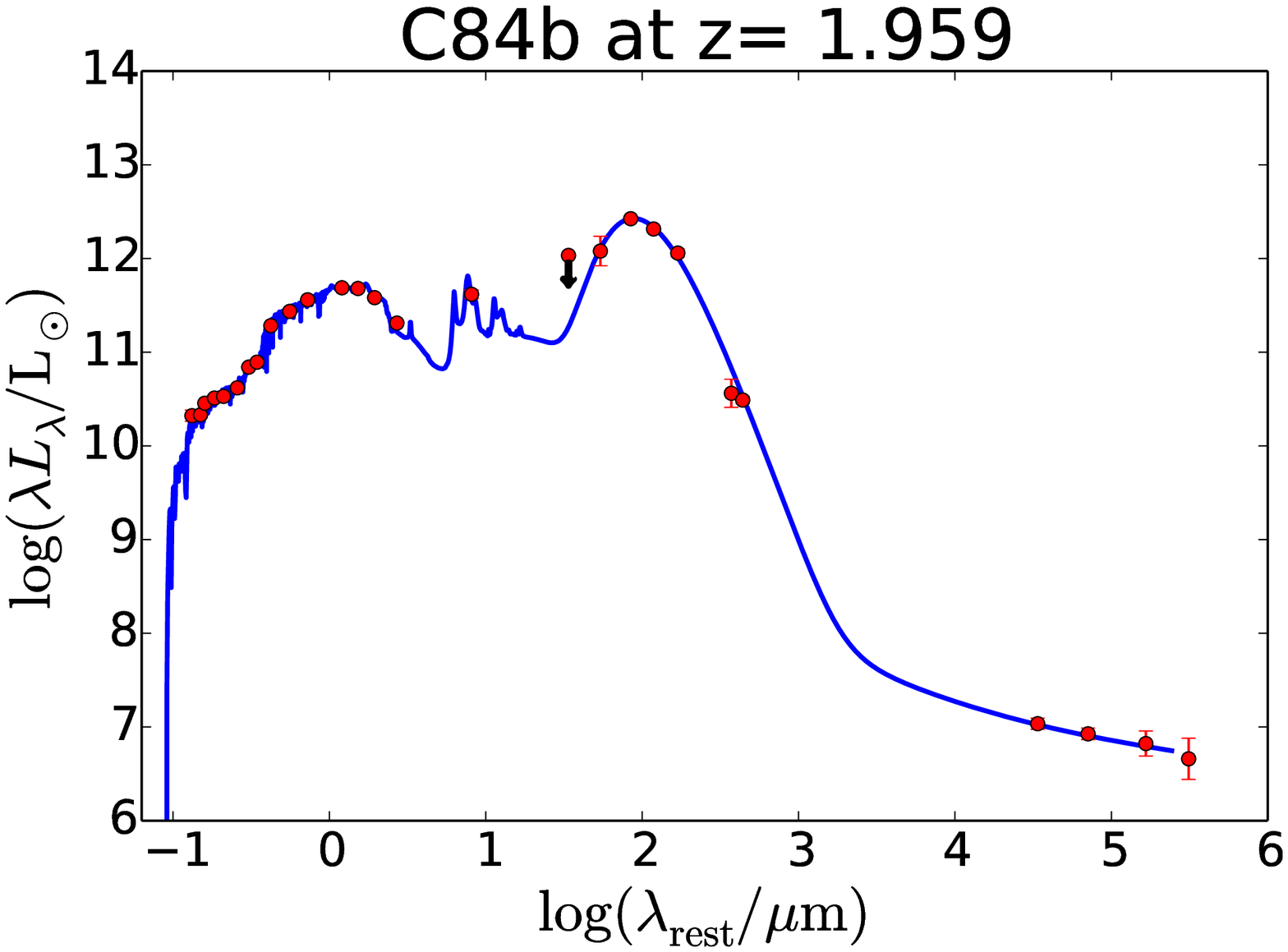}
\includegraphics[width=0.2465\textwidth]{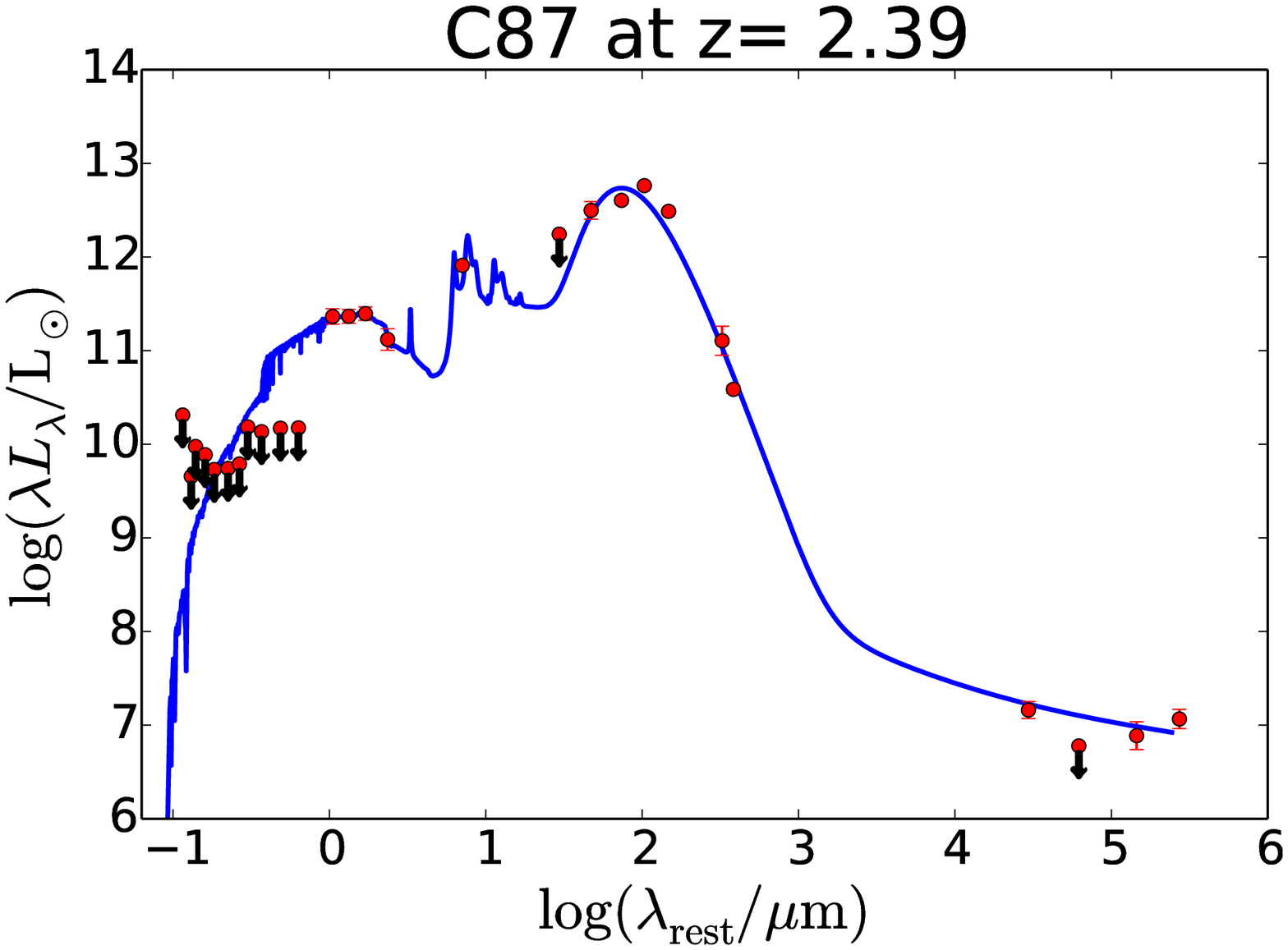}
\includegraphics[width=0.2465\textwidth]{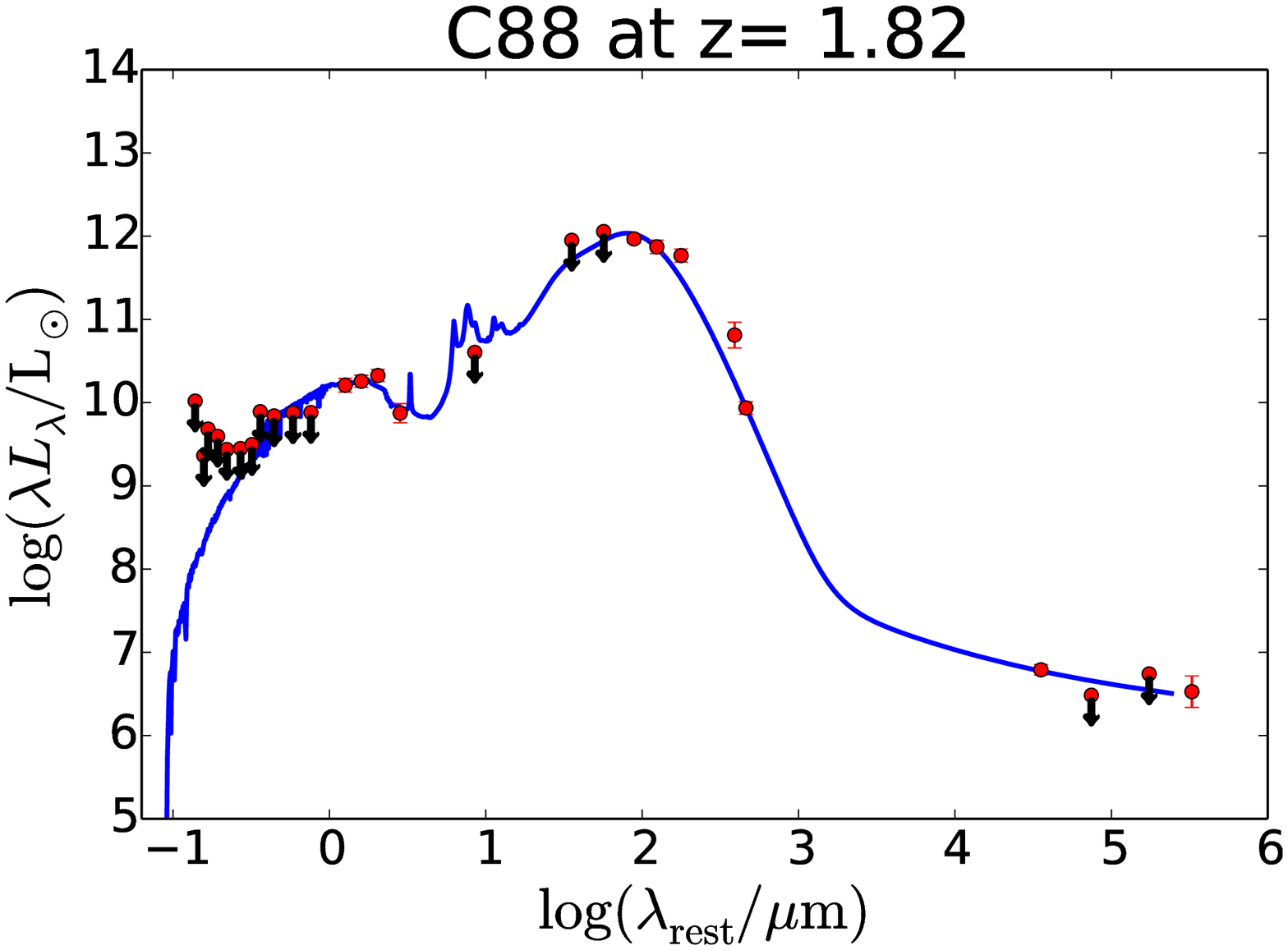}
\includegraphics[width=0.2465\textwidth]{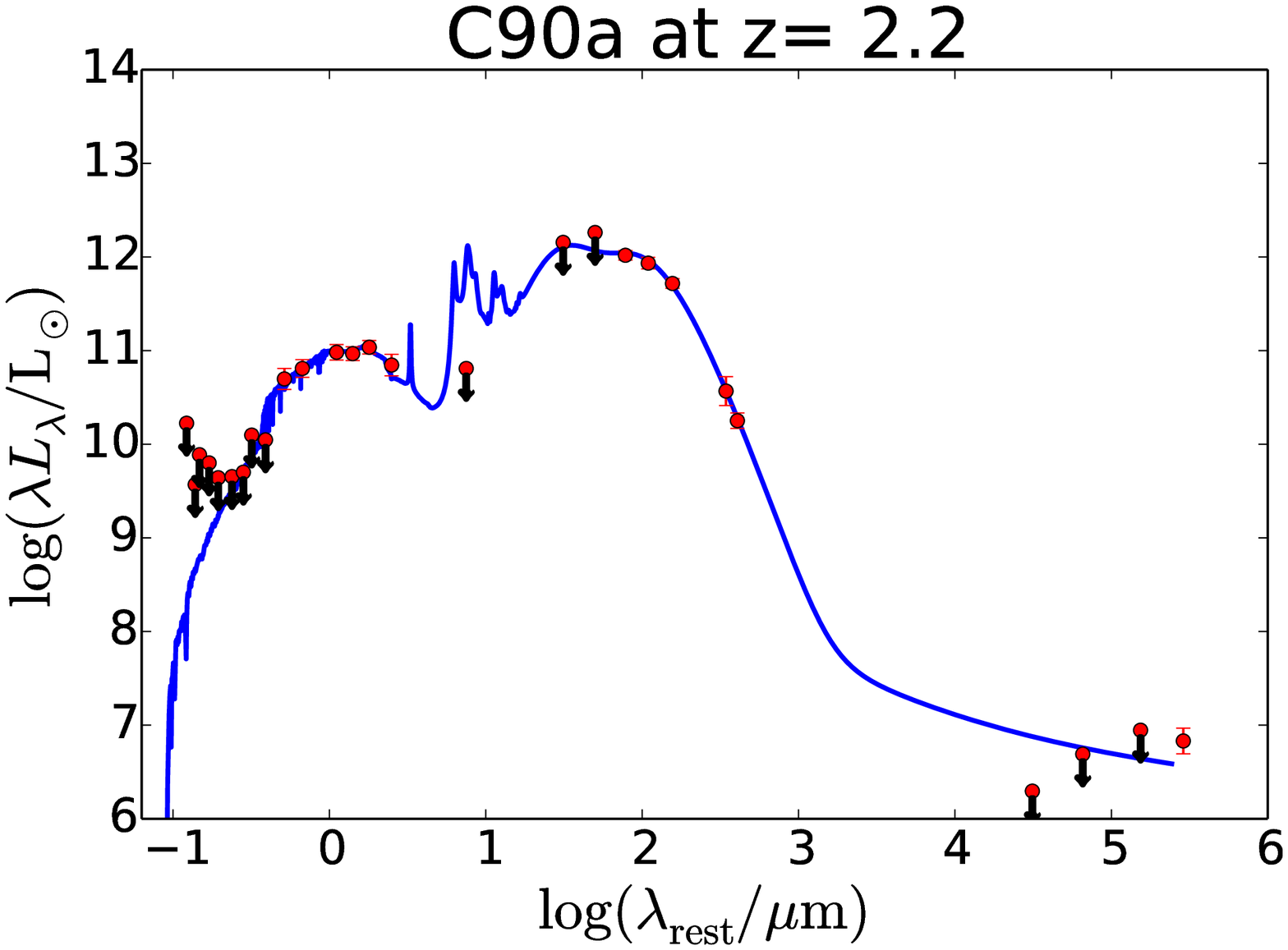}
\includegraphics[width=0.2465\textwidth]{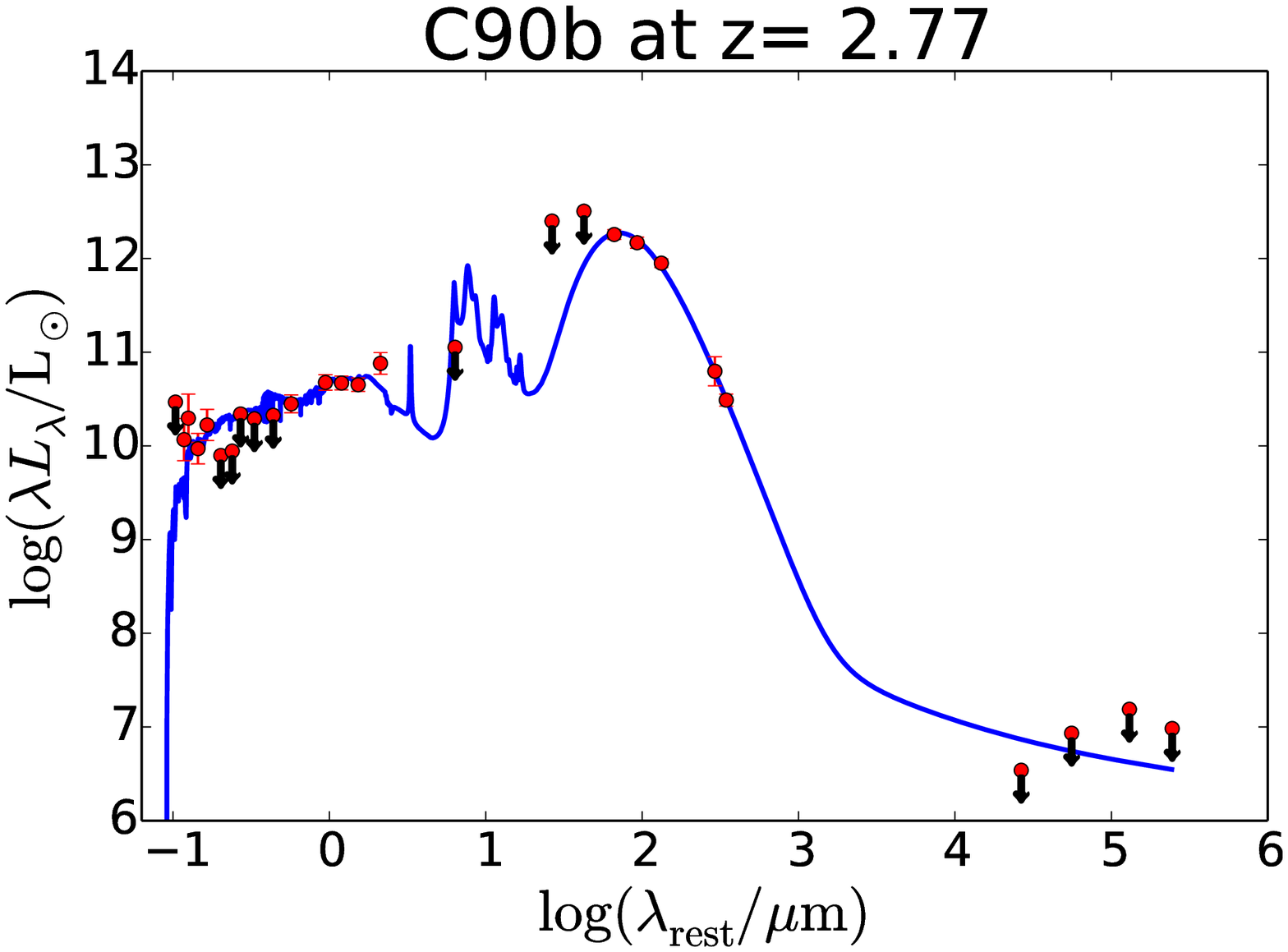}
\includegraphics[width=0.2465\textwidth]{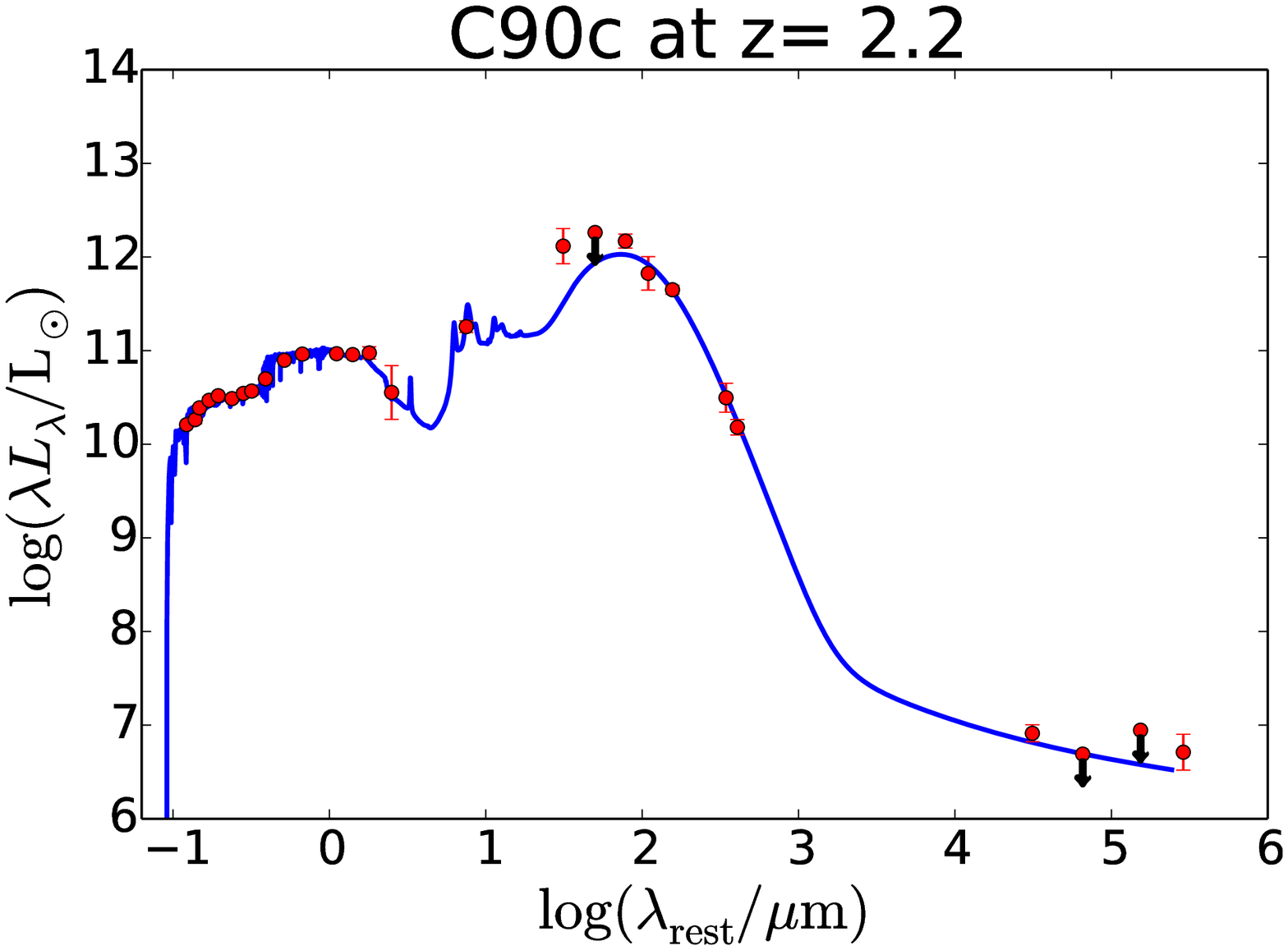}
\includegraphics[width=0.2465\textwidth]{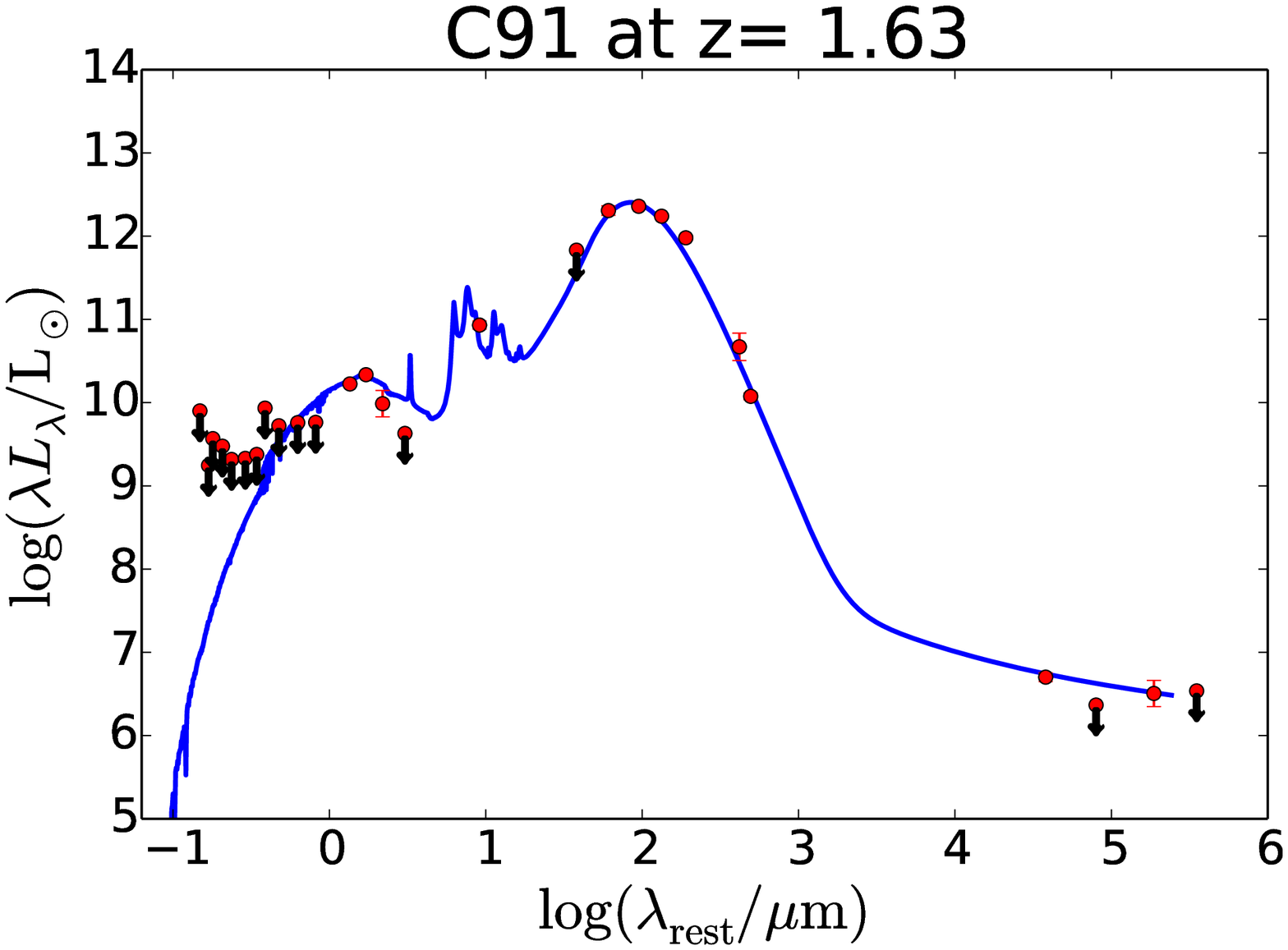}
\includegraphics[width=0.2465\textwidth]{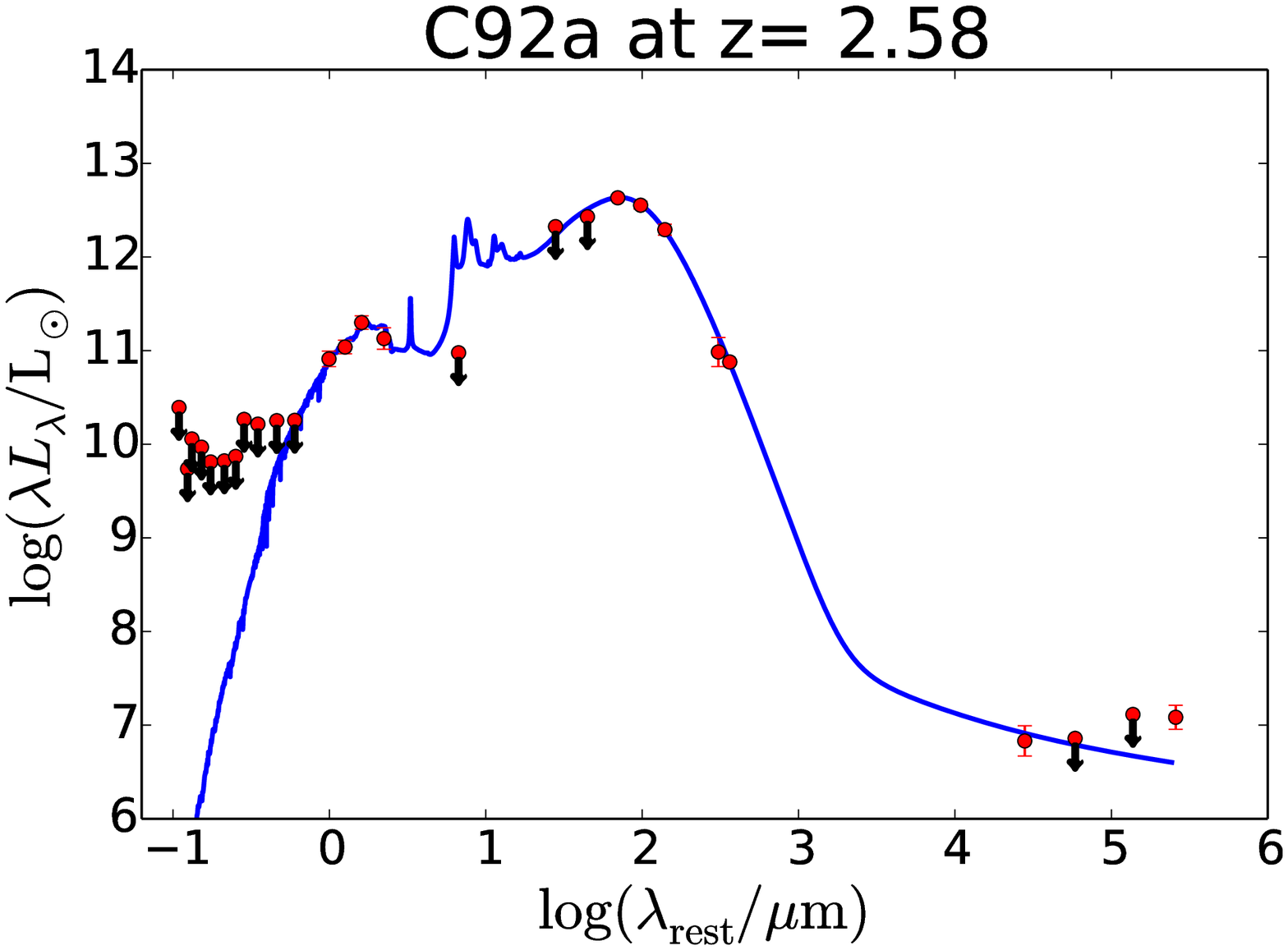}
\includegraphics[width=0.2465\textwidth]{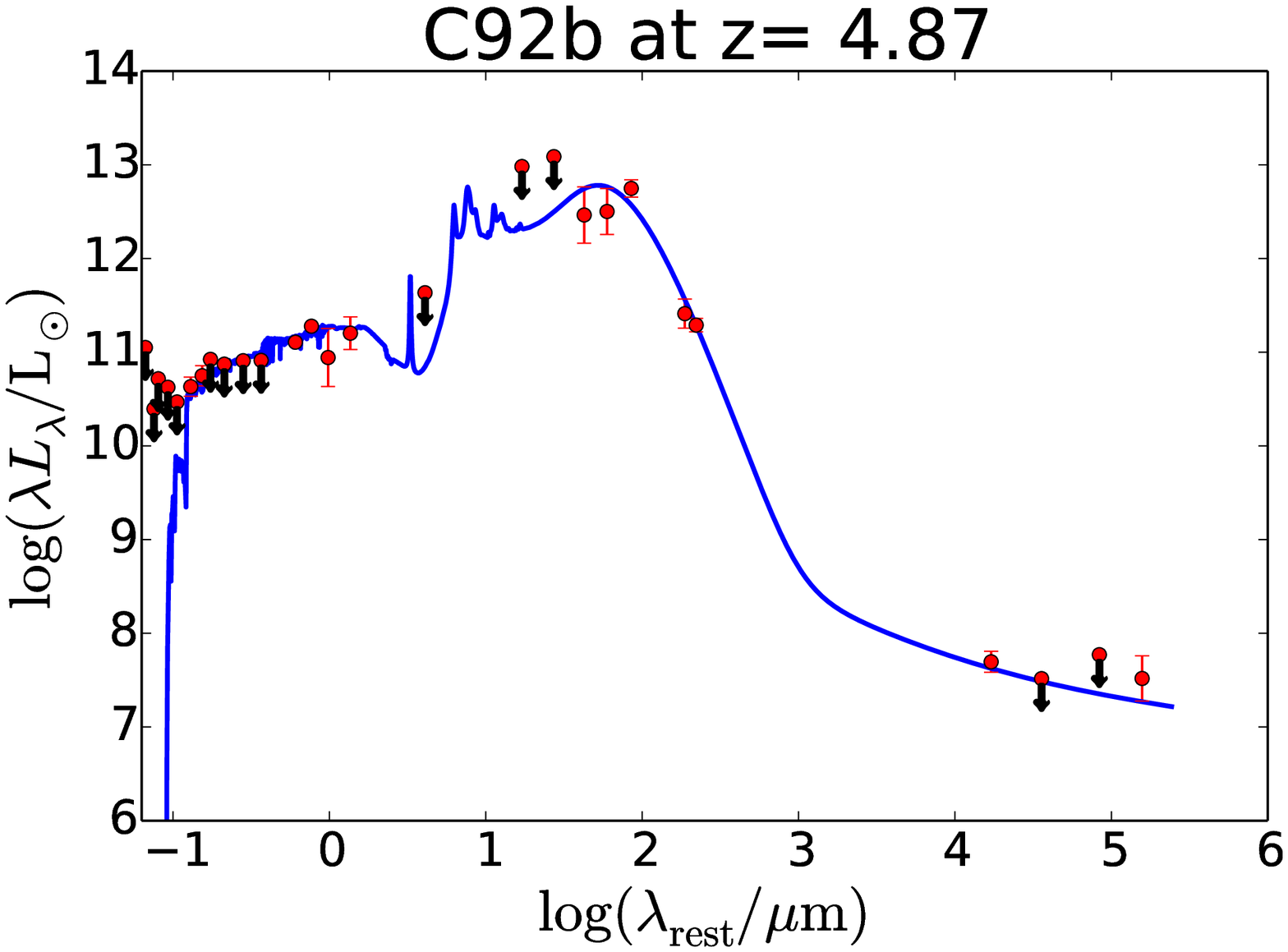}
\includegraphics[width=0.2465\textwidth]{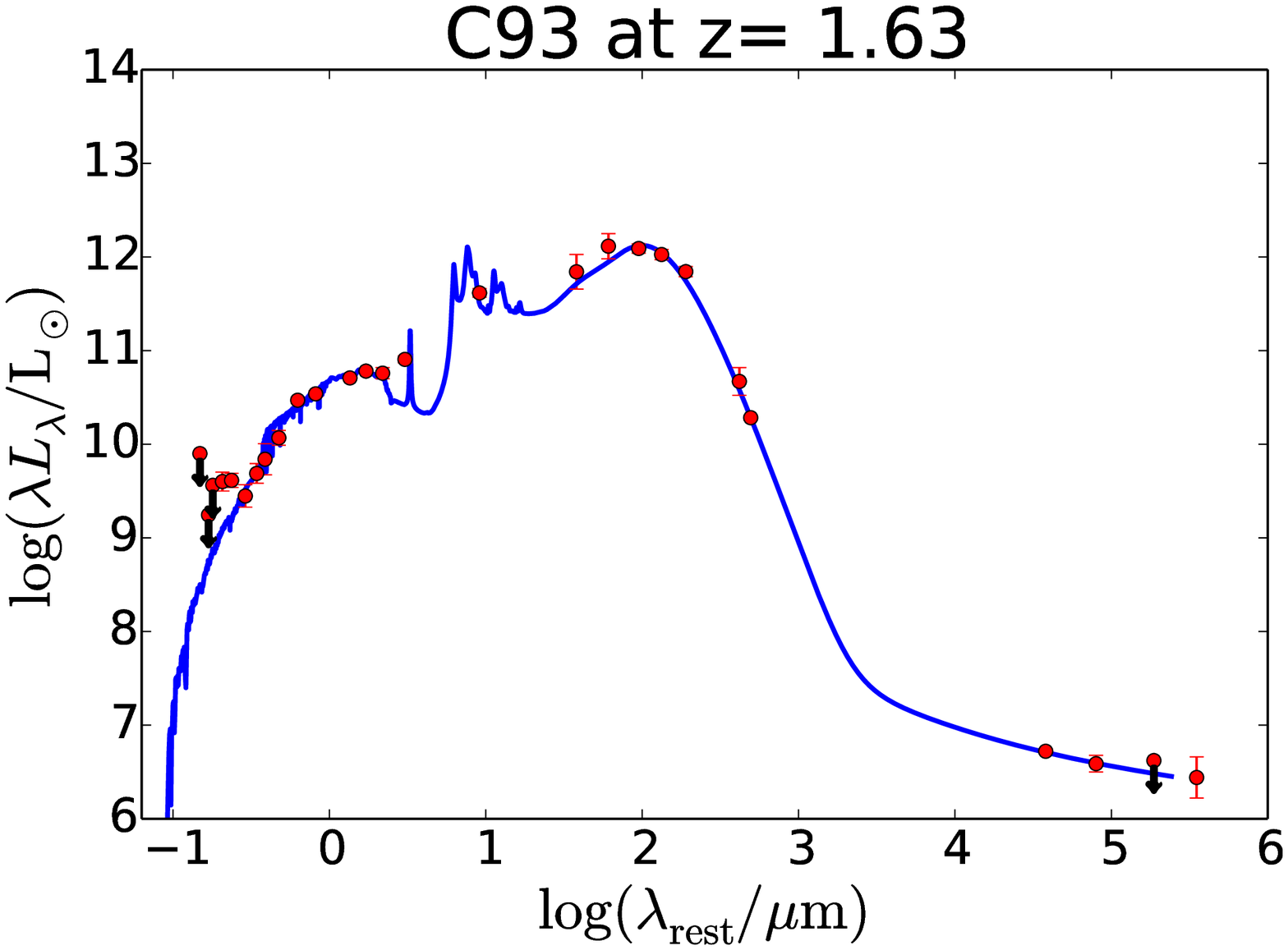}
\includegraphics[width=0.2465\textwidth]{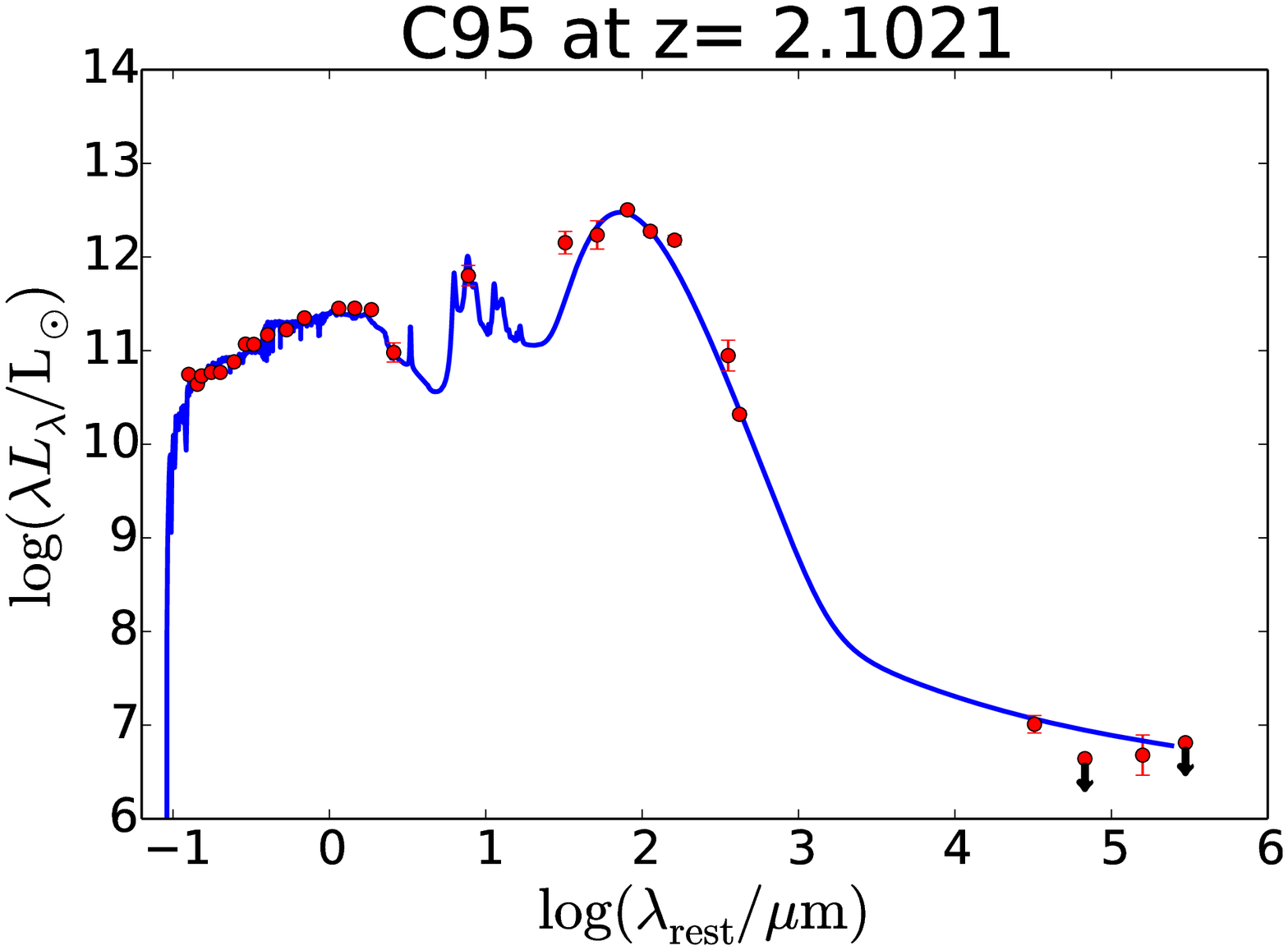}
\includegraphics[width=0.2465\textwidth]{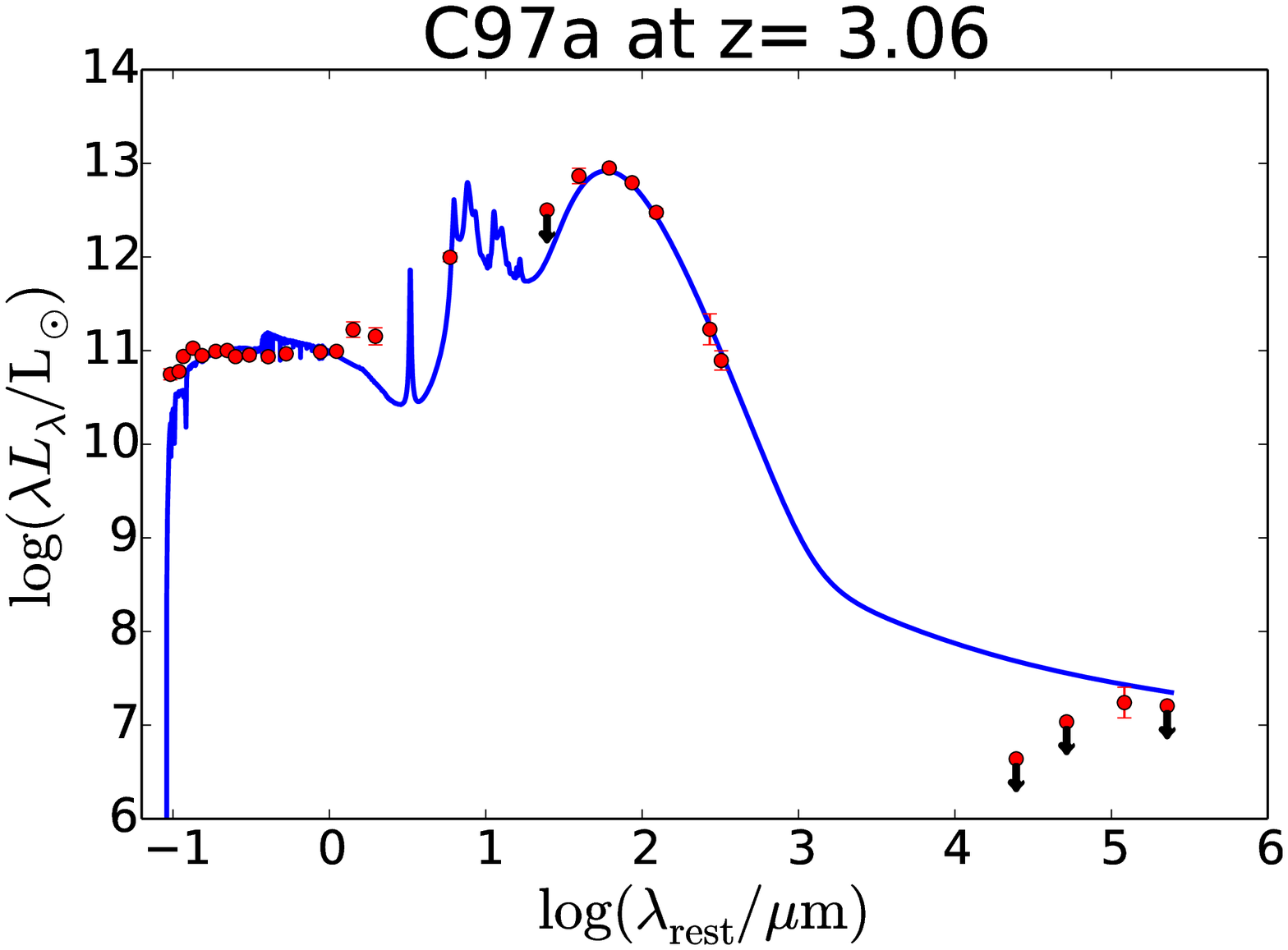}
\includegraphics[width=0.2465\textwidth]{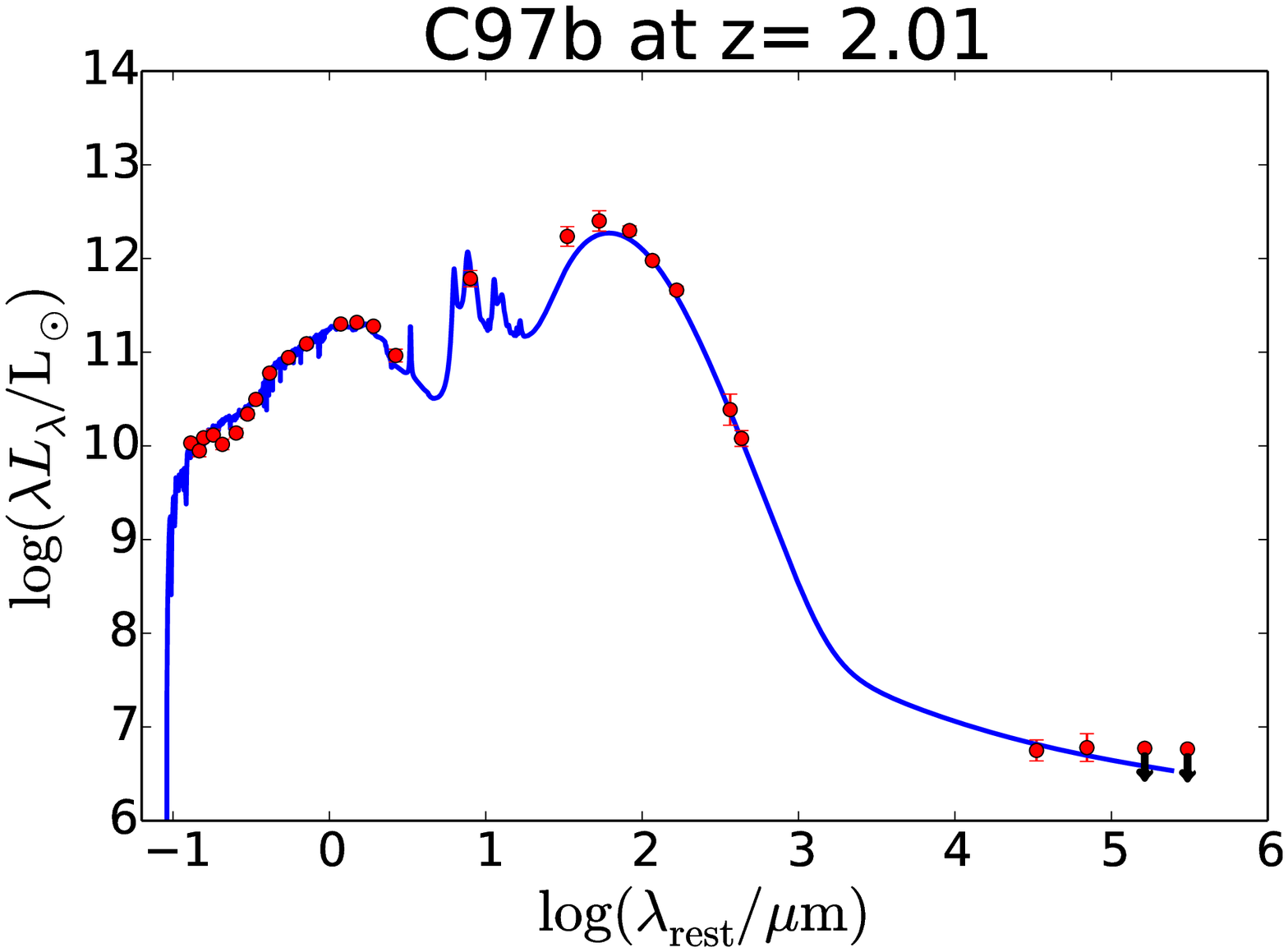}
\includegraphics[width=0.2465\textwidth]{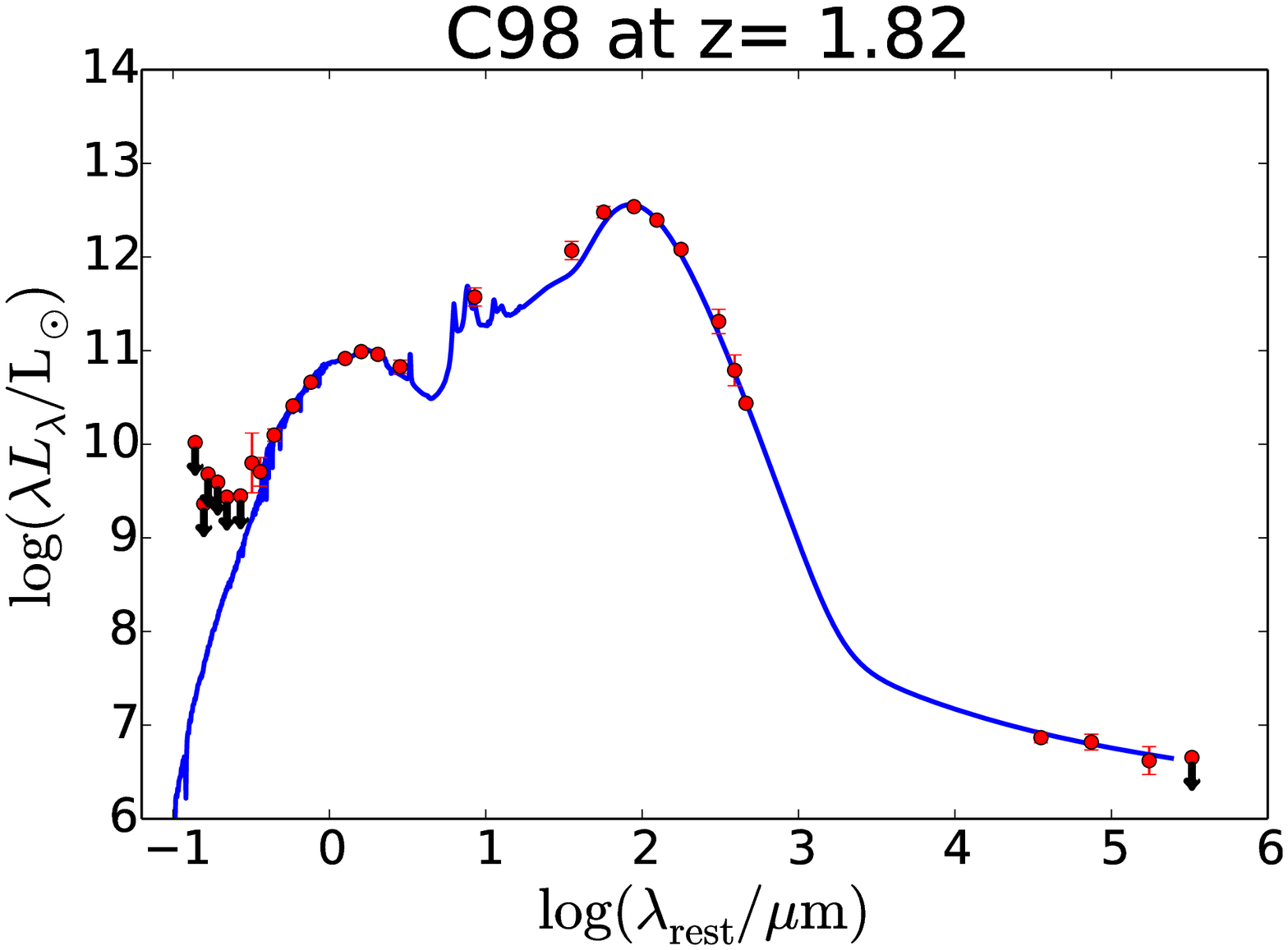}
\includegraphics[width=0.2465\textwidth]{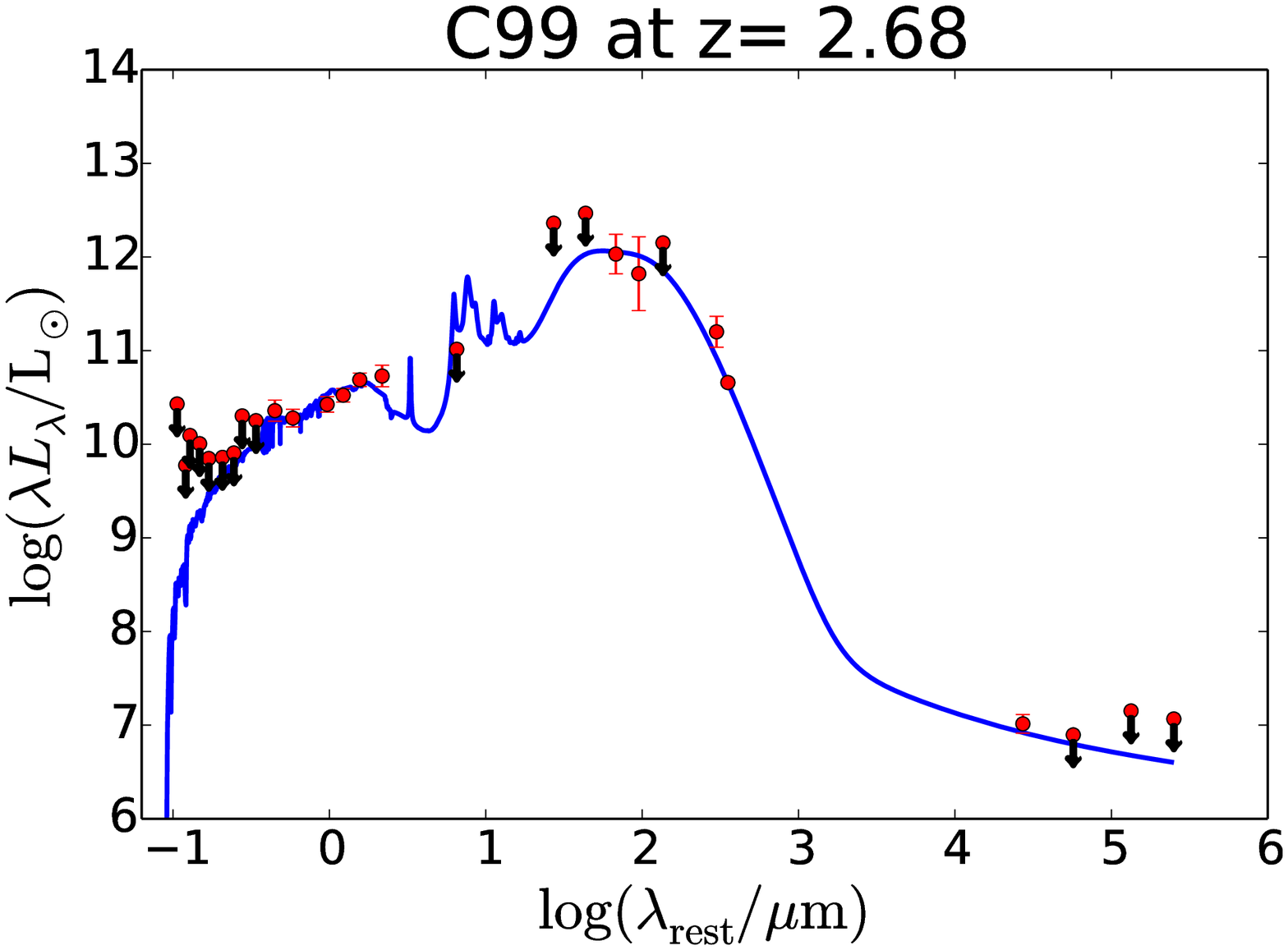}
\includegraphics[width=0.2465\textwidth]{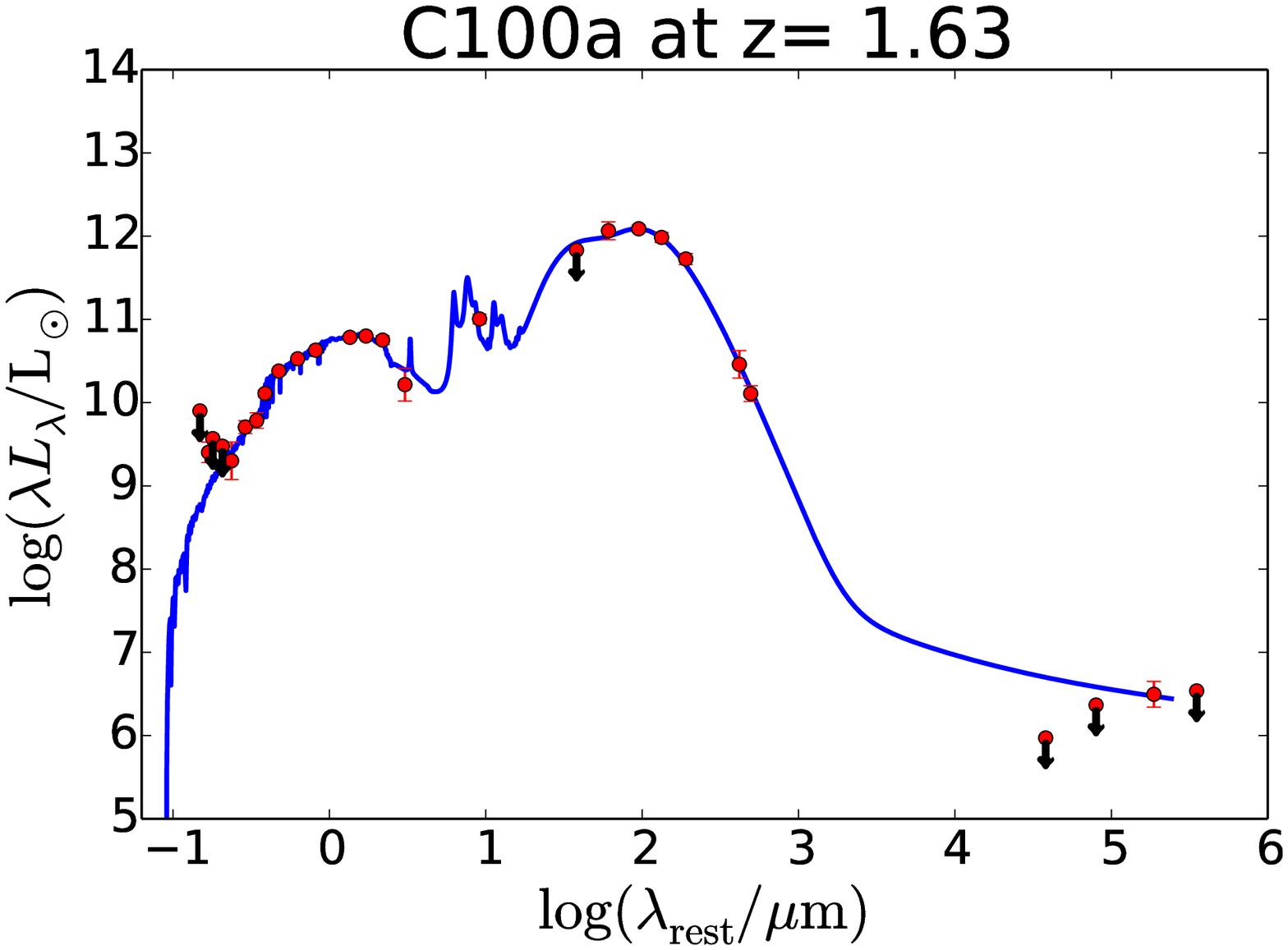}
\includegraphics[width=0.2465\textwidth]{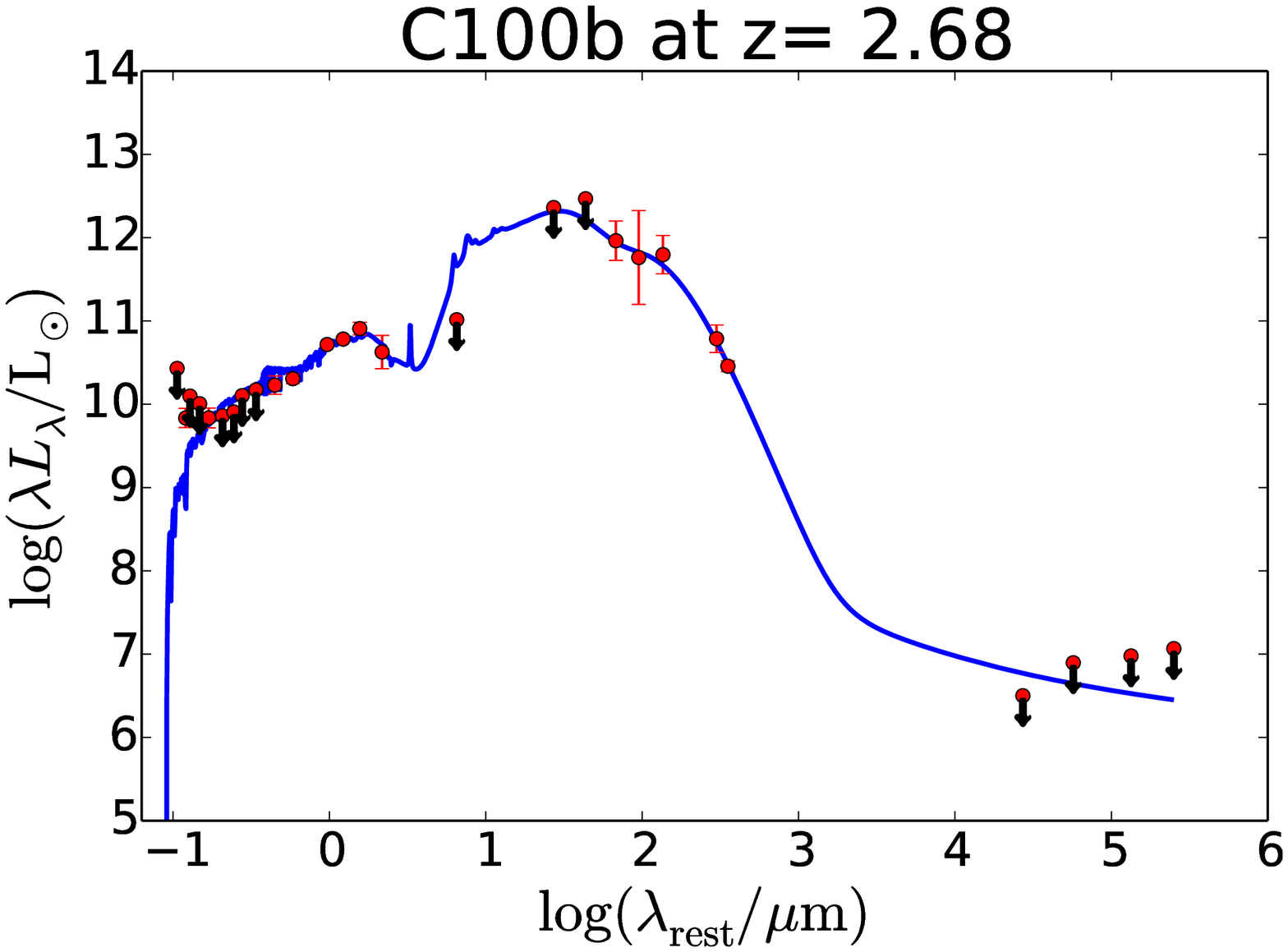}
\includegraphics[width=0.2465\textwidth]{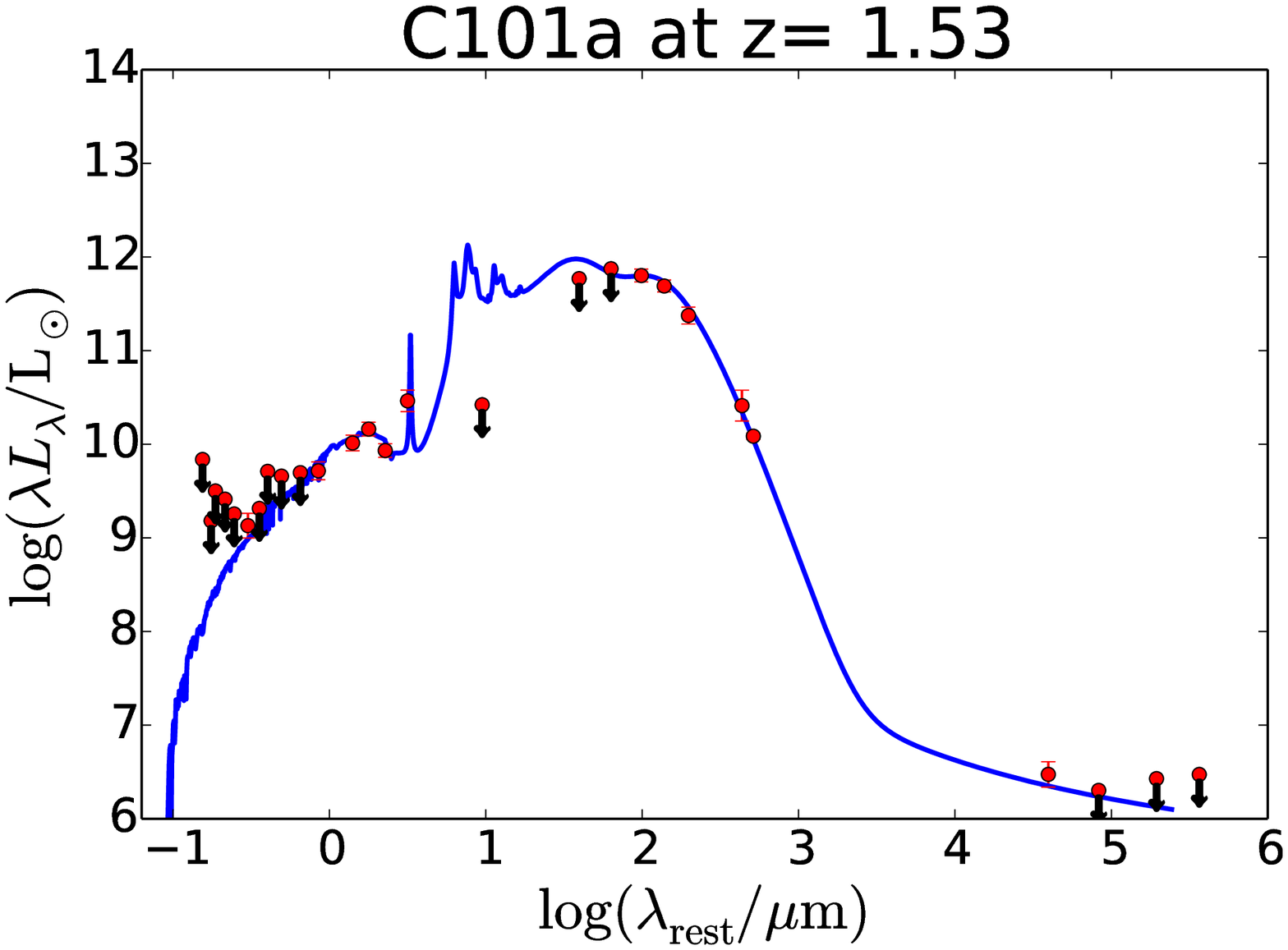}
\includegraphics[width=0.2465\textwidth]{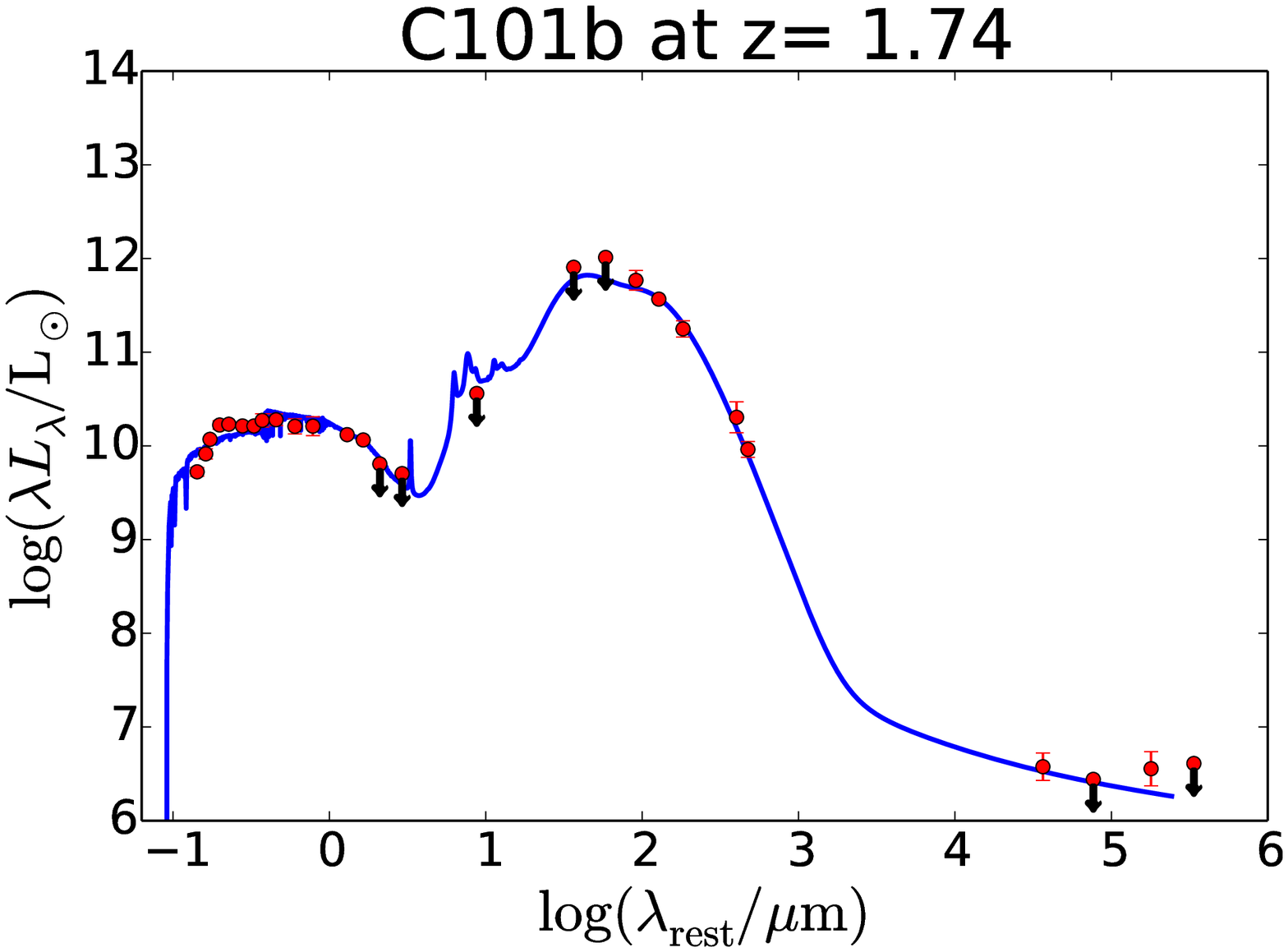}
\caption{continued.}
\label{figure:seds}
\end{center}
\end{figure*} 

\addtocounter{figure}{-1}
\begin{figure*}
\begin{center}
\includegraphics[width=0.2465\textwidth]{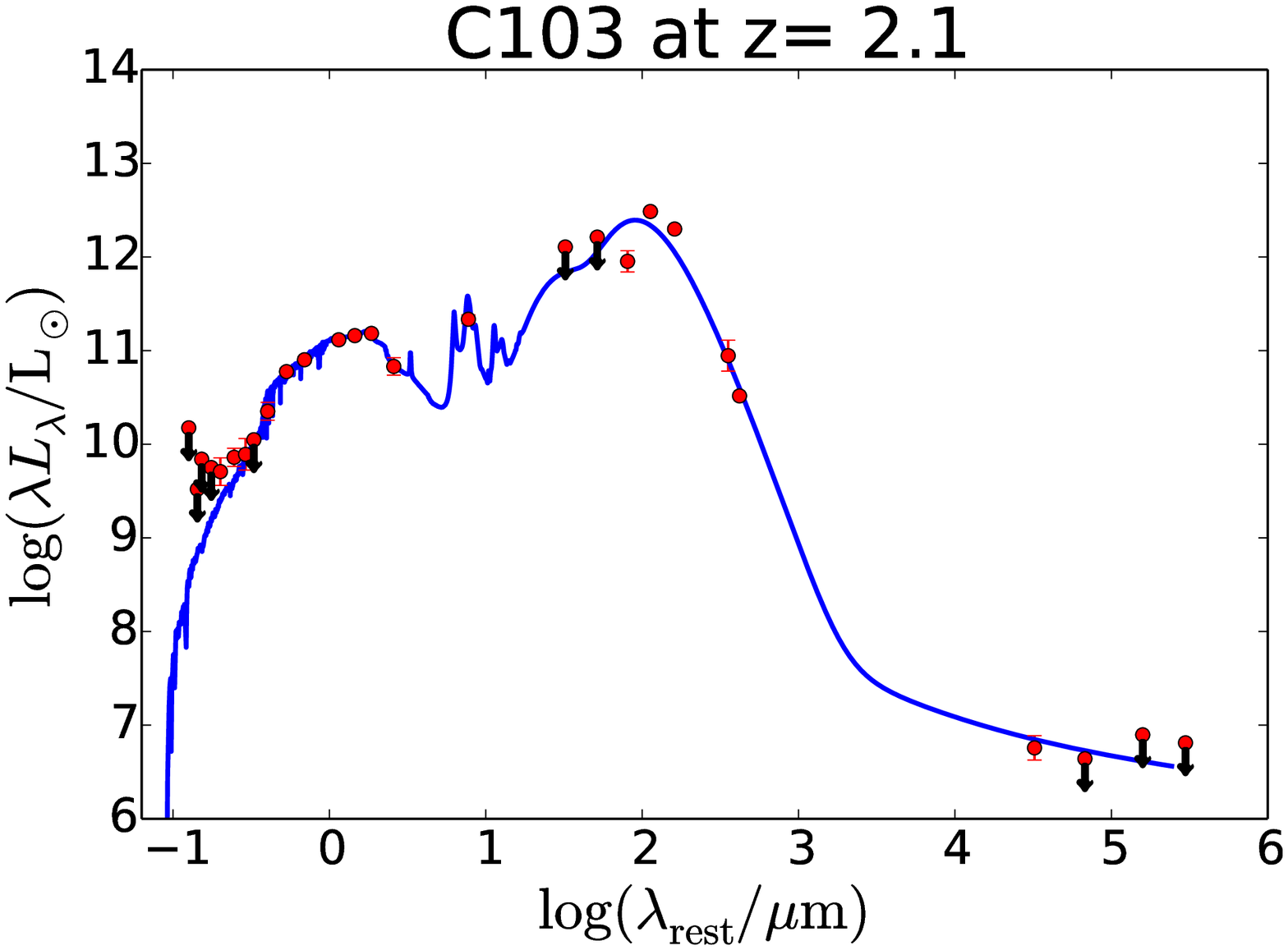} 
\includegraphics[width=0.2465\textwidth]{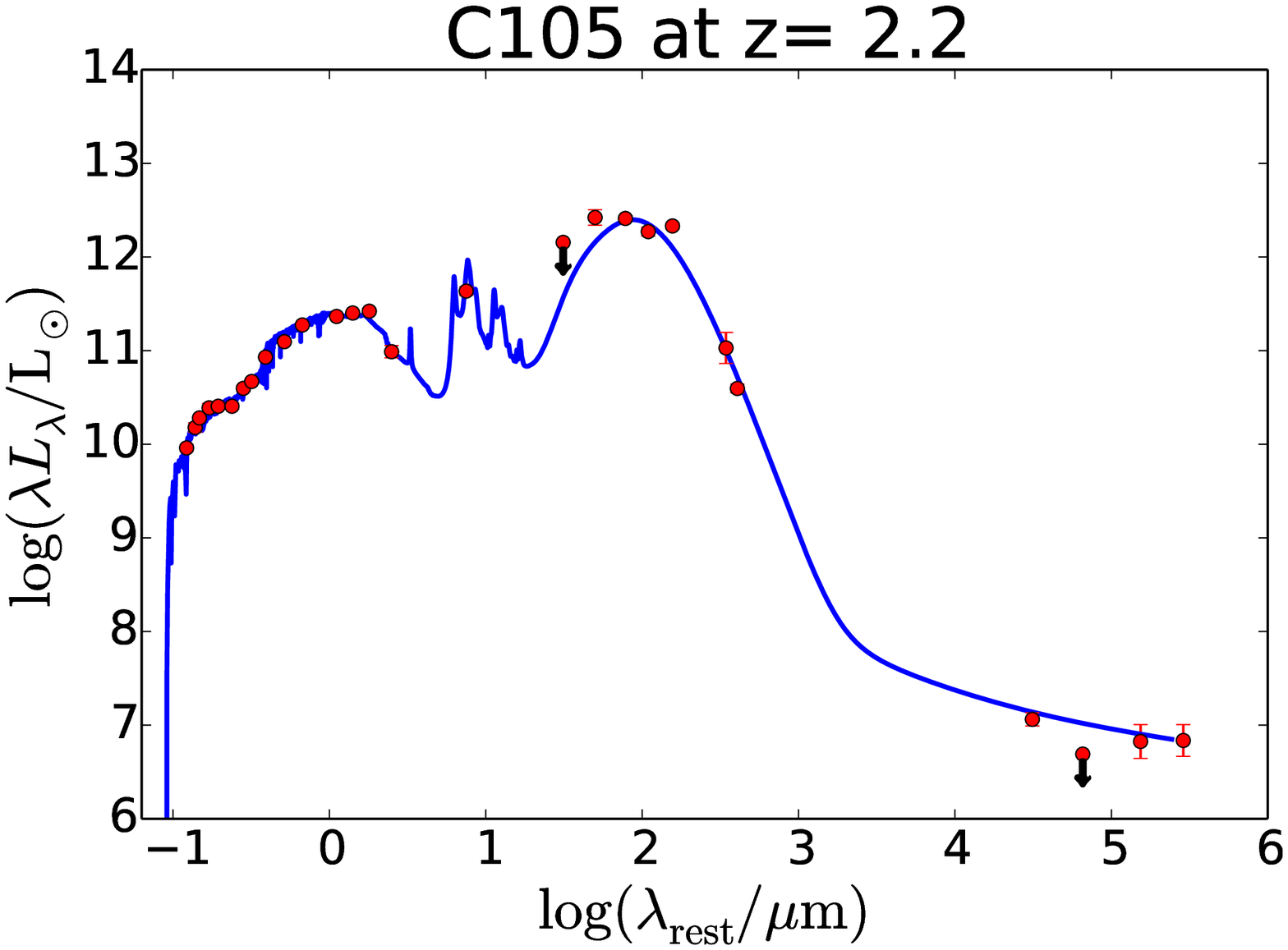}
\includegraphics[width=0.2465\textwidth]{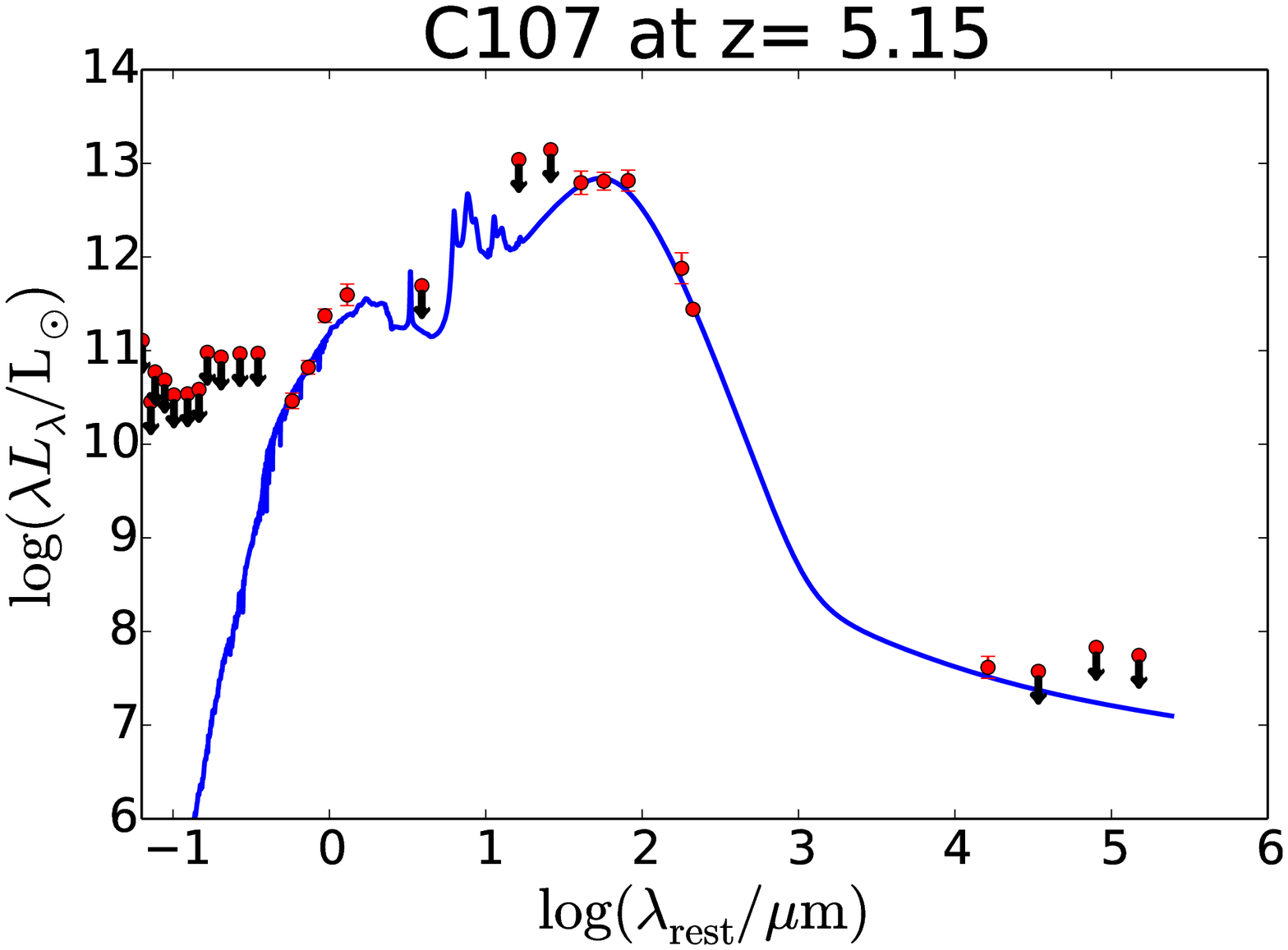}
\includegraphics[width=0.2465\textwidth]{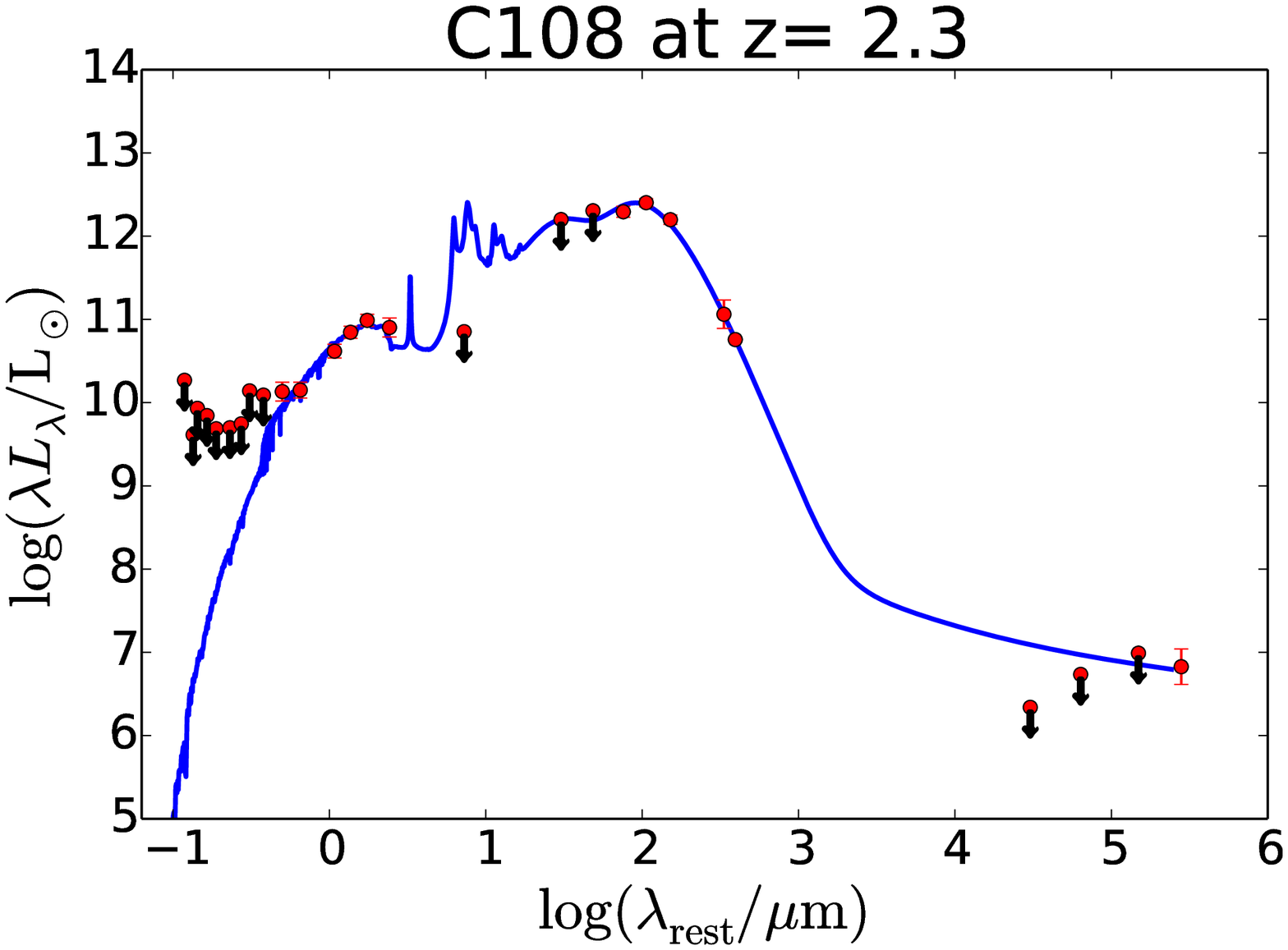}
\includegraphics[width=0.2465\textwidth]{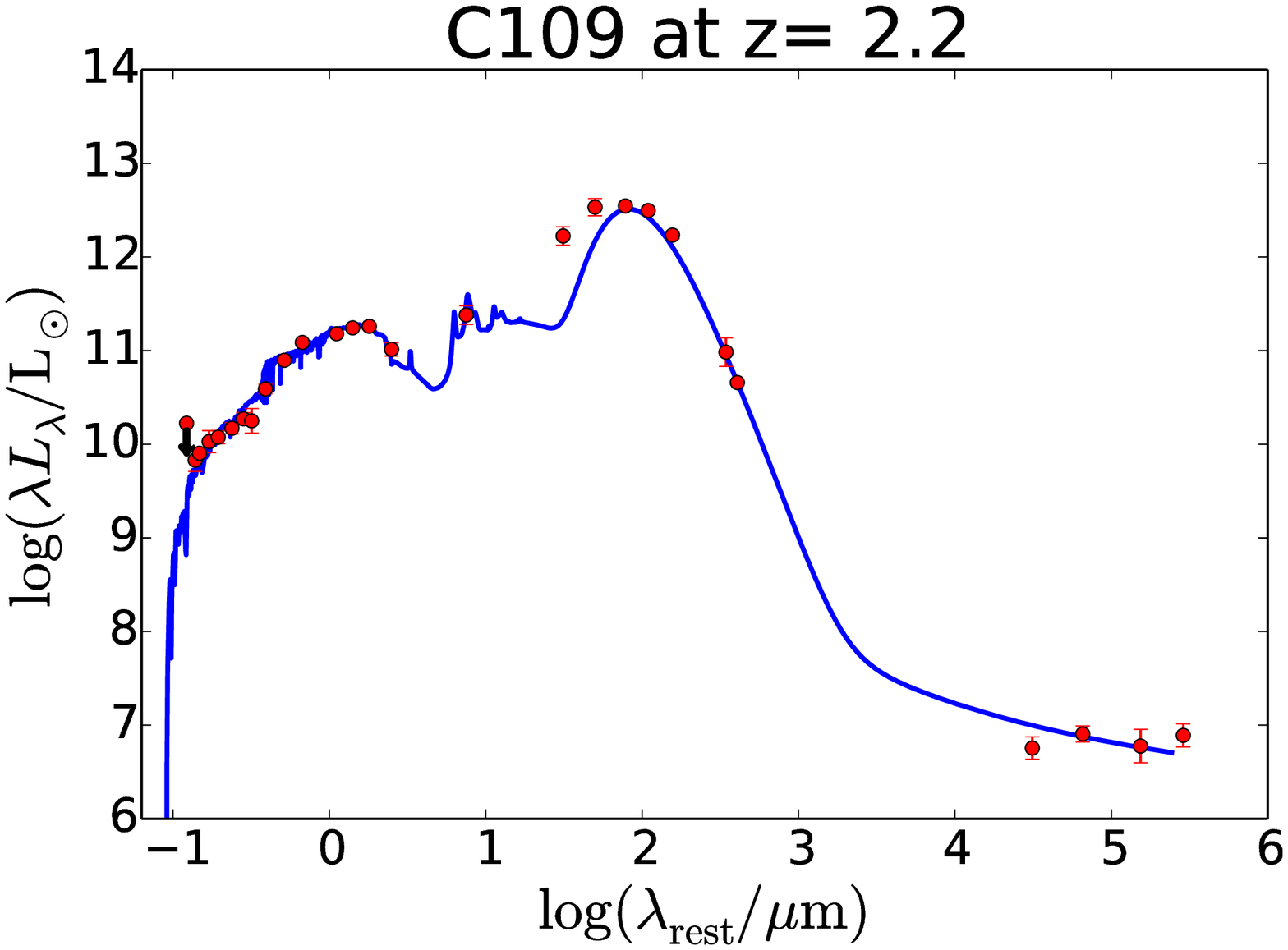}
\includegraphics[width=0.2465\textwidth]{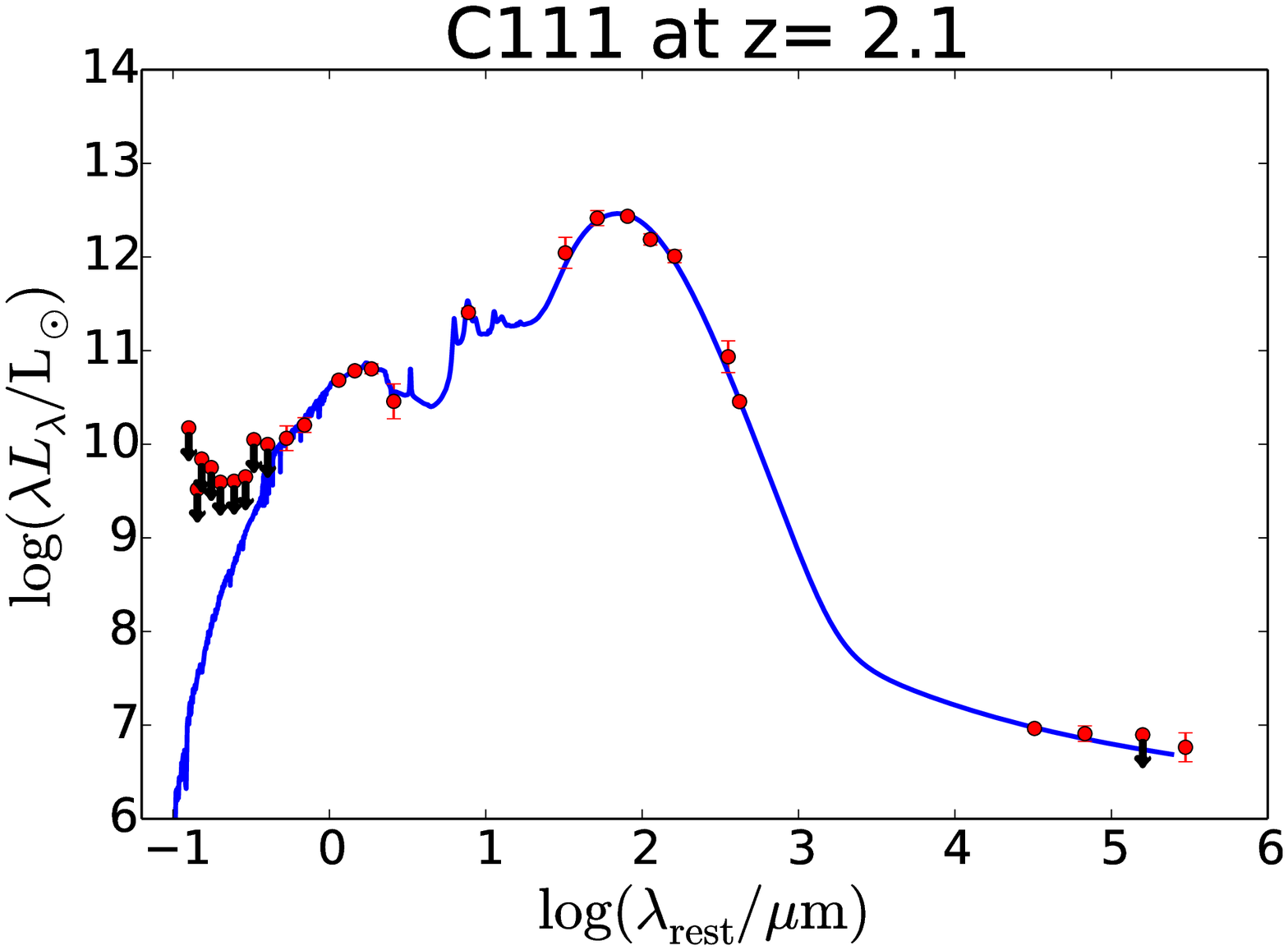}
\includegraphics[width=0.2465\textwidth]{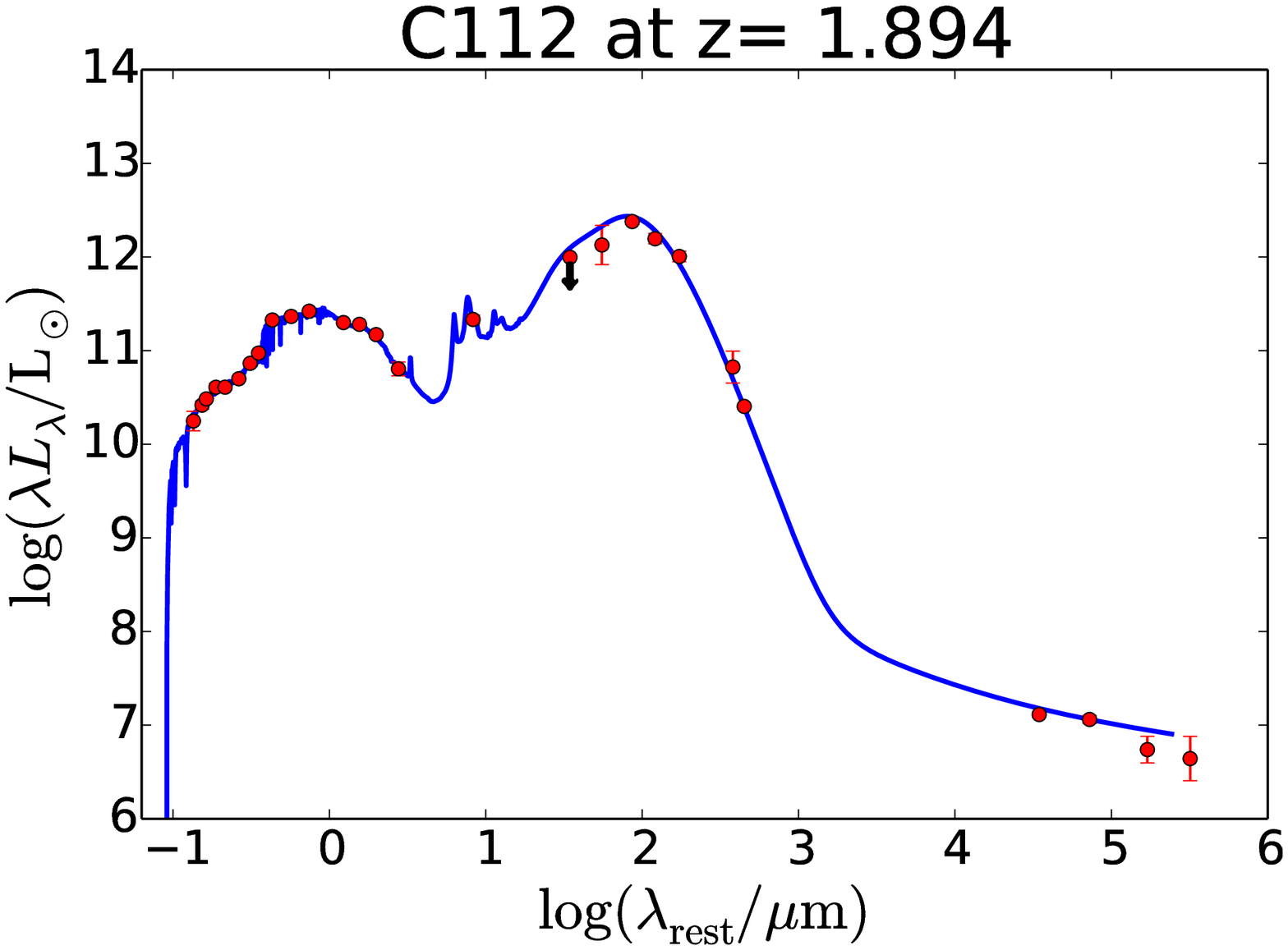}
\includegraphics[width=0.2465\textwidth]{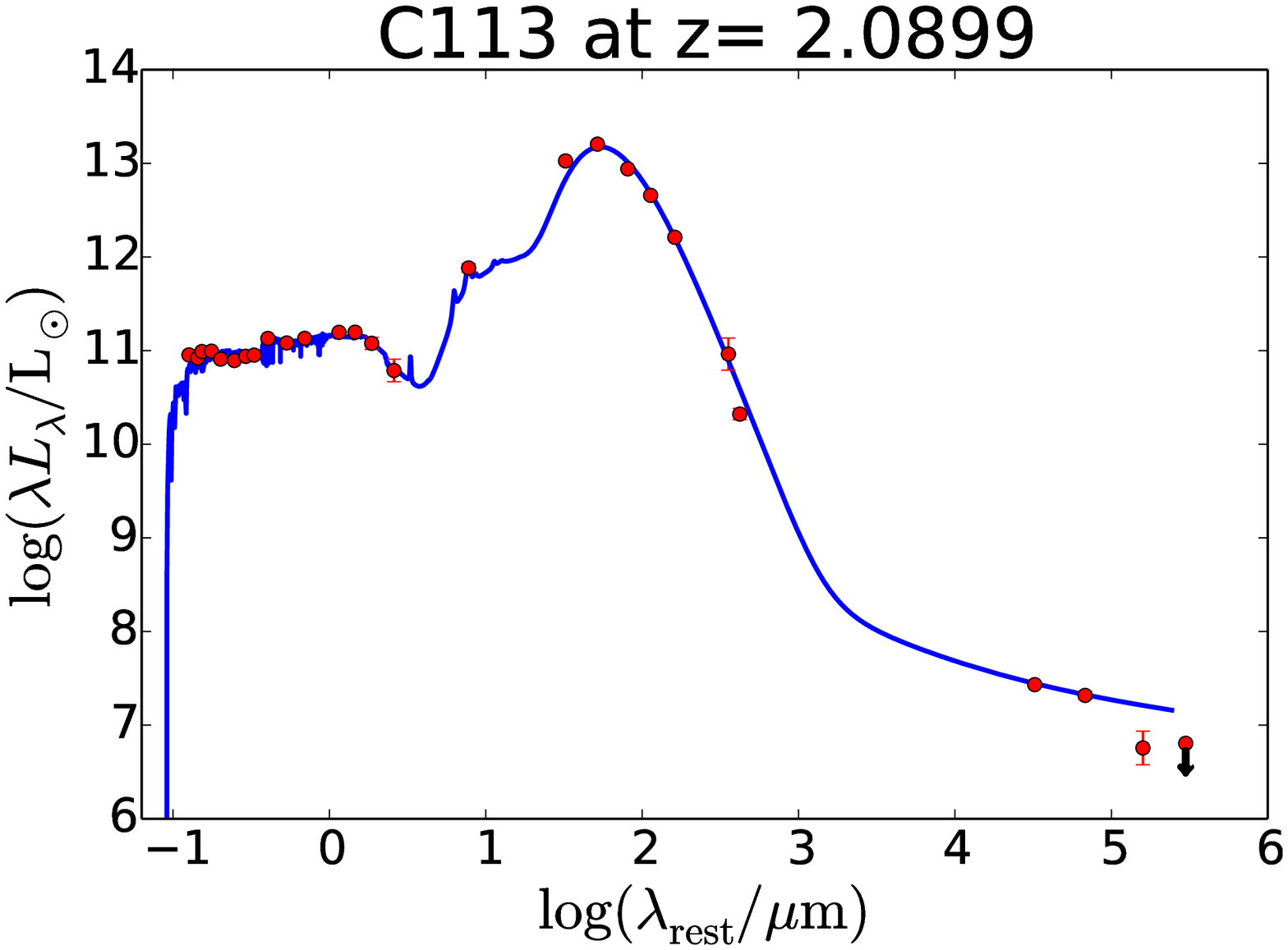}
\includegraphics[width=0.2465\textwidth]{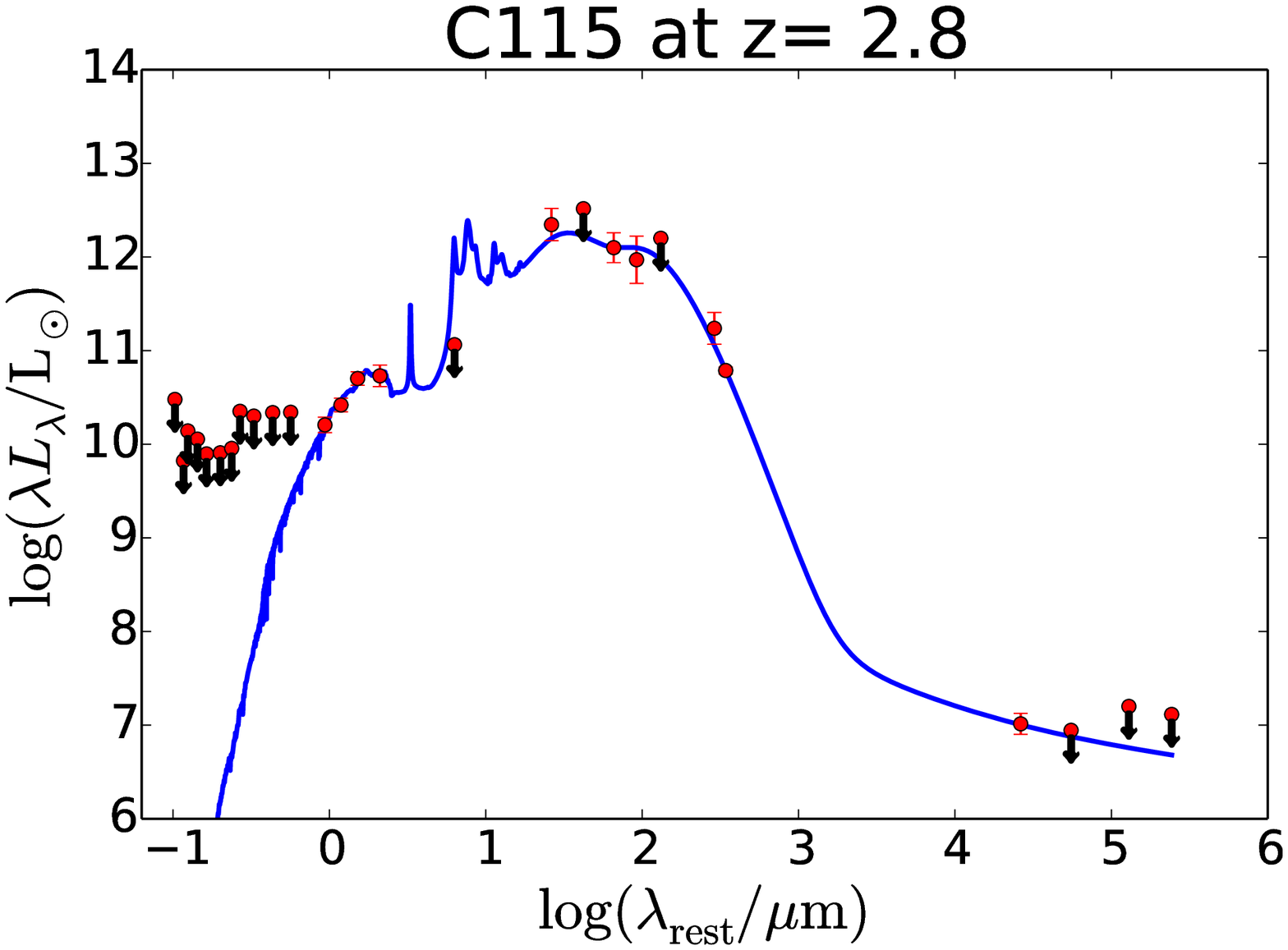}
\includegraphics[width=0.2465\textwidth]{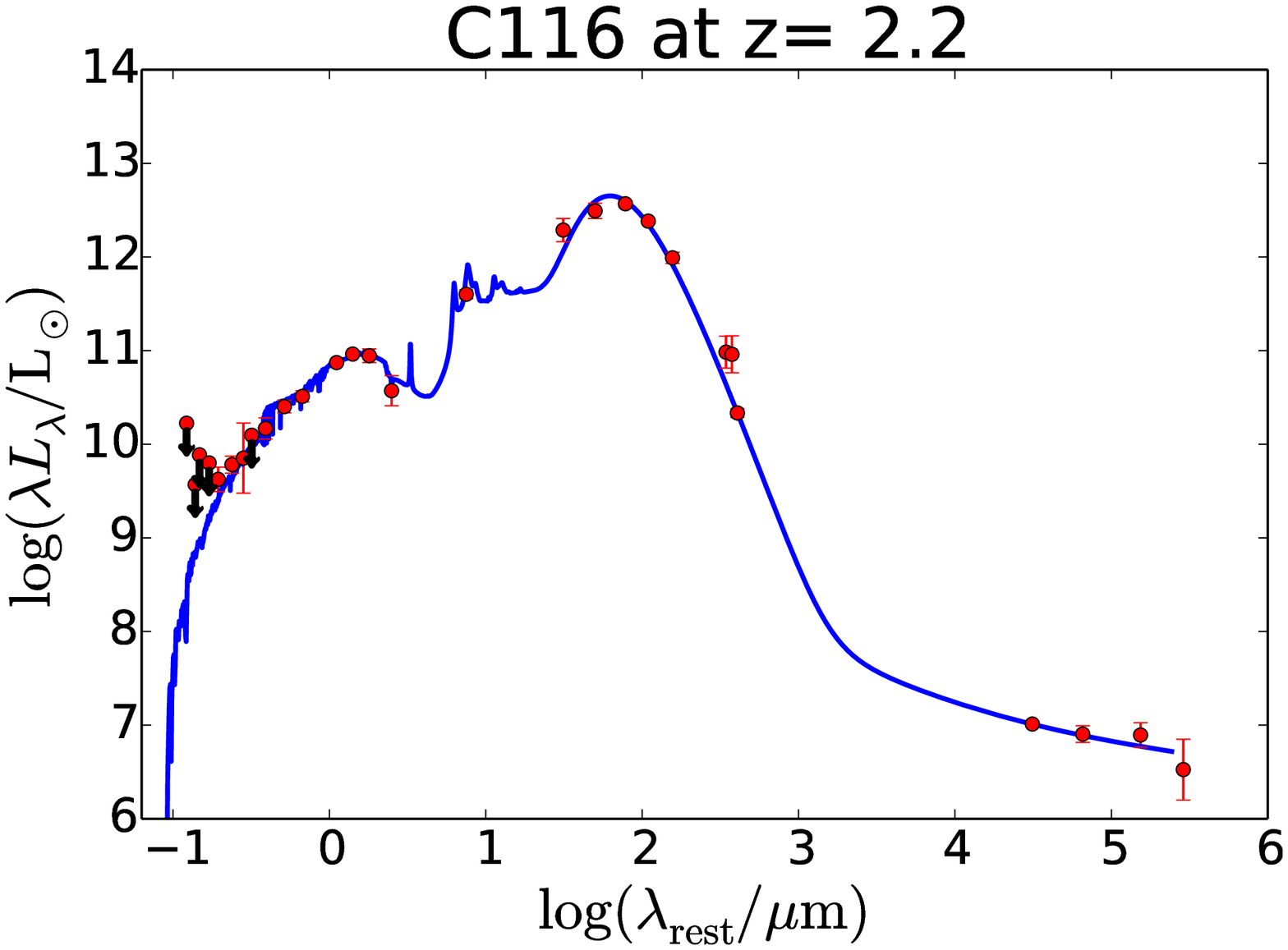}
\includegraphics[width=0.2465\textwidth]{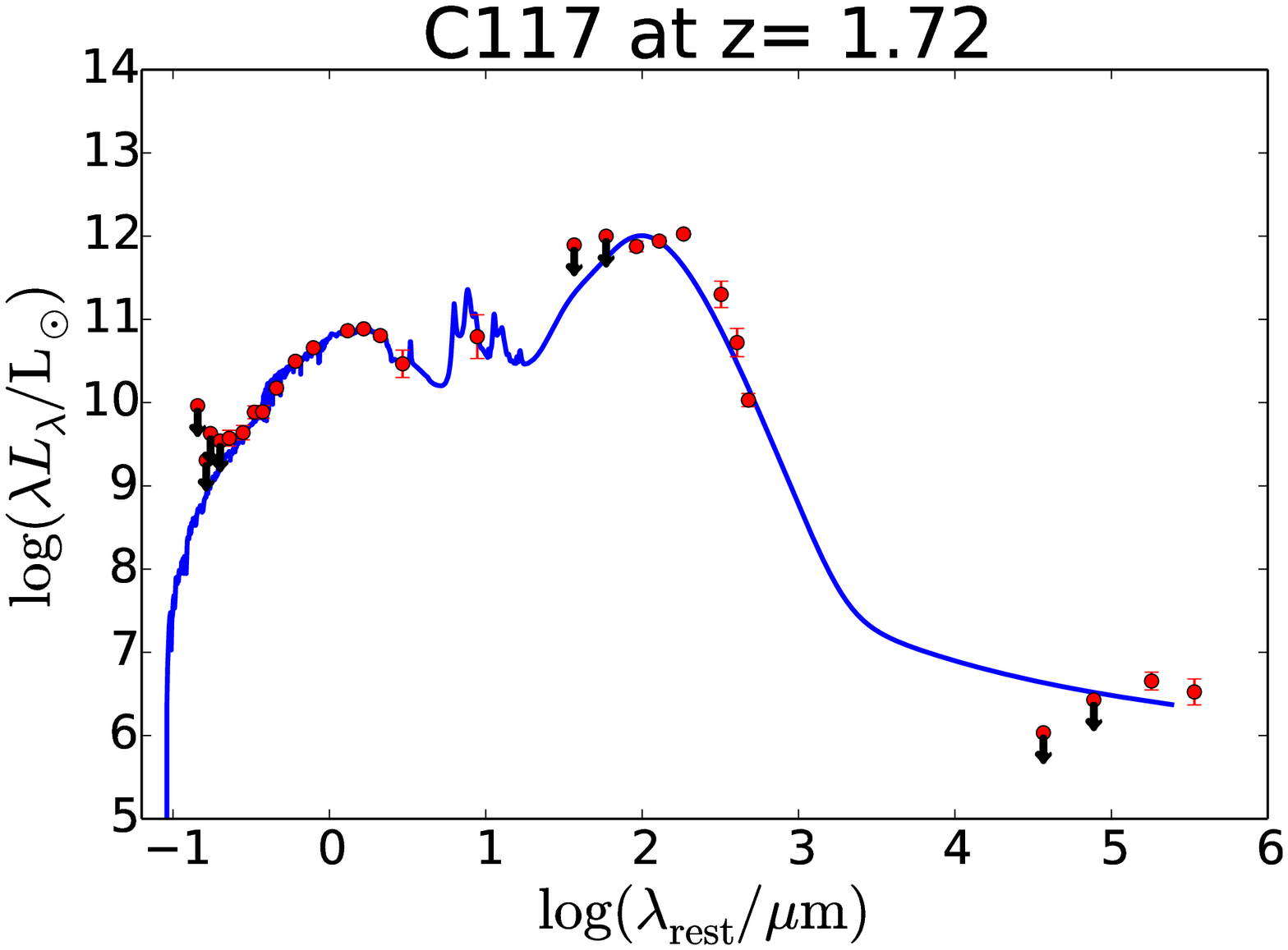}
\includegraphics[width=0.2465\textwidth]{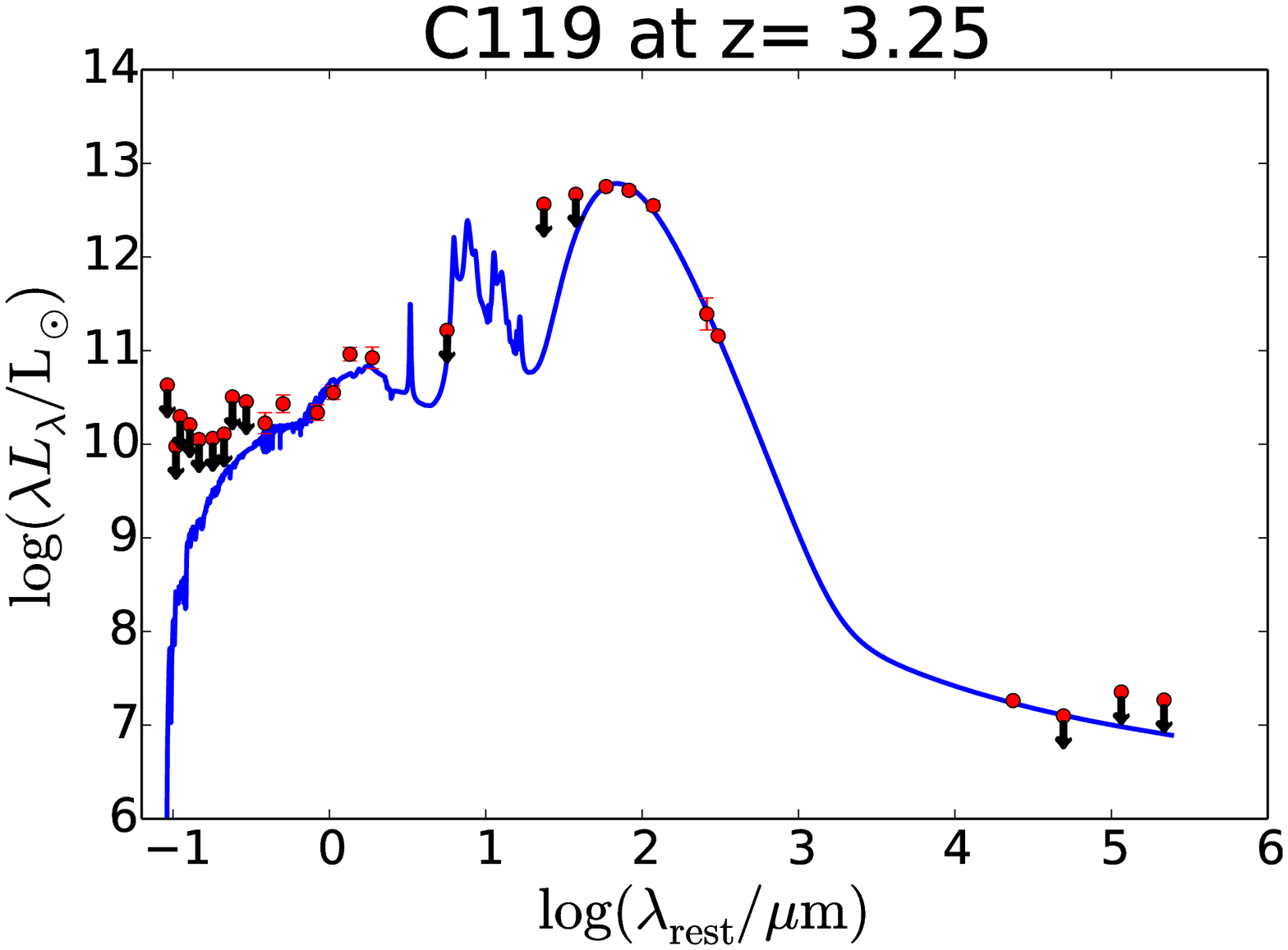}
\includegraphics[width=0.2465\textwidth]{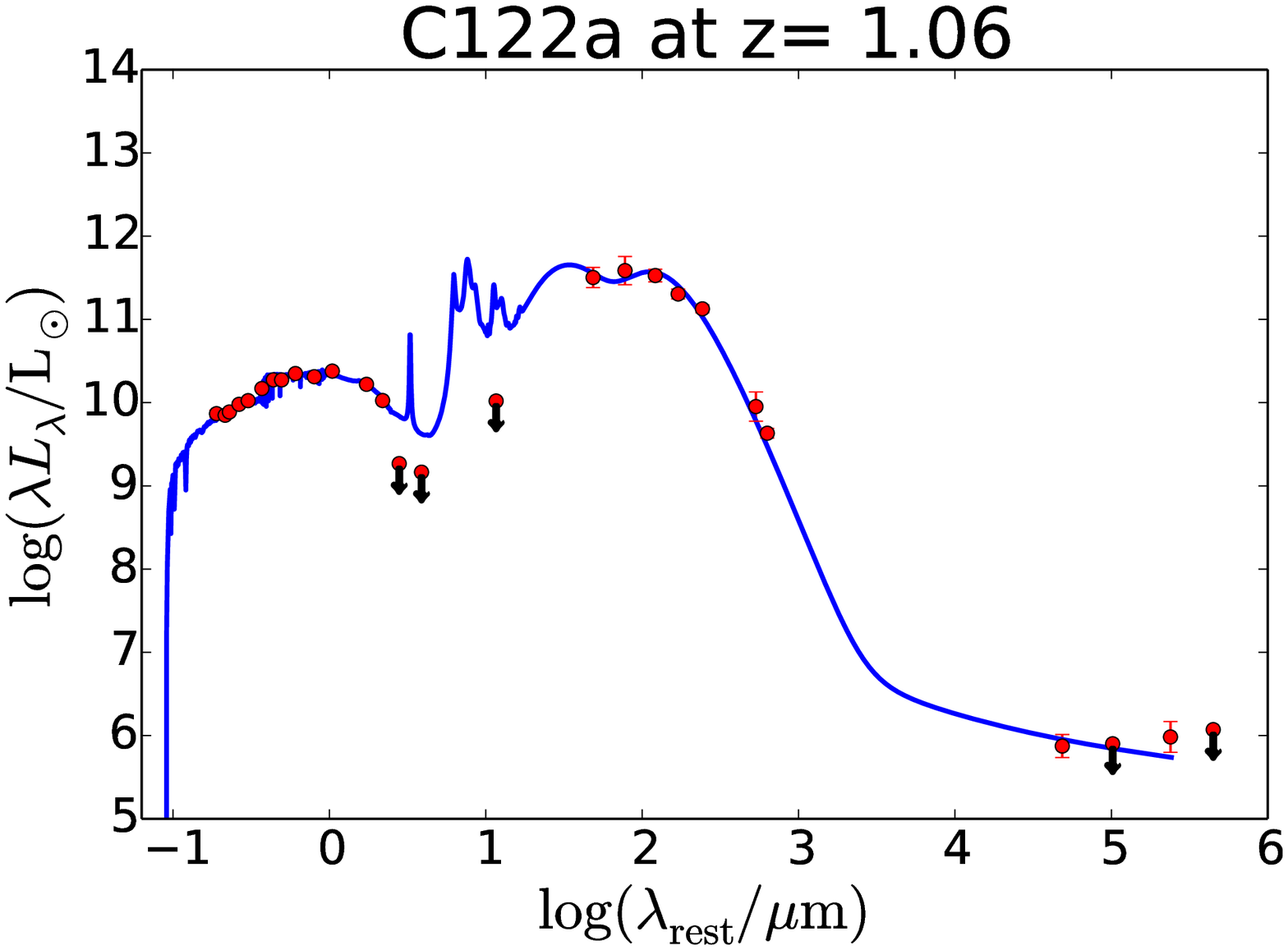}
\includegraphics[width=0.2465\textwidth]{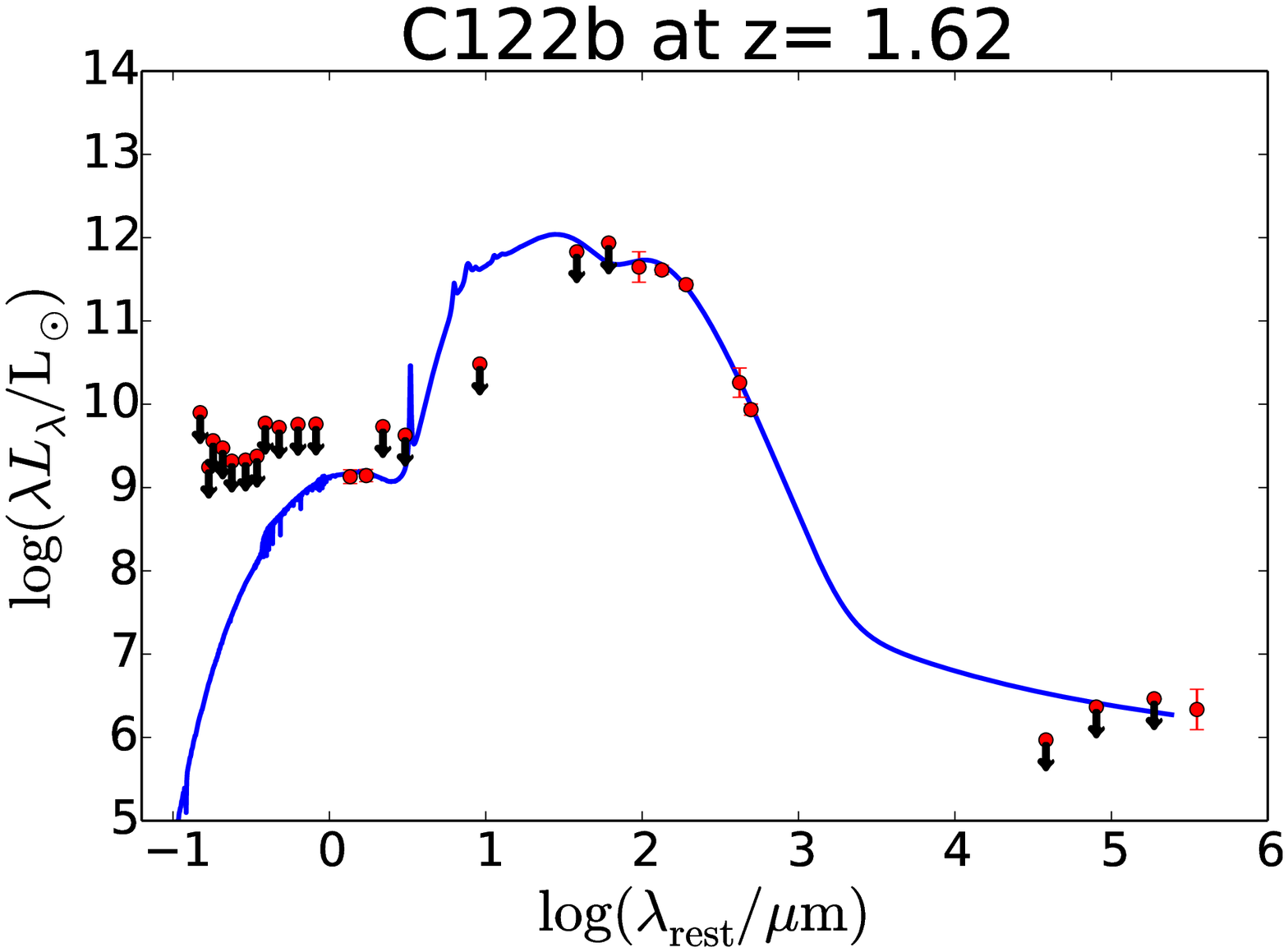}
\includegraphics[width=0.2465\textwidth]{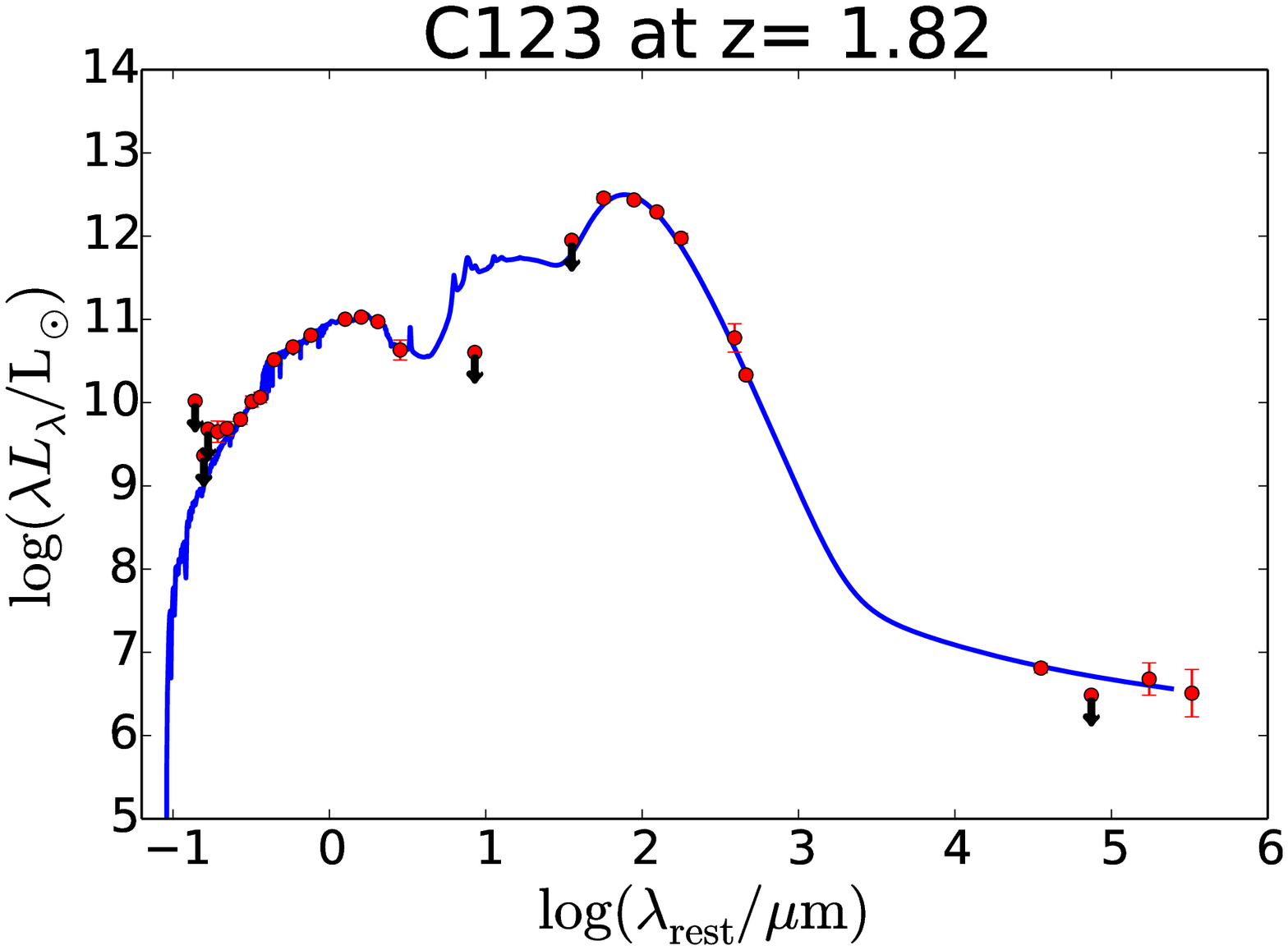}
\includegraphics[width=0.2465\textwidth]{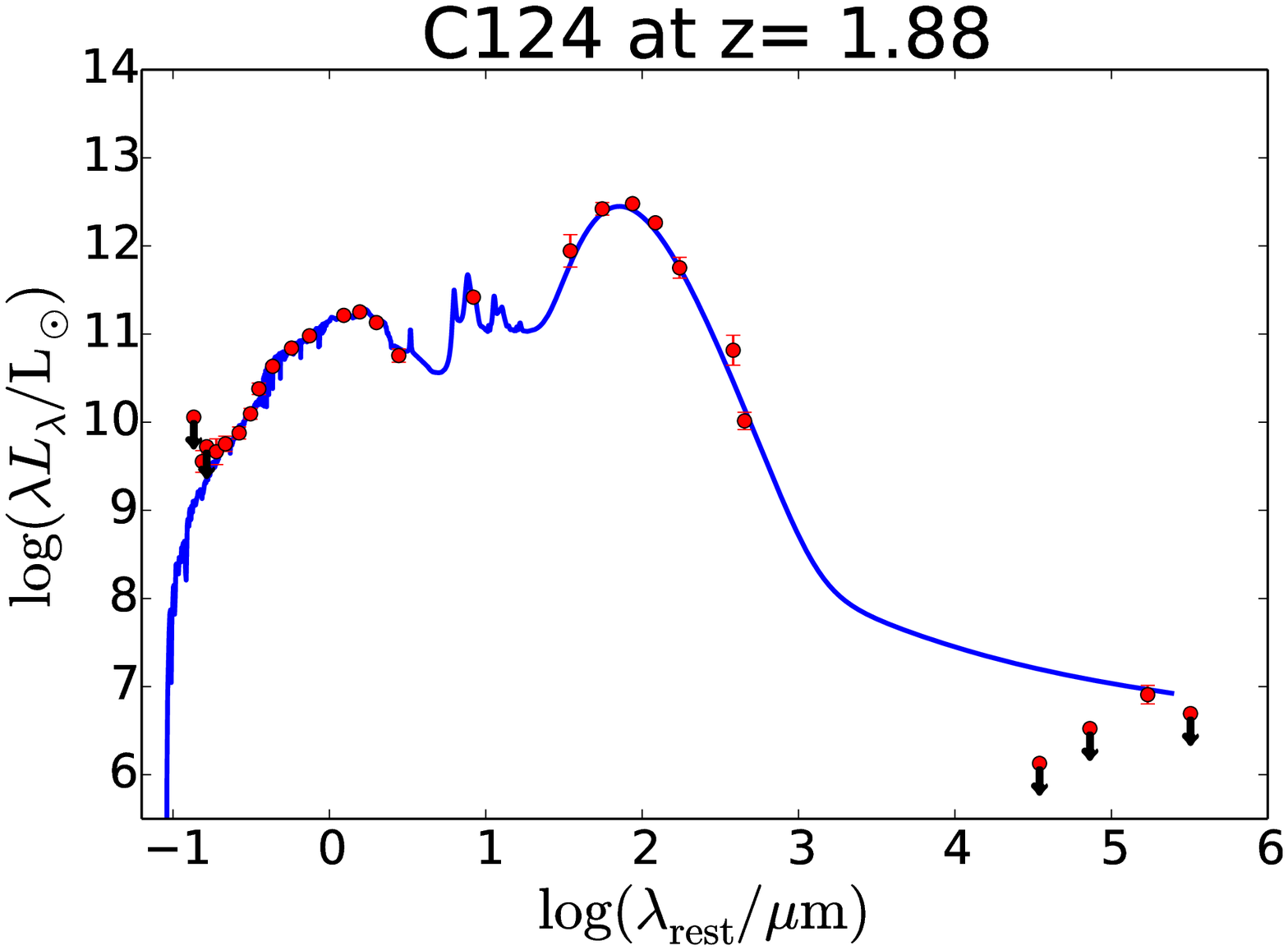}
\includegraphics[width=0.2465\textwidth]{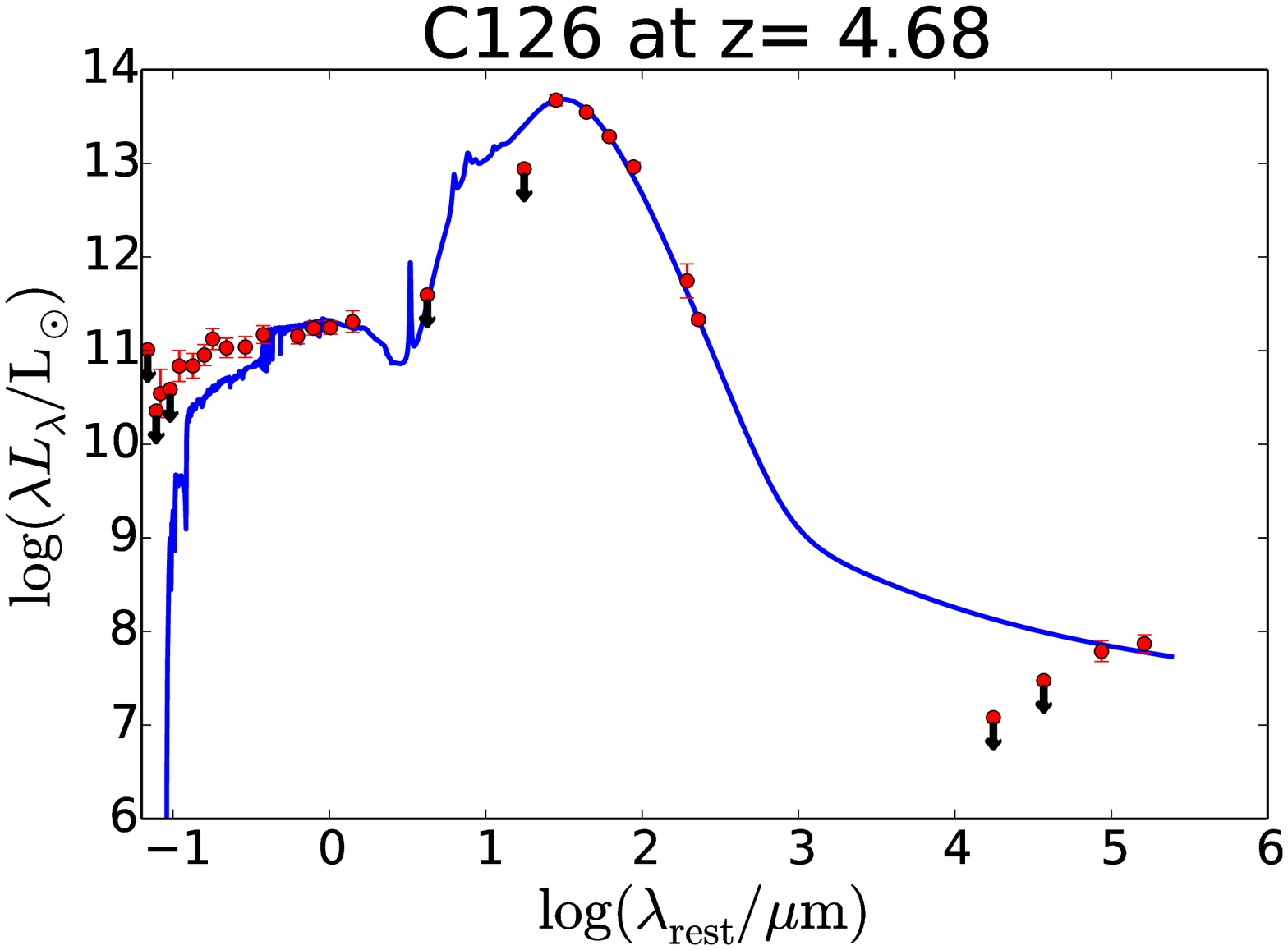}
\includegraphics[width=0.2465\textwidth]{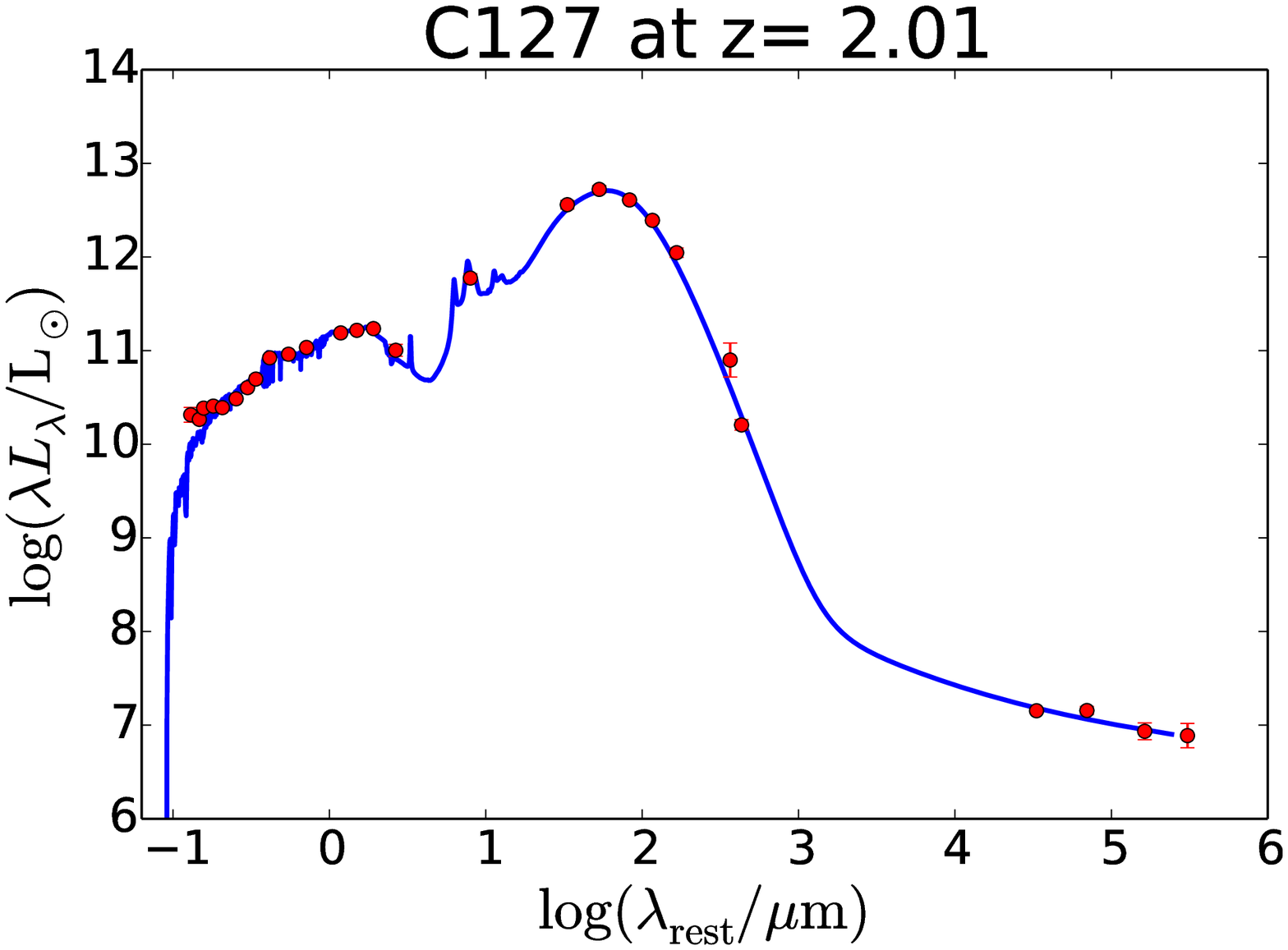}
\includegraphics[width=0.2465\textwidth]{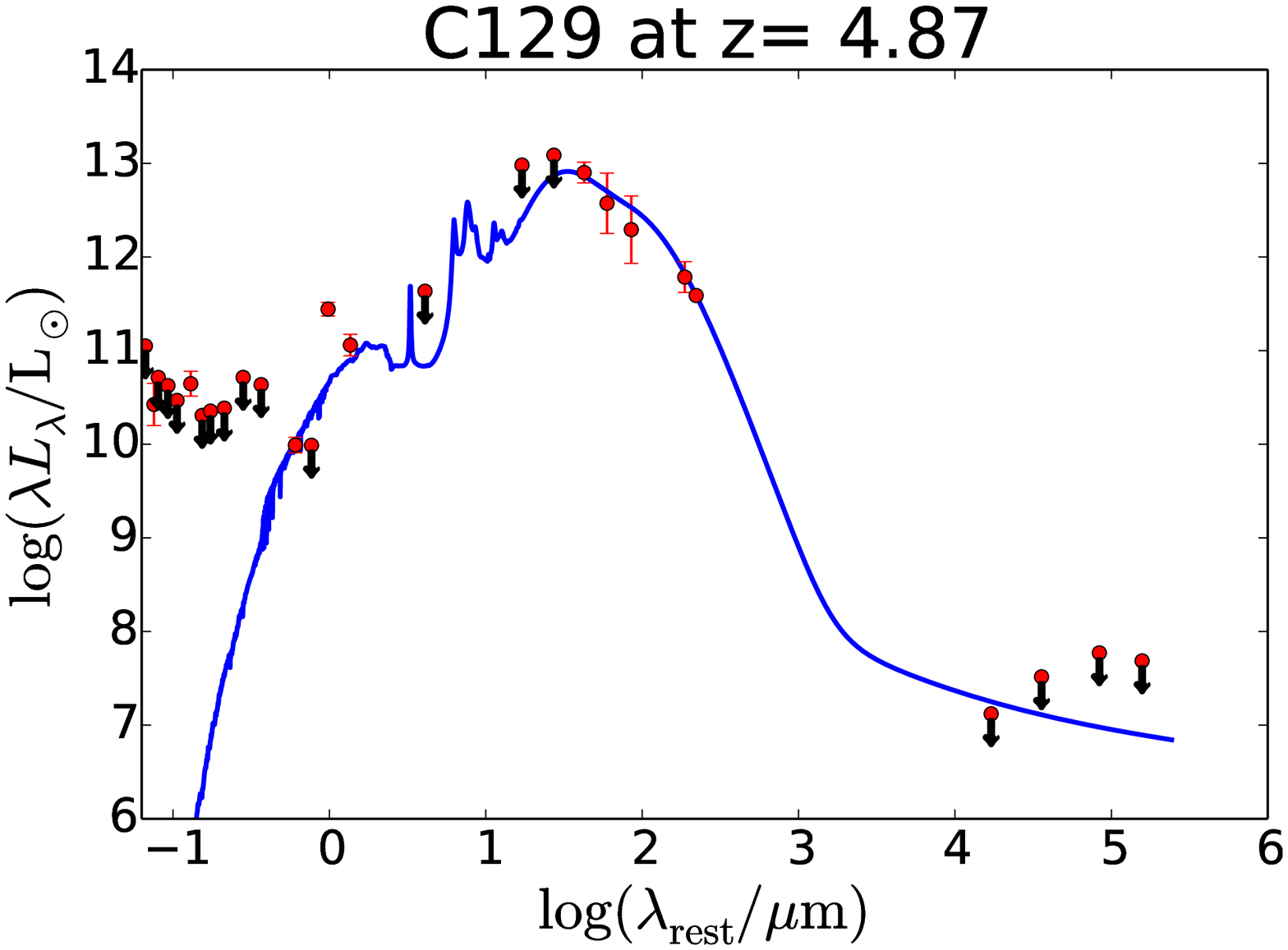}
\caption{continued.}
\label{figure:seds}
\end{center}
\end{figure*}

\begin{figure*}
\begin{center}
\includegraphics[width=0.2465\textwidth]{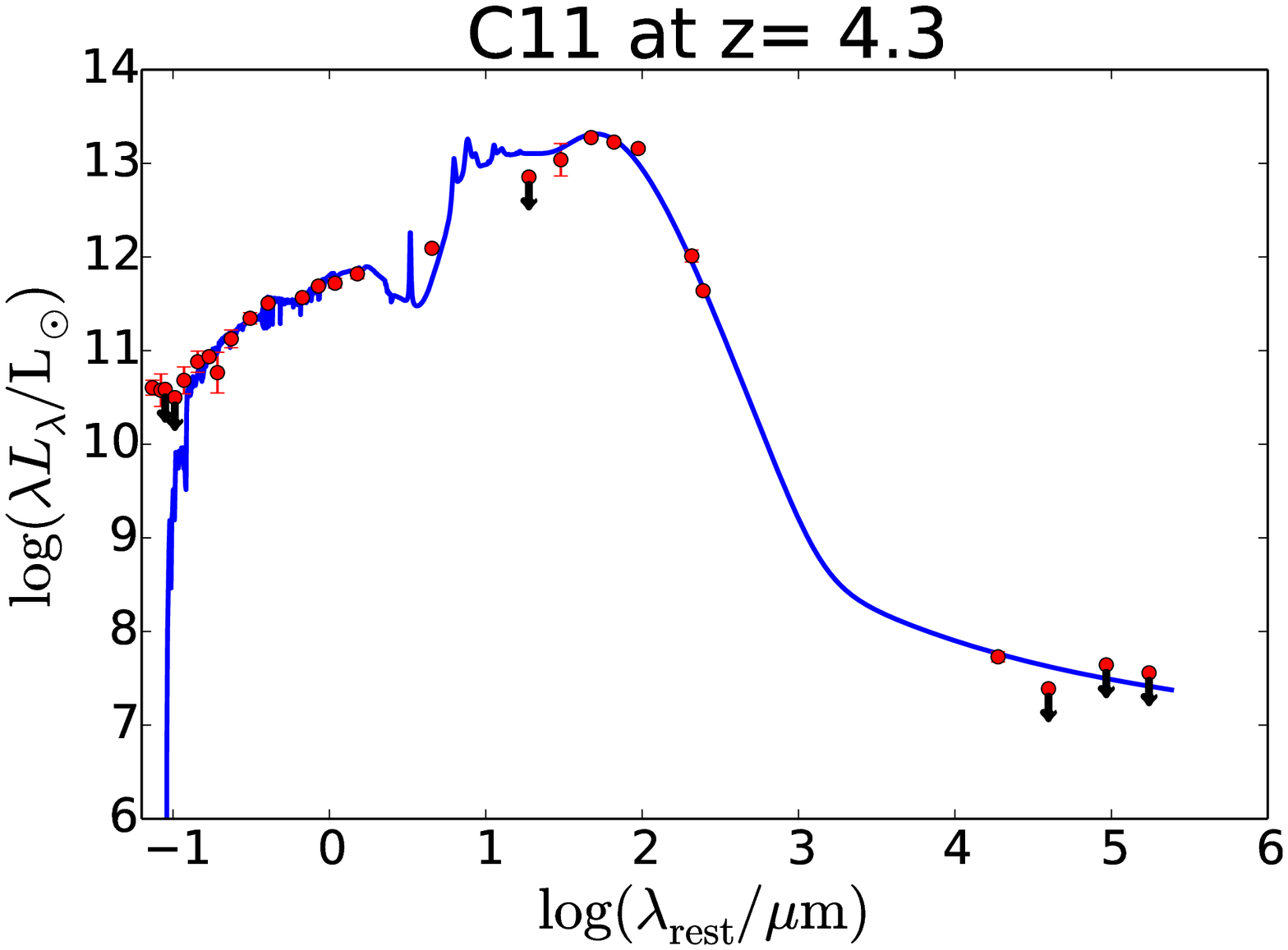}
\includegraphics[width=0.2465\textwidth]{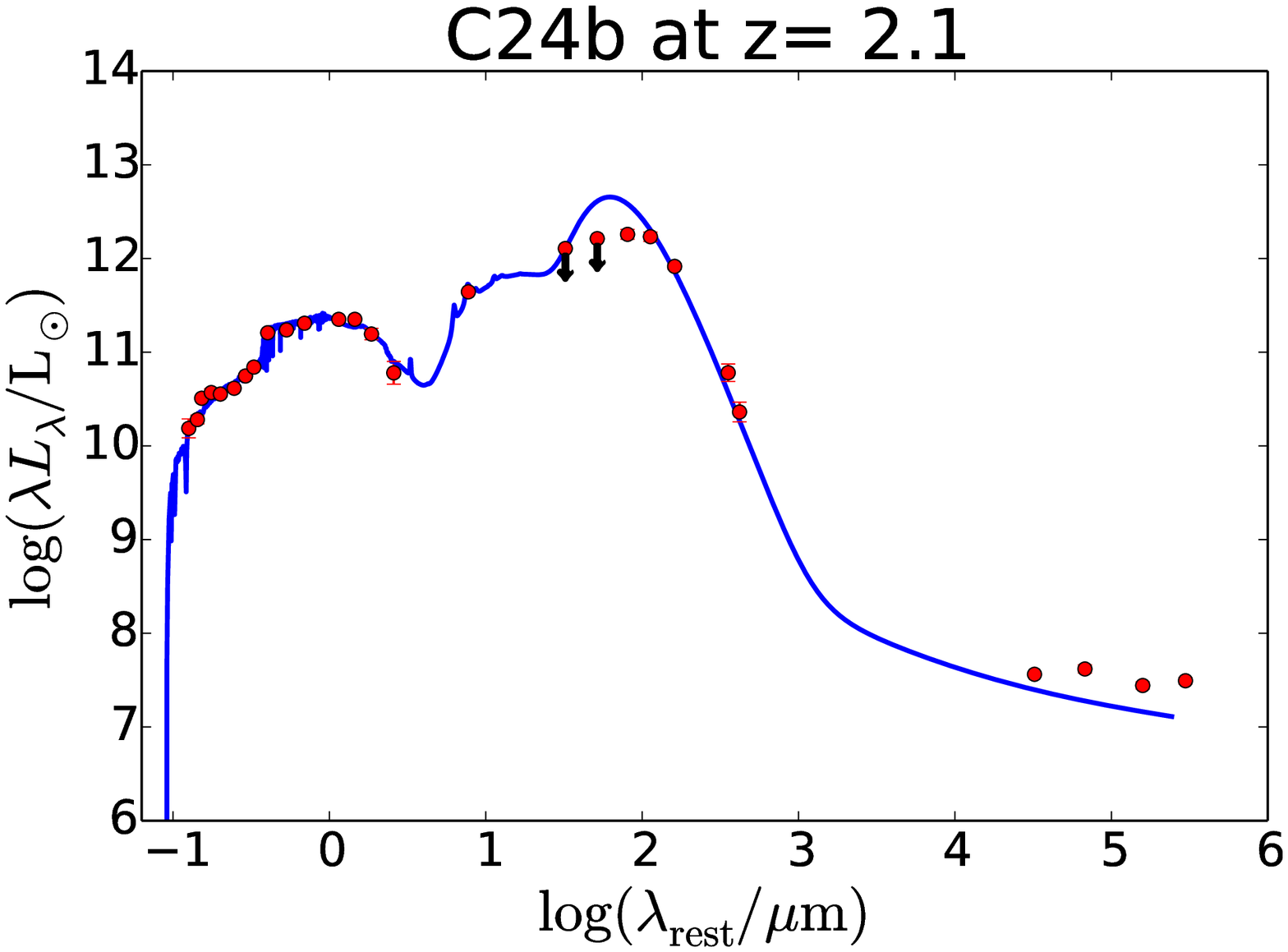}
\includegraphics[width=0.2465\textwidth]{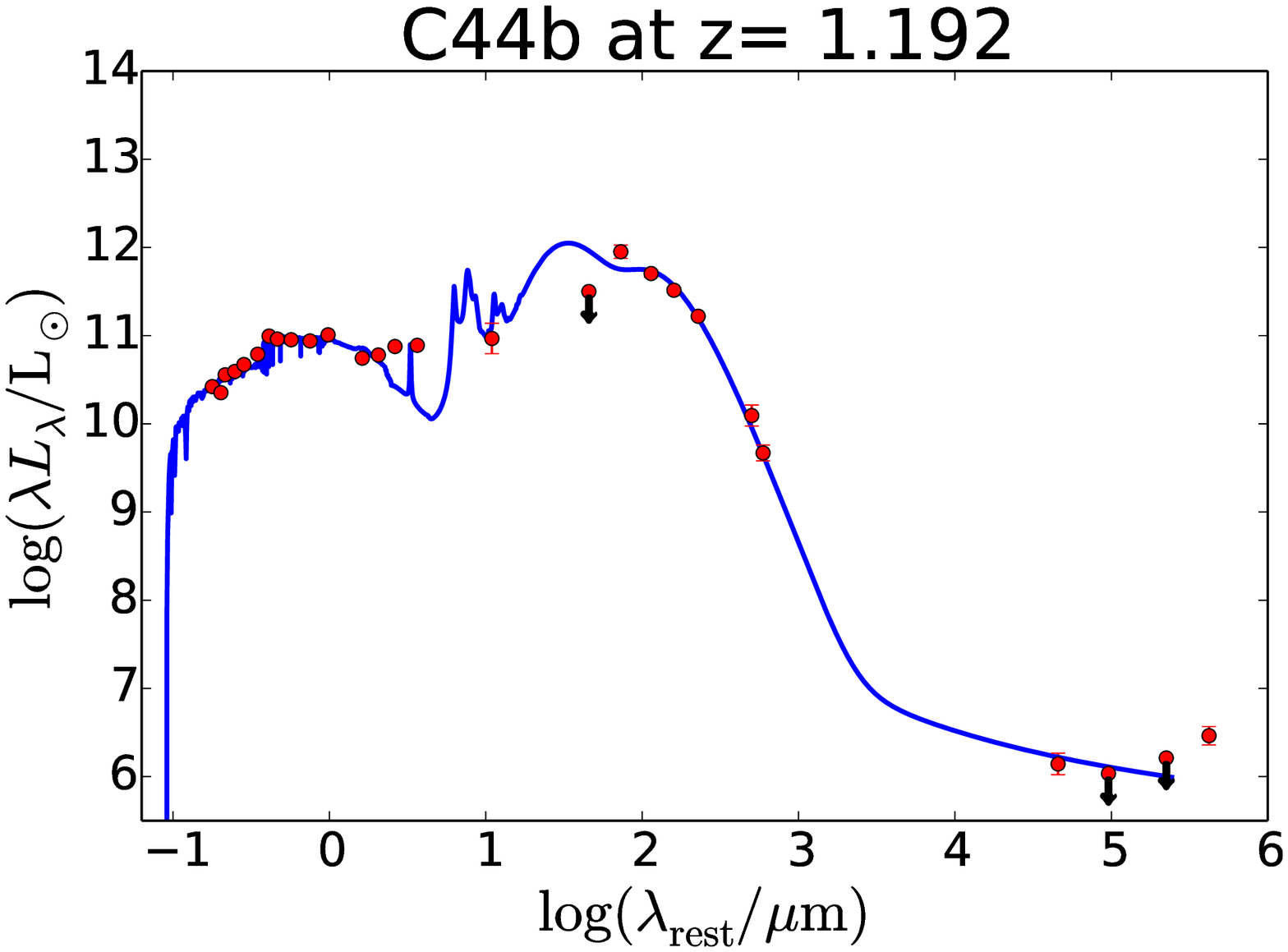}
\includegraphics[width=0.2465\textwidth]{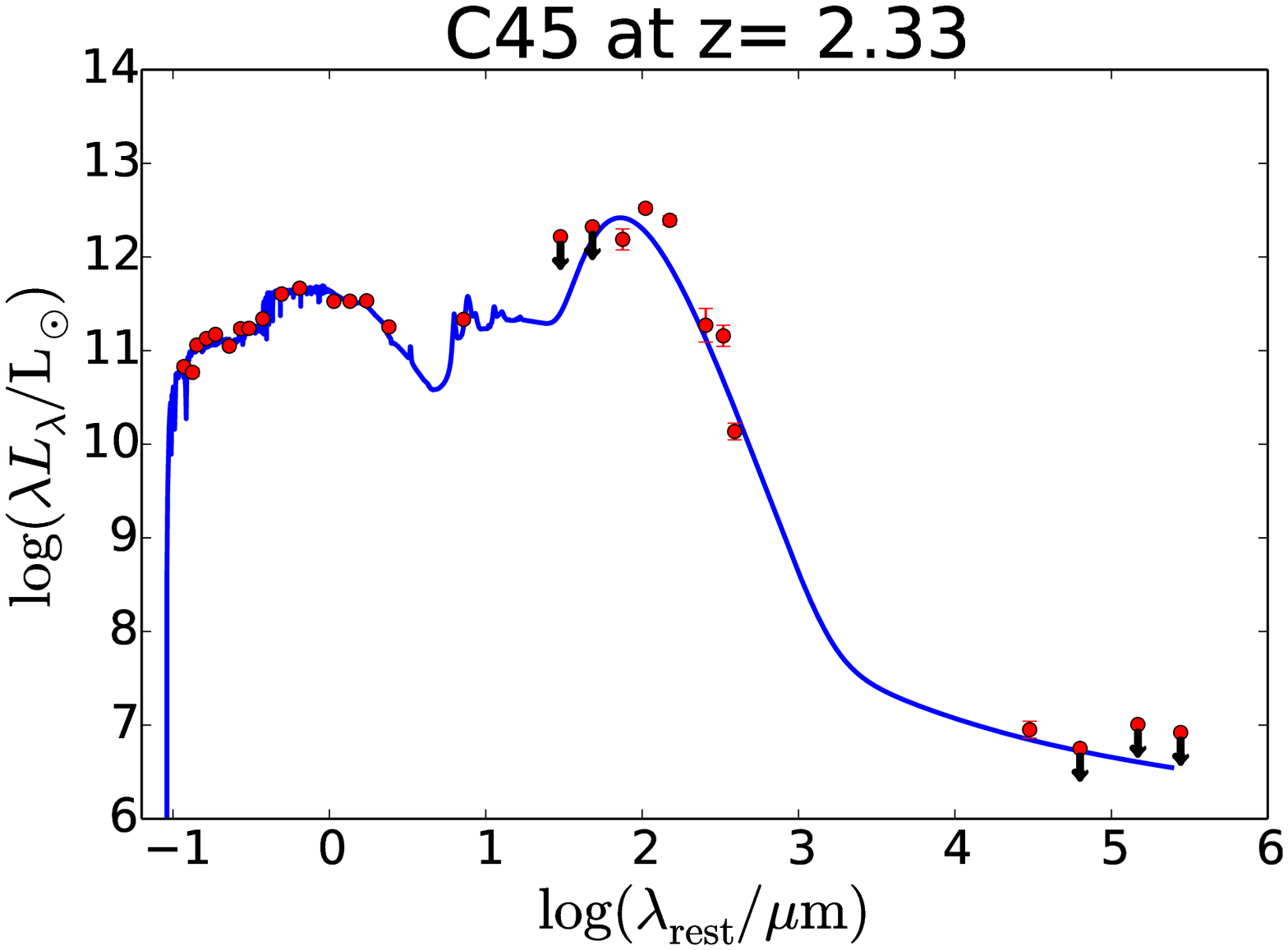}
\includegraphics[width=0.2465\textwidth]{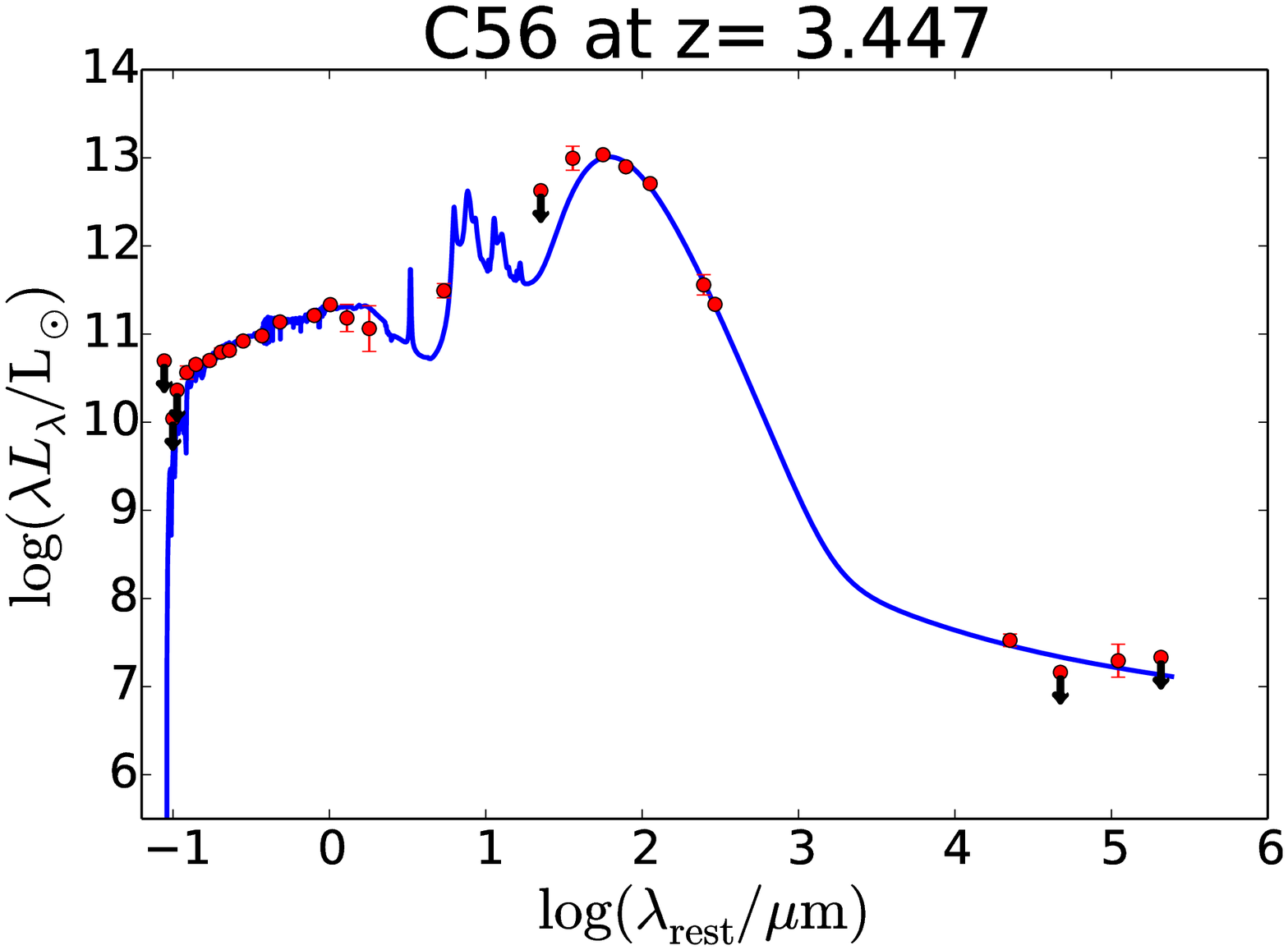}
\includegraphics[width=0.2465\textwidth]{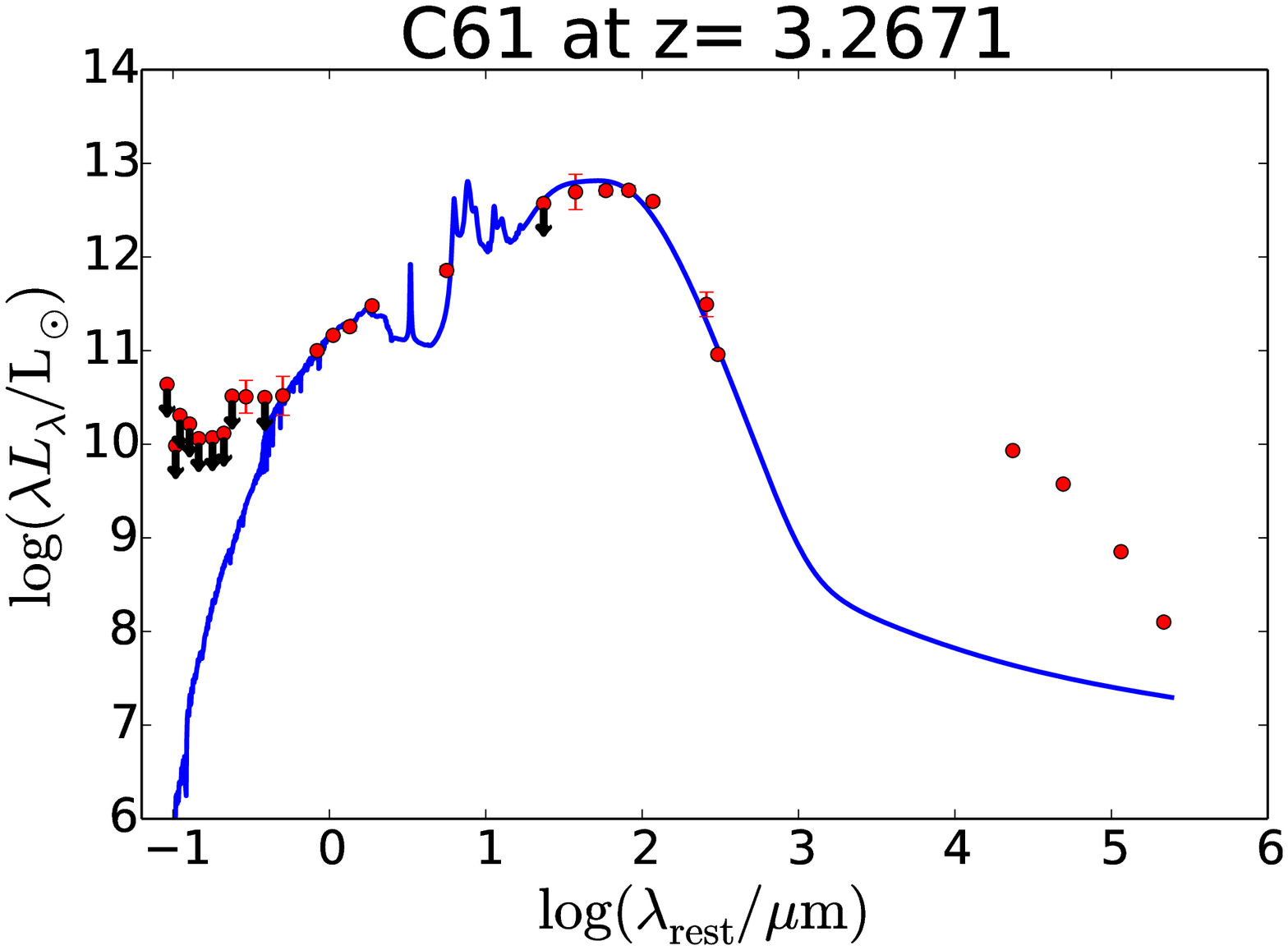}
\includegraphics[width=0.2465\textwidth]{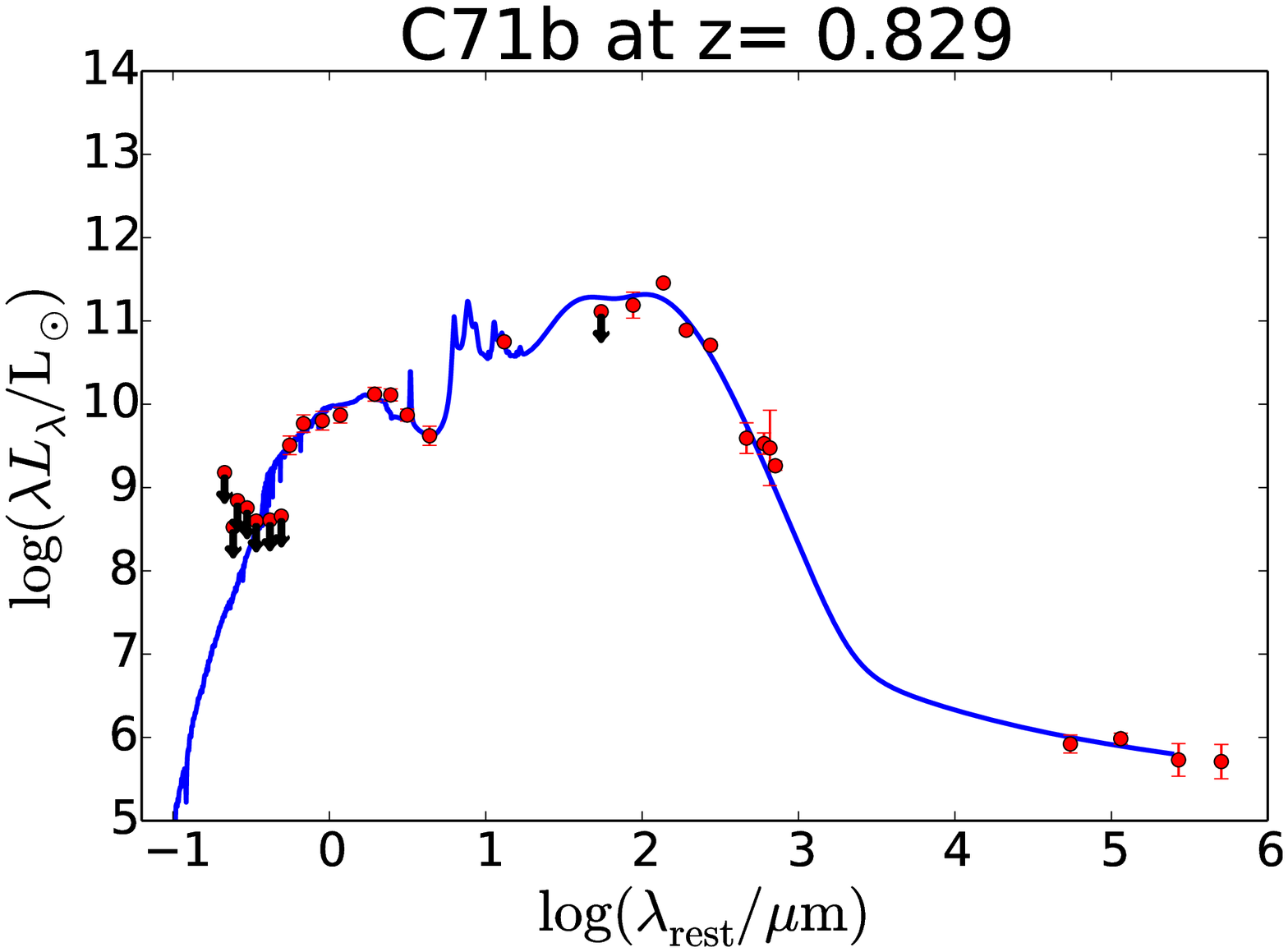}
\includegraphics[width=0.2465\textwidth]{SED_plots/C77a_sed.eps}
\includegraphics[width=0.2465\textwidth]{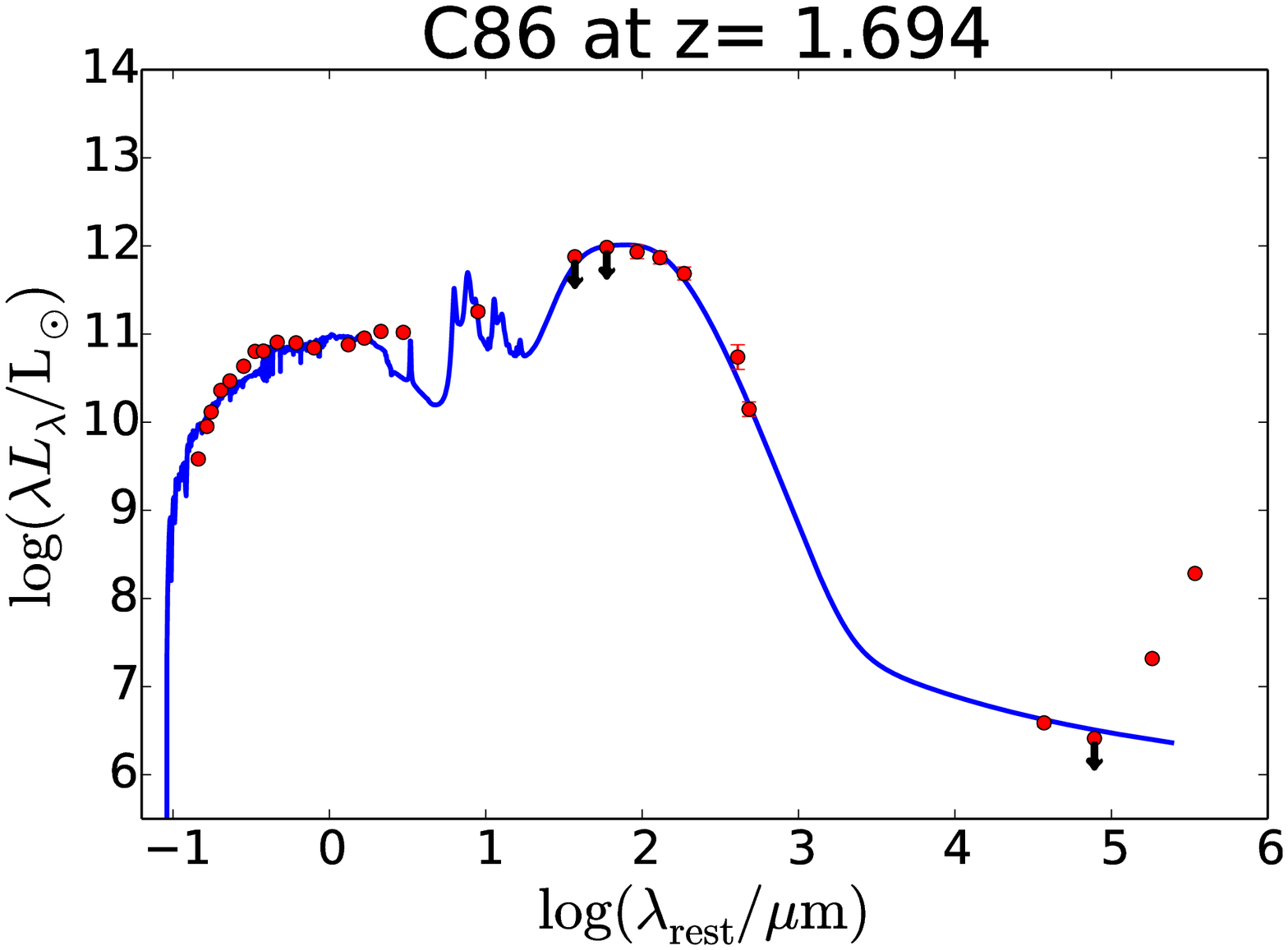}
\includegraphics[width=0.2465\textwidth]{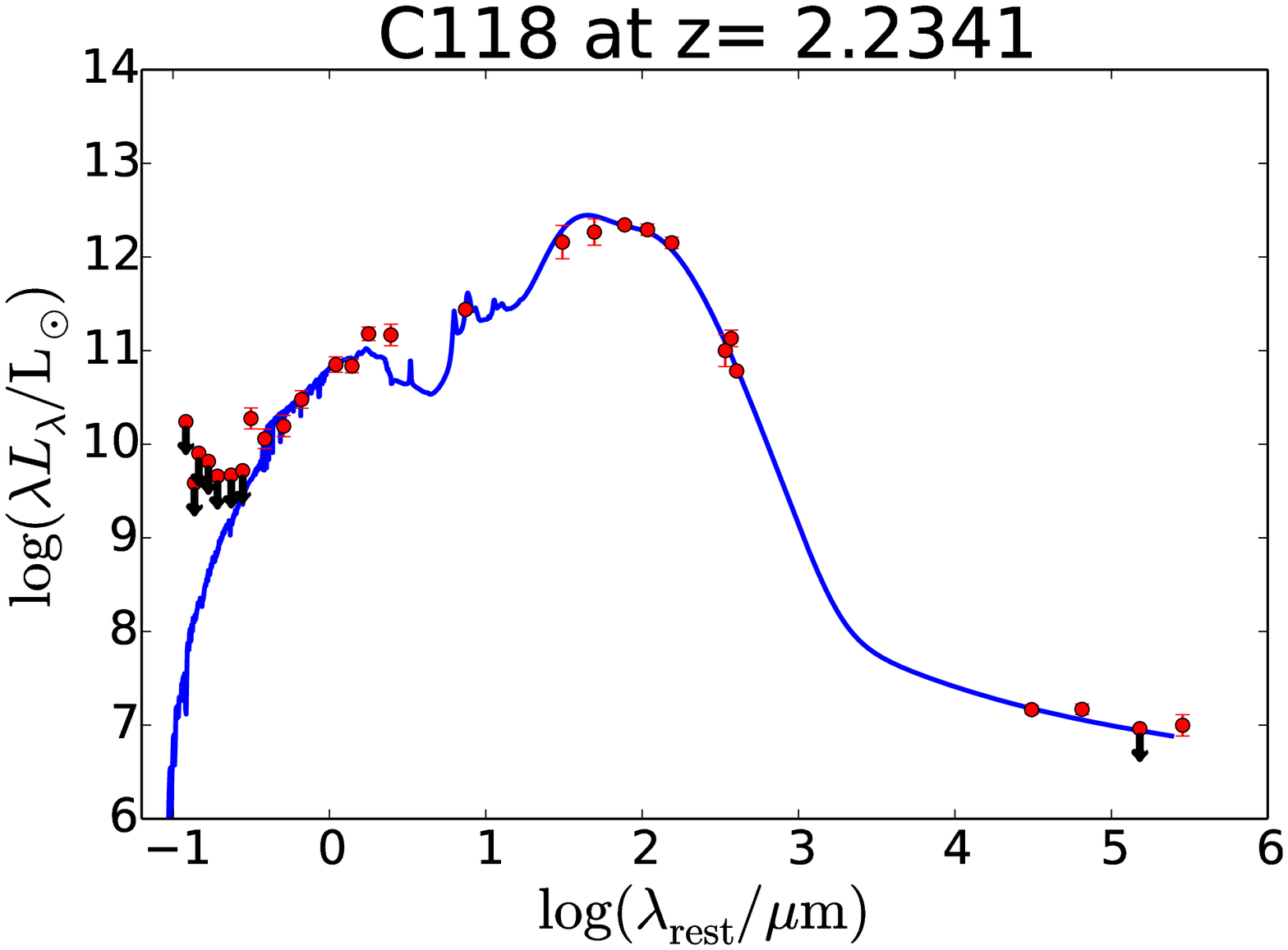}
\caption{Best-fit panchromatic (UV--radio) rest-frame SEDs of our SMGs that are likely to host an AGN. All these sources except AzTEC/C24b and C77a are detected in X-rays. However, AzTEC/C24b and C77a are detected with the VLBA at 1.4~GHz with flux densities of about 0.13~mJy and 0.33~mJy, respectively, and hence both of them are likely to host a radio-emitting AGN.}
\label{figure:seds2}
\end{center}
\end{figure*}

\section{Photometric tables}

A selection of mid-IR to radio (from 24~$\mu$m to 325~MHz) flux densities of our SMGs is provided in Table~\ref{table:fluxes}. 

\begin{landscape}
\begin{table}
\caption{Flux densities at mid-IR to radio wavelengths.}
{\scriptsize
\centering
\renewcommand{\footnoterule}{}
\label{table:fluxes}
\begin{tabular}{c c c c c c c c c c c c c c c}
\hline\hline 
Source ID & $S_{\rm 24\, \mu m}$ & $S_{\rm 100\, \mu m}$ & $S_{\rm 160\, \mu m}$ & $S_{\rm 250\, \mu m}$ & $S_{\rm 350\, \mu m}$ & $S_{\rm 450\, \mu m}$ & $S_{\rm 500\, \mu m}$ & $S_{\rm 850\, \mu m}$ & $S_{\rm 870\, \mu m}$ & $S_{\rm 890\, \mu m}$ & $S_{\rm 1.1\, mm}$ & $S_{\rm 1.3\, mm}$ & $S_{\rm 325\, MHz}$ & $S_{\rm 1.4\, GHz}$\\
          & [mJy] & [mJy] & [mJy] &  [mJy] & [mJy] & [mJy] & [mJy] & [mJy] & [mJy] & [mJy] & [mJy] & [mJy] & [$\mu$Jy] & [$\mu$Jy]\\

         & MIPS & PACS & PACS & SPIRE & SPIRE & SCUBA-2\tablefootmark{a} & SPIRE & SCUBA-2\tablefootmark{a} & LABOCA\tablefootmark{b} & SMA\tablefootmark{c} & AzTEC\tablefootmark{d} & ALMA\tablefootmark{e} & GMRT\tablefootmark{f} & VLA\tablefootmark{f}\\
\hline
C1a & $0.241\pm0.023$ & $<5.0$ & $19.9\pm3.5$ & $23.4\pm1.8$ & $27.0\pm2.3$ & \ldots & $28.5\pm2.6$ & \ldots & $11.5\pm3.4$ & \ldots & $12.1^{+1.0}_{-0.9}$ & $7.24\pm0.10$ & $<234$ & $<36$ \\
C2a &  $<0.054$ & $6.4\pm1.3$ & $30.0\pm3.5$ & $60.3\pm2.4$ & $58.7\pm3.3$ & \ldots & $35.4\pm2.1$ & \ldots & $12.3\pm3.6$ & $21.6\pm2.3$ & $5.9^{+0.6}_{-0.5}$ & $4.07\pm0.15$ & $540.0\pm86.6$ & $102\pm13$\\
C2b & $<0.054$ & $<5.0$ & $<10.2$ & $4.6\pm2.8$ & $<10.7$ & \ldots & $32.3\pm2.0$ & \ldots & \ldots & \ldots & $5.3^{+0.5}_{-0.5}$ & $3.71\pm0.15$ & $<234$ & $<36$ \\
C4 & $<0.054$ & $<5.0$ & $<10.2$ & $11.0\pm2.2$ & $19.3\pm3.0$ & \ldots & $30.5\pm2.8$ & \ldots & \ldots & $14.4\pm1.9$ & $10.5^{+1.0}_{-1.1}$ & $3.87\pm0.10$ & $<234$ & $<36$\\
C5 & $<0.054$ & $<5.0$ & $<10.2$ & $24.1\pm1.8$ & $31.0\pm2.9$ & \ldots & $35.9\pm3.4$ & \ldots & $14.12\pm0.25$ & $15.6\pm1.1$ & $10.0^{+1.1}_{-1.1}$ & $4.40\pm0.11$ & $248.6\pm72.5$ & $48\pm12$\\ 
C6a & $0.288\pm0.022$ & $<5.0$ & $20.8\pm2.6$ & $30.3\pm1.7$ & $77.3\pm2.0$ & \ldots & $67.5\pm2.3$ & \ldots & $5.26\pm0.26$ & \ldots & $6.9^{+0.8}_{-0.7}$ & $2.20\pm0.14$ & $<234$ & $77.8\pm13.0$  \\ 
C6b & $0.216\pm0.033$ & $<5.0$ & $17.5\pm2.8$ & $22.4\pm1.7$ & $30.7\pm0.8$ & \ldots & $26.8\pm0.9$ & \ldots & $3.77\pm0.32$ & \ldots & $2.7^{+0.3}_{-0.3}$ & $0.88\pm0.11$ & $287.4\pm65.5$ & $<36$     \\
C7 & $<0.054$ & $<5.0$ & $<10.2$ & $17.7\pm2.0$ & $27.6\pm3.3$ & $10.67\pm5.79$\tablefootmark{g} & $27.4\pm3.1$ & $11.09\pm1.56$ & $13.8\pm1.5$ & \ldots & $8.9^{+1.1}_{-1.1}$ & $4.20\pm0.11$ & $<234$ & $<36$ \\
C8a & $<0.054$ & $<5.0$ & $<10.2$ & $17.0\pm1.5$ & $35.5\pm2.2$ & \ldots & $28.4\pm2.1$ & \ldots & \ldots & \ldots & $5.3^{+0.7}_{-0.7}$ & $0.96\pm0.11$ & $<234$ & $<36$   \\
C8b & $0.463\pm0.114$ & $<5.0$ & $<10.2$ & $10.8\pm1.0$ & $22.8\pm1.4$ & \ldots & $18.2\pm1.4$ & \ldots & $5.72\pm1.31$ & \ldots & $3.5^{+0.4}_{-0.4}$ & $0.61\pm0.12$ & $<234$ & $<36$     \\
C9a & $0.105\pm0.016$ & $<5.0$ & $<10.2$ & $22.3\pm1.2$ & $26.6\pm1.4$ & \ldots & $30.6\pm1.3$ & \ldots & \ldots & \ldots & $4.5^{+0.6}_{-0.6}$ & $2.82\pm0.12$ & $<216.4$ & $76.9\pm15.9$    \\ 
C9b & $0.074\pm0.016$ & $<5.0$ & $<10.2$ & $9.8\pm0.5$ & $11.7\pm0.6$ & \ldots & $13.5\pm0.6$ & \ldots  & \ldots & \ldots & $2.0^{+0.2}_{-0.3}$ & $1.24\pm0.24$ & $466.7\pm72.1$ & $142.8\pm13.6$ \\ 
C9c & $<0.054$ & $<5.0$ & $<10.2$ & $7.6\pm0.4$ & $9.1\pm0.5$ & \ldots & $10.5\pm0.4$ & \ldots & \ldots & \ldots & $1.6^{+0.2}_{-0.2}$ & $0.97\pm0.11$ & $209.4\pm72.1$ & $<36$ \\
C10a & $<0.054$ & $<5.0$ & $<10.2$ & $<8.1$ & $6.4\pm2.9$ & \ldots & $12.4\pm1.3$ & \ldots & \ldots & \ldots & $3.3^{+0.4}_{-0.4}$ & $2.01\pm0.10$ & $<234$ & $<36$\\
C10b & $0.094\pm0.016$ & $<5.0$ & $<10.2$ & $6.3\pm1.4$ & $10.0\pm1.8$ & \ldots & $10.4\pm1.1$ & \ldots & $5.8\pm1.7$ & \ldots & $3.7^{+0.5}_{-0.5}$ & $1.70\pm0.15$ & $<234$ & $<32$\\
C11\tablefootmark{h} & $0.208\pm0.017$ & $<5.0$ & $12.2\pm4.0$ & $33.0\pm1.9$ & $41.3\pm2.4$ & \ldots & $50.4\pm3.1$ & \ldots & \ldots & \ldots & $7.9^{+1.1}_{-1.1}$ & $3.97\pm0.14$ & $<234$ & $<36$    \\
C12 & $<0.054$ & $<5.0$ & $<10.2$ & $6.7\pm2.4$ & $15.7\pm3.5$ & \ldots & $18.2\pm3.1$ & \ldots & $12.5\pm3.2$ & \ldots & $7.5^{+1.0}_{-1.1}$ & $2.86\pm0.12$ & $216.6\pm70.1$ & $<36$ \\
C13a & $0.327\pm0.041$ & $<5.0$ & $10.9\pm3.9$ & $28.9\pm1.1$ & $37.3\pm1.7$ & \ldots & $35.1\pm1.7$ & \ldots & $8.53\pm3.01$ & \ldots & $6.4^{+1.0}_{-1.0}$ & $3.34\pm0.10$ & $329.1\pm71.8$ & $144.2\pm13.3$  \\ 
C13b & $<0.054$ & $<5.0$ & $<10.2$ & $10.2\pm0.4$ & $13.2\pm0.6$ & \ldots & $12.4\pm0.6$ & \ldots & \ldots & \ldots & $2.3^{+0.3}_{-0.4}$ & $1.18\pm0.12$ & $<215.4$ & $<36$    \\
C14 & $<0.054$ & $<5.0$ & $<10.2$ & $11.8\pm2.6$ & $19.8\pm3.1$ & $11.02\pm4.75$\tablefootmark{g} & $26.7\pm3.4$ & $11.49\pm1.10$ & $16.4\pm3.3$ & $7.4\pm3.0$ & $6.7^{+1.1}_{-1.1}$ & $5.01\pm0.10$ & $<234$ & $68\pm13$  \\
C15 & $<0.054$ & $<5.0$ & $<10.2$ & $40.6\pm2.2$ & $33.5\pm2.5$ & \ldots & $20.7\pm2.3$ & \ldots & \ldots & \ldots & $6.5^{+1.1}_{-1.1}$ & $3.24\pm0.13$ & $<234$ & $<36$      \\
C16a & $0.072\pm0.017$ & $<5.0$ & $13.4\pm3.9$ & $15.9\pm1.7$ & $18.1\pm1.5$ & \ldots & $15.9\pm1.4$ & \ldots & \ldots & \ldots & $3.5^{+0.6}_{-0.6}$ & $1.54\pm0.12$ & $<234$ & $<36$\\
C16b & $0.187\pm0.013$ & $<5.0$ & $11.1\pm3.6$ & $23.2\pm1.7$ & $16.1\pm1.4$ & \ldots & $14.2\pm1.2$ & \ldots & \ldots & \ldots & $3.2^{+0.5}_{-0.5}$ & $1.37\pm0.11$ & $<234$ & $82.1\pm13.8$    \\
C17 & $<0.054$ & $<5.0$ & $<10.2$ & $10.5\pm2.7$ & $<10.7$ & \ldots & $22.5\pm3.1$ & \ldots & \ldots & \ldots & $6.2^{+1.1}_{-1.1}$ & $3.10\pm0.16$ & $<234$ & $<36$    \\
C18 & $0.357\pm0.016$ & $<5.0$ & $23.1\pm2.9$ & $54.9\pm2.4$ & $62.0\pm3.8$ & \ldots & $51.9\pm4.1$ & \ldots & \ldots & $12.8\pm2.9$ & $7.9^{+1.4}_{-1.6}$ & $3.69\pm0.10$ & $254.3\pm75.8$ & $98\pm16$ \\
C19 & $0.073\pm0.015$ & $<5.0$ & $11.0\pm4.1$ & $18.2\pm2.5$ & $19.3\pm2.4$ & \ldots & $19.1\pm2.8$ & \ldots & \ldots & \ldots & $5.9^{+1.1}_{-1.1}$ & $2.44\pm0.21$ & $262.3\pm67.4$ & $<36$ \\
C20 & $<0.054$ & $<5.0$ & $11.9\pm2.7$ & $33.0\pm2.9$ & $44.0\pm3.6$ & \ldots & $33.5\pm2.2$ & \ldots & \ldots & \ldots & $5.7^{+1.2}_{-1.0}$ & $1.62\pm0.10$ & $212.8\pm64.5$ & $<36$ \\
C21 & $0.162\pm0.015$ & $<5.0$ & $<10.2$ & $20.4\pm1.7$ & $26.7\pm2.2$ & \ldots & $24.4\pm2.7$ & \ldots & \ldots & \ldots & $5.9^{+1.0}_{-1.2}$ & $2.19\pm0.11$ & $218.9\pm75.2$ & $<36$     \\
C22a & $0.660\pm0.017$ & $7.8\pm0.9$ & $35.7\pm2.1$ & $56.0\pm1.7$ & $53.9\pm1.4$ & \ldots & $36.7\pm1.9$ & \ldots & \ldots & $4.4\pm2.1$ & $4.7^{+0.9}_{-1.0}$ & $2.22\pm0.10$ & $204.1\pm86.6$ & $132\pm26$  \\
C22b & $<0.054$ & $2.8\pm0.3$ & $12.7\pm0.7$ & $19.9\pm0.6$ & $19.2\pm0.5$ & \ldots & $13.1\pm0.7$ & \ldots & \ldots & $10.0\pm2.1$  & $1.7^{+0.3}_{-0.3}$ & $0.79\pm0.11$ & $213.4\pm86.6$ & $138\pm26$ \\
C23 & $0.463\pm0.019$ & $6.0\pm1.5$ & $30.5\pm3.6$ & $48.3\pm1.6$ & $49.9\pm2.2$ & \ldots & $45.2\pm2.2$ & \ldots & \ldots & \ldots & $5.7^{+1.0}_{-1.1}$ & $1.20\pm0.11$ & $407.3\pm71.1$ & $124.5\pm13.9$     \\
C24a & $0.381\pm0.014$ & $<5.0$ & $12.6\pm3.8$ & $27.4\pm2.1$ & $30.5\pm2.5$ & $12.91\pm4.73$ & $20.6\pm1.4$ & $5.55\pm1.11$ & \ldots & \ldots & $3.1^{+0.6}_{-0.6}$ & $1.50\pm0.10$ & $183.9\pm55.9$ & $99.6\pm12.6$   \\
C24b\tablefootmark{h} & $0.414\pm0.015$ & $<5.0$ & $<10.2$ & $17.7\pm2.0$ & $23.4\pm2.5$ & \ldots & $16.1\pm1.1$ & \ldots & \ldots & \ldots & $2.6^{+0.5}_{-0.5}$ & $1.17\pm0.25$ & $1\,126.5\pm68.0$ & $342.0\pm32.4$  \\
C25 & $0.366\pm0.028$ & $<5.0$ & $17.2\pm4.1$ & $41.4\pm2.0$ & $39.8\pm2.5$ & \ldots & $32.5\pm2.1$ & \ldots & \ldots & \ldots & $5.7^{+1.1}_{-1.1}$ & $1.30\pm0.17$ & $<234$ & $70.2\pm12.2$ \\
C26 & $<0.054$ & $<5.0$ & $<10.2$ & $27.8\pm2.1$ & $28.7\pm3.0$ & \ldots & $34.1\pm3.1$ & \ldots & \ldots & \ldots & $5.7^{+1.2}_{-1.1}$ & $1.93\pm0.11$ & $<234$ & $<36$  \\ 
C27 & $0.140\pm0.017$ & $<5.0$ & $<10.2$ & $12.4\pm1.5$ & $17.2\pm2.2$ & \ldots & $19.8\pm3.1$ & \ldots & \ldots & \ldots & $5.7^{+1.1}_{-1.2}$ & $2.11\pm0.23$ & $<234$ & $<36$\\
C28a & $0.201\pm0.053$ & $<5.0$ & $13.6\pm3.8$ & $25.7\pm1.2$ & $32.4\pm1.6$ & \ldots & $26.4\pm1.5$ & \ldots & \ldots & \ldots & $4.0^{+0.8}_{-0.8}$ & $1.69\pm0.13$ & $1\,161.4\pm98.3$ & $382.0\pm52.1$ \\
C28b & $0.229\pm0.015$ & $<5.0$ & $10.6\pm3.8$ & $12.2\pm0.6$ & $15.4\pm0.8$ & \ldots & $12.5\pm0.7$ & \ldots & \ldots & \ldots & $1.9^{+0.4}_{-0.4}$ & $0.80\pm0.10$ & $<294.8$ & $<36$\\
C29 & $0.129\pm0.015$ & $<5.0$ & $<10.2$ & $18.9\pm2.9$ & $22.0\pm2.7$ & \ldots & $21.3\pm3.1$ & \ldots & \ldots & \ldots & $5.6^{+1.1}_{-1.1}$ & $1.32\pm0.10$ & $272.5\pm65.8$ & $132.8\pm11.6$ \\
C31a  & $<0.054$ & $<5.0$ & $17.2\pm3.2$ & $18.1\pm1.0$ & $18.5\pm1.6$ & \ldots & $18.0\pm1.5$ & \ldots & \ldots & \ldots & $3.1^{+0.6}_{-0.6}$ & $1.28\pm0.12$ & $201.4\pm67.6$ & $<36$ \\
C31b & $<0.054$ & $<5.0$ & $<10.2$ & $13.2\pm0.8$ & $13.5\pm1.2$ & \ldots & $13.2\pm1.1$ & \ldots & \ldots & \ldots & $2.3^{+0.5}_{-0.5}$ & $0.93\pm0.09$ & $<234$ & $<36$   \\
C32 & $0.163\pm0.016$ & $5.4\pm1.5$ & $10.7\pm3.6$ & $33.4\pm1.9$ & $36.3\pm3.1$ & \ldots & $26.9\pm3.6$ & \ldots & \ldots & \ldots & $5.3^{+1.1}_{-1.1}$ & $1.94\pm0.11$ & $356.2\pm 70.8$ & $55.9\pm10.9$ \\
C33a & $0.280\pm0.113$ & $13.1\pm1.5$ & $26.7\pm2.9$ & $42.2\pm1.7$ & $32.1\pm2.7$ & $10.34\pm5.10$\tablefootmark{g} & $19.4\pm1.8$ & $5.15\pm1.28$ & $3.02\pm0.86$ & \ldots & $2.9^{+0.6}_{-0.6}$ & $1.40\pm0.15$ & $309.6\pm74.8$ & $67.3\pm11.7$\\
C34a & $0.094\pm0.037$ & $<5.0$ & $<10.2$ & $31.2\pm1.6$ & $36.3\pm3.8$ &\ldots & $26.1\pm2.4$ & \ldots & $3.46\pm0.93$ & \ldots & $2.7^{+0.6}_{-0.6}$ & $1.80\pm0.11$ & $193.3\pm56.3$ & $<36$ \\
C34b & $0.234\pm0.015$ & $<5.0$ & $<10.2$ & $14.0\pm1.8$ & $18.0\pm3.8$ & \ldots & $18.3\pm2.5$ & \ldots & $3.24\pm0.90$ & \ldots & $2.6^{+0.5}_{-0.6}$ & $1.58\pm0.28$ & $<234$ & $<36$\\
C35 & $<0.054$ & $<5.0$ & $<10.2$ & $13.6\pm1.8$ & $20.6\pm2.5$ & $4.68\pm4.23$\tablefootmark{g} & $35.0\pm2.6$ & $3.25\pm1.00$ & \ldots & \ldots & $5.2^{+1.1}_{-1.1}$ & $1.48\pm0.14$ & $<234$ & $62.0\pm12.0$  \\
C36 &  $0.451\pm0.018$ & $7.3\pm1.8$ & $22.0\pm3.0$ & $54.1\pm1.7$ & $59.8\pm2.2$ & \ldots & $54.8\pm2.1$ & \ldots & \ldots & \ldots & $6.8^{+1.5}_{-1.6}$ & $3.22\pm0.10$ & $417.7\pm83.6$ & $167.6\pm14.9$\\
C37 & $<0.054$ & $<5.0$ & $<10.2$ & $5.6\pm3.3$ & $11.5\pm2.6$ & \ldots & $11.9\pm2.0$ & \ldots & \ldots & \ldots & $5.1^{+1.1}_{-1.1}$ & $3.60\pm0.13$ & $185.3\pm77.9$ & $52.1\pm8.0$\\ 
C38 &  $0.236\pm0.016$ & $6.5\pm1.5$ & $<10.2$ & $21.4\pm2.3$ & $30.7\pm2.8$ & $20.12\pm4.80$ & $29.7\pm3.0$ & $6.60\pm1.12$ & $8.2\pm2.2$ & \ldots & $5.1^{+1.2}_{-1.1}$ & $2.20\pm0.12$ & $<234$ & $43.0\pm11.1$ \\
C39 &  $0.249\pm0.016$ & $5.6\pm1.6$ & $18.9\pm3.2$ & $35.0\pm2.1$ & $36.0\pm3.2$ & \ldots & $25.3\pm2.7$ & \ldots & \ldots & \ldots & $5.1^{+1.1}_{-1.1}$ & $1.66\pm0.11$ & $<234$ & $<36$  \\ 
C41 &  $0.274\pm0.017$ & $<5.0$ & $<10.2$ & $17.9\pm1.6$ & $23.4\pm2.1$ & \ldots & $18.7\pm2.6$ & \ldots & \ldots & \ldots & $4.9^{+1.1}_{-1.1}$ & $1.83\pm0.11$ & $<234$ & $<36$ \\
C42 &  $0.189\pm0.013$ & $<5.0$ & $<10.2$ & $52.3\pm1.8$ & $54.9\pm2.6$ & $25.35\pm6.04$ & $40.3\pm3.1$ & $11.42\pm1.38$ & $9.3\pm1.3$ & \ldots & $4.8^{+1.1}_{-1.1}$ & $2.39\pm0.10$ & $440.8\pm76.3$ & $126\pm15$ \\
C43a &  $0.200\pm0.029$ & $8.6\pm1.2$ & $15.7\pm2.2$ & $21.6\pm0.9$ & $14.7\pm2.2$ & \ldots & $21.1\pm2.4$ & \ldots & \ldots & \ldots & $2.8^{+0.7}_{-0.6}$ & $0.93\pm0.13$ & $283.6\pm62.5$ & $190.0\pm47.9$ \\
C43b &  $0.250\pm0.025$ & $<5.0$ & $10.8\pm1.6$ & $15.5\pm0.7$ & $10.6\pm1.6$ & \ldots & $15.2\pm1.7$ & \ldots & \ldots & \ldots & $2.0^{+0.5}_{-0.5}$ & $0.67\pm0.10$ & $<187.5$ & $<36$\\
C44a &  $0.251\pm0.017$ & $<5.0$ & $<10.2$ & $19.8\pm1.8$ & $24.8\pm1.7$ & \ldots & $18.0\pm2.2$ & \ldots & \ldots & \ldots & $2.8^{+0.7}_{-0.7}$ & $1.31\pm0.13$ & $<234$ & $<36$\\
C44b\tablefootmark{h} & $0.352\pm0.115$ & $<5.0$ & $22.6\pm3.6$ & $20.0\pm1.8$ & $18.1\pm1.2$ & \ldots & $13.1\pm1.6$ & \ldots & \ldots & \ldots & $2.1^{+0.5}_{-0.5}$ & $0.96\pm0.18$ & $424.3\pm90.4$ & $<36$ \\
C45\tablefootmark{h} & $0.157\pm0.013$ & $<5.0$ & $<10.2$ & $11.7\pm2.7$ & $35.2\pm3.2$ & $3.66\pm7.17$\tablefootmark{g} & $37.4\pm4.0$ & $4.82\pm1.63$ &  \ldots & \ldots & $4.8^{+1.1}_{-1.1}$ & $0.54\pm0.10$ & $<234$ & $<36$ \\
\hline
\end{tabular} }
\end{table}
\end{landscape}

\addtocounter{table}{-1}

\begin{landscape}
\begin{table}
\caption{continued.}
{\scriptsize
\centering
\renewcommand{\footnoterule}{}
\label{table:fluxes}
\begin{tabular}{c c c c c c c c c c c c c c c}
\hline\hline 
Source ID & $S_{\rm 24\, \mu m}$ & $S_{\rm 100\, \mu m}$ & $S_{\rm 160\, \mu m}$ & $S_{\rm 250\, \mu m}$ & $S_{\rm 350\, \mu m}$ & $S_{\rm 450\, \mu m}$ & $S_{\rm 500\, \mu m}$ & $S_{\rm 850\, \mu m}$ & $S_{\rm 870\, \mu m}$ & $S_{\rm 890\, \mu m}$ & $S_{\rm 1.1\, mm}$ & $S_{\rm 1.3\, mm}$ & $S_{\rm 325\, MHz}$ & $S_{\rm 1.4\, GHz}$\\
          & [mJy] & [mJy] & [mJy] &  [mJy] & [mJy] & [mJy] & [mJy] & [mJy] & [mJy] & [mJy] & [mJy] & [mJy] & [$\mu$Jy] & [$\mu$Jy]\\

         & MIPS & PACS & PACS & SPIRE & SPIRE & SCUBA-2\tablefootmark{a} & SPIRE & SCUBA-2\tablefootmark{a} & LABOCA\tablefootmark{b} & SMA\tablefootmark{c} & AzTEC\tablefootmark{d} & ALMA\tablefootmark{e} & GMRT\tablefootmark{f} & VLA\tablefootmark{f}\\
\hline
C46 & $0.184\pm0.016$ & $<5.0$ & $9.4\pm3.3$ & $26.5\pm1.6$ & $27.1\pm2.5$ & \ldots & $31.2\pm2.9$ & \ldots & \ldots & \ldots & $4.8^{+1.2}_{-1.1}$ & $1.65\pm0.12$ & $169.1\pm63.8$ & $122.0\pm12.4$  \\
C47 & $0.153\pm0.013$ & $5.5\pm1.3$ & $8.6\pm3.5$ & $21.7\pm1.8$ & $17.5\pm2.2$ & \ldots & $10.4\pm2.9$ & \ldots & \ldots & \ldots & $4.8^{+1.1}_{-1.1}$ & $1.11\pm0.14$ & $527.9\pm72.1$ & $330.0\pm32.4$      \\
C48a & $0.147\pm0.015$ & $<5.0$ & $<10.2$ & $12.7\pm1.5$ & $13.0\pm1.8$ & \ldots & $14.7\pm2.1$ & \ldots & \ldots & \ldots & $3.5^{+0.8}_{-0.9}$ & $1.37\pm0.10$ & $229.4\pm93.6$ & $63\pm13$ \\
C48b & $<0.054$ & $<5.0$ & $<10.2$ & $4.8\pm0.6$ & $4.9\pm0.7$ & \ldots & $5.6\pm0.8$ & \ldots & \ldots & \ldots & $1.4^{+0.3}_{-0.3}$ & $0.52\pm0.11$ & $<280.8$ & $<36$   \\
C49 & $<0.054$ &$<5.0$ & $<10.2$ & $9.0\pm2.0$ & $33.5\pm2.6$ & \ldots & $26.5\pm3.5$ & \ldots & \ldots & \ldots & $5.3^{+1.2}_{-1.3}$ & $1.93\pm0.17$ & $<234$ & $<36$   \\
C50 & $0.112\pm0.018$ & $<5.0$ & $<10.2$ & $20.7\pm1.4$ & $21.8\pm1.9$ & \ldots & $20.2\pm3.0$ & \ldots & \ldots & \ldots & $4.8^{+1.1}_{-1.2}$ & $1.87\pm0.10$ & $179.0\pm61.6$ & $<36$   \\
C51b & $<0.054$ & $<5.0$ & $<10.2$ & $6.8\pm1.6$ & $4.9\pm0.8$ & \ldots & $8.7\pm1.0$ & \ldots & \ldots & \ldots & $1.4^{+0.3}_{-0.3}$ & $0.55\pm0.10$ & $<234$ & $<36$   \\
C52 & $0.385\pm0.036$ & $12.9\pm1.6$ & $37.2\pm3.8$ & $53.5\pm2.7$ & $45.5\pm3.0$ & \ldots & $23.0\pm3.0$ & \ldots & \ldots & \ldots & $4.7^{+1.1}_{-1.1}$ & $1.15\pm0.12$ & $394.2\pm74.8$ & $119.1\pm15.4$      \\
C53 & $0.158\pm0.017$ & $<5.0$ & $<10.2$ & $24.2\pm2.3$ & $34.5\pm2.6$ & \ldots & $28.5\pm2.9$ & \ldots & \ldots & \ldots & $4.6^{+1.1}_{-1.1}$ & $0.55\pm0.10$ & $<234$ & $<36$    \\
C54 & $0.061\pm0.015$ & $<5.0$ & $<10.2$ & $16.5\pm1.8$ & $28.7\pm2.7$ & \ldots & $39.9\pm2.4$ & \ldots & \ldots & \ldots & $4.6^{+1.1}_{-1.1}$ & $2.11\pm0.12$ & $<234$ & $<36$ \\
C55a & $0.262\pm0.052$ & $<5.0$ & $8.7\pm3.6$ & $8.6\pm1.1$ & $18.6\pm1.6$ & \ldots & $15.6\pm1.7$ & \ldots & \ldots & \ldots & $2.3^{+0.6}_{-0.6}$ & $1.71\pm0.09$ & $225.9\pm64.5$ & $<36$    \\
C55b & $0.104\pm0.015$ & $<5.0$ & $<10.2$ & $5.3\pm0.6$ & $11.5\pm1.0$ & \ldots & $9.6\pm1.0$ & \ldots & \ldots & \ldots & $1.5^{+0.3}_{-0.4}$ & $1.05\pm0.19$ & $<234$ & $65.3\pm10.7$ \\
C56\tablefootmark{h} & $0.088\pm0.015$ & $<5.0$ & $18.7\pm5.1$ & $32.0\pm1.6$ & $32.7\pm2.2$ & \ldots & $30.1\pm2.4$ & \ldots & \ldots & \ldots & $4.7^{+1.1}_{-1.1}$ & $3.34\pm0.10$ & $<234$ & $<36$ \\
C58 & $<0.054$ & $<5.0$ & $<10.2$ & $9.8\pm2.3$ & $16.4\pm3.3$ & \ldots & $15.6\pm4.0$ & \ldots & \ldots & \ldots & $5.6^{+1.4}_{-1.5}$ & $2.29\pm0.11$ & $<234$ & $58.9\pm12.5$   \\
C59 & $0.343\pm0.034$ & $16.1\pm1.5$ & $41.0\pm4.1$ & $54.1\pm2.5$ & $43.5\pm3.3$ & \ldots & $48.3\pm3.2$ & \ldots & \ldots & \ldots & $4.6^{+1.1}_{-1.2}$ & $1.03\pm0.11$ & $<234$ & $161.1\pm14.6$  \\
C60a & $<0.054$ & $<5.0$ & $<10.2$ & $4.5\pm1.4$ & $5.7\pm1.6$ & \ldots & $4.9\pm1.5$ & \ldots & \ldots & \ldots & $2.8^{+0.7}_{-0.7}$ & $1.02\pm0.10$ & $283.5\pm52.1$ & $<36$  \\
C60b & $<0.054$ & $<5.0$ & $<10.2$ & $2.7\pm0.8$ & $3.3\pm0.9$ & \ldots & $2.9\pm0.9$ & \ldots & \ldots & \ldots & $1.6^{+0.4}_{-0.4}$ & $0.60\pm0.10$ & $<156.3$ & $<36$ \\
C61\tablefootmark{h} & $0.231\pm0.023$ & $<5.0$ & $10.6\pm3.8$ & $17.2\pm1.8$ & $24.2\pm2.6$ & \ldots & $26.3\pm2.2$ & \ldots & \ldots & \ldots & $4.6^{+1.2}_{-1.1}$ & $1.59\pm0.11$ & $1\,563.4\pm116.2$ & $10\,590\pm127$ \\
C62 & $<0.054$ & $<5.0$ & $<10.2$ & $18.2\pm2.1$ & $28.0\pm3.1$ & \ldots & $21.6\pm3.3$ & \ldots & \ldots & \ldots & $4.7^{+1.2}_{-1.2}$ & $0.71\pm0.15$ & $207.4\pm71.8$ & $<36$ \\
C64 & $0.185\pm0.089$ & $<5.0$ & $21.0\pm3.5$ & $32.8\pm2.2$ & $39.7\pm2.7$ & \ldots & $30.4\pm2.4$ & \ldots & \ldots & \ldots & $4.4^{+1.1}_{-1.1}$ & $2.16\pm0.11$ & $<234$ & $<36$  \\
C65 & $0.401\pm0.040$ & $12.4\pm1.8$ & $32.6\pm3.5$ & $51.3\pm1.9$ & $65.1\pm2.1$ & $23.91\pm7.84$ & $55.7\pm3.2$ & $3.14\pm1.56$ & \ldots & \ldots & $4.4^{+1.2}_{-1.1}$ & $1.25\pm0.15$ & $403.4\pm69.1$ & $153.4\pm12.1$\\
C66 & $0.586\pm0.024$ & $8.7\pm1.4$ & $14.4\pm2.8$ & $38.8\pm2.4$ & $37.1\pm2.9$ & $17.15\pm7.68$ & $25.1\pm3.2$ & $3.06\pm1.70$ & \ldots & \ldots & $4.3^{+1.1}_{-1.1}$ & $1.08\pm0.11$ & $364.6\pm69.7$ & $86.0\pm11.0$  \\
C67 & $0.152\pm0.016$ & $6.1\pm1.4$ & $8.0\pm3.3$ & $27.9\pm3.3$ & $34.2\pm3.3$ & \ldots & $23.6\pm2.8$ & \ldots & \ldots & \ldots & $4.3^{+1.1}_{-1.1}$ & $1.51\pm0.25$ & $208.0\pm64.9$ & $70.0\pm12.3$  \\
C69 & $<0.054$ & $<5.0$ & $<10.2$ & $4.3\pm2.2$ & $19.2\pm2.9$ & \ldots & $23.4\pm3.0$ & \ldots & \ldots & \ldots & $4.3^{+1.1}_{-1.1}$ & $0.61\pm0.09$ & $<234$ & $<36$  \\
C70 & $<0.054$ & $<5.0$ & $21.6\pm4.5$ & $34.5\pm1.9$ & $44.3\pm3.0$ & \ldots & $26.5\pm2.9$ & \ldots & \ldots & \ldots & $4.3^{+1.1}_{-1.1}$ & $1.57\pm0.12$ & $662.9\pm78.9$ & $49.1\pm12.6$ \\
C71b\tablefootmark{h} & $0.523\pm0.017$ & $<5.0$ & $9.6\pm2.9$ & $27.7\pm1.7$ & $10.5\pm0.7$ & $1.33\pm2.03$\tablefootmark{g} & $9.9\pm0.9$ & $1.29\pm0.44$ & \ldots & \ldots & $1.4^{+0.4}_{-0.4}$ & $0.92\pm0.11$ & $183.5\pm69.2$ & $78.5\pm11.1$ \\
C72 & $0.253\pm0.025$ & $<5.0$ & $16.6\pm4.9$ & $31.3\pm2.2$ & $32.9\pm2.6$ & \ldots & $26.0\pm2.2$ & \ldots & \ldots & \ldots & $4.4^{+1.2}_{-1.2}$ & $1.93\pm0.20$ & $<234$ & $95.5\pm13.7$ \\
C73 & $<0.054$ & $<5.0$ & $<10.2$ & $6.13\pm1.8$ & $4.4\pm2.5$ & \ldots & $22.4\pm3.0$ & \ldots & \ldots & \ldots & $4.2^{+1.1}_{-1.1}$ & $0.93\pm0.19$ & $<234$ & $<36$ \\
C74a & $<0.054$ & $<5.0$ & $<10.2$ & $8.3\pm1.9$ & $4.5\pm3.2$ & $0.38\pm6.54$\tablefootmark{g} & $3.9\pm2.3$ & $4.22\pm1.16$ & \ldots & \ldots & $2.8^{+0.7}_{-0.7}$ & $1.54\pm0.10$ & $163.7\pm53.9$ & $<36$ \\
C76 & $0.100\pm0.017$ & $<5.0$ & $<10.2$ & $11.8\pm2.2$ & $11.1\pm3.9$ & \ldots & $41.4\pm2.5$ & \ldots & \ldots & \ldots & $4.2^{+1.1}_{-1.1}$ & $1.62\pm0.12$ & $<234$ & $<36$ \\
C77a\tablefootmark{h} & $0.098\pm0.015$ & $<5.0$ & $9.2\pm2.8$ & $9.2\pm1.5$ & $18.1\pm1.2$ & \ldots & $14.6\pm1.4$ & \ldots & \ldots & \ldots & $2.2^{+0.6}_{-0.6}$ & $1.22\pm0.26$ & $1\,094.6\pm92.8$ & $554.5\pm11.5$ \\
C77b & $0.184\pm0.016$ & $<5.0$ & $12.6\pm2.9$ & $29.0\pm1.6$ & $17.1\pm1.1$ & \ldots & $13.7\pm1.3$ & \ldots & \ldots & \ldots & $2.0^{+0.6}_{-0.5}$ & $1.15\pm0.11$ & $<234$ & $69.1\pm10.7$ \\
C78 & $<0.054$ & $<5.0$ & $<10.2$ & $6.97\pm1.8$ & $7.9\pm3.0$ & \ldots & $9.2\pm3.0$ & \ldots & \ldots & \ldots &  $4.8^{+1.4}_{-1.3}$ & $1.78\pm0.15$ & $<234$ & $<36$ \\
C79 & $0.204\pm0.020$ & $<5.0$ & $9.8\pm3.3$ & $9.4\pm1.4$ & $13.3\pm2.1$ & \ldots & $13.1\pm3.1$ & \ldots & \ldots & \ldots & $4.2^{+1.1}_{-1.2}$ & $1.73\pm0.13$ & $185.1\pm55.9$ & $<36$ \\
C80a & $0.297\pm0.030$ & $9.2\pm1.2$ & $24.6\pm3.4$ & $26.7\pm1.7$ & $26.9\pm2.1$ & $16.79\pm3.25$ & $18.7\pm2.2$ & $5.88\pm0.68$ & $5.90\pm1.83$ & \ldots & $2.7^{+0.7}_{-0.7}$ & $1.74\pm0.13$ & $<234$ &  $<36$ \\
C80b & $0.175\pm0.018$ & $<5.0$ & $<10.2$ & $14.0\pm0.9$ & $14.1\pm1.1$ & \ldots & $9.8\pm1.1$ & \ldots & \ldots & \ldots & $1.4^{+0.4}_{-0.4}$ & $0.91\pm0.10$ & $<234$ & $<36$ \\
C81 & $<0.054$ & $<5.0$ & $<10.2$ & $6.3\pm2.1$ & $18.7\pm3.1$ & \ldots & $15.6\pm3.0$ & \ldots & \ldots & \ldots & $4.1^{+1.2}_{-1.1}$ & $1.06\pm0.13$ & $<234$  & $<36$  \\
C84a & $0.080\pm0.016$ & $<5.0$ & $<10.2$ & $13.8\pm1.6$ & $13.8\pm2.7$ & \ldots & $14.5\pm2.6$ & \ldots & \ldots & \ldots & $2.2^{+0.6}_{-0.6}$ & $2.14\pm0.33$ & $<234$ &  $<36$ \\
C84b & $0.462\pm0.046$ & $<5.0$ & $8.9\pm2.7$ & $30.8\pm1.8$ & $33.5\pm2.5$ & \ldots & $26.5\pm2.5$ & \ldots & \ldots & \ldots & $1.9^{+0.5}_{-0.6}$ & $1.86\pm0.16$ & $196.0\pm77.8$ & $82.4\pm10.9$ \\
C86\tablefootmark{h} & $0.286\pm0.029$ & $<5.0$ & $<10.2$ & $14.2\pm2.3$ & $17.1\pm2.6$ & \ldots & $16.1\pm2.5$ & \ldots & \ldots & \ldots & $4.0^{+1.1}_{-1.1}$ & $1.21\pm0.21$ & $11\,764\pm64.5$ & $<36$  \\
C87 & $0.557\pm0.015$ & $<5.0$ & $14.4\pm2.8$ & $28.7\pm2.2$ & $57.7\pm3.3$ & \ldots & $43.8\pm2.9$ & \ldots & \ldots & \ldots & $4.0^{+1.1}_{-1.2}$ & $1.43\pm0.13$ & $306.8\pm65.2$ & $<36$ \\
C88 & $<0.054$ & $<5.0$ & $<10.2$ & $13.0\pm1.4$ & $14.6\pm2.5$ & \ldots & $16.4\pm2.7$ & \ldots & \ldots & \ldots & $4.0^{+1.1}_{-1.2}$ & $0.63\pm0.10$ & $174.4\pm61.6$ & $<36$ \\
C90a & $<0.054$ & $<5.0$ & $<10.2$ & $9.1\pm1.0$ & $10.5\pm1.4$ & \ldots & $9.1\pm1.0$ & \ldots & \ldots & \ldots & $1.4^{+0.4}_{-0.4}$ & $0.81\pm0.14$ & $218.7\pm59.2$ &  $<36$ \\
C90b & $<0.054$ & $<5.0$ & $<10.2$ & $9.0\pm1.0$ & $10.3\pm1.3$ & \ldots & $8.9\pm0.9$ & \ldots & \ldots & \ldots & $1.4^{+0.4}_{-0.4}$ & $0.80\pm0.11$ & $<177.6$ &  $<36$ \\
C90c & $0.151\pm0.020$ & $4.6\pm1.6$ & $<10.2$ & $12.9\pm2.0$ & $8.2\pm2.8$ & \ldots & $7.8\pm0.8$ & \ldots & \ldots & \ldots & $1.2^{+0.3}_{-0.4}$ & $0.69\pm0.12$ & $165.8\pm59.2$ & $<36$  \\ 
C91 & $0.151\pm0.016$ & $<5.0$ & $23.9\pm2.9$ & $42.1\pm2.3$ & $44.8\pm2.5$ & \ldots & $35.2\pm2.2$ & \ldots & \ldots & \ldots & $3.8^{+1.2}_{-1.1}$ & $1.14\pm0.10$ & $<234$ & $<36$ \\
C92a  & $<0.054$ & $<5.0$ & $<10.2$ & $25.4\pm1.8$ & $29.6\pm2.7$ & \ldots & $23.2\pm2.9$ & \ldots & \ldots & \ldots & $2.5^{+0.7}_{-0.8}$ & $2.33\pm0.14$ & $265.9\pm67.8$ & $<36$  \\
C92b & $<0.054$ & $<5.0$ & $<10.2$ & $3.8\pm1.9$ & $5.8\pm2.5$ & \ldots & $14.6\pm2.8$ & \ldots & \ldots & \ldots & $1.5^{+0.4}_{-0.4}$ & $1.32\pm0.20$ & $158.9\pm67.8$ & $<36$  \\
C93 & $0.730\pm0.073$ & $5.1\pm1.8$ & $15.4\pm4.1$ & $22.7\pm2.5$ & $27.4\pm3.2$ & \ldots & $25.7\pm3.0$ & \ldots & \ldots & \ldots & $3.8^{+1.1}_{-1.1}$ & $1.84\pm0.10$ & $188.0\pm74.7$ & $60.1\pm11.1$ \\
C95 & $0.592\pm0.132$ & $5.6\pm1.3$ & $10.7\pm3.2$ & $31.1\pm1.6$ & $25.7\pm2.5$ & \ldots & $29.5\pm3.3$ & \ldots & \ldots & \ldots & $3.8^{+1.2}_{-1.1}$ & $1.06\pm0.09$ & $<234$ & $<36$  \\
C97a & $0.376\pm0.038$ & $<5.0$ & $18.5\pm3.2$ & $35.2\pm2.4$ & $34.3\pm2.0$ & \ldots & $23.6\pm2.2$ & \ldots & \ldots & \ldots & $2.9^{+0.8}_{-0.9}$ & $1.61\pm0.34$ & $<234$ & $<36$ \\
C97b & $0.637\pm0.115$ & $7.5\pm1.6$ & $17.5\pm3.9$ & $21.6\pm2.4$ & $14.5\pm0.8$ & \ldots & $10.0\pm0.9$ & \ldots & \ldots & \ldots & $1.2^{+0.4}_{-0.4}$ & $0.68\pm0.12$ & $<234$ & $55.1\pm15.9$ \\
C98 & $0.503\pm0.101$ & $6.6\pm1.3$  & $27.0\pm3.6$ & $48.3\pm1.9$ & $48.6\pm3.2$ & \ldots & $33.8\pm2.4$ & \ldots & $10.0\pm2.6$ & \ldots & $3.8^{+1.1}_{-1.2}$ & $2.00\pm0.10$ & $<234$ & $77.8\pm13.7$ \\ 
C99 & $<0.054$ & $<5.0$ & $<10.2$ & $5.8\pm2.3$ & $5.1\pm3.0$ & \ldots & $<15.4$ & \ldots & \ldots & \ldots & $3.8^{+1.1}_{-1.2}$ & $1.29\pm0.09$ & $<234$ & $<36$ \\
C100a & $0.179\pm0.022$ & $<5.0$ & $13.7\pm3.0$ & $22.6\pm2.1$ & $24.9\pm3.2$ & \ldots & $19.6\pm2.8$ & \ldots & \ldots & \ldots & $2.3^{+0.7}_{-0.7}$ & $1.23\pm0.24$ & $<234$ & $<36$ \\
C100b  & $<0.054$ & $<5.0$ & $<10.2$ & $5.0\pm2.1$ & $4.4\pm3.2$ & \ldots & $6.8\pm2.8$ & \ldots & \ldots & \ldots & $1.5^{+0.5}_{-0.4}$ & $0.81\pm0.11$ & $<234$ & $<36$ \\
\hline
\end{tabular} }
\end{table}
\end{landscape}

\addtocounter{table}{-1}

\begin{landscape}
\begin{table}
\caption{continued.}
{\scriptsize
\centering
\renewcommand{\footnoterule}{}
\label{table:fluxes}
\begin{tabular}{c c c c c c c c c c c c c c c}
\hline\hline 
Source ID & $S_{\rm 24\, \mu m}$ & $S_{\rm 100\, \mu m}$ & $S_{\rm 160\, \mu m}$ & $S_{\rm 250\, \mu m}$ & $S_{\rm 350\, \mu m}$ & $S_{\rm 450\, \mu m}$ & $S_{\rm 500\, \mu m}$ & $S_{\rm 850\, \mu m}$ & $S_{\rm 870\, \mu m}$ & $S_{\rm 890\, \mu m}$ & $S_{\rm 1.1\, mm}$ & $S_{\rm 1.3\, mm}$ & $S_{\rm 325\, MHz}$ & $S_{\rm 1.4\, GHz}$\\
          & [mJy] & [mJy] & [mJy] &  [mJy] & [mJy] & [mJy] & [mJy] & [mJy] & [mJy] & [mJy] & [mJy] & [mJy] & [$\mu$Jy] & [$\mu$Jy]\\
         
& MIPS & PACS & PACS & SPIRE & SPIRE & SCUBA-2\tablefootmark{a} & SPIRE & SCUBA-2\tablefootmark{a} & LABOCA\tablefootmark{b} & SMA\tablefootmark{c} & AzTEC\tablefootmark{d} & ALMA\tablefootmark{e} & GMRT\tablefootmark{f} & VLA\tablefootmark{f}\\
\hline
C101a  & $<0.054$ & $<5.0$ & $<10.2$ & $13.5\pm2.0$ & $14.6\pm2.0$ & \ldots & $10.1\pm1.9$ & \ldots & \ldots & \ldots & $2.4^{+0.7}_{-0.8}$ & $1.35\pm0.10$ & $<234$ & $<36$ \\
C101b  & $<0.054$ & $<5.0$ & $<10.2$ & $9.1\pm2.0$ & $8.0\pm1.1$ & \ldots & $5.5\pm1.0$ & \ldots & \ldots & \ldots & $1.4^{+0.4}_{-0.4}$ & $0.74\pm0.13$ & $<234$ & $<36$ \\
C103 & $0.203\pm0.020$ & $<5.0$ & $<10.2$ & $8.79\pm2.0$ & $41.9\pm2.7$ & \ldots & $38.9\pm2.3$ & \ldots & \ldots & \ldots & $3.8^{+1.1}_{-1.2}$ & $1.67\pm0.10$ & $<234$ & $<36$  \\
C105 & $0.362\pm0.036$ & $<5.0$ & $14.8\pm2.6$ & $22.5\pm1.4$ & $22.8\pm2.3$ & \ldots & $37.4\pm2.6$ & \ldots & \ldots & \ldots & $4.1^{+1.2}_{-1.3}$ & $1.79\pm0.18$ & $221.4\pm71.6$ & $<36$\\
C107 & $<0.054$ & $<5.0$ & $<10.2$ & $7.08\pm1.8$ & $10.3\pm2.0$ & \ldots & $14.9\pm3.4$ & \ldots & \ldots & \ldots & $3.8^{+1.1}_{-1.2}$ & $1.64\pm0.11$ & $<234$ & $<36$ \\
C108 & $<0.054$ & $<5.0$ & $<10.2$ & $15.5\pm2.2$ & $27.9\pm3.0$ & \ldots & $24.9\pm3.1$ & \ldots & \ldots & \ldots & $4.0^{+1.2}_{-1.3}$ & $2.35\pm0.12$ & $195.9\pm75.9$ & $<36$ \\
C109 & $0.201\pm0.041$ & $5.9\pm1.2$ & $19.0\pm3.6$ & $30.6\pm2.1$ & $38.3\pm3.5$ & \ldots & $29.9\pm2.6$ & \ldots & \ldots & \ldots & $3.7^{+1.1}_{-1.1}$ & $2.07\pm0.10$ & $251.0\pm62.3$ & $59.1\pm10.6$ \\
C111 & $0.240\pm0.024$ & $4.3\pm1.4$ & $16.3\pm2.8$ & $26.6\pm2.1$ & $21.1\pm2.7$ & \ldots & $19.9\pm2.9$ & \ldots & \ldots & \ldots & $3.7^{+1.2}_{-1.2}$ & $1.45\pm0.12$ & $209.8\pm62.7$ & $66.7\pm11.6$ \\
C112 & $0.260\pm0.026$ & $<5.0$ & $10.8\pm4.1$ & $30.1\pm2.1$ & $27.6\pm3.2$ & \ldots & $25.5\pm3.1$ & \ldots & \ldots & \ldots & $3.7^{+1.1}_{-1.2}$ & $1.66\pm0.11$ & $204.6\pm86.1$ & $121.8\pm11.5$ \\ 
C113 & $0.725\pm0.073$ & $42.0\pm2.1$ & $101.4\pm4.1$ & $86.4\pm1.6$ & $63.0\pm2.5$ & \ldots & $32.1\pm2.3$ & \ldots & \ldots & \ldots & $4.0^{+1.3}_{-1.3}$ & $1.08\pm0.14$ & $<234$ & $173.0\pm15.6$ \\
C115 & $<0.054$ & $4.3\pm1.4$ & $<10.2$ & $6.1\pm1.9$ & $6.4\pm2.8$ & \ldots & $<15.4$ & \ldots & \ldots & \ldots & $3.7^{+1.1}_{-1.2}$ & $1.55\pm0.13$ & $<234$ & $<36$ \\
C116 & $0.335\pm0.034$ & $6.8\pm1.7$ & $17.4\pm3.0$ & $32.3\pm2.2$ & $29.5\pm2.0$ & \ldots & $17.1\pm2.2$ & \ldots & \ldots & \ldots & $3.7^{+1.1}_{-1.2}$ & $0.98\pm0.10$ & $108.2\pm57.0$ & $58.9\pm10.8$ \\
C117 & $0.095\pm0.043$ & $<5.0$ & $<10.2$ & $12.0\pm1.7$ & $19.5\pm2.0$ & \ldots & $33.9\pm2.9$ & \ldots & $11.1\pm3.4$ & \ldots & $3.7^{+1.1}_{-1.2}$ & $0.89\pm0.15$ & $197.2\pm59.4$ & $<36$ \\
C118\tablefootmark{h} & $0.222\pm0.017$ & $4.8\pm1.6$ & $9.9\pm2.7$ & $18.5\pm1.5$ & $23.0\pm3.0$ & \ldots & $23.8\pm3.2$ & \ldots & \ldots & \ldots & $3.7^{+1.1}_{-1.2}$ & $2.65\pm0.12$ & $309.5\pm71.6$ & $104.3\pm12.7$ \\
C119 & $<0.054$ & $<5.0$ & $<10.2$ & $19.3\pm1.8$ & $24.6\pm2.5$ & \ldots & $24.1\pm2.8$ & \ldots & \ldots & \ldots & $3.7^{+1.1}_{-1.2}$ & $2.55\pm0.10$ & $<234$  & $<36$    \\
C122a & $<0.054$ & $6.9\pm1.7$ & $13.3\pm4.3$ & $18.1\pm2.9$ & $15.2\pm1.9$ & \ldots & $14.4\pm1.5$ & \ldots & \ldots & \ldots & $2.1^{+0.6}_{-0.7}$ & $1.20\pm0.15$ & $<234$ & $<36$\\
C123 & $<0.054$ & $<5.0$ & $25.7\pm3.0$ & $38.2\pm2.9$ & $38.4\pm2.9$ & \ldots & $26.5\pm3.3$ & \ldots & \ldots & \ldots & $3.7^{+1.1}_{-1.2}$ & $1.57\pm0.11$ & $167.3\pm80.5$ & $<36$\\
C124 & $0.323\pm0.014$ & $4.5\pm1.6$ & $21.6\pm3.3$ & $38.7\pm2.8$ & $32.9\pm2.8$ & \ldots & $14.5\pm3.5$ & \ldots & \ldots & \ldots & $3.7^{+1.2}_{-1.2}$ & $0.69\pm0.14$ & $<234$ & $<36$\\
C126 & $<0.054$ & $<5.0$ & $43.4\pm5.8$ & $50.4\pm2.3$ & $38.8\pm2.7$ & \ldots & $26.2\pm3.0$ & \ldots & \ldots & \ldots & $3.5^{+1.2}_{-1.2}$ & $1.60\pm0.13$ & $391.9\pm77.8$  & $<36$\\
C127 & $0.622\pm0.062$ & $15.7\pm1.6$ & $36.7\pm3.3$ & $44.2\pm1.8$ & $37.5\pm2.5$ & \ldots & $24.2\pm2.6$ & \ldots & \ldots & \ldots & $3.8^{+1.3}_{-1.3}$ & $0.91\pm0.11$ & $311.0\pm80.5$ & $131.1\pm12.5$ \\
C129  & $<0.054$ & $<5.0$ & $<10.2$ & $10.4\pm2.3$ & $6.8\pm3.6$ & \ldots & $5.1\pm2.9$ & \ldots & \ldots & \ldots & $3.5^{+1.1}_{-1.1}$ & $2.63\pm0.10$ & $<234$ & $<36$\\
\hline
\end{tabular} }
\tablefoot{The third row of the table gives the name of the instrument used to measure the flux density. The quoted \textit{Herschel} flux density uncertainties refer to the total error, that is instrumental plus confusion noise. We placed a $3\sigma$ flux density upper limit for the non-detections. The \textit{Herschel} flux density upper limits include the confusion noise.\tablefoottext{a}{From \cite{casey2013}.}\tablefoottext{b}{From F.~Navarrete et al., in prep.. The quoted 870~$\mu$m flux densities of AzTEC/C5, C6a, and C6b were measured with ALMA (see Sect.~2.2.2 for details).}\tablefoottext{c}{From \cite{younger2007}, 2009.}\tablefoottext{d}{From \cite{aretxaga2011}.}\tablefoottext{e}{From M.~Aravena et al., in prep.}\tablefoottext{f}{See Sect.~2.2.3 for the details of the radio flux densities.}\tablefoottext{g}{A raw 450~$\mu$m SCUBA-2 flux density, which was not used in the SED fit.}\tablefoottext{h}{The SMG is likely to host an AGN, and hence was excluded from the final sample.}}
\end{table}
\end{landscape}

\section{The derived physical properties}

The derived physical properties are tabulated in Table~\ref{table:sed}. 

\begin{table*}[!htb]
\caption{Results of {\tt MAGPHYS} SED modelling of the target SMGs, and the gas mass estimates.}
{\scriptsize
\begin{minipage}{2\columnwidth}
\centering
\renewcommand{\footnoterule}{}
\label{table:sed}
\begin{tabular}{c c c c c c c c c c c}
\hline\hline 
Source ID & $z$ & $\chi^2$ & $\log(M_{\star}/{\rm M}_{\sun})$ & $\log(L_{\rm IR}/{\rm L}_{\sun})$ & SFR [${\rm M}_{\sun}~{\rm yr}^{-1}$] & sSFR [${\rm Gyr}^{-1}$] & $\Delta_{\rm MS}$ & $T_{\rm dust}$ [K] & $\log(M_{\rm dust}/{\rm M}_{\sun})$ & $\log(M_{\rm gas}/{\rm M}_{\sun})$ \\[1ex]
\hline
AzTEC/C1a & 4.7\tablefootmark{a} & 5.973 & $12.15^{+0.12}_{-0.20}$ & $13.57^{+0.02}_{-0.01}$ & $3\,702^{+174}_{-84}$ & $2.6^{+1.7}_{-0.7}$ & $1.3^{+0.1}_{-0.0}$ & $53.0^{+2.9}_{-3.3}$ & $9.45^{+0.04}_{-0.00}$ & $11.95^{+0.10}_{-0.13}$       \\ [1ex]
AzTEC/C2a & 3.179\tablefootmark{a} & 4.148 & $10.91^{+0.01}_{-0.01}$ & $13.41^{+0.01}_{-0.01}$ & $2\,543^{+59}_{-58}$ & $31.3^{+1.5}_{-1.4}$ & $12.4^{+0.3}_{-0.3}$ & $46.2^{+0.9}_{-0.3}$ & $9.19^{+0.09}_{-0.04}$ & $11.75^{+0.10}_{-0.13}$   \\ [1ex]
AzTEC/C2b & $1.10^{+2.60}_{-1.10}$ & 7.188 & $10.84^{+0.35}_{-0.25}$ & $12.08^{+0.11}_{-0.07}$ & $120^{+35}_{-18}$ & $1.7^{+2.2}_{-2.1}$ & $2.5^{+0.7}_{-0.4}$ & $31.5^{+4.9}_{-7.0}$ & $9.36^{+0.02}_{-0.03}$ & $11.77^{+0.10}_{-0.13}$ \\ [1ex]
AzTEC/C4 & $5.30^{+0.70}_{-1.10}$ & 0.359 & $11.59^{+0.12}_{-0.14}$ & $13.31^{+0.06}_{-0.05}$ & $2\,025^{+300}_{-220}$ & $5.2^{+3.0}_{-1.7}$ & $1.9^{+0.3}_{-0.2}$ & $45.8^{+4.1}_{-2.1}$ & $8.94^{+0.04}_{-0.03}$ & $11.67^{+0.10}_{-0.13}$      \\ [1ex]
AzTEC/C5 & 4.3415\tablefootmark{a} & 1.963 & $10.99^{+0.01}_{-0.09}$ & $13.26^{+0.01}_{-0.02}$ & $1\,821^{+42}_{-82}$ & $18.6^{+4.8}_{-1.2}$ & $6.0^{+0.1}_{-0.3}$ & $42.8^{+2.0}_{-2.1}$ & $9.17^{+0.03}_{-0.03}$ & $11.74^{+0.10}_{-0.13}$ \\ [1ex]
AzTEC/C6a & 2.494\tablefootmark{a} & 6.240 & $11.35^{+0.00}_{-0.00}$ & $12.85^{+0.00}_{-0.00}$ & $709^{+0}_{-0}$ & $3.2^{+0.0}_{-0.0}$ & $2.0^{+0.0}_{-0.0}$ & $38.0^{+0.0}_{-0.0}$ & $9.06^{+0.00}_{-0.01}$ & $11.51^{+0.10}_{-0.13}$ \\ [1ex]
AzTEC/C6b & 2.513\tablefootmark{a} & 6.402 & $11.31^{+0.05}_{-0.07}$ & $12.81^{+0.01}_{-0.07}$ & $644^{+15}_{-96}$ & $3.2^{+0.6}_{-0.8}$ & $1.9^{+0.0}_{-0.3}$ & $45.5^{+0.0}_{-5.8}$ & $8.73^{+0.02}_{-0.03}$ & $11.11^{+0.11}_{-0.15}$          \\ [1ex]
AzTEC/C7 & $3.06^{+1.88}_{-1.76}$ & 1.111 & $10.66^{+0.22}_{-0.07}$ & $12.89^{+0.18}_{-0.09}$ & $772^{+396}_{-144}$ & $16.9^{+13.1}_{-8.6}$ & $6.2^{+3.2}_{-1.2}$ & $34.2^{+16.2}_{-2.9}$ & $9.45^{+0.05}_{-0.03}$ & $11.77^{+0.10}_{-0.13}$  \\ [1ex]
AzTEC/C8a & 3.62\tablefootmark{a} & 9.661 & $10.97^{+0.11}_{-0.20}$ & $13.07^{+0.04}_{-0.04}$ & $1\,169^{+113}_{-103}$ & $12.5^{+9.2}_{-3.7}$ & $4.6^{+0.4}_{-0.4}$ & $48.8^{+3.8}_{-3.4}$ & $8.62^{+0.03}_{-0.03}$ & $11.10^{+0.11}_{-0.14}$\\ [1ex]
AzTEC/C8b & $1.10^{+0.30}_{-0.20}$ & 1.987 & $11.27^{+0.06}_{-0.06}$ & $11.92^{+0.04}_{-0.18}$ & $83^{+8}_{-28}$ & $0.4^{+0.1}_{-0.2}$ & $0.9^{+0.1}_{-0.3}$ & $51.3^{+0.0}_{-18.0}$ & $8.97^{+0.02}_{-0.04}$ & $10.98^{+0.12}_{-0.17}$\\ [1ex]
AzTEC/C9a & $2.68^{+0.24}_{-0.51}$ & 3.289 & $11.18^{+0.05}_{-0.04}$ & $12.70^{+0.04}_{-0.00}$ & $505^{+49}_{-0}$ & $3.3^{+0.7}_{-0.4}$ & $1.8^{+0.2}_{-0.0}$ & $31.6^{+4.4}_{-0.0}$ & $9.28^{+0.01}_{-0.05}$ & $11.61^{+0.10}_{-0.13}$   \\ [1ex]
AzTEC/C9b & 2.8837\tablefootmark{a} & 8.972 & $11.40^{+0.02}_{-0.02}$ & $12.54^{+0.05}_{-0.09}$ & $345^{+42}_{-65}$ & $1.4^{+0.2}_{-0.3}$ & $0.8^{+0.1}_{-0.1}$ & $42.8^{+3.5}_{-9.0}$ & $8.89^{+0.14}_{-0.07}$ & $11.24^{+0.12}_{-0.17}$\\ [1ex]
AzTEC/C9c & 2.9219\tablefootmark{a} & 3.222 & $11.21^{+0.05}_{-0.04}$ & $12.48^{+0.03}_{-0.04}$ & $299^{+21}_{-26}$ & $1.8^{+0.3}_{-0.3}$ & $0.9^{+0.1}_{-0.1}$ & $41.3^{+5.0}_{-3.0}$ & $8.77^{+0.04}_{-0.06}$ & $11.13^{+0.11}_{-0.14}$ \\ [1ex]
AzTEC/C10a & $3.40^{+3.60}_{-0.59}$ & 0.639 & $10.93^{+0.30}_{-0.34}$ & $12.55^{+0.12}_{-0.10}$ & $354^{+113}_{-73}$ & $4.2^{+7.8}_{-2.5}$ & $1.6^{+0.5}_{-0.3}$ & $34.2^{+11.1}_{-3.2}$ & $9.07^{+0.06}_{-0.05}$ & $11.43^{+0.10}_{-0.13}$\\ [1ex]
AzTEC/C10b & $2.90^{+0.30}_{-0.90}$ & 0.800 & $11.51^{+0.04}_{-0.08}$ & $12.46^{+0.10}_{-0.06}$ & $289^{+75}_{-37}$ & $0.9^{+0.5}_{-0.2}$ & $0.5^{+0.1}_{-0.1}$ & $39.5^{+8.4}_{-7.6}$ & $9.12^{+0.09}_{-0.07}$ & $11.38^{+0.10}_{-0.14}$ \\ [1ex]
AzTEC/C12 & $3.25^{+0.16}_{-0.51}$ & 3.568 & $10.52^{+0.02}_{-0.26}$ & $13.00^{+0.01}_{-0.40}$ & $991^{+23}_{-596}$ & $29.9^{+25.8}_{-18.5}$ & $9.7^{+0.2}_{-5.8}$ & $47.0^{+14.6}_{-15.9}$ & $9.31^{+0.06}_{-0.06}$ & $11.59^{+0.10}_{-0.13}$ \\ [1ex]
AzTEC/C13a & $2.01^{+0.15}_{-0.49}$ & 2.378 & $11.11^{+0.00}_{-0.00}$ & $12.80^{+0.00}_{-0.00}$ & $637^{+0}_{-0}$ & $4.9^{+0.0}_{-0.0}$ & $3.6^{+0.0}_{-0.0}$ & $47.3^{+0.0}_{-0.0}$ & $9.58^{+0.00}_{-0.00}$ & $11.72^{+0.10}_{-0.13}$ \\ [1ex]
AzTEC/C13b & $2.01^{+0.30}_{-0.50}$ & 2.509 & $9.87^{+0.09}_{-0.01}$ & $12.13^{+0.01}_{-0.03}$ & $136^{+3}_{-9}$ & $18.5^{+0.9}_{-4.5}$ & $6.7^{+0.2}_{-0.4}$ & $28.5^{+1.0}_{-1.2}$ & $9.09^{+0.06}_{-0.04}$ & $11.27^{+0.10}_{-0.14}$ \\ [1ex]
AzTEC/C14 & $4.58^{+0.25}_{-0.68}$ & 0.892 & $10.82^{+0.01}_{-0.10}$ & $13.10^{+0.02}_{-0.03}$ & $1\,246^{+59}_{-83}$ & $18.9^{+6.0}_{-1.7}$ & $5.5^{+0.3}_{-0.4}$ & $38.5^{+1.4}_{-1.2}$ & $9.25^{+0.06}_{-0.05}$ & $11.79^{+0.10}_{-0.13}$ \\ [1ex]
AzTEC/C15 & $3.91^{+0.28}_{-2.35}$ & 2.319 & $11.06^{+0.02}_{-0.06}$ & $13.57^{+0.02}_{-0.05}$ & $3\,752^{+177}_{-408}$ & $32.7^{+6.6}_{-4.9}$ & $11.7^{+0.6}_{-1.3}$ & $69.2^{+3.4}_{-5.0}$ & $9.13^{+0.10}_{-0.07}$ & $11.62^{+0.10}_{-0.13}$\\ [1ex]
AzTEC/C16a & $3.15^{+0.62}_{-1.54}$ & 0.864 & $11.52^{+0.09}_{-0.09}$ & $12.82^{+0.06}_{-0.07}$ & $656^{+97}_{-98}$ & $2.0^{+0.8}_{-0.6}$ & $1.1^{+0.2}_{-0.2}$ & $45.0^{+8.8}_{-6.1}$ & $8.83^{+0.07}_{-0.05}$ & $11.32^{+0.10}_{-0.13}$\\ [1ex]
AzTEC/C16b & $2.39^{+0.27}_{-0.56}$ & 7.994 & $10.79^{+0.00}_{-0.01}$ & $12.42^{+0.00}_{-0.01}$ & $261^{+0}_{-6}$ & $4.2^{+0.1}_{-0.1}$ & $2.1^{+0.0}_{-0.0}$ & $32.2^{+0.0}_{-0.0}$ & $9.01^{+0.00}_{-0.01}$ & $11.31^{+0.10}_{-0.13}$\\ [1ex]
AzTEC/C17 & 4.542\tablefootmark{a} & 3.259 & $10.69^{+0.09}_{-0.10}$ & $12.89^{+0.12}_{-0.14}$ & $773^{+246}_{-213}$ & $15.8^{+10.4}_{-6.5}$ & $4.3^{+1.4}_{-1.2}$ & $38.2^{+9.9}_{-3.7}$ & $9.09^{+0.07}_{-0.06}$ & $11.58^{+0.10}_{-0.13}$    \\ [1ex]
AzTEC/C18 & $3.15^{+0.13}_{-0.44}$ & 2.983 & $11.60^{+0.05}_{-0.02}$ & $13.38^{+0.00}_{-0.10}$ & $2\,421^{+0}_{-498}$ & $6.1^{+0.3}_{-1.8}$ & $3.4^{+0.0}_{-0.7}$ & $42.0^{+9.0}_{-1.5}$ & $9.14^{+0.02}_{-0.01}$ & $11.70^{+0.10}_{-0.13}$ \\ [1ex]
AzTEC/C19 & $2.87^{+0.11}_{-0.41}$ & 2.663 & $10.65^{+0.00}_{-0.00}$ & $12.66^{+0.01}_{-0.00}$ & $454^{+11}_{-0}$ & $10.2^{+0.2}_{-0.0}$ & $3.9^{+0.1}_{-0.0}$ & $32.2^{+2.5}_{-0.0}$ & $9.24^{+0.00}_{-0.00}$ & $11.54^{+0.10}_{-0.14}$\\ [1ex]
AzTEC/C20 & $3.06^{+0.13}_{-0.54}$ & 5.413 & $10.59^{+0.00}_{-0.01}$ & $13.08^{+0.01}_{-0.00}$ & $1\,199^{+28}_{-0}$ & $30.8^{+1.5}_{-0.0}$ & $10.8^{+0.3}_{-0.0}$ & $45.5^{+2.7}_{-1.1}$ & $8.84^{+0.03}_{-0.05}$ & $11.35^{+0.10}_{-0.13}$\\ [1ex]
AzTEC/C21 & $2.70^{+1.30}_{-0.40}$ & 1.509 & $11.84^{+0.21}_{-0.24}$ & $12.74^{+0.03}_{-0.05}$ & $546^{+39}_{-59}$ & $0.8^{+0.7}_{-0.4}$ & $0.6^{+0.0}_{-0.1}$ & $36.2^{+4.6}_{-2.9}$ & $9.08^{+0.05}_{-0.05}$ & $11.50^{+0.10}_{-0.13}$\\ [1ex]
AzTEC/C22a & 1.599\tablefootmark{a} & 9.922 & $10.99^{+0.00}_{-0.01}$ & $12.64^{+0.00}_{-0.01}$ & $434^{+0}_{-10}$ & $4.4^{+0.1}_{-0.1}$ & $4.1^{+0.1}_{-0.1}$ & $33.2^{+0.0}_{-2.3}$ & $9.22^{+0.02}_{-0.00}$ & $11.56^{+0.10}_{-0.13}$     \\ [1ex]
AzTEC/C22b & 1.599\tablefootmark{a} & 3.277 & $10.31^{+0.28}_{-0.21}$ & $12.22^{+0.00}_{-0.06}$ & $165^{+0}_{-21}$ & $8.1^{+5.0}_{-4.4}$ & $4.9^{+0.0}_{-0.6}$ & $32.0^{+0.4}_{-0.0}$ & $8.95^{+0.01}_{-0.10}$ & $11.11^{+0.11}_{-0.15}$ \\ [1ex]
AzTEC/C23 & $2.10^{+0.46}_{-0.41}$ & 6.789 & $11.61^{+0.03}_{-0.10}$ & $12.82^{+0.02}_{-0.00}$ & $663^{+31}_{-0}$ & $1.6^{+0.5}_{-0.1}$ & $1.5^{+0.1}_{-0.0}$ & $36.0^{+3.5}_{-0.5}$ & $8.96^{+0.03}_{-0.04}$ & $11.27^{+0.10}_{-0.14}$\\ [1ex]
AzTEC/C24a & $2.01^{+0.18}_{-0.46}$ & 2.907 & $11.45^{+0.06}_{-0.06}$ & $12.54^{+0.04}_{-0.02}$ & $346^{+33}_{-16}$ & $1.2^{+0.3}_{-0.2}$ & $1.1^{+0.1}_{-0.0}$ & $31.0^{+7.9}_{-0.3}$ & $8.99^{+0.06}_{-0.02}$ & $11.37^{+0.10}_{-0.13}$\\ [1ex]
AzTEC/C25 & 2.51\tablefootmark{a} & 2.193 & $10.73^{+0.01}_{-0.02}$ & $12.98^{+0.01}_{-0.09}$ & $952^{+22}_{-178}$ & $17.7^{+1.3}_{-3.6}$ & $8.1^{+0.2}_{-1.5}$ & $40.2^{+1.8}_{-0.7}$ & $8.87^{+0.14}_{-0.02}$ & $11.28^{+0.11}_{-0.15}$\\ [1ex]
AzTEC/C26 & $5.06^{+0.08}_{-0.90}$ & 9.296 & $10.98^{+0.00}_{-0.01}$ & $13.50^{+0.01}_{-0.01}$ & $3\,140^{+73}_{-71}$ & $32.9^{+1.5}_{-0.7}$ & $9.6^{+0.2}_{-0.2}$ & $65.4^{+2.0}_{-1.6}$ & $8.50^{+0.05}_{-0.02}$ & $11.37^{+0.10}_{-0.13}$\\ [1ex]
AzTEC/C27 & $2.77^{+0.88}_{-0.47}$ & 1.898 & $11.26^{+0.01}_{-0.00}$ & $12.56^{+0.03}_{-0.01}$ & $364^{+26}_{-8}$ & $2.0^{+0.1}_{-0.1}$ & $1.1^{+0.1}_{-0.0}$ & $32.5^{+5.3}_{-1.4}$ & $9.17^{+0.07}_{-0.08}$ & $11.48^{+0.11}_{-0.14}$\\ [1ex]
AzTEC/C28a & 2.319\tablefootmark{a} & 1.880 & $10.95^{+0.00}_{-0.00}$ & $12.67^{+0.00}_{-0.00}$ & $465^{+0}_{-0}$ & $5.2^{+0.0}_{-0.0}$ & $2.9^{+0.0}_{-0.0}$ & $32.2^{+0.0}_{-0.0}$ & $9.25^{+0.00}_{-0.00}$ & $11.41^{+0.10}_{-0.13}$\\ [1ex]
AzTEC/C28b & $2.30^{+0.31}_{-0.48}$ & 7.200 & $10.93^{+0.02}_{-0.10}$ & $12.52^{+0.02}_{-0.18}$ & $329^{+15}_{-112}$ & $3.9^{+1.2}_{-1.4}$ & $2.2^{+0.1}_{-0.7}$ & $50.5^{+1.4}_{-18.3}$ & $8.80^{+0.11}_{-0.01}$ & $11.09^{+0.11}_{-0.14}$\\ [1ex]
AzTEC/C29 & $1.82^{+0.35}_{-0.54}$ & 4.334 & $10.31^{+0.00}_{-0.00}$ & $12.50^{+0.00}_{-0.00}$ & $313^{+0}_{-0}$ & $15.3^{+0.0}_{-0.0}$ & $8.1^{+0.0}_{-0.0}$ & $42.5^{+0.0}_{-0.0}$ & $8.94^{+0.00}_{-0.00}$ & $11.32^{+0.10}_{-0.13}$\\ [1ex]
AzTEC/C31a & $2.10^{+3.20}_{-0.10}$ & 1.735 & $11.28^{+0.26}_{-0.36}$ & $12.64^{+0.06}_{-0.11}$ & $439^{+65}_{-98}$ & $2.3^{+3.8}_{-1.3}$ & $1.7^{+0.3}_{-0.4}$ & $49.2^{+8.5}_{-8.0}$ & $9.03^{+0.07}_{-0.07}$ & $11.30^{+0.10}_{-0.14}$\\ [1ex]
AzTEC/C31b & $2.49^{+2.79}_{-0.51}$ & 1.111 & $11.28^{+0.25}_{-0.33}$ & $12.47^{+0.10}_{-0.07}$ & $292^{+76}_{-43}$ & $1.5^{+2.6}_{-0.8}$ & $0.9^{+0.2}_{-0.1}$ & $37.8^{+8.7}_{-4.2}$ & $8.76^{+0.06}_{-0.06}$ & $11.14^{+0.10}_{-0.14}$\\ [1ex]
AzTEC/C32 & $1.63^{+0.20}_{-0.47}$ & 1.557 & $11.45^{+0.24}_{-0.20}$ & $12.39^{+0.01}_{-0.00}$ & $248^{+6}_{-0}$ & $0.9^{+0.5}_{-0.4}$ & $1.1^{+0.0}_{-0.0}$ & $30.0^{+0.0}_{-0.0}$ & $9.26^{+0.00}_{-0.01}$ & $11.50^{+0.10}_{-0.13}$\\ [1ex]
AzTEC/C33a & $2.30^{+0.16}_{-0.46}$ & 4.255 & $10.99^{+0.01}_{-0.00}$ & $12.82^{+0.01}_{-0.00}$ & $661^{+15}_{-0}$ & $6.8^{+0.2}_{-0.2}$ & $3.9^{+0.1}_{-0.0}$ & $43.2^{+1.6}_{-1.2}$ & $8.77^{+0.00}_{-0.06}$ & $11.33^{+0.11}_{-0.14}$\\ [1ex]
AzTEC/C34a & $3.53^{+0.02}_{-0.52}$ & 3.546 & $10.68^{+0.01}_{-0.00}$ & $13.19^{+0.00}_{-0.00}$ & $1\,542^{+0}_{-0}$ & $32.2^{+0.0}_{-0.7}$ & $10.5^{+0.0}_{-0.0}$ & $51.0^{+1.6}_{-0.0}$ & $8.69^{+0.06}_{-0.00}$ & $11.38^{+0.10}_{-0.13}$\\ [1ex]
AzTEC/C34b & $2.49^{+0.26}_{-0.50}$ & 7.445 & $10.43^{+0.00}_{-0.01}$ & $12.61^{+0.00}_{-0.01}$ & $404^{+0}_{-9}$ & $15.0^{+0.3}_{-0.3}$ & $5.9^{+0.0}_{-0.1}$ & $39.8^{+3.5}_{-6.7}$ & $9.04^{+0.03}_{-0.06}$ & $11.37^{+0.12}_{-0.16}$\\ [1ex]
AzTEC/C35 & $3.91^{+0.18}_{-0.50}$ & 6.073 & $11.07^{+0.09}_{-0.01}$ & $13.04^{+0.01}_{-0.03}$ & $1\,087^{+25}_{-73}$ & $9.2^{+0.4}_{-2.2}$ & $3.3^{+0.1}_{-0.2}$ & $45.7^{+5.1}_{-3.0}$ & $8.68^{+0.03}_{-0.06}$ & $11.28^{+0.10}_{-0.14}$\\ [1ex]
AzTEC/C36 & 2.415\tablefootmark{a} & 8.506 & $11.18^{+0.01}_{-0.00}$ & $13.06^{+0.02}_{-0.00}$ & $1\,154^{+54}_{-0}$ & $7.6^{+0.4}_{-0.2}$ & $4.6^{+0.2}_{-0.0}$ & $36.0^{+8.1}_{-0.0}$ & $9.27^{+0.00}_{-0.06}$ & $11.68^{+0.10}_{-0.13}$\\ [1ex]
AzTEC/C37 & $1.70^{+0.70}_{-0.30}$ & 7.662 & $10.76^{+0.23}_{-0.24}$ & $12.41^{+0.02}_{-0.08}$ & $255^{+12}_{-43}$ & $4.4^{+3.6}_{-2.3}$ & $3.3^{+0.2}_{-0.6}$ & $43.3^{+8.0}_{-5.4}$ & $9.44^{+0.05}_{-0.02}$ & $11.76^{+0.10}_{-0.13}$\\ [1ex]
AzTEC/C38 & $1.91^{+0.52}_{-0.46}$ & 1.105 & $11.52^{+0.01}_{-0.02}$ & $12.45^{+0.01}_{-0.02}$ & $283^{+7}_{-13}$ & $0.9^{+0.2}_{-0.2}$ & $0.8^{+0.0}_{-0.0}$ & $37.7^{+0.7}_{-2.8}$ & $9.29^{+0.03}_{-0.02}$ & $11.54^{+0.10}_{-0.13}$\\ [1ex]
AzTEC/C39 & $2.00^{+0.20}_{-0.40}$ & 1.916 & $11.80^{+0.01}_{-0.78}$ & $12.56^{+0.00}_{-0.01}$ & $362^{+0}_{-8}$ & $0.6^{+2.9}_{-0.0}$ & $0.6^{+0.0}_{-0.0}$ & $31.8^{+0.0}_{-0.6}$ & $9.05^{+0.01}_{-0.07}$ & $11.42^{+0.10}_{-0.13}$\\ [1ex]
AzTEC/C41 & $1.25^{+0.18}_{-0.34}$ & 9.258 & $10.58^{+0.00}_{-0.07}$ & $12.16^{+0.01}_{-0.00}$ & $146^{+3}_{-0}$ & $3.8^{+0.8}_{-0.0}$ & $4.0^{+0.1}_{-0.0}$ & $48.5^{+0.0}_{-0.0}$ & $9.26^{+0.01}_{-0.00}$ & $11.47^{+0.10}_{-0.13}$\\ [1ex]
AzTEC/C42 & $3.63^{+0.37}_{-0.56}$ & 3.160 & $11.46^{+0.00}_{-0.00}$ & $13.58^{+0.00}_{-0.00}$ & $3\,800^{+0}_{-0}$ & $13.2^{+0.0}_{-0.0}$ & $6.1^{+0.0}_{-0.0}$ & $64.0^{+0.0}_{-0.0}$ & $8.86^{+0.00}_{-0.00}$ & $11.49^{+0.10}_{-0.13}$ \\ [1ex]
AzTEC/C43a & $2.01^{+0.23}_{-0.47}$ & 4.070 & $10.96^{+0.37}_{-0.04}$ & $12.52^{+0.05}_{-0.04}$ & $334^{+41}_{-29}$ & $3.7^{+5.1}_{-2.2}$ & $2.4^{+0.3}_{-0.3}$ & $46.3^{+2.2}_{-2.9}$ & $8.71^{+0.14}_{-0.01}$ & $11.17^{+0.11}_{-0.15}$\\ [1ex]
AzTEC/C43b & $1.82^{+0.29}_{-0.36}$ & 3.713 & $11.11^{+0.04}_{-0.04}$ & $12.23^{+0.05}_{-0.07}$ & $169^{+21}_{-25}$ & $1.3^{+0.3}_{-0.3}$ & $1.1^{+0.1}_{-0.2}$ & $36.3^{+6.1}_{-3.6}$ & $8.75^{+0.09}_{-0.10}$ & $11.03^{+0.11}_{-0.15}$\\ [1ex]
AzTEC/C44a & $2.01^{+0.29}_{-0.44}$ & 1.021 & $11.39^{+0.12}_{-0.05}$ & $12.39^{+0.08}_{-0.04}$ & $244^{+49}_{-21}$ & $1.0^{+0.3}_{-0.3}$ & $0.8^{+0.2}_{-0.1}$ & $31.1^{+10.9}_{-1.9}$ & $9.04^{+0.03}_{-0.05}$ & $11.31^{+0.10}_{-0.14}$\\ [1ex]
\hline
\end{tabular} 
\end{minipage} }
\end{table*}

\addtocounter{table}{-1}

\begin{table*}[!htb]
\caption{continued.}
{\scriptsize
\begin{minipage}{2\columnwidth}
\centering
\renewcommand{\footnoterule}{}
\label{table:sed}
\begin{tabular}{c c c c c c c c c c c}
\hline\hline 
Source ID & $z$ & $\chi^2$ & $\log(M_{\star}/{\rm M}_{\sun})$ & $\log(L_{\rm IR}/{\rm L}_{\sun})$ & SFR [${\rm M}_{\sun}~{\rm yr}^{-1}$] & sSFR [${\rm Gyr}^{-1}$] & $\Delta_{\rm MS}$ & $T_{\rm dust}$ [K] & $\log(M_{\rm dust}/{\rm M}_{\sun})$ & $\log(M_{\rm gas}/{\rm M}_{\sun})$ \\[1ex]
\hline
AzTEC/C46 & $1.06^{+1.07}_{-0.41}$ & 8.120 & $10.40^{+0.12}_{-0.06}$ & $11.94^{+0.00}_{-0.00}$ & $87^{+0}_{-0}$ & $3.5^{+0.5}_{-0.8}$ & $4.0^{+0.0}_{-0.0}$ & $32.2^{+0.0}_{-0.0}$ & $9.20^{+0.00}_{-0.00}$ & $11.41^{+0.10}_{-0.13}$  \\ [1ex]
AzTEC/C47 & 2.0468\tablefootmark{a} & 9.065 & $11.31^{+0.00}_{-0.00}$ & $12.61^{+0.00}_{-0.00}$ & $404^{+0}_{-0}$ & $2.0^{+0.0}_{-0.0}$ & $1.6^{+0.0}_{-0.0}$ & $42.5^{+0.0}_{-2.6}$ & $9.05^{+0.00}_{-0.34}$ & $11.24^{+0.11}_{-0.14}$ \\ [1ex]
AzTEC/C48a & $1.91^{+0.18}_{-0.42}$ & 1.795 & $11.20^{+0.00}_{-0.12}$ & $12.22^{+0.14}_{-0.04}$ & $167^{+63}_{-15}$ & $1.1^{+0.9}_{-0.1}$ & $0.9^{+0.3}_{-0.1}$ & $29.9^{+17.5}_{-0.0}$ & $9.16^{+0.01}_{-0.03}$ & $11.34^{+0.10}_{-0.13}$ \\ [1ex]
AzTEC/C48b & $1.82^{+0.21}_{-0.46}$ & 1.521 & $11.23^{+0.00}_{-0.01}$ & $11.86^{+0.00}_{-0.01}$ & $72^{+0}_{-2}$ & $0.4^{+0.0}_{-0.0}$ & $0.4^{+0.0}_{-0.0}$ & $41.5^{+5.5}_{-8.0}$ & $8.77^{+0.09}_{-0.10}$ & $10.92^{+0.12}_{-0.17}$\\ [1ex]
AzTEC/C49 & $0.87^{+0.23}_{-0.33}$ & 9.053 & $10.14^{+0.49}_{-0.41}$ & $11.70^{+0.04}_{-0.08}$ & $50^{+5}_{-8}$ & $3.6^{+6.6}_{-2.7}$ & $4.7^{+0.5}_{-0.8}$ & $36.3^{+1.7}_{-7.7}$ & $8.89^{+0.03}_{-0.03}$ & $11.45^{+0.10}_{-0.14}$ \\ [1ex]
AzTEC/C50 & $3.15^{+0.78}_{-1.32}$ & 1.195 & $11.17^{+0.19}_{-0.31}$ & $12.88^{+0.07}_{-0.04}$ & $750^{+131}_{-66}$ & $5.1^{+7.1}_{-2.1}$ & $2.3^{+0.4}_{-0.2}$ & $41.0^{+7.3}_{-2.7}$ & $8.90^{+0.03}_{-0.05}$ & $11.41^{+0.10}_{-0.13}$\\ [1ex]
AzTEC/C51b & $1.34^{+0.20}_{-0.34}$ & 1.939 & $10.78^{+0.02}_{-0.04}$ & $11.72^{+0.15}_{-0.05}$ & $52^{+22}_{-6}$ & $0.9^{+0.5}_{-0.1}$ & $0.9^{+0.4}_{-0.1}$ & $45.0^{+6.8}_{-6.8}$ & $8.73^{+0.05}_{-0.05}$ & $10.95^{+0.12}_{-0.16}$\\ [1ex]
AzTEC/C52 & 1.1484\tablefootmark{a} & 2.485 & $11.50^{+0.04}_{-0.05}$ & $12.31^{+0.01}_{-0.03}$ & $203^{+5}_{-14}$ & $0.6^{+0.1}_{-0.1}$ & $1.4^{+0.0}_{-0.1}$ & $34.3^{+0.0}_{-2.9}$ & $9.00^{+0.03}_{-0.02}$ & $11.26^{+0.11}_{-0.14}$\\ [1ex]
AzTEC/C53 & $2.20^{+0.60}_{-0.70}$ & 5.667 & $11.05^{+0.19}_{-0.20}$ & $12.61^{+0.03}_{-0.03}$ & $411^{+29}_{-27}$ & $3.7^{+2.6}_{-1.5}$ & $2.3^{+0.2}_{-0.2}$ & $39.2^{+2.6}_{-2.7}$ & $8.69^{+0.06}_{-0.04}$ & $10.93^{+0.12}_{-0.16}$\\ [1ex]
AzTEC/C54 & $3.25^{+0.04}_{-0.52}$ & 5.720 & $10.30^{+0.17}_{-0.00}$ & $12.80^{+0.17}_{-0.00}$ & $630^{+302}_{-0}$ & $31.6^{+15.1}_{-10.2}$ & $9.2^{+4.4}_{-0.0}$ & $36.8^{+3.6}_{-0.0}$ & $9.03^{+0.01}_{-0.05}$ & $11.46^{+0.10}_{-0.13}$\\ [1ex]
AzTEC/C55a & $2.49^{+0.33}_{-0.45}$ & 7.195 & $11.32^{+0.00}_{-0.00}$ & $12.33^{+0.00}_{-0.00}$ & $212^{+0}_{-0}$ & $1.0^{+0.0}_{-0.0}$ & $0.6^{+0.0}_{-0.0}$ & $28.0^{+2.6}_{-0.0}$ & $9.10^{+0.00}_{-0.03}$ & $11.40^{+0.10}_{-0.13}$ \\ [1ex]
AzTEC/C55b & $2.77^{+0.32}_{-0.41}$ & 5.782 & $10.90^{+0.07}_{-0.00}$ & $12.22^{+0.00}_{-0.00}$ & $166^{+0}_{-0}$ & $2.1^{+0.0}_{-0.3}$ & $0.9^{+0.0}_{-0.0}$ & $31.1^{+0.4}_{-3.1}$ & $8.93^{+0.01}_{-0.14}$ & $11.18^{+0.12}_{-0.16}$\\ [1ex]
AzTEC/C58 & $4.10^{+0.32}_{-0.79}$ & 2.060 & $10.51^{+0.08}_{-0.02}$ & $12.90^{+0.09}_{-0.14}$ & $800^{+184}_{-220}$ & $24.7^{+7.1}_{-9.8}$ & $6.7^{+1.5}_{-1.8}$ & $41.0^{+3.6}_{-3.6}$ & $8.93^{+0.06}_{-0.08}$ & $11.46^{+0.10}_{-0.13}$\\ [1ex]
AzTEC/C59 & 1.2802\tablefootmark{a} & 5.450 & $11.37^{+0.00}_{-0.00}$ & $12.39^{+0.01}_{-0.00}$ & $243^{+6}_{-0}$ & $1.0^{+0.0}_{-0.0}$ & $1.7^{+0.0}_{-0.0}$ & $31.5^{+1.3}_{-0.0}$ & $9.08^{+0.00}_{-0.09}$ & $11.22^{+0.11}_{-0.14}$\\ [1ex]
AzTEC/C60a & $0.96^{+0.14}_{-0.40}$ & 9.920 & $9.85^{+0.43}_{-0.32}$ & $11.59^{+0.06}_{-0.08}$ & $39^{+6}_{-7}$ & $5.5^{+7.6}_{-3.8}$ & $4.9^{+0.7}_{-0.8}$ & $51.3^{+5.9}_{-10.2}$ & $8.53^{+0.04}_{-0.04}$ & $11.19^{+0.10}_{-0.14}$ \\ [1ex]
AzTEC/C60b & $4.77^{+0.14}_{-0.75}$ & 0.803 & $10.40^{+0.03}_{-0.09}$ & $12.49^{+0.03}_{-0.13}$ & $308^{+22}_{-80}$ & $12.3^{+3.9}_{-3.8}$ & $2.9^{+0.2}_{-0.7}$ & $50.2^{+8.2}_{-7.3}$ & $8.27^{+0.14}_{-0.13}$ & $10.87^{+0.11}_{-0.16}$  \\ [1ex]
AzTEC/C62 & $3.36^{+0.97}_{-0.97}$ & 4.153 & $10.77^{+0.22}_{-0.15}$ & $12.98^{+0.08}_{-0.11}$ & $966^{+195}_{-216}$ & $16.4^{+11.5}_{-8.7}$ & $5.8^{+1.2}_{-1.3}$ & $50.3^{+6.8}_{-3.8}$ & $8.51^{+0.06}_{-0.06}$ & $10.98^{+0.12}_{-0.18}$   \\ [1ex]
AzTEC/C64 & $2.58^{+0.79}_{-0.63}$ & 1.022 & $11.08^{+0.23}_{-0.24}$ & $12.90^{+0.05}_{-0.04}$ & $801^{+98}_{-70}$ & $6.7^{+6.3}_{-3.1}$ & $3.5^{+0.4}_{-0.3}$ & $40.2^{+6.2}_{-4.1}$ & $9.04^{+0.03}_{-0.03}$ & $11.50^{+0.10}_{-0.13}$\\ [1ex]
AzTEC/C65 & 1.798\tablefootmark{a} & 9.393 & $11.77^{+0.00}_{-0.00}$ & $12.71^{+0.00}_{-0.00}$ & $517^{+0}_{-0}$ & $0.9^{+0.0}_{-0.0}$ & $1.1^{+0.0}_{-0.0}$ & $36.0^{+0.0}_{-0.0}$ & $9.11^{+0.00}_{-0.00}$ & $11.30^{+0.11}_{-0.14}$ \\ [1ex]
AzTEC/C66 & $2.01^{+0.27}_{-0.50}$ & 2.107 & $11.58^{+0.09}_{-0.09}$ & $12.67^{+0.02}_{-0.01}$ & $463^{+22}_{-11}$ & $1.2^{+0.4}_{-0.3}$ & $1.2^{+0.1}_{-0.0}$ & $42.7^{+0.0}_{-2.6}$ & $8.86^{+0.03}_{-0.01}$ & $11.23^{+0.10}_{-0.14}$ \\ [1ex]
AzTEC/C67 & 2.9342\tablefootmark{a} & 6.722 & $10.27^{+0.00}_{-0.00}$ & $12.62^{+0.00}_{-0.01}$ & $420^{+0}_{-10}$ & $22.6^{+0.0}_{-0.5}$ & $7.0^{+0.0}_{-0.2}$ & $37.0^{+0.1}_{-0.3}$ & $9.01^{+0.00}_{-0.04}$ & $11.33^{+0.12}_{-0.16}$ \\ [1ex]
AzTEC/C69 & $3.91^{+0.09}_{-0.50}$ & 4.983 & $10.80^{+0.21}_{-0.03}$ & $12.77^{+0.15}_{-0.13}$ & $590^{+243}_{-153}$ & $9.3^{+4.8}_{-5.1}$ & $3.0^{+1.2}_{-0.8}$ & $49.8^{+7.2}_{-4.3}$ & $8.32^{+0.06}_{-0.07}$ & $10.89^{+0.11}_{-0.15}$\\ [1ex]
AzTEC/C70 & $4.01^{+0.09}_{-0.66}$ & 3.118 & $11.10^{+0.00}_{-0.02}$ & $13.58^{+0.01}_{-0.02}$ & $3\,838^{+89}_{-173}$ & $30.5^{+2.2}_{-1.4}$ & $11.0^{+0.3}_{-0.5}$ & $64.4^{+1.3}_{-3.9}$ & $8.55^{+0.09}_{-0.06}$ & $11.30^{+0.10}_{-0.13}$ \\ [1ex]
AzTEC/C72 & $1.72^{+0.38}_{-0.45}$ & 3.241 & $10.10^{+0.00}_{-0.00}$ & $12.38^{+0.00}_{-0.00}$ & $241^{+0}_{-0}$ & $19.1^{+0.0}_{-0.0}$ & $9.4^{+0.0}_{-0.0}$ & $29.9^{+0.0}_{-0.0}$ & $9.33^{+0.00}_{-0.00}$ & $11.49^{+0.11}_{-0.49}$ \\ [1ex]
AzTEC/C73 & $6.40^{+0.60}_{-1.10}$ & 3.931 & $11.33^{+0.20}_{-0.27}$ & $13.08^{+0.08}_{-0.08}$ & $1\,197^{+242}_{-201}$ & $5.6^{+6.9}_{-2.7}$ & $1.7^{+0.3}_{-0.3}$ & $63.8^{+5.2}_{-4.7}$ & $8.30^{+0.06}_{-0.06}$ & $11.06^{+0.12}_{-0.17}$ \\ [1ex]
AzTEC/C74a & $2.10^{+0.20}_{-0.67}$ & 3.445 & $9.58^{+0.00}_{-0.00}$ & $12.08^{+0.00}_{-0.00}$ & $120^{+0}_{-0}$ & $31.9^{+0.0}_{-0.0}$ & $9.4^{+0.0}_{-0.0}$ & $29.5^{+0.0}_{-0.0}$ & $9.05^{+0.00}_{-0.00}$ & $11.38^{+0.12}_{-0.16}$\\ [1ex]
AzTEC/C76 & $4.01^{+0.07}_{-0.57}$ & 0.192 & $11.60^{+0.04}_{-0.06}$ & $13.21^{+0.07}_{-0.08}$ & $1\,630^{+285}_{-275}$ & $4.1^{+1.4}_{-1.0}$ & $1.9^{+0.3}_{-0.3}$ & $49.8^{+6.4}_{-4.4}$ & $8.70^{+0.04}_{-0.04}$ & $11.31^{+0.10}_{-0.13}$ \\ [1ex]
AzTEC/C77b & $3.06^{+0.59}_{-1.19}$ & 2.631 & $10.78^{+0.23}_{-0.17}$ & $12.98^{+0.04}_{-0.07}$ & $963^{+93}_{-143}$ & $16.0^{+9.9}_{-8.0}$ & $6.2^{+0.6}_{-0.9}$ & $52.7^{+4.3}_{-3.0}$ & $8.58^{+0.14}_{-0.09}$ & $11.20^{+0.10}_{-0.14}$  \\ [1ex]
AzTEC/C78 & $4.77^{+0.09}_{-3.89}$ & 5.469 & $10.83^{+0.11}_{-0.11}$ & $12.93^{+0.08}_{-0.07}$ & $851^{+172}_{-127}$ & $12.6^{+6.9}_{-4.3}$ & $3.6^{+0.7}_{-0.5}$ & $43.8^{+7.9}_{-4.5}$ & $8.87^{+0.07}_{-0.05}$ & $11.34^{+0.10}_{-0.14}$\\ [1ex]
AzTEC/C79 & $2.20^{+0.33}_{-0.96}$ & 1.757 & $11.55^{+0.07}_{-0.04}$ & $12.33^{+0.06}_{-0.07}$ & $212^{+31}_{-32}$ & $0.6^{+0.2}_{-0.2}$ & $0.5^{+0.1}_{-0.1}$ & $34.8^{+8.8}_{-3.0}$ & $9.19^{+0.09}_{-0.05}$ & $11.42^{+0.10}_{-0.13}$\\ [1ex]
AzTEC/C80a & $2.10^{+0.66}_{-0.43}$ & 2.124 & $11.19^{+0.00}_{-0.00}$ & $12.66^{+0.00}_{-0.00}$ & $459^{+0}_{-0}$ & $3.0^{+0.0}_{-0.0}$ & $2.1^{+0.0}_{-0.0}$ & $46.5^{+0.0}_{-0.0}$ & $9.04^{+0.00}_{-0.00}$ & $11.43^{+0.10}_{-0.13}$\\ [1ex]
AzTEC/C80b & $2.01^{+0.68}_{-0.52}$ & 0.885 & $11.36^{+0.11}_{-0.11}$ & $12.23^{+0.05}_{-0.04}$ & $170^{+21}_{-15}$ & $0.7^{+0.3}_{-0.2}$ & $0.6^{+0.1}_{-0.1}$ & $33.5^{+3.1}_{-2.4}$ & $8.78^{+0.06}_{-0.07}$ & $11.15^{+0.11}_{-0.14}$\\ [1ex]
AzTEC/C81 & $4.62^{+1.48}_{-1.48}$ & 4.167 & $11.36^{+0.28}_{-0.36}$ & $12.95^{+0.08}_{-0.08}$ & $893^{+181}_{-150}$ & $3.9^{+6.8}_{-2.2}$ & $1.4^{+0.3}_{-0.2}$ & $52.3^{+6.0}_{-4.4}$ & $8.44^{+0.05}_{-0.05}$ & $11.11^{+0.11}_{-0.14}$\\ [1ex]
AzTEC/C84a & $1.63^{+2.73}_{-0.34}$ & 4.333 & $11.23^{+0.22}_{-0.28}$ & $11.97^{+0.08}_{-0.03}$ & $94^{+19}_{-6}$ & $0.6^{+0.7}_{-0.2}$ & $0.6^{+0.1}_{-0.0}$ & $26.5^{+7.0}_{-2.8}$ & $9.25^{+0.06}_{-0.07}$ & $11.54^{+0.11}_{-0.15}$\\ [1ex] 
AzTEC/C84b & 1.959\tablefootmark{a} & 2.087 & $11.82^{+0.04}_{-0.00}$ & $12.50^{+0.01}_{-0.00}$ & $315^{+7}_{-0}$ & $0.5^{+0.0}_{-0.0}$ & $0.5^{+0.0}_{-0.0}$ & $29.2^{+0.0}_{-0.8}$ & $9.19^{+0.01}_{-0.00}$ & $11.47^{+0.10}_{-0.14}$ \\ [1ex]
AzTEC/C87 & $2.39^{+0.20}_{-0.45}$ & 8.183 & $11.83^{+0.24}_{-0.20}$ & $12.80^{+0.04}_{-0.02}$ & $637^{+61}_{-29}$ & $0.9^{+0.7}_{-0.4}$ & $0.8^{+0.1}_{-0.0}$ & $36.2^{+1.3}_{-1.9}$ & $8.98^{+0.05}_{-0.04}$ & $11.33^{+0.10}_{-0.14}$\\ [1ex]
AzTEC/C88 & $1.82^{+0.38}_{-0.47}$ & 2.156 & $10.52^{+0.29}_{-0.27}$ & $12.16^{+0.09}_{-0.05}$ & $146^{+34}_{-16}$ & $4.4^{+5.7}_{-2.4}$ & $2.6^{+0.6}_{-0.3}$ & $33.8^{+8.6}_{-2.7}$ & $8.71^{+0.12}_{-0.05}$ & $11.00^{+0.11}_{-0.15}$\\ [1ex]
AzTEC/C90a & $2.20^{+2.83}_{-0.46}$ & 0.901 & $11.25^{+0.16}_{-0.20}$ & $12.24^{+0.17}_{-0.10}$ & $174^{+83}_{-36}$ & $1.0^{+1.3}_{-0.4}$ & $0.7^{+0.3}_{-0.1}$ & $37.5^{+12.8}_{-6.3}$ & $8.79^{+0.10}_{-0.09}$ & $11.10^{+0.12}_{-0.16}$\\ [1ex]
AzTEC/C90b & $2.77^{+0.33}_{-1.67}$ & 0.758 & $10.23^{+0.07}_{-0.08}$ & $12.41^{+0.04}_{-0.05}$ & $256^{+25}_{-28}$ & $15.1^{+4.8}_{-3.6}$ & $4.8^{+0.5}_{-0.5}$ & $37.0^{+2.8}_{-1.8}$ & $8.65^{+0.06}_{-0.07}$ & $11.06^{+0.11}_{-0.15}$\\ [1ex]
AzTEC/C90c & $2.20^{+0.23}_{-0.42}$ & 1.657 & $10.90^{+0.00}_{-0.00}$ & $12.22^{+0.00}_{-0.00}$ & $165^{+0}_{-0}$ & $2.1^{+0.0}_{-0.0}$ & $1.2^{+0.0}_{-0.0}$ & $36.8^{+1.6}_{-2.6}$ & $8.63^{+0.16}_{-0.14}$ & $11.03^{+0.12}_{-0.16}$  \\ [1ex]
AzTEC/C91 & $1.63^{+0.29}_{-0.41}$ & 4.856 & $10.75^{+0.90}_{-0.01}$ & $12.46^{+0.03}_{-0.00}$ & $285^{+20}_{-0}$ & $5.1^{+0.5}_{-4.4}$ & $4.0^{+0.3}_{-0.0}$ & $30.8^{+1.3}_{-0.3}$ & $9.05^{+0.07}_{-0.05}$ & $11.27^{+0.10}_{-0.14}$\\ [1ex]
AzTEC/C92a & $2.58^{+2.67}_{-0.46}$ & 1.495 & $11.84^{+0.23}_{-0.25}$ & $12.82^{+0.10}_{-0.07}$ & $656^{+170}_{-98}$ & $0.9^{+1.2}_{-0.5}$ & $0.7^{+0.2}_{-0.1}$ & $38.7^{+9.2}_{-5.2}$ & $9.11^{+0.05}_{-0.05}$ & $11.53^{+0.10}_{-0.13}$\\ [1ex]
AzTEC/C92b & $4.87^{+0.22}_{-0.98}$ & 2.562 & $10.77^{+0.07}_{-0.07}$ & $12.83^{+0.08}_{-0.05}$ & $679^{+137}_{-74}$ & $11.5^{+4.8}_{-2.8}$ & $3.1^{+0.6}_{-0.3}$ & $48.7^{+6.4}_{-3.9}$ & $8.49^{+0.10}_{-0.08}$ & $11.21^{+0.11}_{-0.15}$\\ [1ex]
AzTEC/C93 & $1.63^{+1.10}_{-0.53}$ & 4.235 & $11.01^{+0.06}_{-0.11}$ & $12.38^{+0.03}_{-0.03}$ & $243^{+17}_{-16}$ & $2.4^{+0.9}_{-0.4}$ & $2.2^{+0.2}_{-0.1}$ & $37.3^{+2.3}_{-5.1}$ & $9.30^{+0.03}_{-0.06}$ & $11.48^{+0.10}_{-0.13}$\\ [1ex]
AzTEC/C95 & 2.1021\tablefootmark{a} & 0.095 & $11.28^{+0.00}_{-0.00}$ & $12.55^{+0.00}_{-0.00}$ & $357^{+0}_{-0}$ & $1.9^{+0.0}_{-0.0}$ & $1.4^{+0.0}_{-0.0}$ & $35.2^{+1.4}_{-0.7}$ & $8.83^{+0.06}_{-0.02}$ & $11.22^{+0.10}_{-0.14}$\\ [1ex]
AzTEC/C97a & $3.06^{+0.04}_{-0.52}$ & 3.058 & $10.73^{+0.00}_{-0.01}$ & $13.02^{+0.00}_{-0.01}$ & $1\,059^{+0}_{-24}$ & $19.7^{+0.5}_{-0.4}$ & $7.4^{+0.0}_{-0.2}$ & $44.2^{+1.0}_{-0.9}$ & $8.85^{+0.03}_{-0.08}$ & $11.35^{+0.12}_{-0.17}$ \\ [1ex]
AzTEC/C97b & $2.01^{+0.08}_{-0.48}$ & 5.487 & $11.43^{+0.00}_{-0.00}$ & $12.42^{+0.00}_{-0.00}$ & $265^{+0}_{-0}$ & $1.0^{+0.0}_{-0.0}$ & $0.9^{+0.0}_{-0.0}$ & $41.0^{+3.1}_{-2.2}$ & $8.58^{+0.10}_{-0.10}$ & $11.03^{+0.12}_{-0.16}$\\ [1ex]
AzTEC/C98 & $1.82^{+0.60}_{-0.46}$ & 0.751 & $11.61^{+0.13}_{-0.05}$ & $12.66^{+0.02}_{-0.02}$ & $460^{+22}_{-21}$ & $1.1^{+0.2}_{-0.3}$ & $1.3^{+0.1}_{-0.1}$ & $34.0^{+2.9}_{-0.5}$ & $9.17^{+0.06}_{-0.02}$ & $11.50^{+0.10}_{-0.13}$\\ [1ex]
AzTEC/C99 & $2.68^{+1.37}_{-0.92}$ & 1.094 & $10.67^{+0.23}_{-0.30}$ & $12.39^{+0.11}_{-0.12}$ & $247^{+71}_{-60}$ & $5.3^{+8.3}_{-2.9}$ & $2.2^{+0.6}_{-0.5}$ & $41.8^{+12.6}_{-9.6}$ & $9.03^{+0.08}_{-0.08}$ & $11.27^{+0.10}_{-0.13}$\\ [1ex]
AzTEC/C100a & $1.63^{+0.17}_{-0.44}$ & 1.088 & $11.08^{+0.04}_{-0.05}$ & $12.24^{+0.07}_{-0.04}$ & $174^{+30}_{-15}$ & $1.4^{+0.5}_{-0.2}$ & $1.4^{+0.2}_{-0.1}$ & $32.0^{+8.6}_{-3.1}$ & $9.05^{+0.08}_{-0.05}$ & $11.30^{+0.12}_{-0.17}$\\ [1ex]
AzTEC/C100b & $2.68^{+0.42}_{-0.63}$ & 1.847 & $10.37^{+0.77}_{-0.09}$ & $12.58^{+0.01}_{-0.33}$ & $376^{+9}_{-200}$ & $16.1^{+4.2}_{-14.8}$ & $5.7^{+0.1}_{-3.0}$ & $56.8^{+8.6}_{-17.3}$ & $8.77^{+0.12}_{-0.12}$ & $11.07^{+0.11}_{-0.15}$\\ [1ex]
\hline
\end{tabular} 
\end{minipage} }
\end{table*}

\addtocounter{table}{-1}

\begin{table*}[!htb]
\caption{continued.}
{\scriptsize
\begin{minipage}{2\columnwidth}
\centering
\renewcommand{\footnoterule}{}
\label{table:sed}
\begin{tabular}{c c c c c c c c c c c}
\hline\hline 
Source ID & $z$ & $\chi^2$ & $\log(M_{\star}/{\rm M}_{\sun})$ & $\log(L_{\rm IR}/{\rm L}_{\sun})$ & SFR [${\rm M}_{\sun}~{\rm yr}^{-1}$] & sSFR [${\rm Gyr}^{-1}$] & $\Delta_{\rm MS}$ & $T_{\rm dust}$ [K] & $\log(M_{\rm dust}/{\rm M}_{\sun})$ & $\log(M_{\rm gas}/{\rm M}_{\sun})$ \\[1ex]
\hline
AzTEC/C101a & $1.53^{+0.31}_{-0.51}$ & 0.882 & $10.18^{+0.29}_{-0.15}$ & $12.23^{+0.10}_{-0.13}$ & $169^{+44}_{-44}$ & $11.2^{+8.7}_{-6.9}$ & $6.7^{+1.7}_{-1.7}$ & $49.2^{+7.7}_{-11.6}$ & $9.16^{+0.04}_{-0.03}$ & $11.34^{+0.10}_{-0.13}$\\ [1ex]
AzTEC/C101b & $1.74^{+0.98}_{-0.27}$ & 3.700 & $9.89^{+0.01}_{-0.01}$ & $12.05^{+0.00}_{-0.01}$ & $113^{+0}_{-3}$ & $14.7^{+0.2}_{-0.5}$ & $6.3^{+0.0}_{-0.1}$ & $42.2^{+5.1}_{-5.9}$ & $8.79^{+0.10}_{-0.14}$ & $11.07^{+0.12}_{-0.16}$ \\ [1ex]
AzTEC/C103 & $2.10^{+0.33}_{-0.57}$ & 8.955 & $11.49^{+0.06}_{-0.08}$ & $12.51^{+0.02}_{-0.06}$ & $326^{+15}_{-42}$ & $1.1^{+0.3}_{-0.3}$ & $0.9^{+0.0}_{-0.1}$ & $37.8^{+3.1}_{-6.1}$ & $9.20^{+0.06}_{-0.05}$ & $11.42^{+0.10}_{-0.13}$\\ [1ex]
AzTEC/C105 & $2.20^{+0.08}_{-0.54}$ & 4.427 & $11.59^{+0.01}_{-0.00}$ & $12.53^{+0.00}_{-0.00}$ & $342^{+0}_{-0}$ & $0.9^{+0.0}_{-0.0}$ & $0.7^{+0.0}_{-0.0}$ & $33.5^{+1.3}_{-2.0}$ & $9.27^{+0.00}_{-0.10}$ & $11.44^{+0.10}_{-0.14}$ \\ [1ex]
AzTEC/C107 & $5.15^{+0.93}_{-1.40}$ & 1.718 & $12.25^{+0.07}_{-0.51}$ & $12.96^{+0.10}_{-0.05}$ & $916^{+237}_{-100}$ & $0.5^{+1.6}_{-0.1}$ & $0.3^{+0.1}_{-0.0}$ & $48.5^{+6.2}_{-4.6}$ & $8.53^{+0.03}_{-0.03}$ & $11.30^{+0.10}_{-0.13}$\\ [1ex]
AzTEC/C108 & $2.30^{+1.26}_{-0.47}$ & 1.102 & $11.67^{+0.17}_{-0.19}$ & $12.56^{+0.11}_{-0.08}$ & $361^{+104}_{-61}$ & $0.8^{+0.8}_{-0.3}$ & $0.7^{+0.2}_{-0.1}$ & $33.8^{+8.9}_{-5.0}$ & $9.28^{+0.04}_{-0.05}$ & $11.55^{+0.10}_{-0.13}$\\ [1ex]
AzTEC/C109 & $2.20^{+0.28}_{-0.41}$ & 4.661 & $11.28^{+0.10}_{-0.00}$ & $12.58^{+0.10}_{-0.00}$ & $383^{+99}_{-0}$ & $2.0^{+1.1}_{-0.2}$ & $1.4^{+0.4}_{-0.0}$ & $32.5^{+3.5}_{-1.3}$ & $9.15^{+0.01}_{-0.03}$ & $11.50^{+0.10}_{-0.13}$ \\ [1ex]
AzTEC/C111 & $2.10^{+0.54}_{-0.59}$ & 1.151 & $11.43^{+0.12}_{-0.20}$ & $12.58^{+0.02}_{-0.05}$ & $384^{+18}_{-42}$ & $1.4^{+0.9}_{-0.5}$ & $1.2^{+0.1}_{-0.1}$ & $38.0^{+0.4}_{-2.1}$ & $8.98^{+0.05}_{-0.07}$ & $11.35^{+0.10}_{-0.13}$ \\ [1ex]
AzTEC/C112 & 1.894\tablefootmark{a} & 1.552 & $11.48^{+0.00}_{-0.00}$ & $12.60^{+0.00}_{-0.00}$ & $402^{+0}_{-0}$ & $1.3^{+0.0}_{-0.0}$ & $1.3^{+0.0}_{-0.0}$ & $39.8^{+0.0}_{-0.0}$ & $9.09^{+0.00}_{-0.00}$ & $11.42^{+0.10}_{-0.13}$   \\ [1ex]
AzTEC/C113 & 2.0899\tablefootmark{a} & 9.512 & $10.73^{+0.01}_{-0.00}$ & $13.25^{+0.01}_{-0.00}$ & $1\,764^{+41}_{-0}$ & $32.9^{+0.8}_{-0.8}$ & $18.4^{+0.4}_{-0.0}$ & $47.2^{+8.1}_{-0.0}$ & $8.92^{+0.11}_{-0.00}$ & $11.22^{+0.11}_{-0.15}$\\ [1ex]
AzTEC/C115 & $2.80^{+1.30}_{-0.60}$ & 0.326 & $11.39^{+0.24}_{-0.22}$ & $12.57^{+0.08}_{-0.09}$ & $371^{+75}_{-70}$ & $1.5^{+1.5}_{-0.8}$ & $0.9^{+0.2}_{-0.2}$ & $52.8^{+5.4}_{-8.8}$ & $9.10^{+0.07}_{-0.08}$ & $11.34^{+0.10}_{-0.14}$\\ [1ex]
AzTEC/C116 & $2.20^{+1.75}_{-0.43}$ & 1.400 & $10.90^{+0.13}_{-0.05}$ & $12.74^{+0.01}_{-0.05}$ & $549^{+13}_{-60}$ & $6.9^{+1.0}_{-2.3}$ & $4.0^{+0.1}_{-0.4}$ & $41.2^{+0.0}_{-2.9}$ & $8.65^{+0.16}_{-0.01}$ & $11.18^{+0.10}_{-0.14}$\\ [1ex]
AzTEC/C117 & $1.72^{+0.20}_{-0.68}$ & 5.133 & $11.21^{+0.06}_{-0.04}$ & $12.12^{+0.08}_{-0.02}$ & $132^{+27}_{-6}$ & $0.8^{+0.3}_{-0.1}$ & $0.8^{+0.2}_{-0.0}$ & $31.4^{+9.3}_{-3.7}$ & $9.16^{+0.06}_{-0.11}$ & $11.16^{+0.11}_{-0.16}$\\ [1ex]
AzTEC/C119 & $3.25^{+0.82}_{-0.62}$ & 1.935 & $10.78^{+0.26}_{-0.33}$ & $12.89^{+0.10}_{-0.04}$ & $775^{+201}_{-68}$ & $12.9^{+21.8}_{-6.4}$ & $4.7^{+1.2}_{-0.4}$ & $37.8^{+3.4}_{-1.2}$ & $9.07^{+0.03}_{-0.04}$ & $11.54^{+0.10}_{-0.13}$\\ [1ex]
AzTEC/C122a & $1.06^{+0.12}_{-0.40}$ & 1.909 & $10.12^{+0.01}_{-0.04}$ & $11.97^{+0.01}_{-0.10}$ & $93^{+2}_{-19}$ & $7.0^{+0.9}_{-1.6}$ & $6.7^{+0.2}_{-1.4}$ & $45.0^{+6.7}_{-8.0}$ & $8.97^{+0.05}_{-0.03}$ & $11.27^{+0.11}_{-0.15}$ \\ [1ex]
AzTEC/C123 &  $1.82^{+0.20}_{-0.61}$ & 0.789 & $11.20^{+0.07}_{-0.05}$ & $12.63^{+0.06}_{-0.04}$ & $426^{+63}_{-37}$ & $2.7^{+0.8}_{-0.6}$ & $2.4^{+0.4}_{-0.2}$ & $34.0^{+6.4}_{-1.5}$ & $9.05^{+0.06}_{-0.05}$ & $11.40^{+0.10}_{-0.13}$\\ [1ex]
AzTEC/C124 & 1.88\tablefootmark{a} & 1.808 & $11.48^{+0.05}_{-0.04}$ & $12.57^{+0.01}_{-0.04}$ & $373^{+9}_{-33}$ & $1.2^{+0.2}_{-0.2}$ & $1.2^{+0.0}_{-0.1}$ & $36.2^{+0.0}_{-1.2}$ & $8.75^{+0.02}_{-0.03}$ & $11.04^{+0.12}_{-0.17}$\\ [1ex]
AzTEC/C126 & $4.68^{+0.31}_{-0.64}$ & 3.628 & $11.30^{+0.00}_{-0.00}$ & $13.81^{+0.00}_{-0.00}$ & $6\,501^{+0}_{-0}$ & $32.6^{+0.0}_{-0.0}$ & $11.5^{+0.0}_{-0.0}$ & $79.2^{+0.0}_{-0.0}$ & $8.24^{+0.00}_{-0.00}$ & $11.29^{+0.10}_{-0.13}$ \\ [1ex]
AzTEC/C127 & $2.01^{+0.17}_{-0.51}$ & 5.729 & $10.89^{+0.00}_{-0.00}$ & $12.88^{+0.00}_{-0.00}$ & $761^{+0}_{-0}$ & $9.8^{+0.0}_{-0.0}$ & $6.3^{+0.0}_{-0.0}$ & $48.2^{+0.0}_{-0.0}$ & $8.74^{+0.00}_{-0.00}$ & $11.15^{+0.11}_{-0.14}$ \\ [1ex]
AzTEC/C129 & $4.87^{+0.73}_{-0.97}$ & 5.553 & $11.49^{+0.22}_{-0.38}$ & $13.02^{+0.10}_{-0.12}$ & $1\,047^{+271}_{-253}$ & $3.4^{+6.8}_{-1.8}$ & $1.3^{+0.3}_{-0.3}$ & $57.8^{+7.5}_{-8.1}$ & $9.04^{+0.12}_{-0.11}$ & $11.51^{+0.10}_{-0.13}$\\ [1ex]
\hline 
\end{tabular} 
\tablefoot{The columns are as follows: (1) the name of the SMG; (2) redshift (see \cite{brisbin2017}); (3) chi-square or goodness of fit; (4) stellar mass; (5) IR luminosity calculated by integrating the SED over the rest-frame wavelength range of 
$\lambda_{\rm rest}=8-1\,000$~$\mu$m; (6) SFR calculated using the $L_{\rm IR}-{\rm SFR}$ relationship of Kennicutt (1998); 
(7) specific SFR ($={\rm SFR}/M_{\star}$); (8) ratio of SFR to that of a MS galaxy of the same redshift and stellar mass, ${\rm SFR}/{\rm SFR}_{\rm MS}$ (i.e. offset from the MS); (9) luminosity-weighted dust temperature (see Eq.~(8) in \cite{dacunha2015}); (10) dust mass; (11) gas mass estimated from the dust emission (Sect.~3.2). The quoted values in Cols.~(4)--(10) and their uncertainties represent the median of the likelihood distribution, and its 68\% confidence interval (corresponding to the 16th--84th percentile range).\tablefoottext{a}{Spectroscopic redshift (see \cite{brisbin2017}, and references therein).}}
\end{minipage} }
\end{table*}

\section{Note on the previous studies of the spectral energy distributions of the target submillimetre galaxies}

Miettinen et al. (2017a) analysed the SEDs of a flux-limited sample of JCMT/AzTEC SMGs in COSMOS, and nine of their analysed sources are common with the present work. These are AzTEC~1=AzTEC/C5, AzTEC~4=AzTEC/C4, AzTEC~5=AzTEC/C42, AzTEC~8=AzTEC/C2a, AzTEC~9=AzTEC/C14, AzTEC~11-S=AzTEC/C22a, AzTEC~12=AzTEC/C18, AzTEC~15=AzTEC/C10b, and AzTEC~24b=AzTEC/C48a. Miettinen et al. (2017a) used the same high-$z$ extension of {\tt MAGPHYS} as in the present work. For six of the aforementioned sources (AzTEC/C4, 10b, 14, 18, 42, and 48a), the redshift was revised by Brisbin et al. (2017). The new redshifts are $0.98-2.94$ times the previous ones, with a median factor of only 1.02. The biggest difference applies to AzTEC/C4, for which we previously used a photo-$z$ of $1.80^{+5.18}_{-0.61}$, while in the present work we used a synthetic solution of $z=5.30^{+0.70}_{-1.10}$. Moreover, in the present study, the \textit{Herschel} photometry was extracted by using the ALMA 1.3~mm sources as positional priors, while Miettinen et al. (2017a) used either the 24~$\mu$m priors, or the blind \textit{Herschel} catalogue flux densities in case the source was not detected at 24~$\mu$m. Despite these differences, 
the physical parameters derived here are almost identical to those derived by Miettinen et al. (2017a), even for AzTEC/C4 where the adopted nominal redshift was completely different. We refer to Appendix~C in Miettinen et al. (2017a) for a detailed comparison with earlier literature. 

There is one additional SMG in our sample whose physical properties were determined in previous studies, namely AzTEC/C17 or J1000+0234. This SMG was spectroscopically confirmed ($z_{\rm spec}=4.547$) by Capak et al. (2008), and the authors estimated that $L_{\rm IR}$ is $(0.5-2)\times10^{13}$~L$_{\sun}$. This brackets our value of $L_{\rm IR}=7.8^{+2.4}_{-2.2}\times10^{12}$~L$_{\sun}$. The best-fit SED derived by Capak et al. (2008) under the assumption of a single-burst Bruzual \& Charlot (2003) stellar population model yielded a stellar mass of $M_{\star}\simeq1.2\times10^{10}$~M$_{\sun}$ (scaled to a Chabrier (2003) IMF). This is about $4\pm1$ times lower than our estimate of $M_{\star}=4.9^{+1.1}_{-1.0}\times10^{10}$~M$_{\sun}$. 

Toft et al. (2014) derived a {\tt MAGPHYS} and Chabrier (2003) IMF-based stellar mass of $\log(M_{\star}/{\rm M}_{\sun})=10.9\pm0.1$ for AzTEC/C17. Our stellar mass value is $1.6^{+1.0}_{-0.5}$ times lower. Using a physically motivated dust model of Draine \& Li (2007), Toft et al. (2014) derived a dust mass of $\log(M_{\rm dust}/{\rm M}_{\sun})=9.3\pm0.1$ and IR luminosity of $\log(L_{\rm IR}/{\rm L}_{\sun})=13.17\pm0.09$ for AzTEC/C17. Our values are $1.6^{+0.7}_{-0.5}$ and $1.9^{+1.3}_{-0.7}$ times lower, respectively.

Based on MBB fitting, Huang et al. (2014) derived the values of $T_{\rm dust}=64\pm4$~K, $M_{\rm dust}=(2.2\pm0.5)\times10^9$~M$_{\sun}$, and $\log(L_{\rm FIR}/{\rm L}_{\sun})=12.95\pm0.26$ for AzTEC/C17 (their source Capak4.55; 
see their Table~4). Here, we scaled the dust mass up by a factor of 1.06 from the value reported by Huang et al. (2014) to take 
into account that they adopted a dust opacity of $\kappa_{\rm 250\, \mu m}=5.1$~cm$^2$~g$^{-1}$, while our corresponding value would be 
4.8~cm$^2$~g$^{-1}$ (assuming $\beta = 1.5$). Our dust temperature value of $T_{\rm dust}=38.2^{+9.9}_{-3.7}$~K is clearly lower than the aforementioned Huang et al. (2014) value, but compatible within the uncertainties with their value of $47\pm3$~K derived under the assumption of optically thin emission. Our dust mass estimate is a factor of $1.8^{+0.7}_{-0.6}$ lower than derived by the authors. Finally, we note that the IR luminosity from Huang et al. (2014) refers to the FIR range (although the rest-frame wavelength range was not
specified), and our corresponding value of $L_{\rm FIR} = 0.54 \times L_{\rm IR} = 4.2^{+1.3}_{-1.2}\times10^{12}$~L$_{\sun}$ is $0.5^{+0.6}_{-0.3}$ times their value.

The most direct comparison we can make is with the {\tt MAGPHYS} SED of AzTEC/C17 from Smol{\v c}i{\'c} et al. (2015). However, we note that the prior model libraries used by the authors were those calibrated to reproduce the UV-IR SEDs of local ULIRGs (see \cite{dacunha2010b}). The stellar mass, dust luminosity, and dust mass were found to be $M_{\star}=8.7\times10^{10}$~M$_{\sun}$, $L_{\rm dust}=7\times10^{12}$~L$_{\sun}$, and $M_{\rm dust}=9.1\times10^8$~M$_{\sun}$. The present values are $0.6^{+0.1}_{-0.2}$, $1.2^{+0.4}_{-0.3}$, and $1.4\pm0.2$ times the Smol{\v c}i{\'c} et al. (2015) values. The authors also used a MBB fit to derive a dust temperature of $45.0 \pm 17.5$~K for AzTEC/C17, which, although based on a different method, is consistent with the present estimate (a factor of $1.2\pm0.6$ difference). For comparison, Smol{\v c}i{\'c} et al. (2015) derived a total IR luminosity of $L_{\rm IR} = 4.5^{+13.7}_{-3.4}\times10^{12}$~L$_{\sun}$ and dust mass of $M_{\rm dust}=4.0^{+8.6}_{-2.7}\times10^9$~M$_{\sun}$ for AzTEC/C17 from their best-fit Draine \& Li (2007) dust model. Our values are $1.7^{+7.6}_{-1.4}$ and $0.3^{+0.8}_{-0.2}$ times the latter ones.

\end{document}